\documentclass[a4paper]{book} 

\usepackage{amsmath,amsthm,amsfonts,amssymb,amscd,latexsym,eucal}
\usepackage[a4paper,top=5.5cm,bottom=5.5cm,left=4.0cm,right=4.0cm]{geometry}
\usepackage[utf8]{inputenc}
\usepackage{graphicx} 
\usepackage{nicefrac}
\usepackage{units}

\usepackage{pdfsync} 

\usepackage{lipsum}

\usepackage{imakeidx}
\makeindex
 
\usepackage{relsize}
\usepackage[english]{babel}

\usepackage[T1]{fontenc}
 
 \usepackage{tikz-cd}
\usepackage{graphicx,color}
\usepackage{xcolor}
\colorlet{blu1}{blue!70!black}
\colorlet{blu2}{blue!50!black}
\colorlet{blu3}{blue!70!red}
\colorlet{blu4}{blue!60!green}
\colorlet{red1}{red!80}
\colorlet{red2}{red!50!black}
\colorlet{red3}{red!70!yellow}
\colorlet{red4}{red!50!yellow}
\colorlet{yel1}{yellow!50!black}
\colorlet{yel3}{yellow!20!blue}
\colorlet{gre1}{green!60!blue}
\colorlet{gre2}{green!60!black}
\colorlet{gre3}{green!40!black}

\newtheorem{axiom}{Axiom}
\newtheorem{corollary}{Corollary}
\newtheorem{definition}{Definition}
\newtheorem{example}{Example}

\newtheorem{lemma}{Lemma}
\newtheorem{crit}{Assumption}
\newtheorem{proposition}{Proposition}
\newtheorem{remark}{Remark}

\newtheorem{theorem}{Theorem}
\newtheorem{notation}{Notation}
\newtheorem{problem}{Question}

\newtheorem{citazione}{Quote}
\newtheorem{attenzione}{Warning}
\newtheorem{property}{Property}
\newtheorem{post}{Postulate}

\begin{document}

\title{Statistical Interpretation of the Procedures Measurement of  Physical Quantities}
\author{Carlo Pandiscia
\\
email: pandiscia.carlo@gmail.com }

\maketitle

\newpage
\begin{center}
{\small
© Carlo Pandiscia, 2026

\vspace{0.5cm}

This book is distributed under the Creative Commons 
Attribution -- NonCommercial -- NoDerivatives 4.0 International License 
(CC BY--NC--ND 4.0).

\vspace{0.3cm}

The PDF may be freely downloaded and shared, provided that proper 
attribution to the author is maintained.

\vspace{0.3cm}

No modification, transformation, or creation of derivative works based 
on the content of this volume is permitted. 
No commercial use of the text is allowed.

\vspace{0.3cm}

Full license terms:
\\
\texttt{https://creativecommons.org/licenses/by-nc-nd/4.0/}
}
\end{center}

\frontmatter
 \chapter*{Preface}

\begin{flushright}
\textit{E dunque io esisto  perché tu mi guardi}
\end{flushright}
I reassure the reader immediately: these notes do not develop new concepts nor new axiomatic systems. I lack both the ability and the ambition for such an undertaking. Rather, this is a synthesis project, an attempt to reorganize and connect existing models and to reinterpret their assumptions and basic ideas. I have been guided by the various fundamental works of the pioneers of quantum theory and their successors, and by my Masters whom I have encountered along my life as a student and researcher. In particular, I wish to remember the late Professor John E. Roberts and Professor Luigi Accardi, both from the University of Rome Tor Vergata. Soon, however, I realized how arduous this task was, given the immense volume of published work and the difficulty of navigating it; for this reason the reader may have the sense of a non-exhaustive work --- but how could it be otherwise?

Although I will not delve deeply into philosophical questions of foundations, the nature of the subject will occasionally require brief forays into that territory. I leave the heavier philosophical work to minds more qualified than mine and direct interested readers to the bibliography, where I have cited works I find particularly illuminating on these themes. I will therefore stay away from lengthy philosophical discussions, guided by Dirac's warning \cite{Dirac78}:
\\

\textit{I want to emphasize the necessity for a sound mathematical basis for any fundamental physical theory. Any philosophical ideas that one may have play only a subordinate role. Unless such ideas have a mathematical basis they will be ineffective.}
\\
 
Furthermore, these notes will not repeat standard topics, such as system Hamiltonians, Schrödinger equations, Hilbert spaces, etc., which are covered in any quantum mechanics course. Rather, I intend to build a bridge between the assumptions of algebraic quantum theory and those of quantum probability.

The difficulties encountered in writing these notes have not been mathematical, but conceptual: confronting primitive notions and deceptively simple definitions, whose apparent simplicity often hides subtle pitfalls. It is precisely here, I believe, that one risks self-contradiction or pure nonsense. I hope I have managed to avoid such errors.

The mathematical prerequisites are modest. Readers should be familiar with measure theory (essential for a rigorous theory of probability) and with basic functional analysis, including at least the definition of an algebra of operators.

\newpage

\subsection*{On the use of the AI assistant DeepSeek in this work}

The notes from which this text originates were written many years ago, but I never found the courage to publish them for two main reasons.

First, I do not consider myself a good writer, especially in English: a difficulty that also affects my scientific writing. Second, I never had the opportunity to discuss my hypotheses with colleagues working in quantum physics. Personal circumstances prevented me from meeting them in person, and the absence of direct feedback from experts in the field often left me uncertain about the direction and validity of my research.
\\
In recent months, however, I discovered the virtual assistant DeepSeek and, thanks to its support, I finally managed to complete these notes after years of hesitation.
\\
For grammar checks and text revision, I relied on DeepSeek, which I found particularly effective also in handling and managing physics and mathematics material. DeepSeek proved to be a true research assistant, quickly recalling fundamental mathematical results in a concise manner. Moreover, whenever I encountered interpretative doubts in the theory I was developing, I turned to \textit{her} for clarification, feeling as if I were consulting the Pythia to question Apollo. Sometimes her answers threw me into even deeper confusion, but often her ``evocations'' helped dispel my uncertainties.
\\
Not only that: when I asked the assistant for clarification on a topic discussed in Conway's book, to my surprise she produced a new result, complete with a proof and not present in the text. The proof of this proposition was not well structured, but it worked. Unfortunately, I was unable to identify the precise source from which it might have been drawn. For this reason, Proposition \ref{spec-theorem} on page \pageref{spec-theorem} is not my own work, but a product of artificial intelligence: I merely restructured the proof.

\newpage

\subsection*{ Warning to the reader}  
Throughout the text, the word \textbf{ Question} appears in boldface. It is used to highlight specific issues—sometimes elementary, sometimes conceptually demanding—that require a clear answer. These questions must be addressed within the framework developed in these notes, and not by appealing to the many mathematical solutions that exist outside our minimal scheme.
\\
Chapters and sections marked with an asterisk may also be skipped on a first reading, so as not to interrupt the flow of the main exposition. They contain complementary material which, although not fully developed, outlines a sufficiently clear path for readers who wish to pursue those topics further.
\\

In the present work we adopt the following terminology:

\begin{itemize}
\item \textbf{Axioms}: foundational statements of the logical‑mathematical structure. They are the mathematical rules that define the theory.
\item \textbf{Postulates}: universal physical principles. They are foundational statements about the experimental physical world, which are not derived from other principles.
\item \textbf{Assumptions}: technical conditions, often introduced to simplify or make a mathematical construction well-posed.
\end{itemize}

\newpage

\subsection*{Acknowledgements}

I wish to thank the entire staff of the Haematology Centre at the Santo Spirito Hospital in Rome (ASL Roma 1) for their competence, patience and humanity; without their help I could not have written these notes.

\tableofcontents
\mainmatter

 \addcontentsline{toc}{chapter}{Introduction}
 \chapter*{Introduction}
\markboth{Introduction}{Introduction}

\begin{flushright}
\textit{I am convinced that the philosophers have had a harmful effect upon the progress of scientific thinking,  in removing certain fundamental concepts
from the domain of empiricism, where they are under our control, to the intangible heights of the  a priori  
\\
Albert Einstein 1922 \cite{Einstein45}}.
\end{flushright}

The main aim of this work is to blend two different mathematical approaches for the study of quantum physics: the algebraic and the probabilistic methods. To implement this goal, we will try to follow as much as possible the dictates indicated by Giles in his work \textit{Foundations of Quantum Mechanics} \cite{Giles69} of 1970.\index{Giles}

In this work are highlighted the main characteristics that an ideal physical theory must satisfy, which are:
\begin{itemize}
\item [A.] \textit{The theory should consist of a mathematical structure together with a set of rules of interpretation.}
\item [B.] \textit{The mathematical structure should be expressed in terms of axioms and primitive concepts.}
\item [C.] \textit{The rules of interpretation should give interpretations for all the primitive concepts and only these concepts.}
\item [D.] \textit{The physical concepts referred to in the rules of interpretation should be as direct as possible.}
\end{itemize}

Furthermore\footnote{See Accardi \cite{Accardi97}\index{Accardi}}, \textit{we will avoid deducing assertions about physics starting from a mathematical model of which we cannot clearly isolate the physical presuppositions that justify the main features of the model itself}.

Our starting point is a reworking of von Neumann's measurement theory found in the first paragraph of Chapter 4 of his 1932 book on the foundations of quantum mechanics: \textit{Mathematische Grundlagen der Quantenmechanik} \cite{Neu_32}\index{von Neumann}. Initially we will not deviate much from it, and we will try to introduce only the mathematical structures that are strictly necessary for the management of experimental data obtained through measurements carried out in a hypothetical laboratory. We will see that through the measurement theory of mathematical analysis it is possible to delineate the main properties of a physical system. This approach was introduced by Mackey in 1963 in his book \textit{The Mathematical Foundations of Quantum Mechanics} \cite{Mackey}\index{Mackey} and was developed by Deliyannis in \textit{Theory of Observables} (1969) \cite{Deliyannis}\index{Deliyannis}. The material from this last work will be presented again in these notes, suitably modified to adapt it to our point of view. Furthermore, we will try to describe the conceptual difficulties in associating such a measurement with experimental data as Mackey's formalism requires.

We will only consider experiments that can be physically carried out in the laboratory with measuring instruments and devices capable of experimentally determining the values of the various physical quantities involved. We assume that the experimenter has an idea of the object of his experimental action; he has a theory, a hypothesis, and based on this he designs measuring instruments and devices to explore the physical quantities and to see whether the knowledge acquired experimentally coincides with the hypothesis made a priori about it.\footnote{How such hypothetical assumptions and paradigms have historically had repercussions on scientific experimentation is well described in Kuhn's famous book \cite{kuhn}.} We do not assume the existence of a physical quantity unless we have an effective experimental method capable of measuring it.\footnote{We have underlined our point of view because it might seem that it falls back into the historical problem of \textit{unobservability}, a problem that we can summarize using the following words (see Barone's book \cite{Barone}): \textit{The new physics must deal with entities and quantities susceptible to being observed and measured, leaving aside --- as they are meaningless --- the rest.}

This question was also historically dealt with by Heisenberg, who proposed it to Einstein \cite{Heisenberg}\index{Heisenberg}:

\textit{I told him that this idea of observable quantities was actually taken from his relativity. Then he said, "That may be so, but still it's the wrong principle in philosophy." And he explained that \textbf{it is the theory finally which decides what can be observed} and what cannot, and, therefore, one cannot, before the theory, know what is observable and what not.}

We assume here that to equip the laboratory, our experimenter must possess a hypothesis, a theoretical interpretation of the physical phenomenon studied --- a hypothesis that he will have to verify through experimental action: \textit{he therefore has a provisional idea of what an observable is}, since without it he could not design measuring devices and instruments.} In other words, it is not enough to mentally conceive an experiment and then establish the value of a physical quantity and draw conclusions from it; the experiment must be carried out in practice, in our laboratory.

For example, the management of the experimental data obtained through our tests may not follow a classical probability model dictated by Kolmogorov's axiomatics, as suggested by quantum probability.\footnote{Statistics  models obtained mainly through the analysis of particular mathematical expressions called \textit{statistical invariants}; see, for example, Accardi's works \cite{Accardi85, Accardi97}.\index{Accardi}}

Furthermore, the various equations that link multiple physical quantities must always have a possible experimental confirmation: \textit{mathematical formulae cannot be indiscriminately applied: it is always necessary to specify the conditions of their applicability and then to verify, each time, that such conditions are fulfilled} (Accardi \cite{Accardi97}).
\\
In accordance with the Born-Heisenberg interpretation, we will assume the following obvious point of view (Reichenbach \cite{Reichenbach}, Section 29)\index{Reichenbach}:
\\
\textit{In a physical state not preceded by a measurement of an entity $u$, any statement about a value of the entity $u$ is meaningless.}
\\
Our description of the experimental procedures of the measurement process aims to revisit as much as possible the management --- even if idealized --- of a real laboratory, from the perspective of those who work in experimental physics.\footnote{On this topic, see also Ageno's book \cite{Ageno}.} We will try to describe these operations in simple, as little artificial language as possible, in accordance with Bohr \cite{Bohr}\index{Bohr}\index{Ageno}:
\\
\textit{...however far the phenomena transcend the scope of classical physical explanation, the account of all evidence must be expressed in classical terms. The argument is simply that by the word "experiment" we refer to a situation where we can tell others what we have done and what we have learned, and that, therefore, the account of the experimental arrangement and of the results of the observations must be expressed in unambiguous language with suitable application of the terminology of classical physics.}
\\
In other words\footnote{Cf. Primas \cite{Primas}, p. 101\index{Primas}}:
\\
\textit{The observed result must therefore be registered permanently in objective reality $\ldots$ an observation can be regarded as complete only if there exists a macrophysical document.}\footnote{See also d'Espagnat's book \cite{d'Espagnat}, p. 250.\index{d'Espagnat}
\\
Without entering into this philosophical dispute, we recall that the question of objective reality was first raised by Mach \cite{Mach} at the end of the 19th century, when empiriocriticism attempted to reduce physics to a catalog of immediately observable sensations. This position was criticized by Lenin in \textit{Materialism and Empiriocriticism }\cite{Lenin}, where he warned against the risk of dissolving physical reality into mere subjective impressions. In that work, Lenin anticipated themes that would resurface a few decades later in the subsequent debate between Bohr and Fock \cite{Fock57} \index{Fock}.}
 \\
 After recalling von Neumann's measurement theory and Mackey's formalism, we introduce the algebraic method through the notion of algebraization of a physical system.
\\
As is well known, in his foundations book \cite{Neu_32} von Neumann associates with each observable of the system a self-adjoint operator on a (separable) Hilbert space, initiating the mathematical study of these linear operators. This topic, in collaboration with Jordan and Wigner in 1934 \cite{Jord_34}, led to the first attempt at an algebraic formalization for the observables of a physical system through non-associative real algebras, which subsequently came to be called Jordan algebras.\index{Jordan}\index{Wigner}\index{von Neumann}

Thereafter, von Neumann's works (in collaboration with Murray \cite{Neu_36}) focused on associative algebras of linear operators, while Jordan algebras found themselves again in the foreground in physics in Segal's 1947 work on the postulates of quantum mechanics \cite{Segal_47}, which introduced what would later be called \textit{Segal systems}.\footnote{For a less rapid history, one may consult Primas' book \cite{Primas} and Wightman's work \cite{Wightman76}.}

The von Neumann--Segal algebraic formalization, together with the aforementioned Mackey's theory\footnote{We recall that a synthesis of this formalization was given by Emch in his \textit{Algebraic Methods in Statistical Mechanics and Quantum Field Theory} (1972) \cite{Emch}.}, still constitutes the main toolkit for an axiomatic theory of quantum mechanics.\footnote{See also the works of Accardi \cite{Accardi75, Accardi81b, Accardi97}.}

A further algebraic formalization of quantum physics, due to the joint work of Haag and Kastler, which takes into account the theory of relativity and also absorbs Wightman's field theory \cite{WiSt}, is Local Quantum Physics (LQP) \cite{Haag}\index{Haag}.

The fundamental cornerstone of LQP, formalized for the first time in a rigorous way by the authors in the work \textit{An Algebraic Approach to Quantum Field Theory} (1964) \cite{HaagKastler}, consists in associating with each bounded region $\mathcal O$ of space-time a C*-algebra $\mathfrak A (\mathcal O)$, whose elements are interpreted \textit{as representing} physical operations carried out in the $\mathcal O$ region. This algebra is called the algebra of observables \textit{localized} in $\mathcal O$.

In this work, following the path traced by von Neumann \cite{Neu_32} and Segal \cite{Segal_47}\index{Segal}, states are represented by the positive functionals $\mathfrak A^{(+)^*}$ of the algebra $\mathfrak A$; their value calculated on an element of the algebra is the expected value of the related observable in that state.

In this way we obtain a correspondence $\mathcal O \rightarrow \mathfrak A (\mathcal O)$ that satisfies the well-known properties of isotony, duality and covariance.\footnote{To be exact, in their 1964 article the authors mainly take into consideration two categories of objects in addition to states: operations. Operations are particular linear transformations of the dual set $\mathfrak A^{*}$ that map the set $\mathfrak A^{(+)^*}$ into itself, physically representing the change caused by the measurement apparatus on the initial state of the system. The concept of operation, introduced here for the first time in the literature, even if mathematically well defined, is not easy to interpret (this topic will be addressed later by Davies and Lewis in their work \cite{Davies-Lewis} and taken up by Edwards \cite{Edwards} and Kraus \cite{Kraus}; a modern contribution in the LQP field can be found in \cite{Ozawa015}). In fact, in the subsequent axiomatic versions of LQP (see for example Haag's book \cite{Haag}), this concept has a secondary role while the central role of the observable is re-established, historically defined as a physical quantity measurable in a region of space-time that determines the local algebra. This statement is also reiterated by Horuzhy in \cite{Horu} in the note on p. 2: \textit{However, properties of locality and localizability, fundamental for relativistic quantum theory, are more naturally expressed in terms of observables.}}

The problem with LQP lies in its highly mathematical language, which elegantly hides many apparently banal problems contained in the points outlined by Giles, which in our opinion deserve further study.\footnote{Primas' judgment on this approach is very severe (see \cite{Primas}, p. 178): \textit{"C*-algebraic quantum mechanics uses unashamedly an entirely unjustified operationalistic language."}} In particular:
\begin{itemize}
\item [$\circ$] What should we mean by physically realizable observables in our $\mathcal O$ region?
\item [$\circ$] What are the experimental procedures carried out in space-time, and what role do they have in obtaining the expected value of an observable?
\item [$\circ$] How is it possible to represent physical quantities and localized states in $\mathcal O$ through the elements of an algebra $\mathfrak A (\mathcal O)$ and its dual?\footnote{In LQP the question is often asked in reverse: given the algebra of observables $\mathfrak A$, what are its physically realizable algebraic states? \textit{"The trouble is that only a tiny fraction of the states of our systems have any physical relevance, and one of the basic problems in treating systems with an infinite number of degrees of freedom is to single out and classify the states or representations of physical relevance"} (from Roberts \cite{Roberts04}\index{Roberts}).}
\item [$\circ$] How many possibilities are there for this identification?
\item [$\circ$] What role do simultaneously measurable observables in the $\mathcal O$ region have in this game, and what meaning does this statement have?
\item [$\circ$] Another non-secondary question is to establish what is meant by the temporal evolution of a physical system and what meaning the term "interacting systems" has.
\end{itemize}

Furthermore, in providing clarity, the various introductory statements by Horuzhy in \cite{Horu}\index{Horuzhy} on the algebraic method do not help much. For example, on page 11 we find the following statement:

\textit{...we do not assume that the set of all observable algebras $\mathfrak A (\mathcal O)$ is specified uniquely for a given physical system (although the original set of observables $A_{obs}(\mathcal O)$,\footnote{In our notations such objects will be indicated by $\mathfrak X (\mathcal O)$.} was, of course, unique for a given system). It will be seen that different sets of local algebras corresponding to the same system can have different properties; in particular, a special role is played by the sets of maximal algebras which cannot be embedded into other possible algebras of observables.}

Let us remember that in LQP, through field theory and by introducing a particular compact topological gauge group, the authors Doplicher, Haag and Roberts in \cite{DHR69a, DHR69b}\index{Doplicher} obtain the algebra of observables in an algebraic way. In practice, the algebra $\mathfrak A (\mathcal O)$ is obtained as the fixed points of the action of the gauge group on the field algebra. In this way the annoying problems we have presented are short-circuited, yielding a rigorous mathematical procedure but one totally free from the experimental act, because having the "quantum fields" essentially means already having the algebra of observables.\footnote{See for example the work by Driessler, Summers and Wichmann \cite{DSW}.\index{Wichmann}}
\\

We will adopt the operationalist point of view, well described by Primas in \cite{Primas}\index{Primas}:

\textit{They say that science is a tool for making predictions about what will be observed in different situations, and consider a physical quantity as being defined when the procedures for measuring that quantity are specified.}\footnote{For a philosophical criticism of this approach, see again Primas, p. 147.}

Therefore, the objective of the initial sections will be to establish a \textit{statistical model to be adopted to describe the measurement procedure of a physical quantity} in the laboratory.\footnote{See also Holevo \cite{Holevo}.}

But what should we mean by statistical model?\index{Statistical model}

We have physical quantities to measure at a given instant of time $\tau$. What we can say is that their values will be positioned along the real line.\footnote{We reiterate that for a physical quantity to be such, it must be quantifiable.} So for each physical quantity $a$, each subset $\Delta\subset \mathbb R$ and each time $\tau\geq 0$ we can consider the following logical proposition:
$$ \textsl{A: $a$ takes a value in $\Delta$ at time $\tau$ }$$
Denoting by $\mathcal A$ the set of such propositions, a statistical model on the measurement procedure is given by a family of maps $\left\{ P_\theta \right\}_{\theta\in\Theta}$ with $P_\theta: \mathcal A \rightarrow [0, 1]$, which indicates the truth of the proposition $A$, where the two extremes are:
\begin{equation*}
P_\theta(A)=
\left\{\begin{array}{cc}
  1 & \textit{A is true} \\
  0 & \textit{A is not true}
\end{array}
\right.
\end{equation*}
while the set of parameters $\Theta$ is determined by the experimental procedures and also depends on the physical quantities that we subject to these procedures.

For example, classically a parametrized statistical model consists of a measurable space $(\Omega, \mathcal F)$ (with $\mathcal F$ a sigma-algebra on $\Omega$, the sample space) and a family of probability measures $\left\{ P_\theta \right\}_{\theta\in\Theta}$ on $\mathcal F$.

In this way, for every measurable function $X:\Omega \rightarrow \mathbb R$ we obtain a functional on the space of real functions that vanish at infinity:
\begin{equation}
 \mu_{X,\theta}(f) = \int f\circ X \ d P_\theta \ , \qquad f\in C_o(\mathbb R)
\end{equation}
Therefore, a classical parametrized statistical model is associated with the algebra of real measurable functions denoted by $\Sigma(\Omega, \mathcal F)$ and a map
\begin{equation}\label{mappadual1} 
(X , \theta ) \in \Sigma(\Omega, \mathcal F)\times \Theta \longrightarrow \mu_{X,\theta}\in C_o(\mathbb R)^*
\end{equation}
The elementary propositions (also called questions) given by $\Delta\in\mathcal F$ can be identified with the characteristic function $\mathbf{1}_\Delta\in \Sigma(\Omega, \mathcal F)$ of the set $\Delta$.

In our statistical model we do not initially have any type of mathematical structure associated with the set of propositions $\mathcal A$, but as we will see, by adding further hypotheses on its composition we can also determine in our case a map similar to \eqref{mappadual1}.

As underlined at the beginning of this introduction, the procedure for determining the family of maps $\left\{ P_\theta \right\}_{\theta\in\Theta}$ is the one outlined by von Neumann in \cite{Neu_32}, a procedure which we denote here as \textit{Statistical Ensembles Interpretation}.

In reality, unlike us, Ballentine in \cite{Ballentine}\index{Ballentine} differentiates the measurement procedure treated by von Neumann, defining it "orthodox interpretation" as distinct from the "statistical interpretation" through the following statement:

\textit{"...the basic assumption of the statistical interpretation that a state vector characterizes an ensemble of similarly prepared systems, the orthodox interpretation assumes that a state provides a complete description of an individual system."}

Thus for Ballentine a state of the system is associated with $N$ trials of systems prepared in a similar way (briefly denoted as $N$ copies). We believe that this statement is devoid of experimental sense, because we do not understand how operationally Ballentine can "simultaneously" associate a single state with $N$ copies of our experiment. It is legitimate to ask what the state of $N$ identical trials of our experiment is, and how and where to prepare it.\footnote{See Accardi \cite{Accardi97}, p. 104. However, the reader interested in the philosophical implications, which we will not deal with, can consult the work of Home and Whitaker \cite{H.W.} and the evergreen books of Jammer \cite{Jammer}\index{Jammer} and d'Espagnat \cite{d'Espagnat}.}
\\
We want to underline that in our "orthodox" approach, we leave very little space for the role of the observer. Once the various protagonists (whether people or machines) have completed all the experimental procedures in the laboratory and activated the various instruments, their role will be solely and exclusively that of simple accountants, taking note of the results obtained.\footnote{This is therefore in opposition to the Bayesian QBism approach, much invoked today in the philosophy of science. For a quick background on the topic, the reader may consult Stacey's article \cite{Stacey} and the even more elementary one by von Baeyer \cite{Baeyer}.}
\\
Once the adopted statistical model has been established and analyzed, our program continues by studying the possibility of introducing better equipped mathematical structures in line with the algebraic model described above.
\\

In summary, a physical system will be described by a pair of sets $(\mathcal A, \Theta)$ and by a family of maps $P_\theta : \mathcal A \rightarrow [0,1]$ with $\theta\in \Theta$. The link between the objects $\mathcal A$, $\Theta$ and $P_\theta$ is established by introducing fundamental properties grounded in physical experience, which in the text will be denoted as axioms. In reality, to have an axiomatic formal structure as dictated by Hilbert (see Accardi \cite{Accardi018}), it will be necessary to verify mathematically that these fundamental properties are actually consistent and independent of each other, which requires a further effort to reorganize the material presented.
\\
\textbf{But is a formal axiomatization of physics so important?}
\\
At this point the following remark is useful\footnote{Cf. Accardi \cite{Accardi81}}:

\textit{Nowadays when we speak of "postulates of a physical theory" we simply mean to separate the purely deductive part of the theory from the inductive part. That is, postulates are a set of statements of the theory from which all others can be deduced by purely logical means. In this sense, they represent the conceptual synthesis of empirical knowledge of a given era. However, it should not be forgotten that physics is not a deductive science -- that is, it does not proceed by postulates and deductions -- but it uses these to coordinate and develop the results obtained through experiments and inductions.}

One might think that these problems of axiomatic formulation of physics are recent. In reality, already in the early 17th century the philosopher Francis Bacon had warned the nascent scientific community of the critical issues that could exist in the geometrization of nature. Let us recall a short passage from Amir Alexander's book \cite{Amir}, Chapter 8, relating to this topic:\index{Amir}

\textit{Bacon's suspicion of mathematics as a tool for comprehending the world is not hard to understand. For mathematics to describe nature correctly, nature must be mathematical --- that is, structured according to strict mathematical principles. If that is the case, then all one needs in order to gain insight into the workings of nature is to follow the rules of rigorous mathematics, and all observations and experiments are superfluous.}

\newpage

\part{Laboratories and Measurements}

\chapter{Laboratory Systems}
\begin{flushright}
\textit{It is a capital mistake to theorize before one has data. Insensibly one begins to twist facts to suit theories, instead of theories to suit facts}
\\
Arthur Conan Doyle
\end{flushright}

\begin{flushright}
\textit{Quantum phenomena do not occur in a Hilbert space, they occur in a laboratory}
\\
Asher Peres
\end{flushright}

In this section we will begin the discussion of the experimental procedures that are used to establish the numerical value of a set of physical quantities -- which historically are denoted by the term observables -- through the statistical analysis of the measurements carried out in a well-defined place in space, which we will generically call the laboratory. We will assume that it is limited in space, and for the measurement of time we will rely on a clock supported by the walls of the laboratory itself.

A primary role is played by the various measurement instruments that participate in establishing the numerical value of these observables. Their presence, their method of preparation in the laboratory, and their effectiveness in measurement are characteristics that identify what we will generically call the \textit{physical state of the system}.

The physical states are also determined by other physical quantities of the laboratory system that the experimenter keeps rigidly under his control through his action on the apparatus and on the various instruments when carrying out the experiment.

We will assume a minimal point of view: we will not initially give any mathematical structure to the set of observables and states of the system other than their probable values obtained through measurements.

These trivial statements are the basis of old and new discussions on the foundations of physics.\footnote{For further information, one may consult the works of Accardi cited in the bibliography, as well as those of Ballentine \cite{Ballentine} and Primas \cite{Primas}.} These are topics that we will not cover more than necessary in this section.\index{Primas}\index{Ballentine}\index{Accardi}

\textit{We underline that this chapter is aimed more at the physical motivations of the mathematical formalism that we propose than at its axiomatic presentation.}
\section{Experimental Procedures}\label{proceduresperimentali}
In the model we will develop, an observable corresponds to a physical quantity that we can quantify using devices called measuring instruments. We will not attempt to give a formal definition of measuring instrument, observer, experimenter, etc. The mathematical model we are developing does not need to specify these concepts. However, before moving into the formal discussion, it is useful to make some brief experimental considerations on the measurement of a physical quantity, to better understand the mathematical model we will discuss.

In other words, we will describe the fundamental actions that a hypothetical experimenter must perform when he enters his laboratory to carry out a given physical experiment at a time $\tau$.\footnote{We will assume that the experimenter has the ability to understand where and when he makes the experiment, through the use of rulers and clocks.}

We can say that physically for the measurement we need:
\begin{itemize}
\item [1.] A set of instruments $\textsl{D}$ and a source $\textsl{S}$, the source of the measurement (for example a radioactive material, a steel spring, a sound source, etc.).
\item [2.] Fully specified experimental procedures, i.e., describing step by step all the preparation methods of the instruments $\textsl{D}$ and the environmental conditions to which these procedures are subjected.
\item [3.] A preparation time for our experiment, which must take place in a very specific place $L_o$ of space $\mathcal E$,\footnote{The execution time available to the experimenter, however long it is, is always limited. Moreover, we will consider the space $\mathcal E$ as a locally Euclidean topological space, while space-time $\mathcal M$ is the set given by the Cartesian product
$$\mathcal M = \mathcal E \times \mathbb R$$
As we will discuss later, if we fix a reference system $(K;O)$ with $O\in\mathcal E$ we will have a set (of charts) $\mathcal M^{K;O}=\mathbb R^4$ for the manifold $\mathcal M$.} the spatial region that delimits our laboratory.
So everything happens in a bounded region $\mathcal O_o$ of space-time $\mathcal M$:
$$\mathcal O_o = L_o \times [0, t_p]$$
\item [4.] To assign to the physical quantity $a$ its probable values, which are obtained from the relative frequencies of the numerical values yielded by the $\textsl{D}$ instruments (see von Mises \cite{Mises}) at a measurement time $t_M$.\index{von Mises}
\end{itemize}
Therefore, the knowledge of the physical quantity $a$ requires a preparation time interval $[0, t_p]$, while its values appear on the instrumentation, ready for reading, at the instant $t_L > t_M \geq t_p$.

We remark that the measurement is carried out at a fixed time $t_M$, after the preparation of the experiment, but when we talk about the measurement of the value of the quantity $a$ at a given instant of time $\tau$, we consider the value
$$\tau = t_M - t_p \geq 0$$
In other words, \textit{we begin to establish the temporal evolution of the value of the quantity $a$ after its preparation}.

We summarize our considerations in the following scheme:
\begin{center}
$\begin{array}{ccccc}
\textit{Preparation} & \longrightarrow & \textit{Measurement} & \longrightarrow & \textit{Value reading} \\
\textit{in $[0, t_p]$} & & \textit{at time $t_M > t_p$} & & \textit{at time $t_L > t_M$}
\end{array}$
\end{center}

The measurements of a physical quantity $a$ carried out under appropriate conditions, which we indicate by $\omega$, lead to the study of the following relative frequencies:
\begin{equation}
\label{freq}
 f(a\in\Delta)_\omega = \frac{\textit{number of times that $a$ takes a value in $\Delta\subset \mathbb R$}}{\textit{total number of measurements carried out on $a$ in $L_o$}}
\end{equation} 
Therefore, we can say that the physical system is described through probability laws established via frequencies \eqref{freq}:
\begin{equation}
 (a, \omega) \longrightarrow P(a\in\Delta)_\omega, \qquad \Delta \subset \mathbb R
\label{distribuzio0}
\end{equation}
where $P(a\in\Delta)_\omega$ is the probability that the observable $a$, at time $\tau$, takes a value in a (Borel) subset $\Delta$ of $\mathbb{R}$ conditioned by the state $\omega$.\index{Statistics of the physical system}

In summary, in each state the result of a measurement can be predicted with a certain probability given by \eqref{distribuzio0}, which is called the \textit{statistics of the physical system}.

The statistic that describes a physical system will be called \textsl{exact} (see Accardi \cite{Accardi75}) if for each physical quantity $a$ and state $\omega$ we obtain
$$ P(a\in\Delta)_\omega = 0 \ \text{or} \ 1 $$
for every (Borel) subset $\Delta$ of the real numbers.\index{Exact statistic}\index{Accardi}

The frequentist method "identifies" probabilities -- which are a mathematical notion -- with the relative frequencies that are given by experimental data.

We remind the reader that this identification process is not without conceptual problems; we will try to give a new overview of this topic in Chapter \ref{legge_empirica} of these notes.

\subsection{Some Remarks on States of System}\index{Toller}\index{Peres}\index{Schlegel}\index{Margenau}

It should be noted that in the literature the notion of system state is often used in a slightly different way than the one we propose here.\footnote{For a thorough discussion of this topic, it is advisable to read Accardi's works, in particular \cite{Accardi81}.

Interesting from our point of view is the work of Peres \cite{Peres}, where he talks about preparation as \textit{"a recipe in a good cookbook"}, and that of Toller \cite{Toller75}, where we find the following statement: \textit{"A physical state is defined in terms of the procedure used to prepare it."}

A comment -- in my opinion cryptic -- on preparation and measurement in a given state of the system is due to Margenau \cite{margenau} (see also Schlegel's book \cite{Schlegel}, p. 192):
\textit{"In general, preparation 'determines the state of a physical system but leaves us in ignorance as to the incumbency of that state after preparation,' whereas measurement certifies 'that some system responded to a process, even though we are left in ignorance as to the state of the system after the response.'"}

We want to underline that, unlike Margenau's comment, in our case only after the preparation of the state are the conditions of the laboratory system in which we carry out the actual measurement known.}

So basically, this terminology is used to indicate the situation in which the laboratory finds itself through the knowledge of some physical parameters that the experimenter knows and masters (for example, the temperature of the laboratory, the intensity of the magnetic fields possibly present in it, etc.). It is customary to say that the state of the system changes if these physical parameters change over time. In other words, here the state is identified with the configuration of the system -- i.e., a photograph taken at a given instant of time that completely describes its physical parameters at that instant.

In our case, the state of the system concerns the methodologies with which the observables are measured: not only the physical situation of the laboratory, but also the measurement procedure, the related devices used (for example, any radiation shielding instruments), and the type of measuring instruments used, etc.

\begin{remark}\upshape\label{statoindividuazione}\index{Parametric state}
Let us see, in broad outline, the steps required to identify a state of the physical system.
\\
A state $\omega$ suitable for measuring a physical quantity $a$ can ideally be divided into the following sections:
\\
 - A section, which we denote by $\omega_S$, is related to the instrumentation used and its method of use to determine the various values of $a$ at the established time $\tau$;
\\
- Another section, which we denote by $\omega_C$, is related to the conditions that we set on the various physical parameters $\left\{c_i : i\in T \right\}$,\footnote{This will be denoted by the name \textit{parametric state} of the laboratory. See also the notion of complete set of observables given by Accardi in \cite{Accardi81}, p. 519.} which the experimenter keeps under control during preparation by fixing their values.
\\
Obviously, this component of the state is also influenced by the various instruments for controlling and measuring the variables $c_i$ and the procedures for carrying them out. Therefore, the component $\omega_C$ can also be divided into two parts: one, which we denote by $\omega_{CC}$, related to having fixed the values of the physical parameters $\left\{c_i : i\in T \right\}$; and the other, which we denote by $\omega_{CS}$, relating to the instrumentation and its procedures.\footnote{We are basically stating that even the control and measurement instruments that serve to keep the values of the quantities $c_i$ fixed are important in defining the state of the system (in addition to their physical presence in the laboratory itself). This makes the distinction between parametric $\omega_C$ and instrumental $\omega_S$ states even more nuanced.}
\end{remark}

\subsection{Statistics and Reproducibility}\label{Statistica e Riproducibilità}

Let us now have a brief discussion on how the frequencies defined in \eqref{freq} are determined experimentally. This methodology is a key part of von Neumann's measurement theory, which the reader can find in \cite{Neu_32}, Chapter 4.\index{von Neumann}

We must establish an ensemble, that is, arrange $N$ identical trials of the experiment to be carried out,\footnote{What we will later briefly call the \textsl{copy of the ensemble}.} as highlighted in point 2.

\begin{figure}[htbp]  
	\centering
	\includegraphics[scale=0.4]{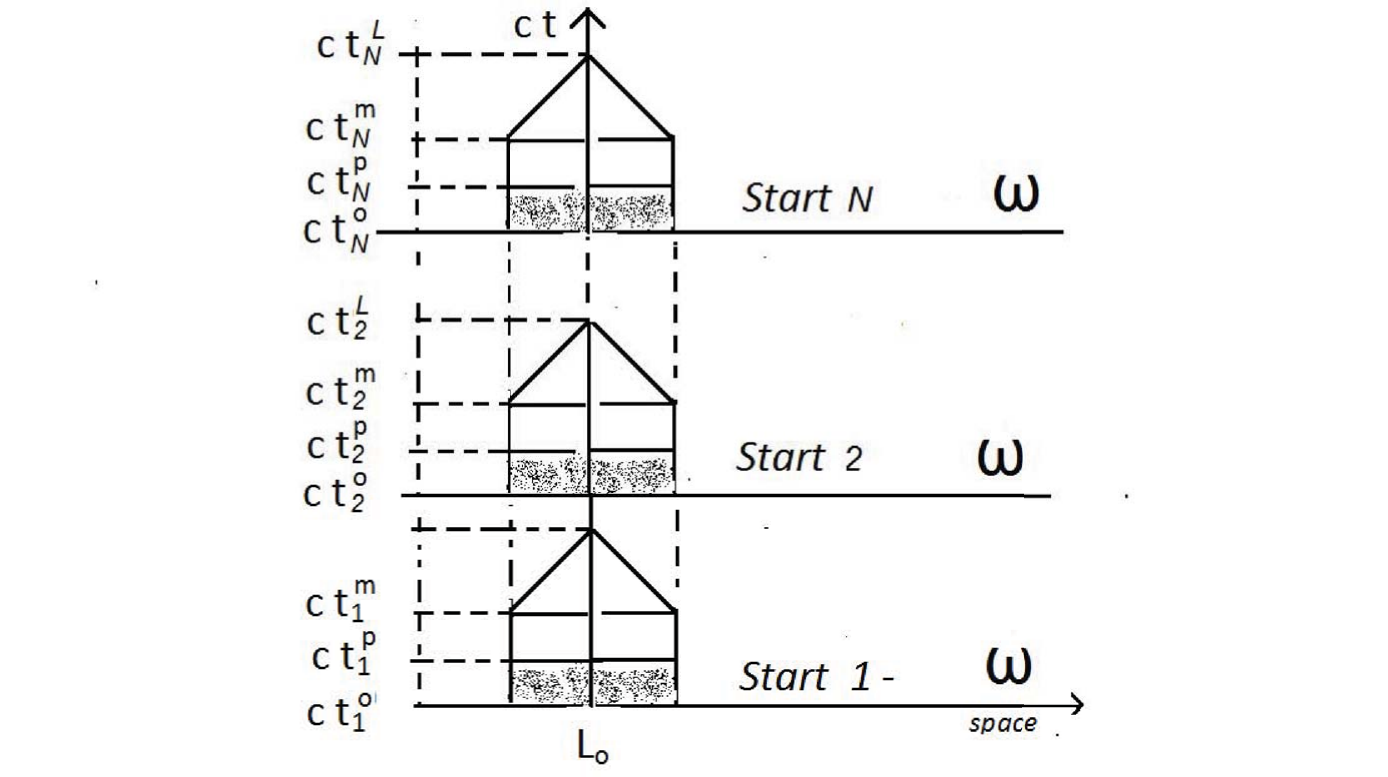}
	\caption{$N$ copies of the ensemble} 
	\label{fig:01}
\end{figure}

Then we must assume that after having carried out the various measurements, the experimenter has the ability to re-establish the initial $\omega$ state, i.e., he has the possibility of eliminating all the effects of the measurements previously carried out in his laboratory (obviously spending time and energy to do so) to prepare a new measurement on the same physical quantity and repeat the experiment. This assumption is referred to as the \textit{reproducibility conditions of the experiment}.

In other words, if we repeat the experiment without resetting the effects of the measurements previously carried out in the laboratory, the parametric state of the system could change.\footnote{After the measurement, in addition to changing the values of the parameters $\left\{c_i\right\}_{i\in I}$ that are specific to the $\omega_C$ state, the conditions of the instruments and devices could also change, such as their calibration and their possible reuse, which the experimenter will have to restore to carry out a new measurement.} Therefore, the subsequent measurements on the ensemble will no longer yield the frequencies  \eqref{freq} relating to the state $\omega$.

Therefore, \textit{the reproducibility of an experiment is the founding element of the experimental method}, and information on the value that a physical quantity possesses is obtained only by analyzing the relative frequencies \eqref{freq} for a very high number of trials (ideally infinite).\footnote{See also Sudbery's comment in \cite{H.W.}, \$5.5 p. 287.\index{Sudbery}}

However, it must be remembered that in statistics, when analyzing a particular characteristic of a given population, since it cannot be done for obvious practical reasons on the entire population, a subset of it is chosen, which is called the sample of the population. From the analysis of the sample we will try to extrapolate information about the particular characteristic of the entire population that we want to examine.

In our case, the repeated trials of the experiment -- which constitute our sample -- will always be finite, however large $N$ may be chosen.

This means that passing from the frequencies of statistical data \eqref{freq} to probability values \eqref{distribuzio0} requires a much more in-depth (and problematic) analysis that goes beyond the content of these notes.\footnote{\label{bayes0} We do not hide the fact that this problem causes our frequentist method to falter, since even if the experimenter has refined statistical and mathematical methodologies to determine the distribution \eqref{distribuzio0} from the results of the frequencies \eqref{freq}, it is necessary to ask ourselves to what extent these methodologies are influenced by the experimenter's choice -- a choice adopted by his instinct and therefore by what the experimenter expects. We will try to resolve this problem later, in Chapter \ref{legge_empirica}.}

The reproducibility of an experiment deserves further investigation, which we will see in Section \ref{riproduzione}. Here we just want to highlight that the state of the system during the preparation of the laboratory is constantly changing.

We start from a parametric state $\omega_{j,C}^\star$, which reflects the initial conditions of the laboratory. This state will be modified during the laboratory preparation interval, assuming a precise identity only once preparation is completed in the state $\omega$, and in this state the measurement physically takes place. Subsequently, when the measurement is carried out, we will have a change in the physical conditions that leads to a new parametric state $\omega'_{j,C}$:
$$ \omega_{j,C}^\star \underbrace{\longrightarrow}_{\text{preparation}} \omega \underbrace{\longrightarrow}_{\text{measurement}} \omega'_{j,C} $$

The starting and arriving (parametric) state depends on the $j$-th copy of the ensemble. We are therefore assuming that the initial laboratory conditions do not hinder or influence the preparation of $\omega$. In this case, we will say that the laboratory has been restored to its initial conditions.

In Figure \ref{fig:01} we have drawn in Minkowskian space-time the preparation of $N$ copies of the experiment after the restoration of the laboratory to its initial conditions (start 1, 2, ..., $N$), with the observer positioned at the center of the laboratory $L_o$.\footnote{Here the observer is the one who collects the data of the various measured values from the instruments; it could itself be a machine.} We have:
\begin{equation}
\label{tempogrand}
0 \leq t_j^m - t_j^p = \tau, \qquad j = 1, 2, \ldots, N
\end{equation}
Furthermore, $t_{j+1}^p - t_j^o$ and $t_j^L - t_j^m$ (which are constant for every $j = 1, 2, \ldots, N$) are, respectively, the preparation time interval and the reading time interval of the $j$-th measurement, while $t_{j+1}^o - t_j^L$, $j = 1, 2, \ldots, N$, is the rest interval from one measurement to the next.

\begin{figure}[htbp]
	\centering
	\includegraphics[scale=0.4]{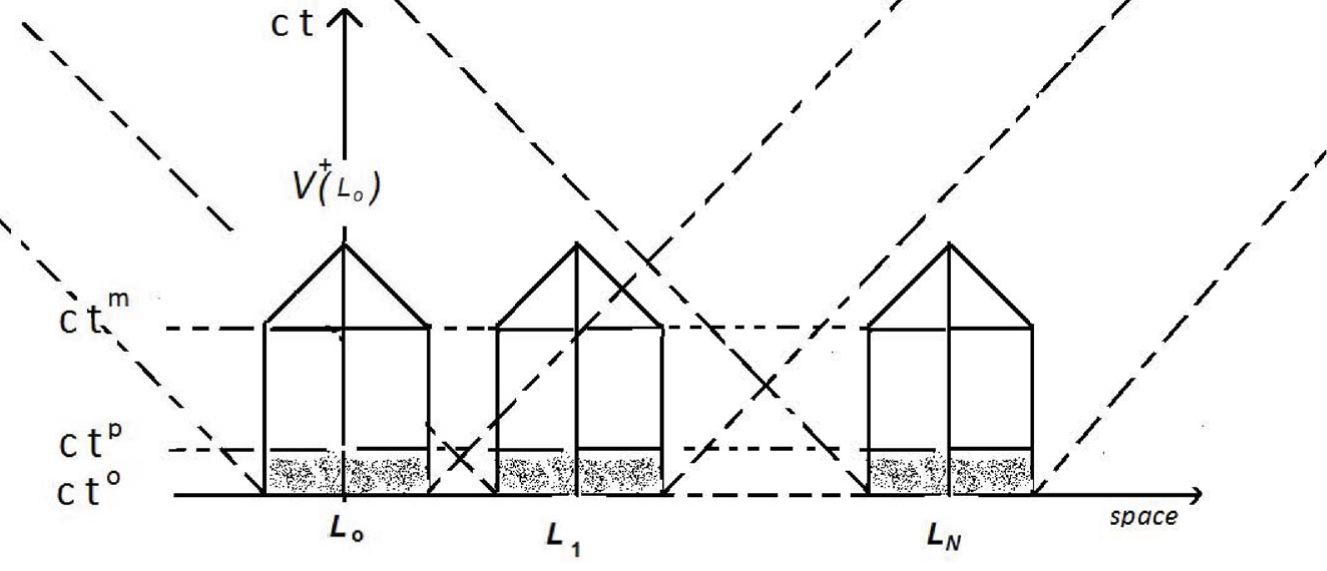}
	\caption{$N$ displaced copies of the ensemble}
	\label{fig:02}
\end{figure}

We observe that we can prepare our copies of the experiment to be carried out as described in Figure \ref{fig:02}. That is, we have located our ensemble in $N$ identical laboratories (not moving with respect to our observer positioned at $L_o$). However, problems arise in this case too: the measurements carried out in the various laboratories influence each other (for example, the future light cone $\mathcal V^+(L_o)$ intersects that of laboratory $L_1$ after a short interval of time), making the frequencies defined in \eqref{freq} ineffective.\footnote{These procedures will be discussed on page \pageref{invarianza-sperimentale}.}
\subsection{Measurement time and state of the physical system}\label{mappastatuale}
As we have underlined, the state of the system is established by the choice of instruments and devices (and in general also by their spatial configuration in the laboratory) that we use to determine the value of a specific physical quantity $a$.

Furthermore, when defining the state of the system, we must explicitly declare in what way and under what conditions these machines are to be used, establishing throughout the preparation time of the state certain physical parameters that the experimenter has decided to keep under his control.

\begin{post}\label{mappastato}\index{Postulate-Time Preparation}
We assume that the experimental procedures carried out in the laboratory to establish the state of the physical system, which require a preparation time
$$t_p = t_j^p - t_j^o, \qquad j = 1, 2, \ldots, N$$
also include the instruction to measure the value of the observable $a$ at every (possible) instant of time\footnote{See Remark \ref{cronos1} on page \pageref{cronos1}.}:
$$\tau = t_j^m - t_j^p \geq 0, \qquad j = 1, 2, \ldots, N$$
\end{post}

Let us study in detail what is the state of the system according to Assumption \ref{mappastato}.

The question is the following:

We want to determine the value of the observable $a$ at various instants of time $\tau$, keeping unchanged the experimental conditions under which these measurements take place, which are:
\begin{itemize}
\item[A.] The use of the same instruments and devices.
\item[B.] The same spatial configuration of the equipment in the laboratory as in point A.
\item[C.] The same physical parameters that must be respected \textit{in the time interval $[0, t_p]$ before carrying out the measurement}\footnote{In other words, the same parametric state.}.
\item[D.] The same procedures are performed, differing only in the instruction of when to make the measurement (at time $\tau$).
\item[E.] The laboratory preparation time $t_p$ is not influenced by the choice of the measurement time $\tau$, as per point D.\footnote{The preparation of the state uses the same procedures and protocols for different measurement times.}
\end{itemize}

Therefore, the experimenter must implement only the instruction "\textit{execute the measurement at time $\tau$}" in the various devices of point A.

In this way, for each value of $\tau \geq 0$, we have a state of the laboratory system that satisfies the five points indicated, which we will denote by $\omega^{(\tau)}$, obtaining a map\footnote{\label{map-temp} Notice that the map \eqref{noevoltemp} is not the temporal evolution of the laboratory state.}:
\begin{equation}
\label{noevoltemp}
\omega : \tau \longrightarrow \omega^{(\tau)}    
\end{equation}

\begin{definition}\upshape\label{statocronos}\index{Chronological state}
The map \eqref{noevoltemp} is said to be the chronological state of the laboratory system.
\end{definition}

\begin{notation}\upshape
By an abuse of language, and when there is no possibility of misunderstanding, the chronological state of the laboratory system \eqref{noevoltemp} is simply called the state of the system and is denoted, unless otherwise stated, still by the symbol $\omega$.
\end{notation}

\subsection{The temporal evolution of the value of an observable}

By changing the value of $\tau\in\mathbb R^+$ in \eqref{noevoltemp}, we obtain \textit{the temporal evolution} of the observable $a$ through the probability distribution law
\begin{equation}
\label{evoltemp0}
\tau \in \mathbb R^+ \longrightarrow P(a\in\Delta)_{\omega^{(\tau)}} \ , \qquad \Delta\subset\mathbb R
\end{equation}
We reiterate that the state $\omega^{(0)}$  differs from the preparation   the state $\omega^{(\tau)}$ only in the instruction: \textit{carry out the measurement at time $\tau$.}

We caution that once the laboratory has been prepared for the measurement of our observable $a$ at time $\tau$, with preparation time $t_p$, by definition of $t_p$ \textit{no further action will be performed on the instruments and devices in the laboratory}, since it is assumed that they have been programmed, during this time interval, to carry out this measurement.

Therefore, the physical parameters initially set in the preparation of the state of the laboratory will, after a certain interval of time $\tau$, most likely not have the same initial values.

\begin{notation} \upshape
For each state $\omega$, the distribution \eqref{evoltemp0} will be denoted as
\begin{equation}\label{evoltemp}
\tau \in \mathbb R^+ \longrightarrow P(a\in\Delta, \tau )_\omega \ , \qquad \Delta\subset\mathbb R
\end{equation}
therefore, in the state $\omega$ we already have the instruction to perform the measurement at time $\tau$. In other words:
$$P(a\in\Delta, 0 )_\omega=P(a\in\Delta )_{\omega^{(0)}} \qquad , \qquad P(a\in\Delta, \tau )_\omega=P(a\in\Delta )_{\omega^{(\tau )}} $$
\end{notation}
\begin{remark}\upshape
Experimentally, measurements can only be performed for a finite number of time values $\left\{\tau_0, \tau_1, \ldots, \tau_n \right\}$; in this way we obtain a discrete family of distribution laws:
\begin{equation}\label{evoltempdis}
\left\{P(a\in\Delta, \tau_j )_\omega : j=0,1, \ldots, n \ , \quad \Delta\subset\mathbb R \right\}
\end{equation}
and therefore, here too we need to establish methodologies that perform the transition from the discrete case \eqref{evoltempdis} to the continuous one \eqref{evoltemp}, a topic that will not be discussed here\footnote{Besides the problem of the discrete-continuous time transition, experimentally we have the non-trivial problem of the rational-real transition, since a measuring instrument always determines fractional values of the physical quantity in question. \\
So one might think that the field of rational numbers is the only one that makes experimental sense. This statement also introduces a series of operational difficulties when we consider the functions of an observable, since only algebraic and non-irrational functions have experimental evidence.}.
\end{remark}

Another notation widely used in the literature for the value $P(a\in\Delta, \tau )_\omega$, which we will not use, is given by $P(a_\tau \in\Delta )_\omega$.
\\
We believe that this notation is misleading, since it would seem that the observable $a$ transforms over time into another observable $a_\tau$, which is obviously not true because what actually changes is the measurement of its value.

\subsubsection{Preparation time and laboratory dimensions}\label{doppio}

\begin{figure}[htbp]
	\centering
		\includegraphics[scale=0.3]{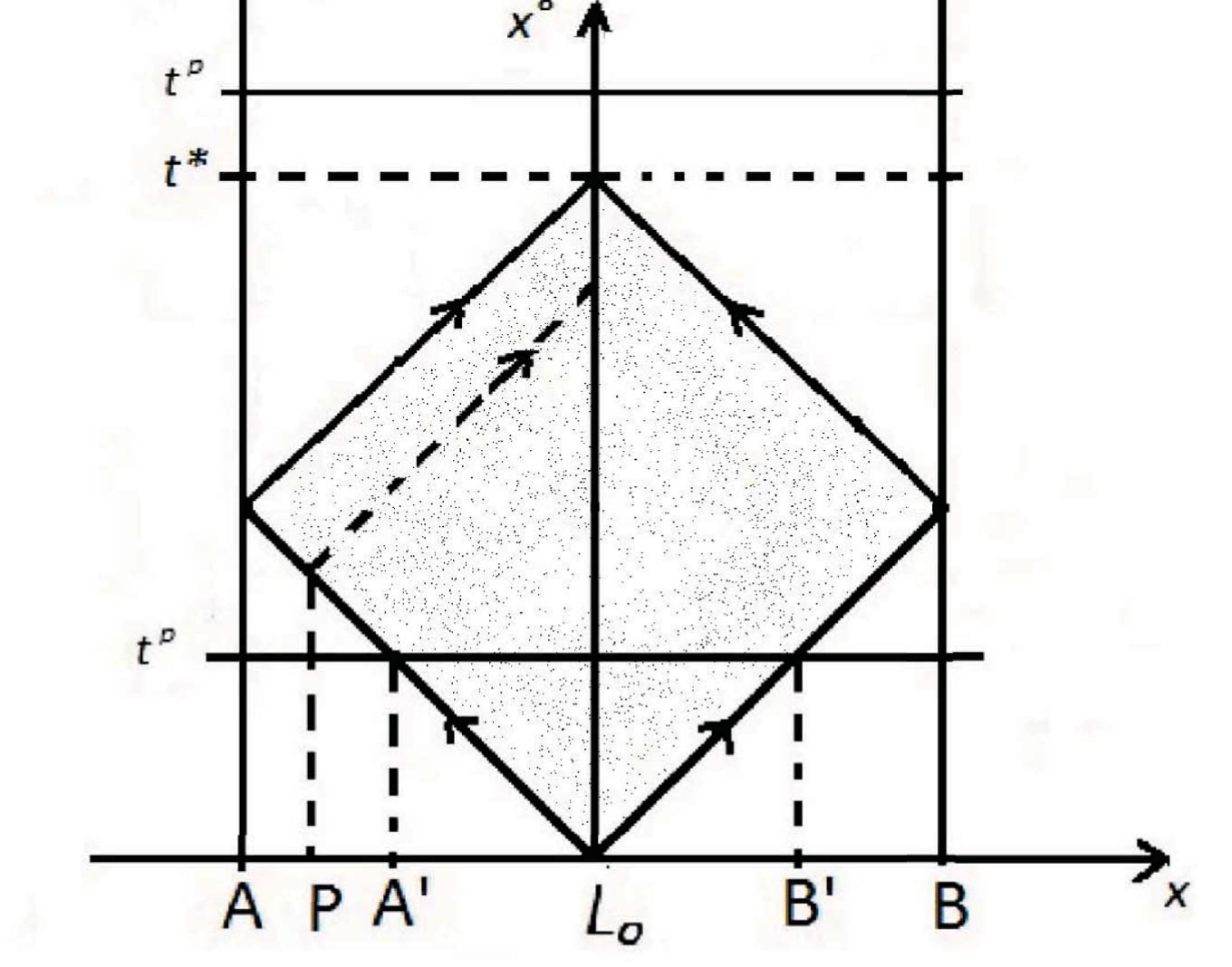}
	\caption{Double cone}
	\label{fig:08bis}
\end{figure}

Before concluding this section, we observe that the speed of signals is always limited by the speed of light $c$, and if the experimenter is positioned at $O$ as in Figure \ref{fig:08bis}, they will be able to control/operate the entire laboratory only after a time interval given by $t^*$. 
Therefore, in our ensembles, it does not make physical sense to take into account preparation times for the instruments $t_p=t^p_j-t^o_j$ lower than the time interval $t^*$ since, for example, during this time interval the instrumentation positioned at point $P$ in Figure \ref{fig:08bis} is not detected; in other words, it is as if it were not there for the purpose of the measurement. Hence, the preparation interval $t_p$ must be no less than the value $t^*$ in each N-copy of the ensemble\footnote{\label{doppio2}Our laboratory-type regions (see definition \ref{regionerettangolo}) are therefore more experimentally ductile, since the preparation time $t_p$, once assumed to be greater than or equal to $t^\ast$ in Figure \ref{fig:08bis}, is independent of the size of the laboratory, which is not the case for double cones. Obviously, for these regions we lose invariance under Lorentz transformations, a property that double cones possess.}. We emphasize that this does not mean that to perform measurements with very short times $\tau$ we need to consider laboratories and therefore instruments of minute dimensions; let us remember once again that the measurement time $\tau$ is given by \eqref{tempogrand}.

\section{External Perturbations and Reproducibility}\label{riproduzione}

Our laboratory, even if it is in optimal isolation conditions, will always be affected by external influences that are physically unavoidable and can disturb the measurement processes, especially for very long times $\tau$.

We underline that bringing the laboratory back to the same initial conditions, re-establishing \textit{all the values of the pre-measurement physical parameters, is practically unachievable}.
\\
So, how can we reconcile reproducibility with these perturbative phenomena? 
\\
Everything depends on the meaning of reproducibility, which for us means:
\begin{center}
\textit{re-establishing the $\omega$ state in each of the $N$ copies of the experiment}.
\end{center}
It follows that we assume that the following steps are carried out in the laboratory:
\begin{itemize}
\item[1.] The experimenter has the possibility of reproducing in each copy of the ensemble the same parametric state, the way of preparation and use of the laboratory equipment, while the infinite values of the physical parameters \textsl{not contained} in the set $\left\{c_i\right\}_{i\in I}$, which we consider the parametric state $\omega_c$, are not under the control of the laboratory apparatus and are free to change their values.
\item[2.] Once the copy of the experiment has been prepared and everything is ready for the measurement at time $\tau$, these \textit{perturbative phenomena are free to influence the measurement procedure} in the time interval $[t^m_j, t^p_j]$ for every $j=1,2,\ldots, n$.
\item[3.] In each of the $N$ copies of the ensemble, the perturbative phenomena occur in the same way as in the previous copies.\footnote{We do not deny that this is a non-negligible problem, since it will be necessary to equip the laboratory to measure such phenomena (and quantify them) and establish that they do not change across the ensemble.}
\end{itemize}
\begin{remark}\upshape
We note that we are not assuming that the laboratory does not interact with the external environment; we affirm that it \textit{always interacts in the same way}, in each copy of our ensemble.
\end{remark}
\begin{figure}[htbp]
	\centering
		\includegraphics[scale=0.4]{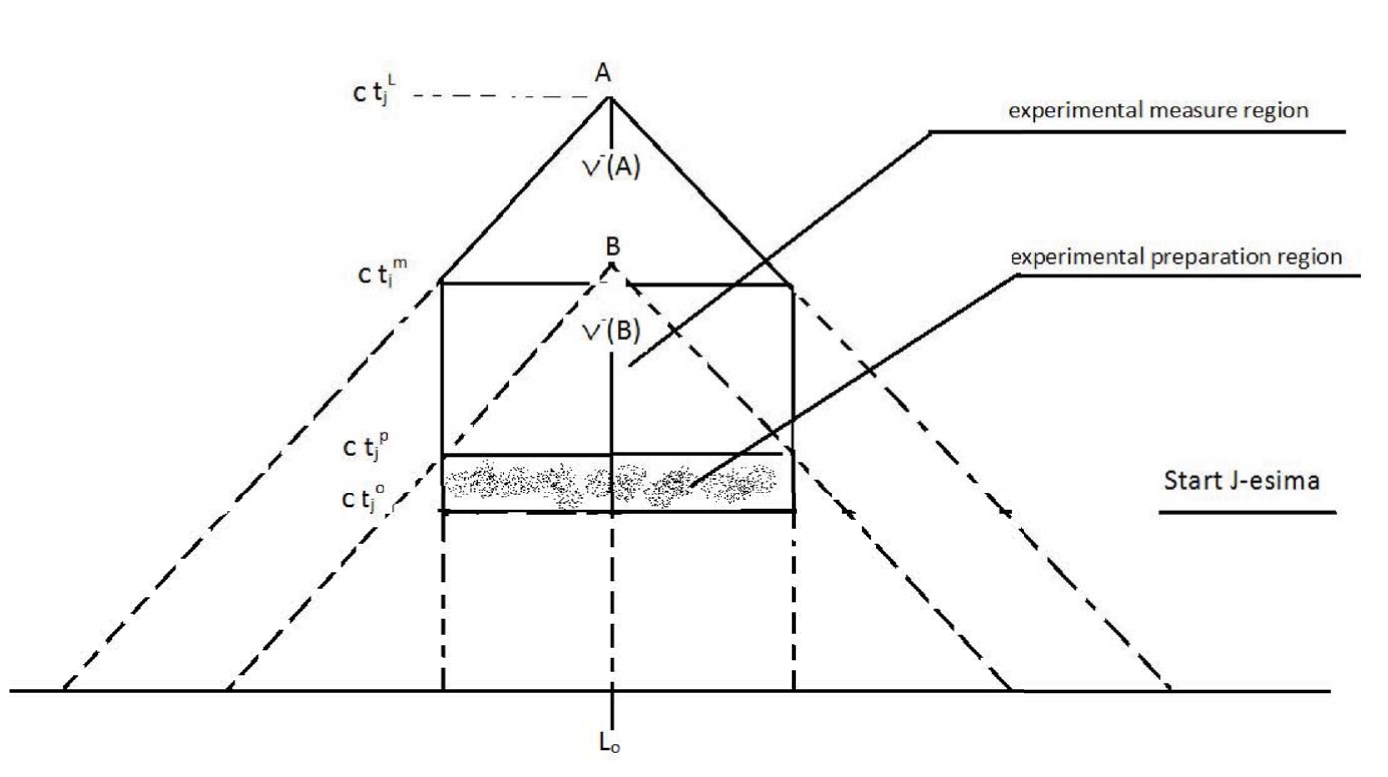}
	\caption{Perturbative region of space}
	\label{fig:05cbis}
\end{figure}
We further clarify the situation through Figure \ref{fig:05cbis}, where we have drawn only the $j$-th copy of the ensemble. We indicate with $\mathcal V^-(A)$ and $\mathcal V^-(B)$ the past light cones of the events $A$ and $B$ as shown in Figure \ref{fig:05cbis}. The entire region of space-time identified by the set $\mathcal V^-(A)$ influences the measurement operations of the $j$-th copy of our ensemble. We observe that the possible perturbations that can be created in the $\mathcal V^-(B)$ region of space-time that influence the preparation of our $j$-th copy of the experiment are recorded and modulated by the experimenter through the state of the system, as established in the first point, while the possible perturbations of the region identified by the set $\mathcal V^-(A)$ are free to influence the operation of the measurement of our observable at time $\tau = t_j^m - t_j^p$.

We made a strong hypothesis about these perturbative regions: at the time of measurement $\tau$, in each copy of the experiment, they influence the act of measurement equally. This essentially happens for small values of $t_N^m$, our measurement time of the last copy of the ensemble; obviously $t_N^m > t_{N-1}^m > \dots > t_1^m > 0$. In practice, the entire region of space-time made up of the $N$ copies of our ensemble is small enough to be subject to the same external influences.

We observe that if $\tau = 0$, the event $A$ coincides with $B$ and the perturbative region $\mathcal V^-(A)$ coincides with $\mathcal V^-(B)$, and we have no external influence that disturbs the \textit{parametric} state of the laboratory.

To sum up, we arrive at the following statement:

The regions "external" to the laboratory can influence the parametric state during the measurement operation of the physical quantity at time $\tau$, since after the preparation of the state, the system is left free to evolve until the measurement time $\tau$ of the value of the physical quantity.

\section{Correlation Between States and Observables}\label{Correlazione-stati-Osservabili}

In the previous section we introduced the observables of the system, i.e., physical quantities that can be measured in the laboratory $L_o$ at certain instants of time $\tau \geq 0$. We have described the ensemble procedures that are used for their measurement in the states that can be achieved in our laboratory:
\begin{equation}
\label{corr1}
 L_o \longrightarrow   \left\{ 
\begin{array}{cc}
\mathfrak S  &  \textit{Set of states of the system} \\ 
\mathfrak X  &  \textit{Set of observables of the system} 
\end{array}%
\right.  
\end{equation}
The pair $(\mathfrak X , \mathfrak S)$ depends on the geometry of $L_o$, since the devices and instruments, which constitute the set $\mathfrak X$, have their own spatial extension.
\\
The experimental effectiveness depends on the instruments contained in the laboratory $L_o$, on their method of use, on the various devices employed, and on the procedures adopted, which will establish the preparation time of the experiment — information all contained in the state $\omega \in \mathfrak S$.
\\
Moreover, we assume the following operational point of view:
\begin{center}
\textit{There is no physical quantity without a device suitable for measuring it}.
\end{center}
Therefore, given an observable $a \in \mathfrak X$, there must exist at least one state $\omega \in \mathfrak S$ of the system in which it can be measured.
\\
We now make a further important clarification of an experimental nature.
\begin{remark}\upshape
Given a state $\omega$ of the system, we cannot measure all observables of our physical system in the $\omega$ state.
\end{remark}
In other words, we arrange our laboratory to measure a specific physical quantity, and this preparation may not be suitable for a possible measurement of another physical quantity different from the previous one.
\begin{definition}\upshape \label{stat-oss}\index{$\mathfrak S_a$}\index{$\mathfrak{X}_\omega$}
Given an observable $a \in \mathfrak X$, we denote by $\mathfrak{S}_a$ the set of states of $\mathfrak S$ for which it is possible to experimentally compute the probability $P(a\in\Delta, \tau)_\omega$ for every time $\tau \geq 0$ and (Borel) subset $\Delta$ of $\mathbb R$.
\\
Conversely, once a state $\omega$ is fixed, we denote by $\mathfrak{X}_\omega$ the set of observables $a$ of $\mathfrak{X}$ for which this probability exists.
\end{definition} 
We observe that
\begin{equation}\label{relazionestati}
a \in \mathfrak X_\omega \ \Longleftrightarrow \ \omega \in \mathfrak S_a
\end{equation}
Moreover, a state $\omega \in \mathfrak S_a$ is said to be \textit{suitable} for the observable $a$ of the physical system, while if $a \in \mathfrak X_\omega$ it is said that $a$ is \textit{measurable in the state $\omega$}.\index{Suitable state}
\bigskip

As mentioned, physical quantities are determined using appropriate devices and instruments; they and their method of use establish what we have defined as a state of the physical system. Each physical quantity $a$ therefore corresponds to its own device/instrument which determines it, and the different measurement procedures establish the state $\omega$ of the system suitable for $a$.
\\

Please note: in this way one might think that given a state $\omega \in \mathfrak S$ there is only a single physical quantity $a$ suitable for it, since the measuring instrument used is specified in $\omega$. 

In actual fact, with the same measuring instrument and the same way of use, more physical quantities can be measured — for example, all the physical quantities derived from the physical quantity $a$. Furthermore, it is possible that a device/instrument can be equipped with multiple functions to measure more than one physical quantity simultaneously or successively (if this is possible)\footnote{In practice it is composed of several instruments and graduated scales, which can be connected to each other.}.

\begin{axiom}[\textbf{States of laboratory}]  \label{fondamentale}\index{Axiom-States of Laboratory}
For each observable $a \in \mathfrak X$ we obtain that
$$ \mathfrak S_a \subset \mathfrak S \qquad , \qquad \mathfrak S_a \neq \mathfrak S $$
in other words, there exists at least one state $\omega \in \mathfrak S$ of the laboratory system that cannot be suitable for the measurement of $a$.
\\
In a specular way, for each state $\omega \in \mathfrak S$ we obtain  
$$ \mathfrak X_\omega \subset \mathfrak X \qquad , \qquad \mathfrak X_\omega \neq \mathfrak X $$
\end{axiom}
\bigskip

We observe that only one condition has been imposed on the laboratory preparation time $t_p$, namely that it is finite. It follows that the set of states $\mathfrak S$ depends only on the geometry of the laboratory $L_o$, while we have tacitly assumed that the set $L_o \subset \mathcal E$ is \textit{connected}; this is to avoid problems that arise when considering multiple laboratories located in space, a topic that we will discuss later.

We want to point out that only after the preparation of our laboratory is it possible to establish the $\omega$ state of the system, which will be associated with the space-time region
\begin{equation} \label{regionerettangolo}
 \mathcal O = L_o \times [0, t_P] \subset \mathcal M
\end{equation}
\begin{definition}\upshape\index{$\mathfrak S (\mathcal O)$}\index{$\mathfrak X (\mathcal O)$}
We denote by $\mathfrak S(\mathcal O) \subset \mathfrak S$ the set of all states $\omega$ of the laboratory $L_o$ with preparation time given by $t_P \in \mathbb R^+$.
\end{definition}
We observe that by decreasing the number of states — and hence the number of various devices and instruments with their experimental procedures — the number of observables could also decrease. In other words, In other words, the states themselves allow observables to be identified through measurement procedures.

\begin{figure}[htbp]
	\centering
		\includegraphics[scale=0.4]{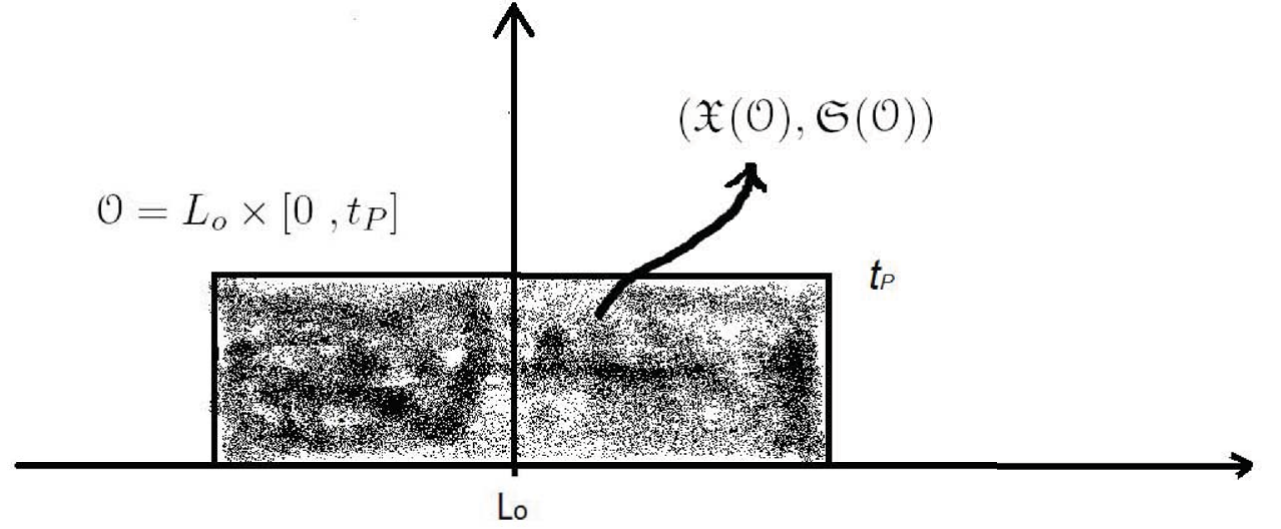}
	\caption{Laboratory-type region}
	\label{fig:regione1}
\end{figure}
\textit{We denote by $\mathfrak X(\mathcal O) \subset \mathfrak X$ the set of observables that can be measured through the set of states $\mathfrak S(\mathcal O)$}:
\begin{equation}\label{statiregionali}
\mathfrak X(\mathcal O) = \bigcup_{\omega \in \mathfrak S(\mathcal O)} \mathfrak X_\omega
\end{equation}

We note that the analogous expression for states is not true, i.e., given the set of observables $\mathfrak X(\mathcal O)$ associated with the region $\mathcal O$, we have
$$
\mathfrak S(\mathcal O) \neq \bigcup_{a \in \mathfrak X(\mathcal O)} \mathfrak S_a
$$
In other words, given an observable $a \in \mathfrak X(\mathcal O)$, there could exist a state $\omega_ \in \mathfrak S_a$ whose preparation is different from that given by the region $\mathcal O$ in \eqref{regionerettangolo}, or which cannot be prepared at all within that region.

We have a simple statement:
\begin{proposition}\upshape
Let $a \in \mathfrak X$. If $\mathfrak S(\mathcal O) \cap \mathfrak S_a \neq \emptyset$, then $a \in \mathfrak X(\mathcal O)$.
\end{proposition}
\begin{proof}
By hypothesis there exists at least one state $\omega_o \in \mathfrak S(\mathcal O) \cap \mathfrak S_a$; this means that $a \in \mathfrak X_{\omega_o}$ with $\omega_o \in \mathfrak S(\mathcal O)$ and therefore from \eqref{statiregionali} we have $a \in \mathfrak X(\mathcal O)$.
\end{proof}
From this proposition it follows that
$$ \mathfrak S(\mathcal O) \cap \mathfrak S_a \neq \emptyset \qquad \Longleftrightarrow \qquad a \in \mathfrak X(\mathcal O)$$

We employ the following notations:
\begin{itemize}
\item For every $a \in \mathfrak X$:
$$ \mathfrak S_a(\mathcal O) = \mathfrak S(\mathcal O) \cap \mathfrak S_a $$   
\item For every $\omega \in \mathfrak S$:
$$ \mathfrak X_\omega(\mathcal O) = \mathfrak X(\mathcal O) \cap \mathfrak X_\omega $$  
\end{itemize}

From \eqref{statiregionali} we obtain the following implication:
$$ \omega \in \mathfrak S(\mathcal O) \qquad \Longrightarrow \qquad \mathfrak X_\omega(\mathcal O) = \mathfrak X_\omega $$
in particular
$$ \omega \in \mathfrak S_x(\mathcal O) \qquad \Longrightarrow \qquad x \in \mathfrak X_\omega(\mathcal O) $$
Moreover
\begin{equation}\label{statiregionali2}
\mathfrak S(\mathcal O) \subset \bigcup_{a \in \mathfrak X(\mathcal O \ )} \mathfrak S_a
\end{equation} 
In fact, if $\omega \in \mathfrak S(\mathcal O)$, given any observable $x \in \mathfrak X_\omega$, then it follows that $x \in \mathfrak X(\mathcal O)$ with $\omega \in \mathfrak S_x$; hence
$$\omega \in \bigcup_{a \in \mathfrak X(\mathcal O \ )} \mathfrak S_a$$ 
\begin{definition}\upshape\index{$\Xi $}
The regions of space-time given in relation \eqref{regionerettangolo} with $L_o$ an open connected subset of $\mathcal E$, are called laboratory-type regions. The set of such regions is denoted by $\Xi$.
\end{definition}
Therefore, each laboratory-type region $\mathcal O$ is associated with a set $\mathfrak S(\mathcal O)$ of all the possible states that can be realized in it and consequently the set of all possible observables $\mathfrak X(\mathcal O)$ measurable in that region.
\\
The pair $(\mathfrak X(\mathcal O), \mathfrak S(\mathcal O))$ is called the physical system associated with the region $\mathcal O$.
\\
In this way we obtain the following correspondence:
\begin{equation}\label{regioni-sistemi}
\mathcal O \in \Xi \longmapsto (\mathfrak X(\mathcal O), \mathfrak S(\mathcal O)) 
\end{equation}
\begin{remark}\upshape\label{dopiconi}\index{Double cones}
In local quantum physics, double cones are considered as laboratory-type regions, mathematically more ductile sets, being a topological basis for space-time and invariant under Lorentz transformations.
\\
Our choice falls on the laboratory-type regions in Figure \ref{fig:regione1}, since physically, the various pieces of equipment can be positioned at every point of our laboratory during the entire preparation time, even if their identification/action by a possible experimenter at a point $O$ of the laboratory leads, as discussed in Figure \ref{fig:08bis}, to a double cone of space-time. Once our experimenter has prepared and operated the laboratory, they no longer act on it; they simply become a data collector.
\end{remark}
\subsection{Chronological state}\label{sezione_cronos}
Let us make some considerations on the definition \ref{statocronos} of chronological state of the laboratory system:
\\
The application \eqref{noevoltemp}, using the new notations, takes the form:
\begin{equation}\label{noevoltemp1}
\omega : \tau \in I \longrightarrow \omega^{(\tau)} \in \mathfrak S_a(\mathcal O), \qquad I \subset [0, \infty]
\end{equation}
this chronological state is said to be suitable for the measurement of $a$\footnote{See note \ref{map-temp}, page \pageref{map-temp}.}.
\begin{remark}\upshape\label{cronos1}
We considered an interval $I \subset [0, \infty]$\footnote{For the topological properties of $\mathbb R^* = [-\infty, \infty]$, consult Kelley's book \cite{Kelley}, page 5.} because in general it is not possible to perform the measurement experimentally for all instants of time $\tau$.
\end{remark}
Indeed, it could happen that the measurement of a value of a given physical quantity can only occur at a single instant of time $\tau$. 
\\
It is assumed that $\omega^{(\infty)}$ corresponds to not performing the measurement, after having prepared the laboratory for the measurement itself.
\begin{notation}\upshape \index{$\mathfrak S_a \vert \tau$}
We denote by $\mathfrak S_a|\tau$ the set of all states $\omega \in \mathfrak S_a$ in which the measurement is carried out at time $\tau \geq 0$.
\\
In the same way, for each laboratory-type region $\mathcal O$ we define the set $\mathfrak S_a(\mathcal O)|\tau$.
\end{notation} 
Furthermore, we assume that for every $\omega_o \in \mathfrak S_a(\mathcal O_{o})| \tau_o$ it is possible to associate a chronological state 
$$\omega : \tau \in I \longrightarrow \omega^{(\tau)} \in \mathfrak S_a(\mathcal O_{o})| \tau, \qquad \tau_o \in I \subset [0, \infty] $$
such that
$$\omega^{(\tau_o)} = \omega_o $$
with the further obvious property
\begin{equation} 
\mathfrak X_{\omega_o} = \mathfrak X_{\omega^{(\tau)}}, \qquad \forall \tau \in I
\end{equation}
We reiterate that the set $I$ could consist of the single element $I = \{\tau_o\}$ or have as a maximal interval $I = [0, \infty]$; in this last case the chronological state is called \textit{globally defined}.

\subsection{States and regions}\label{stati_regioni}

\begin{figure}[htbp]
	\centering
		\includegraphics[scale=0.6]{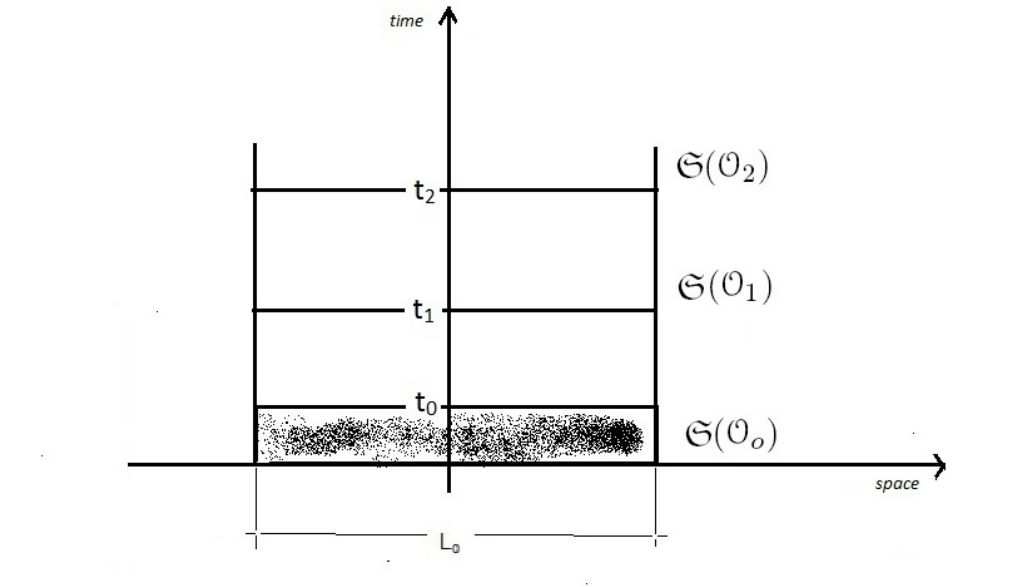}
	\caption{Laboratory with different preparation times}
	\label{fig:09ca}
\end{figure}

We consider different laboratory preparation times to carry out the experiment in our laboratory $L_o$ as in Figure \ref{fig:09ca} and we denote by 
\begin{equation}\label{markovregione_0}
 \mathcal O_k = L_o \times [0, t_k] \ , \qquad k = 0, 1, 2, \ldots, n  
\end{equation}
the related laboratory-type regions at preparation time $t_k$.
We observe that for every value of $t_k$ we have $\mathfrak S(\mathcal O_k) \subset \mathfrak S$, and one could trivially believe that $\mathfrak S(\mathcal O_0)$ is contained in $\mathfrak S(\mathcal O_1)$, but obviously this is not true. We assume that these sets satisfy the following property:
\begin{post}\label{estensione_stato}\index{Postulate-State Extension }
For every $\omega_o \in \mathfrak S(\mathcal O_0)$ there exists $\omega_1 \in \mathfrak S(\mathcal O_1)$ such that
\begin{itemize}
\item [1.] $\mathfrak X_{\omega_o}(\mathcal O_0) = \mathfrak X_{\omega_1}(\mathcal O_1) \cap \mathfrak X(\mathcal O_0)$ 
\item [2.] For each $a \in \mathfrak X_{\omega_o}(\mathcal O_0)$ we have
$$ P(a \in \Delta, \tau)_{\omega_o} = P(a \in \Delta, \tau)_{\omega_1}, \qquad \forall \tau \geq 0 $$
\end{itemize}
\end{post}
\begin{remark}\upshape\label{stessi_strumenti}
Intuitively, property \eqref{estensione_stato} tells us that for the preparation of the state $\omega_1$ we used the same tools and devices as for the state $\omega_o$, since every observable suitable for $\omega_o$ is still suitable for $\omega_1$.\footnote{In practice, once we have finished preparing the state $\omega_o$ for the measurement at time $\tau$, we wait to execute it; the operator in the laboratory \textit{preserves the preparation} of the state until time $t_1$, thus obtaining a new state $\omega_1$.}
\end{remark}
It is useful to anticipate the following notation:\index{$\mathfrak S(\mathcal O_{1} \vert \mathcal O_{0})$ }
$$ \mathfrak S(\mathcal O_1 \mid \mathcal O_0) = \left\{ \omega_1 \in \mathfrak S(\mathcal O_1) : \exists \, \omega_o \in \mathfrak S(\mathcal O_0) \text{ which satisfies the relations in property \ref{estensione_stato}} \right\}$$
Thus, by property \ref{estensione_stato}, the set $\mathfrak S(\mathcal O_1 \mid \mathcal O_0) \subset \mathfrak S(\mathcal O_1)$ is non-empty.\footnote{See proposition \ref{sottostatiprop} on page \pageref{sottostatiprop}, which generalizes this statement.}
\\

Let us provide some further clarification on the states of the system and laboratory-type subregions:
\\
\\
Let $\omega_1 \in \mathfrak S(\mathcal O_1)$ with $\mathfrak X_{\omega_1}(\mathcal O_1) \cap \mathfrak X(\mathcal O_0) \neq \emptyset$ and take any element $x \in \mathfrak X_{\omega_1}(\mathcal O_1) \cap \mathfrak X(\mathcal O_0)$. We obtain that the state $\omega_1$ is suitable for the observable $x$ with $x \in \mathfrak X(\mathcal O_0)$, and this would seem to imply that the preparation of the state $\omega_1$ is carried out in the region $\mathcal O_0$; therefore we would conclude (contradicting previous statements on the subject) that $\omega_1 \in \mathfrak S_x(\mathcal O_0)$.
\\
In reality we can only say that there exists $\omega_o \in \mathfrak S_x(\mathcal O_0)$ that satisfies the relations of property \ref{estensione_stato} and that therefore $\omega_1 \in \mathfrak S(\mathcal O_1 \mid \mathcal O_0)$.\footnote{In other words, $\omega_1$ is the state described in the note of Remark \ref{stessi_strumenti} on page \pageref{stessi_strumenti}.}

We observe that if $\omega \in \mathfrak S_a$, then there exists $t_P \geq 0$ such that $\omega \in \mathfrak S_a(\mathcal O_{t_P})$. Hence,
\begin{equation}\label{stati_lab_prep}
\mathfrak S_a = \bigcup_{t_P \geq 0} \mathfrak S_a(\mathcal O_{t_P})
\end{equation}
where $\mathcal O_{t_P}$ are regions given by relation \eqref{markovregione_0}\footnote{When we need to highlight that the states are associated with the laboratory $L_o$ in question, we write more precisely $\mathfrak S_a(L_o)$ instead of $\mathfrak S_a$, and similarly $\mathfrak X_{\omega}(L_o)$ instead of $\mathfrak X_{\omega}$.}.
\begin{remark}\upshape\label{disjoint}
By definition, to each state $\omega$ of $\mathfrak S$ there corresponds one and only one laboratory preparation time $t_P$; therefore, for every pair of laboratory preparation times $t_1, t_2$ with $t_1 \neq t_2$:
$$\omega \in  \mathfrak S(\mathcal O_{t_1}) \  \Longrightarrow  \ \omega\notin \mathfrak S(\mathcal O_{t_2})   $$
Thus the set-theoretic union in \eqref{stati_lab_prep} is a union of disjoint sets.
\end{remark}

For measurement procedures that are carried out at a time $\tau$, we have the following relations:
$$ \mathfrak S(\mathcal O_t)|\tau \subset \mathfrak S(\mathcal O_t) \ \Longrightarrow \
  \mathfrak S|\tau = \bigcup_{t \geq 0} \mathfrak S(\mathcal O_t)|\tau \subset \bigcup_{t \geq 0} \mathfrak S(\mathcal O_t) = \mathfrak S$$
and by definition
$$ \bigcup_{\tau \geq 0} \mathfrak S(\mathcal O_t)|\tau = \mathfrak S(\mathcal O_t), \qquad \forall t \geq 0 $$
since by definition, if $\omega \in \mathfrak S(\mathcal O_t)$ then there exists a unique $\tau \geq 0$ for which $\omega \in \mathfrak S(\mathcal O_t)|\tau$; the family $\{ \mathfrak S(\mathcal O_t)|\tau \}_{\tau \geq 0}$ consists of disjoint sets.
\\
We will return to the relation between states and regions in Section \ref{sottolaboratorio}.

\section{Achievable and Suitable Systems}\label{attuale}

In the laboratory the experimenter has two alternatives:
\begin{itemize}
\item [A.]  Establish, according to their experimental needs, which physical quantities of the system they want to measure and consequently choose the various instruments, devices, and methods of execution to carry out this measurement.
\item [B.]  Based on the presence of the various instruments and devices in the laboratory, choose the physical quantity $\mathfrak X_o$ to measure.
\end{itemize}
In the first case \textbf{A}, the experimenter initially fixes a set of observables $\mathfrak X_o \subset \mathfrak X$ consisting of the physical quantities that they want to measure\footnote{At this level of discussion we are still considering single measurements on observables, not simultaneous ones, nor subsequent measurements where, as one might suspect, things will be further complicated.}. 
In this way, for each observable $a \in \mathfrak X_o$ there must correspond a state $\omega$ of the system in which it will be possible to carry out a measurement; therefore the experimenter will have to choose a set of states $\mathfrak S_o$ with the following characteristic:
\begin{equation}\label{efficace1}
  \forall a \in \mathfrak X_o \qquad \Longrightarrow \qquad \mathfrak S_a^o = \mathfrak S_o \cap \mathfrak S_a \neq \emptyset
\end{equation}
To avoid an overabundance of states, and therefore to avoid having unused instruments/procedures when measuring the observables in $\mathfrak X_o$, the possible states of the system suitable for each observable $a \in \mathfrak X_o$ will consist of a subset $\mathfrak S_o$ of $\mathfrak S(\mathcal O)$, such that each $\omega \in \mathfrak S_o$ corresponds to an observable $x \in \mathfrak X_o$ with $\omega \in \mathfrak S_x$.
\\
Therefore this set of states must satisfy the following property:
\begin{equation}\label{efficace2}
 \mathfrak S_o \subset \bigcup_{x \in \mathfrak X_o} \mathfrak S_x
\end{equation}
In the second case \textbf{B}, the set of states is physically determined by the various measuring devices actually present in our laboratory and by the various experimental procedures that are adopted for their operation.
\\
In other words, the investigator will initially have a subset $\mathfrak S_o$ of $\mathfrak S(\mathcal O)$. The set of physical quantities $\mathfrak X_o$ that can be measured with this set must satisfy the following properties:
\begin{equation}\label{efficace3}
\forall \omega \in \mathfrak S_o \qquad \Longrightarrow \qquad \mathfrak X_\omega^o = \mathfrak X_o \cap \mathfrak X_\omega \neq \emptyset
\end{equation}
Furthermore, to every observable $x \in \mathfrak X_o$ there must correspond at least one state $\omega \in \mathfrak S_o$ that is suitable for $x$, so
 \begin{equation}\label{efficace4}
 \mathfrak X_o \subset \bigcup_{\omega \in \mathfrak S_o} \mathfrak X_\omega
\end{equation}
We note that \eqref{efficace1} holds if and only if \eqref{efficace4} holds, while \eqref{efficace3} holds if and only if \eqref{efficace2} holds.
\bigskip

Let us reiterate that physically, in order to know the behavior of an observable $a \in \mathfrak X_o$\footnote{This means knowing its values as the state in which the measurement is taken varies.} we have the following family of states of the system at our disposal:
\begin{equation*}
\mathfrak S_a^o = \mathfrak S_o \cap \mathfrak S_a \subset \mathfrak S_a
\end{equation*}
since we cannot have all the instruments and devices that could be implemented in the laboratory. Therefore we affirm:
\\

\textit{With $\mathfrak S_a$ we obtain the maximum degree of knowledge on the observable $a$; it contains all the possible experimental procedures that can be carried out on the observable.}
\\

In summary, given a pair $(\mathfrak X_o, \mathfrak S_o)$ consisting of a family of observables $\mathfrak X_o$ and a family of states $\mathfrak S_o$ of our physical laboratory system, to have experimental evidence it must necessarily satisfy both of the following conditions:
\begin{itemize}
\item [i.] For each element $a \in \mathfrak X_o$ there must exist at least one state $\omega \in \mathfrak S_o$ suitable for the observable $a$; in other words, \eqref{efficace4} must be satisfied.
\item [ii.] For every state $\omega \in \mathfrak S_o$ there must exist at least one observable $a \in \mathfrak X_o$ such that $\omega \in \mathfrak S_a$; therefore,  \eqref{efficace2}	 must be satisfied.
\end{itemize}
In this case we say that the pair $(\mathfrak X_o, \mathfrak S_o)$ is \textbf{physically achievable}.\index{Physically achievable system}
\bigskip

Let us now establish a family of states $\mathfrak S_o$ and consider the set consisting of \textbf{all} the observables of the system that this family allows us to measure:
\begin{equation}\index{$\mathfrak X_\lor ^{\mathfrak S_o} \ , \ \mathfrak S_\lor^ {\mathfrak X_o}$}
\label{osservabilimisurabilimax}
 \mathfrak X_\lor^{\mathfrak S_o} := \bigcup_{\omega \in \mathfrak S_o} \mathfrak X_\omega
\end{equation}
In a symmetric way, once a family of observables $\mathfrak X_o$ is fixed, we define:
\begin{equation}
\label{statimisurabilimax}
 \mathfrak S_\lor^{\mathfrak X_o} = \bigcup_{x \in \mathfrak X_o} \mathfrak S_x
\end{equation}
To summarize, given a pair $(\mathfrak X_o, \mathfrak S_o)$ consisting of a set of observables $\mathfrak X_o$ and a set of states $\mathfrak S_o$, we obtain:
\begin{equation}\label{efficace2bis}
\bigl[ \mathfrak S_a \cap \mathfrak S_o \neq \emptyset \quad \forall a \in \mathfrak X_o \bigr] \qquad \Longleftrightarrow \qquad \mathfrak X_o \subset \mathfrak X_\lor^{\mathfrak S_o} 
\end{equation}
and symmetrically
\begin{equation}\label{efficace4bis}
\bigl[ \mathfrak X_\omega \cap \mathfrak X_o \neq \emptyset \quad \forall \omega \in \mathfrak S_o \bigr] \qquad \Longleftrightarrow \qquad \mathfrak S_o \subset \mathfrak S_\lor^{\mathfrak X_o}
\end{equation}
Thus a pair $(\mathfrak X_o, \mathfrak S_o)$ to be physically achievable must necessarily satisfy the two conditions:
\begin{equation}\label{attuabilis}
(\mathfrak X_o, \mathfrak S_o) \text{ is achievable } \Longleftrightarrow \left\{ 
\begin{array}{ccc}
\mathfrak X_o \subset \mathfrak X_\lor^{\mathfrak S_o}     \\ 
  \\
\mathfrak S_o \subset \mathfrak S_\lor^{\mathfrak X_o}      
\end{array}
\right. 
\end{equation}
By \eqref{efficace4bis}, it is easily verified that given a family of states $\mathfrak S_o$, the pair $(\mathfrak X_\lor^{\mathfrak S_o}, \mathfrak S_o)$ is physically achievable.
\\
Similarly, from \eqref{efficace2bis} it is easily verified that given a family of observables $\mathfrak X_o$, the pair $(\mathfrak X_o, \mathfrak S_\lor^{\mathfrak X_o})$ is physically achievable.
\\

In Section \ref{Sistemi fisici del laboratorio} we will resume the study of physically achievable systems.
\begin{problem}\upshape \label{pb-1}
Which and how many subsets $\mathfrak S_o$ of $\mathfrak S(\mathcal O)$ determine an actually realizable physical system? Is there a method to determine them?
\end{problem}
Let us now focus on a new problem:
\\
Given a set of observables $\mathfrak X_o$ of the system, we want to find a set of states $\mathfrak S_o$ operationally effective for \textbf{all} the observables in $\mathfrak X_o$.
\\
Let us denote 
$$ \mathfrak S_\wedge^{\mathfrak X_o} := \bigcap_{a \in \mathfrak X_o} \mathfrak S_a$$
as a solution to the problem posed,  the set of states $\mathfrak S_o$ must necessarily satisfy the following relation:  
\begin{equation}
\label{adatto1}
\mathfrak S_o \subset \mathfrak S_\wedge^{\mathfrak X_o}  
\end{equation}  
In this case we say that the set $\mathfrak S_o$ is \textbf{suitable} for $\mathfrak X_o$.\index{Suitable set of states}
\begin{definition}\upshape
A pair $(\mathfrak X_o, \mathfrak S_o)$ is called suitable if \eqref{adatto1} is satisfied.
\end{definition}
Given a set of observables $\mathfrak X_o$ we denote by
\begin{equation}\label{osservabilimisurabilimin}
 \mathfrak X_\wedge^{\mathfrak S_o} := \bigcap_{\omega \in \mathfrak S_o} \mathfrak X_\omega
\end{equation}
We now have a simple statement:
\begin{proposition}\upshape\label{adattobis}
Let $\mathfrak X_o \subset \mathfrak X$ be a set of observables and $\mathfrak S_o \subset \mathfrak S$ a set of states. Then we have\footnote{Therefore a suitable pair is physically achievable.}
\begin{equation}
 \mathfrak S_o \subset \mathfrak S_\wedge^{\mathfrak X_o} \qquad \Longleftrightarrow \qquad \mathfrak X_o \subset \mathfrak X_\wedge^{\mathfrak S_o}
\end{equation}
\end{proposition}
\begin{proof}
$(\Rightarrow)$ If $a \in \mathfrak X_o$, then by hypothesis it follows that $\mathfrak S_o \subset \mathfrak S_a$; thus for every $\omega \in \mathfrak S_o$ we have $a \in \mathfrak X_\omega$, hence the thesis.
\\
$(\Leftarrow)$ If $\omega \in \mathfrak S_o$, then $\mathfrak X_o \subset \mathfrak X_\omega$; in other words, for every $a \in \mathfrak X_o$ we have $a \in \mathfrak X_\omega$. It follows that $\omega \in \mathfrak S_a$ for each $a \in \mathfrak X_o$. 
\end{proof}
We observe that if $\mathfrak Y_o \subset \mathfrak X_o$, the set $\mathfrak S_o$ is still suitable for $\mathfrak Y_o$ since
$$ \mathfrak S_o \subset \bigcap_{a \in \mathfrak X_o} \mathfrak S_a \subset \bigcap_{a \in \mathfrak Y_o} \mathfrak S_a $$ 

To summarize:
\\
In case \textbf{A}, we fix a set of observables $\mathfrak X_o$; in this way we obtain that the pair $(\mathfrak X_o, \mathfrak S_\lor^{\mathfrak X_o})$ is the maximum physically achievable, \textit{i.e.}, if $(\mathfrak X_o, \mathfrak S_o)$ is a physically achievable pair, then $\mathfrak S_o \subset \mathfrak S_\lor^{\mathfrak X_o}$. 
\\
Moreover, symmetrically, the pair $(\mathfrak X_o, \mathfrak S_\wedge^{\mathfrak X_o})$ is maximum suitable. Here too, if $(\mathfrak X_o, \mathfrak S_o)$ is a suitable pair, then we have $\mathfrak S_o \subset \mathfrak S_\wedge^{\mathfrak X_o}$.  
\\
In case \textbf{B}, we fix a set of states $\mathfrak S_o$; symmetrically, the pair $(\mathfrak X_\lor^{\mathfrak S_o}, \mathfrak S_o)$ is maximum physically achievable, and the pair $(\mathfrak X_\wedge^{\mathfrak S_o}, \mathfrak S_o)$ is maximum suitable.
\\

From Axiom \ref{fondamentale} we obtain the following:
\begin{remark}\upshape
If $\mathfrak S_o$ is the whole set of states of the laboratory system $\mathfrak S$, then we obtain 
$$ \mathfrak X_\wedge^{\mathfrak S} = \emptyset. $$
\end{remark}

\section{Detected Physical Quantities and Constants}\label{grandezze_rilevate}
Let us make some clarifications on physical quantities.
\\
An observable $a$ is a \textit{constant of the laboratory system} if there exists a real number $\texttt{r}$ such that:
$$P(a \in \{\texttt{r}\}, \tau)_\omega = 1, \qquad \forall \omega \in \mathfrak S_a$$
while an observable $a$ is said to be \textit{null} when
$$P(a \in \{0\}, \tau)_\omega = 1, \qquad \forall \omega \in \mathfrak S_a$$
\begin{notation}\upshape
To indicate that an observable is constantly equal to the real number $\texttt{r}$ in every state of the system suitable for it, we use the notation $a \subset \texttt{r} I$. In particular, we write
$$a \subset 0 \qquad \Longleftrightarrow \qquad P(a \in \{0\}, \tau)_\omega = 1, \qquad \forall \omega \in \mathfrak S_a$$
and
$$a \subset I \qquad \Longleftrightarrow \qquad P(a \in \{1\}, \tau)_\omega = 1, \qquad \forall \omega \in \mathfrak S_a$$
\end{notation} 
One could object that this notation does not take into account the units of measurement of physical quantities, but as we will see in Section \ref{assiomistatici}, we will assume for mathematical reasons\footnote{Basically, to sum non-homogeneous physical quantities to each other.} that every observable of the physical system is dimensionless.
\begin{attenzione}
This should not make us fall into the temptation of treating every real number as a physical quantity of the system.
\end{attenzione}
In other words, \textit{we will not assume} that for every real number $\texttt{r}$ there exists a number observable $r$\footnote{See definition \ref{condition1} on page \pageref{assiomistatici}.} \textit{i.e.}, an observable of the laboratory system, such that
$$P(r \in \{\texttt{r}\}, \tau)_\omega = 1, \qquad \forall \omega \in \mathfrak S$$
since $r$ should be an observable suitable for every $\omega$ state of the system:
$$\mathfrak S_r = \mathfrak S$$
and therefore free from any experimental procedure.    
\\
We will see later, through the Borel functional calculus\footnote{See Axiom \ref{assio3} on page \pageref{assiomistatici}.}, that given any observable $a$ of the laboratory system and any real number $\texttt{r}$, there exists a constant observable $r$ with $\mathfrak S_r = \mathfrak S_a$, such that $r \subset \texttt{r} I$.\footnote{Thus, with our notation,
$$P(r \in \{\texttt{r}\}, \tau)_\omega = 1, \qquad \forall \omega \in \mathfrak S_a.$$}
\begin{definition}[\textbf{Non-detected physical quantity}]\upshape\label{rilevare}\index{Not-detected Quantity}
A physical quantity $a$ is not detected at a given time $\tau$ by the laboratory instruments identified by the set of states $\mathfrak S_o \subset \mathfrak S(\mathcal O)$ where $\mathcal O = L_o \times [0, t_p]$ if we have:
$$P(a \in \{0\}, \tau)_\omega = 1, \qquad \forall \omega \in \mathfrak S_a^o$$
\end{definition}
 
Therefore, the observable $a$ is not detected if our instruments do not indicate the presence of $a$; \textit{this does not mean that $a$ is the null observable}, but only that the observable has a null value established with the devices at our disposal and the various procedures that can be carried out with the laboratory preparation time $t_p$.

\section{Transition Problems}\label{transizioni}
\begin{figure}[htbp]
	\centering
		\includegraphics[scale=0.5]{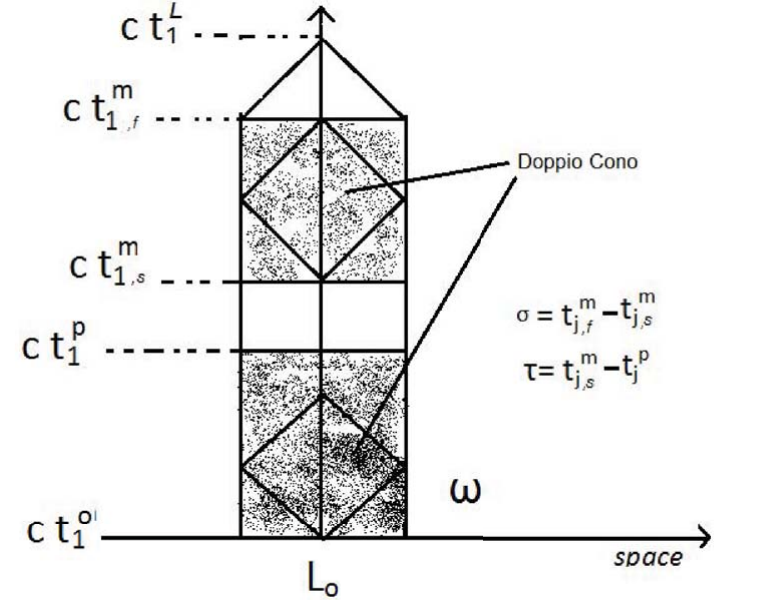}
	\caption{Measurement range}
	\label{fig:misuraintervallo}
\end{figure}

Up to this moment we have not highlighted the fact that the actual act of measurement, which is carried out after the preparation of the laboratory, does not occur instantly at the set time $t^m_j$ but will be carried out, applying the procedures contained in $\omega$, always over a time interval $\sigma > 0$.
\\ 
Since the speed of signals is always finite, this interval must contain a double cone as highlighted in Figure \ref{fig:misuraintervallo}.    
\\
During this time interval 
$$\sigma_j = t^m_{j,f} - t^m_{j,s}$$
the perturbations due to the measurement could cause not only the parametric state $\omega_c$ to change but also the setting and operation of the various apparatuses and devices in the laboratory or the source of the measurement itself. 

Indeed, after the measurement, we cannot say that the same experimental procedures for measuring $a$ contained in the initial state $\omega$ can be applied to this new experimental situation and that they are still suitable for $a$.\footnote{In other words, to still obtain a state suitable for the measurement of $a$.} Furthermore, this new mutated state could depend on the $j$-th copy of the ensemble.
\\
Therefore, from the $N$ copies of our ensemble, after the measurement at time $\tau$, we obtain a set of mutated parametric states which we will denote by 
$$\left\{ \omega_{c,1}', \omega_{c,2}', \ldots, \omega_{c,N}' \right\}$$
which we do not know and which a priori could be different. 
\\
The problem we now have is to establish the link between the mutated parametric states $\omega_{c,k}'$ and the initial parametric state $\omega_c$:
$$ \omega_{\tau,c} \longrightarrow \omega_{j,c}' \ , \qquad j = 1, 2, \ldots, N $$
\textit{Assuming that the experimenter, after having carried out the measurement, is able to recognize whether two states are equal to each other}, we can determine the following relative frequencies:
\begin{equation}
\label{freqtransizione}
 f(\omega_c \rightarrow \omega_c')_a = \frac{\text{number of times that we obtain the state } \omega_c'}{\text{total number of measurements carried out on } a}
\end{equation} 
and in this way obtain the \textit{transition probability} of the parametric state $\omega_c$ passing into the parametric state $\omega_c'$ in the measurement of $a$
\begin{equation}
\label{probtransizione}
\operatorname{Prob}(\omega_c \rightarrow \omega_c')_a
\end{equation}   
The problem we need to solve is that of determining methodologies to characterize the mutated states $\omega_{c,k}'$. In practice, to carry out the measurement we prepare the laboratory as summarized in Remark \ref{statoindividuazione}.  
\\
We remark that we know the parametric state of the system before carrying out the experiment, which is not the case for the changed state that we obtain after perturbing it with the act of measurement. On this issue, the following statement by von Neumann is useful:
\\
\textit{(We) can learn something about the state of $\omega_{c,k}'$ only from the results of measurements.}\footnote{Cf. von Neumann \cite{Neu_32}, p. 337, where instead of our $\omega_{c,k}'$ the state is denoted by \textbf{S}.} \index{von Neumann}
\\
An index of the change in the state after the measurement could be highlighted by carrying out repeated measurements of the same observable in rapid succession (when this is experimentally possible).
\\

In conclusion, in Section~\ref{riproduzione} we saw that external perturbations are unavoidable in any real laboratory and that, even under the assumption that they act identically on each copy of the ensemble, we cannot fully control or manage them during the measurement process. We argue that this inability is not simply a practical drawback, but rather a structural feature of any real laboratory. This characteristic may be the implicit assumption underlying many open problems in the foundations of physics.

\chapter{Preparations and Conditioning} 
In this section we consider the measurement of two or more observables of the laboratory system, studying when it is operationally possible to measure them simultaneously and subsequently.

\section{Jointly Prepareable Observables}   
The experimenter can decide on the basis of experimental considerations the simultaneous or subsequent measurement of two or more physical quantities to be carried out in his laboratory $L_o$. We assume that it is able to distinguish an \textit{earlier} and a \textit{later} in the actions  it performs\footnote{By virtue of the presence of a clock integral with the laboratory that marks the measurement of time.} and  that it is able to carry out the preparations only \textit{in subsequent order} in the laboratory\footnote{Otherwise   one  will have to divide the laboratory into two (or more) separate parts, obtaining two distinct laboratories with the related problems of mutual influence, for example $L_o$ and $L_1$ as in Figure \ref{fig:02}.
\\
\textbf{Warning}: In the real world, the laboratory consists of a room surrounded  by walls delimiting its extent; mathematically it is a connected region $V$ of $\mathbb R^3$.  
\\
In it there may be one or more subjects  preparing and carrying out  various experiments,  whom  \textit{we will denote by the generic name of operators or experimenters}.
\\
The operators in this room  can carry out their experiment autonomously and simultaneously to make two measurements on different physical quantities. Now, this statement may seem to be in contrast with what we have just said above, but it is not so.
\\
In fact, even if the two operators perform the same experiments in the same room, it can never happen that they are in the same place in the room at the same time (you cannot screw two screws simultaneously into the same hole).   
\\
Therefore, the laboratory of operator A, denoted by $L_A$, is  the locus of points of $V \subset \mathbb R^3$ where only operator A carries out the experiment. This set is a connected set of $\mathbb R^3$ (the operator does not dissolve into nothingness and then reappear again).
\\
So, if we have only two operators A and B that act on the room  we obtain  two sets $L_A$ and $L_B$ that can also be non-disjoint.
\\
Let $x^A(t)$ and $x^B(t)$  be  the coordinates of A and B in the laboratory; $x^A(t), x^B(t)\in V$ at  time  $t$.
\\
 We obtain 
$$ V_A =\left\{ (x^A(t),t) : t\in [0, t_p] \right\}   \qquad , \qquad V_B =\left\{ (x^B(t),t) : t\in [0, t_p] \right\}$$ 
 are  disjoint sets of $\mathbb R^4$.
\\
We underline that 
$$V_A=L_A \times [0, t_p] \qquad , \qquad V_B= L_B \times [0, t_p] $$ 
 whether  the two operators, in the laboratory, spend the same amount of time.}.  
\\

This hypothesis is in agreement with Einstein's thoughts \cite{Einstein45}\footnote{See also Section \ref{rif_pb} on page \pageref{rif_pb}.}:\index{Einstein}
\begin{citazione}\label{Ein45}
The experiences of an individual appear to us arranged in a series of events; in this series the single events which we remember appear to be ordered according to the criterion of "earlier" and "later," which cannot be analysed further. 
\\
There exists, therefore, for the individual, a  one-time, or subjective time. This in itself is not measurable. I can, indeed, associate numbers with the events,
in such a way that a greater number is associated with the later event than with an earlier one; but the nature of this association may be quite arbitrary. This association I can define by means of a clock by comparing the order of events furnished by the clock with the order of the given series
of events. We understand by a clock something which provides a series of events which can be counted, and which has other properties of which we shall speak later.
\end{citazione}
For example, if we have to establish the values of two quantities $a$ and $b$, the experimenter must prepare the instrumentation for $a$ and then for $b$ or vice-versa in  its  laboratory and then proceed to measure both quantities and  obtain  their value at time $\tau$.
\\
We observe that not always (even at a macroscopic level)  can two physical quantities $a$ and $b$ that can be measured individually in the laboratory be operationally measurable in succession or simultaneously\footnote{See the preparations given in the example \ref{prepcong2} in section \ref{esempi}}. 
\\
Indeed, it is possible that the preparation of the laboratory for the measurement of the first physical quantity $a$ can negatively influence the subsequent preparation of the measurement of the second physical quantity $b$ or vice-versa,  or  that the preparation of $b$ destroys the possible information that we can obtain on $a$.
\\
We must make a clarification: for the simultaneous measurement of $a$ and $b$ the two preparations may not be totally distinct. For example, it could happen that instruments or devices suitable for the measurement of $b$ are also used for the preparation of $a$; therefore during the preparation of $a$ it could happen that   it  also prepares a part of the experiment that is used for the measurement of $b$ (therefore the measurement of $a$ is interrupted,  it  starts with that of $b$ and  returns  again to the preparation of $a$, etc...). This is obviously always part of the preparation of $a$ which takes place in time $t^p_a$, after which we continue with the actual preparation of $b$ with time $t^p_b$ which ends when everything is ready for its measurement.
\\
Therefore \textit{after} having prepared $a$ before $b$ (if it is possible), we can choose to first measure $a$ at time $\tau_a$ and $b$ at time $\tau_b$ or vice-versa, or carry out a simultaneous measurement (if this is possible) of both at the time $\tau_a=\tau_b$.
\\
We must emphasize that preparing for the simultaneous measurement of two observables may be different from preparing for the subsequent measurement of the two observables\footnote{So  the  laboratory preparation time might also be different.}, therefore the two cases must be distinguished.
\\
In both cases the observables $a$ and $b$ will be said to be \textbf{jointly prepared} (for a simultaneous measurement or for a possible measurement in succession), in the order of preparation that occurred.\index{Jointly prepared observables in a laboratory}
\begin{notation}\upshape
We will use the following notations\index{$a:b$} \index{$a<b$} 
\begin{itemize}
\item $a:b$  for the joint preparation of $a$ and $b$ for their simultaneous measurement where  we prepare  $a$ before $b$.
\item  $a<b$ for the joint preparation of $a$ and $b$ for their subsequent measurement where  we prepare  $a$ before $b$ and measure first $a$ and then $b$.
\item $a>b$  for the joint preparation of $a$ and $b$ for their subsequent measurement where  we prepare  $a$ before $b$ and measure first $b$ and then $a$\footnote{Warning: Unlike arithmetic, here the two writings $a>b$ and $b<a$ obviously have a different meaning.}.
\end{itemize}
\end{notation} 
\textit{We underline that the possibility of jointly preparing two observables for their measurement in succession does not guarantee that the two observables can be jointly prepared for their simultaneous measurement and vice-versa.}

\section{Simultaneous Measurements}\label{misuresimultanee} 
Let's  study  at the \textit{ensemble} level what we mean by the statement: simultaneous measurement of two observables of the system.
\\
The preparation $a:b$ is  carried  out in the same way in all the $N$-copies of our ensemble.
\\
Obviously also in this case the measurement always requires a time interval $[0, t_p]$ and we assume that their value is established simultaneously (\textit{with respect to the laboratory we take as reference}) at the time $t_j^m$ from the device \textsl{D} and read by our experimenter at the time $t^L_j$:
\begin{equation}
 \tau=t_j^m- t^p_{j,b} \qquad , \qquad j=1,2\ldots N
\label{tau2}
\end{equation}
\begin{figure}
 \centering
		\includegraphics[scale=0.5]{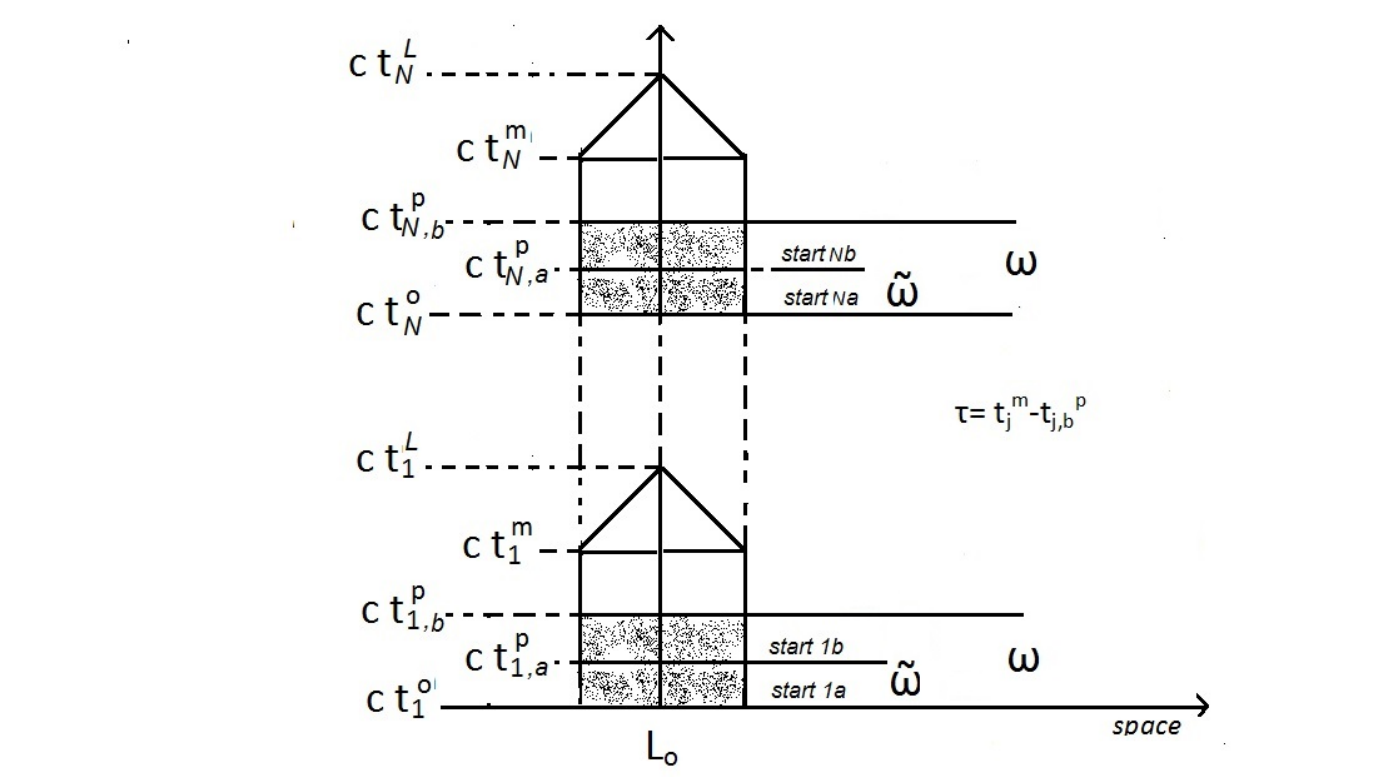}
	\caption{Simultaneous measurements - Ensemble}
	\label{fig:03}
\end{figure} 
Having simultaneous measurability means the possibility of obtaining, for each (Borel) subset $\Delta_0 \ , \Delta_1$ of $\mathbb R$ the following joint relative frequencies\footnote{The notation $(a\in\Delta_0 : b\ in\Delta_1)$ indicates that the measurements are  carried  out at the same instant of time but the preparation of $a$ came before $b$.}:
 \begin{equation}\label{frequenzecongiunte}
 f(a\in\Delta_0 : b\in\Delta_1)_\omega = \frac{n(a\in\Delta_0  : b\in\Delta_1)_\omega}{\textit{Total number of measurements  carried  out}}
\end{equation}
where with 
$$n(a\in\Delta_0 : b\in\Delta_1)_\omega$$
 we have indicated the number of times that $a$ and $b$ take on values in $\Delta_0$ and $\Delta_1$ respectively and therefore   we  obtain the joint probabilities
$$P(a\in\Delta_0 : b\in\Delta_1 , \tau )_\omega $$
at the time $\tau$ given by \eqref{tau2}. \index{$P(a\in\Delta_0 : b\in\Delta_1 , \tau )_\omega $} 
\\
In Figure \ref{fig:03} we have drawn the case of $N-$copies of the experiment to be  carried  out, where with $t^p_{j,b}$ and $t^p_{j,a}$ we have indicated the preparation time for the measurement of the quantities $a$ and $b$ respectively and $ t_j^m$ the time at which their measurement takes place, times obtained from our laboratory clock.
\begin{remark} \upshape
To prepare the simultaneous measurement of $a$ and $b$, we start from an initial parametric state $\omega^*_{c,j}$ of the laboratory, to arrive at an intermediate state $\widetilde\omega\in \mathfrak S_a$, after which starting from this state, it is necessary to obtain a final state $\omega$ that belongs to the set $\mathfrak S_b$ and which preserves in it the preparation of observable $a$ contained in $\widetilde\omega$, a requirement that is not always fulfilled. 
\end{remark}
So we have the following scheme: 
$$  \omega^\star_{c,j} \underbrace{\longmapsto}_{preparation}   \widetilde\omega\in\mathfrak S_a  \underbrace{\longmapsto}_{preparation}   \omega\in\mathfrak S_b\ \underbrace{\longmapsto}_{measure} \omega'_{c,j}  $$
Since the preparation of the observable $a$ must occur in the same way in each copy of the ensemble, the state $\widetilde\omega\in\mathfrak S_a$ relating to the first preparation must be the same in all $N$ copies of the ensemble.
\begin{notation}\upshape\index{$\mathfrak S_{a:b}$ }
We denote by $\mathfrak S_{a:b}$ the set of the states of $\mathfrak S$ for which it is possible to prepare the observables $a$ and $b$ in our laboratory to \textbf{measure them simultaneously}, where obviously we prepared $a$ before $b$.
\end{notation}
\begin{definition}[\textbf{Simultaneous Preparation}]\upshape
The observables $a$ and $b$ are said to be simultaneously prepared \textit{in the order} $a:b$, if 
$$\mathfrak S_{a:b} \ne \emptyset$$
\end{definition}
Repeating the same considerations of section \ref{mappastatuale}, we extend the definition \ref{statocronos} to the case of the preparation of $a:b$: 
\begin{post}\label{mappastato2}{Postulate-Measurement Time}
We assume that in the experimental procedures that are  carried  out in the laboratory to establish the state of the system, which require a preparation time 
$$t_{p,a}=t^p_{j,a}- t^o_j \qquad  and  \qquad t_{p, b}=t^p_{j,b}- t^p_{j,a} \  for    j=1,2 \ldots N$$
 are  included also the instruction to measure the value of observables $a$ and $b$ at each time  
$$\tau= t^m_j - t^p_{j,b} \geq 0 \ , \qquad j=1,2 \ldots N$$
\end{post}
In this case we get an application of the type  \eqref{noevoltemp}:
\begin{equation}\label{noevoltemp2}
\omega \ :  \tau\in I \longrightarrow \omega^{(\tau)} \in \mathfrak S_{a:b} \qquad , \qquad I\subset [0,\infty]
\end{equation}
\begin{definition}[\textbf{Chronological State}]\label{statocronos3}\upshape
The application  \eqref{noevoltemp2}  is called the chronological state of the laboratory system suitable for the simultaneous measurement of $a:b$.
\end{definition}
\textit{The same considerations on the chronological  state  reported immediately after the definition \ref{statocronos}  also apply  in this case.}

\subsubsection{The Joint Distribution Law}
\begin{figure}[htbp]
	\centering
		\includegraphics[scale=0.3]{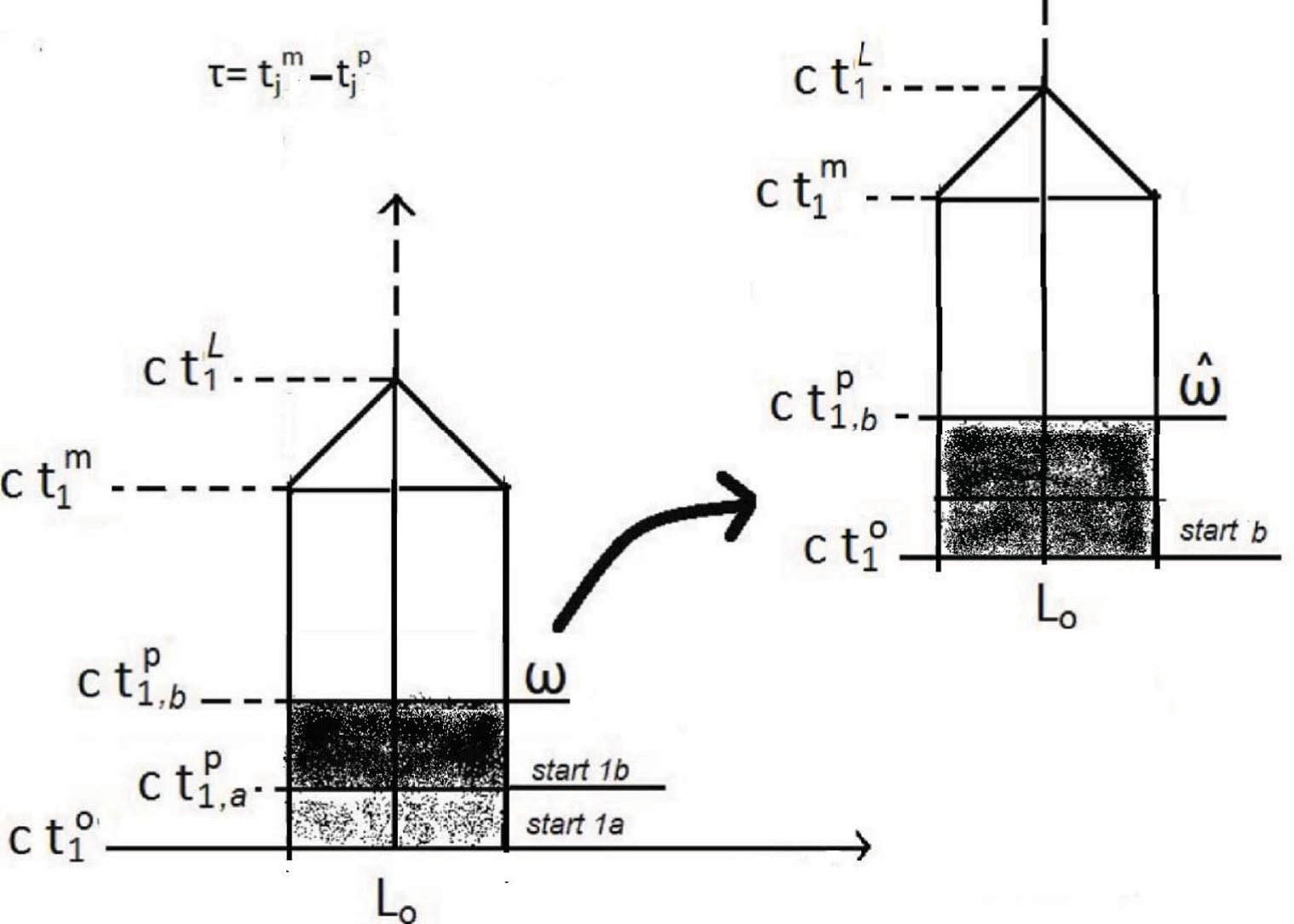}
	\caption{From double to single value}
	\label{fig:04_b}
\end{figure}

We assume that $a$ and $b$ are simultaneously  preparable  in the order $a:b$; for each $\omega\in\mathfrak S_{a:b}$ we define the joint distribution law\footnote{It is useful to note that the following writing $P(a\in\Delta_0 : a\in\Delta_1, \tau )_\omega$ does not make much experimental sense, since it is equivalent to determining the probability law $ P(a\in\Delta_0\cap \Delta_1, \tau )_\omega$.}:
\begin{equation}
\label{leggecongiunta} 
\tau\in \mathbb R^+ \longrightarrow P(a\in\Delta_0 : b\in\Delta_1  \ , \tau)_\omega 
\end{equation}
Let us now ask ourselves what relation exists between the sets $\mathfrak S_{a:b}$, $\mathfrak S_{a}$ and $\mathfrak S_{b}$.
\\
In the state $\omega\in\mathfrak S_{a:b}$, in addition to containing the information that the preparation occurred in the order $a:b$, we also have the instruction of the simultaneous measurement of the two observables at time $\tau$.
\\
Since $a$ was prepared before $b$, we can consider the state $\omega\in\mathfrak S_{a:b}$  as a suitable state for the single measurement of $b$ at time $\tau$.
\\
Practically it is as if in its preparation we also count the time necessary for the preparation of $a$, as shown in Figure \ref{fig:04_b} and the instruction that its measurement is  carried  out simultaneously with $a$.
\\
Therefore we can say that:
$$ \omega \in \mathfrak S_{b}(\mathcal O_o) \ , \qquad \mathcal O_o =L_o \times [0 \ ,t^p_b] $$
Furthermore we will adopt the notation: 
\begin{equation} \label{congiunto2b}
P(b\in\Delta , \tau)_\omega  : =  P(a\in\mathbb R   :   b\in\Delta , \tau)_\omega    
\end{equation} 
which highlights the value of the single observable $b$ that we measured in the $\omega$ state.
\\

The relation \eqref{congiunto2b} at the frequency level,  is  given in \eqref{frequenzecongiunte} and the value 
$$n(a\in\mathbb R : b\in\Delta , \tau)_\omega$$
is the number of times that $b\in\Delta$ when $a$ takes on any value after its measurement occurred simultaneously with the measurement of $b$,  an  instruction contained in $\omega$, therefore
\begin{post}[States of Simultaneous Preparation]  \index{Postulates on States of Simultaneous Preparation} 
If $a$ and $b$ are simultaneously measurable when $b$ is prepared before $a$, we have: 
\begin{equation}\label{congiunzionestati}
\mathfrak S_{a:b}\subset\mathfrak S_{b} 
\end{equation}
\end{post}
\begin{attenzione}\upshape
One might think of implementing the same considerations for the observable $a$, but this is devoid of physical meaning since the preparation of $a$ ends when we start the preparation of $b$.
\\
As we have said several times, the experimenter knows how to distinguish in  its  actions, which take place in the laboratory, \textit{a before and an after} and  it  performs a single action at a time in the laboratory\footnote{Obviously there could exist a state $\omega_o\in\mathfrak S_a$ such that
$$ P(a \in\Delta : b \in\mathbb R, \tau)_\omega = P(a\in\Delta , \tau)_{\omega_o} $$
but $\omega_o$ may have nothing in common (preparation, devices, etc..) with our $\omega\in \mathfrak S_{a:b}$. }.
\end{attenzione}

Let us remember that when the experimenter is preparing the measurement of the observable $b$,  carried  out after the preparation for the measurement of the observable $a$,  it  must not modify (or destroy) this preparation; otherwise, by definition $b$ cannot be simultaneously prepared with $a$  but integrated into the preparation for the measurement of the observable $b$, which could make the state $\omega\in \mathfrak S_{a:b}$ no longer suitable for the sole measurement of $a$.

\begin{remark}\upshape
If $a$ and $b$ are simultaneously measurable when $b$ is prepared before $a$, then it is not necessarily the case that they are still measurable when $a$ is prepared before $b$. 
\\
In other words, \textit{it is not forbidden to obtain $\mathfrak S_{a:b} \neq \emptyset$ with $\mathfrak S_{b:a}=\emptyset$ (or vice-versa).}
\end{remark}
We have another fundamental 
\begin{remark}\upshape
It could happen that $\mathfrak S_{a}\cap\mathfrak S_{b} \neq \emptyset$ but these states, even if they are suitable for the measurement of $a$ and $b$, are not necessarily suitable for their simultaneous measurement.
\end{remark}
\noindent
Let us make the following note on the simultaneous measurement of two observables and on their joint preparation:
\\
Joint preparation refers to the setup of the laboratory in order to carry out the measurements in the required order.
\\
Simultaneous measurement refers to the actual act of measuring in the order established by the preparation, at a specific time determined by the preparation.
\\
\textit{Therefore, a simultaneous measurement occurs if and only if  it is possible to perform a joint preparation (always in the same order).}
\\
\begin{attenzione}
speaking of simultaneous observables is equivalent to speaking of jointly preparable observables (always in the same measurement order), even though the two concepts refer to two experimentally distinct actions.
\end{attenzione}
To summarize:
\begin{itemize}
\item  Joint preparation (setup): A preliminary act of setting up the experimental apparatus. It is a condition of possibility.
\item Simultaneous measurement (execution): The subsequent and single act of data acquisition, made possible by that preparation.
\end{itemize}
We underline that two observables measured simultaneously in a state of the system can have two well-defined values:
\begin{definition}\upshape\label{determinati}\index{Well determined observable}
Let $a,b$ be simultaneously measurable in the order $a:b$ and $\omega\in \mathfrak S_{a:b}$.
If there exist $r_1 , r_2\in\mathbb R$ such that 
$$ P(a\in \left\{r_1 \right\}   :   b\in \left\{r_2 \right\}   , \tau)_\omega =1 $$
 the two observables are said to be  \textbf{well determined} in the state $\omega$ at time $\tau$.
\end{definition} 
\begin{remark}\upshape\label{determinati2}
In our case, saying that the values of two observables cannot be determined simultaneously with absolute precision does not mean that they are not measurable simultaneously, but it implies that the two observables are not well determined.
\end{remark}

\subsection{Complementarity and Compatibility}\label{sez_noncomplementari}
Historically, two observables that are not simultaneously measurable \textit{in every state of the system} are  said  to be \textbf{complementary}.
\\
To be more precise:
\begin{definition}\upshape 
The observables $a$ and $b$ are complementary if we have
\begin{equation}
\label{complementari}
\mathfrak S_{a:b} = \emptyset \qquad \textit{e  } \qquad \mathfrak S_{b:a} =\emptyset
\end{equation}
\end{definition}
Obviously from the negation of this statement we arrive at the notion of non-complementarity:
\\
The observables $a$ and $b$ are \textbf{non-complementary} if 
\begin{equation} \label{noncomplementari}
\mathfrak S_{a:b}\neq \emptyset \qquad \textit{o} \qquad \mathfrak S_{b:a} \neq \emptyset
\end{equation}
\textit{We underline that by non-complementarity  one  does  not  deduce that the set of states $\mathfrak S_{a:b}$ coincides with $\mathfrak S_{b:a}$. }
\\ 
We have the following definition that is found in Accardi \cite{Accardi75}:
\begin{definition}[\textbf{Weak Heisenberg Principle}]\upshape\label{Heisenberg}\index{Weak Heisenberg principle}
A physical system satisfies the (weak) Heisenberg principle if it admits complementary observables.
\end{definition}
Now we give the notion of compatibility of two or more observables. \index{Compatibility}
\\
We explicitly warn that the notion we will give is slightly different from the one we find in many physics texts, where it is formulated in the following way\footnote{See for example Accardi \cite{Accardi88}.}:
\\

\textit{Two observables are compatible if they can be measured simultaneously with arbitrary precision}\footnote{See definition \ref{determinati} and remark \ref{determinati2}.}.
\\

In these notes, compatible observables are observables that can be measured simultaneously (relative to our laboratory clock)  and  whose measurement does not depend on their order of preparation to carry out the experiment itself.
\\
The fact that the preparation procedures of the two observables are interchangeable gives us hope that their preparations have no experimental influence on their measurement;  this  leads us to the following definition of compatibility: 
\begin{definition}[\textbf{Compatibility}]\upshape\label{definizionecompatibile}
Two observables $a$ and $b$ are said to be compatible at  time  $\tau\in\mathbb R^+$ if the following properties are satisfied:
\begin{itemize}
\item [1.] They are not complementary.
\item [2.] The set of states $\mathfrak S_{a:b}$ coincides with $\mathfrak S_{b:a}$.
\item [3.] For every $\omega\in\mathfrak S_{a:b}$ and for every  pair of  Borel subsets $\Delta_0 , \Delta_1$ of $\mathbb R$ we have:
\begin{equation} \label{congiunto0}
P(a\in\Delta_0 : b\in\Delta_1  \ , \tau)_\omega = P(b\in\Delta_1 : a\in\Delta_0  \ , \tau)_\omega 
\end{equation}
\end{itemize}
\end{definition}

An  obvious consequence of the definition of compatibility is the following:
\begin{proposition}\label{congiuntomarginale}\upshape
If two observables $a$ and $b$ are compatible then
$$\mathfrak S_{a:b}=\mathfrak S_{b:a}\subset \mathfrak S_{a} \cap  \mathfrak S_{b}$$
and for the marginal distributions we have
\begin{equation}
 P(a\in\Delta   :   b\in\mathbb R , \tau)_\omega   = P(a\in\Delta,\tau)_\omega  \ , \qquad \forall \omega\in \mathfrak S_{a:b}
\end{equation}
\begin{equation}
 P(a\in \mathbb R  :   b\in\Delta , \tau)_\omega   = P(b\in\Delta,\tau)_\omega   \ , \qquad \forall \omega\in \mathfrak S_{a:b} 
\end{equation}
\end{proposition}
\begin{remark}\upshape
Compatibility between observables is not an equivalence relation in $\mathfrak X$ since it is not transitive.
\end{remark}
We have another fundamental definition:
\begin{definition}\upshape
Two observables $a$ and $b$ are said to be \textit{independent} in the state $\omega\in \mathfrak S_{a:b}$ at time $\tau$ if the following properties are satisfied:
\begin{itemize}
\item they are compatible in this state. 
\item for every  pair of  Borel subsets $\Delta_0 , \Delta_1$ of $\mathbb R$ we have
\begin{equation}\label{indip}
P(a\in\Delta_0 : b\in\Delta_1  \ , \tau)_\omega =P(a\in\Delta_0  \ , \tau)_\omega \cdot\ P(b\in\Delta_1  \ , \tau)_\omega 
\end{equation}
\end{itemize}
\end{definition}
Obviously if the observables are independent, then by \eqref{congiunto0}  and  \eqref{indip}  we obtain the expression of the marginal distributions given in proposition \ref{congiuntomarginale}. 
\\
We highlight that in our case not all observables compatible with each other are independent.
\linebreak

In the definition of jointly  preparable  observable for their simultaneous measurement in the order $a:b$, it could happen that in  \eqref{congiunzionestati}  we obtain  
  $$\mathfrak S_{a:b} = \mathfrak S_b$$
i.e. that in each state $\omega$ of $\mathfrak S_b$ it is possible to jointly prepare $a:b$.
\\
In this way $a$ and $b$ are said to be strongly jointly prepared.
\begin{definition}[\textbf{Strong Compatibility}]\upshape
Let $a$ and $b$ be compatible observables with
$$\mathfrak S_{a:b} = \mathfrak S_b \qquad , \qquad \mathfrak S_{b:a} = \mathfrak S_a$$
 They  are called  strongly compatible \footnote{In this way we have 
$$\mathfrak S_{a} = \mathfrak S_b$$}.
\end{definition}
If we have a \textit{distinct}\footnote{It makes no experimental sense to repeat the same observable, for example to write $a:b:c:a$ etc...} family of observables of the system $\left\{a_1,a_2 , \ldots a_n \right\}$ which  are jointly prepared in the order $a_1:a_2:\ldots a_n$ then we obtain for the set of states the extension of the relation \eqref{congiunzionestati}:
\begin{equation}\label{congiunzionestati2}
\mathfrak S_{a_1:a_2  :\cdots:   a_n} \subset  \mathfrak S_{ a_2  :\cdots:   a_n} \subset \cdots \subset  \mathfrak S_{a_{n-1}  : a_n}\subset \mathfrak S_{ a_n}
\end{equation} 

We now extend the definition of compatibility to a family of observables of the system $\left\{a_1,a_2 , \ldots a_n \right\}$ (at a fixed time $\tau$); these must satisfy the following properties:
\begin{itemize}
\item [A.] All observables $a_i$ and $a_j$ are non-complementary for all $i,j=1,2 \ldots n $.
\item [B.] The set of states $\mathfrak S_{a_1:a_2 :\cdots: a_n}$ coincides with $\mathfrak S_{a_{p(1)}:a_{p(2)} :\cdots: a_{p( n)}}$ for every permutation $p\in S(n)$ with
$$  \mathfrak S_{a_1:a_2  :\cdots:   a_n} \subset \bigcap_{i=1}^n\mathfrak S_{a_i} $$ 
\item [C.] Exchangeability of preparations:
\\
for every $\omega\in\mathfrak S_{a_1:a_2  :\cdots:   a_n}$ and for each family of Borel sets $\left\{ \Delta_i \right\}_{i=1,\ldots n}$  of  $\mathbb R$ we have:
\begin{eqnarray*}
 P(a_1\in\Delta_1 :   :\cdots:a_n\in \Delta_n)_\omega  = P( a_{p(1)}\in\Delta_{p(1)} : :\cdots:  a_{p(n)}\in\Delta_{p(n)})_\omega 
\end{eqnarray*}
\item [D.] Marginal Distributions:
\\
If $\Delta_{i_\alpha} =\mathbb R$ for each $\alpha=1,2 \ldots m <n$ , then for each $\omega\in\mathfrak S_{a_1:a_2  :\cdots:   a_n}$ we have
\begin{eqnarray*}
 P(a_1\in\Delta_1 :  :\cdots:a_n\in \Delta_n)_\omega  =  P( a_{q(1)}\in\Delta_{q(1)} :   :\cdots:  a_{q(n-m)}\in\Delta_{q(n-m)})_\omega   
\end{eqnarray*}  
where $q(1),q(2) \ldots q(n-m) \notin \left\{ i_\alpha \right\}_{\alpha=1,2 \ldots m }$. 
\end{itemize}
\section{Subsequent Measurements}\label{misuresuccessive}
In this section we will analyze the meaning of subsequent measurements on observables at different instants of time; fundamental is the fact that \textit{the laboratory is not rearranged after the first measurement}.
\\
Let's consider two observables $a$ and $b$ of the physical system that we want to measure in succession where  we want to measure $a$ first and $b$ after. 
\\
\begin{figure}[htbp]
	\centering
		\includegraphics[scale=0.5]{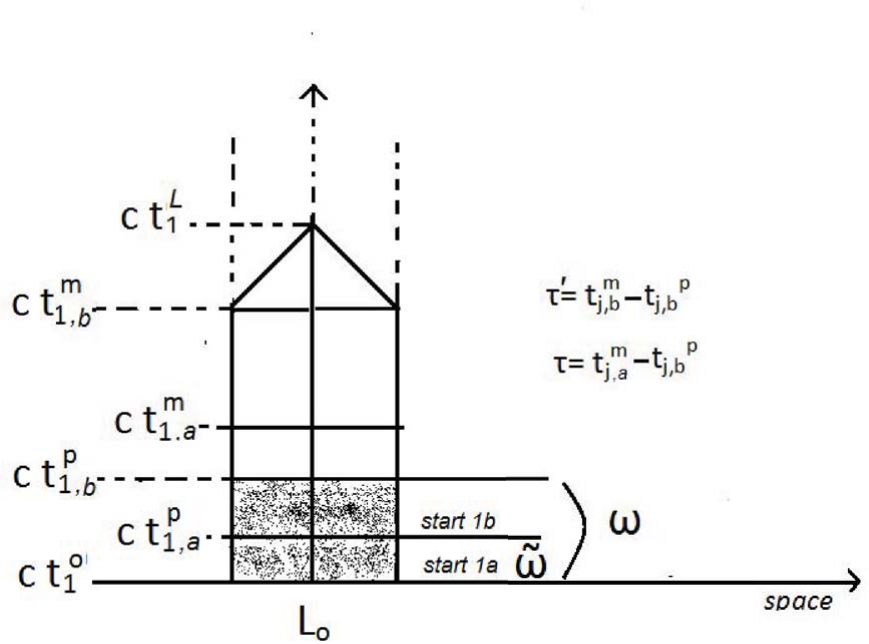}
	\caption{Subsequent Measurements I - Ensemble}
	\label{fig:05}
\end{figure}
We prepare both  observables  $a$ and $b$ before the measurements as represented in Figure \ref{fig:05}\footnote{Where we have represented only a single measurement of the ensemble.}.
\\
So we assume that $a$ and $b$ are jointly prepared and that we prepare $a$ before $b$ in the $\omega$ state of the physical system.
\\
We denote by $t_{j,a}^m, t_{j,b}^m \ , j=1,2\ldots N$ the  respectiv measurement times with different time intervals denoted by $\tau$ and $\tau'$, where
\begin{equation}\label{doppiotempo}
\tau=t_{j,a}^m- t^p_{j,b} \qquad , \qquad  \tau'= t_{j,b}^m - t^p_{j,b}
\end{equation} 
with $\tau' - \tau = t_{j,b}^m -t_{j,a}^m \geq 0$ for every $j=1,2\ldots N$\footnote{Recall that the temporal sequence of  measurements  may not respect that of the preparation; in other words, you may want to measure $b$ first and then $a$.}.
\\
Moreover, even if the two observables are measured at different instants of time, the reading of both results occurs simultaneously, when the laboratory clock shows the time $t_j^L$.
\begin{notation}\upshape\index{$\mathfrak S^{a<b }$}
The set of states where it is possible to  subsequently measure $a$ and then $b$ after having prepared $a$ before $b$\footnote{As in the case of simultaneous measurements, it is not always possible to physically carry out such measurements since the preparation of the observable $b$ could destroy the various information we have obtained about $a$ and the measurement of $a$ could destroy the preparation of $b$.}  is denoted with the symbol $\mathfrak S^{a<b }$ while with the symbol $\mathfrak S^{a>b}$ we indicate the set of states when it is possible to measure $b$ before $a$.
\end{notation}

Also in this case the statement of the property \ref{mappastato} and the definition of the chronological state of the system are proposed again.
\begin{post}\label{mappastato3}
We assume that in the experimental procedures that are  carried  out in the laboratory to establish the state of the system, which require a preparation time $t_{p,a}=t^p_{j,a}- t^o_j$ and $t_{p, b}=t^p_{j,b}- t^p_{j,a}$ for $ j=1,2 \ldots N$,  it  contains the instruction to measure the value of the observables $a$ and $b$ at each time interval $\tau' , \tau$ given by the relation  \eqref{doppiotempo}  with $\tau' \neq \tau$.
\end{post}
The only difference from the cases previously treated is given by the map $\omega$ of the type  \eqref{noevoltemp}; here we have to divide the various cases.  Here  we only deal with the case $a<b$, where $\tau_1$ is the measurement time of $a$ and $\tau_2$ of $b$; the other cases are obtained with easy corrections.
\\
In this case we have a double-valued map:
\begin{equation}\label{noevoltemp3}
\omega: \ (\tau_1,\tau_2) \in \mathbb W^2_u     \longrightarrow \omega^{(\tau_1,\tau_2)} \in \mathfrak S^{a<b} 
\end{equation}
which we assume as the state of the laboratory system, where
$$  \mathbb W^2_u  = \left\{(\tau_1,\tau_2)\in\mathbb R_*^2 : \  \tau_2 > \tau_1\geq 0 \right\} \ , \ \mathbb W^2_d  = \left\{(\tau_1,\tau_2)\in\mathbb R_*^2 : \  \tau_1 > \tau_2\geq 0 \right\} $$
Also in this case, once the experimenter has prepared the laboratory for the two measurements, it  could decide not to carry out the measurement of $b$; this situation corresponds to the case $\tau_2=\infty$.
\\ 
We observe that unlike the case of simultaneous measurements, here we have the following inclusion\footnote{Recall that if $\omega \in \mathfrak S_a \cap \mathfrak S_b$, the laboratory can be prepared in this state both for the measurement of $a$ and $b$ as established in paragraph \ref{Statistica e Riproducibilità}. }: 
$$ \mathfrak S^{a<b} \subset  \mathfrak S_a \cap \mathfrak S_b \qquad , \qquad  \mathfrak S^{a>b} \subset  \mathfrak S_a \cap \mathfrak S_b$$ 
since the state $\omega$ is suitable for both the measurement of $a$ at time $\tau$ and that of $b$ at time $\tau'$. 
\begin{problem}\upshape
If the two observables $a,b$ are not jointly  preparable, simultaneously and/or subsequently,  can the set $\mathfrak S_a \cap \mathfrak S_b$ be non-empty?
\end{problem}
Obviously,
$$ \text{ if \ }  \mathfrak S_a \cap \mathfrak S_b = \emptyset \ \Longrightarrow \  \mathfrak S^{a<b}=\mathfrak S^{a>b}=\emptyset$$ 
 the question remains open in the case of the simultaneous measurement of the observables.
\begin{remark}\upshape
We underline that for every $\tau'\in ]\tau,\infty] $ we have
\begin{equation}\label{noevoltemp3a}
\omega^{\tau'} : \tau \in [0,\tau'[    \longrightarrow \omega^{(\tau,\tau')} \in \mathfrak S^{a} 
\end{equation}
a chronological state suitable for the measurement of $a$, while for each $\tau \in [0,\tau'[$
\begin{equation}\label{noevoltemp3b}
\omega^{\tau} : \tau \in ]\tau,\infty]    \longrightarrow \omega^{(\tau,\tau')} \in \mathfrak S^{b} 
\end{equation}
a chronological state suitable for the measurement of $b$, while the chronological state given in  \eqref{noevoltemp3}  is suitable for $a<b$ and not  individually for the two observables.
\end{remark}
\bigskip
 
For each state $\omega\in\ \mathfrak S^{a<b}$ we obtain the following relative frequencies:
\begin{equation*}
f(a\in\Delta | a<b , \ \tau)_\omega = \frac{n(a\in\Delta | a<b , \ \tau)_\omega}{\textit{Total number of measurements  carried  out}} 
\end{equation*}
where with $n(a\in\Delta \ | a<b , \ \tau)_\omega $ we have denoted the number of times that the observable $a$ has a value in $\Delta$ \textit{after having prepared $a$ before $b$ and having measured $a$ before $b$ at time $\tau$}\footnote{We reiterate that all this information is contained in the $\omega$ state. To avoid making the notation too heavy we have not indicated the time $\tau'$.
\\
One might think that  carrying  out the second measurement at time $\tau'$ has no effect on the first measurement; in reality, having  required  the equipment to carry out the second measurement at time $\tau'$ could influence the first measurement at time $\tau$. }.
\\
In  a  similar way
\begin{equation*}
f(b\in\Delta | a<b ,\ \tau ')_\omega= \frac{n(b\in\Delta |  a<b , \ \tau')_\omega}{\textit{Total number of measurements  carried  out}  } \ , \qquad \tau' > \tau 
\end{equation*}
So we have the following probability distributions:
\begin{equation}\label{primamisura} \index{$P(a\in\Delta_1 |  a<b , \ \tau)_\omega$ }
\tau\in \mathbb R^+ \longrightarrow P(a\in\Delta_1 |  a<b , \ \tau)_\omega  
\end{equation}
\begin{equation}\label{secondamisura}
 \tau' \in  ] \tau , +\infty  [  \longrightarrow P(b\in\Delta_2 |  a<b , \ \tau')_\omega    
\end{equation}
\begin{figure}[htbp]
	\centering
		\includegraphics[scale=0.5]{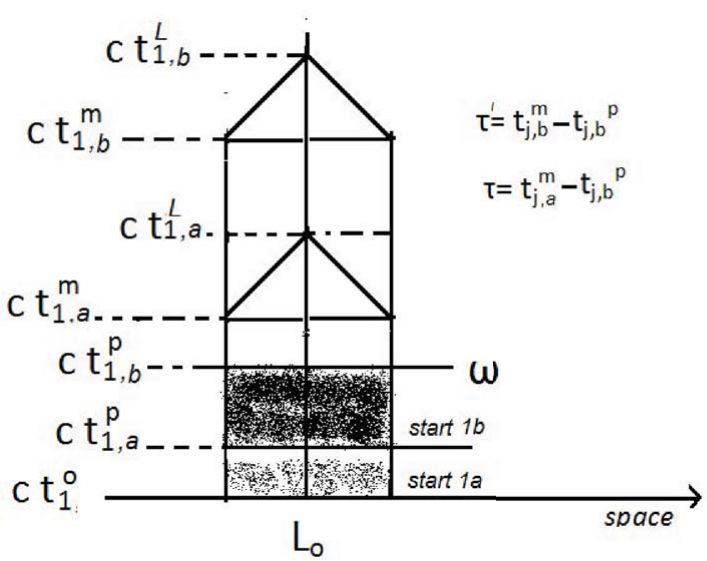}
	\caption{Subsequent measurements with reading}
	\label{fig:06}
\end{figure}
Let us now make some remarks on the notations adopted.
\\ 
In the state $\omega$ there is already information that $a<b$ and that they are measured subsequently at the times $\tau$ and $\tau'$; therefore the notations given in \eqref{primamisura} and \eqref{secondamisura} seem redundant.
\\
It could be written compactly as $P(a\in\Delta_1)_\omega$ for \eqref{primamisura} and $P(b\in\Delta_2)_\omega$ for \eqref{secondamisura}, but in this way the information contained in our $\omega$ state is not explicitly revealed, a notation that will be important for subsequent measurements of the same observable.
\\
Using the chronological state given in \eqref{noevoltemp3}, we can write:
$$P(a\in\Delta_1 | a<b , \tau)_\omega = P(a\in\Delta_1)_{\omega^{(\tau,\tau')}} $$
and
$$   P(b\in\Delta_2 | a<b , \tau')_\omega  = P(b\in\Delta_2)_{\omega^{(\tau,\tau')}} $$
Furthermore, using the chronological states given in \eqref{noevoltemp3a} and \eqref{noevoltemp3b}, we can write:
$$P(a\in\Delta_1 | a<b , \tau)_\omega = P(a\in\Delta_1 ,\tau)_{\omega^{\tau'}} $$
and
$$ P(b\in\Delta_2 | a<b , \tau')_\omega  = P(b\in\Delta_2 , \tau')_{\omega^{\tau}} $$
We remark that the state $\omega\in\mathfrak S^{a<b}$ could contain a different instruction from the previous situation: the instruction to perform the measurements at the times $\tau_1, \tau_2$ as established, but with different reading times (see Figure \ref{fig:06}). In this way, even having all the identical instructions of the previous state for a single reading (the one given by Figure \ref{fig:05}), we obtain a different state since the reading action disturbs the second measurement.

\subsubsection{An alternative method of measurement procedure}
\begin{figure}[htbp]
	\centering
		\includegraphics[scale=0.5]{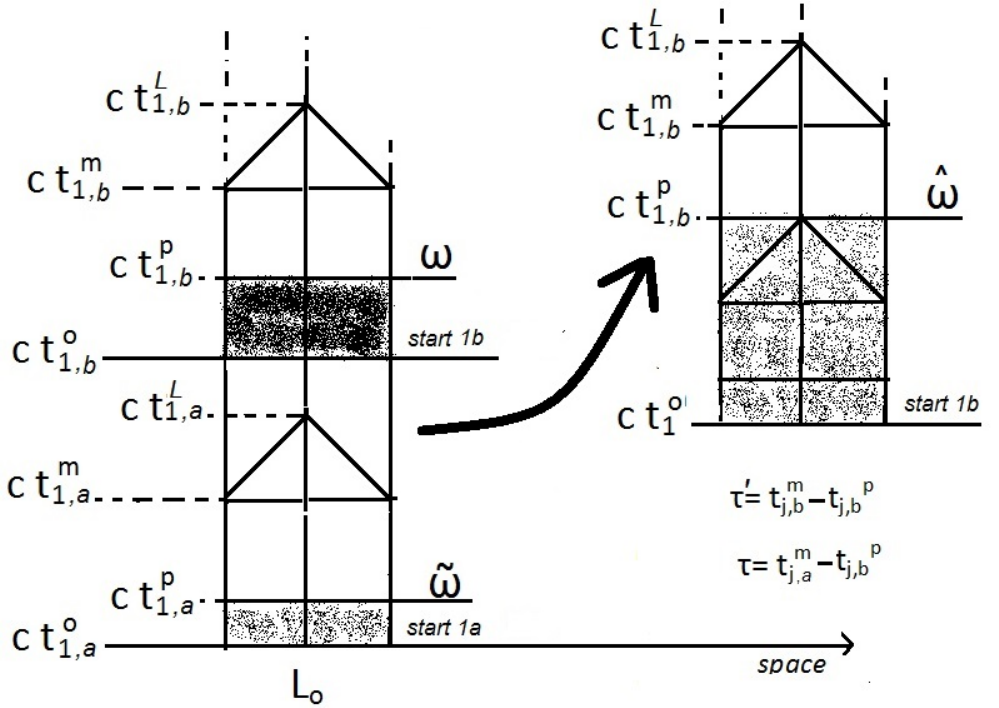}
	\caption{Subsequent measurement II}
	\label{fig:07}
\end{figure}

It could be assumed  that we implement the following alternative method of measurement procedure:
\\
The preparation of the observables $a$ and $b$ occurs after the relative measurements (with or without the readings of their values) as shown in Figure \ref{fig:07}, where for simplicity we have represented only one copy  of the ensemble.
\\
In other words, we prepare the observable $a$ in a state $\widetilde\omega\in\mathfrak S_a$. After its measurement (with or without reading its value) we must prepare the observable $b$  in a way that  takes into account the measurement that occurred on $a$, without intervening on the laboratory devices and equipment or on the source of the measurement; in this way the parametric state of $\widetilde\omega$ is changed.
\\
This   mutated  parametric state $\widetilde\omega_c'$  is not necessarily the same in every copy of our ensemble\footnote{We must underline that this statement is only hypothetical, since to establish how the physical parameters that we indicated  earlier  with $\left\{ c_j \right\}_{j\in I}$ actually change, we must equip the laboratory for  their  measurement.}. 
\\
It follows that this situation does not allow us to establish the same $\widehat\omega$ state in each copy of the ensemble as shown in Figure \ref{fig:07}.	
\\
Let us remember that in our statistical method it has  no  physical value to consider only one measurement of a copy of the ensemble; we should be able to establish a $\omega$ state and repeat the experiment $N$ times in this state and obtain the relative frequencies of type \eqref{freq}.
\bigskip
\\
\textit{
Thus, unless otherwise stated, when we talk about subsequent measurements we will always refer to the first case in which the experimenter prepares both observables in a pre-established order (if this is experimentally possible) for their measurement.}	
\subsection{Repeated Measurements}
We now want to perform two subsequent measurements of the same observable $a$ of the system, respectively at times $\tau$ and $\tau'$. 
\\
Obviously this is not necessarily experimentally possible, since the first measurement could damage the state of the system so that a second measurement of the same observable cannot be  carried  out.
\\

For example, the repeatability of the experiment is possible if the change occurs in the part of the state that we have denoted with the symbol $\omega_C$  in observation \ref{statoindividuazione} and the act of measurement does not damage the instrumentation used and  its  effectiveness (in practice $\omega_S$ is not affected by the measurement) and/or the source of the measurement itself.
\\ 

We assume that this can be done and that therefore there exists  a  $\omega\in \mathfrak S^{a<a}\subset \mathfrak S_a$; in this way, with the notations adopted, for the distribution \eqref{primamisura} we can write that
\begin{equation} 
P(a\in\Delta |  a<a | \tau)_\omega =  P(a\in\Delta  \ , \tau)_\omega 
\end{equation}
and from  \eqref{secondamisura}  
\begin{equation} 
P(a\in\Delta | a<a | \tau')_\omega =  P(a\in\Delta  \ , \tau')_{\omega} \ , \qquad \tau<\tau'
\end{equation} 
since in this case the chronological states given in  \eqref{noevoltemp3a}  and  \eqref{noevoltemp3b}  are both suitable for $a$.
\section{Joint Probabilities}
Let $a$ and $b$ be two jointly prepared and subsequently measurable observables in the state $\omega\in \mathfrak S^{a<b}$ and   let us   consider the joint frequencies of the two events $ A=\left\{a\in\Delta_1 \ , \tau_1 \right\}$ and $B=\left\{b\in\Delta_2 \ , \tau_2 \right\}$:
$$f(a\in\Delta_1 \ , \tau_1 \  \wedge \ b\in\Delta_2 \ , \tau_2 )_\omega : =\frac{ n(a\in\Delta_1 \ , \tau_1\ \wedge  \ b\in\Delta_2 \ , \tau_2)_\omega}{\textit{Total number of measurements carried out} } \ , \  \tau_1 < \tau_2 $$
\\
where $n(a\in\Delta_1 \ , \tau_1 \ \wedge \ b\in\Delta_2 \ , \tau_2)_\omega$ is the count of the number of times that in the two subsequent measurements the value is $a \in\Delta_1 $ at time $\tau_1$ and  then  $b\in\Delta_2 $ at time $\tau_2$.
\begin{remark}\upshape
We specify that here the conjunction of events  refers  to the events that we have indicated with $A$ and $B$ and not to the joint measurement of $a$ and $b$, which for us has the meaning of simultaneous or successive measurement of the observables\footnote{Cox in \cite{Cox}, to avoid generating "temporal" confusion, speaks of propositions and not of events; for example, in our case  one  should speak of the joint probability of the proposition $A$ with $B$, in symbols $A\wedge B$.}.
\end{remark}
From the \textit{joint frequencies} we obtain the \textit{joint probability}:
\begin{equation}
\label{leggecongiunta2}
P(a\in\Delta_1 \ , \tau_1 \  \wedge \ b\in\Delta_2 \ , \tau_2 )_\omega  \ ,  \qquad  \Delta_1 , \Delta_2 \in  B(\mathbb R) \ , \qquad \tau_1 < \tau_2
\end{equation}
As in the simultaneous case, we can give the definition of independence of two observables subjected to successive measurements.
\begin{definition}\upshape\label{leggeindip}
Let $a$ and $b$ be two observables jointly  preparable in the order $a<b$. They are independent in the state $\omega\in \mathfrak S^{a<b}$ at the times $\tau_1 < \tau_2$ if for every $\Delta_1 , \Delta_2 \in B(\mathbb R)$ we obtain 
\begin{equation*}
P(a\in\Delta_1 \ , \tau_1 \  \wedge \ b\in\Delta_2 \ , \tau_2 )_\omega = P(a\in\Delta_1 | a<b ,  \tau_1)_\omega \cdot P(b\in\Delta_2 | a<b , \tau_2)_\omega
\end{equation*}
\end{definition}
\section{Conditional Probability} \label{misurecondizionate}
Usually, for experimental reasons we need to know the value of an observable constrained to some previously established values of other observables.
\\
For example, we want  to  determine the mass of a particle (observable $a$) at time $\tau$ knowing that its velocity (observable $b$) at a certain instant $\tau_o < \tau$ has a well-defined value.
\\
\textit{We must thus determine the statistical law of $a$ at time $\tau$ knowing that $b\in\Delta_o$ at time $\tau_o$.}
\\
What operational meaning should we attribute to this statement?
\\ 
In other words, how do  you experimentally prepare the $N$ copies of the experiment in the laboratory for such measurements?
\\
We need to consider two conditioning options.

\subsection{First Case - Jointly Prepared Observables}
This case falls within the classical definition of conditional probability of two events. We prepare the ensembles for the subsequent measurements of $a$ and $b$, as indicated in section \ref{misuresuccessive}, by choosing among all the measured values obtained for the first observable those established by our conditions.
\\
Let's analyze the situation better:
\\
We prepare the observables $a$ and $b$ for two subsequent measurements in the state $\omega$. We assume that $b$, \textit{the conditioner}, is prepared  before  \textit{the conditioned} $a$; thus $\omega\in\mathfrak S^{ b<a}$ and we obtain the value:
\begin{equation}\label{numerob}
 n(b\in\Delta_o \ | \ b<a \ , \ \tau_o)_\omega
\end{equation}
\textit{i.e.} the number of times that the observable $b$ takes on the value in $\Delta_o$, in symbols $b\in\Delta_o$, after having prepared $b$ before $a$ and \textit{measured $b$ before $a$} (at time $\tau_o$).
\\
 Carrying  out the second measurement, the one on $a$ at time $\tau$, we count the number of times that $a\in\Delta$ when in the first measurement we obtained that $b\in\Delta_o$; in this way we obtain the number
\begin{equation}
\label{condizione}
n(a\in\Delta \ , \tau \ \wedge  \ b\in\Delta_o \ , \tau_o)_\omega
\end{equation}
with
$$ n(a\in\Delta \ , \tau \ \wedge  \ b\in\Delta_o \ , \tau_o)_\omega  \leq n(b\in\Delta_o \ | \  b<a \ , \ \tau_o)_\omega$$
In this way, we consider the frequencies of $a$ at time $\tau$ measured in the state $\omega$, \textit{conditioned} by the values of $b$ measured at time $\tau_o < \tau$:
\begin{equation}
 \label{condiziofreq1}
f(a\in\Delta \ , \tau \ | \ b\in\Delta_o \ , \tau_o )_\omega : =\frac{n(a\in\Delta \ , \tau \ \wedge  \ b\in\Delta_o \ , \tau_o)}{ n(b\in\Delta_o \ | \ b<a \ , \ \tau_o)_\omega}  %
\end{equation} 
Therefore we obtain the distribution (at time $\tau$) of the conditional probabilities:
\begin{equation}
\label{condizio}
\Delta \subset \mathbb R \longrightarrow  P(a\in\Delta  \ , \tau \ |  \ b\in\Delta_o \ , \tau_o )_\omega \ , \qquad \tau_o<\tau
\end{equation}
where
\begin{equation}
P(a\in\Delta  \ , \tau \ |  \ b\in\Delta_o \ , \tau_o )_\omega = \frac{P(a\in\Delta \ , \tau  \  \wedge \ b\in\Delta_o \ , \tau_o )_\omega}{ P(b\in\Delta_o | \ b<a , \tau_o)_\omega} \ , \qquad   \tau_o < \tau 
\end{equation}
The problems of defining the conditional probability in the way just described arise if we consider conditioning due to multiple observables in the system. For example, to  determine the probability that $a\in\Delta$ at time $\tau$, fixing the values of $c\in\Delta_0$ at time $\tau_0$ and $b\in\Delta_1$ at time $\tau_1$:
$$  P(a\in\Delta  \ , \tau \ |  \    b\in\Delta_1  \ , \tau_1 \ | \   c\in\Delta_0  \ , \tau_0 )_\omega \ , \qquad \tau_0 < \tau_1 < \tau$$
We can proceed as established only in the case that $a,b,c$ \textit{are jointly prepared} in the order $c<b<a$ in a state $\omega\in \mathfrak S^{c<b <a}$.
\bigskip

Let's generalize the previous procedure: we assume that $a$ and the observables $\left\{b_1,b_2,\ldots b_n \right\}$ are all jointly prepared in their natural order $b_1<b_2< \cdots < b_n <a$, respectively at times $\tau_1 < \tau_2 < \cdots < \tau_n <\tau$ ; we can consider the conditional probabilities:
\begin{equation}\label{condiziobis}
 P(a\in\Delta  \ , \tau \ |  \    b_1\in\Delta_1  \ , \tau_1 \ |  \cdots  |  \ b_n\in\Delta_n  \ , \tau_n )_\omega 
\end{equation}
with $\omega\in \mathfrak S^{b_1<b_2< \cdots < b_n<a} $.
\\
A further generalization is to assume in \eqref{condiziobis} instead of a single observable $a$ a family of observables $\left\{a_1,a_2,\ldots a_m  \right\}$ jointly prepared in their order $a_1:a_2: \cdots : a_m$ which we measure simultaneously at time $\tau$ and  are jointly prepared with the family of conditioning observables $\left\{b_1,b_2,\ldots b_n \right\}$, obtaining the relation
 \begin{equation} \label{condiziotris}
 P(a_1\in\Delta'_1 : \cdots : a_m\in\Delta_m'  \ , \tau \ |  \    b_1\in\Delta_1  \ , \tau_1 \ |  \cdots  |  \ b_n\in\Delta_n  \ , \tau_n )_\omega 
\end{equation}
where, with the obvious meaning of the symbols, $\omega\in \mathfrak S^{b_1<b_2< \cdots < b_n<(a_1:a_2: \cdots : a_m)}$.
\subsection{Second Case - State Conditioning} \label{cond_statale}
Given a family of observables $\left\{ b_k \ : \ k=1,2, \ldots n \right\}$ \textit{jointly prepared} in their natural order $ b_1<b_2< \cdots < b_n $ respectively at the times $\tau_1 < \tau_2 < \cdots < \tau_n $, we select a set of Borel sets $\left\{ \Delta_k \subset \mathbb R \: \ k=1,2, \ldots n \right\}$.
	\\
Assume that the condition $C$ is given by the following statement:
\begin{center} 
The observables $b_k\in\Delta_k$ at time $\tau_k$ for each $k=1,2, \ldots n$.
\end{center}
We denote with $\mathfrak S^C$ the following  set of  states of our laboratory system:
\begin{equation*}
\mathfrak S^C =\left\{\omega\in \mathfrak S^{b_1< \cdots < b_n} \ :  \ P(b_k\in\Delta_k \ | b_1< \cdots < b_n \ | \tau_k)_\omega=1 \ , \  \forall k=1,2, \ldots n \right\}  
\end{equation*}
Let us now fix any observable of the system $a$; the set 
$$\mathfrak S_a \bigcap \mathfrak S^C$$
 is  composed of the states of observable $a$ \textit{conditioned} by our initial condition $C$\footnote{We cannot exclude the possibility that this set is empty.}. 
\\
We take a state 
$$\omega^C\in \mathfrak S_a \bigcap \mathfrak S^C\subset \mathfrak S_a \bigcap  \mathfrak S^{b_1<b_2< \cdots < b_n}$$
initially selected by our experimenter.
\\
Therefore, we need to prepare the laboratory for subsequent measurements of the family $\left\{ b_k \ : \ k=1,2, \ldots n \right\}$ in our state $\omega^C\in \mathfrak S^{b_1<b_2< \cdots < b_n} $ at the various times $\tau_k \ , k=1,2, \ldots n$ as established in section \ref{misuresuccessive}; after that we proceed with the measurements and determine when
$$P(b_k\in\Delta_k \ | b_1< \cdots < b_n \ | \tau_k)_\omega=1$$ 
\\
and since $\omega^C\in \mathfrak S_a$, we have also prepared the laboratory for the measurement of $a$ at time $\tau$ as established in \ref{proceduresperimentali} and  we determine the  statistical law $P(a\in\Delta , \tau )_{\omega^C}$\footnote{See also the definition of \textit{transition probability} in Accardi's works \cite{Accardi84} p. 307 and in \cite{Accardi97} par. II.10.}.
\\
We note that in this case we have not made any hypotheses on the observable $a$, which could be an observable complementary to the observables of our condition $C$.
\\

Obviously we can consider more than one condition $ \left\{ C_j \ : \ j=1,2, \ldots m \right\}$
\\
For example, we can consider two observables $b$ and $c$ \textit{that are not } jointly  preparable, with the conditions $C_1$ and $C_2$ given respectively by $\left\{ b\in\Delta_o \ , \tau_o \right\}$ and $ \left\{c\in\Delta_1 \ , \tau_1 \right\}$; we  denote with
$$ \mathfrak S^{C_1} =\left\{\omega\in \mathfrak S_{b} \ \textit{t.c.} \ P(b\in\Delta_o , \tau_0)_\omega=1\right\}\subset \mathfrak S_{b}$$
and
$$ \mathfrak S^{C_2} =\left\{\omega\in \mathfrak S_{c} \ \textit{t.c.} \ P(c\in\Delta_1 , \tau_1)_\omega=1\right\} \subset \mathfrak S_{c}$$
It follows that the set of conditioned states of $C_1$ and $C_2$ for $a$ is given by:
$$\mathfrak S_a \bigcap \mathfrak S^{C_1} \bigcap \mathfrak S^{C_2}= \mathfrak S_a \bigcap \mathfrak S^{C_1\wedge C_2}$$
\begin{remark}\upshape \label{doscondizionamenti}
The two conditioning procedures described here are different from each other.
\\
In the first case, to determine the conditioned frequencies given in \eqref{condiziofreq1}, a priori we know nothing about the possible values of $a,b$; the measurements are  carried  out first on $b$ at time $\tau_o$ and in succession those of $a$ at time $\tau > \tau_o$, after which their values are noted and subsequently  analyzed to see  whether these values are  within  $\Delta_o $ for $b$ and in $\Delta$ for $a$; after having  counted  them we obtain  the numbers  \eqref{numerob} and \eqref{condizione}.
\\
In the second case we know a priori that in our state $\omega^C$ the observable $b\in \Delta_o$ at time $\tau_o$.
\\
Furthermore, in this case, as we have defined the conditioning procedures, we have an arbitrariness in choosing the time $\tau_o$ of our condition $C$; nothing prevents us from assuming a future condition by taking the time $\tau_o\geq \tau$\footnote{\label{tempomisura}Recall that here the time $\tau$ is a time interval between the measurement time and the preparation time of the observables.}.  
\end{remark} 
\begin{problem}\upshape
\label{problemacondizionale}
If the observable $a$ and those that establish $C$ are jointly measurable, what relation exists between the conditional probability established in  \eqref{condiziobis}  and the conditional probability $P(a\in\Delta , \tau )_ {\omega^C}$?
\end{problem}
\subsection{Writing Problems}
As we have expressed several times in these notes, to know the value of the observable $a$ I must prepare it for its measurement and this requires knowledge of other observables $\left\{ c_{i,j} \right\}_{ i,j}$ of the laboratory system  that are jointly preparable, previously or simultaneously measured, with the observable $a$, with the necessary instruments and measurement times. Therefore the measurement of $a$ \textit{is always conditioned} by other parameters that contribute to the formation of the state $\omega$ of the system\footnote{Which establishes the parametric state $\omega_c$ of $\omega$.}.
\\
In other words, Whether  we want to focus only on the value of the observable $a$ obtained when we carry out our experiment, instead of writing
\begin{eqnarray*}
 P(a\in\Delta  & | &   c_{1,1}\in\Delta_{1,1}: \cdots  : c_{1,m_1}\in \Delta_{1,m_1} \ |  \cdots \\
& \ldots & |  \ c_{k,1}\in \Delta_{n, 1} : \cdots : c_{k,m_k}\in \Delta_{k,m_k} \ | \cdots )_\omega
\end{eqnarray*}
with 
$$\omega\in \mathfrak S^{ c_{1,1}: \cdots  : c_{1,m_1}<c_{2,1}: \cdots  : c_{2,m_2}< \cdots < c_{k,1}: \cdots  : c_{k,m_k}< \cdots < a } \subset \mathfrak S_a$$
we write $P(a\in\Delta)_\omega$ in a compact way with $\omega\in \mathfrak S_a$\footnote{Obviously in $\omega$ there is  all  the  information  that we omitted in the previous writing, in addition to information on the instruments and methodologies used during the measurement.} and in this case we will talk about \textit{ the single measurement of the observable $a$.}
\\
This also applies to simultaneous measurements of non-complementary observables; when we want to focus on the values of the observables $a_1, a_2, \ldots a_n$ measured simultaneously, instead of writing
\begin{eqnarray*}
P(a_1\in\Delta_1 : \cdots  : a_n\in\Delta_n &:&  |  c_{1,1}\in\Delta_{1,1}: \cdots  : c_{1,m_1}\in \Delta_{1,m_1} \ |  \cdots \\
& \ldots & |  \ c_{k,1}\in \Delta_{n, 1} : \cdots : c_{k,m_k}\in \Delta_{k,m_k} \ | \cdots )_\omega
\end{eqnarray*}
with
$$\omega\in \mathfrak S^{ c_{1,1}: \cdots  : c_{1,m_1}<c_{2,1}: \cdots  : c_{2,m_2}< \cdots < c_{k,1}: \cdots  : c_{k,m_k}< \cdots < a_1:\cdots :a_n } \subset \mathfrak S_{a_1:\cdots :a_n}$$
we write
$$P(a_1\in\Delta_1 : \cdots : a_n\in\Delta_n)_\omega$$  
with $\omega\in \mathfrak S^{ a_1: \cdots  : a_n} $ ;  in this case we will talk about \textit{simultaneous measurements of $a_1: \cdots  : a_n$.}
 \\
Let's summarize everything with the following
\begin{remark}\label{statale}
Given a family of observables $ a_1,a_2, \ldots a_m :a$ jointly  preparable  in their order $ a_1:a_2: \cdots : a_m:a $ which we measure simultaneously at time $\tau$ and which are jointly  preparable  with the family of observables $\left\{b_1,b_2,\ldots b_n \right\}$ and subsequently measurable at the times $\tau_1<\tau_2<\cdots \tau_n<\tau$ in the state $\omega\in \mathfrak S^ {b_1<b_2< \cdots < b_n<(a_1:a_2: \cdots : a_m:a)} $ and by \ref{congiunzionestati2} we have the inclusions:
$$ \mathfrak S^ {b_1<b_2< \cdots < b_n<(a_1:a_2: \cdots : a_m:a)}\subset \mathfrak S^ {a_1:a_2: \cdots : a_m:a}  \subset\mathfrak S_{a}$$
Thus the state $\omega\in\mathfrak S_{a}$ and we can write\footnote{In practice, the observable $a$ that we want to study is prepared last, after having prepared the other physical quantities that determine the parametric state of the system.}
\begin{eqnarray*}
 P(a_1 \in\mathbb R: \cdots : a_m\in\mathbb R : a\in \Delta, \ \tau  & | &  b_1\in\Delta_1, \  \tau_1   |  \cdots   | \ b_n\in\Delta_n, \ \tau_n )_\omega = 
\\
&=&  P(a\in\Delta, \ \tau   )_{\omega}
\end{eqnarray*}
\end{remark}
\section{Examples}\label{esempi}
In this section we collect some simple examples on the feasibility of joint preparations.
\begin{example}\upshape \label{prepcong1}
We inoculate two substances indicated with A (antidote) and B (poison) on a laboratory guinea pig;  we measure the values of their concentrations in the blood.  The concentrations will be called respectively, with the words observables $a$ and $b$.
\\
The poisonous substance $B$ without first inoculating the antidote $A$ leads the poor little animal to certain death.
\\
Therefore it is necessary to prepare the ensemble to first inoculate $A$ and subsequently inoculate the substance $B$, noting all the vital parameters of the guinea pig, the method of administration of the substances etc... before and during the administration of the substances to establish the state $\omega$ of the system.
\\
We can measure the concentration of $A$  and  $B$ in two ways:   either  make two samples and measure the concentration of $A$ and then $B$, or make a single sample and measure the same  concentrations. In both cases, the measurements on the concentration test tubes can take place successively or simultaneously.
\\
Let us remember that in general $\mathfrak S^{a<b}\subset \mathfrak S_{a}\cap \mathfrak S_{b}$ and  that  we have
$$\mathfrak S_{a:b} \subset \mathfrak S_{b} $$
\\
However, if we assume that the substance $A$ is the only existing antidote for $B$, in both cases we must necessarily have:
$$ [ \ \mathfrak S_{a:b} =\mathfrak S_{b} \qquad  and  \qquad \mathfrak S^{a<b}=\mathfrak S_{b} \ ] \qquad \Longrightarrow \qquad \mathfrak S_{a:b} = \mathfrak S^{a<b}$$
we can jointly prepare $a$ and $b$ simultaneously or successively for the measurement, but not vice-versa, therefore
$$\mathfrak S_{b:a} =\emptyset \qquad  and \qquad \mathfrak S^{b<a} =\emptyset$$
We observe that in this case
$$\mathfrak S_{b}= \mathfrak S^{a<b}\subset \mathfrak S_{a} \qquad \Longrightarrow \qquad \omega\in\mathfrak S_{a}$$
\end{example}
\begin{example}\upshape \label{prepcong2}
Let's go back to the previous experiment: this time the two substances indicated with A and B are two drugs that can be inoculated individually but not both into the guinea pig since they lead to the death of the animal.
\\
So individually I can inoculate A and B obtaining a state $\omega_A$ and $\omega_B$ respectively and measure their concentrations $a$ and $b$ in the blood\footnote{In practice we carry out two distinct experiments. }, but I can't jointly prepare $a<b$ or $a>b$ etc...
 $$\mathfrak S^{a<b}=\mathfrak S^{a>b} =\emptyset$$
\end{example}
 
\chapter{Experimental Measures and Radon Measures}
In these notes we are using probability as a synonym for relative frequencies over a large number of trials carried out across copies of the same experiment. We will not analyse the definition of probability of an event more than necessary and we will study the possibility of associating these relative frequencies with a Borel measure in accordance with Kolmogorov probability theory \cite{Kolg}.  
\section{Expected value and Borel measures}\label{misureBorele}
In the previous sections we hypothesized that the relative frequencies $f(a\in \Delta)_\omega $ expressed by  \eqref{freq}  obtained through statistical ensembles stabilized as the total number of measurements carried out increased, around a number denoted by $P(a\in \Delta)_\omega$ which is taken as an index of the probability that the quantity $a$ \textit{(at the time $\tau$)} takes a value in a (Borel) subset $\Delta$ of $\mathbb{R}$, a  measure \textit{conditioned} by the state $\omega$\footnote {We have previously discussed the experimental difficulties of this apparently trivial statement (see also the discussion by Home and Whitaker in \cite{H.W.} paragraph 5.6.); we will  return to  analyzing this problem in chapter \ref{legge_empirica}. }.\index{Home and Whitaker}
\\
From the frequencies  \eqref{freq},  we obtain for each suitable  state $\omega$ of the system, that this index satisfies the following properties:
\begin{itemize}
\item [A.]   $0\leq P(a\in \Delta)_\omega \leq 1$ 
\item [B.]   $ P(a\in \mathbb R)_\omega =1$ 
\item [C.]   $ P(a\in \Delta_1 \cup  \Delta_2)_\omega = P(a\in \Delta_1)_\omega + P(a\in \Delta_2)_\omega$  for each  pair of  disjoint (Borel) subsets $\Delta_1 , \Delta_2$ of $\mathbb R$.
\end{itemize}  
In this way for each physical quantity $a\in \mathfrak X$ a  map is defined:
\begin{equation}
\label{distribuzio1}
 \omega\in\mathfrak S_a \longrightarrow  P(a\in\Delta)_\omega\in [ 0, 1 ]  \ , \qquad \Delta\subset \mathbb R
\end{equation}
As we will see in the next sections, we will assume the following property is true:
\begin{axiom}[\textbf{Axiom of Borel Measure}] \label{Axiom-misura-Borel1}\index{Axiom-Borel Measure}\index{$\mu_{\omega, a}$}
The expression \eqref{distribuzio1} establishes the existence of a Borel measure which we indicate by $\mu_{\omega, a} $ defined as:
\begin{equation}
\mu_{\omega, a}(\Delta)=P(a\in\Delta)_\omega  \ , \qquad \Delta\subset \mathbb R
\end{equation}
\end{axiom}
From axiom \ref{Axiom-misura-Borel1}, for each observable $a$ of the system we have
$$P(a\in\emptyset)_\omega=0$$
furthermore, the following mathematical relation (experimentally non-trivial) of $\sigma$-additivity will be considered valid:
\begin{equation} \label{sigma-addit}
P(a\in\bigcup_{j=1}^{\infty} \Delta_j)_\omega= \sum_{j=1}^{\infty} P(a\in\Delta_j)_\omega
\end{equation}
where $\left\{ \Delta_j \right\}_{j\in\mathbb N} $ is a disjoint family of Borel sets of $\mathbb R$.
\\
As a consequence of our considerations we obtain the following property\footnote{The $\sigma$-additivity is a strong hypothesis of our model; a less restrictive request could be made by considering not measurements on $\sigma$-algebra but on $\pi$–$\lambda$ systems (See \cite{Bobrowski} par.1.2.7) but obviously we would not have the regularity properties stated below.}:
\begin{itemize}
\item \textsl{Regularity of probability measures $\mu _{\omega ,a}$.}\index{Regularity of probability measures}\index{Folland}
	\\
Since $\mathbb R$ is separable and the Borel measures $\mu _{\omega ,a}$ are Radon\footnote{Remember that a Borel measure is Radon if it is finite on every compact set}, they satisfy the properties of internal and external regularity\footnote{See Folland \cite{Folland}, proposition (7.8)}.
	\\
So for every $\Delta\in B\left( \mathbb{R}\right)$ it follows:
	$$\mu _{\omega ,a}(\Delta) = \inf \left\{ \mu _{\omega ,a}(U) : U \ \textit{open with} \  \Delta\subset U \right\}$$ 
while for every open $U$ of $\mathbb R$
		$$\mu _{\omega ,a}(U) = \sup \left\{ \mu _{\omega ,a}(K) : K \ \textit{compact with} \  K \subset  U  \right\}$$ 
	\end{itemize}
Through the probability measure we obtain the average value of the observable $a$ which is defined through the mathematical relation
\begin{equation}
\label{valoremedio0}
\left\langle a \right\rangle_\omega = \int s \ d\mu_{\omega, a} (s)
\end{equation}
As we have previously discussed, the weak point of this procedure is precisely establishing the law experimentally \eqref{distribuzio1}. 
\\
In fact, to have this value we must count the number of times that $a$ takes  a  value in the set $\Delta\in\mathbb R$, a number that we indicated with $n(a\in\Delta)_\omega$ . To do this we must ask the following question every time we do the $N$ trials:
\\
 Is  the value of $a$ in $\Delta$? Yes or No?
\\
 In some way we break down  a question into more basic questions, those whose answer is yes or no (yes-no experiments).
\\
Obviously, however high  the number of  such questions, which must be  carried  out for all the infinite Borel sets
$\Delta$, will never be able to truly cover all the knowledge of the observable $a$ in  \eqref{distribuzio1}.
\begin{problem}\upshape\label{distribuziopb}
In practice it is necessary to identify methodologies which, starting from the knowledge of the value of $P(a\in\Delta)_\omega$ for a subfamily $\mathfrak B_o$ of subsets of $\mathbb R$, allow  us  to establish the measure $\mu_{\omega,a}$.
\end{problem}
\begin{remark}[\textbf{Segal-von Neumann vs. Mackey}]\upshape\label{Segal-von Neumann vs. Mackey}\index{Segal}\index{Mackey}\index{von Neumann} 
As discussed by Accardi in \cite{Accardi75}, the point of view we have described is the one historically adopted by Mackey \cite{Mackey_57, Mackey},  as  contrasted with that of von Neumann \cite{Neu_32} and Segal \cite{ Segal_47} where the relation \eqref{distribuzio1} is replaced by the relation of expected values\footnote{See von Neumann's book \cite{Neu_32} chap.4 p. 306.}: 
\begin{equation}\label{distribuzio2}
  \omega \in   \mathfrak S_a \longrightarrow \left\langle a \right\rangle_\omega \in \mathbb R
\end{equation}
experimentally more sensible since it is an average  over  the $N$ values obtained experimentally by our ensemble (again ideally with the number of  trials  infinite).
\\
However, if we assume for every real continuous function $f$ the existence of an observable $f(a)$ of the system \cite{Neu_32}:
$$ f\in C_o(\mathbb R) \longrightarrow f(a)\in \mathfrak X$$
where $C_o(\mathbb R)$  denotes  the continuous functions that vanish at infinity, we can establish (under appropriate regularity conditions of the previous  map) a probability measure through the Riesz-Markov theorem:
\begin{equation}\label{valoremedio00}
\mu_{\omega, a}(f):= \left\langle f(a)\right\rangle_\omega \ , \qquad f\in C_o(\mathbb R)
\end{equation}
obtaining the reverse procedure of that of Mackey assumed here and which we will discuss in the next sections.
\end{remark}

\section{Simultaneity and Radon Measures}\label{simultanea}
As in the case of a measurement of a single observable $a$ of the physical system, given two non-complementary observables $a$ and $b$ where we assume that $a$ is prepared before $b$ for the measurement, the joint frequencies given by  \eqref{frequenzecongiunte}  define the law of joint probabilities  \eqref{leggecongiunta}:
$$P(a\in\Delta_0 \ : \ b\in\Delta_1)_\omega  \ , \qquad \forall \Delta_0 \times \Delta_1 \subset \mathbb R^2 $$
Now let's make a mathematical digression.
\\
Let's ask ourselves if there exists a probability measure $\nu$ on $B(\mathbb R^2)$ such that\footnote{Obviously here too $\nu$ will depend on $a,b$ and on $\omega$.}
\begin{equation}\label{misuraprodotto}
\nu(\Delta_0 \times  \Delta_1)=P(a\in\Delta_0   :   b\in\Delta_1)_\omega  
\end{equation}
for each (Borel) set $\Delta_0, \Delta_1 \subset \mathbb R$.
\\
The answer is contained in the following theorem on product measures:
\begin{theorem}\upshape\label{meyer}
Let $X$ and $Y$ be metric spaces and $B(X) , B(Y)$ their respective Borel $\sigma$-algebras and let $\beta: B(X)\times B(Y) \rightarrow [0 ,1]$ be a map with the property that $\beta(X\times Y)=1$.
\\
We assume that for each $\Delta_x\in B(X)$ and $\Delta_y\in B(Y)$ the maps
$$\Delta \in B(Y) \rightarrow \beta(\Delta_x \times \Delta)\in [0,1] \quad , \quad \Delta \in B(X) \rightarrow \beta(\Delta \times \Delta_y)\in [0,1]$$
are Radon measures.
\\
Then there exists a unique Radon measure $\nu$ on $B(X \times Y) $ such that
$$\beta(\Delta_1 \times \Delta_2 )= \nu(\Delta_1 \times \Delta_2 ) \ , \qquad \forall \Delta_1 , \Delta_2 \in B(X \times Y)$$
\end{theorem} 
\begin{proof}
For the proof see \cite{D.M.} theorem 74 page 80 III.
\end{proof}
Therefore  if the maps
$$\Delta \rightarrow P( a\in \Delta_o : b\in \Delta)_\omega \qquad ,  \qquad \Delta \rightarrow P(a\in\Delta : b\in \Delta_o)_\omega $$
are Radon  measures, whatever the (Borel) subset $\Delta_o$ of $\mathbb R$, the answer to our question is  affirmative.
\\
Furthermore, if the two observables $a$ and $b$ are independent of each other, the measure $\nu$ can be  factored through the measures induced by the marginal probabilities.
\\
We will assume the following property is true:
\begin{axiom}[\textbf{Product measurement}]\index{Axiom-Product Measurement}
\label{misuraprodotto1}
A family of jointly measurable observables (not necessarily compatible) for their simultaneous measurement at a time $\tau$ admits a probability measure that satisfies \eqref{misuraprodotto}. 
\end{axiom}
Let's analyze this last statement better.
\\
Let $\left\{ a_j \right\}_{j =1,2\ldots n }$ be a generic family of observables of our physical system $(\mathfrak X, \mathfrak S)$ that we want to measure.
\\
We assume that it is possible to prepare them in their natural order $a_1 , a_2 , \ldots a_n$ for their simultaneous measurements at a given instant of time $\tau_o$; in this case the relation \eqref{misuraprodotto} for each state
$$ \omega\in\mathfrak S_{a_1:a_2: \cdots:a_n} \subset\mathfrak S_{a_n}$$
takes the form:
\begin{equation}
  \nu(\Delta_n \times  \Delta_{n-1}\times \cdots \times \Delta_1)=  P( a_n\in\Delta_n : a_{n-1}\in\Delta_{n-1} : \cdots : a_1\in\Delta_1,\tau_o)_\omega
\end{equation}
where $\Delta_j\in B(\mathbb R) \ , j=1,2\ldots n$.
\\

If we change the preparation order of our observables, we obtain  new measurements and a new state\footnote{If we consider the observables  to be  compatible, then by definition \ref{definizionecompatibile} we can change the preparation order of our observables without changing the state.}:
\begin{equation*}
 \nu^p(\Delta_{p(n)} \times  \Delta_{p(n-1)} \times  \cdots \times \Delta_{p(1)}) =  P( a_{p(n)}\in\Delta_{p(n)} :  \cdots :a_{p(1)}\in\Delta_{p(1)},\tau_o)_{\omega'}
\end{equation*}
where $p$ is a permutation of  the $n$ indices and
$$ \omega'\in\mathfrak S_{a_{p(1)}:a_{p(2)}: \cdots:a_{p(n)}} \subset\mathfrak S_{a_{p(n)}}$$
For simplicity of exposition, let us return to the case of only two jointly measurable observables $a$ and $b$; we obtain for each $\Delta_o, \Delta_1 \in B(\mathbb R)$:
$$ \nu(\Delta_o \times \Delta_1) = P( a \in\Delta_o : b\in\Delta_1, \tau_o)_\omega \ , \qquad \omega\in \mathfrak S_{b:a}$$
and 
$$\nu'(\Delta_1 \times  \Delta_o) =  P( b \in\Delta_1 : a\in\Delta_o, \tau_o)_{\omega'} \ , \qquad  \omega'\in \mathfrak S_{a:b}$$
Furthermore, if the observables are compatible then we have
\begin{equation}
\nu(\Delta_o \times  \Delta_1)=\nu'(\Delta_1 \times  \Delta_o)  \ , \qquad \forall  \Delta_o, \Delta_1 \in B(\mathbb R) 
\end{equation}
and we can write that
\begin{equation}
\label{misuraprodotto2a}
P(a \in\Delta_o, \tau_o )_\omega=  \nu(\Delta_o \times  \mathbb R )=\nu'(\mathbb R \times \Delta_o )
\end{equation}
while
\begin{equation}
\label{misuraprodotto2b}
P(b \in\Delta_1, \tau_o )_\omega=  \nu'(\Delta_1 \times  \mathbb R )=\nu(\mathbb R \times \Delta_1 )
\end{equation}
Many authors take into consideration countably infinite families of observables that are simultaneously measurable; even  though   this has no experimental value, the mathematical procedure is easily extendible by considering infinite Cartesian products.
\\ 
Briefly,  if  we denote by $\mathbb R^\mathbb N$ the set of maps $\xi:\mathbb N \rightarrow \mathbb R$  that are zero almost everywhere, i.e. the set $\xi^{-1}( \mathbb R \setminus \{0\})$ has finite cardinality, we can consider the following subsets of $\mathbb R^\mathbb N$ which are called cylinders:
$$C(\Delta, k) = \left\{ \xi\in \mathbb R^\mathbb N : \xi(k)\in\Delta \ , \forall \Delta\subset\mathbb R , k\in \mathbb N \right\} $$
and the $\sigma$-algebra generated by the following family of sets
$$ \left\{ \bigcap_{k\in \mathbb N} C(\Delta, k) : \ \Delta \subset\mathbb R \right\} $$
Having a family of non-complementary observables $\left\{ a_j \right\}_{j\in\mathbb N }$ prepared according to their natural order, we can consider the proposition $A_k$:  
\begin{equation*}
A_{k } =\left\{ 
\begin{array}{cc}
 a_k \in \Delta  & j=k \\ 
 a_j\in \mathbb R  & j\neq k
\end{array}%
\right.  
\end{equation*}
 Then we obtain  
$$ P(A_k)=\nu(C(\Delta, k))  \ ,  \ \forall \Delta\subset\mathbb R  , k\in \mathbb N $$
\section{Joint Average Value}\label{valoremediocongiunto}
Let $a,b$ be two non-complementary (\textit{not necessarily compatible}) observables of the physical system; we assume that they can be prepared in the order $a:b$ in the state $\omega\in \mathfrak S_{a:b }$.
\\
We can treat the pair $a:b$ as a single two-valued observable of our physical system, where its values are expressed by the joint law  \eqref{leggecongiunta}\footnote{Obviously this discussion easily extends to a finite family of non-complementary observables $a_1 : \cdots : a_k$.}:
$$P(a:b\in \Delta_0 \times \Delta_1)_\omega=P(a\in \Delta_0 : b\in\Delta_1)_\omega \ , \qquad \Delta_0, \Delta_1 \in B(\mathbb R)$$
\textit{From assumption \ref{misuraprodotto1} it is possible to  associate to the expression  \eqref{leggecongiunta}  a probability measure  \eqref{misuraprodotto}  which we will indicate with the symbol $\mu_{\omega, a:b}$:}
\begin{equation}\label{misuracongiunta}
\mu_{\omega, a:b}(\Delta_0 \times \Delta_1) : = P (a:b\in \Delta_0 \times \Delta_1)_\omega \ , \qquad \Delta_0, \Delta_1 \in B(\mathbb R)
\end{equation}
Mathematically, from the regularity of the measure $\mu_{\omega, a:b}$, we can extend this definition to every region $\texttt{R}$, a Borel subset of $\mathbb R^2$:
\begin{equation}
\label{misuraduevalori} 
P (a:b\in \texttt{R})_\omega = \mu_{\omega, a:b}(\texttt{R}) \ , \qquad  \forall \ \texttt{R} \in B(\mathbb R^2)
\end{equation}
and we assume that it \textit{is the probability that "the observable" $a:b$ has value in} \texttt{R} in the state $\omega\in \mathfrak S_{a:b}$.
\\
By remark \ref{statale}, for $\omega\in \mathfrak S_{a:b}$ we can write
$$ P (b\in \Delta)_\omega := \mu_{\omega, a:b}(\mathbb R \times \Delta)  \ , \qquad \forall \ \Delta\in B(\mathbb R )$$
Now let's define the joint average value of $a:b$ through the following vector:
\begin{equation}
\left\langle a:b \right\rangle_\omega = \left( \left\langle a \right\rangle_\omega^{a:b} \ , \  \left\langle b \right\rangle_\omega^{a:b}  \right) \in \mathbb R^2
\end{equation}
where with $\left\langle a \right\rangle_\omega^{a:b}$ we denote the average value of the observable $a$ in the joint measurement of $a$ and $b$ in the state $\omega\in \mathfrak S_{a:b}$\footnote{In fact, this information is already contained in the state $\omega$, since it belongs to $\mathfrak S_{a:b}$. Here, however, we want to emphasize the order in which the measurements are performed. }:
\begin{equation}
 \left\langle a \right\rangle_\omega^{a:b} = \int_{\mathbb R^2} s \  d \mu_{\omega, a:b} (s,t)   
\end{equation}
the same  argument applies to the observable $b$:
\begin{equation}
 \left\langle b \right\rangle_\omega^{a:b} = \int_{\mathbb R^2} t \  d \mu_{\omega, a:b} (s,t)   
\end{equation}
In the joint law $P(a\in \Delta_1 : b\in\Delta_2)_\omega$\footnote{Recall that to determine the joint frequencies  \eqref{frequenzecongiunte}  \textit{a priori} we do not know anything about the possible values of $a,b$; we must perform the (infinite) simultaneous measurements after which we note down their values and \textit{a posteriori} we study whether these values are in $\Delta_1$ for $a$ and in $\Delta_2$ for $b$ and calculate the joint frequencies  \eqref{frequenzecongiunte}.} the values of $b$ were influenced by the previous preparation of $a$ and, as argued previously, we can refer to the case of a single measurement of $b$ always in the same state $\omega\in \mathfrak S_b$ since $\mathfrak S_{a:b} \subset\mathfrak S_b $ and for the marginal distribution probability
\begin{equation}
\label{statocond2}
 P(a \in\mathbb R : b\in \Delta)_\omega =  P(b\in \Delta   )_{\omega} = \mu_{\omega , b}( \Delta) 
\end{equation} 
 These arguments, as we have previously argued, do not apply to the measure of $a$; we can only say that there exists a Borel measure $\eta$ defined by
\begin{equation}\label{statocond} 
\eta(\Delta):= P(a\in \Delta  : b\in\mathbb R)_\omega   \ , \qquad \Delta \subset \mathbb R
\end{equation}
and in this case,  it is proved that  we have the following relations\footnote{The proof is not essential for the discussion that follows; however, it can be found on page \pageref{verifiche}.}:
\begin{equation}
\label{statocond3}
\left\langle a \right\rangle_\omega^{a:b} = \int s \  d \eta(s)     \qquad  \ ,  \qquad \left\langle b \right\rangle_\omega^{a:b} = \int t \  d \mu_{\omega,b} (t) \end{equation}
the second relation, written compactly, becomes
$$ \left\langle b \right\rangle_\omega^{a:b} =  \left\langle b \right\rangle_{\omega} $$
Obviously if $a$ and $b$ are compatible then their average values can be "\textit{disentangled}" from the mutual influences of their preparation, and from definition \ref{definizionecompatibile} we can write\footnote{In other words, in the compatible case we have $\eta= \mu_{a,\omega}$ because $\omega\in\mathfrak S_a$.}: 
$$\left\langle a \right\rangle_\omega^{a:b} =  \left\langle a \right\rangle_{\omega} \qquad , \qquad 
\left\langle b \right\rangle_\omega^{a:b} =  \left\langle b \right\rangle_{\omega} $$
 We remark that a physical quantity can be a vector quantity with components given by $a_i , \ i=1,2\ldots n$. These components are treated as if they were different physical quantities and therefore it is not  certain   that they are jointly measurable (or compatible), but when  this  holds, we have a single physical quantity with multiple values, for example $a_1:a_2:\cdots :a_n$.
\\

Recall that if $a$ and $b$ are compatible observables\footnote{The topic will be discussed extensively in section \ref{osservabilicompatibili}.} then we have two observables $a:b$ and $b:a$ connected by  \eqref{congiunto0}:
$$P(a:b\in \Delta_1 \times \Delta_2)_\omega = P( b:a\in \Delta_2 \times \Delta_1)_\omega $$ 
therefore
$$ \mu_{\omega, a:b} (\Delta_1 \times \Delta_2)= \mu_{\omega, b:a} (\Delta_2 \times \Delta_1)$$
\section{Subsequent Measurements and Borel's Property} 
In this section we ask whether subsequent measurements of observables can determine  a  Borel measure, as in the simultaneous case.
\\
We consider a state $\omega\in\mathfrak S^{a<b}$\footnote{Let us remember again that in the state we have set the instruction of the measurements at the times $\tau_1$ and $\tau_2$ of the two observables and the related reading times, which may also be different.},  by  definition we have:
$$ P(a\in \Delta_1 | a<b \ , \tau_1)_\omega = \mu_{\omega , a}(\Delta_1)$$
with 
\begin{equation}\label{marginale_succ_1} 
P(a\in \Delta_1 | a<b \ , \tau_1)_\omega = P(a\in \Delta_1  \wedge b \in \mathbb R)_\omega
\end{equation}
while for the second measure we obtain
$$ P(b\in \Delta_2 | a<b \ , \tau_2)_\omega= \mu_{\omega , b}(\Delta_2)$$
where we can still write
\begin{equation}\label {marginale_succ_2} 
P(b\in \Delta_2 | a<b \ , \tau_2)_\omega = P(a\in \mathbb R \wedge b \in  \Delta_2)_\omega 
\end{equation} 
Practically we determine the following numbers
$$n(a\in \Delta_1 | a<b \ , \tau_1)_\omega$$ 
which indicates the number of times that the observable $a$ takes  a  value in $\Delta_1$ at time $\tau_1$ when we have not yet  carried  out the second scheduled measurement on $b$ at time $\tau_2>\tau_1$; for this second measurement we obtain the values
$$n(a\in \Delta_1 , \tau_1 \wedge b \in  \Delta_2 , \tau_2 )$$ 
where in the first measurement we obtain a value in $\Delta_1$ for $a$ and in the second measurement a value in $\Delta_2$ for $b$.
\\
Moreover, the writing
$$n(a\in \mathbb R , \tau_1 \wedge b \in  \Delta_2 , \tau_2 )$$
 shows  that we are not interested in knowing the value of the observable  $a$  obtained in the first measurement, and in the same way
$$ n(a\in \Delta_1  , \tau_1 \wedge b \in \mathbb R , \tau_2 ) $$ 
 shows  that we are not interested in the value of the observable $b$ obtained in the second measurement.  Thus  the previous equalities \eqref{marginale_succ_1} and \eqref{marginale_succ_2} follow.
\bigskip

Let us now ask ourselves whether the  map
\begin{equation} \label{misura_congiunta}
\Delta_1 \times \Delta_2\in B(\mathbb R^2) \longmapsto P(a\in \Delta_1 , \tau_1 \wedge b \in  \Delta_2 , \tau_2 )_\omega \in [0 , 1]  
\end{equation} 
establishes, as in the simultaneous case, a Borel measure $\mu_{\omega , a<b}$.
\\
\textbf{If we have an affirmative answer}, then we can say that the marginal measures of $\mu_{\omega , a<b}$ are given by $\mu_{\omega,a}$ from \eqref{marginale_succ_1} and by $\mu_{\omega,b}$ from \eqref{marginale_succ_2}.
\bigskip
\\
We underline that these considerations easily generalize to multiple observables in the system.
\bigskip

Differently from the simultaneous case, we have the following definition for joint probabilities from subsequent measurements:
\begin{definition}[\textbf{Kolmogorov's Property}] \upshape\index{Kolmogorov's property}
\label{misuraprodotto1b}
A family of observables $a_1<a_2< \ldots < a_n$ jointly measurable by subsequent   measurements at times $\tau_1 < \tau_2 < \ldots < \tau_n$ in the state $\omega \in\mathfrak S^{a_1 <a_2< \ldots < a_n}$ which admits a probability measure $\mu_{\omega , a_1<a_2< \ldots < a_n}$
that satisfies the relation \eqref{misura_congiunta}:
$$\mu_{\omega , a_1<a_2< \ldots < a_n} (\Delta_1 \times \Delta_2 \times \cdots \times \Delta_n) = P(a_1\in \Delta_1 , \tau_1 \wedge a_2 \in  \Delta_2 , \tau_2 \wedge \cdots  \wedge a_n \in \Delta_n , \tau_n)_\omega$$
is called a family of Kolmogorov observables in the state $\omega$. 
\end{definition}
We observe that for such a family of observables, the Bayes rule for conditional probabilities takes the following form
\begin{eqnarray*}
P(a_n\in \Delta  , \tau_n | a_1 \in  \Delta_1 , \tau_1 | \cdots | a_{n-1} \in \Delta_{n-1} , \tau_{n-1})_\omega   =  
\\  
  =   
\frac{\mu_{\omega , a_1<a_2< \ldots < a_n} (\Delta_1 \times \Delta_2 \times \cdots \times \Delta_{n-1}\times \Delta)}{\mu_{\omega , a_1<a_2< \ldots < a_n} (\Delta_1 \times \Delta_2 \times \cdots \times \Delta_{n-1}\times \mathbb R)}
\end{eqnarray*}

\section{Kolmogorov's Physical Model} 
Experimental data are always measurements conditioned by other factors; in our model we have incorporated these factors into the definition of the state $\omega$ of the system.
\\
In quantum probability, we have the possibility that the statistical data, obtained through experimental measurements (not only relating to the world of physics), may \textit{not follow} a classical probability model established by Kolmogorov's axioms.
\\
This intuition is mainly due to Accardi (see \cite{Accardi84, Accardi88}) and is based on the important notion of Kolmogorov's property of statistical data which we briefly re-elaborate in this section.
\\
Let us consider a family $\mathfrak F$ of sets which we will call \textit{propositions} and we assume that to each proposition $A,B \in\mathfrak F$ we can assign  a conditional probability $P(A|B)$, \textit{i.e.} a number in the interval $[ 0 , 1]$ with the property  
$$ P(A|A)=1 \ , \qquad \forall A \in \mathfrak F $$
We have the following definition:
\begin{definition}\upshape\label{kolmodel}
We consider the pair $( \mathfrak F , P(\cdot | \cdot) )$ consisting of a set of propositions $\mathfrak F$ and a conditional probability $P(\cdot | \cdot)$.
\\
A set of propositions $(A_1, A_2 \ldots A_n)$ of $\mathfrak F$ admits a \textsl{Kolmogorovian model} if there exists a probability space $(\Omega, \Sigma, \nu)$ such that
\begin{itemize}
\item  $A_1, A_2 \ldots A_n\in\Sigma$\footnote{To be more precise, the propositions $A_1, A_2 \ldots A_n$ can be uniquely identified as subsets of $\Omega$. Here we will assume, in order not  to  overload the notations, that they are themselves subsets of $\Omega$. } with 
$$\bigcup_{i=1}^n A_i = \Omega $$
\item Bayes formulas are satisfied:
\begin{equation}\label{kom_bay}
P(A_i|A_j)=\frac{\nu(A_i\cap A_j)}{\nu(A_j)}  \ , \qquad \forall i=1,2\ldots n
\end{equation}
\end{itemize}
\end{definition}
\bigskip

Let's apply these arguments to our experimental data obtained from our hypothetical laboratory.
\\
Given an observable $a\in\mathfrak X$ we can associate the following proposition with it: 
\begin{center}
\textit{$A $ : the value of the observable $a \in\Delta $ at time $\tau $ measured in the state $\omega$.}
\end{center}
We denote by $\mathfrak F$ the set of all these propositions when the parameters $a$, $\Delta$ and $\omega$  vary.
\\
If, on the one hand, we have no problems in defining the negation of the proposition $A$, we have serious operational difficulties in defining the conjunction of two similar propositions. In fact, if $A,B\in\mathfrak F$, as we pointed out in the previous sections, it is not always possible to have a proposition $A\wedge B$ in this family.
\\
We can have a joint proposition if the various preparations of the observables do not destroy the previous preparations for the measurement in the state taken into consideration.
\\
Practically we must assume that our observables are at least jointly preparable.
\\

Let us now consider a generic family of observables $\left\{ a_j \right\}_{j =1,2\ldots n }$ of our physical system $(\mathfrak X, \mathfrak S)$ that we want to measure. 
\\
We assume that they are \textit{jointly preparable for subsequent measurements} and that it is possible to do so in their natural order $a_1 , a_2 , \ldots a_n$ in the state $\omega\in\mathfrak S^{a_1<a_2 < \cdots < a_n} $ and we consider the following propositions:
\begin{equation*}
A_j  =\left\{ 
\begin{array}{ccc }
\textit{the value of the observable} \ a_j\in\Delta_j   & \ \textit{at time} \ \tau_j  & \  \\ 
\textit{the value of the observable} \ a_i\in\mathbb R   & \ \textit{at time} \ \tau_i & \  i \neq j
\end{array}
\right. 
\end{equation*}
As established in section \ref{misurecondizionate}, for each pair of  events $A_i,A_j\in\mathfrak F$ we can assign the  conditional probability $P(A_i|A_j)$ as established by the frequencies given by the relation \eqref{condiziofreq1}:
$$P(A_i|A_j)=P(a_i\in\Delta_i , \tau_i  | \ a_j\in\Delta_j , \tau_j)_\omega  \ , \qquad \tau_j < \tau_i $$
\begin{attenzione}\upshape
You might think that in this case we have only  carried  out two measurements (even though we have prepared all the observables) on $a_i$ and $a_j$ at their respective times, but this is obviously false\footnote{To avoid confusion, we would have had to write the following expression in full:
$$P(A_i|A_j)=P(a_n \in \mathbb R , \tau_n \ldots | \ldots a_i\in\Delta_i , \tau_i \ldots | \ldots \ a_j\in\Delta_j , \tau_j , \ldots | a_1\in\mathbb R , \tau_1 )_\omega $$}.
\end{attenzione}
Continuing with our discussion, we can say that to each proposition $A_k$ we can associate the following subset of $\mathbb R^n$:
$$ A_k= (\mathbb R \times \cdots \times \mathbb R \times \Delta_k \times \mathbb R \times \cdots \times \mathbb R) \in B(\mathbb R^n)$$ 
and if our observables $\left\{ a_j \right\}_{j =1,2\ldots n }$ satisfy Kolmogorov's property in the $\omega$ state,  then  by definition there exists a probability measure $\mu_{\omega , a_1<a_2< \ldots < a_n}$ on $B(\mathbb R^n)$ such that
 \begin{equation}
 \mu_{\omega , a_1<a_2< \ldots < a_n}(\Delta_n \times \cdots \times \Delta_2 \times \Delta_1 ) = P( A_n \wedge \cdots \wedge A_2 \wedge A_1)  
\end{equation}
with
$$ P(A_j| A_i )= \frac{\mu_{\omega , a_1<a_2< \ldots < a_n}(\mathbb R \times \cdots \times \Delta_j \times \mathbb R \times \cdots \times \mathbb R \times \Delta_i \times \cdots \times \mathbb R)}{\mu_{\omega , a_1<a_2< \ldots < a_n}(\mathbb R \times \cdots \times \Delta_i \times \cdots \times \mathbb R)}$$
for each (Borel) set $\Delta_j \subset \mathbb R \ , j=1,2\ldots n$.
\bigskip

Let us now consider the conditional probability defined in section \ref{cond_statale} and let $B$ be the following proposition:
\\

\textit{$B$ : the value of the observable $b \in\Delta_o $ at time $\tau $ measured in the state $\omega\in \mathfrak S_b$.}
\\
 
We assume that 
$$P(b\in\Delta_o , \tau)_\omega=1$$
and \textit{if this state also belongs} to $\mathfrak S_a$, then we can consider the probability 
$$P(A|B)=P(a\in\Delta, \tau)_\omega = \mu_{\omega,a}(\Delta)$$
but in this case we cannot say that a probability space $(\Omega, \Sigma, \nu)$ exists for which relation  \eqref{kom_bay}  is satisfied.
 
\begin{remark}\upshape\label{misurasingola}
 We reiterate that only the study of the frequencies obtained experimentally, as we discussed in the previous sections, can establish whether or not a set of observables has a Kolmogorovian model.\footnote{An important application of these considerations can be found in the works of Accardi \cite{Accardi84, Accardi85, Accardi88, Accardi97} where the author introduces the fundamental notion of \textit{statistical invariants} and the double-slit experiment and Bell localization are described using this notion.}
\end{remark}

In conclusion we have the following
\begin{remark}\upshape
Remark \ref{statale} does not affirm that there is a joint measure\footnote{Otherwise the measurements of all observables would be described by Kolmogorovian models.} but it only tells us that the measurement of an observable, which was last prepared in the laboratory, can be described by Borel measures. 
\end{remark}

\chapter{Frequencies and Probability*}\label{legge_empirica}
In this section we want to study in more detail how the Borel measure of a \textit{single measurement of an observable $a$} is established through the study of the relative frequencies obtained from the experimental data.
\\

The transition from the experimental relative frequencies $f(a\in\Delta)_\omega$ of equation  \eqref{freq}  to the distribution of probabilities $P(a\in\Delta)_\omega$ of equation  \eqref{distribuzio1}  is not without conceptual problems,  difficulties  addressed mainly by von Mises in the 1930s and exposed in his book \textit{The Mathematical Theory of Truth} (see \cite{Mises}).\index{von Mises}
\\
The problem, as is well known, is that the frequencies $f_n(a\in\Delta)_\omega$ do not admit a mathematically rigorous limit when the number of experimental trials $n\rightarrow \infty$;  therefore  we cannot use this   mathematical  tool  to  determine the distribution of probabilities given in \eqref{distribuzio1}.
Von Mises attempts to resolve this problem by introducing the notion of  Kollektiv (collective).
\\
As we will see briefly in section \ref{kollettiv}, the notion of  collective, even if it conceptually solves this problem, does not solve the problem of practically determining the distribution of probabilities given in \eqref{distribuzio1} and therefore the average value of the physical quantity $a$ in the state $\omega$ given by  \eqref{valoremedio0}\footnote{See also von Mises, appendix 2 of \cite{MisesII}.}.
\\
In keeping with this line of thought, we have van Lambalgen's observation in section 4 of \cite{Lamb96}:\index{van Lambalgen}
\\
\textit{Von Mises was aware that collective cannot be explicitly constructed, so that the consistency of the theory can be established only indirectly
[...] Collective are new mathematical objects, not constructible from previously defined objects.}
\\
Here we will try another way: we will use some well-known tools from ergodic theory. In fact, we observe that although experimentally the relative frequencies $f_n(a\in\Delta)_\omega$ do not admit a limit $p$, the numbers $n$ that prevent such convergence to $p$ when the number of trials $n$ increases become more and more \textit{rare}, but  remain always infinite in number.
\section{On Density}
Given any set $F\subset\mathbb N$ we denote its cardinality by $Card(F)$ and on the set $\mathbb N^+$, for every natural number $n>0$, we have the following probability measure:
\begin{equation}\label{misura_natural}
m_n(E) = \frac{Card(E  \cap \mathbb I_n^+ ) }{n}  \ , \qquad \forall E\subset\mathbb N^+
\end{equation}
with support in $\mathbb I_n^+$, where
$$\mathbb I_n^+=\left\{k\in \mathbb N : 1 \leq k \leq n     \right\}\subset \mathbb N^+$$
while with $\mathbb I_n$ we denote the set 
$$\mathbb I_n=\left\{k\in \mathbb N : 0 \leq k \leq n     \right\}\subset \mathbb N$$
Let us briefly recall the definition of density of a set $E\subset\mathbb N^+$ (see N.S.Z. \cite{N.S.Z.}, par. 9 appendix B).
\\
We define the following objects\index{Upper density}\index{Lower density}
\begin{equation}\label{density_natural_up} 
D^*(E ) = \max\lim_{n\rightarrow \infty}m_n(E)     
\end{equation}
 and 
\begin{equation}\label{density_natural_down} 
D_* (E ) = \min\lim_{n\rightarrow \infty}m_n(E)      
\end{equation}
which are respectively called \textit{upper density} and \textit{lower density} of the set $E$.
\\
A set $E\subset\mathbb N^+$ admits density $D(E)$ when the lower density coincides with the upper density; in other words:
\begin{equation}\label{density_natural} 
D(E ) = \lim_{n\rightarrow \infty}m_n(E)    
\end{equation}
A set $E$ is said to have zero density  if $D(E)=0$\footnote{For example, if $E$ is the set of all prime numbers, then by Gauss' law we have
$$Card(E\cap \mathbb I_n^+) \approx \frac{n}{\log(n)}$$
therefore
 $$ m_n(E)\approx \frac{1}{\log(n)} \rightarrow 0$$
and the set of prime numbers has zero density. } .
\bigskip

A sequence $\left\{x_n\right\}_{n\in \mathbb N^+}$ of real numbers converges in density to $0$, in symbols
$$ D-\lim_{n\rightarrow +\infty} x_n =0$$
if there exists a set $E\subset \mathbb N^+$ of zero density such that
$$  \lim_{n\rightarrow +\infty} x_n  \ \mathbf{1}_{\mathbb N \setminus E}=0$$
The sequence $\left\{x_n\right\}_n$ converges in density to $l\in\mathbb R$ if the sequence $\left\{ (x_n - l) \right\}_n$ converges in density to zero; in symbols
$$ D-\lim_{n\rightarrow +\infty} x_n =l$$
We are now ready to state a well-known result of ergodic theory:
\begin{lemma}[\textit{Koopman-von Neumann}]\upshape\label{Koopman-Neumann}\index{Lemma of Koopman-von Neumann}
If $\left\{x_n\right\}_n$  is a bounded sequence of non-negative real numbers, then the following statements are equivalent:
\begin{itemize}
\item [1] The sequence converges in density to zero,
\item [2] The set $F_\alpha$ with $\alpha \in \mathbb R^+$
$$F_\alpha= \left\{n\in\mathbb N^+ : x_n >\alpha\right\}$$
has zero density for each $\alpha$,
\item [3] The Cesàro sum tends to zero:
$$ \lim_{n\rightarrow +\infty} \frac{1}{n} \sum_{k=1}^n x_k =0 $$
\end{itemize}  
\end{lemma}
For the proof see Petersen \cite{Petersen} lemma 2.6.2.

\section{Frequency vs. Probability}
In this section we will establish mathematical methodologies to determine the probability distribution law given in equation  \eqref{distribuzio1}\footnote{And therefore the probability measure $\mu_{a,\omega}$} through the study of the values of the experimental relative frequencies $f_n(a\in\Delta)_\omega$ given in equation \eqref{freq}.
\bigskip
\\
In the previous sections we assumed that there exists a probability measure $\mu_{a,\omega}$ such that the average value of the observable $f(a)$ can be expressed in the following way:
\begin{equation}\label{distribuziopb1}
\left\langle f(a)\right\rangle_{\omega } =\int f(s) \ d \mu_{a,\omega}(s) \qquad , \qquad \forall f\in C_o(\mathbb R)
\end{equation}
Moreover the measure $\mu_{a,\omega}$ is the \textit{unique} probability measure that satisfies equation  \eqref{distribuziopb1}\footnote{In fact, if we have a new probability measure $\mu$ such that  
$$\left\langle f(a)\right\rangle_{\omega } = \mu (f)$$
then $\mu (f)= \mu_{a,\omega}(f)$ for each $f\in C_o(\mathbb R)$.}.
\bigskip

The problem now is to determine, once existence is assumed, the probability measure $\mu_{a,\omega}$.

\begin{figure}[htbp]
	\centering
		\includegraphics[scale=0.5]{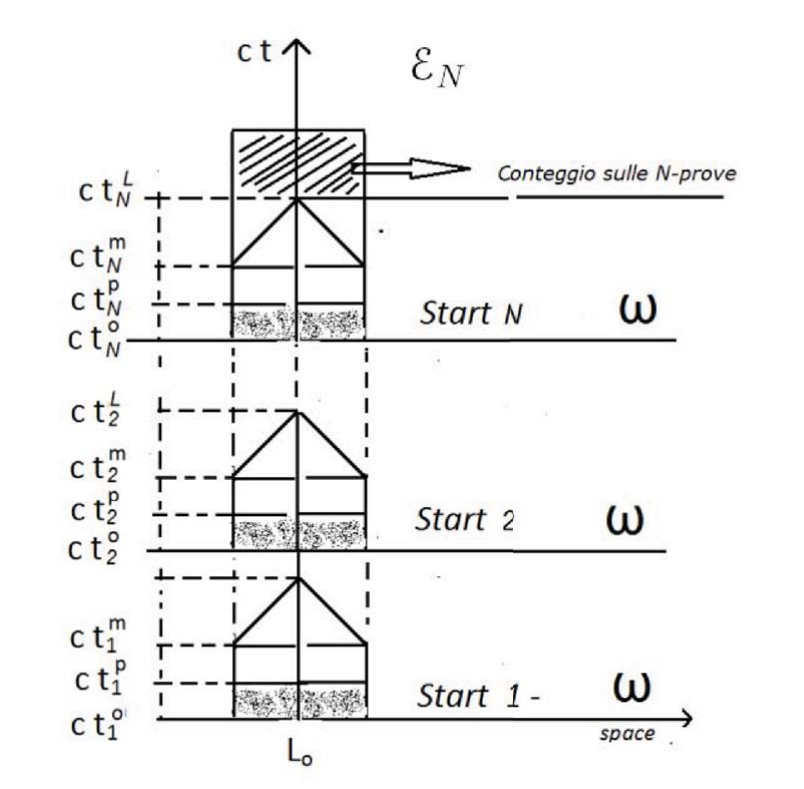}
	\caption{Yes/No counting on $N$ measurements carried out}
	\label{fig:07_contattore}
\end{figure}

Let's give some considerations on ensembles and probability. 
\\
If $\mathcal E_n$ is an ensemble in which the $n$ measurements of the same experiment are  carried  out, briefly referred to as $n$ copies (see Figure \ref{fig:07_contattore}), then we can define the following map:
\begin{equation}\label{vc_idd}
 i\in \mathbb I_n^+  \longrightarrow X_n(i) \in \left\{0,1 \right\}
\end{equation}
where for every $i\in\mathbb I_n^+$:
\begin{equation*}
 X_n(i)  =\left\{ 
\begin{array}{ccc }
1 \ &  \textit{if on the i-th trial we obtain } \ a \in\Delta  \ (\textit{success})  \\ 

0 \ & \textit{if on the i-th trial we obtain} \ a \notin \Delta \ (\textit{failure})
\end{array}
\right. 
\end{equation*}
If we denote with
\begin{equation}\label{frequenze_legge_00}
 \xi(n) = \sum_{i=1}^n  X_n(i) \in \mathbb I_n
\end{equation}
the number of successes obtained in the $n$ copies that constitute the ensemble $\mathcal E_N$, then we can write 
\begin{equation}\label{frequenze_legge_0}
 f_n(a\in\Delta)_\omega = \frac{\xi(n)}{n} \in [0,1]
\end{equation} 

Let's now make some mathematical considerations on frequency and density.
\\
We note that it is possible to give another mathematical expression for our relative frequency. Indeed, let us consider the set
\begin{equation}\label{frequenze_legge_1}
E_n=\left\{j\in\mathbb I_n^+: X_n(j)=1 \right\}\subset \mathbb I_n 
\end{equation}  
we have
$$ Card (E_n) =\xi(n)$$
therefore
$$ E_n= E_n\cap \mathbb I_n^+\subset \mathbb N^+$$
it follows that we can write
\begin{equation}\label{frequenze_legge_2}
 f_n(a\in\Delta)_\omega = \frac{Card (E_n)}{n}=m_n(E_n)
\end{equation}
where $m_n$ is the measure defined in expression  \eqref{misura_natural}.
\\
Obviously $E_m$ being  bounded  has zero density, since
$$m_n(E_m)=\frac{Card (E_m)}{n} \ , \qquad  \forall m<n$$
so for every $m\in\mathbb N$ we obtain
$$D(E_m)= \lim_{n\rightarrow \infty} \frac{Card (E_m)}{n}=0 $$
Let's assume that there exists a map $X_\infty:\mathbb N^+\longrightarrow \left\{0,1 \right\}$ such that\footnote{As we will see later, in the case of \textit{copy added} into an ensemble, we obtain, by definition of copy added, that for every $m,n\in\mathbb N^+$ with $m\leq n$ 
$$X_m (j)= X_n (j) \ , \qquad \forall j\leq m $$ 
Furthermore, in this case, we hypothesize the possibility (only theoretical) to indefinitely increase the number of $n$ copies and explicitly obtain the map $X_\infty$.}
$$X_\infty |_{\mathbb I_n^+}=X_n \ , \qquad \forall n\in\mathbb N^+ $$
and if
$$ E_\infty =\left\{j\in\mathbb N^+: X_\infty(j)=1 \right\}$$
then we have
$$D(E_\infty)=D-\lim_{n\rightarrow +\infty} m_n(E_\infty)=D-\lim_{n\rightarrow +\infty}f_n(a\in\Delta)_\omega$$
since 
$$E_\infty \cap \mathbb I_n^+ = E_n \ , \qquad \forall n\in\mathbb N^+ $$
it follows that if the set $E_\infty$ \textit{admits density}, then 
$$D(E_\infty)=p$$
We must make now a necessary physical remark:\index{Severi}
\\
Experimentally the data obtained through measurements can be \textit{non-predictable and this non-predictability, regardless of its origin, characterizes them as random variables}\footnote{See Severi's book \cite{Sev85} Cap. IX, par. 15.}.
\\
\begin{figure}[htbp]
	\centering
		\includegraphics[scale=0.4]{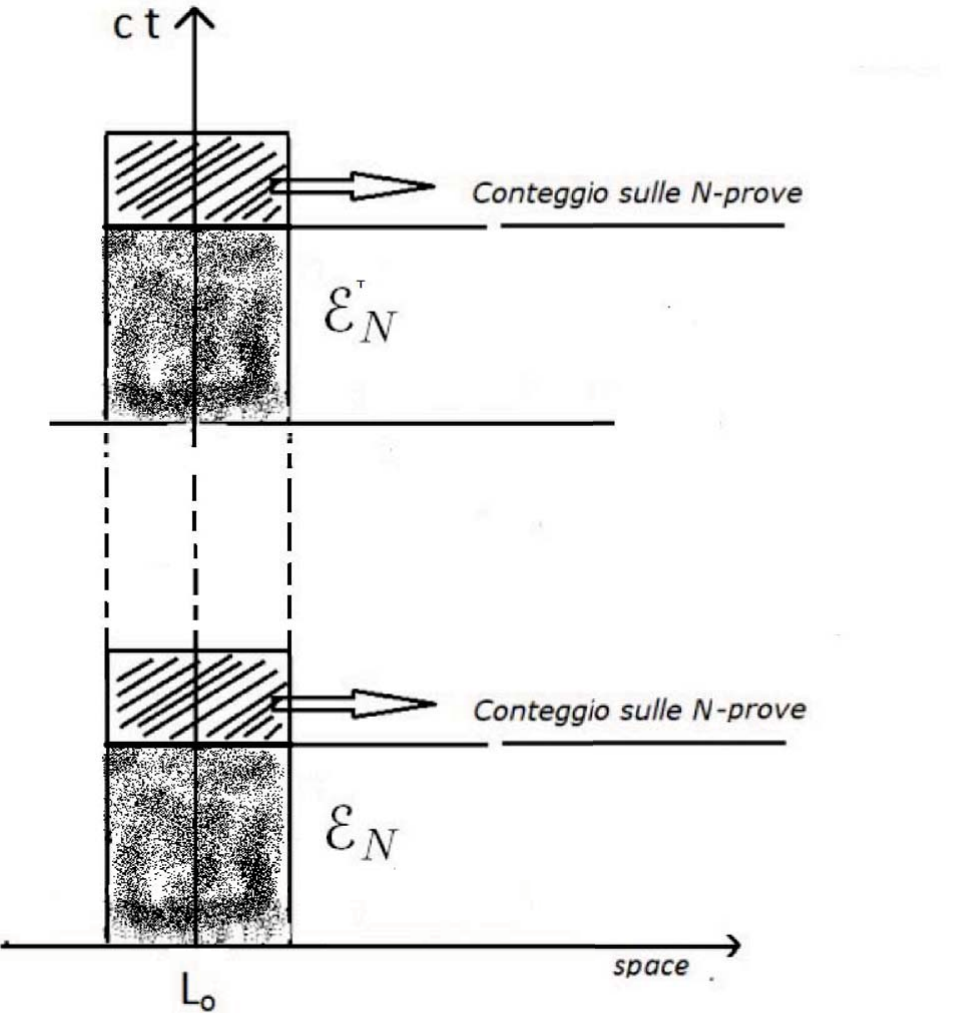}
	\caption{Yes/No count on the $N$ trials  carried  out in each ensemble}
	\label{fig:07_contattore_ripetuto}
\end{figure}
\\
For example, we can repeat the count of how many times we obtain $a\in\Delta$ in the $\omega$ state over $N$ trials, through a new ensemble prepared later\footnote{A young investigator could repeat in the future, in the same laboratory, the same experiment we did previously.}, as in Figure \ref{fig:07_contattore_ripetuto}, at a time interval $T$, which we briefly indicate with $\mathcal E^T_N$ to distinguish it from our $\mathcal E_N$.
\\

In this way we always obtain the random variable
$$ i\in \mathbb I_N  \longrightarrow X^T_N(i) \in \left\{0,1 \right\}$$
but we can have, due to non-predictability, the following result:
$$ X_N(i) \neq X^T_N(i)$$
 for some  same   $i$-th copy in the two ensembles and therefore also obtain a result $\xi(N)\neq \xi^T(N)$ with relative frequencies  \eqref{frequenze_legge_0}  not equal.
\index{Piccinato} 
 We recall that \textit{statistical procedures must be evaluated for their behaviour in hypothetical repetitions of the experiment, which is always assumed to be carried out under the same conditions}\footnote{See Piccinato \cite{Piccinato} par. 4.5, the \textit{principio del condizionamento ripetuto}.}.
\\
These considerations lead to the following
\begin{remark} \upshape
We cannot establish the measure $\mu_{\omega,a}$ with a single relative frequency value $f_N(a\in\Delta)_\omega$ obtained through the $N$ repetitions of our experiment, but we must vary the number of $N$ trials which compose our ensemble. 
\end{remark}
This statement leads to another problem:
\\
How can we experimentally increase the number of $N$ trials that make up our ensemble? 
\\

Furthermore, the following fact of life must be emphasized:
\\
\textit{Even if theoretically we can set up ensembles with any number of trials $N$,  even very high, they will still be limited.}
\subsubsection{First case - the Copy Added}
\begin{figure}[htbp]
	\centering
		\includegraphics[scale=0.5]{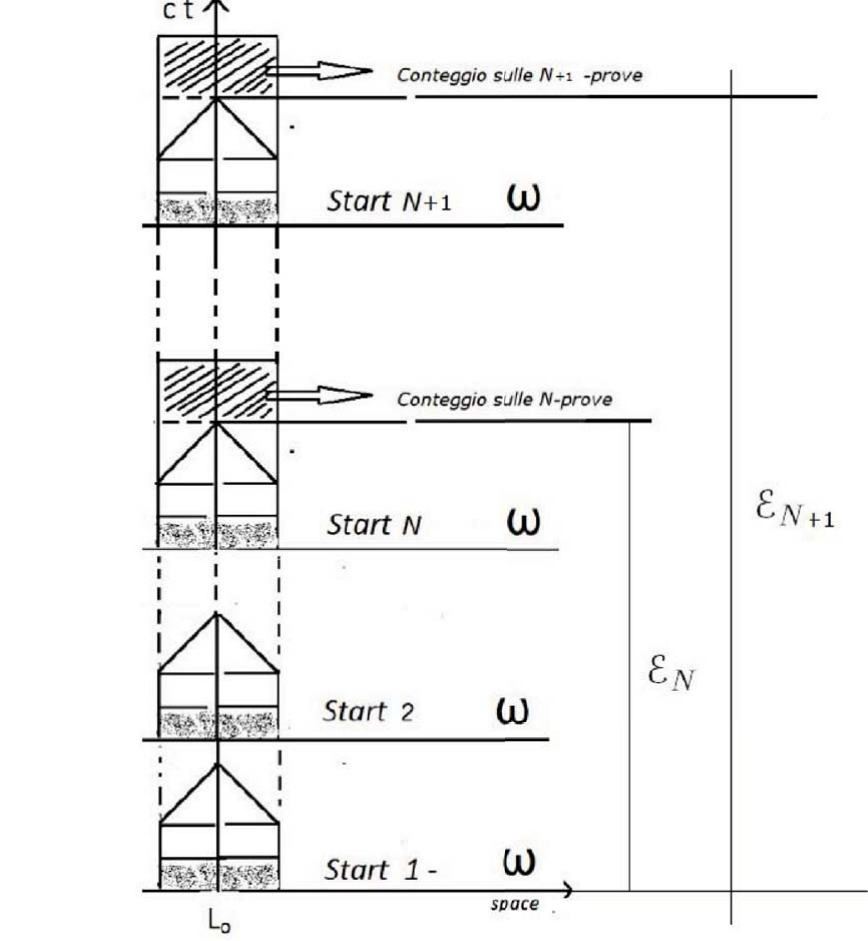}
	\caption{Yes/No count on $N+1$ trials}
	\label{fig:07_contaggio_successivo}
\end{figure}

Let us consider a family of ensembles $\left\{\mathcal E_n \right\}_{n\in\mathbb N^+}$ where
$$  \mathcal E_n \preceq  \mathcal E_{n+1} \ , \qquad \forall n\in\mathbb N $$
The symbol $\preceq$ indicates that $\mathcal E_{n+1}$ is  composed  of the trials contained in the ensemble $\mathcal E_{n}$ plus a new trial of the same experiment prepared identically to the other trials of $\mathcal E_{n}$.
\\
In this way we can define a map $\xi:\mathbb N^+ \longmapsto \mathbb N$ as follows:
\\
The value $\xi(n)$ is the number of times, out of $n$ experimental trials, that we obtain the value $a\in\Delta$; therefore we can write:
\begin{equation}\label{frequenze_legge}
n\in\mathbb N^+ \longrightarrow f_n(a\in\Delta)_\omega = \frac{\xi(n)}{n} \in [0,1]
\end{equation} 
We have for every $n\in\mathbb N$ the following obvious properties:
\\
$1.$ $ \  \xi(n) \leq n \ , \qquad \forall n\in\mathbb N^+  $ 
\\
$2.$ $ \  \xi(n) \leq \xi(n+1) \ , \qquad \forall n\in\mathbb N^+$\footnote{Since we add the   trials  to the ensemble $\mathcal E_N$.}
\\
\begin{remark}\upshape
Counting the results of the $n$ trials does not disturb the outcome of the $(n+1)$-th trial.
\end{remark}
\subsubsection{Second Case - The Repeated Trials}
\begin{figure}[htbp]
	\centering
		\includegraphics[scale=0.35]{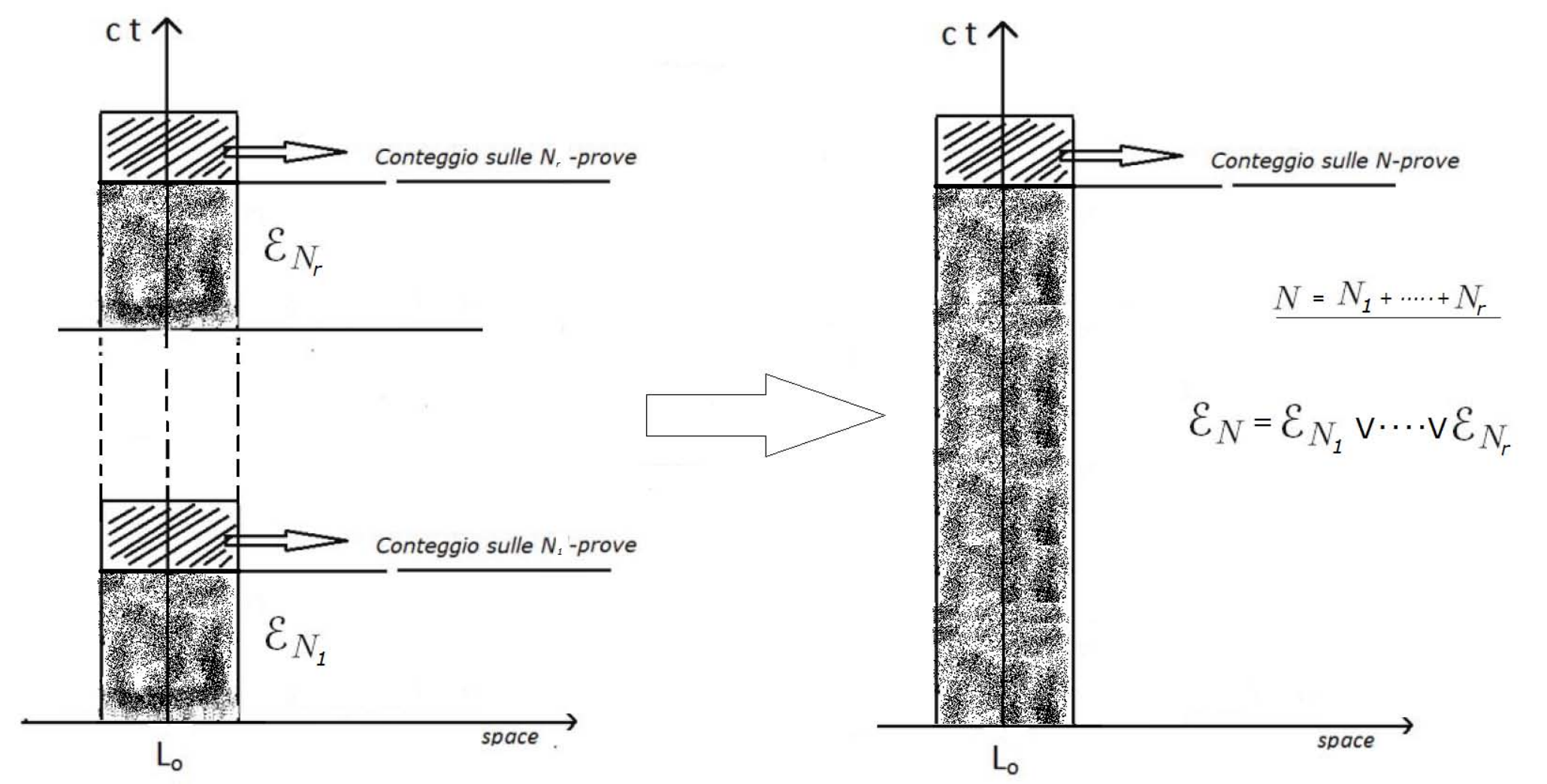}
	\caption{Yes/No count with ensembles}
	\label{fig:07_batteria}
\end{figure}

We can decide to repeat the experiment by setting up (over time) $r$ ensembles 
$$\left\{ \mathcal E_{N_1}, \mathcal E_{N_2} \ldots \mathcal E_{N_r} \right\}$$
 each with $N_1, N_2 \ldots N_r$ trials, all prepared in the same state $\omega$ of the physical system.
This obviously does not bring any additional information, since we can consider everything as an ensemble $\mathcal E_N$  the  union of all $r$ ensembles (see Figure \ref{fig:07_batteria}); in symbols\footnote{As we discussed in previous sections, the act of counting does not influence the preparation of subsequent ensembles.}
 $$\mathcal E_N = \mathcal E_{N_1} \vee  \mathcal E_{N_2} \vee   \ldots  \vee   \mathcal E_{N_r} $$
 composed  of $N = N_1 + N_2 + \ldots + N_r$ copies of the same experiment.
\\
In other words, here we have a generalization of the first case; in practice it is as if we add to our ensemble, here  composed  of $N_1$ copies, the other $N_2 + \ldots + N_r$ copies instead of just one copy as in the previous case.

\section{Ensembles and Collectives}\label{kollettiv}\index{von Mises} \index{Khrennikov}
In this section we will apply the notion of  collective  to our ensemble   composed  of $n$ experimental trials, prepared and  carried  out in the exact same way, described in the previous sections as shown in Figure \ref{fig:07_contattore}.
\\
For completeness, we give a heuristic vision of this notion given by von Mises himself, which is found in Appendix 1 of the book \textit{Mathematical Theory of Probability and Statistics} \cite{MisesII}\footnote{For a quick formal exposition and related problems of this definition, see Khrennikov's book \cite{Khre} par. 2.2. 
Furthermore, for a critical discussion of this notion see the work of Home and Whitaker \cite{H.W.} paragraph 2.6.}:
\begin{citazione}
 Probability calculus as presented in this book is based on some concepts and ideas which may be briefly restated in a non-technical way, as follows:
\\
(1) In probability calculus (or probability theory) we consider aggregates of uniform events, observations which can be repeated over
and over, rather than isolated events; each observation leads to a result which can be expressed by a number (or by several numbers). As the
conceptual counterpart of these observations and results, we introduce an infinite sequence $K = \left\{ x_j \right\} $ of numbers representing the results or
labels of the successive observations. For each label $a_i, i = 1, 2, \ldots $ the limiting value of the relative frequency with which it occurs in $K$
exists and is insensitive to place selections applied to the sequence\footnote{For van Lambalgen \cite{Lamb96}:
\\
\textit{An admissible place selection is a procedure for selecting a subsequence of a given sequence $x$ in such a way that the decision to select a term $x_n$ does not depend on the value of $x_n$.}}.
\\
(2) Such sequences are called collectives and the limiting frequency of a label $a_i$ is the probability $p_i$ of $a_i$ in the collective $K$. The $a_i$ together
with the corresponding $p_i$ form the probability distribution.
\\
(3) By means of the repeated use of certain explicitly defined operations, probability distributions in new collectives are derived from given distributions in a given collective.
\end{citazione}\index{van Lambalgen}
Let $\mathcal L$ be the set of possible results of an experiment, which in our case is  composed  of the Yes/No values; in other words
$$\mathcal L = \left\{ 0,1  \right\}$$
From the $n$ trials we  obtain a series of values
$$\underline{h} = \left\{ h_1,h_2\ldots h_n  \right\} \ , \qquad h_i\in\mathcal L \ \forall i=1,2\ldots n $$
We denote with $K(1,\underline{h} )$ and $K(0,\underline{h} )$ respectively the number of times that the value Yes and the value No are  obtained and we define the relative frequencies\footnote{Therefore: 
$$ f_n(1,\underline{h} )= f_n (a\in\Delta)_\omega$$}:
$$f_n(1,\underline{h} )= \frac{K(1,\underline{h} )}{n} \qquad , \qquad f_n(0,\underline{h} )= \frac{K(0,\underline{h} )}{n} $$
A collective is an infinite sequence of numbers
$$h_\infty=\left\{ h_1,h_2\ldots h_n \ldots  \right\} $$  
which extends our finite series of numbers obtained through the experimental trials $\underline{h}$ and satisfies the existence of the limit of the frequencies:
$$ \lim_{n\rightarrow \infty} f_n(1,\underline{h}) = P(h_\infty,1) $$  
and the \textit{randomness properties}.  
\\
\textit{According to von Mises, every sequence $h_\infty$ established through experimental trials satisfies this property.}
\\
The problem is that we do not have an effective experimental procedure to determine how  this  transition
$$ \underline{h}\longrightarrow h_\infty$$
can happen, since the number of trials, even if  very  high, is always finite. 
\\
Therefore, here too, mathematical/statistical methodologies must be established to determine $h_\infty$ starting from the finite case of $n$ experimental trials (see note \ref{bayes0} in section \ref{Statistica e Riproducibilità}).
\section{The Empirical Law of Chance}
As we have said in the previous sections, the fundamental difficulty in establishing the probability measure $\mu_{a,\omega}$ is the experimental impossibility of establishing the law given in  \eqref{frequenze_legge}, since we can only know a part of it; in other words we have:
\begin{equation}\label{frequenze_legge_tronca}
 n\in \mathbb I_N  \longrightarrow f_n(a\in\Delta)_\omega  \in [0,1]
\end{equation}
Therefore the experimenter will have to establish methodologies to obtain knowledge of the mathematical law reported in \eqref{frequenze_legge}  starting from the experimental sample reported in \eqref{frequenze_legge_tronca} for large values of $N$.
\\
Let's see what the  possible solutions  are  that can be adopted to solve this problem.
\\
Let's initially address the question on a mathematical level;  we will consider the case of infinite repetitions of the same trial and see if this idealization can  serve  as a guide in the finite case.

\subsection{Law of large numbers} 
To solve the problem described above, in this section we will use some probability calculation tools\footnote{See Dall'Aglio's book \cite{Aglio} section V.4.}.\index{Dall'Aglio}
\\
The starting point is the introduction of a probability space $(\Omega, \mathcal A, P)$ with a sequence of independent events $\left\{\texttt{E}_n\right\}_{n\in\mathbb N}$ where $$P(\texttt{E}_n)=p \ , \qquad \forall n\in\mathbb N $$
with $p$ a real number $0<p<1$\footnote{As for us $p$ is our unknown probability $P(a\in\Delta)_\omega$.}.
\\
 
We define a family of random variables
$$A^{(n)} :\Omega \longrightarrow \left\{0,1\right\}$$
such that\footnote{Trivially the inverse image measure of the random variable $A^{(n)}$, called the law of $A^{(n)}$, is given by the probability measure $\mu_n$:
$$\mu_n(\left\{1\right\}) = P(A^{(n)}=1 )= P(\texttt{E}_n)=p \qquad , \qquad \mu_n(\left\{0\right\}) = 1-p=q$$
and
$$\textsc{E}(A^{(n)}) =\int_\Omega A^{(n)}(x) dP(x) = \int_{\left\{0,1\right\} } k d\mu_n (k)= p$$
furthermore, the random variables $A^{(n)}$ have the same law $\mu_n$ since $\mu_n =\mu_m$ for each $n$ and $m$.} 
\begin{equation}\label{v.c.prove0}
\texttt{E}_n=\left\{ x\in \Omega : A^{(n)}(x)=1 \right\}  \ , \qquad \forall n\in\mathbb N 
\end{equation}
The number of successes that occur in $n$ trials is given by the random variable
\begin{equation}\label{v.c.prove1}
A_n  = \sum_{k=1}^n A^{(k)}
\end{equation}
so 
$$A_n: \Omega  \longrightarrow \mathbb{I}_n \ , \qquad \forall n\in\mathbb N $$ 
and by Bernoulli, for every $k\in \mathbb I_n$ we have: 
$$P(\left\{ x\in \Omega  : A_n(x)=k \right\}) =   \binom{n}{k} p^k (1-p)^{n-k} = P^{(n)} (\left\{ k \right\})$$
$P^{(n)}$ is a probability measure (inverse image of $A_n$) on the set $\mathbb I_n$\footnote{We recall that if $m$ is the measure that counts the points:
$$m(\texttt{E})=\text{card}(E) \qquad , \qquad \texttt{E}\subset \mathbb I_n$$
then we obtain
$$dP^{(n)} = \binom{n}{k} p^k (1-p)^{n-k} dm$$
therefore the measure $P^{(n)}$ is the law (or distribution) of $A_n$, while the map
$$k\in\mathbb I_n \longrightarrow \binom{n}{k} p^k (1-p)^{n-k}\in \mathbb R$$
is the distribution function of $A_n$.}.
\\
The average of the variables $A_n$ is given by
$$\textsc{E}(A_n)= \int_\Omega A_n(x) dP(x)= \int_{\mathbb I_n} k \ d P^{(n)} =\sum_{k=1}^n k \binom{n}{k} p^k (1-p)^{n-k}  = np $$
and since the $A_n$ are finite sums of the $A^{(k)}$, which are independent and identically distributed random variables, we can apply Bernoulli's theorem (see Dall'Aglio \cite{Aglio} theorem 4.2).
\\
Therefore, for every $n \in \mathbb N $ and $\alpha >0$ we define the following subset of $\Omega$:
$$\texttt{L}_{n,\alpha}=\left\{ x\in \Omega  : |A_n(x)- np | < \alpha \right\}\in \mathcal A$$
and by the above theorem, we can say that for every $\alpha >0$ we have
$$\lim_{n\rightarrow +\infty}P(\texttt{L}_{n,\alpha}) = 1$$
it follows that  
$$\frac{A_n}{n} \longrightarrow p \ , \ \textit{in probability} $$
and for every $\alpha>0$ and $\epsilon>0$ there exists $\bar{n}=n(\epsilon,\alpha)$ such that
\begin{equation}\label{v.c.prove2}
P(\texttt{L}_{n,\alpha}) > 1-\epsilon \ , \qquad  \forall n> \bar{n}
\end{equation}
Therefore, once  we have determined  a possible value for $p$ through the study of the relative frequencies given in relation  \eqref{frequenze_legge_tronca}, we submit this value to the test of inequality  \eqref{v.c.prove2}.
\\
The problem with this set-up is the arbitrary introduction of the initial probability space $(\Omega , \mathcal A, P)$ with the random variables $\left\{ A^{(n)}\right\}_n$  linked by  relation  \eqref{v.c.prove0}, but this procedure is arbitrary  and  unrelated to the experimental procedures.  In practice, how is this probability space (and the events $\left\{ \texttt{E}_n \right\}_n$) established through the ensemble procedure described above?
\\
We conclude this brief probabilistic review by observing that the variable $\xi:\mathbb N^+ \rightarrow \mathbb N$ of relation \eqref{frequenze_legge_00} is naturally connected to the ensemble procedures; the problem is that we do not know how to introduce an appropriate probability measure on $\mathbb N^+$ so as to use Bernoulli's theorem\footnote{One could think of taking the family of subsets of $\mathbb N^+$ that have density and considering the $\sigma$-algebra generated by this family and taking the value of their density as a candidate for the pre-measurement. The problem is that the set $\mathbb I_n$ has zero density and therefore measure zero.}.
\subsection{Empiricism and ergodicity}
Let's start by defining, for each $p\in \left[ 0, 1 \right]$, the following set of natural numbers:
$$ E(p,\alpha)= \left\{ n \in\mathbb N^+ : \left| f_n - p \right| \geq \alpha \right\} $$
and let us take into consideration the set of real numbers
$$ \mathbb F = \left\{ p\in \left[ 0, 1 \right] : D( E(p,\alpha))=0 \ \ \forall \alpha>0  \right\}\subset \mathbb R $$
Obviously this set could be empty because it is not certain that a generic set $E(p,\alpha)$ admits density and that this density is zero.
\\
Let us remember again that in many experimental physics books it is stated that the values of the relative frequencies $ \left\{f_n(a\in\Delta)_\omega \right\}_n$ "tend to stabilize" towards a numerical value that we denote with $P(a\in\Delta)_\omega$.
\\
This vague statement takes on a precise mathematical meaning if we assume that the following hypothesis is true:
\begin{post}[\textbf{Empirical-Ergodic}]\label{ergo_0}\index{Postulate-Empirical Ergodic}
The set $\mathbb F$ is non-empty.
\end{post}
So in this case there exists at least one $ p\in \left[ 0, 1 \right]$ with $D( E(p,\alpha))=0$.
\bigskip

From hypothesis \eqref{ergo_0}, the following two statements, by the Koopman-von Neumann theorem \ref{Koopman-Neumann}, are equivalent:
\begin{itemize}
\item [1.]  There exists a set $E\subset \mathbb N^+$ of zero density\footnote{Obviously it depends on the Borel set $\Delta$.} such that
$$\lim_{n\rightarrow \infty} \left|f_n - p \right| \mathbf{1}_{\mathbb N \setminus E}(n) =0$$
\item [2.]  \label{relation 2.} 
$$\lim_{n\rightarrow \infty} \frac{1}{n} \sum_{k=1}^n \left|f_k - p \right|  =0$$
\end{itemize}
From  1]   we trivially obtain that 
$$\lim_{n\rightarrow \infty} (f_n - p )\ \mathbf{1}_{\mathbb N \setminus E(p,\alpha)}(n) =0$$
in other words
\begin{equation}\label{legge_emp1}
D-\lim_{n\rightarrow \infty} f_n = p 
\end{equation}
From  2] 
\begin{equation}\label{legge_emp2}
p=\lim_{n\rightarrow \infty} \frac{1}{n} \sum_{k=1}^n f_k
\end{equation}
\begin{remark}\upshape
The second term of equation \eqref{legge_emp2} does not depend on the number $p$ (nor on the set $E$); it follows that if this $p$ exists then it is unique.
\\
Changing the Borel set $\Delta$, we obtain a  value  $p=P(a\in\Delta)_\omega$; this is our candidate for the probability measure $\mu_{\omega,a}$.
\end{remark}
We can rewrite \eqref{legge_emp2} as follows:
\begin{equation}\label{legge_emp3}
P(a\in\Delta)_\omega : =\lim_{n\rightarrow \infty} \frac{1}{n} \sum_{k=1}^n f_k(a\in\Delta)_\omega
\end{equation}
Now we introduce a further assumption:
\begin{post}\label{ergo_0a}
The map $n\in\mathbb N \longrightarrow \xi(n)\in\mathbb N$ defined by \eqref{frequenze_legge_00} depends on the choice of $\Delta\in B(\mathbb R)$; then we take into consideration the Borel measure defined by
$$\nu_n(\Delta)= f_n(a\in\Delta)_\omega \ , \qquad \forall \Delta\in B(\mathbb R)$$
We assume it is regular, as we assumed for our measure $\mu_{\omega,a}$.
\end{post}
With this assumption we can rewrite  \eqref{legge_emp3} as follows:
\begin{equation}\label{legge_emp4}
\mu_{\omega,a}(\Delta) =\lim_{n\rightarrow \infty} \frac{1}{n} \sum_{k=1}^n \nu_k(\Delta)  \ , \qquad \forall \Delta\in B(\mathbb R)
\end{equation}

Let's ask ourselves if we can make the transition from measures to  normal functionals  of $C_o(\mathbb R)$.
\\
In other words, can we write that for every $f\in C_o(\mathbb R)$
\begin{equation}\label{legge_emp5}
\mu_{\omega,a}(f) =\lim_{n\rightarrow \infty} \frac{1}{n} \sum_{k=1}^n \nu_k(f)   
\end{equation}
and for every $f\in C_o(\mathbb R)$ 
\begin{equation}\label{distribuziopb2}
\left\langle f(a)\right\rangle_{\omega } = \lim_{n\rightarrow \infty} \frac{1}{n} \sum_{k=1}^n \left\langle f(a)\right\rangle_{\omega }(k) \ ?  
\end{equation}
where  
\begin{equation}\label{distribuziopb2ab}
\left\langle f(a)\right\rangle_{\omega }(k) = \int f(s) d\nu_k (s)  \ , \qquad \forall  k\in\mathbb N
\end{equation}

If  we denote  by $w_n$ the probability measure on $\mathbb R$:
\begin{equation}\label{misuraw}
w_n(\Delta)=   \frac{1}{n} \sum_{k=1}^n \nu_k(\Delta)  \ , \qquad \forall \Delta\in B(\mathbb R)
\end{equation} 
then we  verify  that the sequence $\left\{w_n\right\}_{n\in\mathbb N}$ converges in the $w^*$-topology to $\mu_{\omega,a}$:
\begin{equation}\label{distribuziopb2abc}
w_n(f)\longrightarrow \mu_{\omega,a}(f)  \qquad , \qquad \forall  f\in C_o(\mathbb R)
\end{equation}
First step: we consider any simple function:
$$\Psi=\sum_{j=1}^m \lambda_j \mathbf{1}_{\Delta_j} \qquad , \qquad \lambda_j \in\mathbb R \ ,\ \Delta_j \in B(\mathbb R) \ \forall j=1,2\ldots m$$
and we evaluate the quantity
$$| \mu_{\omega,a}(\Psi) - w_n(\Psi) |$$
From relation \eqref{legge_emp4}, we obtain:
\begin{eqnarray*}
| \mu_{\omega,a}(\Psi) - w_n(\Psi) | & = & \left|\sum_{j=1}^m \lambda_j \ [ \mu_{\omega,a}(\Delta_j)- w_n(\Delta_j)]\right| \leq
\\
& \leq & \sum_{j=1}^m | \lambda_j| \left| \mu_{\omega,a}(\Delta_j)- w_n(\Delta_j)\right|
\end{eqnarray*}
and for every $\delta>0$ there exists $n(\delta, \Delta_j)$ such that
$$\left| \mu_{\omega,a}(\Delta_j)- w_n(\Delta_j)\right|\leq \delta \qquad , \forall n>n(\delta, \Delta_j)$$
If 
$$ n_o=\max \left\{n(\delta, \Delta_j) \ : \  j=1,2\ldots m \right\}$$
then we can write 
$$ | \mu_{\omega,a}(\Psi) - w_n(\Psi) | \leq \sum_{j=1}^m | \lambda_j| \ \delta  \qquad , \qquad \forall n>n_o $$
It follows that we obtain the convergence \eqref{distribuziopb2abc} for simple functions.  
\\

Second Step: Continuous functions vanishing at infinity.
\\
Let $ f\in C_o(\mathbb R)$. Since it is continuous, it is Borel measurable. As we know (see Folland, Theorem (2.10)), if $f\geq 0$ there exists a countable sequence of simple functions $\Psi_k$ such that 
$$ \left\| f-\Psi_k \right\|_\infty  \to  0$$

By hypothesis, for every natural number $n$ the measure $w_n\in C_o(\mathbb R)^*$, hence 
$$ \left\| w_n(f )- w_n(\Psi_k) \right|\leq  \left\| w_n \right\|   \left\| f-\Psi_k \right\|_\infty \leq     \left\| f-\Psi_k \right\|_\infty $$
and 
$$ \left\| \mu_{\omega,a} (f )- \mu_{\omega,a}(\Psi_k) \right|\leq  \left\| \mu_{\omega,a}  \right\|   \left\| f-\Psi_k \right\|_\infty$$
Thus,
$$ \left\| w_n(f )- w_n(\Psi_k) \right\|+\left\| \mu_{\omega,a} (f )- \mu_{\omega,a}(\Psi_k) \right\| < 2  \left\| f-\Psi_k \right\|_\infty$$

Hence, for every $\delta>0$ there exists a $k_o(\delta)$ independent of $n$ such that for every $k> k_o$,
$$ \left\| w_n(f )- w_n(\Psi_k) \right| +  \left\| \mu_{\omega,a} (f )- \mu_{\omega,a}(\Psi_k) \right\| < 2\delta$$

Then,
$$ \left\| w_n(f )- \mu_{\omega,a} (f ) \right\| \leq 2\delta +  \left\|  w_n(\Psi_k)  - \mu_{\omega,a}(\Psi_k) \right\| $$

By what was said in the first step, for a fixed $k>k_o$ and for every $ \epsilon>0$ there exists an $n(\epsilon, k)$ such that 
$$ \left\|  w_n(\Psi_k)  - \mu_{\omega,a}(\Psi_k) \right\|\leq \epsilon  \quad \forall n > n(\epsilon, k) $$

It follows that 
$$ \left\| w_n(f )- \mu_{\omega,a} (f ) \right\| \leq 2\delta + \epsilon \  \forall n > n(\epsilon, k)$$

From this we obtain the weak $w^*$-convergence.
\\
This means that for every $f\in C_o(\mathbb R)$ and $\epsilon>0$ there exists a natural number $n_o$ such that: 
 $$ \left|\left\langle f(a)\right\rangle_{\omega } - w_n(f)  \right| <\epsilon \  , \qquad \forall n>n_o $$ 
Let us now return to relation  2] on page \pageref{relation 2.} and observe that for this relation we can write:
$$\lim_{n\rightarrow \infty} \frac{1}{n} \sum_{k=1}^n \left|\nu_k(\Delta) - \mu_{\omega,a}(\Delta)  \right|  = 0 \ , \qquad \forall \Delta\in B(\mathbb R)$$
so we can apply the Koopman-von Neumann theorem again to the bounded positive sequence
\begin{equation}\label{distribuziopb2b}
 \left\{\left|\nu_k(\Delta) - \mu_{\omega,a}(\Delta)  \right| \right\}_{k\in\mathbb N} 
\end{equation}
obtaining the existence of a set $\texttt{E} \subset \mathbb N^+$ of zero density\footnote{Dependent on the Borel set $\Delta\in B(\mathbb R)$.} such that
\begin{equation}\label{distribuziopb3}
\lim _{k\rightarrow \infty}\left|\nu_k(\Delta) - \mu_{\omega,a}(\Delta)  \right| \mathbf{1}_{\mathbb N \setminus E}(k)=0
\end{equation}
it follows that we obtain:
$$D-\lim _{k\rightarrow \infty} \nu_k(\Delta)=  \mu_{\omega,a}(\Delta)$$

$$***$$

Using relation  2]  again and adopting the same considerations made up to now, it can be verified that
\begin{equation}\label{distribuziopb5}
\lim_{n\rightarrow \infty} \frac{1}{n} \sum_{k=1}^n \left|\nu_k(f) - \mu_{\omega,a}(f)  \right|  =0 \ , \qquad \forall f\in C_o(\mathbb R)
\end{equation}
and applying again the Koopman-von Neumann theorem to the bounded positive sequence
$$ \left\{\left|\nu_k(f) - \mu_{\omega,a}(f )  \right| \right\}_{k\in\mathbb N} $$

we obtain the existence of a set $\texttt{E} \subset \mathbb N^+$ of zero density\footnote{Which depends on our function $f$.} such that
\begin{equation}\label{distribuziopb6}
\lim _{k\rightarrow \infty}\left|\nu_k(f) - \mu_{\omega,a}(f)  \right| \mathbf{1}_{\mathbb N \setminus E}(k)=0
\end{equation}
and in the same way as before we can write
$$D-\lim _{k\rightarrow \infty} \nu_k(f)=  \mu_{\omega,a}(f)\qquad , \qquad \forall f\in C_o(\mathbb R)  $$
obtaining the following convergence of the average values
$$D-\lim _{k\rightarrow \infty} \left\langle f(a)\right\rangle_{\omega }(k) = \left\langle f(a)\right\rangle_{\omega }  \ , \qquad \forall f\in C_o(\mathbb R)  $$
\subsection{The Finite Case}
As we have already pointed out, the average value of a physical quantity exists and is unique, even if we have problems determining it, but now we can use relation  \eqref{distribuziopb5} to establish this value \textit{up to an "error" made as small as desired}. Indeed, the properties obtained in the case of infinite trials push us to assume the existence of the following mathematical link between the average value of the quantity $a$ and the relative frequencies:
\begin{post}[\textbf{Strong-Hp}]\label{legge_emp_finita1}
For every $\alpha>0$ and $f\in C_o(\mathbb R)$, there exists a natural number $N$ of trials, which depends   on  $\alpha>0$ and on the function $f$, such that
 $$ \left|\left\langle f(a)\right\rangle_{\omega } - \frac{1}{N} \sum_{k=1}^N \left\langle f(a)\right\rangle_{\omega }(k)  \right| <\alpha$$
where the expression $\left\langle f(a)\right\rangle_{\omega }(k)$ is given by  \eqref{distribuziopb2ab}.
\end{post}
Recall that our theoretical measure $\mu_{\omega,a}$ is linked to the average value by the relation:
 $$ \mu_{\omega,a} (f)= \left\langle f(a)\right\rangle_{\omega } \qquad , \qquad \forall f\in C_o(\mathbb R)$$
and from hypothesis \ref{legge_emp_finita1} we obtain the following
\begin{post}[\textbf{Weak-Hp}]\label{legge_emp_finita}
For every $\alpha>0$ and Borel set $\Delta$, there exists a natural number $N$ of trials, which depends  on $\alpha>0$ and on $\Delta$, such that
\begin{equation}\label{legge_emp_finita_0}
 \left| \mu_{\omega,a}(\Delta) - \frac{1}{N} \sum_{k=1}^N f_k(a\in\Delta)_\omega \right| < \alpha
\end{equation}
\end{post}
Therefore this relation indicates the way to determine the measure $\mu_{\omega,a}$\footnote{Warning: it is not true, in general, that taking a number $M$ such that $N<M$, for the value $M$ relation \eqref{distribuziopb4} continues to hold, as in the case of limits of infinite sequences.}:
\begin{equation}\label{distribuziopb4} 
\left|\mu_{\omega,a} (\Delta) - \frac{1}{N} \sum_{k=1}^N \nu_k(\Delta)  \right| <\alpha
\end{equation}
We now fix a Borel set $\Delta$ and let's see how this value of $N$ varies as the parameter $\alpha>0$ varies.
\\
We define for every $\alpha>0$ the set:
$$\Gamma^\Delta_\alpha=  \left\{ n\in \mathbb N^+ : \left| \mu_{\omega,a}(\Delta) - w_n(\Delta) \right|< \alpha\right\} $$
where $w_n$ is the measure given in \eqref{misuraw} and from assumption \ref{legge_emp_finita1} this subset of positive natural numbers is non-empty.
\\

Let
\begin{equation}\label{n-minimo}
N(\alpha,\Delta)=\min\Gamma^\Delta_\alpha
\end{equation}
we have $N(\alpha,\Delta)\geq 0$ and if $\alpha_1 < \alpha_2$ then
$$ \Gamma^\Delta_{\alpha_1} \subset  \Gamma^\Delta_{\alpha_2}  \qquad \Longrightarrow  \qquad N(\alpha_1,\Delta)\geq N(\alpha_2,\Delta)\geq 0 $$
in this way we obtain a map
$$\alpha\in\mathbb R^+ \rightarrow N(\alpha, \Delta)\in \mathbb N$$
and by the properties of  non-increasing  positive functions, it can only be that  
\begin{equation*}
 \lim_{\alpha \rightarrow 0} N(\alpha,\Delta)  =\left\{ 
\begin{array}{ccc }
+\infty \ &       \\ 
N_o\in\mathbb N^+ \ &
\end{array}
\right. 
\end{equation*}

The problem now is to estimate the value of the natural number $N(\alpha,\Delta)$, the main object of property \ref{legge_emp_finita1}; practically we need to find a natural number $N$ such that
$$\mu_{\omega,a} (\Delta) \in \left] w_N (\Delta)-\alpha \ , \ w_N (\Delta)+\alpha \right[ $$
By definition of $N_\alpha$ given in \eqref{n-minimo}, we have:
$$ | w_{N_\alpha} (\Delta) - \mu_{\omega,a} (\Delta) |< \alpha \qquad , \qquad | w_{N_\alpha-1} (\Delta) - \mu_{\omega,a} (\Delta) |\geq \alpha$$
and from these relations it easily follows that
$$ | w_{N_\alpha} (\Delta) - w_{N_\alpha-1} (\Delta) | < 2\alpha$$
If we consider the following set:
$$\Lambda_\alpha= \left\{ m\in\mathbb N : | w_{m} (\Delta) - w_{m-1} (\Delta) | < 2\alpha \right\}$$
then $N_\alpha\in \Lambda_\alpha $ and 
\begin{equation}\label{n-numero1}
\min \Lambda_\alpha \leq N_\alpha
\end{equation}
We observe that 
 $$w_{m} (\Delta) - w_{m-1} (\Delta)= \frac{f_m(a\in\Delta)_\omega - w_{m-1}(\Delta)}{m}$$
so we obtain
\begin{equation}\label{n-numero1b}
\Lambda_\alpha= \left\{ m\in\mathbb N : \frac{1}{m} \left| f_m(a\in\Delta)_\omega - w_{m-1}(\Delta) \right| < 2\alpha \right\} 
\end{equation} 
The quantity
\begin{equation}\label{n-numero2}
\frac{1}{m} \left[ f_m(a\in\Delta)_\omega - w_{m-1}(\Delta) \right]
\end{equation}
\textbf{is an experimental quantity}, therefore \textit{only through $m$ real trials and the related calculation of the quantity \eqref{n-numero2} is it possible to establish whether it belongs to the set $\Lambda_\alpha$}.
\\
We want to underline that the number $N_\alpha$ has the following property:
\begin{equation}\label{n-numero3}
 f_{N_\alpha}(a\in\Delta)_\omega - w_{N_\alpha-1}(\Delta) \neq 0
\end{equation}
Indeed
$$w_{N_\alpha}(\Delta)-\mu_{\omega,a} (\Delta)=w_{N_\alpha-1}(\Delta)-\mu_{\omega,a} (\Delta) + \frac{f_{N_\alpha}(a\in\Delta)_\omega - w_{N_\alpha-1}(\Delta)}{N_\alpha}$$ 
it follows that
\begin{eqnarray*}
\alpha > |w_{N_\alpha}(\Delta)-\mu_{\omega,a} (\Delta)| & \geq & \left| w_{N_\alpha-1}(\Delta)-\mu_{\omega,a} (\Delta) \right| -
\\
&-& \left| \frac{f_{N_\alpha}(a\in\Delta)_\omega - w_{N_\alpha-1}(\Delta)}{N_\alpha} \right| \geq 
\\
& \geq & \alpha - \left| \frac{f_{N_\alpha}(a\in\Delta)_\omega - w_{N_\alpha-1}(\Delta)}{N_\alpha} \right| 
\end{eqnarray*}
and so we can say that 
$$ \left| \frac{f_{N_\alpha}(a\in\Delta)_\omega - w_{N_\alpha-1}(\Delta)}{N_\alpha} \right| > 0 $$

Let us now give a lower estimate of the value of $ w_{n}(\Delta)$.
\\
By definition of empirical frequency we have
$$f_{n}(a\in\Delta)_\omega = \frac{\xi(n)}{n} \qquad \Longrightarrow \qquad (n+1) f_{n+1} \geq n f_{n} $$
since we are in the case of "copy added"
$$\xi(n)\leq\xi(n+1) \ , \qquad \forall n\in\mathbb N$$
so if the number $n_o$ is the largest natural number for which the value 
$$f_{n_o-1}(a\in\Delta)_\omega = 0$$
we have
\begin{equation}\label{n-numero4}
 w_n(\Delta)  \geq  f_{n_o}(a\in\Delta)_\omega \cdot \frac{1}{n}\sum_{k=1}^{n-n_o} \frac{1}{k} 
\end{equation}

++++++++++++

*********
 Let $ f\in C_o(\mathbb R)$. Since it is continuous, it is Borel measurable. As we know (see Folland, Theorem (2.10)), if $f\geq 0$ there exists a countable sequence of simple functions $\Psi_k$ such that 
$$ \left\| f-\Psi_k \right\|_\infty  \to  0$$

By hypothesis, for every natural number $n$ the measure $w_n\in C_o(\mathbb R)^*$, hence 
$$ \left\| w_n(f )- w_n(\Psi_k) \right|\leq  \left\| w_n \right\|   \left\| f-\Psi_k \right\|_\infty \leq     \left\| f-\Psi_k \right\|_\infty $$
and 
$$ \left\| \mu_{\omega,a} (f )- \mu_{\omega,a}(\Psi_k) \right|\leq  \left\| \mu_{\omega,a}  \right\|   \left\| f-\Psi_k \right\|_\infty$$
Thus,
$$ \left\| w_n(f )- w_n(\Psi_k) \right\|+\left\| \mu_{\omega,a} (f )- \mu_{\omega,a}(\Psi_k) \right\| < 2  \left\| f-\Psi_k \right\|_\infty$$

Hence, for every $\delta>0$ there exists a $k_o(\delta)$ independent of $n$ such that for every $k> k_o$,
$$ \left\| w_n(f )- w_n(\Psi_k) \right| +  \left\| \mu_{\omega,a} (f )- \mu_{\omega,a}(\Psi_k) \right\| < 2\delta$$

Then,
$$ \left\| w_n(f )- \mu_{\omega,a} (f ) \right\| \leq 2\delta +  \left\|  w_n(\Psi_k)  - \mu_{\omega,a}(\Psi_k) \right\| $$

By what was said in the first step, for a fixed $k>k_o$ and for every $ \epsilon>0$ there exists an $n(\epsilon, k)$ such that 
$$ \left\|  w_n(\Psi_k)  - \mu_{\omega,a}(\Psi_k) \right\|\leq \epsilon  \quad \forall n > n(\epsilon, k) $$

It follows that 
$$ \left\| w_n(f )- \mu_{\omega,a} (f ) \right\| \leq 2\delta + \epsilon \quad \forall n > n(\epsilon, k)$$

From this we obtain the weak $w^*$-convergence. 
Sia $f\in C_o(\mathbb R)$.

Poiché $f$ è continua e limitata, risulta borelliana.
Dal Teorema 2.10 di Folland, esiste una successione di funzioni semplici
$\{\Psi_k\}_{k\in\mathbb N}$ tale che
$$\|f-\Psi_k\|_\infty \longrightarrow 0$$
Allora, poichè le misure sono di probabilità, per ogni $k$ e per ogni $n$,
$$ |w_n(f)-w_n(\Psi_k)| \le \|w_n\|\,\|f-\Psi_k\|_\infty \le   \|f-\Psi_k\|_\infty $$
e analogamente
$$ |\mu_{\omega,a}(f)-\mu_{\omega,a}(\Psi_k)|
\le  \|f-\Psi_k\|_\infty $$
Dunque
$$  |w_n(f)-w_n(\Psi_k)| + |\mu_{\omega,a}(f)-\mu_{\omega,a}(\Psi_k)| \le 2 \|f-\Psi_k\|_\infty$$
Poiché $\|f-\Psi_k\|_\infty\to 0$, per ogni $\delta>0$ esiste $k_0(\delta)$ tale che,
per ogni $k>k_0(\delta)$ e per ogni $n$,
$$ |w_n(f)-w_n(\Psi_k)| + |\mu_{\omega,a}(f)-\mu_{\omega,a}(\Psi_k)| < 2 \delta $$
Ora fissiamo $k>k_0(\delta)$. 
\\
Poiché $\Psi_k$ è una funzione semplice,
dal primo step otteniamo 
$$w_n(\Psi_k)\longrightarrow \mu_{\omega,a}(\Psi_k)$$
Pertanto, per ogni $\varepsilon>0$ esiste $N(\varepsilon,k)$ tale che, per ogni
$n>N(\varepsilon,k)$,
$$ |w_n(\Psi_k)-\mu_{\omega,a}(\Psi_k)| < \varepsilon $$
Per tali $n$ otteniamo:
$$ \begin{aligned}
|w_n(f)-\mu_{\omega,a}(f)|
&\le  2 \delta  + |w_n(\Psi_k)-\mu_{\omega,a}(\Psi_k)| \le  2 \delta + \varepsilon   
\end{aligned} $$
Poiché $\delta$ ed $\varepsilon$ sono arbitrari, segue che
$$ w_n(f)\longrightarrow \mu_{\omega,a}(f)$$
Questo vale per ogni $f\in C_o(\mathbb R)$, dunque $w_n\to\mu_{\omega,a}$
debolmente-$*$ in $C_o(\mathbb R)^*$.

*************

vedi sezione \S \ref{norm-bound-func}

Consideriamo una famiglia numerabile $ \left\{\mu_n \right\}_n $ di misure borelliane tale che 
$$\mu_n(\mathbb R)< Cost \ \forall n$$
 e sia $B_\infty(\mathbb R)$ funzioni limitate reali, ovviamente otteniamo 
$$B_\infty(\mathbb R) \subset \mathcal L^1(\mu_n) \ , \ \forall n \in\mathbb N$$
dove $\mathcal L^1(\mu_n) $ è lo spazio delle funzioni sommabili con seminorma
$$ \left\| f \right\|_{1,n}= C  \int |f| \ d  \mu_n $$ 
consideriamo per ogni $f\in B_\infty(\mathbb R) $ il seguente oggetto
$$  \left\| f \right\|_{1,s}= \sup_n  \left\| f \right\|_{1,n} \leq \left\| f \right\|_{\infty} $$
è facile verificare che risulta essere una seminorma in $B_\infty(\mathbb R)$.
\\
Abbiamo il seguente risultato: 
\begin{proposition}
Data $f\in B_\infty(\mathbb R)$,  per ogni $\epsilon>0$ esiste una funzione semplice Borelliana  $\Psi\in B_\infty(\mathbb R)$ tale che 
$$ \left\| f -\Psi \right\|_{1,s} < \epsilon$$  
\end{proposition}
\begin{proof}
Per deinizione di estremo superioore, per ogni $\epsilon>0$ esiste un $n$ tale che 
$$|| f||_{i,\infty} < \epsilon + || f ||_{1, n}$$ 
adesso lavoro su questo n fisasto.
\\
Segue che per ogni $\Psi$ funzione semplice borelliana posso scrivere  
$$|| f- \Psi||_{i,\infty} < \epsilon + || f-\Psi ||_{1, n}$$
poichè $f-\Psi\in B_\infty(\mathbb R)$ e l'indice $n$ dipende sia da $\epsilon$ che da $\Psi$.
\\
Ora poichè le funzioni semplici sono dense in $(\mathcal L^1(\mu_n), || \cdot ||_{1, n} ) $ ,  posso  sceglòire la $|Psi$ in modo tale 
$$ || f-\Psi ||_1, n <\epsilon$$ 
quindi ho verificato che esiste una funzione semplice borelliana tale che 
$$|| f- \Psi||_{i,\infty} < 2\epsilon$$
questo mi dice che le funzioni semplici approssimano in tale semibnorma le funzioni  $B_\infty (R)$
\end{proof}

\chapter{Entropies and Temporal Evolutions}
In this section we will briefly recall the notion of Shannon entropy, which we will use as a tool for establishing the quality of a measurement   carried out in a given state at a given time in our laboratory system. Subsequently we will focus on the role played by the time in which the measurement is carried out, on the quality of the measurement itself, and we will determine a family of states of the laboratory system as possible candidates for a temporal evolution of the state.
\section{The Quality of an Experimental Procedure}\label{entropia-I}
Let's ask ourselves if it is possible to determine an index that establishes whether a state of the system is more or less  capable in the measurement of a physical quantity, and  furthermore ask  ourselves what  precise  physical meaning this statement has.
\\
We  now make a simple, apparently harmless observation:
\\ 
To determine the values of a physical quantity we have  some  experimental procedures  that are  more complex than others: some of them give foreseeable results on the value of these quantities, values that do not  hold big surprises, while  with  other procedures (and therefore other states associated with them) we obtain more complex, less  trivial  values of the physical quantity. This statement smacks of Bayesianism, since we are  affirming that the experimenter expects  \textit{a priori}, without carrying  out measurements and in contrast with the Born–Heisenberg interpretation, to obtain certain values for the physical quantities  that are being measured, and hence  there is  surprise when this does not happen after having  carried   out the measurement.
\\

Our approach is \textit{operationalist}: the values of the physical quantities are determined only after calculating the  frequencies given by  \eqref{freq}. Furthermore, the subject who  carries  out the measurement is a cold executor who is not able to conjecture anything about the possible values of the observable; otherwise the experimenter  would know  something more which is not contemplated in our state of the system, which by definition establishes the measurement procedure and all possible boundary conditions\footnote{\textit{The experimenter is not a prophesying oracle}.}.
\\
So let's return to the initial question: how  do we  measure the degree of quality of a state?
\\
 What we can do is  analyse  the values of an observable that we obtained after the measurement. They are (rational) numbers that have  a  distribution along the real line and therefore for every finite partition $\mathcal P$ of the real line into disjoint sets $\left\{ \Delta_j \right\}_{j=1,2\ldots N}$, we obtain from relation \eqref{distribuzio0} a sequence of numbers $ \left\{ p_j \right\}_{j=1,2\ldots N}$ defined by:
$$ p_j =  P\left( a\in\Delta_j\right)_\omega \in\ [0, 1] \ , \qquad    j=1,2\ldots N$$
we can consider the distribution of these numbers as  an index of the complexity of the measurements  carried  out in the $\omega$ state to determine the values of $a$.
\\
In this way we have a finite sequence $(p_1 , p_2, \ldots p_N )$ with $\sum_{j=1}^N p_j=1$, associated with the triple $ (\omega, a , \mathcal P)$ and therefore we can calculate  the  Shannon entropy: \index{Shannon's entropy}
\begin{equation}
 H(\omega,  a ,  \mathcal P) = - \sum_{j=1}^N p_j \log_2 p_j
\end{equation}
We denote with $\mathtt{P}\left( \mathbb{R}\right)$ the set of all possible \textit{finite disjoint partitions} of the set of real numbers $\mathbb{R}$.
\\
We give the following definition of information associated with a state of the laboratory system; we postpone a more in-depth study until section \ref{entropia-II}.  
\begin{definition}\upshape\label{informazionesuperiore}\index{Informative state}
Given an observable $ a\in\mathfrak X$,  let us  consider two states $\omega_1 , \omega_2 \in \mathfrak S_a $. We say that state $\omega_1$ is more informative in the measurement of observable $a$ than state $\omega_2$,  in  symbols $\omega_1 \triangleright\omega_2$, if we have
\begin{equation}\label{infodistato}
 H(\omega_1 , a ,  \mathcal P) \leq H(\omega_2 , a  , \mathcal P)  \ , \qquad \forall  \mathcal P\in  \mathtt{P}\left( \mathbb{R}\right) 
\end{equation}
If in \eqref{infodistato} we have the sign of equality, then the two states are said to be equally informative; in this way we write 
$$\omega_1\sim\omega_2$$
\end{definition}
The relation $\sim$ is an equivalence relation in the set $\mathfrak S_a$; we denote by
$$\kappa_a:\mathfrak S_a \longrightarrow \mathfrak S_a/\sim$$ 
the quotient map and we write
$$ H(\kappa_a(\omega) , \mathcal P) : = H(\omega, a , \mathcal P ) \ , \qquad \forall \mathcal P \in \mathtt{P}( \mathbb{R})$$
We observe that the relation $\triangleright$ induces in $\mathfrak S_a/\sim$ a partial order relation for every observable $a\in\mathfrak X$:
$$ \kappa_a(\omega_1) \triangleright \kappa_a(\omega_2) \qquad \Longleftrightarrow  \qquad  \omega_1 \triangleright \omega_2 $$ 

We now introduce the notion of a purely informational state in the measurement of an observable $a$,
\begin{definition}[\textbf{Purely informational state}]\upshape\label{statopurodelsistema}\index{Purely informational state in the measure of $a$}
A state $\omega_o\in\mathfrak S_a$ is called a purely informational state in the measure of $a$ if the element $\kappa_a(\omega_o)$ is a maximal element of 
$( \mathfrak S_a /\sim \ , \  \triangleright )$.
\\
We denote by $\mathfrak P_a\subset \mathfrak S_a$ the set of such states.
 \\
A state $\omega \in\mathfrak S $ is called a pure state of the physical system $ \left( \mathfrak X , \mathfrak S \right)$ if and only if
$$ \omega  \in\bigcap_{a\in\mathfrak X_\omega} \mathfrak P_a$$
We denote by $\mathfrak{P}\subset \mathfrak S$ the set of such states.
\end{definition}
We underline that the existence of a purely informative state in the measure of $a$ is not ensured, so it could turn out that $\mathfrak P_a = \emptyset$, since the following property is not necessarily satisfied:
\begin{property}\label{statipuri1}
Every linearly ordered family $\mathfrak F$ of $( \mathfrak S_a /\sim,\triangleright)$ admits an upper bound element $\kappa_a(\omega_\sharp)\in \mathfrak F$,  i.e. 
$$   \kappa_a(\omega_\sharp)\triangleright \kappa_a(\omega) \qquad \forall \  \kappa_a(\omega)\in \mathfrak F $$
\end{property}

Moreover, if property \ref{statipuri1} is satisfied, then by Zorn's lemma the partially ordered set $(\mathfrak S_a /\sim \ , \ \triangleright)$ admits a maximal element $\kappa_a(\omega_o)\in\mathfrak S_a/\sim$\footnote{Cf. Folland's book \cite{Folland}.} i.e.
$$ \textit{if } \    \kappa_a(\omega)\triangleright \kappa_a(\omega_o) \qquad  \Longrightarrow \qquad   \omega\sim\omega_o   $$  
\\
We remark that if in $\mathfrak S_a$ there exists a state $\omega_o$ such that
$$P(a\in\ \left\{ \lambda \right\})_{\omega_o} =1 $$ 
for some $\lambda\in\mathbb R$, then we obtain that
 $$H(\omega_o, a , \mathcal P )=0 \ , \qquad \forall \mathcal P \in \mathtt{P}( \mathbb{R} )$$
in this way property \ref{statipuri1} is satisfied for every linearly ordered family $\mathfrak F$ of $( \mathfrak S_a /\sim  , \triangleright)$.
\\

Therefore as defined in these notes, \textit{the pure states of the system are the states of the system with the maximum information}. We want to focus attention on the fact that this concept has nothing to do with the precision of the measurement carried out (in this idealization we will always assume that the measurements carried out take place in an infinitely accurate way, with the instruments and devices that are available), but it is an intrinsic property of our physical system, namely its degree of knowledge of the observable obtained when the system is in that specific state.
\\
We conclude with an Italian statement that we extrapolate from a university textbook on theoretical physics (see Onofri \cite{Onofri} Chap. 7.4):
\\
\textit{We must now observe that pure states, as experimental states that gather the maximum possible information about the system, are in practice not easily prepared. The most general situation is one in which the information obtained is not maximum...}
\\
We must highlight that the authors identify pure states in a conventional way, as orthogonal projectors of rank 1 on a separable Hilbert space; these have zero entropy (von Neumann). We cannot say that states of maximum information, as we have defined them, coincide with the definition of pure state as the extreme point of a convex set.

\section{Information Associated with the Measurement}\label{entropia-II}
We want to briefly introduce the main properties of the entropy function and its meaning as information associated with the state of a physical system.
\\
Let us consider the following set\footnote{The sequence $\xi$ has finite support when there exists a $N\in\mathbb N$ such that $$\xi(j)=0 \ , \forall j > N$$}
\begin{equation}\label{info1}
S_{\infty }=\left\{ \xi :\mathbb{N}^{+}\rightarrow \left[ 0,1\right]
: \ \textsl{with finite support and } \sum\limits_{n=1}^{+\infty }\xi \left( n\right) =1 \right\}
\end{equation}%
It is simple to verify that $S_{\infty }$ is a convex set.
\\
Let us denote for every natural number $k$ by $\delta _{k}$ the elements of $S_{\infty }$ such that 
\begin{equation*}
\delta _{k}\left( n\right) =\left\{ 
\begin{array}{cc}
1 & \text{ }n=k \\ 
0 & \text{ }n\neq k%
\end{array}
\right. 
\end{equation*}
A $k$-schema is an element $\xi $ of $S_{\infty }$ such that $\xi \left(n\right) =0$ for all $n>k$.
\bigskip

A map $H:S_{\infty }\rightarrow \mathbb{R}$ is called \textit{(Shannon) entropy} if it satisfies the following properties (see Khinchin \cite{khinchin}):
\begin{enumerate}
\item[\textbf{K}1.] $H\left( \xi \right) \geq 0$ for every $\xi \in S_{\infty
}$. Furthermore $H\left( \xi \right) =0$ if and only if $\xi =\delta _{k}$ for
some natural $k$.

\item[\textbf{K}2.] $H\left( \xi \right) =H\left( \xi ^{\prime }\right) $ \
where $\ \xi ^{\prime }=\left( 0,\xi \left( 1\right) ,\xi \left( 2\right)
,\xi \left( 3\right) ....\right).$

\item[\textbf{K}3.] $H\left( \xi \right) =H\left( \widehat{\xi }\right) $ \
where $\ \widehat{\xi }=\left( \overset{k}{\overbrace{\frac{\xi \left(
1\right) }{k},\frac{\xi \left( 1\right) }{k}....\frac{\xi \left( 1\right) }{k%
}}},\xi \left( 2\right) ,\xi \left( 3\right) ....\right).$

\item[\textbf{K}4.] The element $\xi _{k}=\left( \overset{k}{\overbrace{p,p...p}},0,0.......\right) $ with $p=1/k$ is a maximal element on the $k$-schemas\footnote{Concept not to be confused with our notion of maximal information.}:
\\
In other words for every $k$-schema $\xi $ we have:
\begin{equation*}
H\left( \xi \right) \leq H\left( \xi _{k}\right) .
\end{equation*}

\item[\textbf{K}5.] The information function is a concave function
\begin{equation*}
H\left( \left( 1-r\right) \xi +r\eta \right) \geq \left( 1-r\right) H\left(
\xi \right) +rH\left( \eta \right)
\end{equation*}%
for each $\xi ,\eta \in S_{\infty }$ and $r\in \left[ 0,1\right] .$

\item[\textbf{K}6.] (Continuity property) For each $\varepsilon
>0$ and $\xi \in S_{\infty }$ there exists a $\theta >0$ such that for every $\eta
\in S_{\infty }$ with $\sum\limits_{n=1}^{+\infty }\left\vert \xi \left(
n\right) -\eta \left( n\right) \right\vert <\theta $ we obtain 
$$\left\vert H\left( \xi \right) -H\left( \eta \right) \right\vert <\varepsilon $$
\end{enumerate}
The function $H$ is an index of the quality of the information
contained in the string of elements belonging to $S_{\infty }$. 
\\
Let's give a simple interpretative example of the six conditions we have given to define the function $H$.
\begin{example}\upshape
We have a box with $k$ compartments and $n$ balls distributed in these $k$ compartments; we indicate with $n_{i}$ the number of balls present in the $i$-th compartment and with $\xi \left( i \right) = n_{i}/n $ their frequency. 
In this way we have a $k$-scheme $\xi =\left( \xi \left( 1\right) ....\xi \left(
k\right) ,0,...0.....\right)$.
\\
We can have different distributions of the balls in the compartments; let's study
the two limiting cases:
\begin{itemize}
\item[I)] We assume that all $n$ balls are all contained in one compartment only; let us assume to fix the ideas that it is the first. Then its $k$-schema is given by $\delta _{1}=\left( 1,0,0...0.....\right) $ and
as has been defined for the entropy function we have $H\left(\delta _{1}\right) =0$.
\item[II)] We assume that the balls are equally distributed in the compartments;
so $n_{i}=m$ for each $i$; it follows that the related $k$-schema is given by   
\begin{equation*}
\xi _{k}=\left( \overset{k}{\overbrace{p,p...p}},0,0.......\right)
\end{equation*}%
with $p=m/n$. In this way by definition the entropy $H\left( \xi _{k}\right) $ is maximal.
\end{itemize}
Suppose that among the $n$ balls there is only one black ball; now we want to know the $k$-scheme relating to the position of the black ball.
\\
In the $k$-scheme given in the first case we can say without any doubt that the black ball is located in the first compartment, while the worst situation, i.e. 
of maximum uncertainty of its position, occurs in the second case; all other $k$-schemes are found in intermediate situations of uncertainty.
\\
Therefore we can say that the higher the value of the entropy function $H$, the more uncertain the \textit{k-schema} is.
 \end{example}
The meaning of the entropy function highlighted in this simple example is reiterated by Khinchin in his book \cite{khinchin}:
\begin{citazione}\index{Khinchin}
Thus, we can say that the information given us by carrying out some experiment consists in removing the uncertainty which existed before the experiment. The larger this uncertainty, the larger we consider the amount of information obtained by removing it.
\end{citazione}
But what does it mean to remove uncertainty?
\\
In our example it was simply looking at where the black ball is placed. If we have to equip the laboratory to verify where the black ball is positioned, we must consider the state of the system in which this recognition occurs. In our example we have prepared the laboratory to measure the observable $q$, the position of the black ball, in two different ways that reflect the state of the system: in the first state we have arranged the balls all in one compartment, in the second state they are all equally distributed in the various compartments.
\\
Our experimenter, unlike Khinchin's, does not give any meaning to the value of the Laplacian probability of favorable cases divided by the total cases taken into consideration previously; it does not expect anything a priori, no surprise effect; the only thing it is authorized to do is to analyse the results obtained in $N$ copies of the ensembles where the black ball is positioned and calculate the relative frequencies of $q$ in the two states of the system described above.
\\
So our information about the measurement of $q$ in the state $\omega$ is lower when the entropy $H(\omega, q, \mathcal P)$ is higher, in line with our definition 
\ref{informazionesuperiore}.
\bigskip

Having clarified this differentiation at the interpretative level of the entropy function\footnote{Substantially between the theory of signals and that of measurement considered by us.} let's resume the mathematical discussion:
\\
We now have a theorem, the proof of which we refer to Khinchin's book
\textit{Mathematical Foundations of Information Theory} \cite{khinchin}, which
defines the form of the information function.
 
\begin{theorem}[Khinchin 1957]\upshape
If $H:S_{\infty }\rightarrow \mathbb{R}^+$ is a function which satisfies the six properties K1–K6 then it follows that
\begin{equation*}
H\left( \xi \right) =-k\sum\limits_{n=1}^{\infty }\xi \left( n\right) \log
\left( \xi \left( n\right) \right)
\end{equation*}%
with $k$ a positive real number and with the convention that $0\log (0)=0$
\end{theorem}
We conclude by recalling that the constant $k$ is fixed by giving a value of the entropy function to certain schemes; for example if for the $2$-scheme
$$\eta=\left( \frac{1}{2},\frac{1}{2},0,0.......\right)$$
we set $H\left( \eta \right) =1$, then we obtain that $k=\frac{1}{\log 2}$ and this trivially implies that
\begin{equation*}
H\left( \xi \right) =-\sum\limits_{k=1}^{\infty }\xi \left( k\right) \log
_{2}\left( \xi \left( k\right) \right)
\end{equation*}
\section{Mutant and Evolutionary Phenomena}\label{mutazionievol}
We prepare our laboratory to carry out a measurement of observable $a$ in the state $\omega \in\mathfrak S_a$ at time $\tau$.
\\
Let us now assume that the experimenter can re-arrange, through a new disposition of the devices/instruments and their preparation time etc., the state of the initial system $\omega\in\mathfrak S_a$ into a new state $\omega_*\in\mathfrak S_a$ such that
\begin{equation}
\label{evol_temp_00}
  P(a\in\Delta, 0)_{\omega_*} = P(a\in\Delta, \tau)_\omega \ , \qquad \forall \Delta\in B(\mathbb R)
\end{equation}
%+
In other words, the experimenter  prepares  the laboratory to test every single element $\omega_*$ of $\mathfrak S_a$ and establish which of these states satisfy  \eqref{evol_temp_00}\footnote{Obviously it is not certain that these states exist.}.
\\
\begin{figure}[htbp]
	\centering
		\includegraphics[scale=0.25]{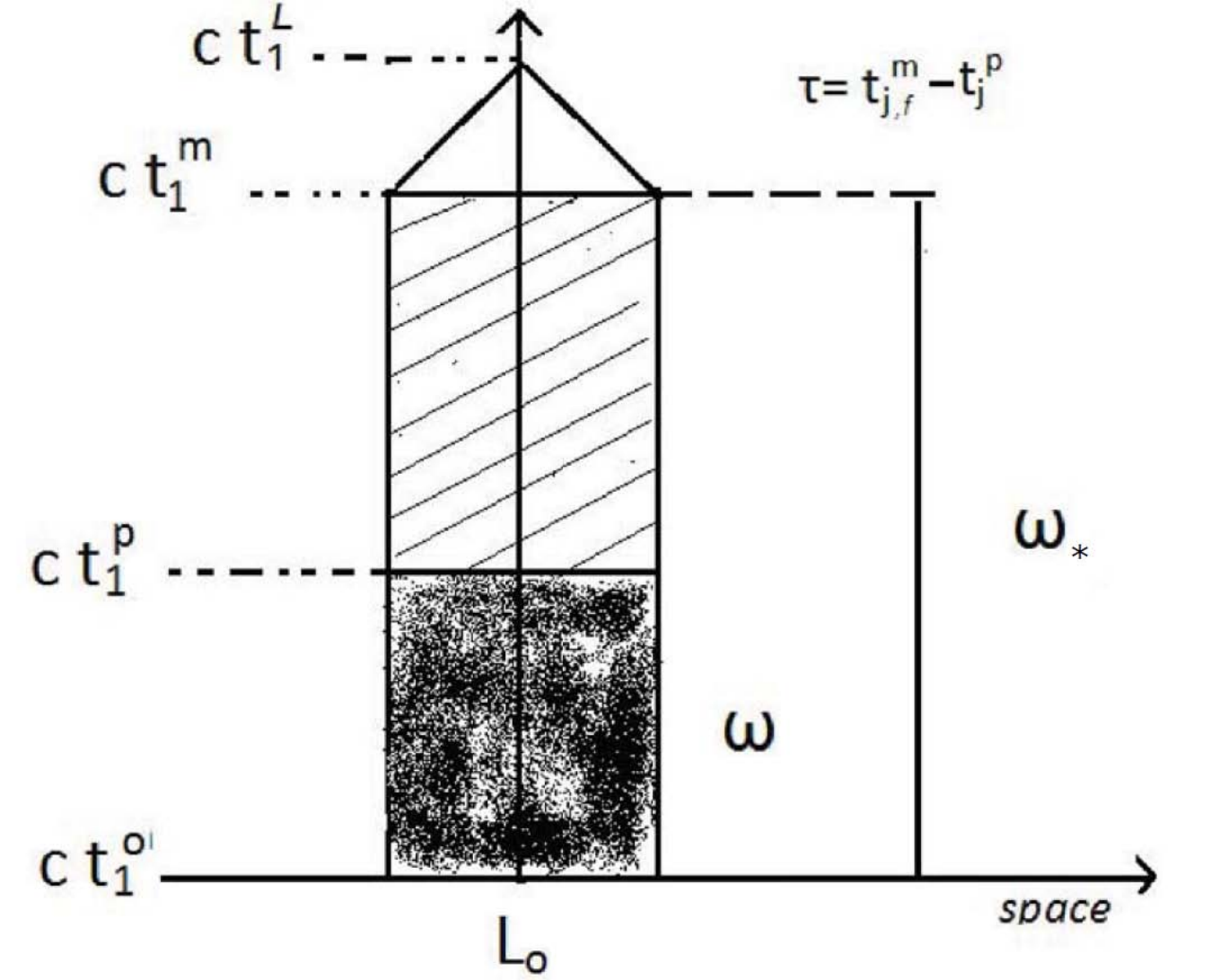}
	\caption{Evolution State}
	\label{fig:statoevol}
\end{figure}
Let's make a further choice on these states:
\\
We consider the two laboratory-type regions (see Figure \ref{fig:statoevol}):
$$ \mathcal O_o =L_o \times [0,t^p] \qquad , \qquad \mathcal O_* =L_o \times [0,t^m]$$
for each $\omega\in\mathfrak S_a(\mathcal O_o)$, we are interested in the following states $\omega_*\in\mathfrak S_a (\mathcal O_*)$ such that:
\begin{itemize}
\item[1.] All physical quantities measurable in the state $\omega$ must also be measurable  in  the state $\omega_*$:
$$\mathfrak X_\omega(\mathcal O_o) = \mathfrak X_{\omega_*}( \mathcal O_*) \bigcap \mathfrak X(\mathcal O_o)$$
\item[2.] The measurement of the observable $a$ in this state occurs at time $\tau'=0$.
\item[3.] They satisfy \eqref{evol_temp_00}.  
\end{itemize}
For every $\omega\in \mathfrak S_a|_\tau$, we define the following subset of $\mathfrak S_a$:
\begin{equation}\index{$\mathfrak S_\tau^{a,\omega}$}
\label{evol_temp_01}
\mathfrak S_\tau^{a,\omega} = \left\{   \omega_*\in\mathfrak S_a (\mathcal O_*): \textit{that satisfy relations (1), (2) and (3)}  \right\}
\end{equation}
Obviously this set  does  not  consist  of a single element and  a priori  we cannot exclude that it is empty.
\\
Therefore, the preparation time of the state $\omega_*$ is given by the entire interval $t^p_{*,j}=t^m_j - t^p_j $ for every $j=1,2\ldots N$ and \textit{its existence will be given axiomatically}.
\\

We underline that for every $ \omega \in \mathfrak S_a (\mathcal O_o)|_\tau$ we have:
\begin{equation} \label{evol_temp_01b}
  \mathfrak S_\tau^{a,\omega} \subset \mathfrak S_a ( \mathcal O_{*})|_{ \tau'=0}
\end{equation}
since, by definition, in the state $\omega_* \in \mathfrak S_a ( \mathcal O_{*})$ the measurement of $a$ occurs at time $\tau'=0$.
\\
We set
$$ \mathfrak S_\tau^{a}:= \bigcup_{\omega \in \mathfrak S_a ( \mathcal O_{o})| \tau} \mathfrak S_\tau^{a,\omega} \qquad \Longrightarrow \qquad \mathfrak S_\tau^{a}\subset \mathfrak S_a ( \mathcal O_{*})|_{ \tau'= 0} $$

\subsection{Evolution of the System State}
We  now  prepare our laboratory to perform a measurement of observable $a$ in the chronological state  \eqref{noevoltemp}:
 $$\omega :  \ \tau\in I \longrightarrow \omega^{(\tau)}\in  \mathfrak S_a (\mathcal O_{t_p})| \tau $$
which establishes the various measurement times of the observable $a$ after having prepared the state of the system, and we obtain for $a$ its temporal evolution expressed by the relation:
$$ \tau\in I \longrightarrow P(a\in\Delta, \tau)_{\omega} \ , \qquad I\subset [0 , \infty[ $$
We have the following
\begin{definition}\upshape\label{evol_temp_02}\index{Temporal evolution of the chronological state}
A temporal evolution of the chronological state $\omega\in \mathfrak S_a (\mathcal O_{t_p})$ is a  map   $\xi:  [0, \infty[ \longrightarrow  \mathfrak S_a $ 
such that 
\begin{itemize}
\item[1.]  $\xi(\tau )\in\mathfrak S_\tau^{a,\omega} \subset\mathfrak S_a (\mathcal O_{t})|_{ \tau'=0}  \qquad \forall \tau=t-t_p\geq 0 \ : \   \mathcal O_t =L_o \times [0,t]$;
\item[2.]  $\xi(0) = \omega^{(0)}$
\end{itemize}
The set of all possible temporal evolutions of our chronological state $\omega$ will be denoted by $\mathcal S_{a,\omega}$.
\end{definition}
By the temporal evolution definition for every $\tau \in I$ there exists $\xi(\tau)\in\mathfrak S_a |_{\tau'=0}$ which satisfies relations \textit{(1)} and \textit{(2)} such that:
$$ P(a\in\Delta)_{\omega^{(\tau)}} = P(a\in\Delta)_{\xi(\tau)}  \ , \qquad \forall \Delta \in B(\mathbb  R)$$
Furthermore 
$$ \xi(0) \in \mathfrak S_0^{a,\omega} \subset \mathfrak S_a (\mathcal O_{t_p})|_{ \tau'=0}$$ 
We underline that the map $\xi$, although very similar to \eqref{noevoltemp}, \textit{is not a chronological state} suitable for the measurement of $a$.
\begin{notation}\upshape\label{evol_temp_s}\index{$S^\tau_a\omega^{(0)}$}
We will use for the temporal evolution of the chronological state $\omega$ the following notation\footnote{In this way the connection between our evolution from the observable $a$ and the state $\omega$ is highlighted.}
$$ \xi(\tau)= S^\tau_a\omega^{(0)} \ , \qquad \forall \tau\in I\subset\mathbb R^+ $$
therefore 
$$S^0_a\omega^{(0)}= \omega^{(0)} $$ 
and for every Borel set $\Delta$ of $\mathbb R$, we can write:
$$ P(a\in\Delta)_{\omega^{(\tau)}}= P(a\in\Delta)_{S^\tau_a\omega^{(0)}} \ , \qquad \forall \tau\in I\subset\mathbb R^+ $$
\end{notation}
\begin{figure}[htbp]
	\centering
		\includegraphics[scale=0.6]{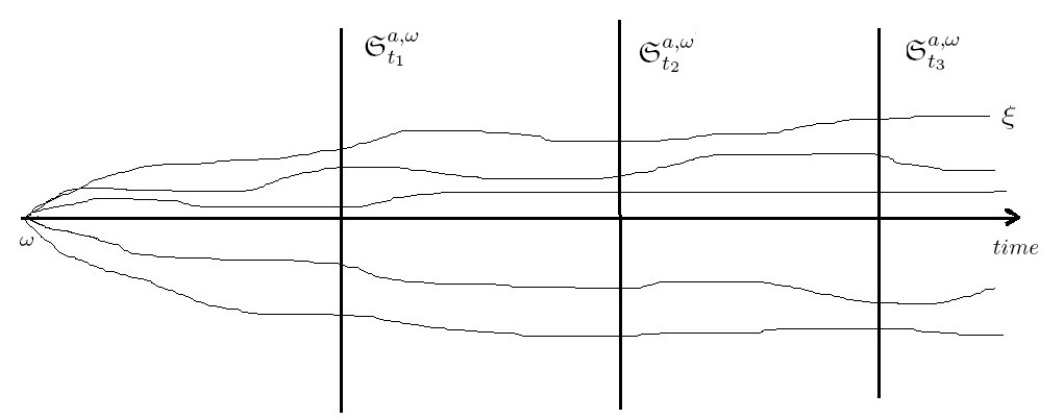}
	\caption{Temporal Evolution}
	\label{fig:05b}
\end{figure}
Let us now consider the set of \textit{possible temporal evolutions of the chronological state $\omega$}:
\begin{equation}
\label{evol_temp_03}
\mathcal S^\omega = \bigcap_{a\in\mathfrak X_\omega} \mathcal S_{a,\omega}
\end{equation}
We remark that $\mathcal S_{a,\omega}$ is a non-empty set, unlike the set $\mathcal S^{\omega}$ which could be.
\\ 
Therefore, for every $\xi\in\mathcal S^{\omega}$ we have:
\begin{equation}
\label{evol_temp_04}
 P(a\in\Delta, \tau)_\omega = P(a\in\Delta, 0)_{\xi(\tau)} \ , \qquad \forall \tau\in\mathbb R^+ \ , \  a\in\mathfrak X_\omega( \mathcal O_o)
\end{equation}
Moreover, if $\xi, \xi'\in\mathcal S^{\omega}$ then we obtain:
 $$P(a\in\Delta, 0)_{\xi(\tau)} = P(a\in\Delta, 0)_{\xi'(\tau)} \ , \qquad \forall \tau\in\mathbb R^+ \ ,\ a\in\mathfrak X_\omega( \mathcal O_o)$$
and we cannot say that $\xi=\xi'$ but   
$$\xi(\tau)  \equiv_{\mathfrak X_\omega( \mathcal O_o)} \xi'(\tau) \ , \qquad \forall \tau\in\mathbb R^+ $$  
\begin{remark}\upshape\label{esistenza-evol}
We reiterate that we do not have experimental methodologies to know which evolution the state of the system has as established in definition \ref{evol_temp_02}; this notion is therefore only theoretical  and  the existence of such a state is assumed. Indeed what is well known experimentally is the chronological state of the system.  
\end{remark}

\subsection{Semi-group Property}
We now prepare our laboratory to perform a measurement of observable $a$ in the chronological state
\begin{equation}\label{statocronos4}
\omega :  \ \tau\in \mathbb R^+ \longrightarrow \omega^{(\tau)}\in  \mathfrak S_a (\mathcal O_{t_p})| \tau 
\end{equation} 
and we set a time $\tau_o=t_o-t_P$; we denote
\begin{equation}\label{statocronos1}
 \xi(\tau_o)=S_a^{\tau_o}\omega^{(0)} \in  \mathfrak S^{a,\omega}_{\tau_o} \subset  \mathfrak S_a(\mathcal O_{t_o})|_{\tau=0} 
\end{equation}
 as its temporal evolution:
$$P (a\in\Delta)_{\omega^{(\tau_o)}}= P(a\in\Delta, 0)_{\xi(\tau_o)} $$
As we hypothesized, the experimenter has the ability to recreate the evolved state $\xi(\tau_o) \in \mathfrak S_a(\mathcal O_{t_o})|_{\tau=0}$.
\\
Moreover we assume that this state generates a chronological state of the system:
 $$\widetilde{\omega} :  \ \tau\in \mathbb R^+ \longrightarrow \widetilde{\omega}^{(\tau)}\in  \mathfrak S_a (\mathcal O_{t_o})| \tau $$ 
where
$$\widetilde{\omega}^{(0)}=\xi(\tau_o)$$
So we set up the various ensembles again to measure the same observable $a$ of the system but using as initial state not our $\omega^{(0)}$ but $\xi(\tau_o)$;  in this way we obtain the temporal evolution:
$$S^{\tau_1}_a  \xi(\tau_o)= \widetilde{\omega}^{(\tau_1)} \qquad , \qquad \tau_1= t_1-t_o$$
so we can write
 $$P(a\in\Delta)_{\widetilde{\omega}^{(\tau_1)}} = P(a\in\Delta)_{S^{\tau_1}_a  \xi(\tau_o)} $$
We have the following  
\begin{property}[\textbf{Semigroup}]
The temporal evolution \eqref{statocronos1} of the state \eqref{statocronos4} satisfies the semigroup property if 
\begin{equation}\label{semigroup}
S^{\tau_1}_a  \xi(\tau_o)=S_a^{\tau_1}\omega^{(0)} \ \qquad , \qquad \forall \tau_1\in\mathbb R^+ 
\end{equation}
\end{property}
We can write the semigroup property as follows:
$$ S^{\tau_1}_a \left( S^{\tau_0}_a \omega^{(0)} \right) =  S^{\tau_1+\tau_o}_a  \omega^{(0)}   $$
where
$$\tau'=\tau_1+\tau_o=t_1-t_o$$
Naturally we ask ourselves the following problem:
\begin{problem}\upshape
\label{pb-1b}
In which cases do we have the validity of property \ref{semigroup}?
\end{problem}
\section{Dissipative Discussion} \label{discussionedissipi} 
Experimentally we note that a state of maximum information over time may not remain so; therefore we have a temporal \textit{wear and tear} of the information qualities of the states of the system; this loss of information occurs due to operationally unavoidable perturbations, which we will denote with the generic term of 
\textit{dissipation}.\index{Dissipation}
\\
We underline that the measures that are carried out in the states $\omega\in\mathfrak S_a | \tau$ with $\tau>0$ are susceptible to dissipative phenomena\footnote{See section \ref{riproduzione}.}.
\\
We have the following question:
\\
If this dissipative phenomenon is quantifiable and therefore measurable\footnote{In other words, if the experimenter notices that something has changed, it means that the mutation is measurable; otherwise, how is it possible to notice the change?}, then it means that we can associate a physical quantity of the laboratory system with it, since by definition physical quantities are such if and only if they can be measured in our laboratory. 
\\
The problem that now arises is to determine this quantity and its properties.
\\
Let's see how it is possible tomanage this problem:
\\
Given a finite partition $\mathcal P = \left\{ \Delta_j \right\}_{j\in I}$ with $I\subset \mathbb N$ of finite cardinality, we consider for each $\tau\geq 0$ the probabilities:
$$ p_j(\tau)=P(a\in\Delta_j , \tau)_\omega \ , \qquad \forall j\in I$$ 
and we calculate the entropy of the measurement of $a$ in the state $\omega\in \mathfrak S_a$ at time $\tau\in\mathbb R^+$:
$$H(\omega, a , \mathcal P , \tau)= - \sum_{j\in I} p_j(\tau)\log_2 (p_j(\tau))$$
We have the following experimental property:
\begin{post}[\textbf{Entropic Property}]\label{prop-entropic}\index{Postulate-Entropic Property}
For every partition $\mathcal P \in \texttt{P}(\mathbb R)$ we have\footnote{Therefore the state $\omega^{(\tau)}$ is more informative than the state $\omega^{(0)}$:
$$\omega^{(\tau)}\triangleright\omega^{(0)} $$ }
\begin{equation}
H(\omega, a , \mathcal P , \tau) \geq H(\omega, a , \mathcal P , 0) \ , \qquad \forall \tau\geq 0
\end{equation}
\end{post}
\begin{remark}\upshape
We highlight that here we are considering the entropy of a measurement and not the entropy of the physical system. 
\end{remark}
We will return to the dissipative topic in section \ref{evol_temp}.

\part{Math Framework for Experimental Procedures}

\chapter{Mackey's Model Revisited}
\label{Mackey_rivisitato}
\begin{flushright}
\textit{The older generation almost always regards the younger as too mathematical.\\ \hfill Wightman 1969}
\end{flushright} 

As explained in the previous sections, we will assume that the measurements of physical quantities are  carried out in a laboratory linked to a reference system where it is possible to establish when and where they occur.
In this chapter we introduce some fundamental properties that connect the sets of physical quantities with those  of their states in relation to the values of the measurements obtained. These properties  induced by laboratory experience  can be related to mathematical axioms of the model we are developing, axioms with all the physical-mathematical limits that we have well described in the introductory chapter of this work.
\\
We will divide the system of axioms into two groups: the first group called "static axioms"  concerns  the definition of the expected value of a physical quantity in a given state; the second group called "dynamic axioms"  establishes  what  we should  understood by the temporal evolution of a physical system. In this chapter we will only cover the static axioms of the system.
\section{Static Axioms}\label{assiomistatici}
In the model we adopted, a physical system of the laboratory is defined through the following objects:
\begin{itemize}
\item [ I. ] A set $\mathfrak{X}$ of observables, where the observables  indicate  all the physical quantities which we can measure in our laboratory.
\\
\textit{We emphasize that we do not make a choice on the set of observables which we want to measure, but we take into account all observables}\footnote{Which we will do in the next sections.}.
\item [ II.] For each observable $a$ of $\mathfrak X$, a set $\mathfrak{S}_a$ of the states of the system\footnote{See definition \ref{stat-oss} on page \pageref{stat-oss}.}, where by system state we indicate the conditions in which the measurement  of  the observable $a$ takes place (the way of preparing the experiment, monitoring procedure, etc.).
\\
We denote by $\mathfrak S$ the set of all possible states of the system: 
$$ \mathfrak S = \bigcup_{a\in\mathfrak X} \mathfrak S_a$$
\item [ III. ] A  map  that associates to each observable $a$ and state $\omega \in\mathfrak S_a$ a \textit{unique} probability measure
\begin{equation}\label{misura_borelliana}
\mu _{\omega ,a}:B\left( \mathbb{R}\right) \rightarrow \left[ 0,1\right] 
\end{equation} 
where by $B\left( \mathbb{R}\right) $ we denote the $\sigma$-algebra of the Borel sets of the real numbers $\mathbb{R}$, such that the expected value
of the observable $a$ in the state $\omega$ denoted by
$\left\langle a\right\rangle _{\omega }$ is given by: 
\begin{equation}
\label{valmedio}
\left\langle   a\right\rangle _{\omega }=\int t\,d\mu _{\omega ,a}(t)\in\mathbb R
\end{equation} 
while the probability that the observable $a$ takes a value in a subset $\Delta$ of $\mathbb{R}$ in the state $\omega$, in symbols
$P\left( a\in \Delta \right)_{\omega }$,  is  
\begin{equation}
P\left( a\in \Delta\right) _{\omega }=\int \mathbf{1}_{\Delta}\,d\mu_{\omega ,a}
\end{equation}
where with $\mathbf{1}_{\Delta }$ we have indicated the characteristic function of the set $\Delta$:
\begin{equation}
\label{funzcar}
\mathbf{1}_{\Delta }(t) =\left\{ 
\begin{array}{cc}
1 & t\in \Delta \\ 
0 & t\notin \Delta
\end{array}%
\right.  
\end{equation}
\end{itemize}
We reiterate that with the symbol $\mathfrak S_a$ we indicate the set of \textbf{all} states of the system $\omega$ for which the probability measure $\mu_{\omega ,a}$ exists; the elements of $\mathfrak S_a$ are called states \textit{suitable} for the observable $a$.
 Moreover, having fixed a state $\omega$ of $\mathfrak S$ we indicate with $\mathfrak X_\omega$ the set of observables $a$ of the system for which the measure $\mu_{\omega ,a}$ exists.
\begin{remark}\upshape
It is worth highlighting that we are assuming that an observable of the physical system located in our laboratory always admits a well-defined average value in any state of the system suitable for it (we do not admit infinite values; the laboratory is limited in space and time). Therefore we can say that the measures $\mu_{\omega, a}$ that have physical validity are those that make the identity function of $\mathbb R$ a $\mu_{\omega, a}$-summable function for every observable $a$ of the system and in every state $\omega$ suitable for it.
\end{remark}
So we obtain the following map:
\begin{equation}
\omega \in \mathfrak{S}_a \longrightarrow \mu_{\omega ,a}\in C_{o}\left( \mathbb{R}\right) ^{\ast}
 \label{mappa mu}
\end{equation}
where with the symbol $\mu_{\omega ,a}$ we indicate both the functional on the Banach algebra of continuous functions that vanish at infinity
$C_{0}\left( \mathbb{R}\right)$ and the corresponding associated Borel measure: 
\begin{equation}
\mu _{\omega ,a}\left( f\right) =\int f\;d\mu _{\omega ,a} \ , \qquad \forall f\in C_{o}\left( \mathbb{R}\right) 
\end{equation}
In summary, a physical system is described by a pair $\left( \mathfrak{X,S}\right)$ where $\mathfrak{X}$ is the set
of the observables and $\mathfrak{S}$ is the set of states; to it there is associated a unique map \eqref{mappa mu} with $\mu _{\omega ,a}$ the associated measure
to the observable $a$ in the state $\omega$ of the laboratory system\footnote{One could object that in reality the values of the measures of a generic observable are not simple real values; more generally Borel measures in $\mathbb{R}^{n}$ should be considered, but this type of generalization, as also observed by Roberts and Roepstroff in \cite{RobertsRoepstorff}, does not lead to any different type of result which has a profound meaning from the one assumed here.}. 
\\
The uniqueness of the map \eqref{mappa mu} derives from the fact that the values assumed by the observable $a$ in the state $\omega$ (at a given fixed time) are uniquely established, experimentally, by the relative frequencies $f_n\left(a\in \Delta\right)_{\omega }$.
\\

Before proceeding with the discussion we must underline the following remarks:
\begin{itemize}\index{Single measurements}\index{Unit of measure}
\item \textbf{Single measurements -} These considerations apply to \textit{single measurements} of the laboratory's physical quantities. We repeat again that experimentally no observable of the system can actually be prepared one at a time, since it is always necessary to take into account other physical quantities that the experimenter knows and keeps under their rigid control; this information contributes to establishing the state of the system $\omega$. So when we talk about average values of an observable we are referring to their \textit{individual measurements} relative to the $\omega$ state\footnote{See remark \ref{statale} on page \pageref{misurecondizionate}.}.
\item \textbf{Unit of measure -} Experimentally, a physical quantity has its own units of measurement in which it is expressed. In this formalism we are assuming that the values of measurements of \textit{physical quantities are dimensionless}. In other words we initially set particular reference values (fixed once and for all by our experimenter) of the physical quantity we want to measure to which we will compare the quantity itself. For example for the electric charge $Q$ we can consider the dimensionless quantity $\widehat{Q}=Q/Q_o$ where $Q_o$ is a test electric charge, etc. This way of proceeding allows us to compare (and we will see later also add) values of the measurements of some physical quantities not homogeneous with each other.
\end{itemize}
For this purpose we give the following definitions\footnote{For further clarification on this definition, see page \pageref{nuovi_axiom}.}:
\begin{definition}\upshape \label{condition1}\index{$a\subset b$}
Let $a$ and $b$ be observables of our physical system; we will say that the observable $b$ extends $a$ and we will write $a\subset b$\footnote{Equivalently we say that the observable $a$ is a restriction of $b$ (see \S \ref{nuovi_axiom}).}, if
\begin{itemize}
\item $\mathfrak S_a \subset \mathfrak S_b \ ,$
\item $ \left\langle a\right\rangle _{\omega }=\left\langle b\right\rangle _{\omega } \ , \qquad \forall \omega\in\mathfrak S_a$
\end{itemize}
\end{definition}
In a symmetrical way we can give the definition for physical states:
\begin{definition} \upshape\label{condition2}\index{$\omega\subset \omega'$}
Let $\omega$ and $\omega'$ be states of our physical system; we will say that the state $\omega'$ extends $\omega$ and we will write $\omega\subset \omega'$\footnote{Equivalently we say that the state $\omega$ is a restriction of $\omega'$.}, if
\begin{itemize}
\item $\mathfrak X_\omega \subset \mathfrak X_{\omega^\prime} \ ,$
\item $ \left\langle a\right\rangle _{\omega }=\left\langle a\right\rangle _{\omega' } \ , \qquad \forall a\in\mathfrak X_\omega$
\end{itemize}
\end{definition}
In the axiomatic formalism we are outlining, the concepts of state and observable of a physical system are primitive concepts having the following fundamental properties:
\begin{axiom}[\textbf{Identity of Observables}]\index{Axiom-Identity of Observables}
\label{assio1} 
Two observables $a$ and $b$ of our physical system are equal if and only if $a \subset b$ and $b \subset a$.
\end{axiom}
For states we have:\index{Mackey}
\begin{axiom}[\textbf{Identity of States}]\index{Axiom-Identity of of States}
\label{assio2}
Two states $\omega_{1}$ and $\omega_{2}$ of our physical system are equal if and only if
$\omega _{1} \subset \omega _{2}$ and $\omega _{2} \subset \omega _{1}$.
\end{axiom}
\begin{remark}\upshape
The axioms we have introduced are slightly different from those adopted by Mackey in \cite{Mackey}, since in addition to considering equality between average values and not between probability measures, we consider a set of states of the system not independent from the set of observables\footnote{See remark \ref{Segal-von Neumann vs. Mackey} on page \pageref{misureBorele}.}.
\end{remark}
Let's make other considerations on the laboratory instruments designed to establish, through measurement, the value of physical quantities.
\\ 
The measurement apparatus by its nature has a graduated scale, a counter where the value of the measurement of our observables can be read, which
in the Mackey scheme is determined by the Borel measure $\mu_{\omega ,a}$.
\\ 
What happens if we rescale (even in a non-linear way) our reading scale of the measuring device?
\\
In practice, we can say that a rescaling of the measurement apparatus is carried out via a Borel function $F:\mathbb{R}\rightarrow \mathbb{R}$; the values of the rescaled measurements are described by the Borel measure \textit{$F$ distribution law} defined as follows:
\begin{equation}
\mu _{\omega ,a}^{F}\left( \Delta \right) =\mu _{\omega ,a}\left(F^{-1}\left( \Delta \right) \right) \ , \qquad \Delta \in B\left( 
\mathbb{R}\right) 
\end{equation}
Before stating the next axiom we give the following definition:
\begin{definition}\upshape\label{a-sommabile}\index{Function a-sommabile}
A Borel function $F:\mathbb R \rightarrow \mathbb R$ is said to be $a$-summable if it is $\mu_{\omega ,a}$-summable for every state $\omega\in\mathfrak S_a$. 
\\
We denote by $L^1(a)$ the set of such functions. \index{$L^1(a)$ }
\end{definition}
Obviously all bounded Borel functions are $a$-summable and as hypothesized the identity function also turns out to be so.
\begin{axiom}[\textbf{Functional Calculus}]\label{assio3}\index{Axiom-Functional Calculus}
For each observable $a$ and Borel function $F:\mathbb{R}\rightarrow \mathbb{R}$ that is $a$-summable, an observable of the system remains associated which we will indicate with $F\left( a\right)$, such that
\begin{equation}
\mathfrak S_{F(a)}=\mathfrak S_a 
\end{equation}
and for every $\omega\in\mathfrak S_a$ it results\footnote{For example, we are stating that directly measuring the square of the velocity $v$ of a particle in the $\omega$ state leads to the same probability law when we measure the observable velocity $v$ in the $\omega$ state (and this state results still suitable for the observable velocity) and then through relation \eqref{funzioni} we determine the probability law of its square $v^2$ (we will see later that the "exponentiation" function is an $a$-summable function for every observable in the system).}
\begin{equation}\label{funzioni}
P\left( F(a)\in \Delta\right) _{\omega }= \mu_{\omega ,a}(F^{-1}(\Delta)) \ , \qquad \forall \ \Delta\in B(\mathbb R) 
\end{equation}
in other words
$$\mu _{\omega ,F\left( a\right) } = \mu_{\omega ,a}^{F}$$
\end{axiom} 
We note that for the properties of the measure \textit{distribution law of $F$} we have the following relation\footnote{
In practice for every Borel function $g:\mathbb{R}\rightarrow \mathbb{R}$ we obtain: 
\begin{equation*}
\int g\left( s\right) \,d\mu _{\omega ,a}^{F}\left( s\right) =
\int g\left(F\left( s\right) \right) \,d\mu _{\omega ,a}\left( s\right)  
\end{equation*}}:
\begin{equation}
\label{eq.a}
\left\langle F\left( a\right) \right\rangle _{\omega }=\int t\,d\mu _{\omega
,F\left( a\right) }=\int F\left( t\right) \,d\mu _{\omega ,a} \in\mathbb R 
\end{equation}
Obviously for every $f \in C_{0}\left(\mathbb{R}\right)$ we obtain the relation established in  \eqref{valoremedio00}.
\\
 
We still want to underline that we have only taken into consideration the $a$-summable Borel functions because, as already said, the observables of our physical system must have an effective average value to be such.
\begin{remark}\upshape
Warning: if $a\subset b$ then it does not mean that $F(a)\subset F(b)$, i.e. that $\mu _{\omega ,a}=\mu _{\omega ,b}$ for every $\omega \in \mathfrak S_a$.
\end{remark}
Let us try to clarify the consequences of this last observation better.

\subsection{Segal-von Neumann vs. Mackey II step}
In definitions \ref{condition1} and \ref{condition2}, the average value of the physical quantities plays a fundamental role in establishing equality between physical quantities and states of the system.
\\
The relations between average values, in the spirit of the Segal–von Neumann formalism,\index{von Neumann} introduce weaker conditions between observables and states of the system, compared to the relations between Borel measures adopted by Mackey, as highlighted by the following observation:
\begin{remark}\upshape\label{sottobs}
If $a,b$ are two observables of the system with $\mathfrak S_a \subset \mathfrak S_b$, then we have the following obvious implication:
$$ [ \ \mu_{\omega,a} = \mu_{\omega,b} \ , \  \forall \omega\in\mathfrak S_a \ ] \qquad \Longrightarrow \qquad a\subset b $$
The reverse implication is not true; to obtain it the following condition must be satisfied:
\begin{equation}
 \left\langle f(a)\right\rangle _{\omega } =\left\langle f(b) \right\rangle _{\omega } \ , \qquad \forall f\in C_o(\mathbb R) \ ,\ \omega\in\mathfrak S_a
\end{equation}
\end{remark}
In this way one could think of replacing condition III, where we introduce the map  \eqref{misura_borelliana}, with the following\footnote{In other words it is the   map  of relation \eqref{distribuzio2} of remark \ref{Segal-von Neumann vs. Mackey} in section \ref{misureBorele}.}:
\begin{itemize}
\item [III.Bis]  \textit{A unique map 
$$\omega\in\mathfrak S_a \longrightarrow  \left\langle a \right\rangle _{\omega } \in\mathbb R$$
 associated with every observable $a$ of the system.}
\end{itemize}
The problem with this substitution arises when we have to introduce the concept of a function of an observable of the system, a fact which becomes problematic with the  sole  management of the average values of the physical quantities, while it is easy to introduce with Borel measures, as was done in axiom \ref{assio3}.
\\
 
Contrary to Mackey's model, which requires the more stringent equality between probability measures, we adopt the Segal–von Neumann model. In other words, for our model, two observables $a$ and $b$ are equal if and only if
$$\mathfrak S_a = \mathfrak S_b \qquad \textit{and} \qquad [ \ \left\langle a \right\rangle _{\omega } =\left\langle b \right\rangle _{\omega } \ , \  \forall \omega\in\mathfrak S_a \ ] $$ 
while for Mackey, we must have equality between the two measures $\mu_{\omega,a}$ and $\mu_{\omega,b}$; therefore
$$\mathfrak S_a = \mathfrak S_b \qquad \textit{and} \qquad [ \ \left\langle f(a) \right\rangle _{\omega } =\left\langle f(b) \right\rangle _{\omega } \ , \  \forall \omega\in\mathfrak S_a \ , \forall f\in C_o(\mathbb R) \ ] $$
Upon closer inspection, our axiomatic conditions do not differ much from those of Mackey, since they assume \textit{the uniqueness} of the Borel measure given in relation \eqref{misura_borelliana} of condition III, together with axioms \ref{assio1} and \ref{assio2}, lead to these simple statements:
\begin{itemize}
\item Let $a,b\in\mathfrak X$; we have:
\begin{equation*}
a=b  \ \Longleftrightarrow \ \left\{ 
\begin{array}{cc}
\mathfrak S_a=\mathfrak S_b  &      \\ 
\mu _{\omega ,a}=\mu _{\omega ,b}  & \forall \omega\in \mathfrak S_a
\end{array}%
\right. 
\end{equation*}
\item Let $\omega,\omega'\in\mathfrak S$; we have
\begin{equation*}
\omega=\omega' \ \Longleftrightarrow \ \left\{ 
\begin{array}{cc}
\mathfrak X_\omega=\mathfrak X_{\omega'}  &      \\ 
\mu _{\omega ,a}=\mu _{\omega' ,a}  & \forall a  \in \mathfrak X_\omega
\end{array}%
\right. 
\end{equation*}
\end{itemize}

\section{Numbers, Constants and Product for a Scalar}\label{numeroecostanti}
In section \ref{grandezze_rilevate} we introduced the constant observables of the system:
\\
Precisely, an observable $r$ is constant if  there exists  a real number $\texttt{r}\in\mathbb R$ such that
$$\mu_{\omega,r} = \delta_\texttt{r} \ , \qquad \forall\omega\in\mathfrak S_r$$
where with $\delta_\texttt{r}$ we have denoted the Dirac measure:\index{Dirac measure}
\begin{equation}
\delta_\texttt{r} \left( \Delta \right)   =\left\{ 
\begin{array}{ccc}
1 & \text{if} & \texttt{r} \in \Delta \\ 
0 & \text{if} & \texttt{r} \notin \Delta%
\end{array}
\right. 
\end{equation}
it follows  
$$\left\langle r \right\rangle _{\omega }= \texttt{r}  , \qquad \forall \omega\in\mathfrak S_r$$
A consequence of axiom \ref{assio3} is the existence of constant observables of our physical system.
\\
Indeed, consider a constant Borel function, for example
\begin{equation}\label{costante0}
C\left(t\right) = \texttt{r} \ , \qquad \forall t\in\mathbb R 
\end{equation}
where $\texttt{r}$ is a real number. In this way, by axiom \ref{assio3}, we obtain an observable $C\left(a\right)$ such that for every state $\omega\in \mathfrak S_a$\footnote{We remark that $\mathfrak S_r = \mathfrak S_a$.} we have:
\begin{equation}\label{costant}
\left\langle C\left( a\right) \right\rangle _{\omega } =\int C\left( t\right)
\,d\mu _{\omega ,a}=\int \texttt{r} \,d\mu _{\omega ,a}= \texttt{r} 
\end{equation}

We want to underline that Mackey in \cite{Mackey}, using this argument, defines the observables $0$ and $I$ as a trivial consequence of axiom \ref{assio3}.	
\\
In our model,  unlike  Mackey, these constant observables \textit{are not defined for each state of the system $\mathfrak S$ but only for  their  subsets}.
\\

Using axiom \ref{assio3} again, we can define another important class of observables of our physical system.
\\
In fact, if for every real number $\lambda $ we  consider  the Borel function defined by 
$$f_{\lambda }\left( t\right) =\lambda t \qquad , \qquad \forall t \in\mathbb{R}$$
we have
\begin{equation*}
\left\langle f_{\lambda }\left( a\right) \right\rangle _{\omega }=
\int f_{\lambda }\left( t\right) \,d\mu _{\omega ,a}=\int \lambda \,t\,d\mu_{\omega ,a}=
\lambda \,\left\langle a\right\rangle_{\omega } 
\end{equation*}
The observable $f_{\lambda }\left( a\right) $ is denoted by the symbol $\lambda a$.
\\
In this way, we obtain the following  map, the product for a scalar:\label{Productscalar} \index{Product for a scalar}
\begin{equation*}
\left( \lambda ,a\right) \in \mathbb{R}\times \mathfrak{X}\rightarrow \lambda a\in \mathfrak{X} 
\end{equation*}
Let us now consider a Borel set $\Delta$ of $\mathbb R$ and its characteristic function $\mathbf{1}_{\Delta }$ defined in  \eqref{funzcar}.
\\
For every observable $a$ of the system we obtain an observable $\mathbf{1}_{\Delta}(a)$ with the following property:
\begin{equation}
\label{funzcar1}
 \left\langle \mathbf 1_\Delta (a) \right\rangle _{\omega } = \mu_{\omega ,a}(\Delta) \ , \qquad  \forall \omega\in\mathfrak S_a
\end{equation} 
Furthermore, for every Borel set $E$ of $\mathbb R$ and $\omega\in \mathfrak S_a$ we have  
\begin{equation}
P\left( \mathbf 1_\Delta (a)\in E \right) _{\omega }=\left\{ 
\begin{array}{cc}
1 & 0,1\in E \\ 
0 & 0,1\notin E\\ 
\mu _{\omega ,a}\left( \mathbb R \setminus \Delta \right) & \qquad 0\in E
,\ 1\notin E \\ 
\mu _{\omega ,a}\left(  \Delta  \right) & \qquad  0\notin E ,\ 1\in E%
\end{array}%
\right.   
\end{equation}
\\
So we can write:
$$ \mu _{\omega ,\mathbf 1_\Delta (a)} = \texttt{r}_0 \delta_0 + \texttt{r}_1 \delta_1 \ , \qquad \texttt{r}_0=\mu _{\omega ,a}\left( \mathbb R \setminus \Delta \right) \ , \ \texttt{r}_1 = \mu _{\omega ,a}\left(  \Delta  \right)  \ , \ \texttt{r}_0 + \texttt{r}_1 =1 $$
where $\delta_0 \ , \delta_1 $ are the respective Dirac delta measures.
\\

We note that if $f,g$ are $a$-summable Borel functions such that
$$0\leq f(t) \leq g(t) \ , \qquad  \forall t\in\mathbb R $$
we have by definition of average value:
\begin{equation}
\label{positive0}
0 \leq \left\langle f(a)\right\rangle_{\omega }  \leq  \left\langle g(a) \right\rangle_{\omega } \ , \qquad \forall \omega\in\mathfrak S_a
\end{equation}
\subsection{Constant Observables and Compatibility}\label{costcomp}
Let's recap the work done so far:
\\
A constant observable of the laboratory system $r\in\mathfrak X$ is such if there exists a set of states $\mathfrak S_r\subset \mathfrak S$ suitable for its measurement with 
\begin{equation}\label{costante1}
  \left\langle f(r)\right\rangle_{\omega } = f(\texttt{r}) \qquad , \qquad \forall \omega\in\mathfrak S_r \ , \forall f\in C_o(\mathbb R)
\end{equation}
for some real number $\texttt{r}\in\mathbb R$.
\\
Through functional calculus, given a real number $\texttt{r}\in\mathbb R$, we can always determine a constant observable $C(a)$ of the system with
\begin{equation}\label{costante2}
\left\langle f(C(a))\right\rangle_{\omega } = f(\texttt{r}) \qquad , \qquad \forall \omega\in\mathfrak S_a \ , \forall f\in C_o(\mathbb R)
\end{equation}
where the function $C$ is defined in relation \eqref{costante0}.
\\
It follows that for every $a\in\mathfrak X$ we have a constant observable of the system that satisfies relation \eqref{costante2}.
\\
Therefore if $x,y\in\mathfrak X$, we have two observables $C(x)$ and $C(y)$ of the laboratory system that satisfy \eqref{costante2} with $\mathfrak S_x$ and $\mathfrak S_y$ as different sets, whose \textit{intersection could also be an empty set}.
\\
Let us assume the following operational point of view:
\begin{axiom}\index{Axiom-Costants}
All system constants are obtained through functional calculus.
\end{axiom}
In other words, the axiom states that if $r$ is a constant observable of the system as per relation \eqref{costante1}, there exists an observable $a\in\mathfrak X$ such that $C(a)=r$ where the function $C$ is defined in relation \eqref{costante0} and by functional calculus $\mathfrak S_r = \mathfrak S_a$. 
\\
Furthermore $r$ is compatible with $a$ and we can say that\footnote{See axiom \ref{assio3bis} on page \pageref{spettrocongiunto}.}  
\begin{equation}
 \mathfrak S_{r:a}=\mathfrak S_{a:r}=\mathfrak S_{a}
\end{equation}
We observe that in this way, having fixed an observable $a\in\mathfrak X$, for every real number $\texttt{r}\in\mathbb R$ we have a set of constant observables compatible with $a$ itself, but it is not certain that they are compatible with another observable $b$ of the system different from $a$, since the laboratory preparation for the measurement of these constant observables occurs in the same way as that of the observable $a$. 
\subsubsection{Problem of the observable number}
Let's ask ourselves if, given a real number $\texttt{r}\in\mathbb R$, there exists an observable of the system, which we indicate again with $r\in\mathfrak X$ and call the observable number\footnote{Not to be confused with the number operator of quantum mechanics.}, such that
\begin{itemize}
\item every state of the system $\omega$ is suitable for $r$\footnote{Therefore $\mathfrak S_r=\mathfrak S$.};
\item  $\mu_{\omega,r} = \delta_r \ , \qquad \forall\omega\in\mathfrak S$
\end{itemize}
So for the observable number $r$ we have
$$\left\langle r \right\rangle _{\omega }= r  , \qquad \forall \omega\in\mathfrak S$$
In this way we obtain the observables $0$ and $rI$, the observables equal to zero and $r$ respectively, \textit{in every state of the system}.
\\
Obviously for every Borel set $\Delta $ of $\mathbb{R}$ we have the following result:
\begin{equation}
\label{funzcost}
P\left( rI \in \Delta \right) _{\omega } = \delta_r(\Delta)  \ , \qquad \forall \omega\in\mathfrak S 
\end{equation}
Furthermore, these number observables do not having a specific laboratory preparation will be jointly preparable with each observable $a$ of the system with  
$$\mathfrak S_{r:a}= \mathfrak S_a \qquad \textit{and} \qquad \mathfrak S_{a:r} =\mathfrak S_a $$
and
$$ P(a\in\Delta_0 : r\in\Delta_1 )_\omega = \delta_\texttt{r}(\Delta_1) P(a\in\Delta_0)_\omega = P(r\in\Delta_1  :  a\in\Delta_0)_\omega $$
and therefore they are compatible with every observable of the system.
\\
\textit{So why not axiomatically introduce the existence of number observables?}
\\
The problem lies in their (non) definition, which as we have said is free from any experimental procedure and therefore these observables are in contrast with our frequentist approach.
\\
We will see in section \ref{Sistemi fisici del laboratorio} that this notion is well defined for some physical subsystems, which are called suitable. 

\section{Norm of an Observable}\label{normaosservabile}
We now give the definition of bounded observable of the laboratory physical system:
\begin{definition}\upshape
An observable $a$ of $\mathfrak{X}$ is said to be bounded if it occurs
\begin{equation*}
\sup \left\{ | \left\langle a\right\rangle _{\omega }|:\omega \in \mathfrak{S}_a%
\right\} <\infty  
\end{equation*}
\end{definition}
From now on, the term observable denotes only the bounded observables of our physical system.
\\
Let us now consider an observable $a$ of $\mathfrak{X}$ and a state $\omega $ of $\mathfrak{S}_a$
and let $f:\mathbb{R}\rightarrow \mathbb{R}$ be a bounded Borel function; by equation \eqref{eq.a} we obtain
\begin{equation}
\underset{\omega \in \mathfrak{S}_a}{\sup }\left\langle f\left( a\right)
\right\rangle _{\omega }\leq 
\underset{t\in \mathbb{R}}{\sup }\left\vert f\left( t\right) \right\vert =
\left\Vert f\right\Vert _{u}
\label{Riscala infty}
\end{equation}
Considering that the observables of the system are all bounded is not an important restriction for our modeling.
\\
 Indeed by \eqref{Riscala infty}, we can use a bounded Borel function
(set by our experimenter) to "rescale" all measurements obtained.
\\
Therefore it is not restrictive to assume the following \textit{fundamental phenomenological principle} (see Segal \cite{Segal_63}):\index{Segal}
\\
\textit{The set of all bounded observables of a physical system defines the system completely, in all its physically observable aspects.} 
\\
In other words we assume:
\begin{axiom}[\textbf{Observable norm}] \index{Axiom-Observable Norm}
\label{assio della norma} 
For each observable $a$ of the system we have
\begin{equation*}
\sup \left\{ | \left\langle a\right\rangle _{\omega }|:\omega \in \mathfrak{S}_a\right\} <\infty  
\end{equation*}%
The value of the upper bound is usually called the norm of the observable $a$ and is indicated
with the symbol $\left\Vert a\right\Vert$:
\begin{equation}
\label{norm}
\left\Vert a\right\Vert =\underset{\omega \in \mathfrak{S}_a}{\sup }\left\vert
\ \left\langle a\right\rangle _{\omega }\right\vert  
\end{equation}
\end{axiom}
\begin{remark}\upshape
If $\left\Vert a\right\Vert =0$ then we have that $\left\langle a\right\rangle _{\omega }=0$ for every state $\omega\in\mathfrak S_a $, so we have
$$  \left\Vert a\right\Vert =0 \qquad \Leftrightarrow \qquad a\subset 0$$
\end{remark}

Given an observable $a$ of $\mathfrak{X}$, for every bounded Borel function $f:\mathbb{R}\rightarrow \mathbb{R}$, we have:
\begin{equation*}
\left\Vert f( a) \right\Vert \leq \left\Vert f\right\Vert_u 
\end{equation*} 
The \textit{functional calculus} defined in this section is a powerful tool for determining and studying the set
of the observables of a physical system.
\\
From Lebesgue's dominated convergence theorem (see Rudin \cite{Rudin}) follows this important statement:
\begin{remark}\upshape
Let $a$ be an observable and $\omega\in\mathfrak S_a$. 
\\
If there exists a net $\left\{f_\alpha\right\}_\alpha$ of equibounded Borel functions such that 
$$f_\alpha \rightarrow f   \ , \ \mu_{\omega,a}\text{-}a.e.$$
then
$$ \left\langle f_\alpha(a)   \right\rangle_\omega \longrightarrow  \left\langle f (a) \right\rangle_\omega$$
\end{remark}
Before proceeding, let us recall a well-known theorem that is a consequence of Egorov's theorem\footnote{See Folland's book \cite{Folland}, \S 2.4, Theorem 2.33 and Bogachev's book \cite{Boga2} \S 7.2 for the detailed version of Lusin.}.
\begin{theorem}[\textbf{Lusin}]\upshape\label{lusin-0}\index{Lusin's theorem}
Let $\mu$ be a finite Borel measure on a compact metric space $K$. For every $\mu$-measurable function $f:K\rightarrow \mathbb R$ and every $\epsilon>0$, there exists a compact $E\subset K$ such that:
\begin{itemize}
\item[1.] $\mu(K \setminus E)<\epsilon$ 
\item[2.] $f|_E$ is continuous.
\end{itemize} 
\end{theorem}
Furthermore: 
\begin{corollary}[\textbf{Lusin's Theorem for Borel Functions}]\upshape\label{lusin} 
Let $\mu$ be a Borel measure on $\mathbb R$. For every Borel function $f:\mathbb R \rightarrow \mathbb R$ and compact subset $K\subset \mathbb R$, there exists an equibounded net of continuous functions $f_\alpha \in C(K)$ that converges $\mu$-a.e. to $f$ on $K$.\footnote{In our setting, for each state $\omega$, there exists a sequence $f_\alpha^\omega \in C(K)$ converging $\mu_{\omega, a}$-a.e. to $f$.}
Moreover, if $f$ is bounded then we obtain
$$ \sup_{x\in K} | f(x)|   \leq \left\| f_\alpha \right\|_\infty$$
\end{corollary}
\begin{proof}
This is a consequence of Lusin's theorem combined with the fact that convergence in measure implies $\mu$-a.e. convergence for some subsequence.
\end{proof}

\section{Spectrum of an Observable}\label{spettro_oss}
The topic that we will develop in this section is broadly the  scheme outlined  by Deliyannis in \cite{Deliyannis}.\index{Deliyannis}
\\
From the first two axioms of the theory, we can define the important concept of \textit{spectrum of an observable} of our physical system, a set
consisting of all the possible values that the observable itself can take on.
\\
Let's take a closer look at the physical meaning of this statement.
\begin{definition}\upshape\label{value}
A real number $\lambda$ is a \textit{possible value} for the observable $a$ if there exists a state $\omega$ of the system such that
$$P(a\in\{ \lambda \}  )_\omega \neq 0 $$
\end{definition}
We fix an observable $a$ of $\mathfrak{X}$; for each state $\omega $ of $\mathfrak{S}_a$ let us denote by $\mathfrak F^\omega (a)$ the family of open sets
\begin{equation}
\label{Fomega}
  \mathfrak F^\omega (a)=\left\{ U \text{ open of } \mathbb{R} : \ \mu _{\omega ,a}\left( U\right) = 0\right\}  
\end{equation}
and with $\rho ^{\omega }\left( a\right) $ the open set 
\begin{equation}
\rho ^{\omega }\left( a\right) = \bigcup_{U\in\mathfrak F^{  \omega} (a)} U  
\end{equation}
The set $\rho ^{\omega }(a)\in\mathfrak F^\omega (a)$, being an open set of the set of real numbers $\mathbb{R}$, is of measure zero.
\\
Indeed, for the regularity of the measure $\mu_{\omega ,a}$ we have
$$ \mu_{\omega ,a}(\rho ^{\omega }(a)) = \sup \left\{\mu_{\omega ,a}(K): K\subset \rho ^{\omega }(a) \ \text{with } K \text{ compact} \right\} $$
From the compactness of $K$ we can say that there is a finite cover of open $U_i$ of $\rho ^{\omega }(a)$; it follows that we obtain
$\mu_{\omega ,a}\left( K \right) =0$, therefore $\mu_{\omega ,a}\left( \rho ^{\omega}\left( a\right) \right) =0$.
\\
Recall that the support of the measure $\mu _{\omega ,a}$ is the closed set
\begin{equation*}
\operatorname{Supp} \mu _{\omega ,a}=\mathbb{R}\setminus \rho ^{\omega }\left(a\right)
\end{equation*}
Obviously, since $\mu _{\omega ,a}$ is a probability measure, $\operatorname{Supp} \mu _{\omega ,a}$ is a non-empty closed set\footnote{Otherwise $\mu _{\omega , a} (\mathbb R)=0$. }:
\begin{equation}
\label{supp}
 \operatorname{Supp} \mu _{\omega ,a} \neq \emptyset
\end{equation} 
\begin{definition}\upshape
An open $U$ of $\mathbb{R}$ is called $a$-null \textit{if for every state }$\omega $ of $\mathfrak{S}_a$ we obtain:
\begin{equation*}
\mu _{\omega ,a}\left( U\right) =0
\end{equation*}
\end{definition}
We denote by $\mathfrak F^\infty (a)$ the family of all $a$-null open sets:
\begin{equation}\index{$\mathfrak F^\infty (a)$}
\label{a-nulli}
\mathfrak F^\infty (a) = \bigcap_{\omega\in\mathfrak S_a} \mathfrak F^\omega (a) 
\end{equation}
and with $\rho \left( a\right) $ the open set 
\begin{equation}\index{$\rho \left( a\right)$}
\rho \left( a\right) =\bigcup _{ U\in\mathfrak F^\infty (a)}U
\end{equation}
Repeating the reasoning done to verify that the set $\rho^\omega (a)$ is of zero measure, it is easily proven that $\mu _{\omega ,a}\left( \rho(a) \right) = 0$ for all $\omega$ states of the system suitable for $a$. 
\begin{definition}[\textbf{Resolvent and Spectrum set}]\upshape\index{Resolvent set}\index{Spectrum}
The open set $\rho \left( a\right) $ is called the resolvent of the observable $a$, while the set
\begin{equation}
\sigma \left( a\right) =\mathbb{R}\setminus \rho \left( a\right)
\end{equation}%
takes the name of the spectrum of the observable $a$.
\end{definition}
The open set $\rho\left( a\right)$, for every state $\omega$ of the system suitable for $a$, belongs to the family $\mathfrak F^\omega (a)$, so 
$$\rho(a)\subset \rho^\omega (a) \ , \qquad \forall \omega\in\mathfrak S_a$$ 
from this it follows
\begin{equation*}
\rho \left( a\right) \subset  \bigcap\limits_{\omega \in \mathfrak{S}_a}\rho^{\omega }\left( a\right)
\end{equation*}
therefore
$$ \underset{\omega \in \mathfrak{S}_a}{\bigcup} \operatorname{Supp}\,\mu_{\omega ,a} \subset \sigma \left( a\right)  $$
We remark that for each state $\omega$ of $\mathfrak{S_a}$ we have that the measure is supported by the spectrum:
\begin{equation}
\operatorname{Supp}\,\mu_{\omega ,a} \subset  \sigma \left( a\right) 
\end{equation}
We can say that the resolvent of $a$ is the largest open set contained in the set $\bigcap\limits_{\omega \in \mathfrak{S}_a}\rho^{\omega }\left( a\right)$, while its spectrum is the smallest closed set containing $\underset{\omega \in \mathfrak{S}_a}{\bigcup} \operatorname{Supp} \mu_{\omega ,a}$.
\begin{remark}\upshape\label{spettrovuoto}
From relation \eqref{supp} we can say that the spectrum of any observable of the system is a closed non-empty set\footnote{Unlike Deliyannis in \cite{Deliyannis}, we do not consider unbounded observables. We will see that the fact that the spectrum of our observables is non-empty will not be a marginal fact. \index{Deliyannis}}. 
\end{remark}
If $\lambda$ is in the resolvent $\rho(a)$ then t results that $\mu_{\omega ,a}(\left\{\lambda \right\})=0$ for every state $\omega$ of the system suitable for $a$.
In fact we can say that there exists an open neighborhood $U$ of $\lambda$ contained in $\rho(a)$, therefore
$$0 \leq \mu_{\omega ,a}(\left\{\lambda \right\})\leq \mu_{\omega ,a}(U)=0$$
In other words we have proved the following proposition:
\begin{proposition}\upshape\label{spettrolambda}
If there exists a state $\omega$ of the system such that\footnote{It should be noted that $\mu_{\omega ,a}(\left\{\lambda\right\})\neq 0$ does not imply that $\lambda\in \operatorname{Supp} \mu_{\omega,a}$.} 
$$\mu_{\omega ,a}(\left\{\lambda\right\})\neq 0 \qquad \Longrightarrow \qquad \lambda\in\sigma(a)$$
\end{proposition}
Furthermore, if $\Delta $ is a Borel set of $\mathbb{R}$ then
\begin{equation*}
P\left( a\in \Delta \right) _{\omega }=
P\left( a\in \Delta \cap \sigma \left( a\right) \right)_{\omega } \ , \qquad  \forall \omega\in\mathfrak{S}_a
\end{equation*}
since
 $ \mu_{\omega ,a}(\Delta)=\mu_{\omega ,a}(\Delta \cap \sigma (a) ) + \mu_{\omega ,a}(\Delta \setminus \sigma (a) )$, 
with $\mu_{\omega ,a}(\Delta \setminus \sigma (a) )=0$.
\\

We have therefore proved the following
\begin{proposition}\upshape
If $\Delta \cap \sigma ( a ) =\emptyset$ then
\begin{equation*}
P\left( a\in \Delta \right) _{\omega }=0 \ , \qquad  \forall \omega \in \mathfrak{S}_a
\end{equation*}
\end{proposition}
In summary, for any observable $a$ in our physical system $\mathfrak{X}$, the spectrum $\sigma ( a )$ constitutes the complete set of possible measurement outcomes.
\begin{remark}\upshape\label{spettros}
For the definition of the spectrum of an observable, we have not used the three axioms \ref{assio1}, \ref{assio2} and \ref{assio3}.
\end{remark}
Let $a$ be a non-null observable of the system and $\Delta$ a Borel set of $\mathbb R$; consider the characteristic function $\mathbf 1_\Delta$ defined in \eqref{funzcar}.
\\
If we have
$$ \Delta \cap \sigma(a)= \emptyset \qquad \Longrightarrow \qquad \mathbf 1_\Delta (a)\subset 0$$
In fact, if $\Delta \cap \sigma(a)= \emptyset$, then we have that $\Delta\subset \mathbb R \setminus \sigma(a)=\rho(a)$ and for every state $\omega$ of the system we obtain 
 $\mu_{\omega ,a}(\Delta)\leq \mu_{\omega ,a}(\rho(a))=0$ and from relation \eqref{funzcar1}:
\begin{equation}
  P\left( a\in \Delta \right) _{\omega }=\left\langle \mathbf 1_\Delta (a) \right\rangle _{\omega } = \mu_{\omega ,a}(\Delta) \ , \qquad  \forall \omega\in\mathfrak S_a
\end{equation}
and by axiom \ref{assio1} it follows that $\mathbf 1_U (a)\subset 0$.
\begin{corollary}\upshape
Let $a$ be a non-null observable of the system and $U$ be an open set of $\mathbb R$; it follows
$$ \mathbf 1_U (a)\subset 0 \qquad \Longleftrightarrow \qquad U \cap \sigma(a)= \emptyset$$
\end{corollary}
\begin{proof}
If $\mathbf 1_U (a)\subset 0$, for every $\omega\in\mathfrak S_a $ we have $\mu_{\omega ,a}(U) =0$ so $U\subset\rho(a)$; it follows that $\sigma(a) \subset \mathbb R \setminus U$.
\end{proof}
Let's pay attention to the following fact:
\begin{remark}\upshape
If $\lambda\in\sigma(a)$ then we cannot yet say that $\mu_{\omega ,a}(\left\{\lambda\right\})\neq0$. Indeed, by definition we can only say that for every neighbourhood $U$ of $\lambda$ there exists a state $\omega$ such that $\mu_{\omega ,a}(U)\neq0$.
\end{remark}
If $\lambda\in\sigma(a)$ and $U_\lambda$ is a neighbourhood of $\lambda$, then we obtain by the previous corollary that 
$$\lambda\in U_\lambda \cap \sigma(a)\neq \emptyset \qquad \Longrightarrow \qquad \mathbf 1_{U_\lambda}(a) \neq 0 $$ 
but we cannot say that the observable $\mathbf 1_{ \left\{\lambda \right\}}(a) \neq 0$.
\\
If there is a neighbourhood of $\lambda$ such that $U_\lambda \cap \sigma(a) =\left\{\lambda \right\}$, then
$$\mathbf 1_{ \left\{\lambda \right\}}(a)= \mathbf 1_{U_\lambda}(a)\neq 0 $$
and there exists at least one state $\omega \in \mathfrak S_a$ such that 
$$\left\langle \mathbf 1_{ \left\{\lambda \right\}}(a) \right\rangle_\omega=\mu_{\omega ,a}(\left\{\lambda\right\})\neq 0$$
We recall that a point $\lambda$ of the spectrum is isolated if there exists an open neighbourhood $U_\lambda$ of $\lambda$ such that
$$ U_\lambda \cap \sigma(a) = \left\{\lambda \right\}$$
the set of isolated points is called the \textit{point discrete spectrum} of the observable $a$, denoted by $\sigma_{pd}(a)$:
$$ \sigma_{pd}(a)\subset\sigma(a) $$
and\footnote{The last inclusion follows from proposition \ref{spettrolambda}.}:
\begin{equation}\label{pd-spettro}\index{$\sigma_{pd}(a)$}\index{Point discrete spectrum}
\sigma_{pd}(a)\subset\left\{ \lambda \in \mathbb R \ : \  \mathbf 1_{ \left\{ \lambda  \right\} }(a) \neq 0 \right\} \subset \sigma(a) 
\end{equation}
\begin{remark}\upshape\label{sp_risol}
From the previous remark, we can say that not all values of the spectrum $\sigma (a)$ are possible values of the observable $a$ as in definition \ref{value}.
\end{remark}
Furthermore we have these simple double implications for $a$ a non-null observable:
\begin{equation}\label{puntispettro-1}
 [ \ \textit{There exists } \ \omega\in \mathfrak S_a \ \textit{such that} \  \mu_{\omega ,a}(\left\{\lambda\right\})\neq0 \  ]  \  \Longleftrightarrow  \ \mathbf 1_{ \left\{ \lambda  \right\} }(a) \neq 0 
\end{equation}
 and 
\begin{equation}\label{puntispettro-2} 
\lambda \in \sigma(a) \qquad \Longleftrightarrow \qquad [ \ \mathbf 1_{ U_\epsilon }(a) \neq 0  \ \forall \  \epsilon>0 \ ]
\end{equation}
where $U_\epsilon = ]\lambda - \epsilon \ , \ \lambda + \epsilon [ $ for some $\epsilon>0$.
\\
The proof of this last statement is simple. 
\\
In fact, it is enough to note that if 
$$ \mathbf 1_{ U_\epsilon }(a) \neq 0    \ ,  \forall \  \epsilon>0 \ $$ 
and $\lambda\notin\sigma(a)$, then it is possible to take an open set $U$ of $\lambda$ such that $\lambda\in U \subset \rho(a)$; it follows that $\mu_{\omega, a} (U)=0$ for every $\omega\in\mathfrak S_a$, so $\mathbf 1_{ U }(a)=0$, which contradicts the initial hypothesis.
\\
If $\lambda \in \sigma(a)$ and 
$$ \exists \  \epsilon>0  :  \mathbf 1_{ U_\epsilon }(a) = 0  $$
 then $\mu_{\omega, a} (U_\epsilon)=0$; it follows that $\lambda\in U_\epsilon  \subset \rho(a)$, so $\lambda \notin \sigma(a)$, which is absurd.
\subsubsection{Mackey Spectrum}\index{Mackey Spectrum}
In Mackey's work \cite{Mackey} we find the following definition of the spectrum of an observable (see also \cite{Deliyannis}).
\\
For every Borel set $\Delta$ of $\mathbb R$, we consider the set of continuous functions:
\begin{equation}\label{nulMac}
\mathfrak{F}_\Delta=\left\{ f\in C(\mathbb R) : 0\leq f \leq   \mathbf 1_\Delta   \right\}
\end{equation}
The Borel set $\Delta$ is said to be $a$-null according to Mackey if $f(a)\subset 0$ for all functions $f\in\mathfrak{F}_\Delta$.
\\
In \cite{Mackey} the resolvent of $a$ is defined as the union of all open sets of $\mathbb R$ that are $a$-null according to Mackey, and the spectrum is its complement.
\\
\textit{Let us now ask ourselves whether our definition of spectrum coincides with Mackey's.}
\\
To prove that this statement is true, it is sufficient to prove that the two definitions of open $a$-null coincide.
\\
If $U$ is an open $a$-null then by definition it follows that $\mu_{\omega ,a}(U) = 0$ for every state $\omega$ and therefore from relation \eqref{funzcar1} we have $\mathbf 1_U (a) \subset 0$.
\\
From relation \eqref{positive0} we obtain that for every $f\in\mathfrak F_U$
$$ 0 \leq \left\langle f(a)\right\rangle_{\omega }  \leq \left\langle \mathbf 1_U (a)\right\rangle_{\omega }=0  \ , \qquad \forall \omega\in\mathfrak S_a$$
therefore $f(a) \subset 0$ and the open set $U$ is $a$-null according to Mackey.
\\
Conversely, if the open set $U$ of $\mathbb R$ is $a$-null according to Mackey then $f(a) \subset 0$ for each $f\in\mathfrak F_U$ and we consider any compact set $K\subset U$.
\\
By Urysohn's lemma\footnote{See Folland \cite{Folland}, proposition 4.32.\index{Folland}}, we can say that there exists a continuous function $g$ with compact support in $U$ with $0 \leq g\leq 1$ and with $g|_K=1$.
\\
It follows that $g\in\mathfrak F_U$ and by hypothesis $\left\langle g(a)\right\rangle_{\omega}=0$ for each $\omega$ suitable for $a$; this implies that $\mu_{\omega ,a}(K) =0$.
\\
From the arbitrariness of the compact $K\subset U$ and from the previously discussed regularity of the Borel measure $\mu_{\omega ,a}$, we obtain that $\mu_{\omega ,a}(U) =0$ for all states $\omega$ suitable for $a$ and therefore $U$ is an open $a$-null.

\subsubsection{Spectral Mapping}
Let's see some simple but important propositions\footnote{See Deliyannis \cite{Deliyannis}\index{Deliyannis}.}:
\begin{proposition}\upshape
Given an observable $a$ and a state $\omega\in\mathfrak S_a$, for every open $U$ of $\mathbb R$ we have:
$$\mu_{\omega ,a}(U) = \sup \left\{ \left\langle f(a)\right\rangle_{\omega } : f\in\mathfrak F_U \right\}    $$           
\end{proposition}
\begin{proof}
Trivial consequence of the regularity of the measure $\mu_{\omega ,a}$ and of Urysohn's lemma.
\end{proof}
\begin{proposition}\upshape
Let $f:\mathbb{R}\rightarrow \mathbb{R}$ be a continuous function; we have $f\left( a\right) \subset f(0)I$ if and only if
$f^{-1}\left( \mathbb{R}\setminus \left\{ 0\right\} \right)$ is an open $a$-null.
\end{proposition}
\begin{proof}
If $f\left( a\right) \subset  f(0)I$ then
$\mu_{\omega ,f \left( a\right)}= \mu_{\omega ,f(0) I}$ for every state $\omega$ of $\mathfrak{S}_a$ ; it follows 
$$
\mu_{\omega , f( a) } ( \mathbb{R}\setminus  \{0\} ) 
= 
\mu _{\omega ,a} ( f^{-1} ( \mathbb{R}\setminus  \{ 0 \} ) ) = 0
$$
Conversely, it is sufficient to note that for each $\omega $ of $\mathfrak S_ a$ the average value is
$$\left\langle f\left( a\right) \right\rangle_{\omega }
=
\int_{f^{-1}( \{ 0 \} ) } f ( t) \ d\mu _{\omega,a}+
\int_{f^{-1} ( \mathbb{R}\setminus  \{ 0 \} )} f( t ) \ d\mu _{\omega ,a}=f(0)
$$
\end{proof}
As a simple consequence of this proposition we have the following
\begin{corollary}\upshape
\label{spettronullo}
The observable $a \subset 0$ \ 
$\iff $ 
\ \ $\mathbb{R}\setminus \left\{
0\right\} $ is an open $a$-null \ \ 
$\iff $ 
\ \ 
$\sigma \left(a\right) \subset \left\{ 0\right\} $.
\end{corollary}
\begin{proof}
In this case it is sufficient to take $f(t)=t$ for every real $t$.
\end{proof}
We observe that given a Borel function $f:\mathbb R \rightarrow \mathbb R$ with $f(r)=0$ for every $r\in \sigma(a)$, we obtain
$f(a) \subset 0$, since for every state $\omega$ suitable for $a$ we have:
$$
\left\langle f\left( a\right) \right\rangle _{\omega }
=
\int_{\sigma(a)} f \left( r\right) \ d\mu _{\omega,a}(r)
+
\int_{\mathbb{R}\setminus \sigma(a)} f\left( r\right) \ d\mu_{\omega ,a}(r) = 0
$$ 
We now give the equivalent of the spectral mapping of self-adjoint operators for our observables
\begin{theorem}[\textbf{Spectral Mapping for Continuous Functions}]\upshape\label{smc}
Let $f:\mathbb{R}\rightarrow \mathbb{R}$ be a continuous function; for each
observable $a$ of $\mathfrak{X}$ we have
\begin{equation*}
\sigma \left( f\left( a\right) \right) =
\overline{f\left( \sigma \left(a\right) \right) } 
\end{equation*}
\end{theorem}
\begin{proof}
We verify that 
$$\overline{f\left( \sigma \left( a\right) \right) }\subseteq
\sigma \left( f\left( a\right) \right)$$
\\
In other words that 
$$\rho \left( f\left( a\right) \right)\subseteq \mathbb{R}\setminus \overline{f\left( \sigma \left( a\right)
\right) }$$
Let $y\in \rho \left( f\left( a\right) \right) $ ; since the set
$\rho\left( f\left( a\right) \right) $ is open, there exists a neighborhood $U_{y}$
of $y$ such that $U_{y}\subset \rho \left( f\left( a\right) \right)$. 
\\
Furthermore we have by definition that for every state $\omega $ of $\mathfrak{S}_a$:
\begin{equation*}
\mu _{\omega ,a}\left( f^{-1}\left( U_{y}\right) \right) =
\mu _{\omega,f\left( a\right) }\left( U_{y}\right) =0
\end{equation*}
Therefore $f^{-1}\left( U_{y}\right) \subset \rho \left( a\right)$ ; it follows that
\begin{equation*}
\sigma \left( a\right) =\mathbb{R}\setminus \rho \left( a\right)
\subset \mathbb{R}\setminus f^{-1}\left( U_{y}\right) \subset
f^{-1}\left( \mathbb{R}\setminus U_{y}\right)
\end{equation*}
and because
$f\left( f^{-1}\left( \mathbb{R}\setminus U_{y}\right)\right) \subset \mathbb{R}\setminus U_{y}$
we have 
\begin{equation*}
f\left( \sigma \left( a\right) \right) \subset
 f\left( f^{-1}\left( \mathbb{R}\setminus U_{y}\right) \right) \subset \mathbb{R}\setminus U_{y}
\quad \Longrightarrow \quad 
\overline{f\left( \sigma \left(a\right) \right) }\subset \overline{\mathbb{R}\setminus U_{y}}=
\mathbb{R}\setminus U_{y} 
\end{equation*}%
so
\begin{equation*}
y\in U_{y}\subset \mathbb{R}\setminus \overline{f\left( \sigma \left(a\right) \right)}
\quad \Longrightarrow \quad 
\rho \left(f\left( a\right) \right) \subset \mathbb{R}\setminus \overline{f\left(
\sigma \left( a\right) \right) }
\end{equation*}
We now verify the inclusion 
$$\sigma \left( f\left( a\right) \right) \subset \overline{f\left( \sigma \left( a\right) \right) }$$
\\
Let $y\in \mathbb{R}\setminus \overline{f\left( \sigma \left( a\right)
\right) }$ ; since $\overline{f\left( \sigma \left( a\right) \right) }$ 
is closed, there exists a neighborhood $U_{y}$ of $y$ such that $U_{y}\subset 
\mathbb{R}\setminus \overline{f\left( \sigma \left( a\right) \right) }%
\subset \mathbb{R}\setminus f\left( \sigma \left( a\right) \right)$.
\\
Moreover, since $\sigma \left( a\right) \subset f^{-1}f\left( \sigma
\left( a\right) \right)$ we have:
\begin{equation*}
f^{-1}\left( U_{y}\right) 
\subseteq 
f^{-1}\left( \mathbb{R}\setminus f\left( \sigma \left( a\right) \right) \right) 
\subseteq
\mathbb{R}\setminus f^{-1}f\left( \sigma \left( a\right) \right) \subset \mathbb{R}\setminus\sigma \left( a\right) 
=
\rho \left( a\right)
\end{equation*}
it follows that for every $\omega $ of $\mathfrak{S}_a$:
\begin{equation*}
\mu _{\omega ,f\left( a\right) }\left( U_{y}\right) 
\leq 
\mu _{\omega,a}\left( \rho \left( a\right) \right) =0 
\end{equation*}
so $y\in U_{y}\subset \rho \left( f\left( a\right)\right) $, 
in other words 
$\mathbb{R}\setminus \overline{f\left(\sigma \left( a\right) \right) }
\subset 
\rho \left( f\left(a\right) \right)$.
\end{proof}
We have the following useful observation for future considerations:
\begin{remark}\upshape
If the spectrum $\sigma(a)$ is bounded then it is closed and bounded, therefore it is a compact subset of $\mathbb R$; from the continuity of the function $f$ it is easy to verify that relation \eqref{smc} becomes:
\begin{equation}
 \sigma \left( f\left( a\right) \right) = f\left( \sigma \left(a\right) \right)
\end{equation}
\end{remark}

\begin{theorem}[\textbf{Spectral Mapping for Borel Functions}]\upshape\label{smb}
Let $a\in\mathfrak{X}$ ; for every Borel function $f:\mathbb{R}\rightarrow \mathbb{R}$ which is $\mu_{\omega,a}$-summable for every state $\omega$, we have
\begin{equation*}
\sigma \left( f\left( a\right) \right) \subset
\overline{f\left( \sigma \left(a\right) \right)} 
\end{equation*}
\end{theorem}
\begin{proof}
Let's take a point $\lambda\in\sigma(f(a))$; let $U$ be any open neighborhood of $\lambda$ ; as mentioned, there exists a state $\omega$ such that $\mu_{ \omega, f(a)}(U)\neq 0$.
\\
By definition we have that $\mu_{\omega, a}(f^{-1}(U))=\mu_{\omega, f(a)}(U) \neq 0$ and therefore we can say that
$ f^{-1}(U) \cap \operatorname{Supp} \mu_{\omega,a} \neq \emptyset$ ; this implies that
$$\emptyset\neq f( f^{-1}(U) \cap \operatorname{Supp} \mu_{\omega,a})\subset f( f^{-1}(U)) \cap f( \operatorname{Supp} \mu_{\omega,a}) \subset U\cap f( \operatorname{Supp} \mu_{\omega,a})$$
it follows that $\lambda$ belongs to the closure of the set $f( \operatorname{Supp} \mu_{\omega,a})$ ; hence the thesis.
\end{proof}
\begin{remark}\upshape
If an observable $a$ has a spectrum with only one element, i.e. $\sigma(a) = \left\{ \lambda_o  \right\}$, then $a\subset \lambda_o I$, because $\mu_{\omega,a} \left\{ \lambda_o \right\}=1$ and
$$ \left\langle a \right\rangle_\omega = \int t \ d\mu_{\omega,a} = \lambda_o \ \mu_{\omega,a}\left\{ \lambda_o \right\} = \lambda_o \ , \qquad \forall \omega \in \mathfrak S_a$$ 
\end{remark}
\subsubsection{Simple consequences of Borel spectral mapping}
Let $\Delta$ be a Borel set of $\mathbb R$; from the spectral mapping for Borel functions, we obtain
\begin{equation*}
 \sigma(\mathbf 1_\Delta (a))\subset \overline{ \mathbf 1_\Delta (\sigma(a))}= \left\{ 
  \begin{array}{ccc}
\left\{ 1  \right\}  & \text{if} & \sigma(a)\subset \Delta \\
 \left\{ 0, 1  \right\}  & \text{if} & \Delta \cap \sigma(a) \neq \emptyset\\
\left\{ 0  \right\}  & \text{if} & \Delta \cap \sigma(a)= \emptyset %
\end{array}
\right.
\end{equation*}
so for each observable $a$ of the physical system we have
$$\sigma(\mathbf{1}_{ \left\{ \lambda \right\} }(a))  \subset \left\{ 0, 1  \right\}  \ , \ \forall \lambda\in \mathbb R$$
Moreover, let $a$ be a non-null observable; we have the following statements:
\begin{itemize}
\item If $\sigma(\mathbf{1}_{ \left\{ \lambda \right\} }(a))=\left\{ 1  \right\}$, then $\sigma(a) = \left\{ \lambda  \right\} $, because $a\subset \lambda I$.  
\\
Indeed, by the previous remark we obtain that $\mathbf{1}_{ \left\{ \lambda \right\} }(a) \subset I$, therefore $\mu_{\omega,a}( \left\{ \lambda \right\} ) =1$ for each state $\omega \in \mathfrak S_a$, so
$$ \left[ \ \left\langle a \right\rangle_\omega = \lambda \ , \qquad \forall \omega \in \mathfrak S_a \ \right] \qquad \Longrightarrow \qquad a\subset \lambda I$$
\item If $\sigma(\mathbf{1}_{ \left\{ \lambda \right\} }(a))=\left\{ 0, 1  \right\}$ then $\lambda\in\sigma(a)$.
\\
In this case the observable $\mathbf 1_{ \left\{ \lambda \right\} }(a)$ cannot be zero by corollary \ref{spettronullo}. It follows that there is at  least one state $\omega$ suitable for $a$ such that $\left\langle \mathbf 1_{ \left\{ \lambda \right\} }(a) \right\rangle _{\omega }\neq 0$, so $\mu_{\omega,a}( \left\{ \lambda \right\} ) \neq 0$.
\item If $\sigma(\mathbf{1}_{ \left\{ \lambda \right\} }(a))=\left\{ 0 \right\}$ then $\mu_{\omega,a}( \left\{ \lambda \right\} ) = 0$ for each state $\omega\in\mathfrak S_a$.
\\
In this case we cannot say whether $\lambda$ is or is not an element of $\sigma(a)$. 
\end{itemize}
It is useful to underline the following statements:
\begin{equation}
\label{spettro_misure}
  \underset{\omega \in \mathfrak{S}_a}{\bigcup} \operatorname{Supp}\,\mu_{\omega ,a}\subset \sigma \left( a\right) \qquad , \qquad \rho \left( a\right) \subset  \underset{\omega \in \mathfrak{S}_a}{\bigcap} \rho^\omega(a)
\end{equation}
We conclude this section with a simple proposition:
\begin{proposition}\upshape\label{modulo-stato}
For every state $\omega\in\mathfrak S$, we have:
$$|\omega|= \sup_{x\in\mathfrak X_{\ \omega}} \frac{\left| \left\langle x \right\rangle_\omega \right|}{\left\| x \right\|} = 1 $$
\end{proposition}
\begin{proof}
By definition, $\left| \left\langle x \right\rangle_\omega \right| \leq \left\| x \right\|$, which implies $|\omega|\leq 1$.
\\
By functional calculus, taking the constant function $C(t)=1$ for all $t\in\mathbb R$, we obtain:
$$\frac{\left| \left\langle C(x) \right\rangle_\omega \right|}{\left\| C(x) \right\|}=1$$
\end{proof}
\section{Spectral Radius}
For each observable $a$ of $\mathfrak{X}$ we define its \textit{spectral radius} $r\left( a\right) $ as
\begin{equation*}
r\left( a\right) =
\sup \left\{ \left\vert \lambda \right\vert :\lambda \in\sigma \left( a\right) \right\}\in  \left[  0 , +\infty \right]
\end{equation*}
\begin{proposition}\upshape
If the observable $a$ of $\mathfrak{X}$ has bounded spectral range, then there exists an element $\lambda_{o}$
of $\sigma \left( a\right) $ such that
\begin{equation*}
r\left( a\right) =\left\vert \lambda _{o}\right\vert
\end{equation*}
\end{proposition}
\begin{proof}
The spectrum of $a$ is a bounded closed subset of $\mathbb{R}$ since $\sigma(a) \subset \left[ -r(a) \ , r(a) \right]$; then it is a compact set.
\\
By definition of the upper bound there exists a natural number $\overline{n}$
such that for every natural number $n>\overline{n}$ there exists an element $\lambda _{n}$ of $\sigma \left( a\right)$
 with the property
\begin{equation*}
0<r\left( a\right) -\frac{1}{n}<\left\vert \lambda _{n}\right\vert
\end{equation*}
The set $\sigma \left( a\right) $ is a compact set of $\mathbb{R}$ ; therefore there exists a subsequence
 $\left\{ \lambda_{n_{k}}\right\} _{k}$ of $\left\{ \lambda_{n}\right\}_{n}$ that is convergent:
\begin{equation*}
\lambda_{n_{k}}\underset{k\rightarrow \infty }{\rightarrow }\lambda_{o}\in \sigma \left( a\right)
\end{equation*}
Obviously $r\left( a\right) \leq \left\vert \lambda _{o}\right\vert$; it follows necessarily that $r\left( a\right) =\left\vert \lambda _{o}\right\vert$.
\\
Therefore if $\lambda _{o}>0$ then $r\left(a\right) \in \sigma \left( a\right)$, while if $\lambda _{o}<0$ then $-r\left(a\right) \in \sigma \left( a\right)$.
\end{proof}
We can give a simple link between the spectral radius and spectral mapping for continuous functions:
\begin{corollary}
Let $a$ be an observable of $\mathfrak{X}$; for every bounded continuous function $f:\mathbb{R}\rightarrow \mathbb{R}$ we have:
\begin{equation*}
r\left( f\left( a\right) \right) =\left\Vert f\right\Vert_u
\end{equation*}
where
\begin{equation*}
\left\Vert f\right\Vert _u
=
\underset{s\in \sigma ( a) }{\sup }\left\vert f( s) \right\vert
\end{equation*}
\end{corollary}
\begin{proof}
By the previous theorem we have
$$
r ( f ( a) ) = \sup \left\{ \left\vert t\right\vert :t\in
\sigma \left( f\left( a\right) \right) \right\} 
=
\sup \left\{ \left\vert t\right\vert :t\in \overline{f( \sigma(a)) }\right\} 
=
\underset{s\in \sigma ( a ) }{\sup }\left\vert f (s) \right\vert
$$
because it is easy to verify that 
$$
\sup \left\{ \left\vert t\right\vert :t\in \overline{f ( \sigma (a)) }\right\} 
=
\sup \left\{ \left\vert f( s)\right\vert :s\in \sigma \left( a\right) \right\}  
$$
\end{proof}
Let's see what link exists between the norm of the observable defined by \eqref{norm} and its spectral radius.
\\
We have:
\begin{equation}
\left\Vert a\right\Vert \leq r\left( a\right)
\end{equation}
Indeed, for every $\varepsilon >0$ there exists a $\omega $ of $\mathfrak S_a$ such that  
\begin{eqnarray*}
\left\Vert a\right\Vert -\varepsilon <\left\vert \left\langle a\right\rangle_{\omega }\right\vert 
 & = &
\left\vert \int t \ d\mu _{\omega ,a}\right\vert 
\leq
\int \left\vert t \right\vert \ d\mu_{\omega ,a} =
\\
& = &
\int\limits_{\sigma(a) }\left\vert t \right\vert \ d\mu_{\omega ,a}
\leq
\int\limits_{\sigma ( a) } r ( a) \ d\mu_{\omega,a}= r( a)
\end{eqnarray*}
Warning: At this level of discussion we cannot yet say that the norm of an observable coincides with its spectral radius (see the spectral property of states)
\\

We conclude this section by recalling the following definition:
\begin{definition}\upshape\index{Quantized observable}
An observable $a$ of the physical system is said to be quantized if its spectrum $\sigma(a)$ is a discrete set.
\end{definition}
\section{Positive Observables}
Let us now define the positive observables of our physical system.
\begin{definition}\upshape
An observable $a\in\mathfrak X$ is positive, in symbols $a>0$, if for every Borel set $\Delta\subset ]-\infty, 0[$ we have
$$P(a\in\Delta)_\omega=0  \qquad \forall \omega\in\mathfrak S_a $$ 
\end{definition}
The following double implication is easily proved
$$a>0  \qquad \Longleftrightarrow  \qquad \sigma(a)\subset [0,  +\infty [ $$
Furthermore, if $a>0$ we obtain that $\left\langle a \right\rangle_\omega > 0$ for every $\omega\in\mathfrak S_a$.
\\

We note that if the function $f$ is $a$-summable and positive, then by spectral mapping we can say that for every observable $a$, the observable $f(a)$ is positive.
\\
Let us now consider the following continuous functions:
\begin{equation} \label{decompositivo}
 f_+(t )= t \ \mathbf{1}_{[0, +\infty[ }(t) \ , \qquad  f_{-}(t)= - t  \ \mathbf{1}_{ ] -\infty, 0[ } (t) 
\end{equation}
with
$$ a_+=f_+(a)>0  \ , \qquad  a_{-} =f_{-} (a) >0 $$
in this way we obtain that each observable of the system is the sum of two positive observables\footnote{See the paragraph dedicated to the sum of compatible observables}:
\begin{equation}
\label{decompositivo1}
a=a_{+} -  a_{-}
\end{equation}
since for every state $\omega\in\mathfrak S_a =\mathfrak S_{f(a)}$ we have
\begin{equation*} 
\left\langle a \right\rangle_\omega = \int\limits_{\sigma ( a) } f_{+} ( t) \ d\mu_{\omega,a}
- \int\limits_{\sigma ( a) } f_{-}(t) \ d\mu_{\omega,a} = \left\langle a_+ \right\rangle_\omega - \left\langle a_{-} \right\rangle_\omega 
\end{equation*}
From spectral mapping for continuous functions we obtain:
\begin{itemize}
\item $\sigma( a_{+}) = \sigma(a) \cap [0, +\infty[$
\item $\sigma( a_{-}) = \sigma(a) \cap ]-\infty, 0[$ 
\end{itemize}
it follows
\begin{equation}
\label{raggio2}
r(a)= \max\left\{ r(a_{+}), r(a_{-})\right\}
\end{equation}
\begin{proposition}\upshape
\label{quadratonullo}
If $a^2\subset 0$ then $a\subset 0$.
\end{proposition}
\begin{proof}
Indeed by \eqref{decompositivo} for every $\omega\in\mathfrak S_a$ we can write:
\begin{eqnarray*}
   \left\langle a^2 \right\rangle_\omega & = & \int [ (f_+ (t) - f_-(t) ]^2 d \mu_{\omega,a} = \int f_+^2 (t) d \mu_{\omega,a} + \int f^2_-(t) d \mu_{\omega,a} = 
\\ 
&=&
\left\langle a_+^2 \right\rangle_\omega + \left\langle a_-^2 \right\rangle_\omega = 0
\end{eqnarray*}
from this it follows that $a_{\pm}^2 \subset 0$ and from the positivity of the observables we obtain that $a_{\pm} \subset 0$.
\end{proof}
We underline that for each $a\in\mathfrak X$ we have
\begin{eqnarray}\label{norm-sum1}
 \left\| a \right\| = \left\| a_+ \right\| + \left\| a_- \right\|
\end{eqnarray}
\section{Spectral Property of States}
In this section we will establish when the spectral radius of an observable coincides with its norm defined by \eqref{norm}.
\\
Let $a$ be an observable; denote by $\underline{\sigma}=\inf \sigma(a)$ and by $\overline{\sigma}=\sup \sigma(a)$; obviously $\sigma(a)\subset \left[ \underline{\sigma} \ , \ \overline{\sigma} \right]$.
\\
In accordance with experimental experience, we obtain that the average value of an observable is always included among its possible values that it can take:
$$ \underline{\sigma} \leq \int t \ d\mu_{\omega ,a} \leq \overline{\sigma} $$
We will assume that a stronger statement than the previous one is true\footnote{See Deliyannis \cite{Deliyannis} \index{Deliyannis}.}, namely that the map
$$\omega\in\mathfrak S_a \rightarrow \left\langle a \right\rangle_\omega \in \left[ \underline{\sigma} \ , \ \overline{\sigma} \right]$$
is surjective.
\\
Precisely we take into consideration the following spectral property of the states of a physical system:

\begin{axiom}[\texttt{SPS}] \label{SPS}\index{Axiom-Spectral Property of the States}\index{\texttt{SPS}} 
Let $a$ be an observable of $\mathfrak{X}$ ; for each element $s_{o} \in \left[ \underline{\sigma} \ , \ \overline{\sigma} \right]$
 there exists a state $\omega _{o}$ of $\mathfrak{S}_a$ such that 
$$\left\langle a \right\rangle_{\omega_o} = s_o$$
\end{axiom}

Let's see a consequence of the \texttt{SPS}:
\\
we have repeatedly said that even if $\mu_{\omega,a}(\left\{ \lambda \right\})=0$, it does not mean that $\lambda$ cannot be an element of the spectrum.
\\
But if $\lambda\in\sigma(a)$, then by the SPS there must exist at least one $\omega_o\in\mathfrak S_a$ such that
$$\left\langle a \right\rangle_{\omega_o} =\lambda$$
in this way there exists a state of the system in which the mean of the observable $a$ is equal to the value $\lambda$, but since by hypothesis
$$P(a\in \left\{ \lambda \right\} )_{\omega} = 0 \qquad \forall \omega \in\mathfrak S_a$$
the observable $a$ will never assume the value $\lambda$.
\\
We see an important consequence of the \texttt{SPS} property.
\begin{proposition}\upshape
If the \texttt{SPS} property holds, then we obtain
\begin{equation*}
\left\Vert a\right\Vert = r\left( a\right)  
\end{equation*} 
furthermore the  map  $f\in C ( \sigma ( a ) ) \rightarrow f ( a ) \in \mathfrak{X}$
where $C ( \sigma ( a ) ) $ is the  linear  space of  real-valued  continuous functions, is an isometry: 
$$
\left\Vert f\left( a\right) \right\Vert 
=
\left\Vert f\right\Vert_u
=
\underset{s\in \sigma ( a ) } {\sup }\left\vert f( s) \right\vert  
$$
\end{proposition}
\begin{proof}
We just need to prove that $\left\Vert a\right\Vert \geq r(a)$.
\\
By SPS, for every $\lambda\in\sigma(a)$ there exists a state $\omega\in\mathfrak S_a$ such that $\left\langle a \right\rangle_\omega =\lambda$; from this it follows that $\left\Vert a\right\Vert \geq \lambda$ and from here the thesis follows.
\\
The observables have finite norm by Axiom \ref{assio della norma}, so the spectrum is a compact non-empty set of $\mathbb R$:
$$\sigma(a) \subset \left[ -\left\Vert a\right\Vert \ , \ \left\Vert a\right\Vert \ \right]$$
The second statement of the proposition is a simple consequence of spectral mapping.
\end{proof}
We have another consequence of SPS\footnote{The converse of this statement is found in the work of Deliyannis \cite{Deliyannis}. Furthermore see also axiom (VIII) of Mackey's book and Wightman's paper \cite{Wightman76}.}:
\begin{corollary}\upshape\label{pss_a}
If the SPS property holds for our physical system, then for every open $U\notin\mathfrak F^\infty (a)$ there exists a state $\omega$ suitable for $a$ such that
$$ \left\langle \mathbf 1_U (a) \right\rangle _{\omega } = 1 $$
\end{corollary}
\begin{proof}
Let $U\notin\mathfrak F^\infty (a)$; by definition we obtain $U \cap \sigma(a) \neq \emptyset$, so we can choose $\lambda\in U \cap \sigma(a)$. By Urysohn's lemma there exists a function $f\in\mathfrak F_U$ with $f(\lambda)=1$. 
\\
By spectral mapping for continuous functions we obtain 
$$ \sup \sigma(f(a)) = \sup \overline{f(\sigma(a))} = 1$$
from SPS there exists a state $\omega\in\mathfrak S_a$ such that $\left\langle f(a) \right\rangle _{\omega } = 1$. Therefore $\mu_{\omega ,a} (U) \geq 1$, hence the thesis.
\end{proof}
\begin{proposition}\upshape
If the SPS property holds then we have
$$ a>0 \qquad \Longleftrightarrow \qquad \left\langle a \right\rangle_\omega \geq 0 \qquad \forall \omega\in\mathfrak S_a $$ 
\end{proposition} 
\begin{proof}
From SPS, for every $\lambda\in\sigma(a)$ there exists $\omega$ such that $\lambda = \left\langle a \right\rangle_\omega \geq 0$; it follows that $\sigma(a) \subset [0, \infty[$.
\end{proof}

The axiom \ref{assio3} and the \texttt{SPS} attribute a meaning to the writing $a^{2}$ (it is the square of the measures of the observable $a$) and in general to the $n$th power $a^{n}$ of our observable $a$.
\\
In fact, from the compactness of $\sigma(a)$ we obtain that the function 
$$f(t)=t^n \qquad , \qquad \forall t\in\mathbb R$$
is $a$-summable for every natural number $n$.
\\
We now have a simple proposition:
\begin{proposition}\upshape
\label{*banach}
Given an observable $a$ of $\mathfrak{X}$, for every state $\omega$ of $\mathfrak{S}_a$ we have:
\begin{equation*}
\left\vert \left\langle a\right\rangle _{\omega }\right\vert ^{2}\leq
\left\langle a^{2}\right\rangle _{\omega }   
\end{equation*}
\end{proposition}
\begin{proof}
This is a simple consequence of H\"{o}lder's inequality (see Folland \cite{Folland} proposition 6.32):
\begin{equation*}
\left\vert \left\langle a\right\rangle _{\omega }\right\vert ^{2}=
\left\vert \int t\,d\mu _{\omega ,a}\right\vert ^{2}\leq \left[ \int \left\vert
t\,\right\vert d\mu _{\omega ,a}\right] ^{2}\leq \int \left\vert
t\right\vert ^{2}\,d\mu _{\omega ,a}=
\left\langle a^{2}\right\rangle_{\omega }
\end{equation*}
\end{proof}
We note that from proposition \ref{*banach} we obtain 
$$ \left\| a \right\| ^2 \leq \left\| a^2 \right\|$$ 
and from spectral mapping for continuous functions we obtain  that 
\begin{equation}
r\left( a^{2}\right) = r\left( a\right) ^{2}
\end{equation}
\begin{remark}[\textbf{C*-norm properties}]\upshape \label{C_star_norma}\index{C*-norm properties}
From the \texttt{SPS} we obtain the following property, which is called \textit{C*-norm properties}:
\begin{equation}
\left\Vert a^{2}\right\Vert = \left\Vert a\right\Vert ^{2}
\end{equation}
\end{remark}
By the C*-norm properties and relation \eqref{norm-sum1}, we obtain
\begin{eqnarray}\label{norm-sum2}
 \left\| a \right\| = \sqrt{\left\| a_+ \right\|^2 + \left\| a_- \right\|^2}
\end{eqnarray}
\begin{attenzione}\upshape
One might think that $a\subset b$ implies the following inclusion $\sigma(a)\subset \sigma(b)$, but from remark \ref{sottobs} we do not have that 
\begin{equation}\label{sottobs2}
\mu_{\omega,a}=\mu_{\omega,b} \ , \qquad \forall \omega\in \mathfrak S_a \subset \mathfrak S_b
\end{equation}
and therefore we cannot say that the family of all $b$-null open sets $\mathfrak F^\infty (b)$ is contained in $\mathfrak F^\infty (a)$\footnote{See page \pageref{a-nulli}.}.
\end{attenzione}
If equality \eqref{sottobs2} holds (in this way we will write $a\subset \subset b$), then $a\subset b$ and moreover $\sigma(a)\subset \sigma(b)$, so $r(a) \leq r(b)$ and $\left\| a \right\| \leq \left\| b \right\|$:\index{$a\subset \subset b$}
$$ a\subset \subset b \qquad \Longrightarrow \qquad \left\| a \right\| \leq \left\| b \right\| $$
\subsubsection{Bayes Measure}
We see another important consequence of the spectral property of states.
\\
Let us fix an observable $a$ and a suitable state $\omega$ for it and we study the following  map: 
$$\Delta_o\in B(\mathbb R) \longrightarrow  \frac{\left\langle \mathbf{1}_{\Delta_o}(a) \ a \ \mathbf{1}_{\Delta_o}(a) \right\rangle_\omega }{\left\langle \mathbf{1}_{\Delta_o}(a) \right\rangle_\omega} \in \mathbb R$$
with $\Delta_o$ a Borel set with $\mu_{\omega,a}( \Delta_o)\neq 0$.
\\
I  state  that we have 
\begin{equation}\label{projection}
 \frac{\left\langle \mathbf{1}_{\Delta_o}(a) \ a \ \mathbf{1}_{\Delta_o}(a) \right\rangle_\omega }{\left\langle \mathbf{1}_{\Delta_o}(a) \right\rangle_\omega} =\int s \ d \eta
\end{equation}
where $\eta$ is the probability measure (Bayes) as defined 
$$\eta(\Delta)= \frac{\mu_{\omega,a}(\Delta \cap \Delta_o) }{\mu_{\omega,a}( \Delta_o)} \qquad , \qquad \forall \Delta\in B(\mathbb R)$$
Obviously
$$\left\langle \mathbf{1}_{\Delta_o}(a) \ a \ \mathbf{1}_{\Delta_o}(a) \right\rangle_\omega = \int f(t) d \mu_{\omega,a}= \int_{\Delta_o} t d \mu_{\omega,a}$$
where
$$f(t)= \mathbf{1}_{\Delta_o}(t) \ t $$
and 
$$f(a)= \mathbf{1}_{\Delta_o}(a) \ a \ \mathbf{1}_{\Delta_o}(a)$$
\\
For every $\Delta\in B(\mathbb R)$ we have 
\begin{equation} 
f^{-1}(\Delta)= \left\{ 
\begin{array}{cccc}
\emptyset & 0\notin\Delta & \Delta \cap \Delta_o=\emptyset \\ 
\Delta \cap \Delta_o & 0\notin\Delta & \Delta \cap \Delta_o \neq \emptyset 
\\
\mathbb R \setminus \Delta_o & 0\in\Delta & \Delta \cap \Delta_o = \emptyset  
\\
\mathbb R \setminus \Delta_o \cup (\Delta \cap \Delta_o) & 0\in\Delta & \Delta \cap \Delta_o \neq \emptyset    
\end{array}
\right. 
\end{equation}
it is easy to realise that
$$ \mu_{\omega,a}(f^{-1}(\Delta)) = \mu_{\omega,a}( \Delta \cap \Delta_o) + \delta_0(\Delta) \mu_{\omega,a}( \mathbb R \setminus \Delta_o ) $$
where $\delta_0$ is the Dirac measure centred at zero, which written in a compact way becomes
\begin{equation}
 \mu_{\omega,f(a)}(\Delta)= c_o \eta (\Delta) + (1-c_o)\delta_0(\Delta) \ , \ \forall \Delta\in B(\mathbb R)
\end{equation}
with $c_0= \mu_{\omega,a}(\Delta_o)$, therefore
\begin{eqnarray*} 
\left\langle \mathbf{1}_{\Delta_o}(a) \ a \ \mathbf{1}_{\Delta_o}(a) \right\rangle_\omega & = & \int s \ d \mu_{\omega,f(a)}(s) =
\\
& = & c_o \int t \ d \eta(t) + (1-c_o)\int t \ d \delta_0(t) = c_o \int t \ d \eta(t)
\end{eqnarray*}
hence expression  \eqref{projection}.
\begin{problem}
Is there $\omega_o\in \mathfrak S_a$  such  that the equation
\begin{equation} \label{projection2}
\left\langle a \right\rangle_{\omega_o} = \frac{\left\langle \mathbf{1}_{\Delta_o}(a) \ a \ \mathbf{1}_{\Delta_o}(a) \right\rangle_\omega }{\left\langle \mathbf{1}_{\Delta_o}(a) \right\rangle_\omega} 
\end{equation}
 holds?
\end{problem}
If we consider the \texttt{SPS} property to be valid, then the answer to this question is positive, since it is an immediate consequence of the mean value theorem of mathematical analysis. 
\\
Indeed, we have that $\mu_{\omega,a} ( \Delta_o \cap \sigma(a) ) = \mu_{\omega,a} ( \Delta_o )$, so
$$\left\langle \mathbf{1}_{\Delta_o}(a) \ a \ \mathbf{1}_{\Delta_o}(a) \right\rangle_\omega = \int_{\Delta_o} t d \mu_{\omega,a} = \int_{\Delta_o \cap \sigma(a)} t d \mu_{\omega,a} = s \ \mu_{\omega,a} ( \Delta_o \cap \sigma(a) )$$
with $s\in \Delta_o \cap \sigma(a) \subset \sigma(a)$; therefore by the \texttt{SPS} property, there exists $\omega_o\in\mathfrak S_a$ such that $s = \left\langle a \right\rangle_{\omega_o}$, hence 
$$ \left\langle \mathbf{1}_{\Delta_o}(a) \ a \ \mathbf{1}_{\Delta_o}(a) \right\rangle_\omega = \left\langle a \right\rangle_{\omega_o} \ \mu_{\omega,a} ( \Delta_o \cap \sigma(a) ) = \left\langle a \right\rangle_{\omega_o} \ \left\langle \mathbf{1}_{\Delta_o}(a) \right\rangle_\omega$$
\begin{problem}\upshape
We now ask ourselves what the state $\omega_o$ given in equation  \eqref{projection2}  physically represents and what physical interpretation we should give to the transition
$$\omega \ \leadsto \ \omega_o $$ 
\end{problem}
We underline that $P(a\in\Delta_o)_{\omega}=c_o$, while
$$ P(a\in\Delta_o)_{\omega_o}= \mu_{\omega_o,a} (\Delta_o)=1 $$
where in this way $\mu_{\omega_o,a}=\eta$.
\subsubsection{Value-States of an Observable}
Let us now make some useful observations on the set of states of a physical system.
\\
If $\lambda\in\sigma(a)$, then by the \texttt{SPS} property there exists at least one state $\omega_o$ suitable for $a$ such that 
$$\left\langle a \right\rangle_{\omega_o} = \lambda$$
The state $\omega_o$ is called a  \textit{value-state}  relative to the value $\lambda$ of the observable $a$ and the set
\begin{equation}
\label{autostati}\index{$\texttt{U}_\lambda (a)$}
  \texttt{U}_\lambda (a) = \left\{ \omega \in \mathfrak S_a \ : \ \left\langle a \right\rangle_\omega = \lambda \right\}
\end{equation}
is conventionally called the value-state set of $\lambda$.
\\
We have the following inclusion:
$$ \bigcup_{\lambda \in \sigma(a)}  \texttt{U}_\lambda (a) \subset \mathfrak S_a $$
Let us now ask ourselves when the equality sign  holds  in the previous inclusion.
\\
This means that for every state $\omega$ suitable for $a$, there exists a $\lambda \in \sigma(a)$ such that 
$$\left\langle a \right\rangle_\omega = \lambda$$
This property is called \textit{dual property of states}.
\section{Questions}
Let us now give the important definition  of a Question or \textit{Yes/No Observable}:
\begin{definition}\upshape
An observable $q$ of $\mathfrak{X}$ is a question if
$$
P\left( q\in \left\{ 0,1\right\} \right) _{\omega } = 1 \ ,\qquad \forall \omega\in\mathfrak{S}_q 
$$
\end{definition}
A \textit{trivial question} is a question $q$ such that $q\subset 0$ or $q\subset I$.\index{Trivial Question observable}
\\
 
We observe that if $q$ is a question with $P( q\in \left\{ 1\right\})_\omega = r$, then we obtain that $P( q\in \left\{0\right\})_\omega = 1 - r$,
since by the well-known properties of measurement we have
$$
\mu _{\omega ,q}( \left\{ 0,1\right\}) 
=
\mu _{\omega ,q}(\left\{ 0\right\}) + \mu_{\omega ,q} ( \left\{ 1\right\} )
$$
Furthermore we can easily verify that for each state $\omega \in\mathfrak S_q$, we have  
\begin{equation}
P\left( q\in \Delta \right) _{\omega }=\left\{ 
\begin{array}{cc}
1 & 0,1\in \Delta \\ 
0 & 0,1\notin \Delta \\ 
\mu _{\omega ,q}\left( \left\{ 0\right\} \right) & \qquad 0\in \Delta
,\ 1\notin \Delta \\ 
\mu _{\omega ,q}\left( \left\{ 1\right\} \right) & \qquad  0\notin\Delta ,\ 1\in \Delta%
\end{array}%
\right.   
\label{osservabile test misura}
\end{equation}
In other words:
\begin{equation}
\label{misuraquestione}
 \mu _{\omega ,q} = r_0 \delta_0 + r_1 \delta_1 \ , \ r_0 = \mu _{\omega ,q} ( \left\{ 0\right\}) \ , \ r_1 = \mu _{\omega ,q} ( \left\{ 1\right\} ) \ , \ r_0 + r_1 = 1 
\end{equation}
with $\delta_0 , \delta_1$ the respective Dirac measures.
\begin{proposition}\upshape \label{spettro test}
If the observable $q$ is a non-trivial question then\footnote{Obviously if $q_o\subset 0$ and $q_1\subset I$ then 
$$
\sigma \left( q_o\right) =\left\{ 0\right\} \ , \qquad \sigma \left(q_1\right) =\left\{ 1\right\}  
$$}
$$
\sigma \left( q\right) = \left\{ 0,1 \right\}
$$ 
\end{proposition}
\begin{proof}
Trivial consequence of relation  \eqref{osservabile test misura}. 
\end{proof}
We have another simple statement.
\begin{proposition}\upshape
An observable $q \in \mathfrak{X}$ is a question if and only if $q^{2} = q$.
\end{proposition}
\begin{proof}
From proposition \ref{spettro test} we have for every $\omega\in\mathfrak S_q$
\begin{equation*}
\left\langle q\right\rangle_{\omega } = \int t \ d\mu_{\omega ,q}
=
0 \cdot \mu_{\omega ,q} ( \left\{ 0\right\} ) + 1 \cdot \mu_{\omega ,q}(\left\{ 1\right\} )
=
\mu_{\omega ,q} ( \left\{ 1\right\})
\end{equation*}
\\
and 
\begin{equation*}
\left\langle q^{2}\right\rangle_{\omega }
=
\int t^{2} \ d\mu_{\omega,q}
=
0^{2}\cdot \mu_{\omega ,q} ( \left\{ 0\right\}) + 1^{2}\cdot \mu_{\omega ,q}( \left\{ 1\right\} )
=
\mu_{\omega ,q} (\left\{ 1\right\})
\end{equation*}
therefore $\left\langle q^{2}\right\rangle _{\omega } = \left\langle q\right\rangle _{\omega }$ for every state $\omega$, so we have $q^{2}=q$.
\\
Conversely, we assume that $q^{2}=q$ and consider the observable $b = f(q)$ where $f$ is the continuous function $f(t) = t^{2} - t \ , \quad t\in\mathbb R $.
\\
We have for every $\omega$ suitable for $q$ 
$$ \left\langle b\right\rangle_{\omega } = \int (t^{2} - t) \ d\mu_{\omega ,q} = \left\langle q^2\right\rangle_{\omega } - \left\langle q\right\rangle_{\omega } = 0 $$
it follows that $b = 0$.
\\
From spectral mapping for continuous functions and from the \texttt{SPS} property, it  follows that
$$ \left\{ 0 \right\} = \sigma(b) = f(\sigma(q))$$
Therefore, if $\lambda\in\sigma(q)$ it must necessarily  hold  that $\lambda^{2} - \lambda = 0$, hence the thesis.
\end{proof}
For every question $q$ of the system  there remains associated  with it the orthogonal question denoted by $q^{\bot}$ defined as follows:
$q^{\bot} = f(q)$ where $f$ is the continuous function $f(t) = 1 - t \ , \ \forall t\in\mathbb R$. In this way we obtain:
\begin{equation}\label{questioneortogonale}
\mu_{\omega, q^{\bot}}(\Delta) = \mu_{\omega, q}( f^{-1} (\Delta)) = \int \mathbf{1}_{\Delta}(1-t) \ d \mu_{\omega, q}(t)
\end{equation}
therefore
\begin{equation*}\label{questioneortogonale1}
P\left( q^{\bot} \in \left\{ 0\right\} \right) _{\omega } = P\left( q\in \left\{ 1\right\} \right) _{\omega } \qquad , \qquad P\left( q^{\bot} \in \left\{ 1\right\} \right) _{\omega } = P\left( q\in \left\{ 0\right\} \right) _{\omega }
\end{equation*}
and of course
$$\left\langle q^{\bot} \right\rangle _{\omega} = 1 - \left\langle q \right\rangle _{\omega } \ , \qquad \forall \omega\in\mathfrak S$$
Let us denote by $\mathcal{P}\left( \mathfrak{X}\right)$ the set of question observables of our physical system and introduce into it the following partial order relation\footnote{See also the work of Jauch and Piron \cite{Jauch63}.}:\index{Jauch}\index{Piron}

\begin{definition}\upshape
Let $p$ and $q$ be two question observables; we say $p$ is contained in $q$, in symbols $p < q$, if and only if the following conditions apply simultaneously:
\begin{itemize}
\item $\mathfrak S_p \subset \mathfrak S_q $ , 
\item $\left\langle p\right\rangle _{\omega } < \left\langle q\right\rangle _{\omega } \ , \qquad \forall \omega\in \mathfrak{S}_p$.
\end{itemize}
\end{definition}
\subsection{Spectral Projectors of an Observable}
Using axiom \ref{assio3} we can construct question observables of our physical system. In fact it is easy to verify that given an observable $a$ of the system and a Borel set $\Delta$ of $B(\mathbb{R})$, the observable $\mathbf{1}_{\Delta} (a)$ is a question.
\\
The question $\mathbf{1}_{\Delta} (a)$ is said to be the \textit{spectral projector} relative to the observable $a$.
\\

Furthermore, it is easy to verify the following double implication:
\begin{equation}
\mathbf{1}_{\Delta_{1}}( a ) 
<
\mathbf{1}_{\Delta_{2}}\left( a\right) 
\qquad \Longleftrightarrow \qquad 
\mu_{\omega ,a}(\Delta_{1}) < \mu_{\omega ,a} ( \Delta_{2} ) \qquad \forall \omega\in\mathfrak S_a
\end{equation}
Can we say that every question observable is a spectral projector?
\\
Let $q$ be a question;  we can write $q = \mathbf{1}_{\Delta}(a)$ for some Borel set $\Delta \supset \left\{ 0,1\right\}$ if there exists an observable $a$ that satisfies the following properties:
\\
\textbf{1} - $\mathfrak S_a = \mathfrak S_q$,
\\
\textbf{2 } - $\mu_{\omega ,a} ( \left\{ 0 \right\} ) = P(q \in \left\{ 0\right\})_\omega \qquad , \qquad \forall \omega \in \mathfrak S_a$.
 \\
 
It is useful to note that two observables $a$ and $b$ of $\mathfrak{X}$ coincide if and only if 
$$\mathbf{1}_\Delta ( a) = \mathbf{1}_\Delta (b)$$ 
for every Borel set $\Delta $ of $B\left(\mathbb{R} \right)$.
\\
Indeed if this equality holds, then we also have $\mathfrak S_a = \mathfrak S_b$ and so for every state $\omega$ suitable for $a$ and $b$ we obtain 
$$\left\langle \mathbf{1}_\Delta ( a ) \right\rangle_{\omega } = \left\langle \mathbf{1}_\Delta ( b )\right\rangle_{\omega }$$ 
therefore $\mu_{\omega ,a} = \mu_{\omega ,b}$ and by axiom \ref{assio1} we have that $a = b$.
\\
Let's now make some simple considerations about spectral projectors and the  \texttt{SPS} property.
\\
We consider an observable $a$ non-null and a Borel set $\Delta$ of $\mathbb R$ in such a way that $\mathbf 1_\Delta (a)$ is a non-trivial spectral projector;  therefore $\sigma( \mathbf 1_\Delta (a)) = \left\{ 0 , 1 \right\}$ and from the \texttt{SPS} property it follows that for every value $s \in [0,1]$ there exists a state $\omega_o \in \mathfrak S_a$ such that
$P(a \in \Delta)_{\omega_o} = s$, since
$$\left\langle \mathbf 1_\Delta (a) \right\rangle_{\omega_o} = P(a \in \Delta)_{\omega_o}$$
so if $\lambda \in \sigma(a)$ and the spectral projector $\mathbf{1}_{ \left\{ \lambda \right\}}(a)$ is non-trivial, then we can modulate the state $\omega \in \mathfrak S_a$ in such a way as to obtain 
$$ P(a \in \left\{ \lambda \right\} )_\omega = 1 $$ 
It follows that if $\lambda \in \sigma_{pd}(a)$ then the set\footnote{We underline that $$  \texttt{V}_\lambda (a) \subset   \texttt{U}_\lambda (a) $$ }  
\begin{equation}\label{autostati1}\index{$\texttt{V}_\lambda (a)$}
   \texttt{V}_\lambda (a) = \left\{ \omega \in \mathfrak S_a \ : \ P(a \in \left\{ \lambda \right\} )_\omega = 1 \right\}
\end{equation} 
is non-empty, because $\mathbf{1}_{ \left\{ \lambda \right\}}(a) \neq 0$.
\\
Furthermore 
$$ \omega \in   \texttt{V}_\lambda (a) \qquad \Longleftrightarrow \qquad \mu_{\omega ,a} = \delta_\lambda $$
and the set $ \texttt{V}_\lambda (a)$ is constituted by purely informational states in the measurement of $a$.
\\
In section \ref{Spectral Decomposition of an Observable} we will use sums of spectral projectors to approximate a generic observable of the system.

\chapter{Pure States and Borel Measures}\label{Pure States and Borel Measures}
In the previous sections we introduced the set of physical states of the system $\mathfrak S$ without providing it with any particular algebraic-topological structure, hence the difficulty of producing mathematical tools to probe its properties. For example, we cannot introduce the definition of a mixed state since we do not have a linear space structure on this set; we will see that to do this we must exploit the properties of the set of probability measures associated with the observables and states of the physical system. 
\section{States and Measures}\label{States and Measures}
This section is a mathematical parenthesis; we will very briefly recall some simple results and definitions of measure theory; for further discussion refer to Folland's book \cite{Folland}.\index{Folland}\index{Folland}
\\
We consider a topological space $X$, locally compact and $T_2$, and denote by $B(X)$ the family of its Borel sets.
\\
We denote by $\Pi(X)$\footnote{If there is no ambiguity about the use of the topological space $X$, we shall denote this set only by the Greek letter $\Pi$.} the set of probability measures defined on the Borel sets of $X$. 
\\
Let $\nu_1$ and $\nu_2$ be two $\sigma$-finite Borel measures on $(X, B(X))$ ; it is said that $\nu_1$ is \textit{absolutely continuous} with respect to $\nu_2$, in symbols $\nu_1 \ll \nu_2$, if we have $\nu_1(\Delta)=0$ for every Borel set $\Delta$ for which $\nu_2(\Delta)=0$. 
\\
If our measures are finite, by the \textit{Radon-Nikodym} theorem, there exists a non-negative function $\rho\in L^1(X , \nu_2)$ such that
$$ \nu_1(\Delta) = \int_\Delta \rho \, d \nu_2 \ , \qquad \forall \Delta\in B(X) $$
The function $\rho$ is called the \textit{Radon-Nikodym derivative}, in symbols:\index{Radon-Nikodym derivative}
$$ d \nu_1 = \rho \, d \nu_2$$
\begin{example}\upshape\label{exempioleb} 
We observe that if $\delta_{t_o}\in\Pi$ is a Dirac delta measure then for every measure $\lambda\in\Pi$ with $\lambda(\left\{ t_o \right\})\neq 0$ we obtain by definition that $\delta_{t_o} \ll \lambda$ and the Radon-Nikodym derivative is given by
\begin{equation}
\label{derivata-RN}
\rho(s)= \frac{1}{\lambda(\left\{ t_o \right\})} \mathbf{1}_{ \left\{t_o  \right\}} (s)  \ , \qquad \lambda\text{-}a.e. 
\end{equation}
\end{example}
Recall that two $\sigma$-finite measures $\nu_1$ and $\nu_2$ \textit{are disjoint}, in symbols $\nu_1 \perp \nu_2$, if there exists a Borel set $\Delta$ such that $\nu_1(\Delta)=0$ and $\nu_2(\mathbb R \setminus \Delta)=0$.
\\

As we have repeated several times, by the Riesz-Markov theorem, the set of probability measures $\Pi$ are in one-to-one correspondence with the states of the C*-algebra $C_o(X)$; we will identify these two sets by writing 
\begin{equation}
\label{insieme-pi}\index{$\Pi$}
 \Pi = \ \left\{ \mu\in C_o(X)^* : \| \mu \| = 1 \right\}
\end{equation}
We denote by $\Pi_p$ \textit{the set of extremal points} of $\Pi$, i.e. the set consisting of the pure states (positive functionals of norm 1) of $C_o(X)^*$.
\\
It is well known that the set $\Pi_p$ is in one-to-one correspondence with the points of the topological space $X$\footnote{See paragraph 9.1 of the book of Hamhalter \cite{Ham}.}:\index{Hamhalter}\index{$\Pi_p$}
$$ \Pi_p = \left\{ \delta_r : r \in X \right\}$$
where $\delta_r$ is the Dirac measure associated with the point $r \in X$.
\\			

Recall that the set $\Pi$ is a W*-compact subset of $C_o(X)^*$. Therefore by the Krein-Milman theorem it is generated by its extremal points; this means that it coincides with the intersection of all the convex sets of $C_o(\mathbb R)^*$ that contain $\Pi_p$.
\bigskip

Let us briefly recall the \textit{Lebesgue decomposition} theorem for Borel probability measures.\index{Lebesgue decomposition}
\begin{theorem}\upshape\index{Lebesgue decomposition theorem}
Let $\nu \in \Pi$; for each $\mu\in\Pi$ we obtain a unique pair of positive Borel measures\footnote{Which are not usually probability measures.} $(\mu_A , \mu_S)$ such that:
$$\nu = \mu_A + \mu_S \qquad \text{with} \qquad \mu_A \ll \mu \ , \ \mu_S \perp \mu $$
\end{theorem}  
\begin{proof}
Let us consider the finite real Borel measure $\lambda = \nu + \mu$; obviously $\nu \ll \lambda$ and $\mu \ll \lambda$ and therefore by the Radon-Nikodym theorem there are two functions $\rho_\nu, \rho_\mu \in L^1(X,\lambda)$ such that
$$ d \nu = \rho_\nu \, d \lambda \qquad \text{and} \qquad d \mu = \rho_\mu \, d \lambda $$
We denote by 
\begin{equation}
\label{Q}
 \mathcal{Q} = \left\{ s\in X : \rho_\mu(s) \neq 0 \right\} \in B(X) \qquad \text{and} \qquad \mathcal{Q}_o = X \setminus \mathcal{Q} 
\end{equation} 
We define for each $\Delta\in B(X)$ the following Borel measures:
$$ \mu_A(\Delta) = \nu(\Delta \cap \mathcal{Q}) \qquad \text{and} \qquad \mu_S(\Delta) = \nu(\Delta \cap \mathcal{Q}_o)$$
since $\mu( \mathcal{Q}_o) = 0$ we can write
$$ \mu_A(\Delta) = \int_{\Delta \cap \mathcal{Q}} \rho_\nu \, d \lambda =  
\int_{\Delta \cap \mathcal{Q}} \frac{\rho_\nu}{\rho_\mu} \, d \mu = \int_{\Delta} \frac{\rho_\nu}{\rho_\mu} \, d \mu $$
it follows that $\mu_A \ll \mu$.
\\
Furthermore, by definition $\mu_S(\mathcal{Q}) = 0$, hence $\mu_S \perp \mu$.
Let us look at the uniqueness of the decomposition.
\\
If $(\mu_A' , \mu_S')$ is another pair of positive measures that decompose the measure $\nu$ according to Lebesgue, then 
$$ \mu_A - \mu_A' = \mu_S - \mu_S'  $$
with $(\mu_A - \mu_A') \ll \mu$ and $(\mu_S - \mu_S') \perp \mu$; it follows that $\mu_A - \mu_A' = 0$.
\end{proof}
Let us now look at some consequences, in the case $X = \mathbb R$, of this famous theorem; for simplicity of writing let us set
$$\chi_{(\nu , \mu)} := \mu_A(\mathbb R) \in [0, 1]$$
If $\chi_{(\nu ,\mu)} \in (0, 1)$ then we can write
\begin{equation}
\label{decoleb0}
\nu = \chi_{(\nu , \mu)} \, \mu_1 + (1 - \chi_{(\nu , \mu)}) \, \mu_2 \qquad \text{with} \qquad \mu_1 \ll \mu \ , \ \mu_2 \perp \mu  
\end{equation}
where $\mu_1 , \mu_2 \in \Pi$ are defined as:
$$\mu_1(\Delta) = \mu_A(\Delta) / \mu_A(\mathbb R)  \qquad , \qquad \forall \Delta\in B(\mathbb R)$$
and
$$\mu_2(\Delta) = \mu_S(\Delta) / \mu_S(\mathbb R) \qquad , \qquad \forall \Delta\in B(\mathbb R)$$
Furthermore, if $\rho_{(\nu ,\mu)} = d \nu / d \mu_A$, then for every $\Delta \in B(\mathbb R)$ we can write:
\begin{equation}
\nu(\Delta) = \chi_{(\nu , \mu)} \int_\Delta \rho_{(\nu , \mu)} (s) \, d \mu_1 (s) + (1 - \chi_{(\nu , \mu)}) \, \mu_2 (\Delta)  
\end{equation}
Summing up:
\\
If there is a $\mu\in\Pi$ such that $\chi_{(\nu , \mu)} \notin \left\{ 0 , 1 \right\}$, then $\nu \notin \Pi_p$.
\\ 
So we can say\footnote{We remark that if $\chi_{(\nu ,\mu)} = 0$ we have $\mu_1 = 0$, while if $\chi_{(\nu ,\mu)} = 1$ we have $\mu_2 = 0$.}
\begin{equation}
\label{pp}
 \nu \in \Pi_p \qquad \Longrightarrow \qquad \left[ \chi_{(\nu , \mu)} \in \left\{ 0 , 1 \right\} \ \ \forall \mu\in\Pi \right]
\end{equation}
Let us give a useful example of Lebesgue's decomposition:
\begin{example}\upshape
Given the measure $\nu\in\Pi$, we find the Lebesgue decomposition with respect to the measure $\mu = \delta_{t_o} \in \Pi$:
$$\nu = \mu_A + \mu_S \qquad \text{with} \qquad \mu_A \ll \delta_{t_o} \ , \ \mu_S \perp \delta_{t_o} $$
In this case, we have that the derivative $\rho_\mu$ is given by expression \eqref{derivata-RN}, so we have $\mathcal{Q} = \left\{ t_o \right\}$ and $\mathcal{Q}_o = \mathbb R \setminus \left\{ t_o \right\}$.
\\
Therefore, for every $\Delta\in B(\mathbb R)$ we obtain 
$$ \mu_A(\Delta) = \nu (\Delta \cap \mathcal{Q}) = \nu ( \left\{ t_o \right\} ) \delta_{t_o}(\Delta) \qquad , \qquad \mu_S(\Delta) = \nu (\Delta \cap \mathcal{Q}_o) $$
Let us now reverse the role of the measures; let $\nu = \delta_{t_o}$ and find the Lebesgue decomposition with respect to any measure $\mu\in\Pi$:
$$ \delta_{t_o} = \mu_A + \mu_S \qquad \text{with} \qquad \mu_A \ll \mu \ , \ \mu_S \perp \mu $$
In this case we have two possibilities\footnote{We highlight that here $\mathcal{Q}$ is in general different from $\left\{ t_o \right\}$. }:
$$\text{if } \ t_o \in \mathcal{Q} \qquad \Longrightarrow \qquad \mu_A = \delta_{t_o} \ , \ \mu_S = 0 $$
while
$$\text{if } \ t_o \notin \mathcal{Q} \qquad \Longrightarrow \qquad \mu_A = 0 \ , \ \mu_S = \delta_{t_o} $$
\end{example}
 $$ \star $$ 
Next, we will focus our attention on the following subsets of the set of probability measures $\Pi$:
\begin{equation}\label{misure_in_tutto}\index{$\mathbb {M}$}
\mathbb M = \left\{\mu_{\omega,a} \in C_o(\mathbb R)^* : a\in\mathfrak X , \   \omega\in\mathfrak S_a\right\}\subset \Pi
\end{equation}
 and for any observable $a$ of $\mathfrak{X}$  
\begin{equation} \label{misure_in_a}\index{$\mathbb {M}\left( a\right)$}
\mathbb {M}\left( a\right) = \left\{ \mu _{\omega ,a} \in C_o(\mathbb R)^* : \omega \in \mathfrak{S}_a \right\}\subset \mathbb M
\end{equation}
\section{Convexity}
Let $\Delta _{o}$ be a Borel set of $B\left( \mathbb{R}\right) $; an observable $a$ of $\mathfrak{X}$ is said to be a \textit{certainty} in $\Delta_{o}$ in the state $\omega $ of $\mathfrak{S}_a$ if we have:
$$ P\left( a\in \Delta_{o}\right)_{\omega }=1$$
For example, a question observable is a certainty in the Borel set $\left\{ 0, 1 \right\}$ for each state suitable for it.
\\
We introduce the following notion of mixed state of the physical system:\index{Mixture of states}
\begin{definition}[\textbf{Mixture}]\upshape
Let $a$ be an observable of the system; a state $\omega $ of $\mathfrak{S}_a$ is a mixture of two
states $\omega _{1}$ and $\omega _{2}$ of $\mathfrak{S}_a$ in the measure of $a$, if there exists $r\in (0,1)$
such that 
\begin{equation}
\label{mistura0}
\mu_{\omega ,a} = (1-r) \mu_{\omega_{1},a} + r \mu _{\omega_{2},a}
\end{equation}
\end{definition}
From this we easily obtain that 
\begin{equation} \label{mistura}
\left\langle a\right\rangle _{\omega } = (1-r) \left\langle a\right\rangle
_{\omega _{1}} + r \left\langle a\right\rangle _{\omega _{2}}
\end{equation}
and by definition and by \eqref{mistura0}, for every $F:\mathbb R \rightarrow \mathbb R$ that is $a$-summable we obtain: 
\begin{equation} \label{mistura2}
\mu_{\omega ,F(a)} = (1-r) \mu_{\omega_{1},F(a)} + r \mu _{\omega_{2},F(a)} 
\end{equation}
We remark that if $\omega $ is a mixture of $\omega _{1}$ and $\omega_{2}$ in the measure of $a$ and the observable $a$
is a certainty in $\Delta_{o}$ in the $\omega $ state, then $a$ still turns out to be a certainty in $\Delta_{o}$ in both the $\omega_{1}$ and $\omega_{2}$ states of the system.
\\
In fact, let $\Delta _{o}$ be a Borel set of $\mathbb{R}$ such that
$$P\left(a\in \Delta_{o}\right)_{\omega } = \mu _{\omega ,a}\left( \Delta_{o}\right) = 1$$
this implies that $\mu_{\omega ,a}\left( \mathbb{R}\setminus \Delta _{o}\right) = 0$ 
and by definition of mixed state 
$$
(1-r) \mu_{\omega _{1},a}\left( \mathbb{R}\setminus \Delta_{o}\right)
+
r \mu_{\omega_{2},a}\left( \mathbb{R}\setminus \Delta _{o}\right) = 0
$$
from this last expression we can say that 
$$
\mu_{\omega_{1},a}\left( \mathbb{R}\setminus \Delta _{o}\right)
=
\mu_{\omega _{2},a}\left( \mathbb{R}\setminus \Delta _{o}\right) = 0
$$
therefore
$\mu _{\omega _{1},a}\left( \Delta _{o}\right) = \mu_{\omega _{2},a}\left( \Delta _{o}\right) = 1$.
\\
The converse of this statement is not true:
\\
If $a$ is a certainty in some Borel set $\Delta_{o}$ for the state $\omega_{1}$, then it is not necessarily still so for the mixed state $\omega$. In other words, \textit{mixtures can obscure certainties.}
\\

In the algebraic theory of quantum mechanics, we have the important definition of \textit{dominated states}. Let us see how this notion transfers to our model:
\begin{definition}\upshape\label{dominato_0}
A state $\omega_o \in \mathfrak{S}_a$ is dominated by a state $\omega \in \mathfrak{S}_a$ in the measure of $a$, if there exists a $\lambda \geq 1$ such that 
\begin{equation}
\label{dominato_1}\index{Domination states}
\mu _{\omega _{o} ,a}\left( \Delta \right) < \lambda \ \mu _{\omega,a}\left(
\Delta \right) \ , \qquad \forall \Delta \in B\left( \mathbb{R}\right)
\end{equation} 
\end{definition}

Given a state $\omega_o$ dominated by $\omega$ we can define the following functional of $C_o(\mathbb R)$:
$$ \nu(f) = \frac{\lambda \mu _{\omega ,a} (f) - \mu _{\omega _{o} ,a}(f)}{\lambda - 1} > 0 \ , \qquad \forall f\in C_o(\mathbb R) $$
Obviously
$$ \left(1 - \frac{1}{\lambda}\right) \nu + \frac{1}{\lambda} \mu _{\omega_o ,a} = \mu _{\omega ,a} $$
Now the functional $\nu \in C_o(\mathbb R)^*$, \textit{but we cannot say} that there exists a state $\omega_1$ of the system such that
$\mu_{\omega_1 ,a} = \nu$; if such a state exists, then the dominant state $\omega$ is a mixture of $\omega_o$ and $\omega_1$. 
\\
Let's see the converse of this statement:
\\
If the state $\omega_o$ is not dominated in the measure of $a$ by any state $\omega \in \mathfrak S_a$, then $\omega_o$ cannot be a mixed state.
\\
In fact, if there exist $\omega_1 , \omega_2 \in \mathfrak S_a$ that satisfy  \eqref{mistura0}  for some $r \in ]0, 1[$, we can write 
$$ \frac{1}{r} \mu _{\omega_o ,a} > \mu _{\omega_1 ,a}$$
contradicting our initial hypothesis.  
\\

We conclude this topic with some simple statements:
\\

If the state $\omega_o \in \mathfrak S_a$ is dominated by $\omega \in \mathfrak S_a$ then $\mu_{\omega_o,a} \ll \mu_{\omega,a}$.
\\
Therefore there exists a function 
$$\rho = \frac{d\mu_{\omega_o,a}}{d\mu_{\omega,a}} \in L^1(\mathbb R, \mu_{\omega,a})$$
such that for each $f \in C_o(\mathbb R)$ we obtain: 
\begin{equation}
 \mu_{\omega_o,a}(f) = \int f(s) \, d\mu_{\omega_o,a}(s) = \int f(s) \rho(s) \, d\mu_{\omega,a}(s)  
\end{equation}
Furthermore, if $a$ and $b$ are observables with $\mu_{\omega,a} \ll \mu_{\omega',b}$ for some state $\omega \in \mathfrak S_a$ and $\omega' \in \mathfrak S_b$, then 
\begin{equation} \mu_{\omega,f(a)} \ll \mu_{\omega',f(b)} \ , \qquad \forall f \in C_o(\mathbb R)
\end{equation}
Indeed for every Borel set $\Delta$ of $\mathbb R$ we have:
$$ \mu_{\omega',f(b)}(\Delta) = \mu_{\omega',b}(f^{-1}(\Delta)) \qquad , \qquad \mu_{\omega,f(a)}(\Delta) = \mu_{\omega,a}(f^{-1}(\Delta)) $$ 
and if $\mu_{\omega',f(b)}(\Delta) = 0$ then $\mu_{\omega,f(a)}(\Delta) = 0$.
\bigskip

Let us now assume that $\omega$ is a mixture in the measure of $a \in \mathfrak X$, as in expression \eqref{mistura0}, then we have:
$$ \mu_{\omega_{i},a} \ll \mu_{\omega ,a} \ , \qquad i = 1,2 $$
and by the Radon-Nikodym theorem, there are two functions $\rho_1, \rho_2 \in L^1(\mathbb R, \mu_{\omega ,a})$, which depend on the observable $a$ and on the states $\omega_1, \omega_2$ and $\omega$, such that
$$ \left\langle f(a)\right\rangle _{\omega_i } = \int f(s) \rho_i(s) \, d \mu_{\omega ,a} (s) \ , \qquad i = 1,2 $$
for each function $f$ which is $\mu_{\omega,a}$-summable.
\\
Again from expression \eqref{mistura0} we obtain    
$$ \int f(s) \, d \mu_{\omega ,a} (s) =
\int f(s) \left[ (1-r) \rho_1(s) + r \rho_2(s) \right] \, d \mu_{\omega ,a} (s) $$
for each function $f$ which is $\mu_{\omega,a}$-summable; therefore for each $r \in (0,1)$ we have
$$ 1 = (1-r) \rho_1(s) + r \rho_2(s) \ , \qquad \mu_{\omega ,a}\text{-}a.e. $$
\subsection{Sectors in the Measure of an Observable}\label{sectors}
In elementary quantum mechanics it is assumed that given two states of the system suitable for the observable $a$ it is always possible to determine an intermediate state that allows us to measure $a$\footnote{See, for example, Emch \cite{Emch84} paragraph 8.3.a.}:\index{Emch}
\begin{property}[\textbf{Convexity}]\label{assio6}\index{Convexity}
Let $a$ be an observable of the system; if $\omega _{1}$ and $\omega _{2}$ are states belonging to $\mathfrak{S}_a$, then
for every real number $r \in \left[ 0,1 \right]$ there exists a state $\omega$\footnote{That is not to say it is unique.} of $\mathfrak{S}_a$ such that
\begin{equation}\label{amistura}
\mu_{\omega ,a} = (1-r) \mu_{\omega_{1},a} + r \mu_{\omega _{2},a}
\end{equation} 
\end{property}
\noindent

In practice this property says that it is always possible experimentally to re-arrange the instruments/devices in the laboratory in order to obtain a state of the system in which it is possible to modulate the measurement between the two values given by the two starting states.
\\

\textit{This statement is very strong and may not have any real experimental feasibility.} 
\\

We will adopt a lighter version; we will assume that in the set $\mathbb {M}\left( a\right)$ there are subsets in which the previous convexity property \ref{assio6} holds, and precisely:
\begin{axiom}[\textbf{Measurement Sectors}]\label{assio6bis}\index{Axiom-Measurement Sectors}\index{$\mathbb {M}_k\left( a\right)$}
The set $\mathbb {M}\left( a\right)$ has a family of convex subsets $ \left\{ \mathbb {M}_k\left( a\right) \right\}_{k\in I}$ of $\mathbb M(a)$ called the  measurement sectors of the observable $a$, with the following properties:
\begin{itemize} 
\item [1.] [\textbf{Weak Closing}]
Each set $\mathbb M_k(a)$ is W*-top. closed in $\Pi$, i.e. for each net $\mu _{\omega_\alpha ,a} \in \mathbb M_k(a)$ such that
$$ [ \ \mu _{\omega_\alpha ,a} (f) \stackrel{\alpha}{\longrightarrow} \mu(f) \ , \ \forall f \in C_o(\mathbb R) \ ] \qquad \Longrightarrow \qquad \mu \in \mathbb M_k(a)$$ 
\item [2.] [\textbf{Upper Bound}]  If there exists a convex subset $S$ of $\mathbb M(a)$ such that
$$\mathbb {M}_k\left( a\right) \subset S \ \Longrightarrow \ \mathbb {M}_k\left( a\right) = S $$  
\item [3.] [\textbf{Covering}] 
$$ \mathbb M(a) = \bigcup_{k \in I} \mathbb {M}_k\left( a\right) \ , \qquad I \subset \mathbb N $$ 
\item [4.] [\textbf{Convex Disjunction}]  For every $h \neq k$ and $\mu _{\omega_h ,a} \in \mathbb M_h(a)$ and $\mu _{\omega_k ,a} \in \mathbb M_k(a)$ property \ref{assio6} does not hold; i.e. there exists $r \in ]0, 1[$ such that
$$ (1-r) \mu_{\omega_{h},a} + r \mu_{\omega _{k},a} \notin \mathbb {M} \left( a\right)$$  
\end{itemize}
\end{axiom}
\index{$\texttt{Ext}(a)$}
Let us denote by $\texttt{Ext}_k(a)$ the extremal points of the convex set $\mathbb {M}_k\left( a\right)$ and 
$$ \texttt{Ext}(a) := \bigcup_{k \in I} \texttt{Ext}_k(a) \subset \Pi $$
\begin{remark}\upshape
The set $\mathbb M_k(a)$ is convex and W*-top. closed; therefore by the Krein-Milman theorem it is generated by its extremal points $\texttt{Ext}_k(a)$.
\end{remark}
\begin{proposition}\upshape\label{disgiunzione-settori} 
If there is more than one measurement sector for $\mathbb M(a)$, then we have
$$ \mathbb M_{h}(a) \cap \mathbb M_{k}(a) = \emptyset \ , \qquad h \neq k $$
\end{proposition}
\begin{proof}
By contradiction, assume that there exists $\mu_{\omega,a} \in \mathbb M_{h}(a) \cap \mathbb M_{k}(a)$; from the convexity of 
$\mathbb M_{h}(a)$, we obtain that for every $\mu_{\omega_h,a} \in \mathbb M_{h}(a)$  
$$ (1-r) \mu_{\omega_{h},a} + r \mu_{\omega,a} \in \mathbb M_{h}(a) \subset \mathbb M(a) \ , \ \forall r \in ]0, 1[ $$
but by hypothesis $\mu_{\omega,a}$ also belongs to $\mathbb M_{k}(a)$; therefore by the convex disjunction property there should exist at least one value $r_o \in ]0, 1[$ such that $(1-r_o) \mu_{\omega_{h},a} + r_o \mu_{\omega, a} \notin \mathbb M(a)$, hence the contradiction.  
\end{proof}
The measurement sectors $\mathbb M_k(a)$ of $\mathbb M(a)$ determine, obviously, a selection in the states suitable for the measurements of $a$, denominated  \textit{sector states in the measurement of $a$}:
\begin{equation}\label{statisettoriali}
\mathfrak S_a^k := \left\{ \omega \in \mathfrak S_a : \mu_{\omega,a} \in \mathbb M_k(a) \right\}
\end{equation}
where for $k \neq h$ we have
$$ \mathfrak S_a^k \cap \mathfrak S_a^h = \emptyset $$
and
$$\mathfrak S_a = \bigcup_k \mathfrak S_a^k $$
Moreover, for every $\omega_1, \omega_2 \in \mathfrak S_a^k$ and for every $r \in [0, 1]$ there exists $\omega \in \mathfrak S_a^k$ for which  \eqref{amistura} holds.
\subsection{Pure States in the Measurement of an Observable}\label{Pure States in the Measurement of an Observable}
\begin{figure}[htbp]
	\centering
		\includegraphics[scale=0.6]{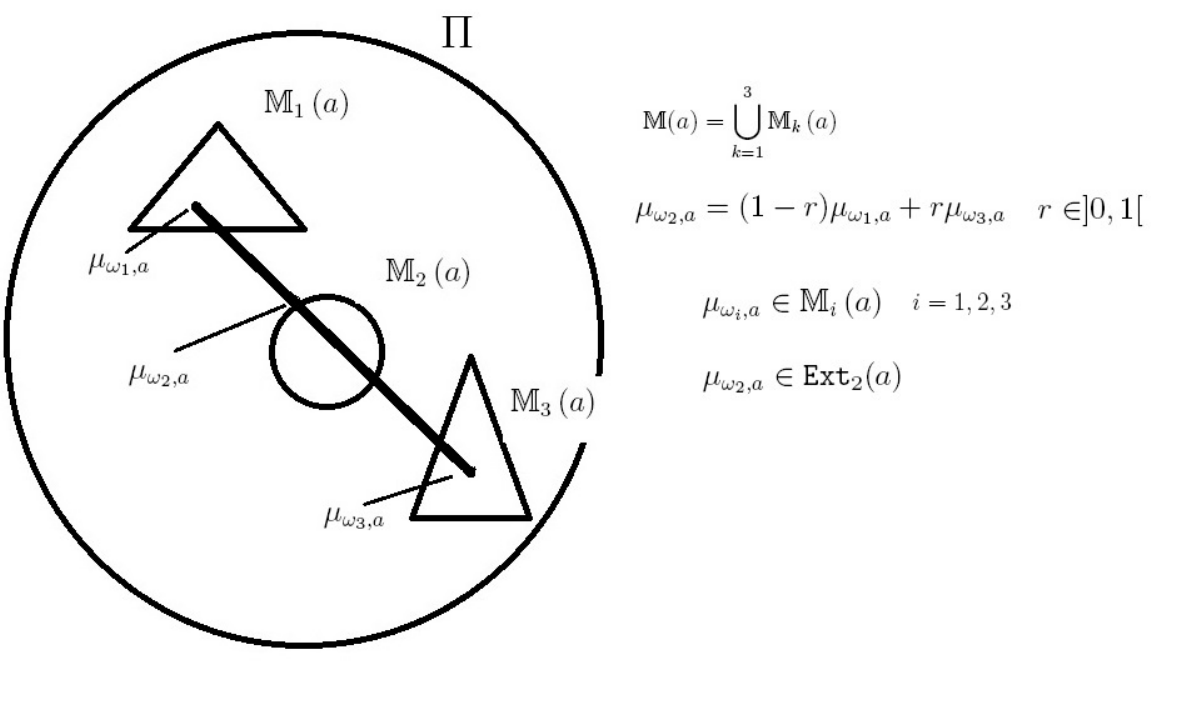}
	\caption{States and purity in the measurement of $a$}
	\label{fig:sector}
\end{figure}

We consider the following sets of states:
\begin{equation} \index{$\mathrm{Ext}(\mathfrak S^k_a)$}
\mathrm{Ext}(\mathfrak S^k_a) = \left\{ \omega \in \mathfrak S^k_a : \mu_{\omega,a} \in \texttt{Ext}_k(a) \right\} \subset \mathfrak S^k_a
\end{equation}
of course
$$\mathrm{Ext}(\mathfrak S^k_a) \cap \mathrm{Ext}(\mathfrak S^h_a) = \emptyset \qquad , \qquad h \neq k $$
With $\mathrm{Ext}(\mathfrak S_a)$ we denote the \textit{pure states} of the system in the measurement of $a$: 
\begin{equation}\label{stati_puri_in_a} 
\mathrm{Ext}(\mathfrak S_a) := \left\{ \omega \in \mathfrak S_a : \mu_{\omega,a} \in \texttt{Ext}(a) \right\} \subset \mathfrak S_a
\end{equation}
where
$$ \mathrm{Ext}(\mathfrak S_a) = \bigcup_{k=1}^n \mathrm{Ext}(\mathfrak S^k_a) $$
Summarizing: 
\begin{definition}\upshape\index{State pure in the measurement of an observable}\index{States pure}
A state $\omega $ of $\mathfrak{S}_a$ is pure in the measurement of $a$ if it is not a mixture of states from the same sector\footnote{We underline that such a state could be a mixture of states of $\mathfrak S_a$ but of different sectors, as represented in Figure \ref{fig:sector}.}.
\end{definition}
it follows that
\begin{center}
\textit{If } $\mu_{\omega,a} \notin \mathrm{Ext}(\mathfrak S_a)$ \textit{ then } $\omega$ \textit{ is not a mixture of states from the same sector}
\end{center}
We have to make a relevant remark:
\begin{remark}\upshape
The notion of pure state differs from that of purely informational state given in definition \ref{statopurodelsistema} on page \pageref{statopurodelsistema}.
\end{remark}
The next step to take is to study the connection between these two notions.
\\
If $\mu_{\omega,a}$ is a pure state of $C_o(\mathbb R)^*$, then the state $\omega$ is pure in the measurement of $a$:
$$ \Pi_p \cap \mathbb{M}(a) \subset \mathrm{Ext}(a) $$ 

\subsubsection{Simple Dual Relations}

Alongside the definition of the measures $\mathbb{M}(a)$, we can define for each state $\omega \in \mathfrak{S}$ the following set:
$$\mathbb{N}(\omega) = \left\{ \mu_{\omega,x} \in \Pi : x \in \mathfrak{X}_\omega \right\}$$
It is straightforward to show that
$$\mathbb{N}(\omega) = \bigcup_{x \in \mathfrak{X}_\omega} \mathbb{M}(x)$$
and dually, for any $a \in \mathfrak{X}$,
$$\mathbb{M}(a) = \bigcup_{\omega \in \mathfrak{S}_a} \mathbb{N}(\omega)$$

As a consequence, if $a \in \mathfrak{X}_\omega$, then for every $k$ the following holds:
$$ \mathbb{N}(\omega) \cap \mathbb{M}_k(a) = \mathbb{M}_k(a)$$

\section{Representations and Physical States}\label{rep-states}
Let $X$ be a locally compact topological space and $T_2$.
\\
For every $\mu\in\Pi(X)$ we have the GNS representation given by the triplet $(\mathcal H_\mu, \pi_\mu, \Omega_\mu)$ where
\begin{itemize}
\item $\mathcal H_\mu$ is the separable Hilbert space $L^2(X ,\mu)$;
\item $\pi_\mu$ is the representation $\pi_\mu: C_b(\mathbb R) \rightarrow B(\mathcal H_\mu)$ defined  by 
$$\pi_\mu(f) \Psi = f \cdot \Psi \ , \qquad \forall f\in C_b(\mathbb R) \ , \ \Psi\in\mathcal H_\mu$$
\item The vector $\Omega_\mu\in\mathcal H_\mu$ is cyclic for $\pi_\mu$ and 
$$ \int f(s) \, d \mu(s) = \left\langle \Omega_\mu , \pi_\mu(f) \ \Omega_\mu \right\rangle \ , \qquad \forall f\in C_b(X)$$ 
\end{itemize}
Recall that a representation $\pi : \mathfrak A \rightarrow B(\mathcal H)$ of a C*-algebra with  unit $\mathfrak A$ is \textit{irreducible} if and only if its  commutant  in $B(\mathcal H)$  is  $\pi(\mathfrak A)^\prime = \mathbb C I$.
\\
Furthermore, if $\varphi$ is a state of $\mathfrak A$ and $(\mathcal H_\varphi, \pi_\varphi, \Omega_\varphi)$ is its GNS  representation, then $\pi_\varphi$ is irreducible if and only if $\varphi$ is a pure state of $\mathfrak A$.
\\
A simple application of the Radon-Nikodym theorem is the following theorem which summarizes the relations between probability measures and representations\footnote{See Arveson's book \cite{Arveson76} \S 2, theorem 2.2.2.}:\index{Arveson}
\begin{proposition}\upshape\label{mis_rep}
Let $\mu$ and $\nu$ be elements of $\Pi(X)$;  we have:
\begin{itemize}
\item $\mu \ll \nu$ if and only if $\pi_\mu \ll \pi_\nu$ ( $\pi_\mu$ is a subrepresentation of $\pi_\nu$);
\item $\mu \approx \nu$ if and only if $\pi_\mu \approx \pi_\nu$ ($\pi_\mu$ is an equivalent representation of $\pi_\nu$);
\item $\mu \perp \nu$ if and only if $\pi_\mu \oslash \pi_\nu$ ( $\pi_\mu$ is a disjoint representation of $\pi_\nu$).
\end{itemize}
\end{proposition}

If $\mu \in \Pi_p$, then the representation $\pi_\mu$ is irreducible \cite{Arveson76} and so
$$ \pi_\mu (f) = \phi (f) I \ , \qquad \forall f \in C_b(X)$$
where $\phi$ is a character (multiplicative functional) of the algebra $C_b(X)$.
\begin{remark}\upshape\label{state-num}
Recall that the set of \textit{states $\mathfrak S$ is a  numerable set}, since the various devices and procedures that can be applied in the laboratory are finite in number. 
\end{remark}
For any observable $a \in \mathfrak X$, we can consider the $k$-sectoral Hilbert space in the measurement of $a$, given by
\begin{equation}
\mathcal H^k_a = \bigoplus_{\omega \in \mathfrak S^{\,k}_a} \mathcal H_\omega
\end{equation} 
and the representation
\begin{equation}\label{rep_di_a}
\pi_{a,k}: C(\sigma(a)) \longrightarrow B(\mathcal H^k_a) \qquad , \qquad \pi_{a,k} = \bigoplus_{\omega \in \mathfrak S^{\,k}_a} \pi_{\mu_{\omega,a}}
\end{equation} 
where $(\mathcal H_\omega, \pi_{\mu_{\omega,a}}, \Omega_\omega)$ is the GNS representation of the functional $\mu_{\omega,a} \in C_o(\mathbb R)^*$, with 
$\mathcal H_\omega := L^2(\sigma(a), \mu_{\omega,a}) $.
\\
Consider the following family $\left\{ \widehat{\Omega}_{\omega} \right\}_{\omega \in \mathfrak S^{\,k}_a}$ of orthonormal vectors of $\mathcal H^k_a$:
\begin{equation}\label{ciclici-sumdir}
\widehat{\Omega}_{\omega} (\omega') = 
 \begin{cases}  
\Omega_{\omega'} & \text{if } \omega = \omega', \\  
0 & \text{if } \omega \neq \omega.  
\end{cases}
\end{equation}
Naturally, for each $\omega \in \mathfrak S_{a}$, we have 
\begin{equation}\label{densit_1}
 \mu_{\omega,a}(f) = \left\langle \widehat{\Omega}_{\omega }, \ \pi_{a,k}(f) \ \widehat{\Omega}_{\omega } \right\rangle_{\mathcal H^{ \ k}_a}   
\end{equation}   
We note that if $\pi_{a,k}(f) = 0$ then $f = 0$ $\mu_{\omega,a}$-a.e. for every $\omega \in \mathfrak S^k_a$.
\\

From representation  \eqref{rep_di_a}  we obtain the following abelian von Neumann algebra associated with the observable $a$ in sector $k$:
$$ \mathfrak R^k (a) = \pi_{a,k}(C(\sigma(a)))'' \subset B(\mathcal H^k_a) $$
while from \eqref{densit_1}, for each function $f \in C(\sigma(a))$, we can write
\begin{equation}\label{densit_2}
 \mu_{\omega,a}(f) = \operatorname{Tr} ( \rho_\omega \ \pi_{a,k}(f) )  
\end{equation}
where
\begin{equation} 
 \rho_\omega = \left| \widehat{\Omega}_{\omega } \right\rangle \left\langle \widehat{\Omega}_{\omega } \right| \in B(\mathcal H^k_a)
\end{equation}
is a rank-1 density matrix.
\\
In this way we obtain a  map 
$$ \mu_{\omega,a} \in \mathbb M(a) \longrightarrow \varphi_\omega \in \mathfrak R^k(a)_*$$
such that
\begin{equation}\label{stati-matrici}
\varphi_\omega(X) = \operatorname{Tr} ( \rho_\omega X ) \ , \qquad \forall X \in \mathfrak R^k(a)
\end{equation}
We explicitly observe that given $\omega \in \mathfrak S^k_a$, the associated vector $\widehat{\Omega}_{\omega } \in \mathcal H^k_a$ is  not  cyclic for the representation $\pi_{a,k}$.

\subsubsection{Operationally Realizable Density Matrices $\mathfrak D^k_{\omega,a}$}
A density matrix in $B(\mathcal H^k_a)$ is called \textit{operationally realizable} if there exists a family of states from the same $k$-sector:
$$\omega_i \in \mathfrak S_a^k \ , \qquad \forall i = 1, 2, \ldots $$
such that 
$$ \widehat{\rho} = \sum_{i=1}^\infty p_i \ \left| \Psi_i \right\rangle \left\langle \Psi_i \right| \in B(\mathcal H^k_a)$$ 
where the vectors $\Psi_i \in \mathcal H^k_a$ are given by \eqref{ciclici-sumdir}, vectors induced via the GNS construction:
\begin{equation} 
\Psi_i(\omega) = 
 \begin{cases}  
 \Omega_{\omega_i} & \text{if } \omega = \omega_i, \\  
0 & \text{if } \omega \neq \omega_i.  
\end{cases}
 \qquad \forall i = 1, 2, \ldots 
\end{equation} 
Moreover, it is straightforward to verify that in this case, for every $f \in C_o(\mathbb R)$, we obtain the relation:
\begin{eqnarray}\label{traccia1}
\operatorname{Tr} ( \widehat{\rho} \ \pi_{a,k}(f) ) = \sum_{j=1}^\infty p_j \ \mu_{\omega_j,a}(f)      
\end{eqnarray}
In fact, the family of vectors of our density matrix $\widehat{\rho}$ is nothing more than a family of the type $\left\{ \widehat{\Omega}_{\omega_j} \right\}_{j=1,2,\ldots}$; it is orthonormal, so we can write by Gram-Schmidt orthogonalization that
\begin{eqnarray*}
\operatorname{Tr} ( \widehat{\rho} \ \pi_{a,k}(f) ) & = & \sum_{j=1}^N p_j \left\langle \widehat{\Omega}_{\omega_j} , \ \pi_{a,k}(f) \ \widehat{\Omega}_{\omega_j} \right\rangle_{\mathcal H_{ \ a}^{\ k}} =
\\
& = & 
\sum_{j=1}^N p_j \left\langle \widehat{\Omega}_{\omega_j} , \pi_{a,k}(f) \ \widehat{\Omega}_{\omega_j} \right\rangle_{\mathcal H_{ \ a}^{\ k}} = 
\\
& = & 
\sum_{j=1}^N p_j \left\langle \Omega_{\omega_j} , \pi_{\mu_{\omega_j,a}}(f) \ \Omega_{\omega_j} \right\rangle_{\mathcal H_{\ \mu_{\omega_j,a}}}= 
\\
& = & 
\sum_{j=1}^N p_j \ \mu_{\omega_j,a}(f)     
\end{eqnarray*}

\begin{definition}\upshape[\textbf{Operationally Realizable Matrices} ]\index{$\mathfrak D^k_{\omega,a}$}\label{Operationally-Realizable-Matrices}
For every $k$-sector and state $\omega \in \mathfrak S_a^k$, we denote by $\mathfrak D^k_{\omega,a}$ the set of operationally realizable density matrices $\rho$ in $B(\mathcal H^k_a)$ such that 
\begin{equation}\label{densit_5}
\mu_{\omega,a}(f) = \operatorname{Tr} ( \widehat{\rho} \ \pi_{a,k}(f) ) \ , \qquad \forall f \in C_o(\mathbb R)
\end{equation}
\end{definition}
%%%
\noindent

We note that for each family of states $\left\{ \omega_1, \omega_2, \ldots, \omega_n \right\}$ that we assume belong to the same sector $\mathfrak S_a^k$ and for numbers $\left\{ t_1, t_2, \ldots, t_n \right\}$ with $t_i > 0$ and $\sum_i t_i = 1$, we can define the following density matrix:
\begin{equation}\label{densit_3}
 \rho = \sum_{i=1}^n t_i \ \left| \widehat{\Omega}_{\omega_i} \right\rangle \left\langle \widehat{\Omega}_{\omega_i} \right| \in B(\mathcal H^k_a)
\end{equation}
By equation \eqref{traccia1} we obtain that
\begin{equation}\label{densit_4}
\operatorname{Tr} ( \rho \ \pi_{a,k}(f) ) = \sum_{i=1}^n t_i \ \mu_{\omega_i,a} (f) \ , \qquad \forall f \in C(\sigma(a))
\end{equation}
Since the family of Borel measures $\left\{ \mu_{\omega_j,a} \right\}_{j=1,2,\ldots,n}$ belongs to the same measurement sector $\mathbb{M}_k(a)$, 
the convexity property implies that there exists a state $\omega \in \mathfrak S^k_a$ such that
$$\mu_{\omega,a} = \sum_{i=1}^n t_i \ \mu_{\omega_i , a} \in \mathbb{M}_k(a)$$ 
Consequently, for this measure we have
\begin{equation*} 
\mu_{\omega,a}(f) = \operatorname{Tr} ( \widehat{\rho} \ \pi_{a,k}(f) ) \ , \qquad \forall f \in C(\sigma(a))
\end{equation*}
where the density matrix is given by \eqref{densit_3}; therefore $\widehat{\rho}$ is an operationally realizable density matrix, i.e., $\widehat{\rho} \in \mathfrak D^k_{\omega,a}$.
\begin{remark}\upshape
For every state $\omega \in \mathfrak S_a^k$ the set $\mathfrak D^k_{\omega,a}$ of its operationally realizable density matrices is constituted as follows:
\begin{itemize}\index{Operational realizable density matrix} \index{$\mathfrak D^k_{\omega,a}$}
\item [1.] a single density matrix of rank 1 given by \eqref{densit_2},
\item [2.] and if $\mu_{\omega ,a} \notin \texttt{Ext}_k(a)$, there are more density matrices of the type \eqref{densit_3}, in addition to the density matrix of rank 1 given in \eqref{densit_2}.
\end{itemize}
\end{remark}

This observation leads to the following definition:

\begin{definition}\upshape[\textbf{Indecomposable Density Matrix Set}]
A set $\mathfrak{D}^k_{\omega,a}$ of operationally realizable density matrices is called indecomposable if every $\rho \in \mathfrak{D}^k_{\omega,a}$ has rank 1.
\end{definition}

The above arguments immediately yield the following result:
\begin{proposition}\upshape
The set $\mathfrak{D}^k_{\omega,a}$ of operationally realizable density matrices is indecomposable if and only if the state $\omega$ is extremal in $\mathfrak{S}^k_a$, i.e.,  
$$\mathfrak{D}^k_{\omega,a} \text{ is indecomposable } \iff \omega \in \mathrm{Ext}(\mathfrak{S}^k_a)$$
\end{proposition}

\subsubsection{The von Neumann Entropy}\index{von Neumann entropy}
We define\index{$h_k(\omega, a )$}
\begin{equation}\label{entropia_neumann}
S_k(\omega, a ):= \sup \left\{ S(\rho) : \ \rho\in \mathfrak D^k_{\omega,a} \right\} 
\end{equation}
with $S(\rho)$ the von Neumann entropy:
$$ S(\rho) = - \operatorname{Tr} ( \rho \log \rho ) $$
\begin{proposition}\upshape\label{entropy_Neumann}[\textbf{Zero von Neumann Entropy Characterization}\footnote{See Wehrl \cite{Wehrl}.} ] 
Let $\mathcal{H}$ be a Hilbert space and $\rho \in B(\mathcal{H})$ a density matrix. 
\\
The von Neumann entropy $S(\rho) = - \operatorname{Tr}(\rho \log \rho)$ satisfies:

$$S(\rho) \geq 0 \qquad \text{with} \qquad S(\rho) = 0 \quad \text{if and only if} \quad \operatorname{rank}(\rho) = 1$$
In other words, $S(\rho)$ vanishes precisely when $\rho$ is a pure state (i.e., $\rho = |\psi\rangle\langle\psi|$ for some unit vector $\psi \in \mathcal{H}$).
\end{proposition}
We have the following easy implication:
$$ \omega \in \mathrm{Ext}(\mathfrak S^k_a) \ \iff\ \mathfrak{D}^k_{\omega,a} \text{ is indecomposable} \ \iff \ S_k(\omega, a) = 0 $$
\begin{attenzione}\upshape
Stating that $\omega \in \mathrm{Ext}(\mathfrak S^k_a)$ does not mean that the measure $\mu_{\omega,a}$ is a Dirac measure, but rather that there exists $\widehat{\Omega} \in \mathcal{H}^k_a$ with $\rho = \left| \widehat{\Omega} \right\rangle \left\langle \widehat{\Omega} \right|$ such that for every $f \in C_o(\mathbb{R})$:
$$\mu_{\omega,a}(f) = \operatorname{Tr} ( \rho \ \pi_{a,k}(f) ) = \left\langle \widehat{\Omega}, \ \pi_{a,k}(f) \ \widehat{\Omega} \right\rangle$$
\end{attenzione}
\begin{attenzione} \upshape
One might conjecture that if there exists a density matrix $\rho \in B(\mathcal H^k_a)$ such that
\begin{equation*} 
 \left\{ 
  \begin{array}{c }
 S(\rho) = 0     \\ 
\mu_{\omega,a}(f) = \operatorname{Tr} ( \rho \ \pi_{a,k}(f) ) \qquad \forall \ f \in C_o(\mathbb R)  
\end{array}%
\right. \qquad \Longrightarrow \qquad \omega \in \mathrm{Ext}(\mathfrak S^k_a)   
\end{equation*} 
\textbf{this implication does not generally hold}. The rank-1 matrix in Proposition \ref{entropy_Neumann} is not necessarily an operationally realizable density matrix. 
\end{attenzione}
\section{Purely Informative States and Extremal Points}\label{entropia}
Let's continue the topic started in sections \ref{entropia-I} and \ref{entropia-II} on information and states.
 \\
Let $\mu\in\Pi$;  for every partition $\mathcal P = \left\{ \Delta_{k} \right\}_{k\in I}$ which belongs to $\mathtt{P}\left( \mathbb{R}\right)$, where $I \subset \mathbb{N}$ is a  set of finite cardinality, we have an element $\xi \in S_{\infty}$ defined as: 
\begin{equation*}
\xi (k) = \mu ( \Delta_{k} ) \ , \qquad k \in I
\end{equation*}
and we define the entropy of the measure $\mu$ on the partition $\mathcal P$\footnote{This definition does not satisfy monotonicity and subadditivity with respect to the partition (see \cite{DNP} paragraph XI).} in the following way:
\begin{equation}
\label{entropymeasure}
H(\mu , \mathcal P) = - \sum_{k=1}^{\infty} \mu( \Delta_{k} ) \log_{2} \left( \mu( \Delta_{k} ) \right) 
\end{equation}
\begin{proposition}\upshape\label{Cantor}
If for each partition $\mathcal P \in \mathtt{P}\left( \mathbb{R}\right)$ we have $H(\mu, \mathcal P) = 0$, then there exists $\lambda \in \mathbb R$ such that $\mu = \delta_{\lambda}$.
\end{proposition}
\begin{proof}
By hypothesis we have that for every $\Delta \in B(\mathbb R)$ it turns out that we can only have two possibilities:
$$\mu(\Delta) = 0 \qquad \text{or} \qquad \mu(\Delta) = 1$$
In this way it is possible to adopt the classic Cantor procedure:
\\
Let $\operatorname{supp} \mu \subset [\alpha, \beta]$ and  consider  the midpoint $\frac{\beta + \alpha}{2}$; by doing so we obtain two intervals
$$ [\alpha , \beta ] = \left[ \alpha, \frac{\beta + \alpha}{2} \right] \cup \left[ \frac{\beta + \alpha}{2} , \beta \right]$$ 
one of the two intervals has measure equal to $1$; let us indicate it by $[ \alpha_1 , \beta_1 ]$. 
\\
Therefore
$$ [ \alpha_1 , \beta_1 ] \subset [\alpha, \beta ] \qquad , \qquad \beta_1 - \alpha_1 \leq \frac{\beta - \alpha}{2} $$
By iterating this procedure we obtain  nested  intervals 
$$ [ \alpha_n , \beta_n ] \subset [ \alpha_{n-1}, \beta_{n-1} ] \subset \cdots \subset [\alpha, \beta ] \qquad , \qquad \mu( [ \alpha_n , \beta_n ] ) = 1 $$ 
Moreover 
$$\alpha \leq \alpha_1 \leq \dots \leq \alpha_n \leq \cdots \qquad , \qquad \beta \geq \beta_1 \geq \dots \geq \beta_n \geq \cdots \ , \ \beta_n - \alpha_n \leq \frac{\beta - \alpha}{2^n}$$
Thus the two sequences converge to the same number $\lambda$ and from this it follows that 
$$ \left\{ \lambda \right\} = \bigcap_{n \in \mathbb N} [ \alpha_n , \beta_n ] $$
and from a well-known theorem of measure theory we obtain:
$$\mu ( \left\{ \lambda \right\} ) = \lim_{n \rightarrow \infty} \mu ( [ \alpha_n , \beta_n ] ) = 1 $$  
\end{proof}
It follows from the latter proposition that 
$$ \mu \in \Pi_p \qquad \Longleftrightarrow \qquad \left[ H(\mu , \mathcal P) = 0 \ , \qquad \forall \mathcal P \in \mathtt{P}\left( \mathbb{R}\right) \right] $$
We now remark that if $\mu$ is a mixture of $\mu_1, \mu_2 \in \Pi$, then we obtain, for each partition $\mathcal P$,
\begin{equation} \label{entropy_miscelata}
H(t \mu_1 + (1-t) \mu_2, \mathcal P) \geq t H( \mu_1, \mathcal P ) + (1-t) H( \mu_2, \mathcal P) 
\end{equation}
Let $a$ be an observable of the system and $\omega$ a state suitable for $a$; by definition it turns out
$$ H(\omega, a, \mathcal P) = H(\mu_{\omega, a}, \mathcal P)$$
since  
\begin{equation*}
\mu _{\omega ,a}\left( \Delta _{k}\right) = P\left( a\in \Delta _{k}\right) _{\omega } \ , \qquad k \in I
\end{equation*}
with 
\begin{equation*}
H(\omega, a, \mathcal P) = - \sum_{k=1}^{\infty} \mu_{\omega, a} ( \Delta_{k} ) \log_{2} \left( \mu_{\omega, a} ( \Delta_{k} ) \right) 
\end{equation*}
We note that the entropy of the partition made up of $\sigma(a)$ and its complement $\mathbb R \setminus \sigma(a)$ is equal to zero; in other words we do not receive any information about the observable $a$, since this partition does not tell us anything new about the possible values of the observable (which, obviously, lie in the spectrum). 
\begin{example}\upshape
Let's calculate the entropy of the measurement of a non-trivial question $q$, carried out in any of its states $\omega$ suitable for it. 
\\
By  \eqref{misuraquestione}, for every partition $\mathcal P \in \mathtt{P}\left( \mathbb{R}\right)$ we obtain  
$$ H(\omega, q, \mathcal P) = - r_0 \log_2 r_0 - r_1 \log_2 r_1 \qquad \text{or} \qquad H(\omega, q, \mathcal P) = 0 $$ 
\end{example}
We have the following simple implications:
$$\mu_{\omega,a} \in \Pi_p \cap \mathbb M(a) \qquad \implies \qquad \left[ H(\mu_{\omega,a}, \mathcal P) = 0 \ , \ \forall \mathcal P \right] \qquad \implies \qquad \omega \in \mathfrak P_a $$
where $\mathfrak{P}_a$ is the set of purely informative states; consequently\footnote{To be more precise, by the \texttt{SPS} property, the set $\mathcal P_a$ is composed of the measurements in $\Pi_p$, so
$$\mathcal P_a = \Pi_p \cap \mathbb M(a)$$ }
$$\Pi_p \cap \mathbb M(a) \subset \left\{ \mu_{\omega,a} \in \Pi : \omega \in \mathfrak P_a \right\}$$
\begin{proposition}\upshape\label{stati-puri-relazione}
Let $\omega \in \mathfrak{S}_a^k$. If $\omega \in \mathfrak{P}_a$, then $\omega \in \mathrm{Ext}(\mathfrak{S}_a^k)$:
$$
\omega \in \mathfrak{P}_a \cap \mathfrak{S}_a^k \ \implies \ \omega \in \mathrm{Ext}(\mathfrak{S}^k_a)
$$
\end{proposition}
\begin{proof}
If $\omega \notin \mathrm{Ext}(\mathfrak{S}_a^k)$, then the measure $\mu_{\omega,a}$ is a mixture of $\mu_{\omega_1}, \mu_{\omega_2} \in \mathbb{M}_k(a)$. Consequently, from  \eqref{entropy_miscelata}  $\mu_{\omega,a}$ cannot be more informative than both constituent states $\omega_1, \omega_2$, and thus cannot be purely informative.
\end{proof}
To summarize:
\begin{equation}\label{rel-puri-stati}
\Pi_p \cap \mathbb{M}_k(a) \subset \left\{ \mu_{\omega,a} \in \Pi : \omega \in \mathfrak{P}_a \cap \mathfrak{S}_a^k \right\} \subset \mathrm{Ext}_k(a)
\end{equation}
We define the set of $k$-eigenstates   of eigenvalue $\lambda$ relative to the observable $a$ as:
$$ \texttt{V}_\lambda^{k} (a) = \left\{ \omega \in \mathfrak S_a^k : P(a \in \left\{ \lambda \right\})_\omega = 1 \right\} $$ 
We have
\begin{equation}\label{autostatifisici}\index{$\texttt{V}_\lambda^{k}(a)$}
\texttt{V}_\lambda^{k}(a) \subset \mathfrak{P}_a \subset \mathrm{Ext} (\mathfrak S_a^k)
\end{equation}
The set $\texttt{V}_\lambda^{k}(a)$ is non-empty\footnote{By the \texttt{SPS} property, see Axiom \ref{SPS} on page \pageref{SPS}.}, and its cardinality satisfies: 
$$ \operatorname{Card} \left\{ \texttt{V}_\lambda^{k}(a) \right\} \geq 1 $$
In other words, there can exist multiple pure $k$-eigenstates $\omega_i \in \mathfrak S_a^k$ for the same eigenvalue $\lambda$ of the observable $a$.

\subsubsection{von Neumann entropy and Partitions}

We now establish fundamental relationships between probability measures, density matrices, and information-theoretic quantities for observable measurements.
\\
Let $\mu \in \Pi$, and suppose there exists a non-trivial partition $\mathcal P_o \in \mathtt{P}\left( \mathbb{R}\right)$ (i.e., $\Delta_k \subsetneq \sigma(a)$ with $\mu(\Delta_k) \neq 0$ for all $\Delta_k \in \mathcal{P}_o$)\footnote{Dirac measures $\mu = \delta_\lambda$ admit no non-trivial partitions since $\delta_\lambda(\Delta_k) \in \left\{ 0, 1 \right\}$.}.
\\
For such $\mu$ we define the conditional probability measures:
\begin{equation}\label{misurepartzizionate}
 \mu_k(\Delta)= \frac{\mu(\Delta \cap \Delta_k)}{\mu(\Delta_k)} \ , \qquad k = 1, 2, \ldots, n
\end{equation}
and in this way
$$ \mu = \sum_{k=1}^n t_k \mu_k \ , \qquad t_k = \mu(\Delta_k) \ , \qquad k = 1, 2, \ldots, n \ , \ \ \sum_{k=1}^n t_k = 1 $$
Obviously we have $\mu_k \ll \mu$ for each $k = 1, 2, \ldots, n$ and from proposition \ref{mis_rep} there exists an isometry $V_k : \mathcal H_{\mu_k} \to \mathcal H_\mu$ defined by
\begin{equation}\label{isometria-V}
 V_k \Psi = \sqrt{F_k} \ \Psi \ , \qquad \forall \Psi \in \mathcal H_{\mu_k} 
\end{equation}
where
\begin{equation} 
F_k = \frac{d \mu_k}{d \mu} = \frac{1}{\mu(\Delta_k)} \mathbf{1}_{\Delta_k} \ , \qquad k = 1, 2, \ldots, n
\end{equation}
and since the partition $\mathcal P_o$ is disjoint we obtain:
\begin{equation}\label{partizionedisgiunta} 
F_k F_h = \delta_{h,k} \mathbf{1}_{\Delta_k}
\end{equation}
Consequently the vectors $\left\{ V_k \Omega_{\mu_k} \right\}_{k=1,2,\ldots,n}$ are orthonormal in $\mathcal H_\mu$. 
\\
Let us now consider the density matrix
$$\rho = \sum_{k=1}^n t_k \ \left| V_k \Omega_{\mu_k} \right\rangle \left\langle V_k \Omega_{\mu_k} \right| \in B(\mathcal H_\mu) $$

\begin{proposition}\upshape\label{entropy-part}
The von Neumann entropy of $\rho$ coincides with the measurement entropy relative to the non-trivial partition $\mathcal P_o$ defined in \eqref{entropymeasure}:
$$S(\rho) = H(\mu, \mathcal P_o)$$ 
\end{proposition}
\begin{proof}
Using the Gram-Schmidt orthogonalization, the von Neumann entropy can be expanded as:
\begin{align*}
S(\rho) = - \operatorname{Tr}( \rho \log \rho ) 
&= - \sum_{j=1}^N \left\langle V_j \Omega_{\mu_j}, \ \rho \log \rho \ V_j \Omega_{\mu_j} \right\rangle \\
&= - \sum_{j=1}^N \left\langle \rho V_j \Omega_{\mu_j}, \ \log \rho \ V_j \Omega_{\mu_j} \right\rangle \quad \text{(since $\rho$ is self-adjoint)} \\
&= - \sum_{j=1}^N \left\langle t_j V_j \Omega_{\mu_j}, \ \log(t_j) V_j \Omega_{\mu_j} \right\rangle \quad \text{(using the eigenvalue equation below)} \\
&= - \sum_{j=1}^N t_j \log(t_j) \quad \text{(by normalization $\| V_j \Omega_{\mu_j} \| = 1$)} \\
&= - \sum_{j=1}^N \mu (\Delta_j) \log( \mu (\Delta_j) ) = H(\mu, \mathcal{P}_o)
\end{align*}
where we have used the key property that $\rho$ acts on the vectors $V_j \Omega_{\mu_j}$ as\footnote{Therefore for the functional calculus
$$ f(\rho) V_k \Omega_{\mu_k} = f(t_k) V_k \Omega_{\mu_k} \ , \qquad k = 1, 2, \ldots, N $$}:
$$ \rho \ V_j \Omega_{\mu_j} = t_j V_j \Omega_{\mu_j} \ , \quad \forall j = 1, 2, \ldots, N$$
with $t_j = \mu (\Delta_j)$ being the measurement probabilities.
\end{proof}
Let's see the connection between von Neumann entropy and the entropy of a measurement more clearly, repeating the previous reasoning, adapting it for the measure $\mu_{\omega,a} \in \mathbb M_k(a)$.
\\
We consider again the partition $\mathcal P_o \in \mathtt{P}\left( \mathbb{R}\right)$ such that $\mu_{\omega,a}(\Delta_j) \neq 0$ for each $\Delta_j \in \mathcal P_o$ and the probability measures $\left\{ \mu_j \right\}_j$ defined by \eqref{misurepartzizionate}.
\\
Let $(\mathcal{H}_{\mu_j}, \pi_j, \Omega_{\mu_j})$ be the GNS representation associated with the functionals $\mu_j$ on $C_o(\mathbb{R})$.
\\
Since the $\mu_j$ do not belong to the set $\mathbb{M}_k(a)$, the vectors $\pi_j(f) \Omega_{\mu_j}$ are not contained in $\mathcal{H}_a^k$. We must therefore consider the operators $\hat{V}_j : \mathcal{H}_{\mu_j} \to \mathcal{H}_a^k$ from equation \eqref{isometria-V}, where for each $j = 1, 2, \ldots, n$
$$ \hat{V}_j \pi_{\mu_j} (f) \Omega_{\mu_j} = \sqrt{F_j} \ \pi_{a,k}(f) \ \widehat{\Omega}_{\omega} \ , \qquad \forall f \in C(\sigma(a))$$ 
the operator $\hat{V}_j$ is an isometry:
\begin{eqnarray*}
\left\| \hat{V}_j \pi_{\mu_j} (f) \Omega_{\mu_j} \right\|^2 & = & \left\langle \hat{V}_j \pi_{\mu_j} (f) \Omega_{\mu_j} , \ \hat{V}_j \pi_{\mu_j} (f) \Omega_{\mu_j} \right\rangle_{\mathcal H_a^k} = \\
& = & \left\langle \sqrt{F_j} \ \pi_{a,k}(f) \ \widehat{\Omega}_{\omega} , \ \sqrt{F_j} \ \pi_{a,k}(f) \ \widehat{\Omega}_{\omega} \right\rangle_{\mathcal H_a^k} = \\
& = & \left\langle \sqrt{F_j} \ \pi_{\mu_{\omega,a}}(f) \ \Omega_{\omega} , \ \sqrt{F_j} \ \pi_{\mu_{\omega,a}}(f) \ \Omega_{\omega} \right\rangle_{\mathcal H_\omega} = \\
& = & \int F_j f^2 \, d\mu_{\omega,a} = \int f^2 \, d\mu_j = \left\| \pi_{\mu_j} (f) \Omega_{\mu_j} \right\|^2
\end{eqnarray*}
From \eqref{partizionedisgiunta} the vectors $\left\{ \hat{V}_j \pi_{\mu_j} (f) \Omega_{\mu_j} \right\}_j$, all belonging to the same subspace $\mathcal H_\omega$ of $\mathcal H_a^k$, are orthogonal:
$$\left\langle \hat{V}_i \pi_{\mu_i} (f) \Omega_{\mu_i} , \ \hat{V}_j \pi_{\mu_j} (f) \Omega_{\mu_j} \right\rangle_{\mathcal H_a^k} = 0 \ , \qquad \forall i \neq j $$
We now consider the density matrix in $B(\mathcal H_a^k)$:
\begin{equation}\label{matrixd0}
\widehat{\rho} = \sum_{j=1}^N t_j \ \left| \hat{V}_j \Omega_{\mu_j} \right\rangle \left\langle \hat{V}_j \Omega_{\mu_j} \right|
\end{equation}
where
$$ t_j = \mu_{\omega,a}(\Delta_j) > 0 \ , \qquad \forall j = 1, 2, \ldots, N $$
We have
\begin{equation}\label{matrixd}
 \mu_{\omega,a}(f) = \operatorname{Tr} ( \widehat{\rho} \ \pi_{a,k}(f) ) \ , \qquad \forall f \in C(\sigma(a))
\end{equation}
Indeed, the family of vectors $\left\{ \hat{V}_j \Omega_{\mu_j} \right\}_j$ is orthonormal, so we can write by Gram-Schmidt orthogonalization that
\begin{eqnarray*}
\operatorname{Tr} ( \widehat{\rho} \ \pi_{a,k}(f) ) & = & \sum_{j=1}^N t_j \left\langle \hat{V}_j \Omega_{\mu_j} , \ \pi_{a,k}(f) \ \hat{V}_j \Omega_{\mu_j} \right\rangle_{\mathcal H_{\ a}^{\ k} }= \\
& = & \sum_{j=1}^N t_j \left\langle \sqrt{F_j} \ \widehat{\Omega}_{\omega} , \ \pi_{a,k}(f) \ \sqrt{F_j} \ \widehat{\Omega}_{\omega} \right\rangle_{\mathcal H_{\ a}^{\ k} } = \\
& = & \sum_{j=1}^N t_j \left\langle \sqrt{F_j} \ \Omega_{\omega} , \ \pi_{\mu_{\omega,a}}(f) \ \sqrt{F_j} \ \Omega_{\omega} \right\rangle_{\mathcal H_{ \ \omega} } = \\
& = & \sum_{j=1}^N t_j \int F_j f \, d\mu_{\omega,a} = \sum_{j=1}^N t_j \int \frac{1}{\mu_{\omega,a}(\Delta_j)} \mathbf{1}_{\Delta_j} f \, d\mu_{\omega,a} = \\
& = & \sum_{j=1}^N \int \mathbf{1}_{\Delta_j} f \, d\mu_{\omega,a} = \mu_{\omega,a}(f)
\end{eqnarray*}

From proposition \ref{entropy-part} it follows:  
 \begin{equation}
S(\widehat{\rho}) = - \operatorname{Tr}(\widehat{\rho} \log \widehat{\rho}) = H(\omega, a , \mathcal P_o) 
 \end{equation}
\begin{attenzione}\upshape
The density matrix is  therefore not derived from physically realizable measurements, i.e., those belonging to $\mathbb{M}(a)$, since in general $\mu_i \notin \mathbb{M}_k(a)$.
\\
The set $\mathbb{M}(a)$ contains measurement outcomes attainable in experiments, while the measures $\mu_i$ used to construct $\widehat{\rho}$ are mathematical artifacts (from the GNS representation) that may lack physical realizability.
Thus, while $\widehat{\rho}$ correctly computes entropies via $H(\omega, a, \mathcal{P}_o)$, it represents an \textit{idealized  object} that includes non-physical configurations.   
\end{attenzione} 
\begin{remark}\upshape
Even when the measure $\mu_{\omega,a}$ satisfies relation \eqref{matrixd}, we cannot conclude that
\begin{equation}\label{entropy-part2}
S_k(\omega,a) \geq H(\omega, a , \mathcal P_o)
\end{equation}
since the density matrix $\widehat{\rho}$ is not guaranteed to be operationally realizable, i.e., $\widehat{\rho} \in \mathfrak{D}^k_{\omega,a}$.
\end{remark}
 \begin{remark}\upshape
All the considerations made so far can be developed within the framework of the universal representation of the algebra $C_o(\mathbb R)$:
$$\pi_u: C_o(\mathbb R) \longrightarrow B(\mathcal H_u) \ , \qquad \pi_u := \bigoplus_{\mu \in \Pi} \pi_\mu \ , \qquad \mathcal H_u := \bigoplus_{\mu \in \Pi} \mathcal H_\mu$$
where
$$\mathcal H_a^k \hookrightarrow \mathcal H_u \qquad , \qquad \pi_{a,k} \ll \pi_u$$
Let $\omega \in \mathfrak S_a$; by repeating the above reasoning, for each partition $\mathcal P$, we obtain a density matrix $\widehat{\rho} \in B(\mathcal H_u)$\footnote{Naturally, $\widehat{\rho}$ depends on both $\omega$ and $\mathcal P$:
$$\widehat{\rho} = \widehat{\rho}(\omega, \mathcal P)$$}: 
$$ \text{Partition } \mathcal P \in \mathtt{P}\left( \mathbb{R}\right) \qquad \Longrightarrow \qquad \text{Density Matrix } \widehat{\rho} \in B(\mathcal H_u) $$
such that   
$$ \mu_{\omega,a}(f) = \operatorname{Tr} ( \widehat{\rho} \ \pi_u(f) ) \qquad \forall f \in C_o(\mathbb R)$$
and 
$$S(\widehat{\rho}) = H(\omega, a, \mathcal P)$$
\end{remark}
\section{States, Measures and Domination*}
Let us conclude the discussion on Borel measures begun in the first section of this paragraph by noting that not all probability measures $\mu \in \Pi$ are actually measures induced by an experimental action.
\\
In other words, it is not necessarily possible to determine an observable $a$ of $\mathfrak X$ and a state $\omega$ in $\mathfrak S_a$ such that $\mu_{\omega,a}$ is  equivalent  to $\mu$.
\\
Furthermore, if we have a measure $\mu \in \Pi$ and an experimental measure $\mu_{\omega,a} \in \mathbb M$ with $\mu \ll \mu_{\omega,a}$, then 
$$ \operatorname{supp} \mu \subset \sigma(a)$$
\\
Indeed, we have that $\mu_{\omega,a}(\rho(a)) = 0$; it follows that $\mu(\rho(a)) = 0$, therefore $\rho(a) \subset \mathbb R \setminus \operatorname{supp} \mu$ and 
hence the thesis. 
\\
The support of the dominated measure lies in the set of values assumed by the observable $a$; this observation leads to the introduction of a new axiom of the model: 
\begin{crit}\upshape 
\label{dominato}
Let $\omega \in \mathfrak S_a^k$ and $\mu \in \Pi$. If $\mu \ll \mu_{\omega,a}$ then there exists a state $\omega_o \in \mathfrak S_a^k$ such that $\mu \approx \mu_{\omega_o ,a}$.
\end{crit}
If the Assumption is true, then for every $\rho \in L^1(\mathbb R, \mu_{\omega,a})$ we have a state $\omega_\rho \in \mathfrak S_a^k$ such that 
$$ \mu_{\omega_\rho ,a} (\Delta) = \int_\Delta \rho(s) \, d \mu_{\omega,a} \ , \qquad \forall \Delta \in B(\mathbb R)$$ 
so we have a map
\begin{equation}
\rho \in L^1(\mathbb R, \mu_{\omega,a}) \to  \omega_\rho \in \mathfrak S_a^k  
\end{equation}
We now observe that if the measure $\mu_{\omega,a} \notin \Pi_p$ then we can write
$$ \mu_{\omega,a} = (1-r) \mu_1 + r \mu_2 \ , \qquad \mu_1, \mu_2 \in \Pi$$
and from Assumption \ref{dominato}, we have the existence of two states $\omega_1, \omega_2 \in \mathfrak S_a$ such that 
$$\mu_i \approx \mu_{\omega_i,a} \ , \qquad i = 1, 2$$
 
This obviously does not tell us that $\omega$ is a mixture of $\omega_1$ and $\omega_2$ in the measurement of $a$, but rather that
$$ \mu_{\omega,a} (f) = (1-r) \int f(s) \rho_1(s) \, d \mu_{\omega_1,a}(s) + r \int f(s) \rho_2(s) \, d \mu_{\omega_2,a}(s) \ , \qquad \forall f \in C_o(\mathbb R) $$
where 
$$\rho_i = \frac{d \mu_i}{d \mu_{\omega_i,a}} \ , \qquad i = 1, 2$$
\begin{remark} \upshape
If in Assumption \ref{dominato} we strengthen the assumption to  
$$ \mu \ll \mu_{\omega,a} \quad \implies \quad \mu \in \mathbb{M}_k(a) $$  
then this leads to the following characterization of extremal states:  
$$ \mathrm{Ext}_k(a) = \Pi_p \cap \mathbb{M}_k(a)$$  
\textbf{Note that this equality may fail in experimental settings.}
\end{remark}
Indeed, the measures $\mu_i$ defined in \eqref{misurepartzizionate} lie in $\mathbb{M}_k(a)$. Consequently, there exists a corresponding family of states $\{ \omega_i \}_{i \in \mathbb{N}} \subset \mathfrak{S}_a^k$ with $\mu_i = \mu_{\omega_i, a}$ for each $i$.
\\
The density matrix $\widehat{\rho}$ constructed in  \eqref{matrixd0}  therefore satisfies $\widehat{\rho} \in \mathfrak{D}^k_{\omega,a}$. From inequality \eqref{entropy-part2}, we observe that for any extremal state $\omega \in \mathrm{Ext}(\mathfrak{S}_a^k)$, the entropy $S_k(\omega, a)$ vanishes.
\\
This yields the implication\footnote{Note that for a trivial partition $\mathcal{P}_o$ we have $H(\omega, a, \mathcal{P}_o) = 0$.}:  
$$ \left[ H(\omega, a, \mathcal{P}) = 0 \qquad \forall \mathcal{P} \in \mathtt{P}(\mathbb{R}) \right] \quad \implies \quad \mu_{\omega,a} \in \Pi_p$$
\subsection{Partitions and Related Issues}

Given a fixed Borel set $\Delta \subset \mathbb{R}$, we determine the measure associated with the observable $\mathbf{1}_\Delta(a) a \mathbf{1}_\Delta(a)$ and its relation to the measure defined in  \eqref{misurepartzizionate}.  
\\
For every $f \in C_o(\mathbb{R})$, we have  
\begin{align*}  
\mu_{\omega, \mathbf{1}_\Delta(a) a \mathbf{1}_\Delta(a)}(f)  
&= \int f(s) \, d\mu_{\omega, \mathbf{1}_\Delta(a) a \mathbf{1}_\Delta(a)}(s) \\  
&= \int f\left(\mathbf{1}_\Delta(s) s \mathbf{1}_\Delta(s)\right) \, d\mu_{\omega, a}(s) \\  
&= f(0) \, \mu_{\omega, a}(\mathbb{R} \setminus \Delta) + \int f(s) \mathbf{1}_\Delta(s) \, d\mu_{\omega, a}(s),  
\end{align*}  
since  
\begin{equation}  
\mathbf{1}_\Delta(s) s \mathbf{1}_\Delta(s) =
 \begin{cases}  
s & \text{if } s \in \Delta, \\  
0 & \text{if } s \notin \Delta.  
\end{cases}  
\end{equation}  
Moreover,  
$$ \left\langle f\left(\mathbf{1}_\Delta(a) a \mathbf{1}_\Delta(a)\right) \right\rangle_\omega = f(0) \left(1 - \mu_{\omega, a}(\Delta)\right) + \left\langle f(a) \mathbf{1}_\Delta(a) \right\rangle_\omega$$  
By applying well-known measure-theoretic results (previously cited), we may consider $f(s) = \mathbf{1}_{\texttt{E}}(s)$, yielding  
\begin{align*}  
\mu_{\omega, \mathbf{1}_\Delta(a) a \mathbf{1}_\Delta(a)}(\texttt{E})  
&= \mathbf{1}_{\texttt{E}}(0) \, \mu_{\omega, a}(\mathbb{R} \setminus \Delta) + \int \mathbf{1}_{\texttt{E}}(s) \mathbf{1}_\Delta(s) \, d\mu_{\omega, a}(s) \\  
&= \mathbf{1}_{\texttt{E}}(0) \, \mu_{\omega, a}(\mathbb{R} \setminus \Delta) + \mu_{\omega, a}(\texttt{E} \cap \Delta).  
\end{align*}  

Thus, for every Borel set $\texttt{E} \subset \mathbb{R}$, we explicitly obtain  
\begin{equation}  
\mu_{\omega, \mathbf{1}_\Delta(a) a \mathbf{1}_\Delta(a)}(\texttt{E}) = \begin{cases}  
1 - \mu_{\omega, a}(\Delta) + \mu_{\omega, a}(\texttt{E} \cap \Delta) & \text{if } 0 \in \texttt{E}, \\  
\mu_{\omega, a}(\texttt{E} \cap \Delta) & \text{if } 0 \notin \texttt{E}.  
\end{cases}  
\end{equation}

\section{Free Dispersion States}\label{FD-states}
We analyze, within our model, the notion of \textit{free dispersion states}\footnote{For a review of this topic, see Emch's book \cite{Emch} and the work of Plymen \cite{Plymen}.}\index{Emch}\index{Plymen}.  
\\
The \textit{dispersion} of an observable $a$ in the state $\omega \in \mathfrak{S}_a$ is defined by  
\begin{equation}\label{varianza}\index{variance} \index{$\Delta _{\omega }\left( a\right)$}  
\Delta _{\omega }\left( a\right) = \left\langle a^{2} \right\rangle_{\omega } - \left\langle a \right\rangle_{\omega }^{2}.  
\end{equation}  
Furthermore, we have\footnote{Recall that the square of the quantity $\Delta _{\omega }\left( a\right)$ in statistical mathematics is called the variance.}   
\begin{equation*}  
\Delta _{\omega }\left( a\right) = \left\langle \left( a - \left\langle a \right\rangle_{\omega } \mathbf{1} \right)^{2} \right\rangle_{\omega } \geq 0.  
\end{equation*}  
\begin{definition}\upshape\index{State of free dispersion}  
A state $\omega$ of $\mathfrak{S}_a$ is said to be free of dispersion on the observable $a$ if  
$$ \Delta _{\omega }\left( a\right) = 0$$ 
The state $\omega$ is called free of dispersion or deterministic state if it is free of dispersion for every observable $a \in \mathfrak{X}_\omega$:  
$$ \Delta _{\omega }\left( a\right) = 0 \qquad \forall a \in \mathfrak{X}_\omega$$  
\end{definition}  

We now have a well-known result from measure theory:
\begin{proposition}\upshape\label{freeds1}
Let $a$ be an observable and $\omega \in \mathfrak{S}_a$ a state such that $\Delta_{\omega}(a) = 0$. Then the spectral measure $\mu_{\omega,a}$ is a point mass, i.e., $\mu_{\omega,a} = \delta_{\lambda_0}$ for some $\lambda_0 \in \sigma(a)$.
\end{proposition}
\begin{proof}
Let $m = \langle a \rangle_{\omega}$ be the expectation value of $a$ in the state $\omega$.
\\
By definition, the variance is:
$$ \Delta_{\omega}(a) = \langle (a - m \mathbf{1})^2 \rangle_{\omega} = \int_{\sigma(a)} (t - m)^2 \, d\mu_{\omega,a}(t) $$
The hypothesis $\Delta_{\omega}(a) = 0$ is equivalent to:
$$\int_{\sigma(a)} (t - m)^2 \, d\mu_{\omega,a}(t) = 0$$
Since the integrand $(t - m)^2$ is a continuous, non-negative function, and $\mu_{\omega,a}$ is a positive Borel probability measure, the integral vanishes if and only if the integrand is zero $\mu_{\omega,a}$-almost everywhere. That is,
$$\mu_{\omega,a}\left( \left\{ t \in \sigma(a) : (t - m)^2 \neq 0 \right\} \right) = 0$$
The set $\{ t \in \sigma(a) : (t - m)^2 \neq 0 \}$ is exactly $\sigma(a) \setminus \{m\}$. 
\\
Therefore,
$$\mu_{\omega,a}( \sigma(a) \setminus \{m\} ) = 0$$
Since $\mu_{\omega,a}$ is a probability measure $(\mu_{\omega,a}(\sigma(a)) = 1)$, it follows that:
$$\mu_{\omega,a}( \{m\} ) = 1$$
This proves that $\mu_{\omega,a}$ is the point mass (Dirac measure) concentrated at $m$, denoted $\delta_m$. Furthermore, since the support of the spectral measure is contained in the spectrum, we must have $m \in \sigma(a)$.
\end{proof}

Consequently we have the following result\footnote{See also Plymen \cite{Plymen}, Lemma 3.1.}:  
\begin{corollary} \upshape
A state $\omega \in \mathfrak{S}$ is free of dispersion if and only if the measure $\mu_{\omega,a} \in \Pi_p$ for every $a \in \mathfrak{X}_\omega$.  
\end{corollary}  
\begin{remark}\upshape
We can say more: from the \texttt{SPS} property  it follows that   every state free of dispersion in the measurement of $a$ is a purely informative state in the measurement of $a$:
$$\mathcal P_a = \left\{ \omega \in \mathfrak S_a : \Delta_{\omega}(a) = 0 \right\}$$
\end{remark}
Therefore every state free of dispersion in the measurement of $a$ is a pure state in the measurement of $a$: 
$$\Delta _{\omega }\left( a\right) = 0 \qquad \Longrightarrow \qquad \omega \text{ is a pure state in the measurement of } a$$

We conclude this section with the following definition:  
\begin{definition}[\textbf{Fluctuations}]\upshape\index{Fluctuations}  
We say there are fluctuations in the measurement of an observable $a$ in the state $\omega \in \mathfrak{S}_a$ if  
$$\left\langle a \right\rangle_{\omega } = 0 \quad \text{and} \quad \left\langle a^2 \right\rangle_{\omega } > 0$$ 
and hence,  
$$\Delta _{\omega }\left( a\right) = \left\langle a^2 \right\rangle_{\omega } > 0$$ 
\end{definition}  
The topic will be revisited in Section \ref{Physic-math-states} when we compare it with the definitions in the algebraic case.
 \section{Operational Vacuum}
In this section we will try to define a particular state of the laboratory that possesses an analog of the vacuum state in quantum field theory, namely the quantum state of minimum energy of our physical system.
\\
In our model we can define this particular state as follows:
\\
\textit{The operational vacuum state} is the state established by a specific procedure (and by the experimental instruments that realize it) such that, for every physical quantity that this procedure allows us to measure, its value is minimal with respect to any other experimental procedure (i.e., with respect to any other physical state).
\\
In our symbols:
\begin{definition}\upshape
The state $\omega_o \in \mathfrak S$ is an operational vacuum if for every $x \in \mathfrak X_{\omega_o}$ we have:
$$ \langle x \rangle_{\omega_o} \leq \langle x \rangle_{\omega} \qquad \forall \omega \in \mathfrak S_x$$
\end{definition}
Heuristically, we can think of this state as "white noise", a reference state defined by measurement procedures, instruments, and the operator.
\\
Recall that in an experimental procedure, \textit{white background noise} is:
\begin{itemize}
  \item[-] The signal measured when no source is active;
  \item[-] It depends on the instrumentation;
  \item[-] It is subtracted to reveal the signal;
  \item[-] It is an \textit{operational convention} and different laboratories have different background noises.
\end{itemize}

Thus, in the preparation of this state, the role of the operator performing the measurement is fundamental, since:
\begin{itemize}
 \item[-] they choose the sensitivity of the instruments (detection thresholds);
 \item[-] they choose the calibration procedure (background subtraction);
 \item[-] they define the threshold between "background" and "signal".
\end{itemize}

\noindent
However, once the experimental procedures are fixed (choice of instrumentation, settings, calibration, detection thresholds), the actual execution of the measurements can be carried out by any operator (or even by an automatic device) who strictly follows the established instructions. In this sense, during the execution phase, the operator is a cold executor, interchangeable, and plays no active role in modifying the measurement outcome. Nevertheless, the definition of the operational vacuum state depends on the initial choices made by the operator (or the research team) before the experiment begins.

We can assign a numerical value to the background noise through the variance of an observable, previously defined, namely the value
$$R_f(\omega_o) := \inf \left\{ \Delta_{\omega_o}(a) : a \in \mathfrak X_{\omega_o} \right\}$$

It should be emphasized that the existence of such states in our model is not guaranteed a priori; it must be postulated:

\begin{post}[Existence and Positive Background Noise]\index{Operational vacuum state}\index{Postulate-Existence of Vacuum State}
The laboratory system defined by the pair $(\mathfrak X, \mathfrak S)$ admits at least one operational vacuum state.
\\
For every operational vacuum state $\omega_o \in \mathfrak S$ we have
$$ R_f(\omega_o) > 0$$
\end{post} 
In other words, no physical quantity accessible through the procedure that defines $\omega_o$ can be measured with zero variance in this state; that is, $\Delta_{\omega_o}(a) > 0$. This implies that the state $\omega_o$ might not be a pure state with respect to the measurement of $a$.

\chapter{Dynamic Axioms}
Building upon the static axioms introduced earlier, which remain largely consistent with Mackey's foundational work \cite{Mackey_57, Mackey}, we now turn to the investigation initiated in Section \ref{mutazionievol} concerning the temporal evolution of physical quantities. This section establishes a new set of principles governing these dynamical changes, which we shall refer to as the dynamic axioms.
 
\section{Foundations of Temporal Evolution}
Before proceeding, let us summarize our current framework while emphasizing the notational conventions, as these will prove crucial for the subsequent development:
\\
As extensively established in our preceding analysis, for any state $\omega \in \mathfrak S_a(\mathcal O_o)$ associated with observable $a$, we can construct, via the frequency interpretation formalized in \eqref{freq}, a probability measure:
$$\mu_{\omega,a}(\Delta) = P(a \in \Delta)_\omega, \quad \forall \Delta \in \mathcal{B}(\mathbb{R})$$
This construction induces a canonical mapping for each observable $a \in \mathfrak X(\mathcal O_o)$:
\begin{equation} \label{misura_cronos}
\omega \in \mathfrak S_a(\mathcal O_o) \longrightarrow \mu_{\omega , a} \in C_o(\mathbb R)^*
\end{equation} 
Recall that in $\omega$ there is a specified time instruction indicating when to perform a measurement at a particular instant $\tau_o$ (where $\tau_o$ is a fixed or reference time):
\\
If $\omega_o \in \mathfrak S_a(\mathcal O_o)$, then there exists a unique $\tau_o \geq 0$ such that $\omega_o \in \mathfrak S_a(\mathcal O_o) | \tau_o$ and we have set that
$$ P(a \in \Delta, \tau_o)_{\omega_o} = \mu_{\omega_o , a}(\Delta) \ , \qquad \Delta \in B(\mathbb R) $$
Let us consider  the  chronological state $\omega$ related to $\omega_o \in \mathfrak S_a(\mathcal O_o) | \tau_o$\footnote{See \S \ref{sezione_cronos} on page \pageref{sezione_cronos}.}:
$$ \omega : \tau \in I \longrightarrow \omega^{(\tau)} \in \mathfrak S_a(\mathcal O_o) | \tau $$
where $0 \in I \subset [0, \infty)$ and 
$$ \omega^{(0)} = \omega_o$$ 
in this case $\mu_{\omega , a}$ given in  \eqref{misura_cronos}  is itself a  map:
\begin{equation} \label{evol_temp_1}
\mu_{\omega , a} : \tau \in I \longrightarrow \mu_{\omega , a}^\tau \in \Pi \subset C_o(\mathbb R)^*
\end{equation} 
where\footnote{In other words 
$$\mu_{\omega , a}^\tau = \mu_{\omega^{(\tau)} , a} \ , \qquad \tau \in I$$
This is a subtle point in our notation:
\\
- When $\omega$ represents a state of the system, the symbol $\mu_{\omega, a}$ denotes a measure.
\\
- When $\omega$ represents a chronological state, it is a mapping as defined in equation  \eqref{evol_temp_1}.}
$$ \mu_{\omega , a}^\tau (\Delta) = P(a \in \Delta, \tau)_\omega \qquad , \qquad \Delta \in B(\mathbb R) $$
with
$$ \mu_{\omega , a}^0 (\Delta) = P(a \in \Delta, \tau_0)_\omega \qquad , \qquad \Delta \in B(\mathbb R) $$
Thus, relation \eqref{evol_temp_1} determines how the possible values of an observable $a$ of the physical system in the state $\omega$ change, obtaining the time-dependent average value of $a$ in the state $\omega$:
\begin{equation}\label{axio_evol_2}
\tau \in I \longrightarrow \langle a \rangle_\omega (\tau) \in \mathbb{R},
\end{equation}
where
$$ \left\langle a \right\rangle_\omega (\tau) = \int s \, d \mu^\tau_{\omega , a} (s) $$
We now proceed with the axiomatization by postulating the following property, which prevents time leaps (discontinuous jumps in the evolution):
\begin{axiom}[\textbf{Continuous Time Evolution (No Time Leap)}]\label{axio_evol_c}\index{Axiom-No Time Leap}
For every state $\omega \in \mathfrak S$, $a \in \mathfrak X_\omega$ and Borel set $\Delta$ of $\mathbb R$, the probability measure $P(a \in \Delta, \tau)_\omega$ evolves continuously in time:
$$\lim_{\tau \rightarrow \tau_\bullet} P(a \in \Delta, \tau)_\omega = P(a \in \Delta, \tau_\bullet)_\omega \ , \qquad \forall \tau_\bullet \in I$$
\end{axiom}
The following axiom is introduced for mathematical completeness:
\begin{axiom}[\textbf{Measurable Time Evolution}]\label{axio_evol_d}\index{Axiom-Measurable Time Evolution}
For every $\omega \in \mathfrak S$, $a \in \mathfrak X_\omega$ and  Borel set  $\Delta$  of  $\mathbb R$, the mapping
$$ \tau \in I \longrightarrow \mu_{\omega,a}^{\tau}(\Delta) \in [0, 1] \ , \qquad I \subset [0, \infty[ $$
is a Borel-measurable function.
\end{axiom}
We emphasize that \textit{we have not assumed} norm-continuity for the mapping:
$$ \tau \in I \longrightarrow \mu_{\omega,a}^{\tau} \in \Pi \subset C_o(\mathbb R)^* $$ 
where $\Pi$ denotes the set of probability measures. This continuity property would be strictly stronger than the requirements imposed by Axioms \ref{axio_evol_c} (time continuity of probabilities) and \ref{axio_evol_d} (measurability).
\begin{axiom}[\textbf{Existence for Evolutionary States}]\label{esistenzastatoevol}\index{Axiom-Existence of Evolutionary States}
For every initial state $\omega \in \mathfrak S$, observable $a \in \mathfrak X_\omega$, and time parameter $\tau \in I$, the set of evolved states $\mathfrak S^{a, \omega}_\tau$ defined in \eqref{evol_temp_01} is non-empty. 
\end{axiom}

\subsection{Spectrum and Measurement Time}
We emphasize that all measurements of physical quantities $a$ are performed in our laboratory $L_o$ at specific time instances $\tau$.
\\
To properly characterize the system's states, we consider:
 $$ \mathfrak S_a = \bigcup_{t_o \geq 0} \mathfrak S_a(\mathcal O_{t_o} ) \ : \qquad \mathcal O_{t_o} = L_o \times [0, t_o]$$
this represents the complete set of physically preparable states for observable $a$ in $L_o$, in a finite time interval preparation $[0, t_o]$.
 \\
To select states corresponding specifically to measurements performed at time $\tau$, we define the restricted set:
$$ \mathfrak S_a | \tau \ \text{ where the measurement of $a$ occurs at time } \tau$$ 
The spectrum $\sigma(a)$ of observable $a$, as defined in Section \ref{spettro_oss} on page \pageref{spettro_oss}, satisfies the fundamental property\footnote{
The equality holds by proposition \ref{spettrolambda} and the \texttt{SPS} property.
\\
We underline that if $\lambda = \left\langle a \right\rangle_\omega^{(\tau)}$, this does not imply that $\mu_{\omega,a}^\tau$ is concentrated entirely at $\lambda$, i.e., $\mu_{\omega,a}^\tau(\left\{\lambda\right\}) = 1$.
\\
Furthermore, if $f$ is a real bounded Borel function, then the statement 
$$ \lambda = \left\langle a \right\rangle_\omega (\tau) \ \Longrightarrow \ f(\lambda) = \left\langle f(a) \right\rangle_\omega (\tau)$$
is \textit{not generally true} for arbitrary observables $a$ and arbitrary time-averaging processes.}
$$\sigma(a) = \left\{ \lambda \in \mathbb R : \exists \tau \in I, \ \omega \in \mathfrak S_a | \tau \text{ such that } \lambda \text{ is realizable as } \left\langle a \right\rangle_\omega^{(\tau)} \text{ and } \mu_{\omega,a}^\tau(\left\{\lambda\right\}) \neq 0 \right\}$$
This means $\sigma(a)$ contains exactly those values that:
\begin{itemize} 
   \item[-] Can be physically realized through measurements of $a$,
   \item[-] Occur at some time $\tau$,
    \item[-]  Are obtainable in some admissible state $\omega\in\mathfrak S_a|\tau$.
 \end{itemize}

If we want the possible values of $a$ only at a given time $\tau$, then we must redefine the concepts introduced in Section \ref{spettro_oss} as follows:
\\
We denote by $\sigma(a)^\tau$ the possible values of the observable $a$ at time $\tau$:
$$ \sigma(a)^\tau = \mathbb{R} \setminus \rho(a)^\tau $$
where
\begin{equation}
\rho(a)^\tau = \bigcup_{U \in \mathfrak F^\infty_\tau (a)} U
\end{equation}
and\footnote{See formula \eqref{a-nulli}.}
\begin{equation}
\label{a-nulli-tau}
\mathfrak F^\infty_\tau (a) = \bigcap_{\omega \in \mathfrak S_a | \tau} \mathfrak F^\omega (a) 
\end{equation}
Since for every $\tau \geq 0$ we have $\mathfrak S_a | \tau \subset \mathfrak S_a$, it readily follows that
$$ \mathfrak F^\infty(a) \subset \mathfrak F^\infty_\tau (a) \qquad \Longrightarrow \qquad \rho(a) \subset \rho(a)^\tau \qquad \Longrightarrow \qquad \sigma(a)^\tau \subset \sigma(a) $$ 
Moreover, as established in earlier sections, the spectrum of an observable is always non-empty, so
$$ \sigma(a)^\tau \neq \emptyset \ , \qquad \forall \tau \in I $$
and
\begin{equation*}
 \operatorname{Supp} \mu_{\omega ,a}^\tau \subset \sigma(a)^\tau \ , \qquad \forall \omega \in \mathfrak S_a
\end{equation*}
\subsection{Dissipative Spectrum}
By repeating all the steps indicated in Section \ref{spettro_oss}, we define a new spectrum of an observable through the set of states $\mathfrak S^{a, \omega}_\tau$:
$$ \mathfrak S_\tau^a = \bigcup_{\omega \in \mathfrak S_a | \tau} \mathfrak S^{a, \omega}_\tau \qquad ,
 \qquad \mathfrak F^\tau_d := \bigcap_{\omega \in \mathfrak S_{\ \tau}^{ \ a} } \mathfrak F^\omega(a) $$  
with
$$ \sigma(a)^\tau_d = \mathbb{R} \setminus \rho_d(a)^\tau \qquad , \qquad \rho_d(a)^\tau = \bigcup_{U \in \mathfrak F^{ \ \tau}_{ \ d} (a)} U $$
Moreover we have
\begin{equation} \label{spettrodis}
  \sigma(a)^\tau_d \subset \sigma(a)^\tau 
\end{equation}
because 
$$\rho(a)^\tau \subset \rho_d(a)^\tau$$
The set $\sigma(a)^\tau_d$ is called the \textit{dissipative spectrum} at the measurement time $\tau$.
\\
We observe the following:
\\
If  for  every $\tau_2 \geq \tau_1 \geq 0$  we obtain  $\mathfrak S_{\tau_2}^a \subset \mathfrak S_{\tau_1}^a$,  then it  is easy to prove 
$$ \sigma(a)^{\tau_2}_d \subset \sigma(a)^{\tau_1}_d $$
In other words, \textit{as the measurement time $\tau$ increases, system states are lost, so the possible values that the observable $a$ can take on decrease.}
\section{Time Evolution}
In this section, we will consider, for simplicity, only globally defined chronological states of the laboratory system.
\\
Experimentally, what we observe is the temporal evolution of the probability measures \eqref{evol_temp_1}. However, regarding the temporal evolution of states, we have no experimental method to prefer one temporal evolution $\xi \in \mathcal S_{a,\omega}$ of the state $\omega$ over another $\xi' \in \mathcal S_{a,\omega}$ in the measurement of $a$\footnote{See definition \ref{evol_temp_02}.}. We can only assert that
$$ \mu_{\xi(\tau), a} = \mu_{\xi'(\tau), a} \ , \qquad \forall \tau \in [0, \infty) $$
Furthermore, for every $\omega_* \in \mathfrak S^{\omega,a}_\tau$ we can write:
$$ \left\langle a \right\rangle_{\omega}(\tau) = \left\langle a \right\rangle_{\omega_*} = \int s \, d \mu_{\omega_* , a} (s) $$
Let us denote by $\mathbb M(a)$ the set of mappings (defined by \eqref{evol_temp_1}) associated with each globally defined chronological state $\omega$, suitable for the measurement of $a$\footnote{We emphasize that in Definition \ref{misure_in_a} on page \pageref{misure_in_a}, $\mathbb M(a)$ is a set of Borel measures, whereas here it is a set of mappings whose images are Borel measures.}.
\\
We now introduce the following mapping\footnote{Not to be confused with the one given in Notation \ref{evol_temp_s}.} 
$$\texttt{S}_a^\tau: \mathbb M(a) \longrightarrow \mathbb M(a)$$
such that for every $\mu_{\omega , a} \in \mathbb M(a)$
\begin{equation}\label{evol_temp_2}
 ( \texttt{S}_a^\tau(\mu_{\omega , a}) )^{\tau_o} = \mu_{\omega, a}^{\tau + \tau_o} \ , \qquad \tau, \tau_o \geq 0
\end{equation}
This mapping is called the temporal evolution of the measure of $a$.

\section{Evolution and Dissipation}\label{evol_temp}
In this section, we resume the discussion initiated in Section \ref{discussionedissipi}. Empirical observations demonstrate that dissipative processes typically transform pure states into mixed states, while the reverse transition from mixed states to pure states does not generally occur. That is, states that become mixed through temporal evolution remain mixed indefinitely. To formalize this behavior, we introduce the following fundamental property:
\begin{definition}\upshape\label{prop_affine_0}
The temporal evolution \eqref{evol_temp_2} is said to possess the affine property if it satisfies the following condition:
\\
For any mixed state $\omega$ at initial time $\tau_o = 0$ that is a convex combination of states $\omega_1, \omega_2 \in \mathfrak S_a$ in the measurement of observable $a$:
$$\mu_{\omega,a} = r \mu_{\omega_1,a} + (1-r) \mu_{\omega_2,a} $$
the time-evolved state at any $\tau \in \mathbb R^+$ maintains the same convex combination:
$$ \mu_{\omega,a}^\tau = r \mu_{\omega_1 , a}^\tau + (1-r) \mu_{\omega_2,a}^\tau $$
\end{definition} 
It follows that: 
\begin{property}[\textbf{Entropy Inequality under Affine Evolution}]
The temporal evolution map \eqref{evol_temp_2} is affine, and consequently satisfies the entropy monotonicity relation:
$$ H( \mu_{\omega,a}^\tau, \mathcal P ) \geq H( \mu_{\omega, a} , \mathcal P ) \ , \qquad \mathcal P \in \texttt{P}(\mathbb R) \ , \ \tau \in I_\bullet \subset I $$ 
where the inequality follows from Postulate \ref{prop-entropic}.
\end{property}
\begin{definition}[\textbf{Dissipation-Free Evolution}]\upshape\label{def:dissipation_free}
A temporal evolution \eqref{evol_temp_2} is called \emph{dissipation-free} in the measurement of the observable $a$ on the time interval $I_\bullet$ if it preserves the measurement entropy for all partitions:
$$ H( \mu_{\omega,a}^\tau, \mathcal P ) = H( \mu_{\omega, a} , \mathcal P ) \ , \qquad \mathcal P \in \texttt{P}(\mathbb R) \ , \ \tau \in I_\bullet \subset I $$
\end{definition}
\begin{remark}[\textbf{Constant Observables and Dissipation}]\upshape\label{rem:constant_obs}
For a constant observable $c$ where there exists a real number $\texttt{r} \in \mathbb R$ such that for all admissible states $\omega \in \mathfrak S_c$, the measurement probability is deterministic:
$$P(c \in \left\{ \texttt{r} \right\}, \tau)_\omega = 1 \qquad , \qquad \forall \omega \in \mathfrak S_c$$
the temporal evolution \eqref{evol_temp_00} yields Dirac measures:
$$\texttt{S}_c^\tau(\mu_{\omega,c}) = \delta_{\texttt{r}}, \quad \forall \tau \geq 0$$
and the entropy vanishes identically:
$$H(\mu_{\omega,c}^\tau, \mathcal{P}) = 0, \quad \forall \mathcal{P} \in \mathtt{P}(\mathbb{R}), \ \tau \geq 0$$
Consequently, no dissipative effects occur in measurements of any constant observable $c$ for states $\omega \in \mathfrak S_c$.
\end{remark}
\subsubsection{Dissipation Indicator via Pure State Preservation}
The extent of dissipative effects in the temporal evolution of our system can be quantified by examining the cardinality of the pure state preservation sets:
$$ \mathcal D_a(\tau) := \mathrm{Ext}(\mathfrak S_a|_{\tau=0}) \cap \mathrm{Ext}(\mathfrak S_a|_{\tau}) \ , \qquad \tau \geq 0 $$
 where:
\begin{itemize}
\item $\mathrm{Ext}(\mathfrak S_a|_{\tau=0})$ denotes the pure states in the measurement of $a$ (defined in \eqref{stati_puri_in_a}) at initial time $\tau = 0$
\item $\mathrm{Ext}(\mathfrak S_a|_{\tau})$ represents the pure states in the measurement of $a$ at evolution time $\tau > 0$
\end{itemize}
The decreasing cardinality $|\mathcal D_a(\tau)|$ provides a quantitative measure of increasing dissipation.
\begin{definition}\upshape\label{def:rev_evolution}
A physical system $(\mathfrak X , \mathfrak S)$ exhibits:
\begin{enumerate}
\item \textbf{Reversible (non-dissipative) evolution} up to time $T > 0$ if it preserves pure states for all observables:
$$\mathrm{Ext}(\mathfrak S_a|\tau) = \mathrm{Ext}(\mathfrak S_a|_{\tau=0}) , \qquad \forall a \in \mathfrak X \ , \qquad \forall \tau \in [0, T]$$
\item \textbf{Irreversible (dissipative) evolution} if there exists an $a \in \mathfrak X$ and a $\tau_* > 0$ such that
$$ \mathrm{Ext}(\mathfrak S_a|_{\tau_*}) \subsetneq \mathrm{Ext}(\mathfrak S_a|_{\tau=0}) $$ 
\item \textbf{Completely irreversible evolution} if for every $a \in \mathfrak X$ there exists a $\tau_* > 0$ such that
$$
\mathrm{Ext}(\mathfrak{S}_a|_{\tau_*}) \cap \mathrm{Ext}(\mathfrak{S}_a|_{\tau=0}) = \emptyset
$$
\end{enumerate}
\end{definition}
 $$ \star \star \star $$
The same considerations naturally extend to a family $\mathfrak X_C$ of simultaneously measurable observables $a_1 : a_2 : \ldots : a_n$:
\\
For every $\omega \in \mathfrak S_{a_1: a_2 : \ldots : a_n}$ and $\tau \geq 0$, we define
$$\mu_{\omega, a_1:a_2:\cdots:a_n}^\tau( \Delta_1 \times \Delta_2 \times \cdots \times \Delta_n) = P(a_1 \in \Delta_1 : a_2 \in \Delta_2 : \cdots : a_n \in \Delta_n, \tau)_{\omega}$$
where $P$ denotes the joint probability measure at time $\tau$.
\\
This yields a joint temporal evolution:
$$ ( \texttt{S}^\tau_{\mathfrak X_C} \mu_{\omega, a_1:a_2:\cdots:a_n} )^{\tau_o} = \mu_{\omega, a_1:a_2:\cdots:a_n}^{\tau + \tau_o} \ , \qquad \tau, \tau_o \geq 0$$
More generally, for any subset $\mathcal O$ of the space-time $\mathcal M$, there exists a map
$$S_{\mathcal O}^t : \mathbb M \rightarrow \mathbb M$$
where $\mathbb M \subset \Pi$ is defined by  \eqref{misure_in_tutto}, satisfying the following properties:
\begin{itemize} 
\item Consistency with individual observables:
\\
 For every $a \in \mathfrak X(\mathcal O)$ and $\omega \in \mathfrak S_a(\mathcal O)$ 
$$\texttt{S}_{\mathcal O}^t (\mu_{\omega , a}) = \mu_{\omega ,a}^t \ , \qquad t \in \mathbb R^+$$ 
\item Linearity:
\\
For every $r \in [0, 1]$ and $\omega_1, \omega_2 \in \mathfrak S_a$     
$$ \texttt{S}_{\mathcal O}^t ( r \mu_{\omega_1 , a} + (1-r) \mu_{\omega_2,a} ) = r \texttt{S}_{\mathcal O}^t (\mu_{\omega_1 , a}) + (1-r) \texttt{S}_{\mathcal O}^t ( \mu_{\omega_2,a} ) $$ 
\end{itemize}
\section{The Irreversible Part*}
Let us recall that at the measurement instant $t = 0$, no dissipative phenomena are present. We now compare the evolution $\left\{ \mu_{\omega,a}^t \right\}_{t \in \mathbb R^+}$ through the Lebesgue decomposition relative to the initial measure $\mu_{\omega,a}$:
\begin{equation}\label{decoleb1a}
 \mu_{\omega , a}^t = \mu_A^t + \mu_S^t \ , \qquad \text{with} \qquad \mu_A^t \ll \mu_{\omega , a} \ \text{and} \ \mu_S^t \perp \mu_{\omega , a} 
\end{equation}
we define
$$\chi^t_\omega = \mu_A^t(\mathbb R) \ , \qquad t \geq 0$$
and
$$D^+ = \left\{ t \in \mathbb R^+ : \chi^t_\omega \notin \left\{ 0, 1 \right\} \right\}$$
For every $t \in D^+$, we obtain from equation \eqref{decoleb0} the decomposition:
\begin{equation}
\label{decoleb1}
 \mu_{\omega , a}^t = \chi^t_\omega \mu_1^t + (1 - \chi^t_\omega) \mu_2^t  
\end{equation}
where
$$ \mu_1^t \ll \mu_{\omega , a} \ \text{and} \ \mu_2^t \perp \mu_{\omega , a} $$
Consequently, for every Borel set $\Delta \in B(\mathbb R)$ and $t \in D^+$, we have 
\begin{equation}
\label{decoleb2} \mu_1^t (\Delta) = \frac{\mu_A^t(\Delta)}{\chi^t_\omega} \in \Pi \qquad , \qquad \mu_2^t (\Delta) = \frac{\mu_S^t(\Delta)}{1 - \chi^t_\omega} \in \Pi
\end{equation}
\begin{remark}\upshape
 If $t_1 \notin D^+$ then $\mu_{\omega,a}^{t_1} \perp \mu_{\omega,a}$.
\end{remark}
Let
$$K_{\omega,a}(s,t) = \frac{d \mu_1^t}{d \mu_{\omega , a}} \ \mathbf{1}_{D_*}(t,s) \ \in L^1(\mu_{\omega , a}) $$
where
$$D_* = D^+ \times \mathbb R \subset \mathbb R^2$$
For every Borel set $\Delta \in \mathbb R$ and $t \in \mathbb R^+$, we have
\begin{equation}
\mu_{\omega,a}^t (\Delta) = \chi^t_\omega \int_\Delta K_{\omega,a}(s,t) \, d\mu_{\omega , a}(s) + (1 - \chi^t_\omega) \mu_2^t (\Delta)
\end{equation}
By the definition of singular measures, for every $\lambda \in \operatorname{supp} \mu_{\omega,a} \subset \sigma(a)$ it follows that
$$ \mu_{\omega,a}^t ( \left\{ \lambda \right\} ) = \chi^t_\omega \ K_{\omega,a}(\lambda,t) \ \mu_{\omega,a}( \left\{ \lambda \right\} ) \ , \qquad t \in \mathbb R^+$$
We define the positive \textit{Koopman operator}:
$$ f \in L^1(\mu_{\omega,a}) \longrightarrow T_{\omega,a} (f) \in L^1(\mu_{\omega,a}) $$
where
\begin{equation}\label{opkoop}
T_{\omega,a} (f)(t) = \int f(s) K_{\omega,a}(s,t) \, d\mu_{\omega , a}(s)
\end{equation} 
\begin{example}\upshape[\textbf{Special Case: Delta Initial Measure}] 
\\
Consider an observable $x$ that admits a state $\omega \in \mathfrak S_x$ such that for $\tau = 0$,
$$ \mu_{\omega , x} = \delta_r $$
(i.e., $P(x \in \left\{ r \right\}, 0)_\omega = 1$).
 \\
For dissipative phenomena, in general, for $t > 0$, the evolved measure satisfies
$$ \mu_{\omega , x}^t \neq \delta_r $$ 
Applying the Lebesgue decomposition (see Example \ref{exempioleb} in Section \ref{States and Measures}), we obtain:
$$\mu_{\omega,x}^t = \mu_A^t + \mu_S^t \ , \qquad \mu_A^t \ll \delta_r \ , \ \mu_S^t \perp \delta_r$$
where for any Borel set $\Delta$ of $\mathbb R$  
$$ \mu_A^t (\Delta) = \mu_{\omega,x}^t(\left\{ r \right\} ) \delta_r(\Delta) \qquad , \qquad \mu_S^t (\Delta) = \mu_{\omega,x}^t( \Delta \setminus \left\{ r \right\} )$$
Decomposing according to Lebesgue we have:
$$\mu_{\omega,x}^t = \mu_A^t + \mu_S^t \ , \qquad \mu_A^t \ll \delta_r \ , \ \mu_S^t \perp \delta_r$$
where for every Borel set $\Delta$ of $\mathbb R$  
$$ \mu_A^t (\Delta) = \mu_{\omega,x}^t(\left\{ r \right\} ) \delta_r(\Delta) \qquad , \qquad \mu_S^t (\Delta) = \mu_{\omega,x}^t( \Delta \setminus \left\{ r \right\} )$$
Letting
 $$\chi_{\omega}^t = \mu_{\omega,x}^t(\left\{ r \right\} )$$
we can express the measure as
\begin{equation}
\mu_{\omega,x}^t = \chi_{\omega}^t \ \delta_r + (1 - \chi_{\omega}^t) \ \mu_2^t \ , \qquad \mu_2^t \perp \delta_r
\end{equation}
with 
$$ \mu_1^t = \delta_r \ , \qquad t \geq 0 $$
\end{example}
$$ $$
For any partition $\mathcal P \in \texttt{P}(\mathbb R)$, the entropy satisfies:
$$H( \mu_{\omega,x}^t , \mathcal P ) \geq \chi_{\omega}^t H( \mu_{1}^t , \mathcal P ) + (1 - \chi_{\omega}^t) H( \mu_2^t , \mathcal P ) = (1 - \chi_{\omega}^t) H( \mu_2^t , \mathcal P )$$ 
where the equality follows since $H(\mu_{1}^t , \mathcal P) = 0$ (as $\mu_{1}^t = \delta_r$ is a Dirac measure).
\begin{remark}\upshape[\textbf{Dissipative Part}]\label{Parte Dissipativa}
The preceding analysis reveals that dissipation is characterized by the map
\begin{equation}\label{mappadissi}
\omega \in \mathfrak S_a \longrightarrow \chi_{\omega}^t \in [0, 1] \ , \qquad \chi^t_\omega = \mu_A^t(\mathbb R)  
\end{equation} 
since if $\chi_{\omega}^t = 1$ for each $t \in [0, t_1]$, then the system exhibits no dissipation during this interval. This suggests that the measure $\mu_1^t$ defined in \eqref{decoleb2} represents the non-dissipative component of the temporal evolution, and this suggests a natural decomposition of the dynamics into dissipative and conservative parts.
\end{remark} 
To rigorously characterize a non-dissipative temporal evolution, the following entropy conservation condition must hold:
\begin{equation}
H(\mu_1^t, \mathcal{P}) = H(\mu_{\omega,a}, \mathcal{P}) \quad \forall \mathcal{P} \in \texttt{P}(\mathbb{R}), \ \forall t \geq 0,
\end{equation}
where $\mu_1^t$ is the absolutely continuous component from decomposition \eqref{decoleb1}.
\\
This equality represents the fundamental criterion for the absence of dissipation in the system's evolution.
 %%%
\subsection{Theoretical Statistics for Non-Dissipative Evolution}
Under the hypothesis of Remark \ref{Parte Dissipativa}, where $\mu_1^t$ represents the non-dissipative component of the temporal evolution, we define the \textit{theoretical statistic}\footnote{For the reader interested in the theory of von Neumann algebras, the article \cite{pan18} may be consulted.} :
$$ P_{\texttt{theor}}(a \in \Delta, \tau)_\omega = \mu_1^\tau(\Delta) \ , \qquad \forall \Delta \in B(\mathbb R) \ , \ \tau \in \mathbb R^+$$
For the observable $c$ previously discussed, this yields a degenerate theoretical statistic:
$$P_{\texttt{theor}}(c \in \Delta, \tau)_\omega = \delta_r(\Delta) \ , \qquad \forall \Delta \in B(\mathbb R) \ , \ \tau \in \mathbb R^+$$
Suppose that the \ref{dominato}  Assumption  holds. Then there exist two states (generally non-unique) $\omega_t^r , \omega_t^d \in \mathfrak S_t^{a,\omega}$ such that:
$$ \mu_1^t \approx \mu_{\omega_t^r , a} \qquad \text{and} \qquad \mu_2^t \approx \mu_{\omega_t^d , a}$$
where:
\\
- $\omega_t^r$ represents the non-dissipative component of the evolution with $\omega_0^r = \omega$ 
\\
- $\omega_t^d$ captures the purely dissipative effects.
\\
 
We further assume the following condition is satisfied:
\begin{crit} \label{Parte DissipativaB}
For all $\tau \in \mathbb R^+$ the theoretical statistic satisfies:
\begin{equation}
P_{\texttt{theor}}(a \in \Delta, \tau)_\omega = P(a \in \Delta, 0)_{\omega_{\tau}^r}
\end{equation}
\end{crit}
\noindent

Under Assumption \ref{Parte DissipativaB}, the following fundamental relations hold:
\\
The Radon-Nikodym Derivative Characterization:
$$ [ \ \mu_1^t (\Delta) = \mu_{\omega_t^r , a} (\Delta) \ , \ \forall \Delta \in B(\mathbb R) \ ] \qquad \Longrightarrow \qquad [ \ \frac{d \mu_1^t}{d \mu_{\omega_t^r , a}} = 1 \ , \ \forall t \geq 0 \ ] $$
and the Integral Representation:
$$ \mu_{\omega_t^r , a} (\Delta) = \int_\Delta K_{\omega,a}(s,t) \, d\mu_{\omega , a}(s) \ , \qquad \forall \Delta \in B(\mathbb R) $$
Let $f$ be a bounded Borel function; the expectation value of the observable $f(a)$ at time $\tau$, in the dissipation-free regime, is given by the Koopman operator:
\begin{equation}
\langle f(a) \rangle_{\text{no-diss}}^\tau = T_{\omega,a}(f)(\tau)
\end{equation}
where the operator $T_{\omega,a} (f)(\tau)$ is defined in \eqref{opkoop}.
 
Let's make some considerations on dynamic decomposition.
\\
The measure evolution \eqref{decoleb1} admits the following structure:
 $$ \mu_{\omega , a}^t = \chi^t_\omega \ \mu_{\omega_t^r , a} + (1 - \chi^t_\omega) \mu_2^t \ , \qquad \mu_2^t \approx \mu_{\omega_t^d , a}$$
 where $\mu_{\omega_t^r , a}$ is the non-dissipative component and $\mu_{\omega_t^d , a}$ the dissipative component.
\\
The temporal evolution \eqref{evol_temp_2} splits affinely as:
 \begin{equation}\label{decoleb3}
S_a^t = \chi^t S^t_{a,r} + (1 - \chi^t) S^t_{a,d}
\end{equation}
with:
\begin{itemize}
\item $S^t_{a,r}$ represents the non-dissipative evolution component
\item $S^t_{a,d}$ represents the dissipative evolution component
\end{itemize}
The evolution components act on the initial state measure as:
$$S^t_{a,r} \mu_{\omega,a} = \mu_1^t \qquad , \qquad S^t_{a,d} \mu_{\omega,a} = \mu_2^t $$ 
where $\chi^t$ is the dissipation coefficient from Definition \ref{mappadissi}.

\chapter{Compatible Observables}\label{osservabilicompatibili}
As we have highlighted, for experimental reasons it is often necessary to perform simultaneous measurements of two or more observables; the meaning of this statement has been widely discussed in the previous sections. Here we want to resume the discussion to establish the meaning of the sum and product of observables performed \textit{in the same state $\omega$ at an instant of time $\tau$}. 
\\
The simultaneity of the measurement of some observables of the physical system allows us to introduce particular algebraic operations of sum and product in the set of observables. The first attempts to give a valid algebraic structure associated to a quantum system are due to various works by von Neumann and Jordan\footnote{For example, see \cite{Jord_34, Neu_32}.} and subsequently by Segal in \cite{Segal_47}. 
\\
For a historical and epistemological discussion on the algebraization of quantum physics, the interested reader can certainly consult the books by Primas \cite{Primas} and by Emch \cite{Emch}.
\section{Function of an Observable and Compatibility}\label{funzioni_compatibil}
Let's consider a generic observable $a$ of our physical system; we have seen in the previous sections what we mean by $F(a)$ where $F: \mathbb R \rightarrow \mathbb R$ is a generic Borel function. Precisely, it is an observable such that for every Borel set $\Delta$ of $\mathbb R$ we have:
\begin{equation}
\label{calcolofunz2}
P(F(a) \in \Delta)_\omega = P(a \in F^{-1}(\Delta))_\omega \ , \qquad \forall \omega \in \mathfrak S_a
\end{equation}
Furthermore, the observable $a$ and $F(a)$ are simultaneously measurable in every state $\omega$ suitable for $a$, since by measuring $a$ we can know through \eqref{calcolofunz2} the value of $F(a)$ (and vice versa), through the equality
 $$P(F(a) \in \Delta_0 : a \in \Delta_1)_\omega = P(a \in F^{-1}(\Delta_0) \cap \Delta_1)_\omega $$
so
$$P(F(a) \in \Delta_0 : a \in \Delta_1)_\omega = P(a \in \Delta_1 : F(a) \in \Delta_0)_\omega$$
it follows that they are also compatible observables.
\\
Given any two real Borel functions $F_1$ and $F_2$, we can define the following observable product:
\begin{equation}
\label{calcolofunz3}
 F_1(a) \cdot F_2(a) = F_2(a) \cdot F_1(a) = (F_1 \cdot F_2)(a)
\end{equation}
with $(F_1 \cdot F_2)(t) = F_1(t) F_2(t)$ for each real number $t$.
\\
Obviously
$$ \left\langle F_1(a) \cdot F_2(a) \right\rangle_\omega = \int F_1(t) F_2(t) \, d \mu_{\omega,a}(t)$$
and we can write $F(a) \cdot a = \widetilde{F}(a)$ where $\widetilde{F}(t) = F(t) t$.
\\
Therefore we can affirm that for every state $\omega$ of the system suitable for $a$ we obtain
$$ \left\langle 1 \cdot a \right\rangle_\omega = \left\langle a \cdot 1 \right\rangle_\omega = \left\langle a \right\rangle_\omega$$
while
$$ \left\langle 0 \cdot a \right\rangle_\omega = \left\langle a \cdot 0 \right\rangle_\omega = 0$$
where the observables $0$ and $1$ are given by the following relations:
$$ \left\langle 1 \right\rangle_\omega = 1 \qquad \forall \omega \in \mathfrak S_a \qquad , \qquad \left\langle 0 \right\rangle_\omega = 0 \qquad \forall \omega \in \mathfrak S_a$$
We observe that by definition, given two Borel sets $\Delta_1$ and $\Delta_2$ of $B(\mathbb R)$, we obtain
\begin{equation*}
\mathbf{1}_{\Delta_1}(a) \cdot \mathbf{1}_{\Delta_2}(a) = \mathbf{1}_{\Delta_1 \cap \Delta_2}(a)
\end{equation*}
\noindent
We give the following
\begin{definition}\upshape
A finite set of observables $\left\{ b_1, b_2, \ldots, b_n \right\}$ of the physical system $(\mathfrak X , \mathfrak S)$ consists of functionally dependent observables if there exists an observable $a$ and $n$ Borel functions $f_j: \mathbb R \rightarrow \mathbb R \ , \ j = 1, 2, \ldots, n$  that are  $a$-summable, such that  
$$b_j = f_j(a) \ , \qquad j = 1, 2, \ldots, n$$
\end{definition}
\noindent
We explicitly note that functionally dependent observables are compatible with each other. 
\\
In fact, in the case of only two functionally dependent observables $b_1$ and $b_2$ we have that
$b_1 = f_1(a)$ and $b_2 = f_2(a)$ with $f_1, f_2$ $a$-summable functions.
\\
So for every $\Delta_1, \Delta_2 \in B(\mathbb R)$, we have:
\begin{eqnarray*}
P(b_1 \in \Delta_1 : b_2 \in \Delta_2)_\omega & = & P( a \in f_1^{-1}(\Delta_1) : a \in f_2^{-1}(\Delta_2))_\omega =
\\
&=& \mu_{\omega,a} (f_1^{-1}(\Delta_1) \cap f_2^{-1}(\Delta_2)) =
\\
&=& P( b_2 \in \Delta_2 : b_1 \in \Delta_1)_\omega
\end{eqnarray*}
while
\begin{eqnarray*}
P(b_1 \in \Delta_1 : b_2 \in \mathbb R)_\omega & = & P( a \in f_1^{-1}(\Delta_1) : a \in f_2^{-1}(\mathbb R))_\omega =
\\
&=& \mu_{\omega,a} (f_1^{-1}(\Delta_1) \cap \mathbb R) = \mu_{\omega,b_1} (\Delta_1) 
\end{eqnarray*}
We remark that for each observable $a \in \mathfrak X$ we have the functional calculus map
\begin{equation}
\label{calculusfun}
f \in L^1(a) \stackrel{\Upsilon}{\longrightarrow} f(a) \in \mathfrak X_a := \bigcap_{\omega \in \mathfrak S_a} \mathfrak X_\omega
\end{equation}
where $L^1(a)$ is the set of definition \ref{a-sommabile}.
\begin{attenzione}\upshape
From our definition of compatibility, it does not follow that compatible observables are functionally dependent.
\end{attenzione}
Recall that if $b_1$ and $b_2$ are compatible observables we obtain that the probability measures $\mu_{\omega,b_1}$ and $\mu_{\omega,b_2}$ are the marginal measures of the product measure defined in  \eqref{misuraprodotto}, which we have subsequently denoted by $\mu_{\omega,b_1:b_2}$:
$$\mu_{\omega,b_1:b_2}(\Delta_1 \times \Delta_2) = P(b_1 \in \Delta_1 : b_2 \in \Delta_2)_\omega \ , \qquad \Delta_1, \Delta_2 \in B(\mathbb R) $$
in other words by \eqref{misuraprodotto2a} and \eqref{misuraprodotto2b} we have:
\begin{equation}\label{comp}
\mu_{\omega,b_1}(\Delta_1) = \nu(\Delta_1 \times \mathbb R) \qquad , \qquad \mu_{\omega,b_2}(\Delta_2) = \nu (\mathbb R \times \Delta_2)
\end{equation}
It is proved that for each bounded Borel function $f: \mathbb R \rightarrow \mathbb R$ we obtain\footnote{The proof is found in section \ref{verifiche}.}:
\begin{equation}
\label{comp1} \left\langle f(b_i) \right\rangle_\omega = \int_{\mathbb R^2} f(s_i) \, d \mu_{\omega,b_1:b_2}(s_1, s_2) \ , \qquad i = 1, 2
\end{equation}

\begin{remark}\upshape
For every bounded Borel function $f, g$ the observables $f(b_1)$ and $g(b_2)$ are still compatible\footnote{Warning: this statement is not as trivial as it seems since it involves the set of jointly prepared states $\mathfrak S_{b_1 : b_2}$ and $\mathfrak S_{f(b_1) : g(b_2)}$ and will have to be postulated with axiom \ref{compat_calc_func1}.}.
\end{remark}
In fact, the experimental procedures to be carried out for the measurement of $f(b_1)$ are the same as for the observable $b_1$, since $\mathfrak S_{f(b_1)} = \mathfrak S_{b_1}$ and, as we have repeatedly said, once the statistics for $b_1$ (always at a fixed time) given by
$$\Delta \in B(\mathbb R) \longrightarrow P(b_1 \in \Delta)_\omega \in [0,1]$$  has been established, we derive that of $f(b_1)$, since by definition
$$P(f(b_1) \in \Delta)_\omega = P(a \in f^{-1}(\Delta))_\omega $$ 
and vice-versa. 
 \\
Furthermore we can write the following expression:
\begin{eqnarray*}
P(f(b_1) \in \Delta_1 : g(b_2) \in \Delta_2)_\omega & = & P( b_1 \in f^{-1}(\Delta_1) : b_2 \in g^{-1}(\Delta_2))_\omega =
\\
&=& 
P( b_2 \in g^{-1}(\Delta_2) : b_1 \in f^{-1}(\Delta_1) )_\omega =
\\
&=& 
P( g(b_2) \in \Delta_2 : f(b_1) \in \Delta_1 )_\omega
\end{eqnarray*} 
and
\begin{eqnarray*}
\mu_{\omega,f(b_1)}(\Delta_1) & = & \mu_{\omega,b_1} (f^{-1}(\Delta_1)) = P( b_1 \in f^{-1}(\Delta_1) : b_2 \in \mathbb R )_\omega =
\\
& = & 
P( f(b_1) \in \Delta_1 : b_2 \in \mathbb R )_\omega 
\end{eqnarray*}
We return to the study of the relation between compatibility and functional dependence of observables with the following proposition:
\begin{proposition}\upshape
\label{condizionamentocompatibile}
If $a$ and $b$ are compatible observables, then there exists a Borel function $F$ which depends on the state $\omega$ and is $\mu_{\omega,a}$-summable such that  
$$ \left\langle b \right\rangle_\omega = \int F \, d \mu_{\omega,a}$$ 
\end{proposition}
\begin{proof}
We observe that the product measure $\mu_{\omega,a} \otimes \mu_{\omega,b}$ is absolutely continuous with respect to the measure $\nu$ defined in \eqref{misuraprodotto}:
$$ \nu \ll \mu_{\omega,a} \otimes \mu_{\omega,b}$$
given that if $(\mu_{\omega,a} \otimes \mu_{\omega,b})(\Delta_1 \times \Delta_2) = 0$, then by \eqref{comp} we obtain that $\nu(\Delta_1 \times \Delta_2) = 0$.
\\
It follows by the Radon-Nikodym theorem that there exists a probability density $K(s,t)$ which is $\mu_{\omega,a} \otimes \mu_{\omega,b}$-summable; therefore by Fubini's theorem we can write
\begin{eqnarray*}
\left\langle b \right\rangle_\omega & = & \int_{\mathbb R^2} t \, d\nu(s,t) = \int_{\mathbb R^2} t K(s,t) \, d (\mu_{\omega,a} \otimes \mu_{\omega,b})(s,t) =
\\
&=&
\int \left[ \int t K(s,t) \, d \mu_{\omega,b}(t) \right] \, d \mu_{\omega,a}(s) = \int F(s) \, d \mu_{\omega,a}(s)
\end{eqnarray*}
\end{proof}
Obviously, in the proposition we can swap the roles of $a$ and $b$ and write 
$$\left\langle a \right\rangle_\omega = \int G \, d \mu_{\omega,b}$$ 
with $G$ a $\mu_{\omega,b}$-summable function.
\section{Sum and Product of Compatible Observables} \label{compatibili-algebre}
The real number sum of two observables is not always a value that can be related to an observable that is actually measurable.
\\
Intuitively, compatible observables, since they are not influenced by the mutual preparations that we can do on them\footnote{Preparing them individually or jointly does not change their probability measure $\mu_{\omega ,a}$.} can be added, so the sum of the expected values of two compatible observables can be derived from an expected value of a new observable of our physical system. This statement introduces a new axiom:
\begin{axiom}[\textbf{Sum of compatible observables}] \label{assioma somma}\index{Axiom- Sum of Compatible Observables}
Given two compatible observables $a$ and $b$ of $\mathfrak{X}$, there always exists an observable $c$ of $\mathfrak{X}$ compatible with $a$ and $b$ such that
for each state $\omega$ of $\mathfrak S_{a:b}$ we have:
\begin{equation}
\label{assioma somma1}
\left\langle c\right\rangle _{\omega } = \left\langle a\right\rangle _{\omega} + \left\langle b\right\rangle _{\omega}.
\end{equation}
with
\begin{equation}
\mathfrak S_c = \mathfrak S_{a:b}
\end{equation}
\end{axiom}
For every compatible $a$ and $b$ of $\mathfrak{X}$ we can define their sum $a+b$ as the observable $c$ enabled by the previous axiom:  
\begin{equation*}
\left\langle a+b\right\rangle _{\omega } = \left\langle a\right\rangle_{\omega } + \left\langle b\right\rangle_{\omega}
\end{equation*}%
for all states $\omega$ of $\mathfrak S_{a:b}$: 
$$\mathfrak S_{a+b} = \mathfrak S_{a:b} \subset \mathfrak S_{a} \cap \mathfrak S_{b}$$
This definition is obviously well-posed, i.e., we have a unique element $c$ of $\mathfrak{X}$ by axiom \ref{assio1}.
\\
As we shall see in section \ref{funzioneosservabile2}, if $\mu_{\omega,a:b}$ is the product measure defined in \eqref{misuraprodotto}, then 
\begin{equation*}
\left\langle c\right\rangle _{\omega } = \int_{\mathbb R^2} (s+t) \, d \mu_{\omega,a:b} (s,t) \qquad , \ \ \omega \in \mathfrak S_{a:b}
\end{equation*}%
Therefore for all compatible $a$ and $b$ of $\mathfrak{X}$ we obtain 
\begin{equation}
\label{norma della addi}
\left\Vert a+b\right\Vert \leq \left\Vert a\right\Vert + \left\Vert b\right\Vert  
\end{equation}
given that 
\begin{equation*}
\left\Vert a+b\right\Vert = \underset{\omega \in \mathfrak S_{a+b}}{\sup} \left( \left\langle a\right\rangle _{\omega} + \left\langle b\right\rangle _{\omega} \right) \leq \underset{\omega \in \mathfrak S_a }{\sup} \left\langle a\right\rangle _{\omega} + \underset{\omega \in \mathfrak S_b }{\sup} \left\langle b\right\rangle _{\omega} = \left\Vert a\right\Vert + \left\Vert b\right\Vert  
\end{equation*}
furthermore for every $\lambda \in \mathbb C$ 
\begin{equation}
\left\| \lambda a \right\| = |\lambda| \left\| a \right\|
\end{equation}
\begin{proposition}\upshape
We have the following properties of the sum of mutually compatible observables:
\begin{itemize}
\item Commutativity:
\begin{equation*}
a + b = b + a
\end{equation*}
\item Associativity\footnote{Actually here we have anticipated a result that will be discussed in detail in section \ref{calcolo_funz_n_dim}, in particular that it turns out
$$\mathfrak S_{(a+b)+c} = \mathfrak S_{(a+b):c} = \mathfrak S_{a:b:c} \qquad , \qquad \mathfrak S_{c+(a+b)} = \mathfrak S_{c:(a+b)} = \mathfrak S_{c:a:b}$$}:
\begin{equation*}
(a + b) + c = a + (b + c)
\end{equation*}

\item Neutrality of zero\footnote{Warning: here the observable $0$ is the observable compatible with $a$ given by
$c_0(a)$ where $c_0(t) = 0$ for every $t \in \mathbb R$. }:
\begin{equation*}
a + 0 = 0 + a = a
\end{equation*}

\item Cancellation law: 
\begin{equation*}
a + b = a + c \quad \implies \quad b = c
\end{equation*}
\end{itemize}
\end{proposition}
\begin{proof}
It's trivial.
\end{proof}
As in the case of the sum, given any two observables $a$ and $b$ of the physical system and individually establishing their distribution laws at the time $\tau$ (marked by the clock in our laboratory), $P(a \in \Delta, \tau)_\omega$ and $P(b \in \Delta, \tau)_\omega$, one could think of carrying out the numerical multiplication of the various experimentally determined values of the two observables to establish the numerical value of the product of the two observables in question. This way of operating is experimentally incorrect, since the numerical value established in this way does not always actually correspond to an observable of the physical system. We can give a physical (operational) sense only to the product of compatible observables; to do this we use the sum and square operation of an observable established in the previous sections\footnote{In \textit{Grundlagen}, von Neumann \cite{Neu_32}\index{von Neumann} asserts that the sum is well defined even for incompatible observables, since in his framework he assumes that the observables of a system are identified with the self-adjoint operators of a separable Hilbert space and the sum of self-adjoint operators still results in a self-adjoint operator. This assertion is taken up by Segal in his \textit{Postulates for General Quantum Mechanics} \cite{Segal_47} but subsequently in his \textit{Mathematical Problems of Relativistic Physics} \cite{Segal_63} we explicitly find the statement that the sum and the product of observables makes sense only for simultaneously observable observables. }.
\\ 
\textit{We want to recall that if $a$ and $b$ are compatible observables, then $f(a)$ and $g(b)$ for every $f, g \in \mathfrak B_\infty (\mathbb R)$ are compatible observables.}
\subsubsection{Jordan Product}
For every compatible $a$ and $b$ of $\mathfrak{X}$ we can define the following \textit{Jordan product}
\begin{equation}\label{prodotto jordan}
a \cdot b = \frac{1}{2} \left[ (a + b)^2 - a^2 - b^2 \right]   
\end{equation}
with
$$ \mathfrak S_{a \cdot b} = \mathfrak S_{a:b} $$

We note that for every natural number $m, n$ we have
$$a^m \cdot a^n = a^{m+n}$$
Since  taking the real Borel function $f(t) = \frac{1}{2} \left[ (t^m + t^n)^2 - t^{2m} - t^{2n} \right]$ we obtain $f(a) = a^m \cdot a^n$ and $f(t) = t^{m+n}$.
\begin{remark}\upshape
By definition of the power of an observable it turns out that
$$a^0 = c_1(a) \qquad \Longrightarrow \qquad a^0 \subset I$$
where the function $c_1(t) = 1$ for each $t \in \mathbb R$
\end{remark}
\begin{proposition}\upshape\label{norma_banach}
For every pair of compatible observables $a$ and $b$ of $\mathfrak{X}$ we obtain:
\begin{equation}
  \left\| a \cdot b \right\| \leq \left\| a \right\| \left\| b \right\| 
\end{equation}
\end{proposition}
\begin{proof}
By definition
$$\left\| a \cdot b \right\| = \sup_{\omega \in \mathfrak S_{a:b}} \left| \left\langle a \cdot b \right\rangle_\omega \right| = 
\frac{1}{2} \sup_{\omega \in \mathfrak S_{a:b}} \left\{ \left| \left\langle (a + b)^2 \right\rangle_\omega - \left\langle a^2 \right\rangle_\omega - \left\langle b^2 \right\rangle_\omega \right| \right\}$$
From proposition \ref{*banach} we obtain:
$$ -\left\langle a^2 \right\rangle_\omega - \left\langle b^2 \right\rangle_\omega \leq 
 -\left\langle a \right\rangle_\omega^2 - \left\langle b \right\rangle_\omega^2 $$
and given that
$$( \left\langle a \right\rangle_\omega + \left\langle b \right\rangle_\omega )^2 = \left\langle (a + b) \right\rangle_\omega ^2$$
we can write:
$$ \left| \left\langle (a + b)^2 \right\rangle_\omega - \left\langle a^2 \right\rangle_\omega - \left\langle b^2 \right\rangle_\omega \right| \leq 
   \left| \left\langle (a + b)^2 \right\rangle_\omega - \left\langle (a + b) \right\rangle_\omega ^2 + 2 \left\langle a \right\rangle_\omega \left\langle b \right\rangle_\omega \right| $$
Going to the least upper bound and by remark \ref{C_star_norma}, we have:
$$ 0 \leq \sup_{\omega \in \mathfrak S_{a:b}} \left\{ \left\langle (a + b)^2 \right\rangle_\omega - \left\langle (a + b) \right\rangle_\omega ^2 \right\} = \left\| a + b \right\|^2 - \left\| (a + b)^2 \right\| = 0$$
and thus the thesis.
\end{proof}
If $a$ and $b$ are compatible observables of the system, then as said previously, for every pair of bounded Borel functions $f$ and $g$ we obtain
$$f(a) \cdot g(b) = g(b) \cdot f(a)$$  
therefore for every pair of Borel sets $\Delta_1, \Delta_2$ we have
\begin{equation*}
\mathbf{1}_{\Delta_1}(a) \cdot \mathbf{1}_{\Delta_2}(b) = 
 \mathbf{1}_{\Delta_2}(b) \cdot \mathbf{1}_{\Delta_1}(a)
\end{equation*} 
\\
Let's now make some simple considerations on the product defined in \eqref{prodotto jordan}.
\\
From  \eqref{decompositivo1}  we obtain that every $a \in \mathfrak X$ decomposes into $a = a_+ - a_-$ with the observables $a, a_+, a_-$ compatible with each other; now it is easy to prove that 
$$a_+ \cdot a_- = 0 \qquad \Longrightarrow \qquad a^2 = a_+^2 + a_-^2$$
\begin{proposition}\upshape
\label{formalreal}
If $a$ and $b$ are compatible observables such that $a^2 + b^2 = 0$, then we obtain that $a, b \subset 0$.
\end{proposition}
\begin{proof}
Indeed, for every $\omega \in \mathfrak S_{a:b}$ we have:
$$\left\langle a^2 + b^2 \right\rangle_{\omega} = \left\langle a^2 \right\rangle_{\omega} + \left\langle b^2 \right\rangle_{\omega} = 0 \qquad \Longrightarrow \qquad a^2, b^2 \subset 0$$
and from proposition \ref{quadratonullo} the thesis follows.
\end{proof}
The next step in the mathematical modelling of observables is to determine particular families of observables that possess a well-defined algebraic structure, to which end the next section is devoted.
\section{Notes on Jordan Algebras*}
Let us now briefly recall the theory of Jordan algebras\footnote{For more details see the text by Hanche-Olsen and Størmer \cite{Ol-St}.}.
\begin{definition}\upshape[\textbf{Jordan Algebra}]
A real Jordan algebra $\texttt{B}$ is a real linear space with a product, called the Jordan product, which satisfies the following properties:
\\
For every $A, B, C \in \texttt{B}$  
\begin{itemize}
\item Commutativity:
$A \circ B = B \circ A$
\item Distributivity:
$A \circ (B + C) = A \circ B + A \circ C$
\item Weak Associativity:
$A^2 \circ (B \circ A) = (A^2 \circ B) \circ A$
\end{itemize}
\end{definition}
Following Jordan et al. (see Jordan \cite{Jord_34} p. 32), the algebra $\texttt{B}$ is \textit{formally real} if 
$$A^2 + B^2 = 0 \qquad \Longrightarrow \qquad A = B = 0$$
Let $\texttt{B}_1$ and $\texttt{B}_2$ be two Jordan algebras; a $\mathbb R$-linear map $\Phi: \texttt{B}_1 \rightarrow \texttt{B}_2$ is a Jordan morphism if it preserves the Jordan product:
$$\Phi(A \circ B) = \Phi(A) \circ \Phi(B) \ , \qquad \forall A, B \in \texttt{B}_1 $$
We denote by $L(\texttt{B})$ the set of linear operators from $\texttt{B}$ to itself and for each $A \in \texttt{B}$ we define \textit{the multiplication operator}\index{$L(\texttt{B})$}:
\begin{equation}\label{op-multi}
T_A X = A \circ X \ , \qquad \forall X \in \texttt{B}
\end{equation}
We note that $T_A T_B = T_B T_A$ if and only if
\begin{equation}
\label{commutante-J}
 B \circ (X \circ A) = (B \circ X) \circ A \ , \qquad \forall X \in \texttt{B} 
\end{equation}
For every $A \in \texttt{B}$ we define the set\index{$\mathfrak Z(A)$}
$$ \mathfrak Z(A) = \left\{ B \in \texttt{B} \ : \ T_A T_B = T_B T_A \right\} $$
while the \textit{center} of the Jordan algebra is given by the set:\index{Center of the Jordan algebra}
\begin{equation}\label{centro}
 \mathfrak{Z}(\texttt{B}) = \bigcap_{A \in \texttt{B}} \mathfrak{Z}(A) \subset \texttt{B} 
\end{equation}
The linear subspace $\mathfrak{Z}(\texttt{B})$ of $\texttt{B}$ is an associative algebra (see \cite{Ol-St} Lemma 2.5.3):
\begin{itemize}
\item[1.] $x \circ y \in \mathfrak{Z}(\texttt{B}) \ , \ \forall x, y \in \mathfrak{Z}(\texttt{B})$ ;
\item[2.] $x \circ (y \circ z) = (x \circ y) \circ z \ , \ \forall x, y, z \in \mathfrak{Z}(\texttt{B})$ .
\end{itemize}
Let's briefly focus on associative algebras and analyse how they relate to Jordan algebras.
\subsubsection{Real associative algebra}
Let $\mathcal A$ be a real associative algebra; we can define in it a so-called Jordan product, as
\begin{equation}\label{prodotto jordan2}
A \circ B = \frac{1}{2} (AB + BA) \ , \qquad \forall A, B \in \mathcal A
\end{equation}
which makes $\mathcal A$ a real Jordan algebra, conventionally denoted by $\mathcal A^{(+)}$\footnote{Not to be confused with the positive elements of the associative algebra.}.
\\
In this case it is easy to see that we obtain:
$$A \circ B = \frac{1}{2} \left[ (A + B)^2 - A^2 - B^2 \right]$$
\begin{remark}\upshape Let $A, B$ be elements of a generic associative algebra $\mathcal A$; the following standard notation is used:
$$ [A, B] = AB - BA \qquad , \qquad \{A, B\} = AB + BA $$
which are called the commutator and the anti-commutator of the algebra, respectively.
\\
Furthermore, if we consider the product
$$A \ast B = \frac{1}{2} (AB - BA) \qquad , \qquad \forall A, B \in \mathcal A $$
the algebra $\mathcal A$ becomes a Lie algebra, which is denoted by the symbol $\mathcal A^{(-)}$.
\\
We observe that for the associative product of $\mathcal A$ we have a decomposition into a Jordan part and a Lie part:
\begin{equation}
AB = A \circ B + A \ast B \ , \qquad \forall A, B \in \mathcal A
\end{equation}
\end{remark}
After this brief reminder we return to the study of the link between the center of the Jordan algebras and the associative algebra.
\\

We have a fundamental definition (see Kalisch \cite{kali}):
\begin{definition}\upshape
A real Jordan algebra $\texttt{B}$ is called \textit{special} if there exists a subalgebra $\mathcal R$ of $\mathcal A^{(+)}$ isomorphic to $\texttt{B}$\footnote{In other words, $\mathcal R$ is a subspace of $\mathcal A$ closed with respect to the Jordan product \eqref{prodotto jordan2}. }  
\end{definition}
\begin{definition}\upshape[\textbf{Commuting Relative}] \index{Commuting relative in Jordan algebra}
Let $\mathfrak D$ be any subset of the real associative algebra $\mathcal A$ ; we define its  commuting
relative  in $\mathcal A$ as the set\footnote{Therefore $\mathfrak A^c \subset \mathfrak A$ and $\mathfrak A = \mathfrak A^{cc}$.}
$$ \mathfrak D^{c} = \left\{ A \in \mathcal A : AX = XA \ \ \forall X \in \mathfrak D \right\} \subset \mathcal A$$ 
\end{definition}
It is easily verified that the elements of $\mathfrak D^c$ constitute an abelian subalgebra of the associative algebra $\mathcal A$.
\\
Let us take into consideration the Jordan algebra $\mathcal A^{(+)}$ induced by the real associative algebra $\mathcal A$ with the product given in \eqref{prodotto jordan2}.
\\
In this case, from \eqref{commutante-J} it is easily proved that we have the following equality:
\begin{equation}
\label{prodotto jordan3}
\mathcal A^c := \left\{ A \in \mathcal A : AB = BA \ , \ \forall B \in \mathcal A \right\} \subset \mathfrak{Z}(\mathcal A^{(+)}) 
\end{equation}
\begin{remark}\upshape\label{commutante0}
Let $\mathcal A$ be a subalgebra of $B(\mathcal H_{\mathbb R})$ of bounded operators on a real Hilbert space $\mathcal H_{\mathbb R}$. We have   
 $$Z(\mathcal A) := \mathcal A' \cap \mathcal A = \mathcal A^c$$
where $\mathcal A'$ denotes the commutant of the algebra $\mathcal A$ in $B(\mathcal H_{\mathbb R})$:
$$\mathcal A' = \left\{ T \in B(\mathcal H_{\mathbb R}) : AT = TA \ , \ \forall A \in \mathcal A \right\}$$
Thus we obtain:
 $$Z(\mathcal A) \subset \mathfrak{Z}(\mathcal A^{(+)}) $$
\end{remark}
In other words, this remark tells us that the center $Z(\mathcal A)$ of the subalgebra of bounded operators on a Hilbert space $\mathcal A$ is included in the center $\mathfrak{Z}(\mathcal A^{(+)})$ of the Jordan algebra $\mathcal A^{(+)}$, but we cannot say that it coincides\footnote{See equality \eqref{centro_bis}.}
\\

The Jordan algebras $\texttt B$ of interest to us will all be real, unital and Banach, i.e., $\texttt B$ is a real Banach space with the property 
\begin{equation}\label{disBanach}
\| A \circ B \| \leq \| A \| \cdot \| B \| \ , \qquad \forall A, B \in \texttt B
\end{equation}
and among these we will consider those denoted by \textit{JB} algebras whose norm satisfies these two further conditions\footnote{It is proved in \cite{Sh79} that \eqref{disBanach} is a consequence of these two conditions.}:
\begin{itemize}
\item [1.] $\| A^2 \| = \| A \|^2 \ , \qquad \forall A \in \texttt{B}$ (it follows that $\| I \| = 1$)
\item [2.] $\| A^2 \| \leq \| A^2 + B^2 \| \ , \qquad \forall A, B \in \texttt{B}$
\end{itemize}
\begin{remark}\upshape\label{commutante001}
Let $\mathcal A$ be a real Banach algebra; then $\mathcal A^{(+)}$ is a real JB algebra because relation \eqref{disBanach} holds.
\end{remark}

\begin{remark}\upshape\label{commutante01}
Given a complex Banach *-algebra $\mathfrak B$ we can obtain a real JB-algebra by considering the set of its self-adjoint elements $\mathfrak B_{s.a.}$ with the Jordan product given in \eqref{prodotto jordan2}.
\end{remark}

In the following we will consider real Jordan algebras contained in the set of self-adjoint operators of the algebra $B(\mathcal H)$ of bounded operators on a Hilbert space $\mathcal H$. In this regard, please note that\footnote{See Størmer and Topping \cite{stormer,topp}.} a Jordan algebra \textit{JC} [\textit{JW}] is a Jordan algebra $\texttt{B}$ contained in $B(\mathcal H)_{s.a.}$, closed in norm [\textit{weakly closed}] with the product given by equation \eqref{prodotto jordan2}. 
\\
Obviously a \textit{JC} algebra is a \textit{JB} algebra.
\\
\noindent
We have the following statement\footnote{See Topping \cite{topp} Proposition 3.1.}:
\begin{proposition}\upshape[\textbf{Topping}]\label{centro_commutante}
Let $A, B \in B(\mathcal H)_{s.a.}$; we have:
$$ AB = BA \quad \Longleftrightarrow \quad T_A T_B = T_B T_A $$
with
$$ T_A X = A \circ X = \frac{1}{2} (AX + XA) \ , \qquad X \in B(\mathcal H)_{s.a.}$$
\end{proposition}
Therefore, if $\texttt{B} \subset B(\mathcal H)_{s.a.}$ is a Jordan JC algebra, then we have\footnote{We note that the set $Z(\texttt{B})$ is not an associative subalgebra of $B(\mathcal H)$. }
\begin{equation} \label{centro_bis}
 \mathfrak{Z}(\texttt{B}) = \left\{ A \in \texttt{B} : AB = BA \ , \ \forall B \in \texttt{B} \right\} = \texttt{B} \cap \texttt{B}' = Z(\texttt{B})  
 \end{equation}
\begin{attenzione}\upshape
Let us remember that a JB-algebra generally cannot be represented as an algebra of operators on a Hilbert space, and therefore for these algebras it makes no sense to talk about closure in the weak topology \cite{Sh79}.
\end{attenzione}
The analogue of real W*-algebras in Jordan algebras is given by JBW-algebras:
\\
A real JBW-algebra is a Jordan algebra isomorphic to the dual of a real Banach vector space.
\\
We underline that a Jordan JW-algebra is also a Jordan JBW-algebra.
\\
We have the following result which is found in \cite{Sh79} Theorem 3.9\footnote{I invite you to also consult the works of Alfsen and Shultz \cite{AS78, AS80} where a necessary and sufficient condition for a Jordan JB-algebra to be the self-adjoint part of a C*-algebra is given.}:
\begin{theorem}\upshape[\textbf{Alfsen-Shultz}]
A JBW-Jordan algebra $\texttt{B}$ admits a unique direct sum decomposition:
$$\texttt{B} = \texttt{B}_{sp} \oplus \texttt{B}_{ex} $$
where $\texttt{B}_{sp}$ is isomorphic to a Jordan JW-algebra, while $\texttt{B}_{ex}$ is purely exceptional\footnote{See the book Hanche-Olsen and Størmer \cite{Ol-St}, paragraph 7.2, for the definition of a purely exceptional algebra.}.
\end{theorem}
\section{Center of a Set of Observables}
After this brief excursion into the world of Jordan algebras, we return to our physical system $(\mathfrak X, \mathfrak S)$.
\begin{definition}\upshape\label{centro_obs}\index{$\mathcal C(a)$}
Let $a$ be an observable of the system; we denote by $\mathcal C(a)$ the set of observables compatible with $a$.
\end{definition}
We remark that from section \ref{funzioni_compatibil}, we have that if a function $f$ is $a$-summable, then the observable $f(a)$ is still in $\mathcal C(a)$.
\\
So, if we take the constant function $c(t) = \texttt{r} , \ \forall t \in \mathbb R$, then the constant observable $c(a)$ is still in $\mathcal C(a)$; it follows that the constant observables of the system belong to the set
$$ \left\{ c \in \mathfrak X : c \subset \texttt{r} I \ , \ \texttt{r} \in \mathbb R \right\} \subset \mathcal C(a) \subset \mathfrak X$$
We emphasize that if $x \in \mathcal C(a)$, then by definition $\mathfrak S_x \cap \mathfrak S_a \neq \emptyset$; it follows that
\begin{equation}\label{centrosemplice}
 \mathcal C(a) \subset \mathfrak X_\vee^{\mathfrak S_a} = \bigcup_{\omega \in \mathfrak S_a} \mathfrak X_\omega
\end{equation}
We will denote by $\mathcal C^F(a)$ the set of observables \textit{strongly compatible} with $a$; obviously:
$$ \mathcal C^F(a) \subset \mathcal C(a) $$
furthermore 
$$x \in \mathcal C^F(a) \qquad \Longrightarrow \qquad \mathfrak S_x = \mathfrak S_a $$
Let's now select a family of observables of the system $\mathfrak X_o$; the following sets of observables are associated with it:
\begin{definition}\upshape[\textbf{Center of a Set of Observables}]\label{commutante}\index{Center of a set of observables}\index{$\mathcal{Z}(\mathfrak X_o ) $}
Let $\mathfrak X_o \subset \mathfrak X$; we denote by $\mathcal{C}(\mathfrak{X_o})$ the set of all observables compatible with every observable of $\mathfrak X_o$:
\begin{equation} 
\mathcal{C}(\mathfrak X_o ) := \bigcap_{a \in \mathfrak X_o} \mathcal C(a) \subset \mathfrak X
\end{equation}
The set $\mathcal{C}(\mathfrak X_o )$ is said to be the \textit{commutant }of $\mathfrak X_o$, while the \textit{center }of the observables $\mathfrak X_o$ is given by  
\begin{equation} \index{Commutant}
\mathcal{Z}(\mathfrak X_o ) := \mathcal{C}(\mathfrak X_o) \cap \mathfrak X_o
\end{equation}
\end{definition}
Let's see some simple properties of this set:
\begin{remark}\upshape
Let $a \in \mathfrak X_{oo} \subset \mathfrak X_o \subset \mathfrak X$; by definition we obtain that: 
$$ \mathcal{C}(\mathfrak X) \subset \mathcal{C}(\mathfrak X_o) \subset \mathcal{C}(\mathfrak X_{oo}) \subset \mathcal{C}(a) $$
Furthermore, the set $\mathcal{C}(\mathfrak X_o)$ may be empty, since, unlike the algebraic case, we have not assumed the existence of number observables.
\end{remark}
\begin{remark}\upshape\label{sottosist}
Let $x, y \in \mathcal{C}(\mathfrak X_o)$; it is not necessarily true that $x$ and $y$ are compatible with each other, so it is not necessarily true that the sum observable $x + y$ exists, and even if it did, it is not necessarily true that it belongs to the set $\mathcal{C}(\mathfrak X_o)$.
\end{remark}
We set 
$$\mathfrak X_o' := \left\{ x \in \mathfrak X : \mathfrak X_o \subset \mathcal C(x) \right\} \subset \mathfrak X $$ 
we then have the following  
\begin{proposition}\upshape
For the commutant set related to $\mathfrak X_o$, it turns out that  
$$ \mathcal C(\mathfrak X_o) = \mathfrak X_o' $$
and thus we can write 
$$ \mathcal{Z}(\mathfrak X_o) = \mathfrak X_o' \cap \mathfrak X_o$$   
\end{proposition}
\begin{proof}
If $x \in \mathcal C(\mathfrak X_o)$, then by definition $x$ is compatible with every observable $a \in \mathfrak X_o$; therefore:
$$ [ a \in \mathcal C(x) \qquad \forall a \in \mathfrak X_o ] \qquad \Longrightarrow \ \mathfrak X_o \subset \mathcal C(x) \qquad \Longrightarrow \ x \in \mathfrak X_o'$$ 
Conversely, if $x \in \mathfrak X_o'$, then by definition $\mathfrak X_o \subset \mathcal C(x)$, so every observable $a \in \mathfrak X_o$ is compatible with
 $x$; therefore: 
$$ [ x \in \mathcal C(a) \qquad \forall a \in \mathfrak X_o ] \qquad \Longrightarrow \ x \in \bigcap_{a \in \mathfrak X_o} \mathcal C(a) = \mathcal C(\mathfrak X_o)$$
\end{proof}
Given a family of observables $\mathfrak X_o$ of our physical system, we can define its strong commutator:
$$ \mathcal C^F(\mathfrak X_o) = \bigcap_{a \in \mathfrak X_o} \mathcal C^F(a)$$ 
and the strong center:
 $$ \mathcal Z^F(\mathfrak X_o) = \mathcal C^F(\mathfrak X_o) \cap \mathfrak X_o$$ 
We underline that if $\mathcal C^F(\mathfrak X_o) \neq \emptyset$, then every observable of $\mathfrak X_o$ admits the same set of suitable states:
$$\mathfrak S_o = \mathfrak S_a \ , \qquad \forall a \in \mathfrak X_o$$
Indeed, let $x \in \mathcal C^F(\mathfrak X_o)$; by definition of strong compatibility we obtain the following identity:
$$ \mathfrak S_{a:x} = \mathfrak S_{x:a} = \mathfrak S_x = \mathfrak S_a \ , \qquad \forall a \in \mathfrak X_o$$
so we fix as a set of states $\mathfrak S_o$ the set $\mathfrak S_x$.  
\\
Thus, if $\mathcal C^F(\mathfrak X_o) \neq \emptyset$, then we have a suitable pair $(\mathfrak X_o , \mathfrak S_o)$.
\bigskip

In the next sections we want to equip a subset of the observables of our physical system $\mathfrak X_o \subset \mathfrak X$ with an algebraic structure induced by the Jordan product given in \eqref{prodotto jordan}. It is natural to consider a Jordan algebra $\texttt{B}$ where it is possible to embed the set $\mathfrak X_o$ in $\texttt{B}$ appropriately, i.e., which maintains the product of compatible observables of the system, thus obtaining an algebraic inclusion:
$$ (\mathcal Z(\mathfrak X_o), \cdot ) \hookrightarrow ( \texttt{B}, \circ ) $$
To pursue this aim, we will \textit{focus only on special Jordan algebras}, which in addition to being formally real are also more mathematically tractable. Nothing prevents us from remaining in more general settings and also considering exceptional Jordan algebras; we reiterate that ours is a choice of a mathematical nature.
\\

So our goal is to identify the observables of the physical system with the elements of a real Banach algebra $\mathfrak B$ and therefore with a Jordan subalgebra $\mathcal R$ of $\mathfrak B^{(+)}$ with the further property that
\begin{equation}
\label{condizionecentrum0}
\mathcal{Z}(\mathfrak{X_o}) \hookrightarrow \mathfrak{Z}(\mathcal R)
\end{equation}
and to establish the role that compatible observables of the physical system have in determining the properties of $\mathcal R$.

\chapter{Spectral Decomposition}
Every experimental apparatus exhibits an intrinsic resolution limit $\delta$, such that for any physical observable $a$:
\begin{itemize}
\item Measurements cannot distinguish values within intervals smaller than $\delta$.  
\item The spectral projection $E_{[t-\delta/2 , t +\delta/2]}$ represents the minimal detectable event,
where
$$E_\Delta := \mathbf{1}_{\Delta}(a) \qquad \text{for each Borel set } \Delta \subset \mathbb R.$$
\item Expectation values $\left\langle f(a) \right\rangle_{\omega}$ are empirically indistinguishable from coarse-grained averages:
$$\left\langle f(a) \right\rangle_{\omega, \delta} := \sum_k f(t_k) \, \mu_{\omega, a}([t_k - \delta/2, t_k + \delta/2])$$ 
where $\left\{ t_k \right\}_k$ forms a $\delta$-spaced grid of $\sigma(a)$\footnote{Therefore
$$\left\langle f(a) \right\rangle_{\omega, \delta} = \left\langle \sum_k f(t_k) E_{[t_k - \delta/2, t_k + \delta/2]} \right\rangle_{\omega}$$ }.
\\
This means that we can discretize the observable's spectrum into small intervals $[t_k - \delta/2, t_k + \delta/2]$, where $f$ is nearly constant.
\end{itemize}	
In this chapter, we will analyze the meaning of such statements mathematically.5
\section{Spectral Decomposition of a Discrete Observable}\label{Spectral Decomposition of a Discrete Observable}
Let $a$ be a discrete observable, i.e., an observable with a point discrete spectrum\footnote{The spectrum is a set of points such that each point in the set is an isolated point.}:
$$\sigma(a) = \left\{ \lambda_n \right\}_{n \in \mathbb N} \subset \mathbb R$$
By the properties of functional calculus, we can write the following sum of compatible observables, which is called the spectral decomposition of $a$:
\begin{equation}
\label{decosp}
a = \sum_{n \in \mathbb N} \lambda_n \ \mathbf{1}_{\left\{\lambda_n\right\}}(a)
\end{equation}
and $\mathbf{1}_{\left\{\lambda_n\right\}}(a)$ is called the spectral projection of $a$ associated with $\lambda_n$.
\\
The function
 $$\Psi(t) = \sum_{n \in \mathbb N} \lambda_n \ \mathbf{1}_{\left\{\lambda_n\right\}}(t)$$
is a simple Borel function and, as established in the preceding discussion, this allows us to define the corresponding observable $\Psi(a)$ via the Borel functional calculus.
\\ 
Moreover, for every state $\omega \in \mathfrak S_a$, the following holds:
\begin{eqnarray*}
 \left\langle \sum_{n \in \mathbb N} \lambda_n \mathbf{1}_{\left\{\lambda_n\right\}}(a) \right\rangle_{\omega} & = & 
 \sum_{n \in \mathbb N} \lambda_n \left\langle \mathbf{1}_{\left\{\lambda_n\right\}}(a) \right\rangle_{\omega} =
 \sum_{n \in \mathbb N} \lambda_n \mu_{\omega, a} (\left\{\lambda_n \right\}) =
\\
& = & \int t \, d \mu_{\omega,a} (t) = \left\langle a \right\rangle_{\omega}  
\end{eqnarray*}
For every Borel measurable function $f: \mathbb R \rightarrow \mathbb R$, the following operator identity holds:
$$f(a) = \sum_{n \in \mathbb N} f(\lambda_n) \ \mathbf{1}_{\left\{\lambda_n\right\}}(a)$$
Convergence is understood in the weak sense:
$$\left\langle f(a) \right\rangle_{\omega} = \lim_{n \rightarrow \infty} \left\langle \sum_{k=1}^n f(\lambda_k) \mathbf{1}_{\left\{\lambda_n\right\}}(a) \right\rangle_{\omega} \ , \qquad \forall \omega \in \mathfrak S_a$$
In the point discrete spectrum case we have the existence of pure states in the measurement of $a$:
\begin{proposition}\upshape
If $a$ is a discrete observable, then $\mathfrak S_a$ admits a pure state in the measurement of $a$.   
\end{proposition}
\begin{proof}
Let $U_i$ be an open set of $\mathbb R$ such that $U_i \cap \sigma(a) = \left\{ \lambda_i \right\}$; then from corollary \ref{pss_a} there exists a state $\omega \in \mathfrak S_a$ for which $\left\langle \mathbf{1}_{U_i}(a) \right\rangle_{\omega} = 1$; it follows that 
$$ \mu_{\omega, a} (\left\{\lambda_n \right\}) = \delta_{n,i} \ , \qquad \forall n \in \mathbb N$$
in other words, $\mu_{\omega, a}$ is a Dirac measure, so $\omega$ is a pure state and from decomposition \eqref{decosp} we have\footnote{For the definition of the set $\texttt{V}(\lambda_i)$, see \eqref{autostati1}.}:
$$ P(a \in \left\{ \lambda_i \right\})_{\omega} = 1 \qquad \Longrightarrow \qquad \omega \in \texttt{V}(\lambda_i) $$
\end{proof}
We have another simple proposition:
\begin{proposition}\upshape
If $\sigma(a)$ is a discrete set and bounded below\footnote{The same reasoning applies if it is bounded above.} by the element $\lambda_0 \in \sigma(a)$, then every state $\omega_o \in \texttt{U(}\lambda_0)$ is a pure state in the measurement of $a$\footnote{For the definition of the set $\texttt{U}(\lambda_i)$, see \eqref{autostati}.}.
\end{proposition}
\begin{proof}
Since $\omega_o \in \texttt{U}(\lambda_0)$, by definition 
$$\lambda_0 = \left\langle a \right\rangle_{\omega_o}$$
and by the spectral decomposition of $a$ we have
$$ \left\langle a \right\rangle_{\omega_o} = \sum_{n \in \mathbb N} \lambda_n \ \mu_{\omega_o, a} (\left\{\lambda_n \right\}) $$
it follows that
$$ \sum_{n > 0} (\lambda_n - \lambda_0) \ \mu_{\omega_o, a} (\left\{\lambda_n \right\}) = 0$$
because 
$$ \sum_{n \in \mathbb N} \mu_{\omega_o, a} (\left\{\lambda_n \right\}) = 1 $$
and $\lambda_n - \lambda_0 > 0$ for each $n \neq 0$; thus 
$$\mu_{\omega_o, a} (\left\{\lambda_n \right\}) = 0 \ , \ \forall n > 0 \qquad \Longrightarrow \qquad \mu_{\omega_o, a} (\left\{\lambda_0 \right\}) = 1$$
In other words, $U(\lambda_0) = V(\lambda_0)$.
\end{proof}

\section{A norm on the linear space $\mathcal B(\mathbb R)$*}\index{$\mathcal B(\mathbb R)$}\label{norm-bound-func}
Before proceeding with the discussion, we need to make some mathematical considerations.
\\
We denote by $\mathcal B(\mathbb R)$ the linear space of real-valued bounded Borel functions.
\\
It is evident that  
$$ \mathcal B(\mathbb R) \subset L^1(\mu_{\omega,a}) \qquad , \qquad \forall \omega \in \mathfrak S_a $$
and we denote by   
$$\left\| f \right\|_{1,\omega} = \int_{\sigma(a)} |f(t)| \, d\mu_{\omega,a}(t) \qquad , \qquad \forall f \in L^1(\mu_{\omega,a})$$
the norm in $L^1(\mu_{\omega,a})$.
\\
Recall that the linear space $L^1(a)$ is defined as follows:
$$ L^1(a) = \bigcap_{\omega \in \mathfrak S_a} L^1(\mu_{\omega,a}) \qquad \Longrightarrow \qquad \mathcal B(\mathbb R) \subset L^1(a) $$
For every $f \in \mathcal B(\mathbb R)$, we define
$$\left\| f \right\|_{a,1} = \sup_{\omega \in \mathfrak S_a} \left\| f \right\|_{1,\omega} \qquad , \qquad [ \ \left\| f \right\|_{a,1} \leq \left\| f \right\|_\infty \ ] $$
where $\left\| \cdot \right\|_\infty$ denotes the supremum norm.
\\
It is straightforward to verify that $\left\| \cdot \right\|_{a,1}$ defines a norm on $\mathcal B(\mathbb R)$.
\\
We define the following set function on Borel sets:
$$\mu_a(\Delta) = \sup_{\omega \in \mathfrak S_a} \mu_{\omega,a}(\Delta) \ , \qquad \forall \Delta \in B(\mathbb R)$$

Let $\left\{ \Delta_j \right\}_{j \in \mathbb N}$ be any countable disjoint partition of a Borel set. From the positivity and $\sigma$-additivity of each $\mu_{\omega, a}$, we deduce:   
$$\mu_a \left( \bigcup_{j \in \mathbb N} \Delta_j \right) = \sum_{j \in \mathbb N} \mu_a( \Delta_j)$$
\textit{Thus $\mu_a$ is a $\sigma$-additive measure.}
\\
Since each $\mu_{\omega,a}$ is a regular probability measure, it follows that $\mu_a$ inherits this regularity property.
\\
For positive simple functions of the form:
$$\Psi(t) = \sum_{j=1}^{N} \texttt{k}_j \ \mathbf{1}_{\Delta_j}(t) \qquad (k_j \geq 0, \ \Delta_j \text{ disjoint Borel sets})$$
we have the equality:
$$\left\| \Psi \right\|_{a,1} = \int_{\sigma(a)} |\Psi(t)| \, d\mu_a(t) = \sum_{j=1}^{N} k_j \mu_a(\Delta_j)$$
By standard approximation arguments in measure theory, this equality extends to all bounded Borel functions $f \in \mathcal B(\mathbb R)$: 
$$\left\| f \right\|_{a,1} = \int_{\sigma(a)} |f(t)| \, d\mu_a(t)$$

We define the Banach space $\mathcal{L}^1(a)$ as the $\left\| \cdot \right\|_{a,1}$-completion of bounded Borel functions:
$$\mathcal{L}^1(a) := \overline{\mathcal B(\mathbb R)}^{\left\| \cdot \right\|_{a,1}} \subset L^1(a)$$
For each state $\omega \in \mathfrak S_a$, we have:
$$ L^1(\mu_{\omega, a}) = \overline{\mathcal B(\mathbb R)}^{\left\| \cdot \right\|_{1,\omega}}$$
The inclusion chain:
$$ \mathcal{L}^1(a) \subset L^1(a) \subset L^1(\mu_{\omega, a}) $$
holds for all $\omega \in \mathfrak S_a$.
\section{Spectral Decomposition of an Observable}\label{Spectral Decomposition of an Observable}
It is a standard result in measure theory that every bounded Borel function
$$F: \sigma(a) \rightarrow \mathbb R$$
can be approximated in the $L^1$-norm by simple functions. Specifically, for the Borel measure $\mu_a$, there exists a sequence of simple functions
$$\Psi_m (t) = \sum_{j=1}^{N(m)} \texttt{k}_{j,m} \ \mathbf{1}_{\Delta_{j,m}}(t) $$
where $\left\{ \Delta_{j,m} \right\}_{j=1,2,\ldots, N(m)}$ is a finite disjoint partition of $\sigma(a)$ and $\texttt{k}_{j,m} \in \mathbb R$, such that
$$\lim_{m \rightarrow \infty} \left\| F - \Psi_m \right\|_{a,1} = 0 \ , \qquad \forall \omega \in \mathfrak S_a$$
In other words, for every $\epsilon > 0$ there exists a simple function $\Psi_\epsilon$ such that
$$\left\| F - \Psi_\epsilon \right\|_{a,1} = \int_{\sigma(a)} \left| F(t) - \Psi_\epsilon(t) \right| \, d \mu_a(t) < \epsilon $$
The expectation of $F(a)$ in the state $\omega$ satisfies
$$ \left| \left\langle F(a) - \Psi_\epsilon(a) \right\rangle_{\omega} \right| = \left| \int_{\sigma(a)} \left( F(t) - \Psi_\epsilon(t) \right) \, d \mu_a(t) \right| \leq \left\| F - \Psi_\epsilon \right\|_{a,1} < \epsilon$$
\\
By definition of the norm of an observable, we have:
$$\left\| F(a) - \Psi_\epsilon(a) \right\| = \sup_{\omega \in \mathfrak S_a} \left| \left\langle F(a) - \Psi_\epsilon(a) \right\rangle_{\omega} \right| $$ 
From this, it follows that:
$$ \left\| F(a) - \Psi_\epsilon(a) \right\| = \left\| F - \Psi_\epsilon \right\|_{a,1} \leq \epsilon$$
 Consequently, for every $\omega \in \mathfrak S_a$ we obtain 
$$ \left\langle F(a) \right\rangle_{\omega} = \lim_{m \rightarrow \infty} \left\langle \Psi_m(a) \right\rangle_{\omega} $$
where $\Psi_m \rightarrow F$ in the $\left\| \cdot \right\|_{a,1}$-norm.
\\

Many results of the spectral theory of self-adjoint operators on Hilbert spaces can be repeated without particular difficulty in the case of observables of a physical system (see Conway \cite{Conway} Chapter IX.1).
\begin{theorem} \upshape \label{spec-theorem}
Let $a$ be an observable with spectral measure $E_\Delta = \mathbf{1}_{\Delta}(a)$. For any bounded Borel function $f \in \mathcal B(\mathbb R)$ and $\epsilon > 0$, there exists a finite partition $\left\{ \Delta_k \right\}_{k=1}^n$ of $\sigma(a)$ such that:
$$ \operatorname{diam}(f(\Delta_k)) < \epsilon \qquad \forall k = 1, 2, \ldots, n$$
and the following approximation holds
$$\left\| f(a) - \sum_{k=1}^n f(t_k) E_{\Delta_k} \right\| < \epsilon$$ 
\end{theorem}
\begin{proof}
We prove the existence of the partition.
\\
We apply Lusin's theorem  \ref{lusin-0}  to our case, where:
\\
The set $K = \sigma(a)$ (a compact spectrum in $\mathbb R$), the Borel measure $\mu = \mu_a$ and the function $f$ is bounded and thus $\mu_a$-measurable.
 \\
Taking $\epsilon/2 > 0$, there exists a compact set $C \subset K$ such that: 
\begin{itemize}
\item[1.] $\mu(\sigma(a) \setminus C) < \epsilon/2$ (measure control)
\item[2.] $f|_C$ is continuous on $C$, therefore uniformly continuous. 
\end{itemize} 
We exploit the uniform continuity on $C$:
$$ \text{For } \delta = \delta(\epsilon) > 0, \ |t - s| < \delta \ \Longrightarrow \ |f(t) - f(s)| < \epsilon/2$$ 
Now, we partition the set $C$ as follows:
\\
We choose a partition $\left\{ C_k \right\}_{k=1}^n$ of $C$ with $\operatorname{diam}(C_k) < \delta$
\\
We then extend this partition to all of $\sigma(a)$ by setting $\Delta_k = C_k$ and $\Delta_0 = \sigma(a) \setminus C$. 
\\
Thus, the final partition is given by: 
$$\Delta_0 \cup \left\{ \Delta_k \right\}_{k=1}^n$$
On each $\Delta_k$ ($k \geq 1$), $\operatorname{diam}(f(\Delta_k)) < \epsilon/2$ by construction, while on $\Delta_0$, we have $\mu_a(\Delta_0) < \epsilon/2$ by measure control.
\\
In our estimate:
$$ \left\| f(a) - \sum_{k=1}^n f(t_k) E_{\Delta_k}(a) \right\| = \left\| f - \sum_{k=1}^n f(t_k) \mathbf{1}_{\Delta_k} \right\|_{a,1}$$
and
$$ \left\| f - \sum_{k=1}^n f(t_k) \mathbf{1}_{\Delta_k} \right\|_{a,1} \leq \sum \int_{\Delta_k} |f(t) - f(t_k)| \, d\mu_a(t)$$
For the terms $k \geq 1$ (on $C$):   
$$\int_{\Delta_k} |f(t) - f(t_k)| \, d\mu_a(t) < \frac{\epsilon}{2} \mu_a(\Delta_k)$$
For the term $k = 0$ (on $\Delta_0$): 
$$\int_{\Delta_0} |f(t) - f(t_k)| \, d\mu_a(t) \leq 2 \| f \|_\infty \mu_a(\Delta_0) \leq 2 \| f \|_\infty \cdot \frac{\epsilon}{2} = \| f \|_\infty \epsilon$$
Thus
 $$ \left\| f(a) - \sum_{k=1}^n f(t_k) E_{\Delta_k}(a) \right\| < \frac{\epsilon}{2} \mu_a(C) + \| f \|_\infty \epsilon < \epsilon \left( \frac{1}{2} + \| f \|_\infty \right)$$ 
since $\mu_a(C) < 1$.
 \\
We observe that to obtain:
 $$\left\| f(a) - \sum_{k=1}^n f(t_k) E_{\Delta_k} \right\| < \epsilon$$ 
it suffices to reapply Lusin's theorem with $\epsilon_* = \frac{\epsilon}{\frac{1}{2} + \| f \|_\infty}$, which guarantees:
$$ \epsilon_* \left( \frac{1}{2} + \| f \|_\infty \right) = \epsilon$$
\end{proof}

For any state $\omega \in \mathfrak S_a$, the expectation values converge: 
\begin{equation}\label{Statistical-Convergence}
\lim_{n \rightarrow \infty} \left\langle \sum_{k=1}^n f(t_k) E_{\Delta_k} \right\rangle_{\omega} = \left\langle f(a) \right\rangle_{\omega}
\end{equation}
This justifies the compact notation:
$$ f(a) = \int_{\sigma(a)} f(t) \, dE_t$$
so we can write
$$ a = \int_{\sigma(a)} t \, dE_t$$
\subsubsection{Physical Interpretation} 
In summary, the spectral decomposition theorem establishes two operating principles:
\begin{itemize}
\item \textit{Measurement Granularity:}
\\
The partition $\left\{ \Delta_k \right\}_{k=1}^n$ corresponds to detector resolution limits in experiments (through the value of $\epsilon > 0$).
\item \textit{Statistical Convergence:}
\\
The state convergence \eqref{Statistical-Convergence} reflects how finite-precision measurements approach ideal expectations.
\end{itemize}
\section{Questions and States}\label{questions-states}
Let $\Delta$ be a Borel set decomposed into a countable union of pairwise disjoint Borel sets:
\begin{equation}\label{partizioe-00}
\Delta = \bigcup_{k \in \mathbb N} \Delta_k \quad \text{with} \quad \Delta_i \cap \Delta_j = \emptyset \ \forall i \neq j
\end{equation}
Then, for every state $\omega \in \mathfrak S_a$, the expectation of the question $\mathbf{1}_\Delta(a)$ satisfies:  
$$\left\langle \mathbf{1}_\Delta(a) \right\rangle_{\omega} = \int \mathbf{1}_\Delta(s) \, d \mu_{\omega,a}(s) = \int \sum_{k \in \mathbb N} \mathbf{1}_{\Delta_k}(s) \, d \mu_{\omega,a}(s) = \sum_{k \in \mathbb N} \left\langle \mathbf{1}_{\Delta_k}(a) \right\rangle_{\omega} $$
where the interchange of the integral and the sum is justified by the monotone convergence theorem (since $\mathbf{1}_{\Delta_k} \geq 0$).
This establishes the $\sigma$-additivity of the spectral measure $\mu_{\omega,a}$ as referenced in equation \eqref{sigma-addit} on page \pageref{sigma-addit}:
$$ \mu_{\omega,a}(\Delta) = \sum_{k \in \mathbb N} \mu_{\omega,a}(\Delta_k) = \mu_{\omega,a} \left( \bigcup_{k \in \mathbb N} \Delta_k \right) = \lim_{N \rightarrow \infty} \mu_{\omega,a} \left( \bigcup_{k=1}^N \Delta_k \right) $$
The key point is the convergence of the state of partial sums:
\begin{equation}\label{proiettori}
\left\langle \mathbf{1}_\Delta(a) \right\rangle_{\omega} = \lim_{N \rightarrow \infty} \sum_{k=1}^N \left\langle \mathbf{1}_{\Delta_k}(a) \right\rangle_{\omega} \ , \qquad \forall \omega \in \mathfrak S_a
\end{equation}
This last relation will be important when we discuss the algebraization of a physical system.
\begin{remark}\upshape
The partition $\left\{ \Delta_k \right\}_{k \in \mathbb N}$ of equation \eqref{partizioe-00} is chosen arbitrarily, countable, and not constrained by the assumptions made in Theorem \ref{spec-theorem}.
\end{remark}

\chapter{Function of Several Observables}
In this section we introduce the functional calculus for simultaneously preparable and compatible observables . We will define a particular subset of $\mathbb R^n$ of all possible values that a family of $n$ jointly preparable observables in a given order can assume simultaneously, which will be called the joint spectrum. 
\section{Joint Spectrum}\label{spettrocongiunto}
\begin{figure}[htbp]
	\centering
		\includegraphics[scale=0.5]{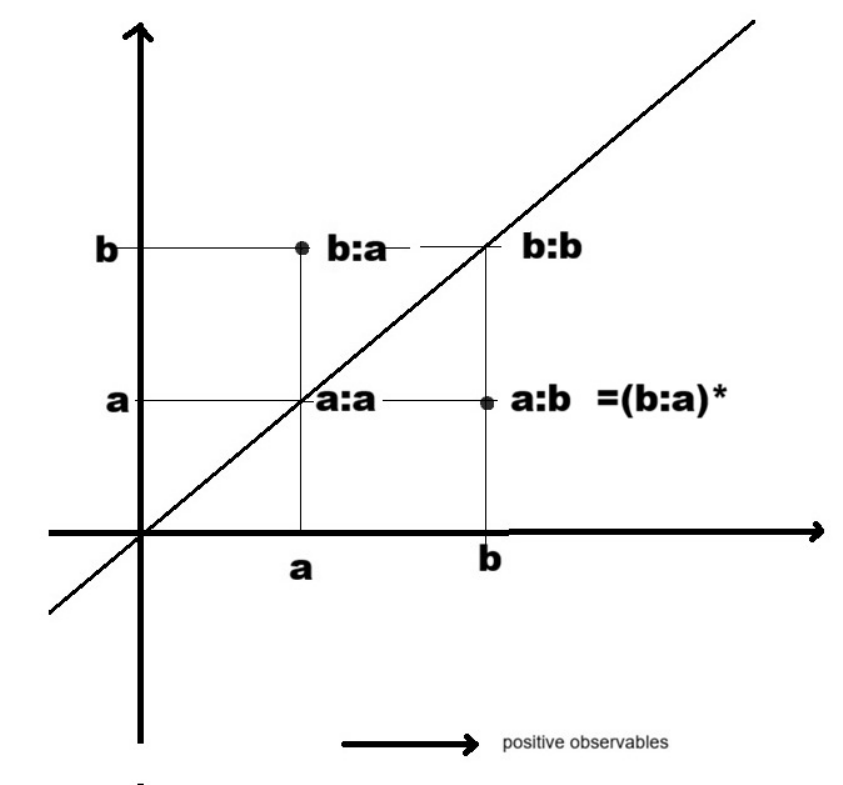}
	\caption{Simultaneous Cartesian Product $\mathfrak X \times_{\mathfrak S} \mathfrak X$}
	\label{fig:Simultaneous-Cart--prod}
	\end{figure}

Let us now resume the study carried out in section \ref{valoremediocongiunto}. Let $a$ and $b$ be two \textit{non-complementary} (not necessarily compatible) observables that we want to measure simultaneously; furthermore we assume that the preparation is $a:b$ ($a$ prepared before $b$). As we have seen in equation \eqref{misuraprodotto}, page \pageref{misuraprodotto}, there is a probability measure, which we have denoted by $\mu_{\omega, a:b} \in \Pi(\mathbb R^2)$, such that:
\begin{equation}\label{misuracongiuntasimultanea}
 P(a \in \Delta_1 : b \in \Delta_2)_\omega = \mu_{\omega, a:b}(\Delta_1 \times \Delta_2) \ , \qquad \Delta_1, \Delta_2 \in B(\mathbb R)
\end{equation}
As we have already said, we treat the pair $a:b$ as a single observable of the \textit{two-valued} system:
\begin{equation*}
 P(a:b \in \Delta_1 \times \Delta_2)_\omega = \mu_{\omega, a:b}(\Delta_1 \times \Delta_2) \ , \qquad \omega \in \mathfrak S_{a:b}
\end{equation*}
\\
We can regard $a:b$ as an element of the $\mathfrak S$-simultaneous Cartesian product (see Figure \ref{fig:Simultaneous-Cart--prod}), defined as follows:
$$\mathfrak X \times_{\mathfrak S} \mathfrak X := \left\{ (a,b) \in \mathfrak X \times \mathfrak X : \text{$a$ is jointly preparable with $b$ in the order $a:b$} \right\}$$
Clearly, the compatible observables are positioned symmetrically with respect to the bisector axis of $\mathfrak X \times \mathfrak X$\footnote{This set will consist of a discrete collection of points, with
$$\mathfrak X \times_{\mathfrak S} \mathfrak X \subset \mathfrak X \times \mathfrak X$$ }.
\\
\\
As in the case of a single observable, for each $\omega \in \mathfrak S_{a:b}$ we can define the following family of subsets of $\mathbb R^2$: 
$$\mathfrak F^\omega(a:b) = \left\{ V \text{ open set of } \mathbb R^2 : \mu_{\omega, a:b}(V) = 0 \right\} $$
and the related open set:
$$\rho^\omega(a:b) = \bigcup_{V \in \mathfrak F^{ \ \omega}(a:b)} V \subset \mathbb R^2$$
here too, by definition, we have that the support of the measure is given by
$$\operatorname{Supp} \mu_{\omega, a:b} = \mathbb R^2 \setminus \rho^\omega(a:b) $$
We define
$$\mathfrak F^\infty(a:b) = \left\{ V \text{ open set of } \mathbb R^2 : \mu_{\omega, a:b}(V) = 0 \ \forall \omega \in \mathfrak S_{a:b} \right\} $$
and the resolvent of $a:b$ as
$$ \rho^\infty(a:b) = \bigcup_{V \in \mathfrak F^\infty(a:b)} V $$
while its \textit{joint spectrum} is given by the set:
$$\sigma(a:b) = \mathbb R^2 \setminus \rho^\infty(a:b) $$
\subsubsection{Adjoint of Jointly Preparable Observables}\label{aggiunto-accardi}
Let $a, b$ be non-complementary observables of the system with the property that it is possible to prepare $a$ before $b$  and  $b$ before $a$ for the experiment, and such that
$$ \mathfrak S_{a:b} = \mathfrak S_{b:a}$$
These observables are not necessarily compatible, but only jointly preparable — i.e., preparable in both orders — and this class of observables may be denoted as \textit{non-orthogonal observables} of the system\footnote{Thus, $\mathfrak S_{a:b} = \mathfrak S_{b:a} \subset \mathfrak S_a \cap \mathfrak S_b$.}. 
\\ 
We emphasize that if $a, b$ are non-orthogonal, the observables may be simultaneously measurable independently but not compatible according to our definitions, because the associated measures $\mu_{\omega, a:b}$ and $\mu_{\omega, b:a}$ may differ, and the observable $a:b$ need not coincide with $b:a$.  
For this class of observables, we define a $*$ operation of \textit{time reversal} (cf. Accardi \cite{Accardi018}):
\[
(a:b)^* = b:a 
\]
These considerations extend straightforwardly to a family of non-orthogonal observables of the system.
\subsubsection{Joint Spectrum and Functional Calculus}
We introduce a new property that generalizes axiom \ref{assio3} of the functional calculus:
\begin{axiom}\label{assio3bis}\index{Axiom-Functional Calculus II}
For each pair of observables $a, b$ jointly preparable in the order $a:b$ and for every bounded Borel function $F: \mathbb R^2 \rightarrow \mathbb R$, we have an observable of the laboratory, which we will denote by $F(a:b)$, such that
\begin{equation}
\mathfrak S_{F(a:b)} = \mathfrak S_{a:b} 
\end{equation}
and
\begin{equation}
\mu_{\omega, F(a:b)}(\Delta) = \mu_{\omega, a:b}(F^{-1}(\Delta)) \ , \qquad \forall \Delta \in B(\mathbb R)
\end{equation}
\end{axiom}
So by definition it turns out
$$ P(F(a:b) \in \Delta)_\omega := P(a:b \in F^{-1}(\Delta))_\omega = \mu_{\omega, a:b}(F^{-1}(\Delta)) $$
The next step is to understand, given bounded Borel functions $F: \mathbb R^2 \rightarrow \mathbb R$, what the observables $F(a:b) \in \mathfrak X$ of the system are. 
\\
By definition of the average value of an observable we have
$$
\left\langle F(a:b) \right\rangle_{\omega} = \int_{\mathbb R} r \, d\mu_{\omega, F(a:b)}(r)
$$
in other words
\begin{equation}\label{functiondiab}
\left\langle F(a:b) \right\rangle_{\omega} = \int_{\mathbb R^2} F(s,t) \, d\mu_{\omega, a:b}(s,t)
\end{equation}
\begin{remark}\upshape
As we will see later, it could happen that for each $\omega \in \mathfrak S_{a:b}$ we have
$$ \left\langle F(a:b) \right\rangle_{\omega} = \left\langle b \right\rangle_{\omega}$$
but this does not mean that the observable $F(a:b)$ coincides with $b$, but only that $F(a:b) \subset b$, since $\mathfrak S_{F(a:b)} \subset \mathfrak S_b$\footnote{If $a:b$ are strongly simultaneously preparable, then we have that the set $\mathfrak S_{a:b}$ coincides with $\mathfrak S_b$; it follows that in this case $F(a:b) = b$. }.
\end{remark}
Using the same tools as in Theorem \ref{smc} and Theorem \ref{smb}, we prove that
\begin{theorem}\upshape
For every $F \in C_b(\mathbb R^2)$ we have:
$$\sigma(F(a:b)) = F(\sigma(a:b))$$
while for every Borel function $F: \mathbb R^2 \rightarrow \mathbb R$
$$\sigma(F(a:b)) \subset \overline{F(\sigma(a:b))}$$
\end{theorem}

Also in this case, using similar reasoning to that of Proposition \ref{spettrolambda} on page \pageref{spettrolambda}, we obtain:  
\begin{equation}
 [ \ \exists \ \omega \in \mathfrak S_{a:b} \text{ such that } \mu_{\omega, a:b}(\{\lambda_1, \lambda_2\}) \neq 0 \ ] \quad \Longrightarrow \quad (\lambda_1, \lambda_2) \in \sigma(a:b)
\end{equation}
Recall that if $a$ and $b$ are compatible, then from \eqref{comp} we obtain:
\begin{equation}
\mu_{\omega,a}(\Delta_1) = \mu_{\omega, a:b}(\Delta_1 \times \mathbb R) \qquad , \qquad \mu_{\omega,b}(\Delta_2) = \mu_{\omega, a:b}(\mathbb R \times \Delta_2)
\end{equation}
 
Let us now see what connection exists between the joint spectrum of two compatible observables and their respective spectra.
\begin{proposition}\upshape\label{spettrocongiunto1}
If $a$ and $b$ are compatible observables, then we obtain
$$\sigma(a:b) \subset \sigma(a) \times \sigma(b)$$
\end{proposition}
\begin{proof}
Let $(\lambda_1, \lambda_2) \in \rho(a) \times \rho(b)$. By definition, there exist open neighbourhoods $U_{\lambda_1}$ and $U_{\lambda_2}$ such that 
$$\mu_{\omega,a}(U_{\lambda_1}) = \mu_{\omega,b}(U_{\lambda_2}) = 0$$
From compatibility we obtain:
\begin{eqnarray*}
0 & = & \mu_{\omega,a}(U_{\lambda_1}) = \mu_{\omega, a:b}(U_{\lambda_1} \times \mathbb R) \geq
\\
& \geq & \mu_{\omega, a:b}(U_{\lambda_1} \times U_{\lambda_2}) \ \Longrightarrow \ \mu_{\omega, a:b}(U_{\lambda_1} \times U_{\lambda_2}) = 0
\end{eqnarray*}
Therefore 
$$\rho(a) \times \mathbb R \subset \rho(a:b) \ \Longrightarrow \ \sigma(a:b) \subset \mathbb R^2 \setminus (\rho(a) \times \mathbb R) = \sigma(a) \times \mathbb R $$
and the same reasoning applied to $\lambda_2$ yields   
$$ \sigma(a:b) \subset \mathbb R^2 \setminus (\mathbb R \times \rho(b)) = \mathbb R \times \sigma(b) $$
it follows that
$$ \sigma(a:b) \subset \sigma(a) \times \mathbb R \cap \mathbb R \times \sigma(b) \subset \sigma(a) \times \sigma(b)$$
\end{proof}   

One might think that if the two observables are also independent as well as compatible, then we have 
\begin{equation}\label{jointspectrumindip}
\sigma(a:b) = \sigma(a) \times \sigma(b)
\end{equation}
In practice, by Proposition \ref{spettrolambda} on page \pageref{spettrolambda} it should be proved that given $\lambda_1 \in \sigma(a)$ and $\lambda_2 \in \sigma(b)$ there exists a state $\omega \in \mathfrak S_{a:b}$ such that    
$$ \mu_{\omega, a:b} ( \left\{ \lambda_1 \right\} \times \left\{ \lambda_2 \right\}) \neq 0$$ 
and from the independence property we can write
$$ \mu_{\omega, a:b} ( \left\{ \lambda_1 \right\} \times \left\{ \lambda_2 \right\}) = \mu_{\omega, a}(\left\{ \lambda_1 \right\}) \cdot \mu_{\omega, b}(\left\{ \lambda_2 \right\}) $$
but reapplying Proposition \ref{spettrolambda}, we can only say that there exists $\omega_1 \in \mathfrak S_a$ and $\omega_2 \in \mathfrak S_b$ such that
$$\mu_{\omega_1,a}(\left\{ \lambda_1 \right\}) \neq 0 \qquad , \qquad \mu_{\omega_2,b}(\left\{ \lambda_2 \right\}) \neq 0$$
therefore we  cannot  say that  \eqref{jointspectrumindip}  holds.
\begin{remark}\upshape\label{compconsestesso}
The observable $a:a$ is defined only formally\footnote{Formally, because the simultaneous measurement of $a$ and $a$ itself (we are not making a repeated measurement of $a$) does not, of course, make physical sense.} through the expression
$$P(a:a \in \Delta_1 \times \Delta_2, \tau)_\omega = P(a \in \Delta_1 \cap \Delta_2, \tau)_\omega$$
in other words
$$\mu_{\omega, a:a}(\Delta_1 \times \Delta_2) = \mu_{\omega, a}(\Delta_1 \cap \Delta_2)$$
and so
 $$\sigma(a:a) = \sigma(a) \times \sigma(a)$$
\end{remark}
\section{Functional Calculus and Compatibility}\label{calcolo_funz_comp}
Let us return to the study of the main properties of the functional calculus.
\begin{axiom}\label{compat_calc_func1}\index{Axiom-Functional Calculus III}
Given a pair of observables $a, b$ jointly preparable in the order $a:b$, for each pair of bounded Borel functions $f, g: \mathbb R \rightarrow \mathbb R$ we have:
\begin{itemize}
\item the observables $f(a)$ and $g(b)$ are jointly preparable in the same order:
$$ a:b \ \Longrightarrow \ f(a):g(b)$$ 
\item they have the same joint states:
$$ \mathfrak S_{a:b} = \mathfrak S_{f(a):g(b)} $$
\end{itemize} 
\end{axiom}
Let us see how the following probability is calculated
$$ P(f(a):g(b) \in \Delta_0 \times \Delta_1)_\omega = P(f(a) \in \Delta_0 : g(b) \in \Delta_1)_\omega$$
and to do so we prove the following
\begin{lemma}\upshape\label{prodottodifunzioni}
Let $a, b$ be two compatible observables of the system. For each pair of bounded Borel functions $f, g: \mathbb R \rightarrow \mathbb R$, we have:
$$\mu_{\omega, f(a):g(b)} = \mu_{\omega, a:b}^V $$
where $V: \mathbb R^2 \rightarrow \mathbb R^2$ is the function defined by
$$V(s,t) = (f(s), g(t)) \ , \qquad \forall s, t \in \mathbb R$$
\end{lemma}
\begin{proof}
For every $\Delta_0, \Delta_1 \in B(\mathbb R)$ and $\omega \in \mathfrak S_{a:b}$ we obtain
$$P(f(a) \in \Delta_0 : g(b) \in \Delta_1)_\omega = \mu_{\omega, f(a):g(b)}(\Delta_0 \times \Delta_1)$$
and by definition
\begin{eqnarray*}
P(f(a) \in \Delta_0 : g(b) \in \Delta_1)_\omega & = & \mu_{\omega, a:b}(f^{-1}(\Delta_0) \times g^{-1}(\Delta_1)) =
\\
& = & \mu_{\omega, a:b}(V^{-1}(\Delta_0 \times \Delta_1)) 
\end{eqnarray*}
\end{proof}
It follows that for every bounded Borel function $F: \mathbb R^2 \rightarrow \mathbb R$ we can write
\begin{equation}
\int F(s,t) \, d \mu_{\omega, f(a):g(b)}(s,t) = \int F(V(s,t)) \, d \mu_{\omega, a:b}(s,t)
\end{equation}
\\
We give a simple consequence of the previous results concerning the centre of an observable:
\begin{corollary}\upshape\label{compat_calc_func2}
Let $b \in \mathcal C(a)$ . For every bounded Borel function $f: \mathbb R \rightarrow \mathbb R$, the observable $f(b)$ is still in $\mathcal C(a)$.
\end{corollary}
\begin{proof}
From the previous axiom we obtain
$$ \mathfrak S_{a:b} = \mathfrak S_{a:f(b)} \qquad , \qquad \mathfrak S_{b:a} = \mathfrak S_{f(b):a} $$ 
The proof follows trivially from Lemma \ref{prodottodifunzioni}.
\end{proof}
Now inevitably we need to be able to answer the following question:
\begin{problem}\upshape
Given two observables $a$ and $b$ simultaneously preparable in the order $a:b$, what relation exists between these observables and the observable of the laboratory system obtained through the Borel functional calculus $F(a:b)$? 
\end{problem}
Let's try to give a satisfactory answer to this question.
\\
Let $\omega \in \mathfrak S_{a:b}$. Given two Borel sets $\Delta_0, \Delta_1 \subset \mathbb R$, through the simultaneous measurements of $a$ and $b$ we can determine when the values of $b$ are in $\Delta_0 \subset \mathbb R$ and when their simultaneous values are in $F^{-1}(\Delta) \subset \mathbb R^2$, i.e.,
$$ a:b \in F^{-1}(\Delta) \qquad \Longleftrightarrow \qquad F(a:b) \in \Delta \subset \mathbb R $$   
and therefore in the state $\omega \in \mathfrak S_{a:b}$ we can establish both statistics:
$$ P( b \in \Delta_0)_\omega \qquad \text{and} \qquad P( F(a:b) \in \Delta)_\omega$$
in this way we can say that the observable $b$ is simultaneously preparable with the observable $F(a:b)$ in both orders:
$$ b : F(a:b) \qquad , \qquad F(a:b) : b$$
with
\begin{equation}\label{stati_e_funzioni}
 \mathfrak S_{b : F(a:b)} = \mathfrak S_{a:b} = \mathfrak S_{F(a:b) : b}
\end{equation}
Furthermore we can write the following equation:
\begin{eqnarray*}
P \left( b \in \Delta_0 : F(a:b) \in \Delta \right)_\omega & = & P \left( (a:b) \in \mathbb R \times \Delta_0 : (a:b) \in F^{-1}(\Delta) \right)_\omega =
\\
& = & P \left( (a:b) \in \left( \mathbb R \times \Delta_0 \right) \cap F^{-1}(\Delta) \right)_\omega =
\\
& = & \mu_{\omega, a:b} \left( \left( \mathbb R \times \Delta_0 \right) \cap F^{-1}(\Delta) \right)
\end{eqnarray*}

\subsection{Several Observables: $n$-Dimensional Case}\label{calcolo_funz_n_dim}
Even for a family of compatible observables $\left\{ a_1, a_2, \ldots, a_n \right\}$ of the system we can extend the notions given for two observables; in this way we can define their joint spectrum in a similar way:
$$\sigma(a_1 : a_2 : \cdots : a_n) \subset \mathbb R^n$$
and for every Borel function $F: \mathbb R^n \rightarrow \mathbb R$ the observable $F(a_1 : a_2 : \cdots : a_n)$ is defined by 
$$\left\langle F(a_1 : a_2 : \cdots : a_n) \right\rangle_{\omega} = \int F(t) \, d \mu_{\omega, a_1 : a_2 : \cdots : a_n}(t)$$
where $\mu_{\omega, a_1 : a_2 : \cdots : a_n}$ is the measure established by 
$$P(a_1 \in \Delta_1 : a_2 \in \Delta_2 : \cdots : a_n \in \Delta_n)_\omega = \mu_{\omega, a_1 : a_2 : \cdots : a_n}( \Delta_1 \times \Delta_2 \times \cdots \times \Delta_n)$$
Furthermore, as in the two-dimensional case we obtain a result contained in Proposition \ref{spettrolambda} of section \ref{spettro_oss}:
\\
We have that $(\lambda_1, \lambda_2, \ldots, \lambda_n) \in \sigma(a_1 : a_2 : \cdots : a_n)$ if and only if there exists a state $\omega \in \mathfrak S_{a_1 : a_2 : \dots : a_n}$ such that
$$\mu_{\omega, a_1 : a_2 : \cdots : a_n} \left( \left\{ \lambda_1 \right\} \times \left\{ \lambda_2 \right\} \times \cdots \times \left\{ \lambda_n \right\} \right) \neq 0$$ 
\\
and from this result we derive that
$$\sigma(a_1 : a_2 : \cdots : a_n) \subset \sigma(a_1) \times \sigma(a_2) \times \cdots \times \sigma(a_n) \subset \mathbb R^n $$
Let's make some considerations that will be important to verify the associativity of the product of compatible observables given in equation \eqref{prodotto jordan}.
\\
We examine the case of three observables $\left\{ a, b, c \right\}$ simultaneously preparable in the order $a : b : c$ and let $G: \mathbb R^2 \rightarrow \mathbb R$ be a bounded Borel function. For each state $\omega \in \mathfrak S_{a:b:c}$ we have
$$ G(a:b) \in \Delta \subset \mathbb R \qquad \Longleftrightarrow \qquad a:b \in G^{-1}(\Delta) \subset \mathbb R^2$$
With the same motivations that led to the establishment of the set relation given in \eqref{stati_e_funzioni}, we can establish the simultaneous values of the three observables $a:b:c$ and therefore also of the observable $G(a:b)$ and determine the statistical law:
\begin{equation}\label{calcolo}
P(G(a:b) \in \Delta_o : c \in \Delta)_\omega = \mu_{\omega, G(a:b):c} (\Delta_o \times \Delta) \qquad \forall \Delta_o, \Delta \in B(\mathbb R)
\end{equation}
and we necessarily need to introduce a new model axiom establishing the relation between observables and their functions:
\begin{axiom}\label{compat_calc_func1c}\index{Axiom-Functional Calculus IV}
Let $a, b, c$ be three simultaneously preparable observables in the order $a:b:c$. We have
\begin{itemize}
\item[1.] $a:b:c \ \Longrightarrow \ G(a:b):c$
\item[2.] $\mathfrak S_{G(a:b):c} = \mathfrak S_{a:b:c}$
\end{itemize}
\end{axiom}
Let's go back to calculating the probability given in equation \eqref{calcolo}; it is equivalent to calculating the following probability
$$ P( a:b:c \in G^{-1}(\Delta_o) \times \mathbb R \ \wedge \ a:b:c \in \mathbb R \times \mathbb R \times \Delta )_\omega $$
which obviously coincides with
$$ P( a:b:c \in G^{-1}(\Delta_o) \times \Delta )_\omega $$
therefore
\begin{eqnarray}\label{perlassociativa}
P(G(a:b) \in \Delta_o : c \in \Delta)_\omega = \mu_{\omega, a:b:c} ( G^{-1}(\Delta_o) \times \Delta) 
\end{eqnarray}
Furthermore we have the following  
\begin{proposition}\upshape
For every pair of bounded Borel functions $F, G: \mathbb R^2 \rightarrow \mathbb R$ we have:
\begin{equation}
\label{perlassociativa1}
\int F(r,s) \, d \mu_{\omega, G(a:b):c}(r,s) = \int F(G(t_1, t_2), r) \, d \mu_{\omega, a:b:c}(t_1, t_2, r)
\end{equation}
\end{proposition}
\begin{proof}
The proposition is verified by using relation \eqref{perlassociativa} and noting that 
$$\mu_{\omega, a:b:c} ( G^{-1}(\Delta_o) \times \Delta ) = \mu_{\omega, a:b:c} ( \hat{G}^{-1}(\Delta_o \times \Delta) ) \qquad \forall \Delta_o, \Delta \in B(\mathbb R) $$
where $\hat{G}: \mathbb R^3 \rightarrow \mathbb R^2$ is a bounded Borel function defined as follows:
\begin{equation}
\label{perlassociativa2}
\hat{G}(t_1, t_2, r) = (G(t_1, t_2), r) \ , \qquad \forall t_1, t_2, r \in \mathbb R
\end{equation}
In other words, we have proved that
$$ \mu_{\omega, G(a:b):c} = \mu_{\omega, a:b:c}^{\hat{G}} $$
and from this we obtain \eqref{perlassociativa1}.
\end{proof}
If $a, b, c$ are compatible observables, then with the same considerations, it occurs that                   
$$ \mu_{\omega, a:G(b:c)} = \mu_{\omega, a:b:c}^{\check{G}} $$
where this time $\check{G}: \mathbb R^3 \rightarrow \mathbb R^2$ is the function
$$\check{G}(t, r_1, r_2) = (t, G(r_1, r_2)) \ , \qquad \forall t, r_1, r_2 \in \mathbb R$$

\subsubsection{The Spectral Property of States and Compatibility}
Let's try to establish the \texttt{SPS} property of Axiom \ref{SPS} in the case of a family of compatible observables:
\begin{axiom}[\texttt{SPS2}]\label{PSS2}\index{Axiom-Spectral Property of States II}\index{\texttt{SPS2}}
Given a finite set of compatible observables $\left\{ a_1, a_2, \ldots, a_n \right\}$ belonging to $\mathfrak X$, for every $s_i \in \left[ \underline{\sigma_i}, \overline{\sigma_i} \right]$, $i = 1, 2, \ldots, n$, where:
$$ \underline{\sigma_i} = \inf \sigma(a_i) \qquad , \qquad \overline{\sigma_i} = \sup \sigma(a_i) $$
there exists a state $\omega_*$ belonging to $\mathfrak S_{a_1 : a_2 : \ldots : a_n}$ such that 
$$\left\langle a_i \right\rangle_{\omega_*} = s_i \ , \qquad \forall i = 1, 2, \ldots, n$$
\end{axiom}
In other words, the family of compatible observables $\left\{ a_1, a_2, \ldots, a_n \right\}$, for a set of values $\left\{ s_1, s_2, \ldots, s_n \right\}$, admits a common eigenstate given by $\omega_*$.
\section{Conditioning of Simultaneous Measurements}\label{condizio_misure_simult}
In this section we give some measurement theory properties found in \cite{Cinlar, Nielsen} applied to our Borel probability measures $\Pi(\mathbb R^n)$.
\\
Recall that we have the following decomposition:
$$ B(\mathbb R^n) = B(\mathbb R^m) \otimes B(\mathbb R^{n-m}) \ , \qquad m \leq n $$  
where $\otimes$ denotes the generated $\sigma$-algebra.
\begin{definition}\upshape
A real Markov kernel\footnote{Another definition is that of \textit{transition probability kernel}.} is a family $\left\{ P_s \right\}_{s \in \mathbb R^m}$ of probability measures of $\Pi(\mathbb R^{n-m})$ such that for every $\Delta \in B(\mathbb R^{n-m})$ the map
$$ s \in \mathbb R^m \longrightarrow P_s(\Delta) \in [0, 1] $$ 
is Borel-measurable.
\end{definition}

It is not difficult to verify that given a real Markov kernel 
\begin{equation}
 s \in \mathbb R^m \longrightarrow P_s \in \Pi(\mathbb R^{n-m})
\end{equation}
for each measure $\mu \in \Pi(\mathbb R^m)$ there is a unique measure $\pi \in \Pi(\mathbb R^n)$ defined as follows:
$$ \pi(\Delta_0 \times \Delta_1) = \int_{\Delta_0} P_s(\Delta_1) \, d \mu(s) \ , \qquad \forall \Delta_0 \in B(\mathbb R^m) \ , \ \Delta_1 \in B(\mathbb R^{n-m})$$
The converse of this statement is given by the following proposition:
\begin{proposition}\upshape\label{markovkernel0}
Let $n > m$. For every measure $\pi \in \Pi(\mathbb R^n)$ there exists a measure $\mu \in \Pi(\mathbb R^m)$ and a real Markov kernel $\left\{ P_s \right\}_{s \in \mathbb R^m}$ of probability measures of $\Pi(\mathbb R^{n-m})$ such that for each set $\Delta_0 \in B(\mathbb R^m)$ and $\Delta_1 \in B(\mathbb R^{n-m})$ we have
$$ \pi(\Delta_0 \times \Delta_1) = \int_{\Delta_0} P_s(\Delta_1) \, d \mu(s) $$
\end{proposition}
\begin{proof}
See \cite{Cinlar} Theorem 2.18, page 154.
\end{proof}
We consider two compatible observables $a$ and $b$ with $\omega\in\mathfrak S_{a:b}$, and the measure $\mu_{\omega, a:b}$ defined in \eqref{misuracongiuntasimultanea} and let $\left\{ P_s^{a:b} \right\}_{s \in \mathbb R}$ and $\mu \in \Pi(\mathbb R)$ be, respectively, the real Markov kernel and the Borel measure established by the previous proposition:
$$\mu_{\omega, a:b}(\Delta_0 \times \Delta_1) = \int_{\Delta_0} P_s^{a:b}(\Delta_1) \, d \mu(s) \ , \qquad \Delta_0, \Delta_1 \in B(\mathbb R)$$
From the compatibility properties we obtain:
\begin{equation}\label{int_a}
\mu_{\omega, a}(\Delta_0) = \mu_{\omega, a:b}(\Delta_0 \times \mathbb R) = \int_{\Delta_0} P_s^{a:b}(\mathbb R) \, d \mu(s) = \mu(\Delta_0)
\end{equation}
it follows that $\mu = \mu_{\omega, a}$ and therefore
\begin{equation}\label{a1}
\mu_{\omega, a:b}(\Delta_0 \times \Delta_1) = \int_{\Delta_0} P_s^{a:b}(\Delta_1) \, d \mu_{\omega, a}(s) \ , \qquad \Delta_0, \Delta_1 \in B(\mathbb R)
\end{equation}
Adopting the same reasoning we obtain the following link between the two measurements relating to the two observables:
\begin{equation}\label{b}
\mu_{\omega, b}(\Delta_1) = \int P_s^{a:b}(\Delta_1) \, d \mu_{\omega, a}(s)
\end{equation}
As is well known, for each function $f \in L^1(\mu_{\omega, b})$ it is possible to approximate in the $\left\| \cdot \right\|_1$-norm by summable simple functions\footnote{See Folland \cite{Folland} Proposition 6.7}, so we have a linear map
\begin{equation}\label{applicazioneP}
P^{a:b}: L^1(\mu_{\omega, b}) \longrightarrow L^1(\mu_{\omega, a})
\end{equation}
such that 
\begin{equation}
\label{c0}
\mu_{\omega, b}(f) = \int P_s^{a:b}(f) \, d \mu_{\omega, a}(s) \ , \qquad f \in L^1(\mu_{\omega, b})
\end{equation}
and for each $s \in \mathbb R$ it turns out that 
\begin{equation}\label{applicazione_P2}
P_s^{a:b}(f) = \int f(t) \, d P_s^{a:b}(t) 
\end{equation}
We note that for each $\Delta \in B(\mathbb R)$ we have
$$P_s^{a:b}(\mathbf{1}_{\Delta}) = P_s^{a:b}(\Delta)$$
with
$$ 0 \leq P_s^{a:b}(\Delta) \leq 1 \ , \qquad s \in \mathbb R $$
and by \eqref{c0} it is easy to prove
$$\operatorname{Supp} P_s^{a:b} = \operatorname{Supp} \mu_{\omega, a:b} \qquad \mu_{\omega,a}\text{-}a.e. $$
The map \eqref{applicazioneP} is a positive map:
$$ f \geq 0 \qquad \Longrightarrow \qquad P^{a:b}(f) \geq 0 $$
because by \eqref{applicazione_P2} we have 
$$P_s^{a:b}(f) \geq 0 \ , \qquad \forall s \in \mathbb R $$
Furthermore, for each positive function $f \in L^1(\mu_{\omega, b})$ we have
$$\| P^{a:b}(f) \|_1 = \int |P_s^{a:b}(f)| \, d \mu_{\omega,a}(s) = \mu_{\omega,b}(f) \leq \| f \|_1$$
and from this follows the continuity, in the norm topology, of the map \eqref{applicazioneP}.
\\
Since from \eqref{int_a} we obtain:
$$\int (1 - P_s^{a:b}(\sigma(b))) \, d \mu_{\omega,a} = 0 $$ 
it follows that
\begin{equation}\label{applicazione_P3}
 P_s^{a:b}(\sigma(b)) = 1 \qquad \mu_{\omega,a}\text{-}a.e. 
\end{equation}
From \eqref{c0} we have\footnote{See also Proposition \ref{condizionamentocompatibile}.}: 
$$ \left\langle b \right\rangle_\omega = \mu_{\omega,b}(\theta_1) = \int P_s^{a:b}(\theta_1) \, d \mu_{\omega,a}(s) $$ 
where for every natural number $n$ we have defined
$$\theta_n(t) = t^n \ , \qquad \forall t \in \mathbb R$$

\subsection{$N$-Dimensional Case}
Let's try to apply the same considerations to a finite family of compatible observables of the system $\left\{ a_1, a_2, \ldots, a_n \right\}$.
\\
For every $\Delta_0 \in B(\mathbb R^m)$ and $\Delta_1 \in B(\mathbb R^{n-m})$ we have:
$$ \mu_{a_1 : a_2 : \cdots : a_n} (\Delta_0 \times \Delta_1) = \int_{\Delta_0} P_s^{(a_1 : \cdots : a_m) : a_{m+1} : \cdots : a_n}(\Delta_1) \, d \mu_{\omega, a_1, a_2, \ldots, a_m}(s) $$
and also in this case, from the compatibility properties, for each $\Delta \in B(\mathbb R^{n-m})$ we obtain 
$$ \mu_{\omega, a_{m+1} : a_{m+2} : \cdots : a_n} (\Delta) = \int P_s^{(a_1 : \cdots : a_m) : a_{m+1} : \cdots : a_n}(\Delta) \, d \mu_{\omega, a_1 : a_2 : \cdots : a_m}(s) $$
and obtain a continuous linear map:
$$P_s^{(a_1 : \cdots : a_m) : a_{m+1} : \cdots : a_n} : L^1( \mu_{\omega, a_{m+1} : a_{m+2} : \cdots : a_n}) \longrightarrow L^1(\mu_{\omega, a_1 : a_2 : \cdots : a_m})$$
such that for every $F \in L^1( \mu_{\omega, a_{m+1} : a_{m+2} : \cdots : a_n})$ it turns out
$$ P_s^{(a_1 : \cdots : a_m) : a_{m+1} : \cdots : a_n}(F) = \int_{\mathbb R^{n-m}} F(t) \, d P_s^{a_1 : a_2 : \cdots : a_n}(t)$$ 
and
$$ \mu_{\omega, a_{m+1} : a_{m+2} : \cdots : a_n} (F) = \int P_s^{(a_1 : \cdots : a_m) : a_{m+1} : \cdots : a_n}(F) \, d \mu_{\omega, a_1 : a_2 : \cdots : a_m}(s) $$
\section{Functional Calculus and Compatibility - II Step}\label{funzioneosservabile2}
Given two observables $a$ and $b$ \textbf{compatible}, let's ask ourselves what the observables $f(a:b)$ defined by relation \eqref{functiondiab} are.
\\
To answer this question we must recall some well-known results of integration theory\footnote{See \cite{Nielsen}.}.
\begin{proposition}\upshape\label{markovkernel}
Let $\pi \in \Pi(\mathbb R^2)$ be the integration of the real Markov kernel $\left\{ P_s \right\}_{s \in \mathbb R}$ with respect to $\mu \in \Pi(\mathbb R)$\footnote{ i.e., the measure established by Proposition \ref{markovkernel0}.}.
\\
For every Borel function $f: \mathbb R^2 \rightarrow \mathbb R$ we have:
\begin{itemize}
\item [1.] that the function
$$ s \in \mathbb R \longrightarrow \int f(s,t) \, dP_s(t) $$   
is Borel-measurable.
\item [2.] (Extension of Tonelli's theorem): 
$$\int f(s,t) \, d \pi(s,t) = \int \left( \int f(s,t) \, d P_s(t) \right) d \mu(s) $$
\item [3.] (Extension of Fubini's theorem):
\\
If we denote
$$\Delta_f = \left\{ s \in \mathbb R : \int |f(s,t)| \, dP_s(t) < \infty \right\}$$
then we obtain 
\item [a.] $\mu(\Delta_f) = 1$
\item [b.] the function
$$ s \in \mathbb R \longrightarrow \left( \int f(s,t) \, dP_s(t) \right) \mathbf{1}_{\Delta_f}(s) $$ 
is Borel-measurable and $\mu$-summable;
\item [c.] and 
$$\int f(s,t) \, d \pi(s,t) = \int_{\Delta_f} \left( \int f(s,t) \, d P_s(t) \right) d \mu(s) $$
\end{itemize}
\end{proposition}
Let us now apply these results to our measure $\mu_{\omega, a:b}$, where $a$ and $b$ are compatible and $\omega \in \mathfrak S_{a:b}$\footnote{Obviously, by the compatibility of the observables, this is equal to $\mu_{\omega, b:a}$.} and let $\left\{ P_s \right\}_{s \in \mathbb R}$ be its Markov kernel with respect to $\mu_{\omega,a}$.
\\
For every $m, n \in \mathbb N$ we denote:
\begin{equation}\label{theta_func}
 \Theta_{m,n}(s,t) = s^m t^n \cdot \mathbf{1}_{\sigma(a) \times \sigma(b)}(s,t)
\end{equation} 
therefore in this way it turns out that
$$ \Delta_{\Theta_{m,n}} = \mathbb R $$

Let us now study the various types of functions given by \eqref{theta_func}.
\\

$\bullet$ - \textbf{First Step}
\\
Here we consider the function:
$$ \Theta_{m,0}(s,t) = s^m \cdot \mathbf{1}_{\sigma(a) \times \sigma(b)}(s,t)$$
Then
$$\int \Theta_{m,0}(s,t) \, d \mu_{\omega,a:b}(s,t) = \int \left( \int s^m \cdot \mathbf{1}_{\sigma(a) \times \sigma(b)}(s,t) \, d P_s(t) \right) d \mu_{\omega,a}(s) $$
and
\begin{eqnarray*}
 \int s^m \cdot \mathbf{1}_{\sigma(a) \times \sigma(b)}(s,t) \, d P_s(t) & = & s^m \cdot \int \mathbf{1}_{\sigma(a) \times \sigma(b)}(s,t) \, d P_s(t) \\
& = & s^m \cdot \int \mathbf{1}_{\sigma(b)}(t) \, d P_s(t) \\
& = & s^m \cdot P_s(\sigma(b)) = s^m
\end{eqnarray*}
It follows that
$$\int \Theta_{m,0}(s,t) \, d \mu_{\omega,a:b}(s,t) = \int s^m \, d \mu_{\omega,a}(s) = \left\langle a^m \right\rangle_\omega$$
So in this case we can write
$$ \left\langle \Theta_{m,0}(a:b) \right\rangle_\omega = \left\langle a^m \right\rangle_\omega \ , \qquad \forall \omega \in \mathfrak S_a $$
therefore 
$$ \Theta_{m,0}(a:b) = a^m $$
$\bullet$ - \textbf{Second Step}
 \\
Let's now consider the function:
$$ \Theta_{0,n}(s,t) = t^n \cdot \mathbf{1}_{\sigma(a) \times \sigma(b)}(s,t)$$
then
$$\int \Theta_{0,n}(s,t) \, d \mu_{\omega,a:b}(s,t) = \int \left( \int t^n \cdot \mathbf{1}_{\sigma(a) \times \sigma(b)}(s,t) \, d P_s(t) \right) d \mu_{\omega,a}(s) $$
from \eqref{applicazione_P2} we have
\begin{eqnarray*}
 \int t^n \cdot \mathbf{1}_{\sigma(a) \times \sigma(b)}(s,t) \, d P_s(t) & = & \left( \int t^n \cdot \mathbf{1}_{\sigma(b)}(t) \, d P_s(t) \right) \cdot \mathbf{1}_{\sigma(a)}(s) \\
& = & P_s(\theta_n) \cdot \mathbf{1}_{\sigma(a)}(s)  
\end{eqnarray*}
where
$$\theta_n(t) = t^n \cdot \mathbf{1}_{\sigma(b)}(t) $$
therefore
\begin{eqnarray*}
\int \Theta_{0,n}(s,t) \, d \mu_{\omega,a:b}(s,t) = \int P_s(\theta_n) \, d \mu_{\omega,a}(s) = \mu_{\omega,b}(\theta_n) = \left\langle b^n \right\rangle_\omega 
\end{eqnarray*}
So even in this case we can write 
$$ \Theta_{0,n}(a:b) = b^n $$
$\bullet$ - \textbf{Third Step}
\\
Let's consider the function:
$$ v(s,t) = (s + t) \ \mathbf{1}_{\sigma(a) \times \sigma(b)}(s,t)$$
Also in this case we obtain with a simple calculation that 
$$\Delta_v = \mathbb R$$
Therefore
$$\int v(s,t) \, d \mu_{\omega,a:b}(s,t) = \int \left( \int (s + t) \ \mathbf{1}_{\sigma(a) \times \sigma(b)}(s,t) \, d P_s(t) \right) d \mu_{\omega,a}(s) $$
with
\begin{eqnarray*}
\int (s + t) \ \mathbf{1}_{\sigma(a) \times \sigma(b)}(s,t) \, d P_s(t) = s P_s(\sigma(b)) + P_s(\theta_1) \cdot \mathbf{1}_{\sigma(a)}(s) 
\end{eqnarray*}
it follows that
\begin{eqnarray*}
\int v(s,t) \, d \mu_{\omega,a:b}(s,t) & = & \int \left[ s P_s(\sigma(b)) + P_s(\theta_1) \cdot \mathbf{1}_{\sigma(a)}(s) \right] d \mu_{\omega,a}(s) \\
& = & \int s \, d \mu_{\omega,a}(s) + \int P_s(\theta_1) \cdot \mathbf{1}_{\sigma(a)}(s) \, d \mu_{\omega,a}(s) \\
& = & \left\langle a \right\rangle_\omega + \mu_{\omega,b}(\theta_1) \\
& = & \left\langle a \right\rangle_\omega + \left\langle b \right\rangle_\omega 
\end{eqnarray*}
So we can write 
$$ 
 \left\langle v(a:b) \right\rangle_\omega = \left\langle a \right\rangle_\omega + \left\langle b \right\rangle_\omega $$
then 
$$ v(a:b) = a + b $$
In other words
$$\mu_{\omega, a+b} (\Delta) = \mu_{\omega, a:b} (v^{-1}(\Delta)) \ , \qquad \Delta \in B(\mathbb R) $$
therefore this measure is nothing more than the convolution of the two measures\footnote{Recall that for every Borel set $\Delta$ of $\mathbb R$ we have:
$$ (\mu_{\omega,a} \ast \mu_{\omega,b})(\Delta) = \int_{\mathbb R^2} \mathbf{1}_{\Delta}(s + t) \, d \mu_{\omega,a}(s) \, d \mu_{\omega,b}(t)$$
}
\begin{eqnarray}\label{sum_conv}
 \mu_{\omega, a+b} = \mu_{\omega,a} \ast \mu_{\omega,b}
\end{eqnarray}
$\bullet$ - \textbf{Fourth Step}
\\
We note that we have the following identity:
 
$$ \Theta_{1,1}(s,t) = \frac{1}{2} \left[ (s + t)^2 - s^2 - t^2 \right] \cdot \mathbf{1}_{\sigma(a) \times \sigma(b)}(s,t) $$
and using the previous results it is easy to prove that
$$ \Theta_{1,1}(a:b) = a \cdot b$$
 
$\bullet$ - \textbf{Fifth Step} [The Product of Powers] 
\\
We set $V_{m,n}: \mathbb R^2 \rightarrow \mathbb R^2$ with $V_{m,n}(s,t) = (s^m, t^n)$ . We have 
$$ \Theta_{m,n}(s,t) = \Theta_{1,1}(V_{m,n}(s,t))$$  
and from Proposition \ref{prodottodifunzioni} it turns out that
$$ \mu_{\omega, a^m : b^n} = \mu_{\omega, a:b}^{V_{m,n}}$$
therefore
\begin{eqnarray*}
\left\langle \Theta_{m,n}(a:b) \right\rangle_{\omega} & = & \int \Theta_{m,n}(s,t) \, d \mu_{\omega, a:b}(s,t) \\
& = & \int \Theta_{1,1}(V_{m,n}(s,t)) \, d \mu_{\omega, a:b}(s,t) \\
& = & \int \Theta_{1,1}(r, t) \, d \mu_{\omega, a:b}^{V_{m,n}}(r, t) \\
& = & \int \Theta_{1,1}(r, t) \, d \mu_{\omega, a^m : b^n}(r, t) \\
& = & \left\langle \Theta_{1,1}(a^m : b^n) \right\rangle_{\omega} = \left\langle a^m \cdot b^n \right\rangle_{\omega} 
\end{eqnarray*}
It follows that 
$$\Theta_{m,n}(a:b) = a^m \cdot b^n $$
$\bullet$ - \textbf{Sixth Step}
\\
We assert that the relation 
\begin{equation}\label{sestocaso}
\mu_{\omega, a:b} (\Delta_1 \times \Delta_2) = \left\langle \mathbf{1}_{\Delta_1}(a) \cdot \mathbf{1}_{\Delta_2}(b) \right\rangle_{\omega}
\end{equation}
holds.
\\
By definition it turns out
\begin{eqnarray*}
P(a \in \Delta_1 : b \in \Delta_2)_\omega & = & \mu_{\omega, a:b} (\Delta_1 \times \Delta_2) \\
& = & \int \mathbf{1}_{\Delta_1 \times \Delta_2}(s,t) \, d \mu_{\omega, a:b}(s,t) \\
& = & \left\langle \mathbf{1}_{\Delta_1 \times \Delta_2}(a:b) \right\rangle_{\omega}
\end{eqnarray*}
and to verify equation \eqref{sestocaso} we use Proposition \ref{prodottodifunzioni} again, since
$$\mathbf{1}_{\Delta_1 \times \Delta_2}(s,t) = \Theta_{1,1}(V_{\Delta_1 \times \Delta_2}(s,t))$$
where
$$V_{\Delta_1 \times \Delta_2}(s,t) = ( \mathbf{1}_{\Delta_1}(s), \mathbf{1}_{\Delta_2}(t) )$$
it follows that:
\begin{eqnarray*}
& = & \int \Theta_{1,1}(s,t) \, d \mu_{\omega, \mathbf{1}_{\Delta_1}(a) : \mathbf{1}_{\Delta_2}(b)}(s,t) \\
& = & \int \Theta_{1,1}(V_{\Delta_1 \times \Delta_2}(s,t)) \, d \mu_{\omega, a:b}(s,t) \\
& = & \int \mathbf{1}_{\Delta_1}(s) \mathbf{1}_{\Delta_2}(t) \, d \mu_{\omega, a:b}(s,t) \\
& = & \int \mathbf{1}_{\Delta_1 \times \Delta_2}(s,t) \, d \mu_{\omega, a:b}(s,t)
\end{eqnarray*}
In summary, from the Sixth Step, for every $\omega \in \mathfrak S_{a:b}$ we have:
$$\left\langle \mathbf{1}_{\Delta_1 \times \Delta_2}(a:b) \right\rangle_{\omega} = \left\langle \mathbf{1}_{\Delta_1}(a) \cdot \mathbf{1}_{\Delta_2}(b) \right\rangle_{\omega} $$
therefore  
\begin{equation}
\mathbf{1}_{\Delta_1 \times \Delta_2}(a:b) = \mathbf{1}_{\Delta_1}(a) \cdot \mathbf{1}_{\Delta_2}(b) 
\end{equation}
for real polynomials, from the Fifth Step, we have  
\begin{equation}
p(a:b) = \sum_{i,j} r_{i,j} \, a^i b^j 
\end{equation}
where
$$p(s,t) = \sum_{i,j} r_{i,j} \, s^i t^j = \sum_{i,j} r_{i,j} \, \Theta_{i,j}(s,t) $$
and by the Stone–Weierstrass theorem, if
$$F(s,t) = f(s) g(t) \ , \qquad f, g \in C(\sigma(a:b))$$
then 
\begin{equation}
\label{prodottofunzioni}
F(a:b) = f(a) g(b)
\end{equation}

\section{Product of Compatible Observable }\label{associati}
 
We consider three observables $\left\{ a,b,c \right\}$ compatible with each other and verify that their Jordan product given by  \eqref{prodotto jordan} is associative and distributive.
\begin{itemize}
\item \textbf{Associative Property}
\end{itemize}
We need to prove the relation:
\begin{equation}\label{associaa}
a\cdot(b\cdot c)= (a\cdot b) \cdot c
\end{equation}
and for the functional calculus,  it is enough to prove  that for each $\omega \in\mathfrak S_{a:b:c}$ we have:
\begin{equation}\label{associab}
 \left\langle \Theta_{1,1}(a:\Theta_{1,1}(b:c) ) \right\rangle _{\omega }=  \left\langle \Theta_{1,1}( \Theta_{1,1}(a:b):c) \right\rangle _{\omega }
\end{equation}
By definition and by \eqref{perlassociativa1} we have
\begin{eqnarray*}
\left\langle \Theta_{1,1}(a:\Theta_{1,1}(b:c) ) \right\rangle _{\omega } &=& \int_{\mathbb R^2}  \Theta_{1,1}(r,s) d\mu_{\omega, a:\Theta_{1,1}(b:c)}(r,s)=
\\
& = &
\int_{\mathbb R^3}  \Theta_{1,1}(r,\Theta_{1,1}(t_1,t_2)) d\mu_{\omega, a:b:c}(r,t_1,t_2) 
\end{eqnarray*} 
while
\begin{eqnarray*}
 \left\langle \Theta_{1,1}( \Theta_{1,1}(a:b):c) \right\rangle _{\omega } &=& \int \Theta_{1,1}(r,s) d\mu_{\omega, \Theta_{1,1}(a:b):c}(r,s)=
\\
& = &
\int \Theta_{1,1}( \Theta_{1,1}(r_1,r_2),t) d\mu_{\omega, a:b:c}(r_1,r_2,t) 
\end{eqnarray*} 
obviously for every $r,s,t\in \mathbb R$ we have
$$\Theta_{1,1}( \Theta_{1,1}(r ,s ),t)=\Theta_{1,1}( r,\Theta_{1,1}(s,t))= r s t $$
therefore \eqref{associab} holds.
\\
We will see that \eqref{associaa} will have an important role in the algebraization of a physical system.
\begin{itemize}
\item \textbf{Distributive Property}  
\end{itemize}
Let us now prove the following relation: 
\begin{equation}
\label{distribuitia}
(a + b) \cdot c = a \cdot c + b \cdot c
\end{equation} 
For every $\omega \in \mathfrak S_{a:b:c}$ we can write
\begin{eqnarray*}
\left\langle (a + b) \cdot c \right\rangle_{\omega} & = & \left\langle \Theta_{1,1}(v(a:b) : c) \right\rangle_{\omega} \\
& = & \int_{\mathbb R^2} \Theta_{1,1}(r,s) \, d \mu_{\omega, v(a:b) : c}(r,s) \\
& = & \int_{\mathbb R^3} (r t + s t) \, d \mu_{\omega, a:b:c}(r,s,t) \\
& = & \int_{\mathbb R^3} r t \, d \mu_{\omega, a:b:c}(r,s,t) + \int_{\mathbb R^3} s t \, d \mu_{\omega, a:b:c}(r,s,t) 
\end{eqnarray*}
and from \eqref{prodottofunzioni} we obtain:
\begin{eqnarray*}
\left\langle a \cdot c \right\rangle_{\omega} = \int_{\mathbb R^3} r t \, d \mu_{\omega, a:b:c}(r,s,t) \qquad , \qquad \left\langle b \cdot c \right\rangle_{\omega} = \int_{\mathbb R^3} s t \, d \mu_{\omega, a:b:c}(r,s,t)
\end{eqnarray*}
 $$ \star \star \star $$
We apply the above considerations to the study of the observable of the type $p \cdot a$ where $p$ is a non-trivial question of the system, therefore with $\sigma(p) = \left\{ 0, 1 \right\}$ and which is compatible with the observable $a$.
 \\
Let us now study the associated Borel measure $\mu_{\omega, p \cdot a}$.
\\
By expression \eqref{a1}, we can write\footnote{We remark that the Markov kernel $\left\{ P^{p:a}_s \right\}_{s \in \mathbb R}$ obviously also depends on the state $\omega$.}: 
\begin{eqnarray*}
\mu_{\omega, p : a} (\Delta_0 \times \Delta_1) & = & \int_{\Delta_0} P^{p:a}_s(\Delta_1) \, d \mu_{\omega, p}(s) \\
& = & \texttt{r}_0 \delta_0(\Delta_0) P^{p:a}_0(\Delta_1) + \texttt{r}_1 \delta_1(\Delta_0) P^{p:a}_1(\Delta_1)    
\end{eqnarray*}
where 
$$ \texttt{r}_0 = \mu_{\omega, p} ( \left\{ 0 \right\} ) > 0 \qquad , \qquad \texttt{r}_1 = \mu_{\omega, p} ( \left\{ 1 \right\} ) > 0 \qquad , \qquad \texttt{r}_0 + \texttt{r}_1 = 1 $$  
with $\delta_0, \delta_1$ we have indicated the respective Dirac measures. 
\\
By definition, for every Borel set $\Delta$ of $\mathbb R$ we obtain:
\begin{eqnarray*}
\mu_{\omega, p \cdot a} (\Delta) & = & \int_{\mathbb R^2} \mathbf{1}_{\Delta}(r s) \, d \mu_{\omega, p:a}(r,s) \\
& = & \texttt{r}_0 \int_{\mathbb R^2} \mathbf{1}_{\Delta}(r s) \, d \delta_0(r) \, d P^{p:a}_0(s) \\
& & + \texttt{r}_1 \int_{\mathbb R^2} \mathbf{1}_{\Delta}(r s) \, d \delta_1(r) \, d P^{p:a}_1(s) \\
& = & \texttt{r}_0 \delta_0(\Delta) P^{p:a}_0(\mathbb R) + \texttt{r}_1 P^{p:a}_1(\Delta) \\
& = & \texttt{r}_0 \delta_0(\Delta) + \texttt{r}_1 P^{p:a}_1(\Delta)
\end{eqnarray*}
which is written in a concise way
\begin{equation}\label{misura-pa-1}
 \mu_{\omega, p \cdot a} = \texttt{r}_0 \, \delta_0 + \texttt{r}_1 \, P^{p:a}_1
\end{equation}
Let us study the link between $\mu_{\omega,a}$ and the Markov kernel $\left\{ P^{p:a}_s \right\}_{s \in \mathbb R}$.
\\
Using expression \eqref{b} we have:
$$\mu_{\omega,a} (\Delta_1) = \int P^{p:a}_s(\Delta_1) \, d \mu_{\omega,p}(s) \ , \qquad \forall \Delta_1 \in B(\mathbb R) $$ 
therefore
$$\mu_{\omega,a} (\Delta_1) = \texttt{r}_0 P^{p:a}_0(\Delta_1) + \texttt{r}_1 P^{p:a}_1(\Delta_1)$$
it follows that 
\begin{equation}
\label{misura-pa-2}
 \mu_{\omega, p \cdot a} (\Delta) = \texttt{r}_0 \left( \delta_0(\Delta) - P^{p:a}_0(\Delta) \right) + \mu_{\omega,a} (\Delta) 
\end{equation}

We summarize the considerations made so far in the following 
\begin{proposition}\upshape
If $p$ is a non-trivial question compatible with the observable $a$, then
\begin{equation}\label{spettro-pa}
 \left\{ 0 \right\} \subset \sigma(p \cdot a) \subset \left\{ 0 \right\} \cup \sigma(a)
\end{equation}
\end{proposition}
\begin{proof}
From \eqref{misura-pa-1} we obtain
$$ \mu_{\omega, p \cdot a} ( \left\{ 0 \right\}) = \texttt{r}_0 + \texttt{r}_1 P^{p:a}_1(\left\{ 0 \right\}) \geq r_0 > 0 \qquad \Longrightarrow \qquad 0 \in \sigma(p \cdot a) $$
If $\lambda \in \sigma(p \cdot a)$ with $\lambda \neq 0$, from \eqref{misura-pa-2} we have:
$$ \mu_{\omega, p \cdot a} ( \left\{ \lambda \right\}) = - \texttt{r}_0 P^{p:a}_0(\left\{ \lambda \right\}) + \mu_{\omega, a} ( \left\{ \lambda \right\}) > 0$$ 
therefore 
$$ \mu_{\omega, a} ( \left\{ \lambda \right\}) > \texttt{r}_0 P^{p:a}_0(\left\{ \lambda \right\}) \geq 0 \qquad \Longrightarrow \qquad \lambda \in \sigma(a) $$ 
\end{proof}
From the distributive property, the following decomposition is obtained:
\begin{equation}\label{decomposizione}
 a = p \cdot a + p^\bot \cdot a 
\end{equation}
since the question $p^\bot$ is compatible with the observable $a$ and by the above\footnote{See \eqref{sum_conv}.}, for every state $\omega \in \mathfrak S_a$ we can write
\begin{equation}
\mu_{\omega, a} = \mu_{\omega, p \cdot a} \ast \mu_{\omega, p^\bot \cdot a} 
\end{equation}
\subsubsection{Properties of the measure $\mu_{\omega, p^\bot \cdot a}$.}
Given $f(t) = 1 - t$, from \eqref{questioneortogonale} it is easily deduced that
$$ \mathbf{1}_{\Delta}(p^\bot) = \mathbf{1}_{f^{-1}(\Delta)}(p) \ , \qquad \forall \Delta \in B(\mathbb R) $$  
and using \eqref{sestocaso} we can write:
$$ \mu_{\omega, p^\bot : a} (\Delta_0 \times \Delta_1) = \mu_{\omega, p : a} (f^{-1}(\Delta_0) \times \Delta_1) = \mu_{\omega, p : a}^F (\Delta_0 \times \Delta_1) $$ 
where
$$ (r,s) \in \mathbb R^2 \longrightarrow F(r,s) = (f(r), s) \in \mathbb R^2$$
therefore
\begin{eqnarray*}
\mu_{\omega, p^\bot \cdot a} (\Delta) & = & \int \mathbf{1}_{\Delta}(r s) \, d \mu_{\omega, p : a}^F (r, s) \\
& = & \int \mathbf{1}_{\Delta}((1 - r) s) \, d \mu_{\omega, p : a}^F (r, s) \\
& = & \int \mathbf{1}_{\Delta}(r s) \, d P_s^{p:a}(r) \, d \mu_{\omega, p}(s)
\end{eqnarray*}
and also in this case we can write a relation similar to the previous one
\begin{eqnarray}  
\mu_{\omega, p^\bot \cdot a} (\Delta) = \texttt{r}_0 P_0^{p:a}(\Delta) + \texttt{r}_1 \delta_0(\Delta) P_1^{p:a}(\Delta)
\end{eqnarray}
Applying these results to the calculation of the mean value, we obtain:
\begin{eqnarray*} 
\left\langle p \cdot a \right\rangle_{\omega} & = & \int t \, d \mu_{\omega, p \cdot a}(t) \\
& = & \texttt{r}_1 \int t \, d P_1^{p:a}(t) 
\end{eqnarray*}
and
\begin{eqnarray*} 
\left\langle p^\bot \cdot a \right\rangle_{\omega} & = & \int t \, d \mu_{\omega, p^\bot \cdot a}(t) \\
& = & \texttt{r}_0 \int t \, d P_0^{p:a}(t) 
\end{eqnarray*}

If the commutant $\mathcal C(a)$ is non-trivial and $x$ is one of its elements, then for every disjoint partition $\left\{ \Delta_k \right\}_{k \in \mathbb N}$ of $\sigma(x)$, we have a family of questions $\left\{ q_k \right\}_{k \in \mathbb N}$ mutually orthogonal\footnote{In fact, by the Borel calculus, for each $i \neq j$ it turns out that 
$$ q_i \cdot q_j = 0 $$ } such that
$$ I = \sum_{k \in \mathbb N} q_k \qquad , \qquad q_k \in \mathcal C(a) $$
and we can write\footnote{See also equation \eqref{decosp}.}
\begin{equation}
\label{decomposizionedue}
a = \sum_{k \in \mathbb N} q_k \cdot a \cdot q_k 
\end{equation}
 $$ \star \star \star $$
\begin{remark}[\textbf{Kolmogorov's Observables}]\upshape\label{Kolmobservab}
For subsequently jointly preparable observables which are Kolmogorov, it is possible to retrace the notions and definitions given in \ref{valoremediocongiunto} and \ref{spettrocongiunto}, since in this case we have a probability measure as described by Definition \ref{misuraprodotto1b}. 
\end{remark}

\section{Proofwriter*}\label{verifiche}
In this section we collect the proofs of some results that we presented in the previous sections but without having verified them.
\\
We recall that $\omega \in \mathfrak S_{a:b} \subset \mathfrak S_b$.
\begin{itemize}
\item [$\star$] \textbf{Proof of relation \eqref{statocond3} in section \ref{valoremediocongiunto}, page \pageref{valoremediocongiunto}:}
$$ \left\langle a \right\rangle_\omega^{a:b} = \int s \, d \eta(s) \qquad , \qquad \left\langle b \right\rangle_\omega^{a:b} = \int t \, d \mu_{\omega_2,b}(t) $$
From Proposition \ref{markovkernel0} we obtain that there is a Borel measure $\mu \in \Pi$ such that:
\begin{equation*}
\mu_{\omega, a:b}(\Delta_1 \times \Delta_2) = \int_{\Delta_1} P_s^{a:b}(\Delta_2) \, d \mu(s)
\end{equation*}
and from \eqref{statocond} it follows that in this case $\mu = \eta$, while from Proposition \ref{markovkernel} we obtain
\begin{eqnarray*}
\int_{\mathbb R^2} s \, d \mu_{\omega, a:b}(s,t) & = & \int_{\mathbb R^2} s \mathbf{1}_{\mathbb R}(t) \, d \mu_{\omega, a:b}(s,t) \\
& = & \int_{\mathbb R^2} s \mathbf{1}_{\mathbb R}(t) \, d P_s^{a:b}(t) \, d \eta(s) \\
& = & \int s \, d P_s^{a:b}(\mathbb R) \, d \eta(s) = \int s \, d \eta(s) 
\end{eqnarray*}
Therefore
$$\left\langle a \right\rangle_\omega^{a:b} = \int s \, d \eta(s) $$
Similarly, for the calculation of the average value of $b$ we can write:
$$\mu_{\omega, a:b}( \mathbb R \times \Delta_2 ) = \int_{\mathbb R} P_s^{a:b}(\Delta_2) \, d \eta(s) = \mu_{\omega, b}(\Delta_2) $$
and from \eqref{c0} we have that for every $\eta$-summable Borel function $f$ 
$$\mu_{\omega, b}(f) = \int_{\mathbb R} P_s^{a:b}(f) \, d \eta(s)$$
If $f(t) = t$, then
\begin{eqnarray*}
\mu_{\omega, b}(f) & = & \int_{\mathbb R} t \, d \mu_{\omega, b}(t) = \int_{\mathbb R} P_s^{a:b}(f) \, d \eta(s) \\
& = & \int_{\mathbb R^2} f(t) \, d P_s^{a:b}(t) \, d \eta(s) = \int_{\mathbb R^2} t \, d \mu_{\omega, a:b}(s,t)  
\end{eqnarray*}
therefore
$$ \left\langle b \right\rangle_\omega^{a:b} = \left\langle b \right\rangle_{\omega} $$
\item [$\star$] \textbf{Proof of relation \eqref{comp1} in section \ref{funzioni_compatibil}, page \pageref{funzioni_compatibil}}:
\\
We have to prove that for every bounded Borel function $f: \mathbb R \rightarrow \mathbb R$ it holds that
$$ \left\langle f(b_1) \right\rangle_\omega = \int f(s) \, d \mu_{\omega, b_1}(s) = \int_{\mathbb R^2} f(s) \, d \nu(s, t)$$
where $\nu$ is the product measure given in \eqref{misuraprodotto}, which coincides with our measure $\mu_{\omega, b_1:b_2}$.
\\
Therefore, using the results described in \ref{condizio_misure_simult} and \ref{funzioneosservabile2} we can write:
\begin{eqnarray*}
\int_{\mathbb R^2} f(s) \, d \nu(s, t) & = & \int_{\mathbb R^2} f(s) \mathbf{1}_{\mathbb R}(t) \, d \nu(s, t) \\
& = & \int_{\mathbb R^2} f(s) \mathbf{1}_{\mathbb R}(t) \, d \mu_{\omega, b_1:b_2}(s, t) \\
& = & \int_{\mathbb R^2} f(s) \mathbf{1}_{\mathbb R}(t) \, d P^{b_1:b_2}_s(t) \, d \mu_{\omega, b_1}(s) \\ 
& = & \int f(s) \, d \mu_{\omega, b_1}(s) \cdot \int \mathbf{1}_{\mathbb R}(t) \, d P_s(t) = \int f(s) \, d \mu_{\omega, b_1}(s) 
\end{eqnarray*}
from here follows relation \eqref{comp1}.
\end{itemize}

\chapter{Selection of States and Observables}
As argued in detail in the previous sections, to our laboratory system, located in $\mathcal O$, we have associated a pair $\left( \mathfrak X , \mathfrak S \right)$ consisting of all the physical quantities that it is potentially possible to measure in the laboratory in a given preparation time and all the possible states in which these measurements are made. However, in the work done so far we have not taken the following fact into account:
\\
\textit{It makes no physical sense to take into account all the various devices and measuring equipment of the laboratory system, because we can only use a finite family of them (or at most a countable limit) and thus only really make measurements on a finite family of observables (or at most a countable limit) $\mathfrak S_o$ of $\mathfrak S$}.
\\
In this section we will highlight the minimal properties which this family of states (and observables) must satisfy to determine a physical subsystem of the system.
\section{Physical Laboratory Systems} \label{Sistemi fisici del laboratorio} 
Let's make a direct choice on the states of the physical system $\left( \mathfrak X , \mathfrak S \right)$ of our laboratory, considering (for example for experimental reasons) a subset $\mathfrak S_o \subset \mathfrak S$ and we study the consequences of this selection.
\\
The first question we ask ourselves is the following:
\\
\textit{What subset $\mathfrak X_o \subset \mathfrak X$ can be associated with the state set $\mathfrak S_o$?}
\\
An initial answer to this question was given in section \ref{attuale} on page \pageref{attuale}: 
\\
To make physical sense, the pair $\left( \mathfrak X_o , \mathfrak S_o \right)$ must necessarily satisfy at least the physically achievable conditions:
\begin{itemize}
\item [1.] For every observable $a$ belonging to $\mathfrak X_o$, there must be at least one state $\omega \in \mathfrak S_o$ suitable for it, so:
$$ \mathfrak X_o \subset \mathfrak X_\lor^{\mathfrak S_o} = \bigcup_{\omega \in \mathfrak S_o} \mathfrak X_\omega $$
\item [2.] For every state $\omega$ belonging to $\mathfrak S_o$ there must be at least one observable $a \in \mathfrak X_o$ suitable for measurement in this state $\omega$, so: 
$$ \mathfrak S_o \subset \mathfrak S_\lor^{\mathfrak X_o} = \bigcup_{x \in \mathfrak X_o} \mathfrak S_x $$
\end{itemize}
Given the physically achievable pair $\left( \mathfrak X_o , \mathfrak S_o \right)$, for each observable $a \in \mathfrak X_o$ we define the following set of real Borel functions
$$ L^1(a, \mathfrak S_o) = \left\{ f : \mu_{\omega,a}(f) < \infty \ \forall \omega \in \mathfrak S^o_a \right\} $$
which are called $a$-$\mathfrak S_o$-summable functions\footnote{Recall that 
$$\mathfrak S^o_a = \mathfrak S_o \cap \mathfrak S_a$$ }.
Our physically achievable pair $\left( \mathfrak X_o , \mathfrak S_o \right)$, to be a good candidate for a physical subsystem of the laboratory system $\left( \mathfrak X , \mathfrak S \right)$, must satisfy the following functional closure property:
\begin{property}[\textbf{Functional Closure}]\label{c.f.}\index{Property of functional closure}
If $a \in \mathfrak X_o$, then $F(a) \in \mathfrak X_o$ for each function $F \in L^1(a, \mathfrak S_o)$.
\end{property}
Note that the set of $a$-$\mathfrak S_o$-summable functions contains the set of $a$-summable functions:
$$ L^1(a) \subset L^1(a, \mathfrak S_o)$$
In this way, if the pair $\left( \mathfrak X_o , \mathfrak S_o \right)$ satisfies property \ref{c.f.}, then for each Borel function $f \in L^1(a, \mathfrak S_o)$ we obtain an observable $f^o(a) \in \mathfrak X_o$ such that  
\begin{equation}\label{c.f.2} 
\left\langle f^o(a) \right\rangle_\omega = \mu_{\omega,a}(f) \qquad \forall \omega \in \mathfrak S^o_a 
\end{equation}
\begin{attenzione}\upshape
Given that $\mathfrak X_o \subset \mathfrak X$, we obtain that $f^o(a)$ is also an observable of the laboratory physical system $\left( \mathfrak X , \mathfrak S \right)$, where its associated states are
$$
\mathfrak S_{f^o(a)} = \mathfrak S^o_a \subset \mathfrak S_a
$$
Therefore, if the function $f \in L^1(a)$, then we obtain, through the Borel functional calculus, an observable $f(a)$ of $\mathfrak X$, with $\mathfrak S_{f(a)} = \mathfrak S_a$. Obviously $f$ is also an element of $L^1(a, \mathfrak S_o)$; then we obtain an observable of $\mathfrak X_o$, denoted by $f^o(a)$, which satisfies relation \eqref{c.f.2}. In this way  
$$f^o(a) \subset f(a)$$
because 
$$\mathfrak S_{f^o(a)} = \mathfrak S^o_a \subset \mathfrak S_a = \mathfrak S_{f(a)}$$
\end{attenzione}
We introduce the equality of observables of a physically achievable pair $\left( \mathfrak X_o , \mathfrak S_o \right)$:
\begin{definition}[\textbf{Equality in Selection States}]\upshape\index{$a\stackrel{\mathfrak S_o}=b$}\index{Equality in selection states}
Two observables $a, b \in \mathfrak X_o$ are $\mathfrak S_o$-equal, in symbols
$$a \stackrel{\mathfrak S_o}{=} b$$
if
\begin{itemize}
\item $\mathfrak S_a^o = \mathfrak S_b^o$  
\item $\mu_{\omega,a} = \mu_{\omega,b} \ , \qquad \forall \omega \in \mathfrak S_a^o$ 
\end{itemize}
\end{definition}
Obviously if $a, b \in \mathfrak X_o$ are equal, then they are $\mathfrak S_o$-equal.
\\

Let us now make some simple considerations.
\\
We remark that, having fixed a state $\omega \in \mathfrak S$ and any $a \in \mathfrak X_\omega$, we obtain
$$ f(a) \in \mathfrak X_\omega \ , \qquad \forall f \in L^1(a)$$
Furthermore, having fixed a set of states $\mathfrak S_o$ of the system, we have for each function $f \in L^1(a, \mathfrak S_o)$
$$ [ \ a \in \mathfrak X_\omega \ \text{and} \ \omega \in \mathfrak S^o_a \ ] \qquad \Longrightarrow \qquad f^o(a) \in \mathfrak X_\omega $$
Obviously
$$ [ \ a \in \mathfrak X_\omega \ \forall \omega \in \mathfrak S^o_a \ ] \qquad \Longrightarrow \qquad [ \ f^o(a) \in \mathfrak X_\omega \ \forall \omega \in \mathfrak S_a^o \ ] $$
It follows that we have two physically achievable pairs which satisfy property \ref{c.f.} of functional closure: 
$(\mathfrak X_\lor^{\mathfrak S_o}, \mathfrak S_o)$, where the set of observables $\mathfrak X_\lor^{\mathfrak S_o}$ is defined by equation \eqref{osservabilimisurabilimax}:
$$ \mathfrak X_\lor^{\mathfrak S_o} = \bigcup_{\omega \in \mathfrak S_o} \mathfrak X_\omega $$
and $(\mathfrak X_\wedge^{\mathfrak S_o}, \mathfrak S_o)$, where $\mathfrak X_{\land}^{\mathfrak S_o}$ are the observables given in equation \eqref{osservabilimisurabilimin}:
$$ \mathfrak X_\wedge^{\mathfrak S_o} = \bigcap_{\omega \in \mathfrak S_o} \mathfrak X_\omega $$
\begin{proposition}\upshape \label{numero} 
If $a \in \mathfrak X_\wedge^{\mathfrak S_o}$, then
$$ \mathfrak S_a^o = \mathfrak S_o $$
\end{proposition}
\begin{proof}
If $\omega' \in \mathfrak S_o$, from the hypothesis $a \in \mathfrak X_{\omega'}$ it follows that $\omega' \in \mathfrak S_a$; therefore we have $\omega \in \mathfrak S_o \cap \mathfrak S_a$.
\end{proof}
From these trivial considerations it follows that the pair $(\mathfrak X_{\land}^{\mathfrak S_o}, \mathfrak S_o)$ admits constant observables, i.e., observables $c$ such that:
\begin{equation} \label{costatntisottosistem}
\left\langle \texttt{c} \right\rangle_{\omega} = c \ , \qquad \forall \omega \in \mathfrak S_o
\end{equation}
In fact, if $C(t) = c \ , \forall t \in \mathbb R$, then for each $a \in \mathfrak X_{\land}^{\mathfrak S_o}$ we have 
$$ \left\langle C^o(a) \right\rangle_{\omega} = c \ , \qquad \forall \omega \in \mathfrak S_a^o = \mathfrak S_o$$
It follows that $C^o(a) = \texttt{c}$ with $\mathfrak S_{\texttt{c}} = \mathfrak S_o$.
\subsection{Relative Spectrum} 
We observed in section \ref{Correlazione-stati-Osservabili} that fixing a set of states $\mathfrak S_o$ of the system influences the set of possible observables, since a state of the system is synonymous with an experimental procedure, a procedure implemented with instruments and devices present in the laboratory\footnote{We again point out that the states of the system also depend on the type of the various devices used in the measurements and these may be more or less effective, exploratory, in obtaining information on the values of our physical quantity.}. This led us to the notion of a physically achievable pair $(\mathfrak X_o , \mathfrak S_o)$.
\\
One effect of this choice is to have the spectrum of an observable $a \in \mathfrak X_o$ smaller than the totality of the states, in symbols\footnote{See remark \ref{spettros} of section \ref{spettro_oss} on page \pageref{spettros}.}:\index{$\sigma_{\mathfrak S_o} (a)$}
$$ \sigma_{\mathfrak S_o}(a) \subset \sigma(a) $$
since for every observable $a$, for the set of open $a$-null sets given in equation \eqref{a-nulli}, we obtain: 
$$\mathfrak F^\infty (a) \subset \mathfrak F^{\infty}_{ \ \mathfrak S_o}(a) := \bigcap_{\omega \in \mathfrak S^{ \ o }_{a}} \mathfrak F^\omega (a)$$
and therefore for the resolvent we have by definition:
$$\rho(a) \subset \rho_{\mathfrak S_o}(a) := \mathbb{R} \setminus \sigma_{\mathfrak S_o}(a) $$
Increasing (or decreasing) the possible states of the system means increasing (or reducing) the complexity of the measurements that can be carried out on the observables of the system, thereby increasing (or decreasing) their knowledge (for example, their spectrum becomes larger or smaller) and even limiting what is physically impossible to know about the observables themselves\footnote{Therefore we have a limit on their knowledge.}:
$$
\text{Experimenter} \quad \Longrightarrow \quad \text{choice of states} \quad \Longrightarrow \quad \text{information on the physical quantity}
$$
\subsection{Compatibility and Complementarity in Selection}
For a physically achievable pair $(\mathfrak X_o , \mathfrak S_o)$, we introduce the notions of complementary, jointly preparable and compatible observables.
\begin{definition}\upshape\index{Compatibility in Selection}
The observables $a, b \in \mathfrak X_o$ can be jointly prepared for their simultaneous measurement (at a given time $\tau$) in the order $a:b$ with respect to the family of states $\mathfrak S_o$ if it holds that
$$\mathfrak S_{a:b}^o := \mathfrak S_{a:b} \cap \mathfrak S_o \neq \emptyset $$
\end{definition}
As it is easy to see, to obtain the definitions in selection, it is sufficient to make the following substitutions in the various definitions given on page \pageref{sez_noncomplementari}:
$$
\begin{array}{ccc}
\mathfrak S_a & \longmapsto & \mathfrak S_a^o = \mathfrak S_o \cap \mathfrak S_a 
\\ 
  \\
\mathfrak S_{a:b} & \longmapsto & \mathfrak S_{a:b}^o = \mathfrak S_o \cap \mathfrak S_{a:b}  
\\
\\
\mathfrak S_{b:a} & \longmapsto & \mathfrak S_{b:a}^o = \mathfrak S_o \cap \mathfrak S_{b:a}     
\end{array}
$$
In this way we will talk about $\mathfrak S_o$-complementarity, $\mathfrak S_o$-compatibility, etc.
\\
For example, if $a, b \in \mathfrak X_o$ are $\mathfrak S_o$-compatible, then we have
\begin{itemize}
\item [$\star$] $\mathfrak S_{a:b}^o = \mathfrak S_{b:a}^o \neq \emptyset$,
\item [$\star$] For each $\omega \in \mathfrak S_{a:b}^o$ 
$$ P(a \in \Delta_0 : b \in \Delta_1)_\omega = P(b \in \Delta_1 : a \in \Delta_0)_\omega \ , \qquad \forall \Delta_0, \Delta_1 \in B(\mathbb R)$$
\end{itemize}
We must call attention to the following fact:
\\
Following the same lines of reasoning as in section \ref{sez_noncomplementari} on page \pageref{sez_noncomplementari}, two observables $a$ and $b$ belonging to $\mathfrak X_o$ are $\mathfrak S_o$-strongly compatible if they are $\mathfrak S_o$-compatible and
$$ \mathfrak S_{a:b}^o = \mathfrak S_b^o \qquad , \qquad \mathfrak S_{b:a}^o = \mathfrak S_a^o $$
Therefore if $a, b \in \mathfrak X_o$ are strongly compatible, then they are, as is easy to verify, still $\mathfrak S_o$-strongly compatible, but this \textit{is not true} for simple compatibility since it is not guaranteed that $\mathfrak S_{a:b}^o$ or $\mathfrak S_{b:a}^o$ are non-empty.
\\ 
However, if $a, b \in \mathfrak X_o$ are $\mathfrak S_o$-jointly preparable in the order $a:b$, then we have $\mathfrak S_{a:b}^o \neq \emptyset$, so it also follows that $\mathfrak S_{a:b} \neq \emptyset$, i.e., that $a, b$ are still jointly preparable.  
\begin{remark}\upshape
Let us consider the pair $(\mathfrak X_{\land}^{\mathfrak S_o}, \mathfrak S_o)$. From the properties of functional calculus, we have that the observable $c$ is $\mathfrak S_o$-compatible with every observable belonging to $\mathfrak X_{\land}^{\mathfrak S_o}$, since $C(a) \stackrel{\mathfrak S_o}{=} c$ is $\mathfrak S_o$-compatible with $a$. 
\end{remark}
\subsection{Sum of $\mathfrak S_o$-compatible Observables}
Let $(\mathfrak X_o , \mathfrak S_o)$ be an achievable pair, and consider two observables $x, y$ of $\mathfrak X_o$ which are $\mathfrak S_o$-compatible. What can we say about their sum?
\\
In general, we cannot affirm that there exists the sum of $x$ and $y$ such that 
\begin{equation}\label{somma_comp}
\left\langle x + y \right\rangle_{\omega} = \left\langle x \right\rangle_{\omega} + \left\langle y \right\rangle_{\omega} \ , \qquad \forall \omega \in \mathfrak S_o
\end{equation}
and even if that observable exists, it does not necessarily belong to the set $\mathfrak X_o$ and is $\mathfrak S_o$-compatible with $x$ and $y$.
\\
To solve this problem, we introduce a generalization of property \ref{c.f.}  in the case of $\mathfrak S_o$-joint preparation (see Axiom \ref{assio3bis} on page \pageref{assio3bis}):
\begin{property}[\textbf{Multivariable Functional Closure}]\label{c.f.m.}\index{Property of multivariable functional closure}
The physically achievable pair $(\mathfrak X_o , \mathfrak S_o)$ satisfies the property of multivariable functional closure if for every pair of observables $a, b \in \mathfrak X_o$ which are $\mathfrak S_o$-simultaneously preparable in the order $a:b$ and for every bounded Borel function $F: \mathbb R^2 \rightarrow \mathbb R$, there remains associated an observable, which we denote by $F^o(a:b) \in \mathfrak X_o$, such that
\begin{equation}
\mathfrak S_{F^o(a:b)} = \mathfrak S_{a:b}^o
\end{equation}
and 
\begin{equation}
\mu_{\omega, F^o(a:b)}(\Delta) = \mu_{\omega, a:b}(F^{-1}(\Delta)) \ , \qquad \forall \Delta \in B(\mathbb R)
\end{equation}
\end{property}
We can extend, as we have done for the general case, this property to the case of three or more observables that are $\mathfrak S_o$-simultaneously preparable. 
\\

For example, if $a, b, c$ are observables of $\mathfrak X_o$ which are $\mathfrak S_o$-simultaneously preparable in the order $a:b:c$, then we have the following generalization of property \ref{c.f.m.}:
\begin{itemize}
\item If $a:b:c$ are $\mathfrak S_o$-simultaneously preparable, then $G(a:b):c$ are $\mathfrak S_o$-simultaneously preparable. 
\item $\mathfrak S_{G^o(a:b):c}^o = \mathfrak S_{a:b:c}^o$ 
\item For every $\omega \in \mathfrak S_{a:b:c}^o$ it holds that
$$ \mu_{\omega, G^o(a:b):c} = \mu_{\omega, a:b:c}^{\hat{G}}$$
where $\widehat{G}: \mathbb R^3 \rightarrow \mathbb R^2$ is the function defined in \eqref{perlassociativa2}. 
\end{itemize} 
To recapitulate, if the physically achievable pair $(\mathfrak X_o , \mathfrak S_o)$ satisfies the multivariable functional closure, then we have the following results:
\\
If $a, b$ are $\mathfrak S_o$-compatible observables, then the observable $v^o(a,b)$ introduced in section \ref{funzioneosservabile2} on page \pageref{funzioneosservabile2} — third step — is still $\mathfrak S_o$-compatible with the observables $a$ and $b$, with 
$$ \left\langle v^o(a:b) \right\rangle_\omega = \left\langle a \right\rangle_\omega + \left\langle b \right\rangle_\omega \ , \qquad \forall \omega \in \mathfrak S_{a:b}^o$$
Furthermore, always by property \ref{c.f.m.}, also for the Jordan product we obtain $a \cdot b \in \mathfrak X_o$, because the observable $\Theta^o_{1,1}(a:b)$ introduced in section \ref{funzioneosservabile2} — fourth step — is still $\mathfrak S_o$-compatible with $a$ and $b$, with 
$$\left\langle \Theta^o_{1,1}(a:b) \right\rangle_\omega = \left\langle a \cdot b \right\rangle_\omega \ , \qquad \forall \omega \in \mathfrak S_{a:b}^o$$
\subsection{Center of a Set of Observables in Selection}
Let $\mathfrak S_o$ be a subset of $\mathfrak S$. For every observable $a \in \mathfrak X_\lor^{\mathfrak S_o}$\footnote{Since from the hypothesis $a \in \mathfrak X_\lor^{\mathfrak S_o}$ we obtain that $\mathfrak S_a^o \neq \emptyset$.}, we denote by $\mathcal C_{\mathfrak S_o}(a)$ the commutant of $a$ relative to the family of states $\mathfrak S_o$: the set of all observables $x \in \mathfrak X$ which are $\mathfrak S_o$-compatible with $a$\footnote{By remark \ref{compconsestesso} on page \pageref{spettrocongiunto}, we implicitly assume that any observable $a$ is compatible with itself; then $a$ belongs to the set $\mathcal C(a)$ defined in \eqref{centrosemplice} on page \pageref{centrosemplice} and so $a \in \mathcal C_{\mathfrak S_o}(a)$ because $a \in \mathfrak X_\lor^{\mathfrak S_o}$. }.\index{$\mathcal C_{\mathfrak S_o}(a)$}\index{Commutant of a Set of Observables in Selection}
\\
We recall that $x \in \mathcal C_{\mathfrak S_o}(a)$ if and only if these three conditions hold simultaneously:
\begin{itemize} 
\item [1.] $x \in \mathfrak X_\lor^{\mathfrak S_o}$,
\item [2.] $\mathfrak S_{a:x}^o = \mathfrak S_{x:a}^o \neq \emptyset$,
\item [3.] $\mu_{\omega, a:x} = \mu_{\omega, x:a} \ , \qquad \forall \omega \in \mathfrak S_{a:x}^o$ 
\end{itemize}
Therefore the relative $\mathfrak S_o$-commutant satisfies the following relation:
$$\mathcal C_{\mathfrak S_o}(a) \subset \mathfrak X_\lor^{\mathfrak S_o} $$
We must make the following
\begin{remark}\upshape
If $\mathfrak S_{oo} \subset \mathfrak S_o \subset \mathfrak S$ and $a \in \mathfrak X_\lor^{\mathfrak S_{oo}}$, then 
 $$ \left\{ x \in \mathcal C_{\mathfrak S_o}(a) : \mathfrak S_{a:x}^{oo} \neq \emptyset \right\} \subset \mathcal C_{\mathfrak S_{oo}}(a)$$ 
Indeed, if $x \in \mathcal C_{\mathfrak S_o}(a)$, the observable $x$ is compatible with $a$; then from the hypothesis we obtain that 
$\mathfrak S_{a:x}^{oo} = \mathfrak S_{x:a}^{oo}$, but we cannot say that they are non-empty. 
 \end{remark}
Intuitively, as the number of states increases, the possibility of compatibility decreases while complementarity increases.
\\

Let $\mathfrak X_o \subset \mathfrak X_\lor^{\mathfrak S_o}$ be a family of observables of the laboratory system. We define the \textit{commutant} of $\mathfrak X_o$ as the set:
\begin{equation}\label{centronosemplice}\index{$\mathcal C_{\mathfrak S_o }(\mathfrak X_o) $}
\mathcal C_{\mathfrak S_o}(\mathfrak X_o) := \bigcap_{a \in \mathfrak X_o} \mathcal C_{\mathfrak S_o}(a) \subset \mathfrak X_\lor^{\mathfrak S_o}
\end{equation}
while the set
\begin{equation}\label{commutanteobs}\index{$\mathcal Z_{\mathfrak S_o }(\mathfrak X_o)$ }\index{Center in selection of states}
\mathcal Z_{\mathfrak S_o}(\mathfrak X_o) := \mathcal C_{\mathfrak S_o}(\mathfrak X_o) \cap \mathfrak X_o 
\end{equation}
is called the \textit{center} of the family of observables $\mathfrak X_o$.
\\

Obviously, we can introduce the notion of the strong commutant of a family of observables $\mathfrak X_o$, which we denote by $\mathcal C^F_{\mathfrak S_o}(\mathfrak X_o)$; it is the set of all $x \in \mathfrak X$ that are strongly $\mathfrak S_o$-compatible with our observable $a$. Similarly we define their strong center $\mathcal Z^F_{\mathfrak S_o}(\mathfrak X_o)$.
\\

We transfer these definitions to the case of a physically achievable pair $(\mathfrak X_o , \mathfrak S_o)$, pairs that are the prototypes of our physical subsystems in the laboratory\footnote{See section \ref{Sottosistemi_Fisici} on page \pageref{Sottosistemi_Fisici}.}.
\\
Let us consider a physically achievable pair $(\mathfrak X_o , \mathfrak S_o)$. We define the commutant and center of a set of observables $\mathfrak Y \subset \mathfrak X_o$ as the following sets:
$$\mathcal C_{(\mathfrak X_o, \mathfrak S_o)}(\mathfrak Y) := \mathcal C_{\mathfrak S_o}(\mathfrak Y) \cap \mathfrak X_o$$ 
 and
$$\mathcal Z_{(\mathfrak X_o, \mathfrak S_o)}(\mathfrak Y) := \mathcal C_{\mathfrak S_o}(\mathfrak Y) \cap \mathfrak Y$$ 
It follows that for each $a \in \mathfrak X_o$ 
$$\mathcal C_{(\mathfrak X_o, \mathfrak S_o)}(a) = \mathcal C_{\mathfrak S_o}(a) \cap \mathfrak X_o $$

\subsection{Sectors in the Measurement of an Observable in Selection}\label{Sectorsbis}
In this section we remodel the notion of sectors in the measurement of an observable for physically achievable pairs $(\mathfrak X_o , \mathfrak S_o)$.
\\

Let $a \in \mathfrak X_o$ and let us define the following set of probability measures:
$$\mathbb M_o(a) = \left\{ \mu_{\omega,a} \in \Pi : \omega \in \mathfrak S_a^o \right\} \subset \mathbb M(a)$$
As we established in section \ref{sectors}, the set $\mathbb M(a)$ admits a covering of convex $k$-sets $\mathbb M_k(a)$ that satisfy the properties contained in Axiom \ref{assio6bis} on page \pageref{assio6bis}.
\\
One might think that the $k$-sets 
$$\mathbb M_k(a) \cap \mathbb M_o(a)$$
satisfy the conditions dictated by Axiom \ref{assio6bis} for the set $\mathbb M_o(a)$, but this is obviously not true, since the convexity property for such a set is not even guaranteed.
\begin{property}[\textbf{Sectors of Measurement in the Selection of States}]\label{assio6tris}\index{Property of sectors of measurement in the selection of states}
For every $a \in \mathfrak X_o$, the set $\mathbb M_o(a)$ has a family of convex sets $\left\{ \mathbb M^k_o(a) \right\}_k$ which satisfy the four conditions contained in Axiom \ref{assio6bis} of section \ref{sectors}.\footnote{We reiterate that they may not coincide with any set of the form $\mathbb M_k(a) \cap \mathbb M_o(a)$.}.  
\end{property}
\begin{remark}\upshape\label{settoreoperativo}
Only experimentally is it possible to check whether a family of states $\mathfrak S_o$ satisfies property \ref{assio6tris} on page \pageref{assio6tris} for each observable $a \in \mathfrak X_o$. 
\end{remark}
Assuming this property to be true, we can also in this case define the set of sectorial states in the measurement of $a$ in the selection of states $\mathfrak S_o$:
$$\mathfrak S^{o,k}_a = \left\{ \omega \in \mathfrak S_a^o : \mu_{\omega,a} \in \mathbb M^k_o(a) \right\} $$ 
and repeating step by step the considerations made in section \ref{sectors} on page \pageref{sectors}, we obtain 
$$ \mathfrak S_a^o = \bigcup_k \mathfrak S^{o,k}_a \qquad \text{with} \qquad \mathfrak S^{o,k}_a \cap \mathfrak S^{o,h}_a = \emptyset \ , \ h \neq k $$
\section{Physical Subsystems}\label{Sottosistemi_Fisici}
Let us take up again the initial problem of determining when a pair $(\mathfrak X_o , \mathfrak S_o)$ can be considered a physical subsystem of our laboratory as described in Problem \ref{pb-1} of section \ref{attuale} on page \pageref{attuale}. 
\\
The pair $(\mathfrak X_o , \mathfrak S_o)$, to be a physical subsystem, must satisfy the following conditions:
\begin{itemize}
\item [A.] Physical achievability, as defined in section \ref{attuale}.  
\item [B.] The functional closure property \ref{c.f.}.
\item [C.] The property of temporal evolution, i.e., for every $\omega \in \mathfrak S_o$ we obtain\footnote{We underline that this property is the least trivial to verify experimentally.}:  
$$ \mathfrak S^{x,\omega}_t \subset \mathfrak S_x^o \ , \qquad \forall x \in \mathfrak X_\omega \cap \mathfrak X_o$$
\end{itemize}
\begin{remark}\upshape
We emphasise that a physical subsystem of the laboratory system $(\mathfrak X , \mathfrak S)$ does not necessarily require that the sector measurement property \ref{assio6tris} be satisfied\footnote{See previous remark \ref{settoreoperativo}.}.
\end{remark}
Let us now see when two different choices of state families can give physically equivalent systems.
\begin{definition}[\textbf{Equivalent Physical Systems}]\upshape\label{equivalenzafisica}\index{Equivalent physical systems}
Two physical subsystems $(\mathfrak X_o , \mathfrak S_o)$ and $(\mathfrak X_{oo} , \mathfrak S_{oo})$ of the laboratory system $(\mathfrak X , \mathfrak S)$
are said to be physically equivalent if there exists a pair of bijections 
$$\Lambda: \mathfrak X_o \rightarrow \mathfrak X_{oo} \qquad , \qquad \Lambda^\natural: \mathfrak S_o \rightarrow \mathfrak S_{oo}$$
such that
\begin{itemize}
\item[0.] for every $\omega \in \mathfrak S_a^o$ with $a \in \mathfrak X_o$ it holds that
$$ \Lambda^\natural(\omega^{(\tau)}) \in \mathfrak S_a^o|_\tau \qquad , \ \forall \tau \in \mathbb R^+$$
\item[1.] for every $a \in \mathfrak X_o$  
$$ \Lambda^\natural(\mathfrak S_a^o) = \mathfrak S_{\Lambda(a)}^{oo} $$
\item[2.] for every $\omega \in \mathfrak S_o$
$$ \Lambda(\mathfrak X_\omega^o) = \mathfrak X_{\Lambda^\natural(\omega)}^{oo} $$
\item[3.] for every $a \in \mathfrak X_o$ and $\omega \in \mathfrak S_o$ we obtain
$$P(\Lambda(a) \in \Delta, \tau)_{\Lambda^\natural(\omega)} = P(a \in \Delta, \tau)_{\omega} \ , \qquad \forall \tau \geq 0$$ 
\item[4.] the maps carry each jointly preparable observable pair of $(\mathfrak X_o , \mathfrak S_o)$ to a jointly preparable observable pair of $(\mathfrak X_{oo}, \mathfrak S_{oo})$.
\\
In particular, if $a, b \in \mathfrak X_o$ are jointly preparable in $\mathfrak S_o$ in the order $a:b$, then we have $\Lambda(a):\Lambda(b)$ with
$$ \Lambda^\natural (\mathfrak S_{a:b}^o) = \mathfrak S_{\Lambda(a):\Lambda(b)}^{oo}$$
and
$$P(\Lambda(a) \in \Delta_1 : \Lambda(b) \in \Delta_2, \tau)_{\Lambda^\natural(\omega)} = P(a \in \Delta_1 : b \in \Delta_2, \tau)_{\omega} \ , \qquad \forall \tau \geq 0$$ 
\item[5.] for every $a \in \mathfrak X_o$ we obtain:
$$ \Lambda(\mathcal Z_{(\mathfrak X_o , \mathfrak S_o)}(a)) = \mathcal Z_{(\mathfrak X_{oo} , \mathfrak S_{oo})}(\Lambda(a))$$
\end{itemize}
\end{definition}
We note that there are some redundancies in the definition, namely:
\\ 
Relation [1.] holds if and only if relation [2.] holds, while relation [5.] is a consequence of [4.].
\subsection{Abelian System}\label{sist_abel}
Let's now look at a particular class of physical subsystems: the Abelian ones.
\begin{definition}[\textbf{Abelian System}]\upshape\label{sist_abeliano}\index{Abelian System}
The physical subsystem $(\mathfrak X_o , \mathfrak S_o)$ is said to be Abelian if it satisfies the following relation\footnote{Of course we have
$$ \mathfrak X_o \subset \mathcal C_{\mathfrak S_o} (\mathfrak X_o) \qquad \Longleftrightarrow \qquad \mathcal Z_{\mathfrak S_o} (\mathfrak X_o) = \mathfrak X_o $$}:
\begin{equation}
\label{classic}
 \mathfrak X_o \subset \mathcal C_{\mathfrak S_o} (\mathfrak X_o) 
\end{equation}
Furthermore, if 
\begin{equation}\label{mass} 
\mathfrak X_o = \mathcal C_{\mathfrak S_o} (\mathfrak X_o)  
\end{equation}
it will be called a \textit{maximal abelian subsystem} (\texttt{MASS}).\index{Mass}
\end{definition} 

\subsubsection{A Particular Abelian System} 
Let us consider a finite family $\mathfrak X_c \subset \mathfrak X$ of compatible observables of our physical system $(\mathfrak X , \mathfrak S)$.
\\
From compatibility it follows that for every distinct $a_1, a_2, \ldots, a_l \in \mathfrak X_c$, by definition we have\footnote{Recall that $S(l)$ is the group of permutations on $l$ elements; here $l \leq n = \operatorname{Card}(\mathfrak X_c)$.}:
$$\mathfrak S_{a_1, a_2, \ldots, a_l} = \mathfrak S_{a_{p(1)}, a_{p(2)}, \ldots, a_{p(l)}} \ , \qquad \forall p \in S(l) $$
and to make writing easier, we denote by $\mathfrak S_o$ the following set of states:
$$\mathfrak S_o := \bigcup_{l=1}^n \bigcup_{p \in S(n)} \mathfrak S_{a_{p(1)}, a_{p(2)}, \ldots, a_{p(l)}} $$ 
and since for each permutation $p \in S(n)$ we have:
$$ \mathfrak S_{a_{p(1)}, a_{p(2)}, \ldots, a_{p(l)}} \subset \bigcap_{j=1}^l \mathfrak S_{a_{p(j)}} $$ 
therefore
$$ \mathfrak S_o \subset \bigcup_{p \in S(n)} \bigcap_{j=1}^l \mathfrak S_{a_{p(j)}} \subset \bigcup_{j=1}^{n} \mathfrak S_{a_j} = \mathfrak S_\lor^{\mathfrak X_c}$$ 
Thus the pair $(\mathfrak X_c , \mathfrak S_o)$ is physically achievable, but it is not yet a physical subsystem, because it does not satisfy property \ref{c.f.m.} on page \pageref{c.f.m.}. For this purpose we must consider a set of observables $\mathfrak X_o$ obtained from the family $\mathfrak X_c$ closed under functional calculus; hence we consider the set of observables:
\begin{equation}\label{abel}
\mathcal G(\mathfrak X_c) = \left\{ f(a_{p(1)}: \cdots : a_{p(l)}) : \forall f \in L^1(a_{p(1)}: \cdots : a_{p(l)}) \ , \ p \in S(n) \right\}
\end{equation}
In this way it is easy to verify that the pair $(\mathfrak X_o , \mathfrak S_o)$ is a physical subsystem with  
\begin{equation*}
\mathcal G(\mathfrak X_c) \subset \mathcal C_{\mathfrak S_o} (\mathcal G(\mathfrak X_c)) \ \Longrightarrow \ \text{the physical system is Abelian}
\end{equation*}
and $\mathcal G(\mathfrak X_c)$ is said to be the Abelian system generated by the set $\mathfrak X_c$.
\section{Restricted States and Observables}\label{nuovi_axiom}
In section \ref{assiomistatici} on page \pageref{assiomistatici}, we introduced the axioms governing the relation that exists between states and observables and their average values obtained through measurements made using the ensemble procedure.   
\\
In this section we introduce additional properties\footnote{See definitions \ref{condition1} and \ref{condition2} on page \pageref{condition2}.} that link the set of observables with the set of states of a physical system, properties that are in line with experimental procedures carried out in the laboratory.
\begin{property}\textbf{[Restricted State]} \label{esistenza_stato_ristretto}\index{Property of restricted states}
Let $\omega \in \mathfrak S$ and $\mathfrak X_o \subset \mathfrak X_\omega$. A state $\omega_o \in \mathfrak S$\footnote{If this state exists, it is not necessarily unique.} is called a restricted state of $\omega$ to the set $\mathfrak X_o$ if it satisfies the following properties:
\begin{itemize}
\item[1.] $\omega_o \subset \omega$;
\item[2.] $\mathfrak X_o = \mathfrak X_{\omega_o}$.
\end{itemize} 
The restriction $\omega_o$ of $\omega$ to $\mathfrak X_o$ is denoted by $\omega|_{\mathfrak X_o}$\footnote{Obviously it turns out that $\omega_o \in \mathfrak S_\wedge^{\mathfrak X_o}$.}.
\end{property}
This property affirms that the instruments, and thus the related procedures included in $\omega$, are limited in their use, being directed only to the measurement of the observables in the family $\mathfrak X_o$.
\\
In particular, if the set $\mathfrak X_o$ consists only of the observable $a$, then we have a restricted state of $\omega$ suitable only for its measurement\footnote{Hence the instruments/procedures are used only for the measurement of $a$.}.
\\
We denote by $\mathfrak S_\odot$ the set of states which admit the restricted state $\omega|_{\mathfrak X_o}$; it necessarily follows that
$$ \mathfrak S_\odot \subset \mathfrak S_\wedge^{\mathfrak X_o} $$
and we may consider the non-empty set 
$$ \mathfrak S_o = \left\{ \omega|_{\mathfrak X_o} : \omega \in \mathfrak S_\odot \right\} \subset \mathfrak S_\wedge^{\mathfrak X_o} $$
Therefore, the pair $(\mathfrak X_o , \mathfrak S_o)$ is a suitable system, but not necessarily a physical subsystem.
\\

Similarly, we introduce the following property concerning observables.
\begin{property}\textbf{[Restricted Observable]} \label{esistenza_osservabile_ristretto}\index{Property of restricted observables}
Let $a \in \mathfrak X$ and $\mathfrak S_o \subset \mathfrak S_a$. An observable $a_o \in \mathfrak X$ is called a \emph{restricted observable} of $a$ to the set $\mathfrak S_o$ if it satisfies the following properties:
\begin{itemize}
\item[1.] $a_o \subset a$;
\item[2.] $\mathfrak S_o = \mathfrak S_{a_o}$.
\end{itemize} 
\end{property}

Here too we can introduce the sets $\mathfrak X_\odot \subset \mathfrak X_\wedge^{\mathfrak S_o}$ and $\mathfrak X_o$ in a similar way to the case of restricted states.

Therefore, the fact that $a_o$ is a restriction of $a$ has the following physical interpretation:
\\ 
The experimental procedures identified with the set of states $\mathfrak S_{a_o}$ do not fully reveal the nature (i.e., the properties) of the physical quantity represented by $a_o$.
\\ 
Through technological and theoretical progress, we may develop new experimental procedures $\mathfrak S_b$ that extend the old ones and reveal those properties.
\subsection{$\mathfrak S_o$-Complete Observables}\label{completi_sottosist}
Let us consider a physical subsystem $(\mathfrak X_o , \mathfrak S_o)$ of our laboratory $L_o$ and a family of observables $\mathfrak X_\oslash \subset \mathfrak X_o$ mutually $\mathfrak S_o$-compatible, i.e., 
$$ \mathfrak X_\oslash \subset \mathcal C_{\mathfrak S_o}(\mathfrak X_\oslash) \subset \mathfrak X$$
We recall that 
$$ \mathcal Z_{(\mathfrak X_o , \mathfrak S_o)}(\mathfrak X_\oslash) = \mathcal C_{\mathfrak S_o}(\mathfrak X_\oslash) \cap \mathfrak X_\oslash $$
so
\begin{equation} \label{obs_abel_comp}
\mathfrak X_\oslash \subset \mathcal Z_{(\mathfrak X_o , \mathfrak S_o)}(\mathfrak X_\oslash)  
\end{equation}
and since $\mathcal C_{\mathfrak S_o}(\mathfrak X_o) \subset \mathcal C_{\mathfrak S_o}(\mathfrak X_\oslash)$ it turns out that  
$$ \mathcal Z_{\mathfrak S_o}(\mathfrak X_o) \subset \mathcal Z_{(\mathfrak X_o , \mathfrak S_o)}(\mathfrak X_\oslash) \subset \mathfrak X_o $$
in other words
\begin{equation} \label{obs_abel_comp1}
\mathfrak X_\oslash \subset \mathcal Z_{\mathfrak S_o}(\mathfrak X_o) \subset \mathfrak X_o 
\end{equation}
\begin{definition}[\textbf{Set of Complete Observables}]\upshape\index{Set of complete observables}
Let $\mathfrak X_\oslash \subset \mathfrak X_o$. It is called a set of complete observables of the subsystem $(\mathfrak X_o , \mathfrak S_o)$ if we have \footnote{Furthermore we also have
 $$ \mathfrak X_\oslash = \mathcal Z_{\mathfrak S_o}(\mathfrak X_\oslash) $$ }
$$ \mathfrak X_\oslash = \mathcal Z_{(\mathfrak X_o , \mathfrak S_o)}(\mathfrak X_\oslash) $$ 
 \end{definition}
\begin{remark}\upshape
If we consider the entire laboratory system $(\mathfrak X , \mathfrak S)$, a set of observables $\mathfrak X_o$ is complete if 
$$ \mathfrak X_o = \mathcal C(\mathfrak X_o) $$
since in this case we have
$$\mathcal Z_{(\mathfrak X , \mathfrak S)}(\mathfrak X_o) = \mathcal C(\mathfrak X_o)$$
\end{remark}
\begin{remark}\upshape
If $(\mathfrak X_o , \mathfrak S_o)$ is a\texttt{ MASS}, then the set of observables $\mathfrak X_o$ is not necessarily complete in $(\mathfrak X , \mathfrak S)$.
\\
In fact we cannot generally say that $\mathfrak X_o \subset \mathcal C(\mathfrak X_o)$, because it is not necessarily true that the observables of $\mathfrak X_o$ are mutually $\mathfrak S$-compatible with each other; we can only say that 
$$\mathfrak X_o = \mathcal C_{\mathfrak S_o}(\mathfrak X_o)$$ 
therefore 
$$ \mathcal C(\mathfrak X_o) \subset \mathcal C_{\mathfrak S_o}(\mathfrak X_o) \ \Longrightarrow \ \mathcal C(\mathfrak X_o) \subset \mathfrak X_o $$
In other words, from the \texttt{MASS} property we obtain that every observable of $\mathfrak X$ which is $\mathfrak S$-compatible with the elements of $\mathfrak X_o$ belongs to $\mathfrak X_o$. 
\end{remark}
\section{Topologies in Suitable Systems}
Let us consider a suitable physical system $(\mathfrak X_o , \mathfrak S_o)$ in which we can define particular topologies on the set of its observables and states.

\subsection{Topologies on the Set $\mathfrak X_o$}
On the set of observables $\mathfrak X_o$ we can introduce a metric space structure defined by the following metric:
\begin{equation}\label{distanza}
d(a, b) := \sup_{\omega \in \mathfrak S_o} \left| \left\langle a \right\rangle_{\omega} - \left\langle b \right\rangle_{\omega} \right| \ , \qquad \forall a, b \in \mathfrak X_o
\end{equation}
and for every $a \in \mathfrak X_o$ it results:
\begin{equation} \label{normainselezione}
d(a, 0) = \| a \|_{\mathfrak S_o} := \sup_{\omega \in \mathfrak S_o} | \left\langle a \right\rangle_{\omega} | 
\end{equation}
where the observable $0$ is defined by functional calculus, with $C(t) = 0$:  
$$\left\langle C(x) \right\rangle_{\omega} = 0 \qquad \forall x \in \mathfrak X_o, \ \omega \in \mathfrak S_o$$
and $0 = C(x)$ for every observable $x$ of $\mathfrak X_o$. 
\\
Thus we will say that $a_\alpha \stackrel{\alpha}{\longrightarrow} a_o$ in the norm topology if $d(a_\alpha, a_o) \stackrel{\alpha}{\longrightarrow} 0$.
\\
Note that if we have a sequence of continuous real bounded functions $f_\alpha$ which converges in uniform norm to $f$, then we have
$$d(f_\alpha(a), f(a)) \stackrel{\alpha}{\longrightarrow} 0$$
We introduce another topology, weaker than the previous one, induced by the following subbase of open sets:
$$U_{\epsilon,\omega}(a_o) = \left\{ a \in \mathfrak X_o : | \left\langle a \right\rangle_{\omega} - \left\langle a_o \right\rangle_{\omega} | < \epsilon \right\} \ , \qquad \forall \omega \in \mathfrak S_o, \ \epsilon \in \mathbb R^+ $$
This topology will be denoted by \textit{$w$-top}.
\\
A net $a_\alpha \stackrel{\alpha}{\longrightarrow} a_o$ in the $w$-top if and only if 
$$\left\langle a_\alpha \right\rangle_{\omega} \stackrel{\alpha}{\longrightarrow} \left\langle a_o \right\rangle_{\omega} \ , \qquad \forall \omega \in \mathfrak S_o$$
Obviously, if $a_\alpha \stackrel{\alpha}{\longrightarrow} a_o$ in the norm topology, then it implies that $a_\alpha \stackrel{\alpha}{\longrightarrow} a_o$ in the $w$-top.
\subsection{Topologies on the Set $\mathfrak S_o$}
On the set of states $\mathfrak S_o$ we can introduce the following metric:
\begin{equation}
\label{normstate}
d(\omega_1 , \omega_2) := \sup_{a \in \mathfrak X_o} \frac{| \left\langle a \right\rangle_{\omega_1} - \left\langle a \right\rangle_{\omega_2} |}{\| a \|_{\mathfrak S_o}} \ , \qquad \omega_1, \omega_2 \in \mathfrak S_o
\end{equation}
where $\| a \|_{\mathfrak S_o}$ is the norm of the observable defined in \eqref{normainselezione}.
\\
Furthermore we can also give the following weak topology on $\mathfrak S_o$:
\\
A net $\omega_\alpha$ in $\mathfrak S_o$ converges in the *-weak topology ($\textit{w*-top}$) to $\omega_o$ if
$$ \left\langle a \right\rangle_{\omega_\alpha} \stackrel{\alpha}{\longrightarrow} \left\langle a \right\rangle_{\omega_o} \ , \qquad \forall a \in \mathfrak X_o$$
A subbase for this topology is given by the family of open sets:
$$V_{\epsilon,a}(\omega_o) = \left\{ \omega \in \mathfrak S_o : | \left\langle a \right\rangle_{\omega} - \left\langle a \right\rangle_{\omega_o} | < \epsilon \right\} \ , \qquad \forall a \in \mathfrak X_o, \ \epsilon \in \mathbb R^+ $$
Furthermore it is easy to verify that if $\left\langle a \right\rangle_{\omega_\alpha} \stackrel{\alpha}{\longrightarrow} \left\langle a \right\rangle_{\omega_o}$ in the $w*$-top, then we have the convergence $\mu_{\omega_\alpha,a} \stackrel{\alpha}{\longrightarrow} \mu_{\omega_o,a}$ in the $w*$-top of $C_o(\mathbb R)^*$ for every $a \in \mathfrak X_o$.
\\
 
In addition to the weak notion of convergence we can also introduce the following notion of strong convergence ($\tau_F$-top):
\\
We have that $\omega_\alpha \stackrel{\tau_F}{\longrightarrow} \omega_o$ if $d(\omega_\alpha , \omega_o) \rightarrow 0$.
\\
Obviously, strong convergence implies *-weak convergence.
\section{Mackey's Systems}\label{Mack_syst}
Let us remember again that a pair $(\mathfrak X_o , \mathfrak S_o)$ is suitable if:
$$ \mathfrak S_o \subset \bigcap_{x \in \mathfrak X_o} \mathfrak S_x \qquad \qquad \qquad \qquad \left( \text{equivalently} \qquad \mathfrak X_o \subset \bigcap_{\omega \in \mathfrak S_o} \mathfrak X_\omega \right) $$
\begin{definition}\upshape
Let $(\mathfrak X_o , \mathfrak S_o)$ be a physical subsystem of the laboratory $L$\footnote{See definition on page \pageref{Sottosistemi_Fisici}}. It is said to be a Mackey system if
\begin{itemize}
\item[a.] It is a suitable subsystem;
\item [b.] for every $a, b \in \mathfrak X_o$ such that
$$ \left[ \ \mu_{\omega,a} = \mu_{\omega,b} \qquad \forall \omega \in \mathfrak S_o \ \right] \qquad \Longrightarrow \qquad a = b$$ 
\item [c.] for every $\omega, \omega' \in \mathfrak S_o$ such that
$$ \left[ \ \mu_{\omega,a} = \mu_{\omega',a} \qquad \forall a \in \mathfrak X_o \ \right] \qquad \Longrightarrow \qquad \omega = \omega' $$ 
\item [d.] [\textbf{Convexity}] For every $\omega_1$ and $\omega_2$ belonging to $\mathfrak S_o$ and for every real number $r \in [0, 1]$, there exists a state $\omega \in \mathfrak S_o$ such that, for each $a \in \mathfrak X_o$,  
\begin{equation}\label{convex_mack}
\mu_{\omega ,a} = (1 - r) \mu_{\omega_1 ,a} + r \mu_{\omega_2 ,a} 
\end{equation}
\end{itemize}
\end{definition}
In this way, the state $\omega$ in \eqref{convex_mack} is unique, and for this we will adopt the following notation\footnote{Obviously we can generalize this writing in the following way:
$$\omega = \sum_{i} r_i \omega_i \ , \qquad \sum_{i} r_i = 1 \ , \qquad r_i \geq 0 \ \forall i$$
where
$$ \mu_{\omega, a} = \sum_{i} r_i \mu_{\omega_i, a} \ , \qquad \forall a \in \mathfrak X$$ }:
\begin{equation}
\label{mistura1}
\omega = (1 - r) \omega_1 + r \omega_2
\end{equation}
We denote by $\mathfrak S_o^p$ the pure states of the Mackey system, i.e., the states of $\mathfrak S_o$ which are not mixtures expressed by \eqref{mistura1} for $r \in ]0, 1[$.
\\
Thus, by definition, a Mackey system, for each of its observables $a$, admits \textit{only one measurement sector} given by 
$$\mathbb M_o(a) = \left\{ \mu_{\omega,a} \in \Pi : \omega \in \mathfrak S_o \right\}$$

Let us now return briefly to a problem left open on page \pageref{spettrocongiunto1} concerning the joint spectrum of two compatible observables, with the following 
\begin{proposition}\upshape \label{spettrocongiunto2}
If $a$ and $b$ are independent observables of the Mackey system $(\mathfrak X_o , \mathfrak S_o)$, then 
$$\sigma_{\mathfrak S_o}(a:b) = \sigma_{\mathfrak S_o}(a) \times \sigma_{\mathfrak S_o}(b)$$
\end{proposition}
\begin{proof}
By Proposition \ref{spettrocongiunto1} on page \pageref{spettrocongiunto1}, we only need to prove the inclusion $\sigma_{\mathfrak S_o}(a) \times \sigma_{\mathfrak S_o}(b) \subset \sigma_{\mathfrak S_o}(a:b)$.
\\
Let $\lambda_1 \in \sigma_{\mathfrak S_o}(a)$ and $\lambda_2 \in \sigma_{\mathfrak S_o}(b)$. By relation \eqref{puntispettro-2} on page \pageref{puntispettro-2}, for each open set $U_1$ containing $\lambda_1$ and $U_2$ containing $\lambda_2$, respectively, there exist $\omega_1, \omega_2 \in \mathfrak S_o$ such that $\mu_{\omega_1,a}(U_1) \neq 0$ and $\mu_{\omega_2,b}(U_2) \neq 0$. 
\\
By relation \eqref{mistura1}, we obtain $\omega = (1 - r) \omega_1 + r \omega_2 \in \mathfrak S_o$; it follows that
$$ \mu_{\omega, a:b} (U_1 \times U_2) = \mu_{\omega, a}(U_1) \cdot \mu_{\omega, b}(U_2) \neq 0$$ 
so we have 
$$ \mathbf{1}_{U_1 \times U_2}(a:b) \neq 0 \ \Longrightarrow \ (\lambda_1, \lambda_2) \in \sigma_{\mathfrak S_o}(a:b)$$
\end{proof}
%
%%%
\section{Classical and Quantum Systems}
As we have pointed out, the compatibility of two or more observables depends on the set of states that we use to make our measurements, a set fixed by the experimenter through the measuring instruments that he uses to prepare and measure physical quantities. Increasing the possible states of the system means increasing the complexity of the experiment, of the measurements that can be made on the observables of the system, thus increasing our knowledge of them. Therefore, the center of the observables of the system defined by Definition \ref{commutante} on page \pageref{commutante} depends on the set of states that we take into consideration.
\\
Let us reiterate the concept: \textit{increasing the number of measuring instruments and their preparations essentially means increasing the possibility of determining a state for which two observables can be complementary. }
\subsection{The Classical System}
Let us analyze the following set of states of our physical laboratory system:
\begin{equation}\label{set_abeliano}
\mathfrak F_{cl} = \left\{ \mathfrak S_o \subset \mathfrak S \ , \ Z_{\mathfrak S_o} (\mathfrak X^{\mathfrak S_o}_\wedge ) = \mathfrak X^{\mathfrak S_o}_\wedge \right\} 
\end{equation} 
For each $\mathfrak S_o \in \mathfrak F_{cl}$ we consider the pair $(\mathfrak X^{\mathfrak S_o}_\wedge , \mathfrak S_o)$, which is physically achievable and suitable\footnote{Warning: It is not certain that condition [C.] of temporal evolution is satisfied; even if it were, it is not guaranteed that the physical subsystem obtained is of Mackey type.}. 
 \\
Furthermore, the set $\mathfrak F_{cl}$ is partially ordered by inclusion.
\begin{proposition}\upshape
If the set $\mathfrak F_{cl}$ is non-empty\footnote{We note that the existence of a set of compatible observables of the laboratory physical system, as per section \ref{sist_abel} on page \pageref{sist_abel}, does not guarantee that the set $\mathfrak F_{cl}$ is different from the empty set.}, then it admits at least one maximal element.
\end{proposition}
\begin{proof}
Consider any linearly ordered subset
$$\mathfrak F_0 = \left\{ \mathfrak S_k : k \in \mathbb N \right\} \subset \mathfrak F_{cl}$$
and assume that $\mathfrak S_i \subset \mathfrak S_j$ for $i < j$. Consider the following set of states
\begin{equation}\label{stati_maxi}
 \mathfrak S_\infty = \bigcup_{k \in \mathbb N} \mathfrak S_k 
\end{equation}
We have to check that $Z_{\mathfrak S_\infty} (\mathfrak X^{\mathfrak S_\infty}_\wedge ) = \mathfrak X^{\mathfrak S_\infty}_\wedge$; since by definition we have 
$$Z_{\mathfrak S_\infty} (\mathfrak X^{\mathfrak S_\infty}_\wedge ) \subset \mathfrak X^{\mathfrak S_\infty}_\wedge$$
only the reverse inclusion will need to be checked.
\\
We observe that for each natural number $k$, it holds that
$$\mathfrak S_k \subset \mathfrak S_\infty \qquad \Longrightarrow \qquad \mathfrak X^{\mathfrak S_\infty}_\wedge \subset \mathfrak X^{\mathfrak S_k}_\wedge$$
furthermore
$$ Z_{\mathfrak S_\infty} (\mathfrak X^{\mathfrak S_\infty}_\wedge ) \subset \mathfrak X^{\mathfrak S_\infty}_\wedge \subset \mathfrak X^{\mathfrak S_k}_\wedge = Z_{\mathfrak S_k} (\mathfrak X^{\mathfrak S_k}_\wedge ) $$
Let us now assume that there exists an element $a \in \mathfrak X^{\mathfrak S_\infty}_\wedge$ that does not belong to the central set $Z_{\mathfrak S_\infty} (\mathfrak X^{\mathfrak S_\infty}_\wedge )$:
$$a \in \mathfrak X^{\mathfrak S_\infty}_\wedge \qquad \text{such that } \qquad a \notin Z_{\mathfrak S_\infty} (\mathfrak X^{\mathfrak S_\infty}_\wedge )$$
This means that there must exist at least one state $\omega \in \mathfrak S_\infty$ and an observable $x \in \mathfrak X^{\mathfrak S_\infty}_\wedge$ that is not compatible with $a$ in state $\omega$.
\\
By definition there must exist a $k$ for which $\omega \in \mathfrak S_k$ with $a, x \in \mathfrak X^{\mathfrak S_k}_\wedge$; in other words, we have determined an observable $x$ of $\mathfrak X^{\mathfrak S_k}_\wedge$ not compatible in the state $\omega$ with an element $a$ of $Z_{\mathfrak S_k} (\mathfrak X^{\mathfrak S_k}_\wedge )$, contradicting the hypothesis that such a set coincides with all of $\mathfrak X^{\mathfrak S_k}_\wedge$.
\\
The thesis follows from Zorn's lemma.
\end{proof}
\begin{definition}\upshape
For every maximal element $\mathfrak S_o \in \mathfrak F_{cl}$ whose associated pair $(\mathfrak X^{\mathfrak S_o}_\wedge , \mathfrak S_o)$ turns out to be a Mackey subsystem, it is called a classical system.
\end{definition} 
From physical experience we can say that
\begin{axiom}[\textbf{Classical World}]\index{Axiom-Classical World}
The laboratory system $(\mathfrak X , \mathfrak S)$ admits at least one classical physical system $(\mathfrak X_{cl}, \mathfrak S_{cl})$.  
\end{axiom}
In summary, a classical system $(\mathfrak X_{cl}, \mathfrak S_{cl})$ is an Abelian Mackey system consisting of a maximal set of states for which all observables of the system are compatible:
$$ Z_{\mathfrak S_{cl}}(\mathfrak X_{cl}) = \mathfrak X_{cl}$$
\subsection{The Pure Quantum System}
Similar to the classical case, consider the following family of states of the physical laboratory system:
\begin{equation}\label{set_quantistico}
\mathfrak F_{mq} = \left\{ \mathfrak S_o \subset \mathfrak S \ : \ Z_{\mathfrak S_o} (\mathfrak X^{\mathfrak S_o}_\wedge ) = \mathbb R I \right\} 
\end{equation} 
Here, too, for each $\mathfrak S_o \in \mathfrak F_{mq}$ we consider the pair $(\mathfrak X^{\mathfrak S_o}_\wedge , \mathfrak S_o)$, which is physically achievable and suitable.
\\
As in the classical case, if $\mathfrak F_{mq}$ is non-empty, then every linearly ordered family of states $\mathfrak F_o \subset \mathfrak F_{mq}$ admits a maximal element, given by the set of states defined in \eqref{stati_maxi}, since for each natural number $k$ we obtain
$$ \mathfrak X^{\mathfrak S_\infty}_\wedge \subset \mathfrak X^{\mathfrak S_k}_\wedge \qquad , \qquad Z_{\mathfrak S_\infty} (\mathfrak X^{\mathfrak S_\infty}_\wedge) \subset Z_{\mathfrak S_k} (\mathfrak X^{\mathfrak S_k}_\wedge) = \mathbb R I $$
\begin{definition}\upshape
Each maximal element $\mathfrak S_o \in \mathfrak F_{mq}$ whose associated pair $(\mathfrak X^{\mathfrak S_o}_\wedge , \mathfrak S_o)$ turns out to be a Mackey subsystem is called a purely quantum system.
\end{definition} 
\begin{axiom}[\textbf{Purely Quantum World}]\index{Axiom-Purely Quantum World}
The laboratory system $(\mathfrak X , \mathfrak S)$ admits at least one purely quantum physical system $(\mathfrak X_{mq}, \mathfrak S_{mq})$.  
\end{axiom}
Thus, an elementary quantum system $(\mathfrak X_{mq}, \mathfrak S_{mq})$ is a suitable physical subsystem consisting of a maximal set of states for which all observables in the system are mutually incompatible:
$$ Z_{\mathfrak S_{mq}}(\mathfrak X_{mq}) = \mathbb R I$$
Experimentally we have that classical observables are also quantum observables, so 
$$\mathfrak X_{cl} \subset \mathfrak X_{mq}$$
Mind you, the observables $\mathfrak X_{cl}$ are not, in general, compatible observables of the system $(\mathfrak X_{mq}, \mathfrak S_{mq})$. 
\begin{remark}\upshape
If we have two physically achievable pairs $(\mathfrak X_o, \mathfrak S_o)$ and $(\mathfrak X_\sharp, \mathfrak S_\sharp)$ with $\mathfrak X_o \subset \mathfrak X_\sharp$, then by definition, for every observable $x \in \mathfrak X_o$ there exists a state $\omega \in \mathfrak S_o$ suitable for $x$; but since $x$ is also in $\mathfrak X_\sharp$, there will exist a state $\omega' \in \mathfrak S_\sharp$ suitable for it\footnote{Not unique and dependent on $x$.}. 
\end{remark}
This last observation authorizes us to affirm that the states $\mathfrak S_{mq}$ are more numerous than $\mathfrak S_{cl}$, since each state $\omega$ of $\mathfrak S_{cl}$ corresponds to at least one state $\omega'$ of $\mathfrak S_{mq}$.  
 \\ 
Furthermore, with the $\omega'$ state of $\mathfrak S_{mq}$ being more experimentally structured, I can measure \textit{more things} compared to its corresponding $\omega$ state of the classical case:
\begin{equation*}
\mathfrak X_\omega \subset \mathfrak X_{cl} \qquad , \qquad \mathfrak X_\omega \subset \mathfrak X_{\omega'} \subset \mathfrak X_{mq}
\end{equation*}
\subsection{Hidden Variables}
Let us ask ourselves the following question:
\\
When is it possible to describe a generic physical system $(\mathfrak X_o, \mathfrak S_o)$ in a classical way?
\\
Let us see exactly what this question means.
\\

There exists a \textit{classical system} described by the pair $(\mathfrak X_{\ast}, \mathfrak S_{\ast})$, not necessarily induced by our laboratory\footnote{In other words, it is not a physical subsystem of our laboratory, i.e., the observables $\mathfrak X_{\ast}$ and states $\mathfrak S_{\ast}$ are not observables of $\mathfrak X$ and $\mathfrak S$, respectively, of our laboratory $L$.}, and surjective maps $(\texttt{v}, \texttt{v}^\circ)$:
$$\texttt{v}: \mathfrak X_o \hookrightarrow \mathfrak X_{\ast} \qquad , \qquad \texttt{v}^\circ: \mathfrak S_o \hookrightarrow \mathfrak S_{\ast}$$
such that for each $a \in \mathfrak X_o$ and $\omega \in \mathfrak S_o$
\begin{itemize}
\item[a.] $\qquad \texttt{v}^\circ\left( \mathfrak S^o_a \right) \subset \mathfrak S^{\ast}_{\texttt{v}(a)} \qquad , \qquad \texttt{v}\left( \mathfrak X^o_\omega \right) \subset \mathfrak X^{\ast}_{\texttt{v}^\circ(\omega)}$  
\item[b.] $\qquad \widetilde{\mu}_{\texttt{v}^\circ\omega, \texttt{v}a} = \mu_{\omega,a}$,
\\
where $\widetilde{\mu}_{\omega, x}$ is the probability measure related to the classical system $(\mathfrak X_{\ast}, \mathfrak S_{\ast})$. 
\item[c.] $\qquad \texttt{v}(f(a)) = f(\texttt{v}(a))$ \qquad for each $f \in L^1(a)$ 
\end{itemize}
is said to be a \textit{parametric pair of maps} for our physical system. 
\\
If there exists such a parametric pair of maps, we have
$$ \left\langle f(a) \right\rangle_{\omega} = \left\langle \texttt{v}(f(a)) \right\rangle_{\texttt{v}^\circ\omega} = \int f(t) \, d \widetilde{\mu}_{\texttt{v}^\circ\omega, \texttt{v}a}(t) $$
We consider the variance $\Delta_{\omega}(a)$ of the observable $a$ in the state $\omega \in \mathfrak S_a$\footnote{See section \ref{FD-states} on page \pageref{FD-states}.}.
\\
We note that 
$$\Delta_{\texttt{v}^\circ \omega}(\texttt{v}(a)) = \Delta_{\omega}(a)$$
Therefore if $\omega$ is a free dispersion state in the measurement of $a$, then $\texttt{v}^\circ \omega$ is a free dispersion state in the measurement of $\texttt{v}(a)$.
\\
We have the following open question\footnote{To learn more about hidden variables, see Jammer's book, Chapter 7.}:
\begin{problem}\upshape
For which physical subsystems $(\mathfrak X_o , \mathfrak S_o)$ does such a classical parametrization exist?
\end{problem}
\section{Inferences}\label{inferenze}
Paraphrasing the definition in the encyclopaedia \cite{Treccani}, inference can be considered as a generalization of the results obtained through a partial sample survey, i.e., limited to the consideration of a few individual cases of the phenomenon under study, to the totality of the cases of the phenomenon itself, on the basis of \textit{plausible hypotheses}\footnote{See also footnote \ref{bayes0} on page \pageref{Statistica e Riproducibilità}.}. 
\\

In this section we will briefly study this notion by applying it to our interpretative model of values obtained through experiments.
\\

Let $a$ be an observable of the system, and let us assume that the experimenter does not have at his disposal all the states suitable for $a$ but only a part of them, $\mathfrak S_a^o \subset \mathfrak S_a$. In this case we only have partial information on the observable due to the distributions 
\begin{equation}
\label{distribuzio3}
 \left\{ P(a \in \Delta, \tau)_\omega : \omega \in \mathfrak S_a^o \right\}
\end{equation}
as the Borel set $\Delta$ varies.\index{States of sufficiently informative} 
\\
Let us ask when the set $\mathfrak S_a^o$ is \textit{sufficiently informative} about the possible values of the observable $a$ of the system, i.e., if there exists a map $E_a : \mathfrak S_a \rightarrow \mathfrak S_a$ such that
\begin{itemize}
\item $E_a(\mathfrak S_a) \subset \mathfrak S_a^o$
\item For every $\Delta \in B(\mathbb R)$ it satisfies 
$$P(a \in \Delta, \tau)_\omega = P(a \in \Delta, \tau)_{E_a(\omega)} \qquad , \qquad \forall \omega \in \mathfrak S_a $$
in other words
$$\mu_{\omega,a} = \mu_{E_a(\omega),a} \qquad , \qquad \forall \omega \in \mathfrak S_a $$ 
\end{itemize}
In this way for the average value of the observables we obtain:
$$\left\langle a \right\rangle_\omega = \left\langle a \right\rangle_{E_a(\omega)} \ , \qquad \omega \in \mathfrak S_a $$
The map $E_a : \mathfrak S_a \rightarrow \mathfrak S_a$ is called the \textit{projector of the states in the measurement of $a$}\footnote{Warning: This map does not necessarily always exist.}.
 $$ \star \star \star $$
As discussed in section \ref{misureBorele} on page \pageref{misureBorele}, for experimental reasons we can establish the values of relation \eqref{distribuzio1} only for some Borel sets $\Delta$ of $\mathbb R$; let $\mathfrak F_o \subset B(\mathbb R)$ be such a family of subsets. 
\\
We now weaken the hypotheses on the family $\mathfrak F_o$ by considering it not as a $\sigma$-algebra but as a $\pi$–$\lambda$ system. By the monotone class theorem, if $\mathfrak F \subset B(\mathbb R)$ is the $\sigma$-algebra generated by $\mathfrak F_o$, then there exists a unique measure $\tilde{\mu}_{\omega,a}$ on $\mathfrak F$\footnote{See Bobrowski's book \cite{Bobrowski} par. 1.2.7.\index{Bobrowski}} such that 
$$\tilde{\mu}_{\omega,a}(\Delta) = P(a \in \Delta, \tau)_\omega \ , \qquad \Delta \in \mathfrak F $$
Thus we have two probability measure spaces $(\mathbb R, \mu_{\omega,a}, B(\mathbb R))$ and $(\mathbb R, \tilde{\mu}_{\omega,a}, \mathfrak F)$, and as is known there exists a positive map ( \textit{conditional expectation})\index{Conditional expectation}
$$\mathcal{E}_{a,\omega} : L^1(\mu_{\omega,a}) \rightarrow L^1(\tilde{\mu}_{\omega,a})$$
such that for every $f \in L^1(\mu_{\omega,a})$ we have:
$$\int_\Delta f \, d\mu_{\omega,a} = \int_\Delta \mathcal{E}_{\omega,a}(f) \, d\tilde{\mu}_{\omega,a} \qquad , \qquad \forall \Delta \in \mathfrak F$$
in other words
$$ \mu_{\omega,a}(f) = \tilde{\mu}_{\omega,a}(\mathcal{E}_{\omega,a}(f)) \qquad , \qquad \forall f \in L^1(\mu_{\omega,a})$$
and in particular
$$ P(a \in \Delta, \tau)_\omega = \left\langle \mathbf{1}_\Delta(a) \right\rangle_{\omega} = \int \mathcal{E}_{\omega,a}(\mathbf{1}_\Delta)(t) \, d\tilde{\mu}_{\omega,a}(t) \qquad , \qquad \forall \Delta \in B(\mathbb R) $$ 
Thus for every function $f$ that is $a$-summable we obtain:
$$ \left\langle f(a) \right\rangle_{\omega} = \int \mathcal{E}_{\omega,a}(f) \, d\tilde{\mu}_{\omega,a} $$

\part{Measurement Procedures in More Laboratories}

\chapter{Reference and States}
In this chapter we resume the study of the physical laboratory system from the operational point of view, addressing the various problems avoided in section \ref{Correlazione-stati-Osservabili} of the initial chapter.
\\
Let's start the discussion by remembering that in our case \textit{to observe} is synonymous with \textit{to measure}, and to measure you must have instruments at hand. It makes no experimental sense to say that we observe a physical phenomenon in a given reference system without saying how to carry it out; to do so, you need to have a laboratory available to carry out measurements, equipped at least with rulers for distances and clocks for time. In this way, a reference system \textit{is a real measurement apparatus} to be positioned at a chosen point $O$ of the laboratory; without it we cannot establish the physical state of the system, hence its role as the primary instrument of the laboratory follows.
\section{Clocks, Rulers and States}\label{rif_pb}
In this section we will highlight what role the adopted reference system plays in the measurement process of physical quantities. We assume that the laboratory has a spatial extension given by a bounded, open and connected set $L_o$ of a Euclidean topological space $\mathfrak E$\footnote{In general it may turn out not to be simply connected.}.
\\
In agreement with Einstein (see \cite{Einstein}), we assume that at \textit{every point of the laboratory} it is possible to associate clocks that are all synchronized with a clock located at a well-fixed point $O$ of the laboratory that the experimenter uses to establish when events happen, and we assume that
\begin{post}\index{Einstein}\index{Postulate-Synchronized Clocks}\label{ipotesiorologiosincro}
All clocks positioned anywhere in the laboratory, once synchronized, keep the same time as our reference clock positioned at $O$.
\end{post}
We note that this postulate is not always true; for example, clocks run more or less slowly in the presence of gravity, which must be taken into account if we consider very large laboratories or have instruments with a large mass\footnote{In general we must also ask ourselves the question of whether the graduated scales of the various instruments depend on their location in the laboratory.}.
\\
From this hypothesis, it follows that every instrument in the laboratory must be subject to the same acceleration (which can even be zero) and this \textit{excludes the possibility of there being laboratory devices in relative motion with each other}.
\\
Let's make a further clarification on this last statement:
\\
A measuring instrument occupies a more or less large space in the laboratory; inside it we can have moving mechanisms that allow the instrument to carry out its measuring function. What we are stating is that \textit{the reading of the value of the measurement given by the instrument at a given time $\tau$} is carried out with a clock that satisfies our postulate \ref{ipotesiorologiosincro}.
\\

For example, let us assume we have a geostationary satellite with various instruments and an operations centre at an Earth base where we place our $O$ clock. One might think of extending the Earth laboratory to include the satellite and the base itself. In our definition of a laboratory, this cannot happen, since, by definition, if we place a clock anywhere in the laboratory, as established by postulate \ref{ipotesiorologiosincro}, it must always mark the same time as indicated by the clock at $O$. In this case, the satellite is moving and therefore time flows differently from our reference clock, and the reading of the measurements taken by the satellite is carried out in the satellite (and then possibly communicated to the ground base)\footnote{We will see that what we can do is to consider two distinct laboratories: the satellite laboratory and the ground-base laboratory.}.
\\ 

So far we have only used a clock fixed at a point $O$ in space, thus obtaining the laboratory-type region:  
\begin{equation}
\label{regionelab0}
\mathcal O_o = L_o \times [0, t_p] \subset \mathfrak E \times \mathbb R
\end{equation}
\textit{But when do spatial coordinates come into play?}
\\

In order to establish the state of the laboratory system, the experimenter must have the ability to say where the various instruments are operating and when they are activated. For this purpose we establish a methodology to label the points of the laboratory environment space; to do this we will use 
oriented rulers\footnote{If desired, we can also use rulers and goniometers, but it is always better to limit the use of other types of instruments with their associated physical quantities, such as angles.}, which we will denote by $K = (e^1, e^2, e^3)$, and thus obtain spatial coordinates with origin at $O$, where our reference clock resides. We denote this space-time reference by $(K, O)$\footnote{In practice, with $K$ we also indicate the three numbers that are used to identify the points in space with respect to the point $O$, plus the time value of the laboratory indicated by the clock positioned at $O$.}.
\\
The question we ask ourselves now is the following: \textit{Does the state of the system depend on the choice of the spatial reference $K$ that we have taken into consideration?}
\\
Physically, in establishing the state $\omega$ of the system in the measurement of its observable $a$, what is important is the position, and hence the mutual position, of the various devices in the laboratory, which would seem to be independent of the spatial coordinates we adopt to indicate them. \textit{Is this sufficient to affirm that the state does not depend on the choice of the adopted coordinate reference $(K, O)$?}
\\
The answer is no, because clocks and rulers are themselves devices that the experimenter uses in the laboratory, and changing their position could change the state of the system\footnote{Also because in order to note down in my notebook the procedures, etc., that I carry out in the laboratory, I have to rely on a reference system that is also noted down, so as to indicate to a new experimenter the operations to be carried out in order to reproduce the same experiment.}.
\begin{remark}\upshape
The chosen point $O$ must be taken inside the laboratory $L_o$, because all measuring instruments/equipment must be in the laboratory, and placing the origin of the reference system outside the laboratory would mean changing the very extent of the laboratory.
\end{remark}
In summary, we imagined positioning our experimenter at a point $O$ of $L_o \subset \mathfrak E$ and fixing a reference system $K$ centered at $O$ to identify the relative position of the instruments/devices for measurement.
\\
In other words, at $O$ we have positioned our operations center (which we have briefly denoted by the name of operator, experimenter, observer, etc.) which prepares, activates the various devices present in the laboratory, records and analyzes and transmits their data.
\\
In this way, once the state of the system has been established, we can consider the $L_o$ laboratory as a single measuring instrument, an instrument equally spread across the entire $L_o$ region which detects the value of our observable to be measured in our state of the system at a given time $\tau$ established by our clock positioned at $O$.
\\
  \begin{figure}[htbp]
	\centering
		\includegraphics[scale=0.3]{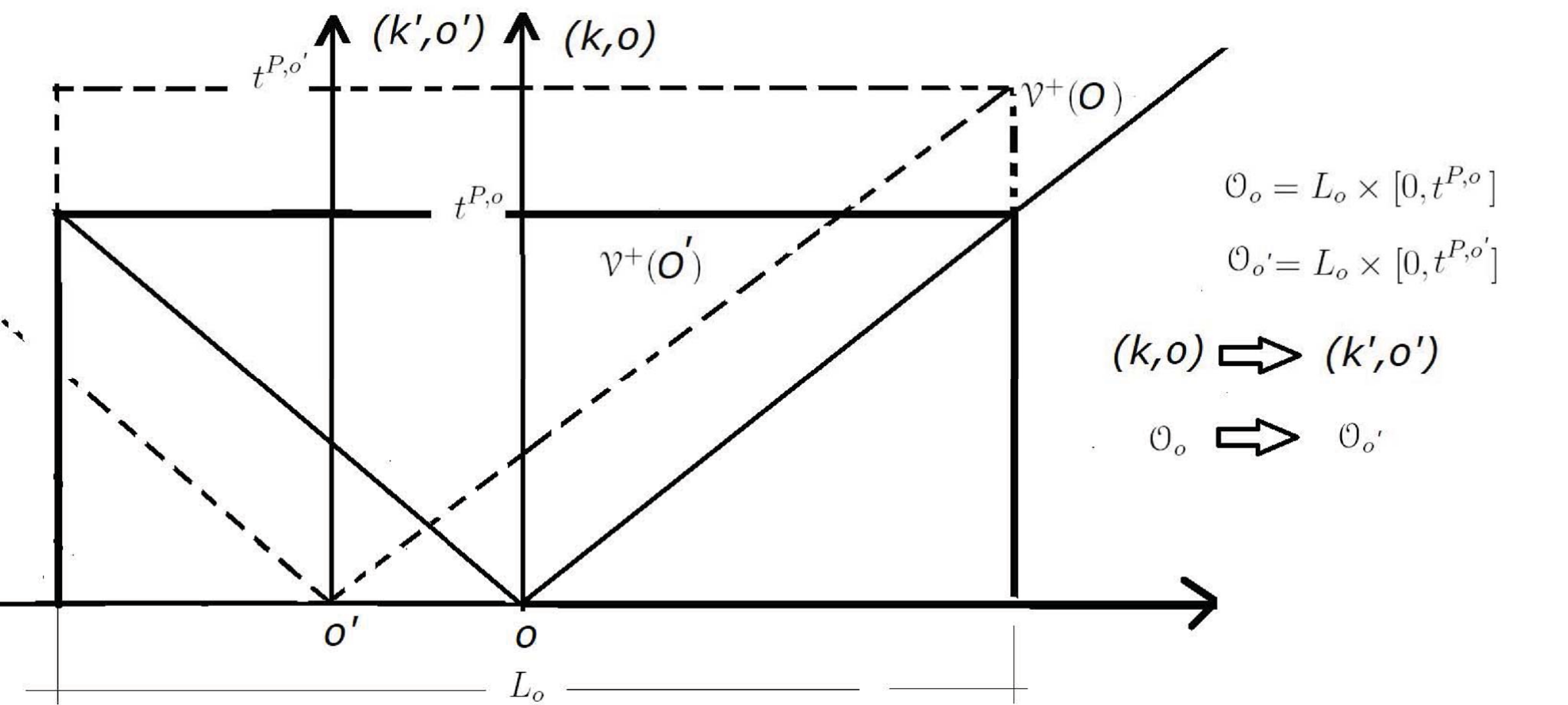}
	\caption{R.S. change and preparation time}
	\label{fig:disle.8}
\end{figure}
\subsection{Reference Problems}\label{Reference Problems}
Let us ask ourselves what happens if we change our reference system $(K,O)$ to another $(K',O')$ with $O, O' \in L_o$.
\\
What changes in the passage of coordinates
\begin{equation}\label{transform0}
 S = (K,O) \longmapsto S' = (K',O')
\end{equation}
is the state of the laboratory system $\omega$.
\\
Furthermore, in addition to the problems of the delay in readings, we must also consider the modification of the activation time of the various devices, which the operator must take into account as shown in Figure \ref{fig:08bis}.
\\
For example, if we change the coordinate system from $(K, O)$ to $(K', O')$ as shown in Figure \ref{fig:disle.8}, the "minimal" times to perform the various experimental procedures change, and this leads to the following
\begin{definition}[\textbf{Illuminated Region}]\upshape\label{illuminata} \index{Illuminated region}
A laboratory-type region
$$\mathcal O_o = L_o \times [0, t_p] \subset \mathbb R^3 \times \mathbb R$$
is said to be illuminated if, in each fixed reference system $(K, O)$, the experimenter at $O \in L_o$ manages to "illuminate" the entire laboratory with the fixed preparation time $t_p$\footnote{This means that each ray of light from a torch fixed at $O$ manages to illuminate our entire laboratory in the time interval $[0, t_p]$; this is possible because our laboratory $L_o$, no matter how large, is always spatially limited.
\\
We note that in Figure \ref{fig:disle.8} the laboratory-type region $\mathcal O_o$ is obviously not illuminated.}.
\end{definition}
We assume that the coordinate change \eqref{transform0} induces a bijective transformation
\begin{equation}\label{transform1}
\Theta : \mathfrak S(L_o) \longrightarrow \mathfrak S(L_o)
\end{equation}
with
\begin{equation}\label{transform2}
 \Theta(\mathfrak S_a(L_o)) = \mathfrak S_a(L_o) \ , \qquad \forall a \in \mathfrak X(L_o)
\end{equation}
and for each $a \in \mathfrak X(L_o)$ determines a map
\begin{equation} 
 \mu_{\omega, a} \in \mathbb{M}(a) \longmapsto \mu_{\Theta(\omega), a} \in \mathbb{M}(a) 
\end{equation}
where the set $\mathbb{M}(a)$ is defined in \eqref{misure_in_a}.
\begin{remark}\upshape
Obviously nothing prevents us from placing the experimenter outside the laboratory $L_o$; this would mean extending the laboratory, a situation we will study in the next sections.
\end{remark}
 \begin{figure}[htbp]
	\centering
		\includegraphics[scale=0.7]{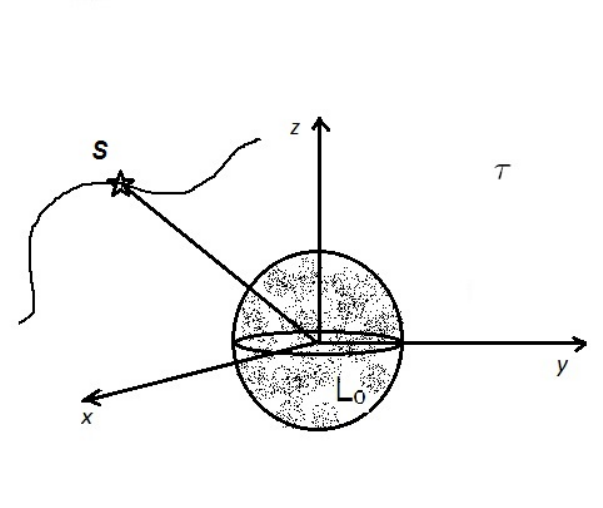}
	\caption{Source}
	\label{fig:08aa}
\end{figure}
Before concluding we must make some banal but important remarks on the source $S$, referred to in the first chapter as the source of the measurement, which is nothing other than the object of investigation of our laboratory $L_o$.
\\
Unlike the measuring devices/equipment $D$, which must reside in the laboratory $L_o$, the source $S$ may not. For example, if we want to determine the velocity or position of an object $S$ moving in space, as depicted in Figure \ref{fig:08aa}, we have that $S$ is intercepted by the instruments of $L_o$ at time $\tau$, identified through a reference system.
\\
We reiterate the concept with the following
\begin{remark}\upshape\label{misure_oss}
When we speak of observables of the laboratory system $L_o$ located in the laboratory-type region $\mathcal O_o$, we mean that the measurement of the physical quantity takes place in this region of space-time, in a precise reference system\footnote{Information on the reference system adopted is contained in the $\omega$ state of the system we have chosen for our measurements.}, even if the source $S$, the object of the measurement, is not necessarily contained in it, as in the case of Figure \ref{fig:08aa}. 
\end{remark}
\section{Time Observable} 
Let's start by underlining that we have taken the \textit{clock as a real laboratory measuring instrument}\footnote{See also Fabri's book \cite{Fabri} - section: \textit{gli orologi come strumenti fisici}.}\index{Fabri}.    
\\
Time is a fundamental unit in many systems of units of measurement, and in the International System the definition of standard time is well established thanks to the advent of atomic clocks\footnote{See Muga et al. \cite{Muga1}, Chap. 1.}. It follows that time has the same dignity as a physical quantity as length and mass; therefore, as such, it is an observable of the physical laboratory system.
\\
For the time observable $\texttt{t}$ of our laboratory $L_o$, we can detect the following features:
\begin{itemize}
\item It needs no preparation time, as the clock is already present and functioning, and it is the \textit{first instrument} to be activated in the laboratory.
\item The set of its suitable states $\mathfrak S_\texttt{t}(\mathcal O_o)$ is made up of the various types of clocks and the way they are used, etc.
\item It is prepared before each observable $a$ of the system, so it is jointly prepared in the order $\texttt{t}:a$, since $a$ is measured at a time $\tau$:
$$ \mathfrak S_{\texttt{t}:a}(\mathcal O_o) \subset \mathfrak S_a(\mathcal O_o) \qquad , \qquad \forall a \in \mathfrak X$$
\end{itemize}

Although it sounds like a play on words, we must be careful not to confuse the time observable $\texttt{t}$ with its value $t_O$ marked by the clock positioned in our laboratory at $O \in L_o$\footnote{We will often forget about the clock at $O$ and denote $t_O$ simply as $t$.}. 
\\
We note that for each state $\omega \in \mathfrak S_\texttt{t}(\mathcal O_o)$ which measures the time indicated by our laboratory clock located at $O \in L_o$, we have:
$$ P(\texttt{t} \in \left\{ \tau \right\} : t_O)_\omega = 1 \qquad \Longleftrightarrow \qquad \tau = t_O $$
We observe that this statement is tautological because the clock is the first measuring instrument that our experimenter makes operational in his laboratory.
\\ 
For the spectrum of the observable $\texttt{t}$ the following property is assumed\footnote{We have already discussed that for measurements it would be physically more appropriate to consider only rational values; moreover, we have an \textit{experimental} limitation in determining shorter and shorter time intervals. To date, we are around $3 \times 10^{-19}$ seconds. In other words, assuming that one can take arbitrarily small time intervals is a strong assumption.}: 
\begin{post}[\textbf{Continuous Time}]\label{Conti_time}\index{Postulate-Continuous Time}
$$\sigma(\texttt{t}) = \mathbb R$$ 
\end{post}
Therefore its spectral measure is given by
$$ \mu_{\omega, \texttt{t}}^{t} = \delta_{t} \ , \qquad \forall t \geq 0$$
In summary, the time observable $\texttt{t}$ is jointly preparable with every observable $a$ of the system in the order $\texttt{t}:a$, but it is not compatible with any observable of the laboratory system\footnote{According to our definition given in Definition \ref{definizionecompatibile}.} since it must necessarily be prepared before each observable.
\\
This last statement could be a methodological solution to Pauli's old problem of \textit{time as a parameter}\footnote{See Muga et al. \cite{Muga1}, Chap. 3.}.
\bigskip

To summarize, we have redefined the concept of a reference system as a measurement apparatus located in the laboratory. It follows that:
\begin{itemize}
\item [-]  Time is not an external parameter, but an observable $\texttt{t}$ with a continuous spectrum.
\item [-]  Clock synchronization is a hypothesis, not a mathematical axiom.
\item [-]  A change of reference transforms the state of the system $\Theta$, not just the coordinates.
\item [-]  The source may lie outside the laboratory, but the measurement takes place inside.
\end{itemize}
\section{Inclusion Problems}\label{pbinclus}
In the previous sections we established that each laboratory region $L_o \subset \mathfrak E$ is associated with a set of states $\mathfrak S(L_o)$ which describes \textbf{all} the possible equipment/instruments and their procedures/modes of use, etc., that can be implemented in $L_o$.
\\
We also pointed out that in the real case, not all the devices will be available to those who carry out experiments in the laboratory; however, this experimenter must always have a clock available, positioned somewhere in the laboratory for time, and some oriented rulers to establish the distance and position of each individual device in the laboratory.
\\
We have noticed that changing the reference system leads to a reshuffling of the set of states through the map given in \eqref{transform1}; for example, if we establish that the state of the system with respect to $(K,O)$ is given by $\omega$, then changing the laboratory system from $(K,O)$ to $(K',O')$, the state will change to $\Theta(\omega)$:
$$\omega \in \mathfrak S(L_o) \ \longrightarrow \ \Theta(\omega) \in \mathfrak S(L_o) $$
\begin{figure}[htbp]
	\centering
		\includegraphics[scale=0.3]{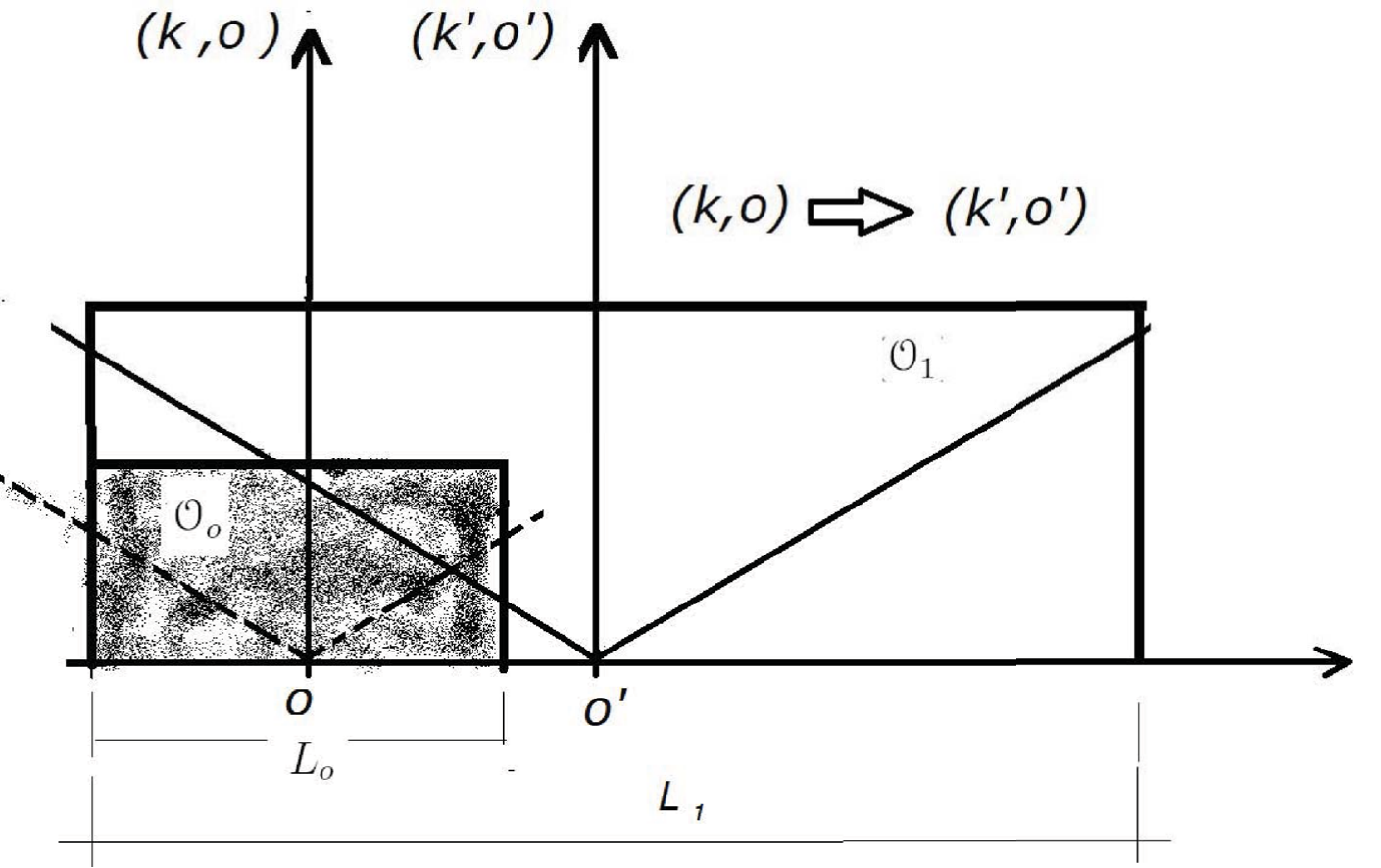}
	\caption{Reference Change - Illuminated Regions}
	\label{fig:disle.9}
\end{figure}
Given a laboratory $L_o$, we can decide to extend its "walls" and consider a larger laboratory $L_1$ that contains it (see Figure \ref{fig:disle.9}). In the laboratory $L_1$, you will have the possibility of having more states of the system than those that can be obtained in $L_o$\footnote{In section \ref{sottolaboratorio}, we will analyze more carefully the relationship existing between the set of states $\mathfrak S(\mathcal O_o)$ and $\mathfrak S(\mathcal O_1)$ when one laboratory-type region contains the other.}.
\\

As we have repeatedly reiterated, the experimenter decides, based on his objectives, which states to choose in the set $\mathfrak S(L_1)$. For example, even if he has extended his laboratory, he could decide to use only that part of the instruments located in $L_o$.
\\
\textit{But how does this selection happen starting from $\mathfrak S(L_1)$?}
 \\
One method to achieve this selection is to choose a set of states $\mathfrak S_{\bullet}$ of $\mathfrak S(L_1)$ given by\footnote{Obviously $\mathfrak S_{\bullet}$ also depends on $L_1$ and $L_o$, so we should have written $\mathfrak S_{\bullet}(L_1, L_o)$, but that would be heavy notation.}:
\begin{equation}\label{labristretto}
 \mathfrak S_{\bullet} = \left\{ \omega \in \mathfrak S(L_1) \ : \ \mathfrak X_\omega(L_1) \subset \mathfrak X(L_o) \right\} \subset \mathfrak S(L_1) 
\end{equation} 
Let us now ask ourselves another question:
\\
\textit{Using the set of states $\mathfrak S_{\bullet}$, can we obtain all the observables of $\mathfrak X(L_o)$?}
\\
The answer to this question is affirmative if the following statement is assumed to be true:
\begin{property}\label{misurabil_sotto}
Given $a \in \mathfrak X(L_o)$, for every $\omega \in \mathfrak S_a(L_o)$ there exists a state (not necessarily unique) $\widehat{\omega} \in \mathfrak S_{\bullet}$ suitable for $a$\footnote{Subsequently we will axiomatically assume that every quantity measurable in $L_o$ is also measurable in $L_1$ (see Axioms \ref{subreg} and \ref{subregB} in section \ref{sottolaboratorio}).}, such that 
$$\mu_{\omega,a} = \mu_{\widehat{\omega},a} $$
 \end{property} 
We now assume property \ref{misurabil_sotto} to be true and fix a laboratory reference system $(K,O)$ for $L_o$; in this way it also becomes a reference for $L_1$, since $O \in L_o \subset L_1$.
\\
Let $a \in \mathfrak X(L_o)$ and $\omega \in \mathfrak S(L_o)$. Using this reference system, we relate a state $\omega_1 \in \mathfrak S_{\bullet} \cap \mathfrak S_a(L_1)$. 
\\
Now if we change the system from $(K,O)$ to $(K',O')$ with $O' \in L_1$ as shown in Figure \ref{fig:disle.9}, the reference system of the laboratory is only $L_1$.
\\
The state $\omega_1 \in \mathfrak S(L_1)$ previously identified will have mutated into the new state $\Theta(\omega_1) \in \mathfrak S(L_1)$. 
\\
We observe that in this case we cannot say that $\Theta(\omega_1)$ satisfies relation \eqref{labristretto} and is therefore still a state of $\mathfrak S_{\bullet}$.
\\
In the next sections we will study the problems related to the inclusion of laboratory $L_o$ with its own reference system into that of $L_1$.
\section{Groups of Transformations*}  
Let us imagine that we have a laboratory centered at a point $O \in \mathfrak E$ and we establish its reference system $(K,O)$. In this way, each laboratory system is associated with a 4-dimensional Euclidean space
$$(K,O) \longmapsto \mathbb R^3 \times \mathbb R  $$
Without going into specifics about the properties of space-time, it is enough to note that once a point $O \in \mathfrak E$ has been fixed, a physical quantity time $\texttt{t}_O$ is associated with it, whose value is indicated by the clock at $O$, and once the rulers $K$ are fixed, we obtain the values of lengths with respect to the chosen point $O$\footnote{In other words, for the four-dimensional manifold $\mathfrak E \times \mathbb R \subset \mathbb R^5$, we have a system of local charts.}.
\\
\begin{figure}[htbp]
	\centering
		\includegraphics[scale=0.6]{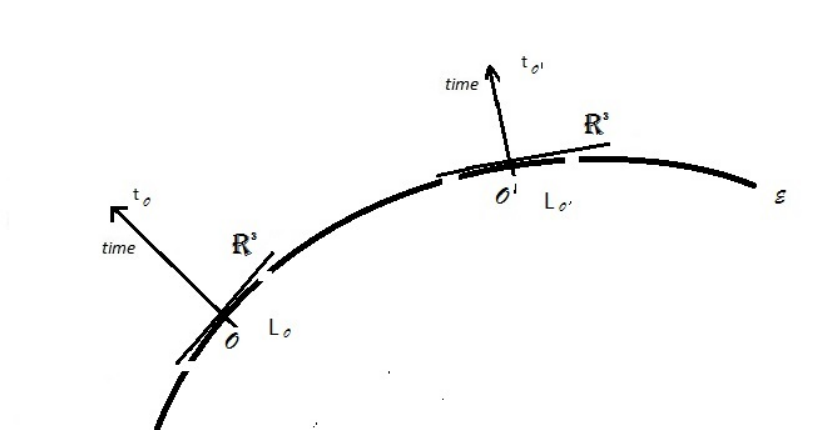}
	\caption{Space-Time}
	\label{varieta}
\end{figure}
Intuitively we can think of the Euclidean space $\mathfrak E$ as a three-dimensional submanifold of $\mathbb R^4$ "curved by gravitational time" (see schematic Figure \ref{varieta}). 
\begin{remark}\upshape
Since we have considered that time in the laboratory $L_o$ flows in the same way at each of its points, the region given by $L_o$ can be considered a subset of a three-dimensional Euclidean space $\mathbb R^3$.
\end{remark}
\subsection{Change of Reference System}
We observe that the transition from one laboratory reference system of $L_o$ to another 
\begin{equation}\label{cambiosist}
(K,O) \longmapsto (K',O')
\end{equation}
is carried out by a bijective transformation 
\begin{equation}
\label{cambiocoord}
T^{K',O'}_{K,O}: \mathbb R^3 \times \mathbb R \longrightarrow \mathbb R^3 \times \mathbb R
\end{equation}
What properties do the transformations \eqref{cambiocoord} have?
\\
If we make the following reference changes:
$$(K,O) \longmapsto (K',O') \longmapsto (K'',O'') \qquad , \qquad (K,O) \longmapsto (K'',O'') $$
we obtain that the transformations \eqref{cambiocoord} satisfy the group property:
\begin{equation}
T^{K'',O''}_{K',O'} \circ T^{K',O'}_{K,O} = T^{K'',O''}_{K,O}
\end{equation}
\begin{attenzione}\upshape 
In these considerations, nothing prevents us from assuming that the clock positioned at $O'$ is not at rest with respect to our clock at $O$.
\end{attenzione}

If we denote by $\operatorname{Aut}(\mathbb R^4)$ the set of bijective maps of $\mathbb R^4$ onto itself that are continuous in the Euclidean topology, then a subgroup $\mathcal G \subset \operatorname{Aut}(\mathbb R^4)$ is called a group of \textit{kinematic transformations} that govern the laws of coordinate transformations when passing from one reference to another.
\\
\begin{figure}[htbp]
	\centering
		\includegraphics[scale=0.8]{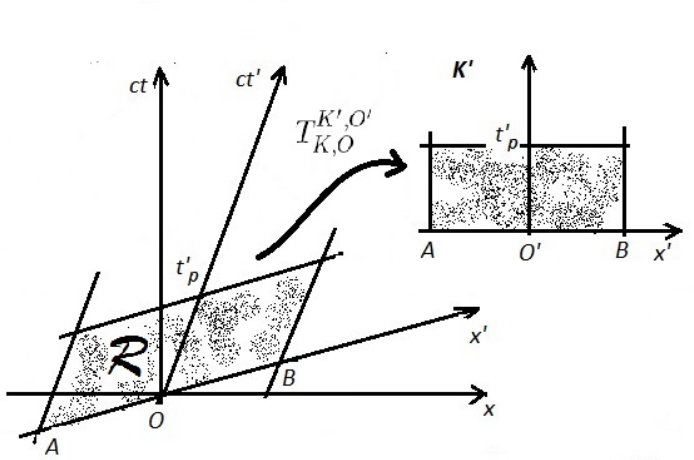}
	\caption{Laboratory Regions}
	\label{fig:08tris}
\end{figure}
We recall that the properties of space-time depend on the group of transformations $\mathcal G$, which describes the reference changes given in relation \eqref{cambiosist}, which we choose according to the conditions dictated by our operational needs.
\bigskip

Let us now consider a group of transformations $\mathcal G$ and take a point $O$ of $\mathfrak E$ and any reference system $(K,O)$\footnote{Even if we omit clearly naming the laboratory, in reality it is always present since we can always consider a "globular laboratory" centered at $O$ with the given reference system.}.
\begin{definition}\upshape\label{G-regione}
An open set $\mathcal R$ of $\mathbb R^3 \times \mathbb R$ is called a $\mathcal G$-laboratory region if there exists a new reference $(K',O')$ such that the transformation $T^{K',O'}_{K,O} \in \mathcal G$ transforms the set $\mathcal R$ into a standard laboratory-type region:
$$T^{K',O'}_{K,O} (\mathcal R) = L_o \times [0, t_o] \ , \qquad L_o \subset \mathbb R^3 $$ 
\end{definition}
For example, consider as a group of transformations $\mathcal G$ those induced by Lorentz transformations. A laboratory centered at $O'$ moving with respect to another centered at $O$, with their respective reference systems, transforms the set $\mathcal R$ considered by the experimenter in the laboratory centered at $O$ (as given in Figure \ref{fig:08tris}) into a laboratory-type region (see also note \ref{doppio2} in section \ref{doppio}). 
\\
We underline that the $\mathcal G$-region $\mathcal R$ is not a subset of $\mathfrak E \times \mathbb R$ but of $\mathbb R^4$, since in order to establish it we must have fixed a laboratory system $(K,O)$ and determined a new laboratory system $(K',O')$ and the related transformation, which satisfies Definition \ref{G-regione}.   
 \\

Before concluding this short section on changing coordinates, it is useful to note the following facts.
\\
We introduced the choice of the group of transformations $\mathcal G$ in a purely mathematical way, independent of the experimental act. In reality, to determine a coordinate transformation \eqref{cambiosist}, experimental results must be taken into account. For example, consider the study of the decay time of an elementary particle in two distinct laboratories $L_O$ and $L_{O'}$ not at rest with respect to each other. As is known\footnote{For example, one could see the nice video document on physics teaching by the PSSC \cite{PSSC}.}, if this elementary particle is at rest with respect to the laboratory $L_{O'}$ with reference system $(K',O')$ centered at $O'$, then its decay time, measured with the instruments and procedures of laboratory $L_O$\footnote{Thus the measurement takes place in a state $\omega \in \mathfrak S(L_O)$, and this state of the system is also established through the use of mirrors and light signals, as indicated by Einstein in \cite{Einstein}.}, is dilated.
\\
This leads us to consider the Poincaré group as the transformation group $\mathcal G$\footnote{For this topic, see the book by Costa and Fogli \cite{Costa}.}, and if the velocity and acceleration of $O'$ relative to $O$ are small (compared to the speed of light), these values of the decay time differ slightly; in this case we can assume the Galilean group as the group of transformations.
\section{Symmetries in the Laboratory}
Recall that given a reference system $(K, O)$ of the laboratory $L_o$, we can write $L_o \subset \mathbb R^3$, and the pair $\left( \mathfrak X(L_o), \mathfrak S(L_o) \right)$ consists of the physical quantities and the relative states that can be measured and prepared in this laboratory, respectively.
\\
From Roberts and Roepstorff \cite{RobertsRoepstorff} we have:\index{Roberts} \index{Roepstorff}
\\
\textit{A symmetry of a physical system is intuitively a transformation of the system leaving all physically significant features invariant.}
 \\
Let us see what mathematical meaning the term transformation has, and let us adapt the definition given in \cite{RobertsRoepstorff} for the algebraic case to our model:
\begin{definition} \upshape \label{sym_mis}\index{Semi-symmetry}
A semi-symmetry of the physical system $\left( \mathfrak X(L_o), \mathfrak S(L_o) \right)$ is a pair of maps $(\alpha, \alpha^\natural)$ 
$$\alpha : \mathfrak X(L_o) \rightarrow \mathfrak X(L_o) \qquad , \qquad \alpha^\natural : \mathfrak S(L_o) \rightarrow \mathfrak S(L_o)$$
such that
\begin{itemize}
\item [a.] $\alpha(\mathfrak X_\omega) \subset \mathfrak X_{\alpha^\natural(\omega)} \ , \qquad \forall \omega \in \mathfrak S$
\item [b.] $\alpha^\natural (\mathfrak S_a) \subset \mathfrak S_{\alpha(a)} \ , \qquad \forall a \in \mathfrak X$
\item [c.] for each $a \in \mathfrak X$ and $\omega \in \mathfrak S_a$ we obtain:
$$P(a \in \Delta, \tau)_\omega = P(\alpha(a) \in \Delta, \tau)_{\alpha^\natural(\omega)} \ , \qquad \forall \Delta \in B(\mathbb R) $$
\end{itemize}
\end{definition}
We underline that conditions [a.] and [b.] are equivalent:
$$ \left[ \alpha^\natural (\mathfrak S_a) \subset \mathfrak S_{\alpha(a)} \ , \ \forall a \in \mathfrak X \right] \qquad \Longleftrightarrow \qquad \left[ \alpha (\mathfrak X_\omega) \subset \mathfrak X_{\alpha^\natural(\omega)} \ , \ \forall \omega \in \mathfrak S \right] $$
Furthermore, from [c.] we obtain
\begin{equation}
\label{sym_mis2}
\mu_{\omega, a} = \mu_{\alpha^\natural(\omega), \alpha(a)} 
\end{equation}
In this way, for every $a \in \mathfrak X(\mathcal O_o)$ and $\omega \in \mathfrak S_a(\mathcal O_o)$:  
\begin{equation}
\left\langle \alpha(a) \right\rangle_{\alpha^\natural(\omega)} = \left\langle a \right\rangle_{\omega} 
\end{equation}
\begin{definition}
A semi-symmetry consisting of bijective maps is called a symmetry of our system.
\end{definition}
We denote by $\operatorname{Sym}(L_o)$ the set of symmetries of the physical system $\left( \mathfrak X(L_o), \mathfrak S(L_o) \right)$. 
\\
We point out that with the composition of maps, the set $\operatorname{Sym}(L_o)$ turns out to be a group.
\begin{proposition}\upshape
If $(\alpha, \alpha^\natural) \in \operatorname{Sym}(L_o)$, then we have 
$$\| \alpha(a) \| = \| a \| \ , \qquad \forall a \in \mathfrak X $$
and
$$\| \alpha^\natural(\omega) \| = \| \omega \| \ , \qquad \forall \omega \in \mathfrak S $$
where we define the norm of an observable and a state by the well-known expressions:
$$\| a \| = \sup_{\omega \in \mathfrak S_a} \left| \left\langle a \right\rangle_\omega \right| \qquad , \qquad \| \omega \| = \sup_{a \in \mathfrak X_\omega} \left| \left\langle a \right\rangle_\omega \right|$$
\end{proposition}
\begin{proof}
Trivial consequence of the bijectivity of the two maps.
\end{proof}
\begin{proposition}
Let $a \in \mathfrak X$; for every function $f: \mathbb R \rightarrow \mathbb R$ that is $a$-summable, we have:
$$ \alpha(f(a)) = f(\alpha(a))$$
Specifically,
$$ \alpha(a^2) = \alpha(a)^2$$
\end{proposition}
\begin{proof}
For every $\omega \in \mathfrak S_a(L_o)$ we obtain:
\begin{eqnarray*}
\left\langle \alpha(f(a)) \right\rangle_{\alpha^\natural(\omega)} & = & \int s \, d \mu_{\alpha(f(a)), \alpha^\natural(\omega)} \\
& = & \int s \, d \mu_{f(a), \omega} = \left\langle f(a) \right\rangle_{\omega}  
\end{eqnarray*}
while
\begin{eqnarray*}
\left\langle f(\alpha(a)) \right\rangle_{\alpha^\natural(\omega)} & = & \int f(s) \, d \mu_{\alpha(a), \alpha^\natural(\omega)} \\
& = & \int f(s) \, d \mu_{a, \omega} = \left\langle f(a) \right\rangle_{\omega}  
\end{eqnarray*}
From bijectivity we have that $\alpha^\natural(\mathfrak S_a(L_o)) = \mathfrak S_{\alpha(a)}(L_o)$, and we can write
$$ \left\langle \alpha(f(a)) \right\rangle_{\omega} = \left\langle f(\alpha(a)) \right\rangle_{\omega} \ , \qquad \forall \omega \in \mathfrak S_a(L_o) $$
hence the thesis.
\end{proof}
\begin{proposition}
Symmetries preserve the spectrum of observables of the system:
$$ \sigma(\alpha(a)) = \sigma(a) \ , \qquad \forall a \in \mathfrak X(L_o) $$
\end{proposition}
\begin{proof}
Trivial consequence of the relations \eqref{Fomega} and \eqref{sym_mis2} on pages \pageref{Fomega} and \pageref{sym_mis2}, respectively.
\end{proof}
We underline that the definition of symmetry that we have given is independent of the notion of compatibility of observables:
\begin{remark}\upshape
If $x \in \mathcal C(a)$, then it does not necessarily follow that $\alpha(x) \in \mathcal C(\alpha(a))$, where $\mathcal C(a)$ is the set of observables compatible with $a$.
\end{remark}
As we discussed in section \ref{proceduresperimentali}, observables are physical quantities that remain so in time and space; what changes are their values, but not their typology. To have a broader mathematical definition, in the definition of symmetry we considered a map $\alpha$ that transforms the observables of the system, since the symmetries of a physical system could be induced by processes internal to the system itself, independent of space-time\footnote{By internal symmetries one usually means, for elementary particles, symmetries that are independent of the space-time structure of the world (see Nuyts in \cite{BBG}).}.
\subsection{Operations}\index{Group of operations}
A group $\mathbb{G}$ is called a \textit{group of operations} on the physical system of the laboratory $(\mathfrak X(L_o), \mathfrak S(L_o))$ if $\mathbb{G}$ acts on the set $\mathfrak S(L_o)$; in other words, if for every $g$ belonging to $\mathbb{G}$ there is a map $\gamma_g : \mathfrak S(L_o) \longrightarrow \mathfrak S(L_o)$ such that for each $g$ and $h$ in $\mathbb{G}$ we have
\begin{equation*}
\gamma_{gh} = \gamma_g \circ \gamma_h 
 \end{equation*}%
The map $\gamma$ is called an \textit{operation on the physical system}.\index{Operation on the physical system}
 \\
We now have a set of definitions.
\\
Let $\mathbb{H}$ be a subgroup of the operation group $\mathbb{G}$. A set $\mathfrak S_o \subset \mathfrak S$ is said to be $\mathbb{H}$-stable if
$$ \gamma_h (\mathfrak S_o) \subset \mathfrak S_o \ , \qquad \forall h \in \mathbb{H}$$
and a state $\omega_o$ is $\mathbb{G}$-stable if 
\begin{equation*}
\gamma_g(\omega_o) = \omega_o \ , \qquad \forall g \in \mathbb{G}
 \end{equation*}
while an observable $a$ of the laboratory system is said to be $\mathbb{G}$-invariant if for every $g \in \mathbb{G}$ we have
$$ \gamma_g(\mathfrak S_a) \subset \mathfrak S_a $$
and
$$ \left\langle a \right\rangle_\omega = \left\langle a \right\rangle_{\gamma_g(\omega)} \ , \qquad \forall \omega \in \mathfrak S_a(L_o) $$ 
\begin{problem}\upshape
If $\mathbb{G}$ is a group of operations on the system and it is a Lie group, then what connection exists between the generators of its Lie algebra and the $\mathbb{G}$-invariant observables of the system?
\end{problem}
The group $\mathbb{G}$ of operations on the physical system is said to be \textit{symmetric} if the operations leave the behaviour of the system unchanged. In mathematical terms, this means that there is a group homomorphism
\begin{equation}
g \in \mathbb{G} \longrightarrow (\operatorname{id}, \gamma_g) \in \operatorname{Sym}(L_o)
\end{equation}
Therefore for every $g \in \mathbb{G}$ and $a \in \mathfrak X$ we have:
$$ \gamma_g(\mathfrak S_a) \subset \mathfrak S_a $$
and   
$$ \mu_{a, \omega} = \mu_{a, \gamma_g(\omega)} \ , \qquad \forall \omega \in \mathfrak S_a(L_o) $$  
\chapter{Reference and Laboratory Inclusions}
In this section we will address the issue of sublaboratories of our physical laboratory and its division into two or more parts. We observe that in circumscribing a part of our laboratory, a particular selection of states (and thus observables) is made, which falls under the arguments covered in section \ref{Sistemi fisici del laboratorio} of our mother laboratory.
\\
In addition, we will consider two separate laboratories at rest with respect to each other and study their mutual independence. 
\\
We emphasize that these problems find their proper territory in the general framework given by the theory of relativity, a topic that we will analyze only briefly in this section.
\section{Centered Sublaboratories}\label{sottolaboratorio}
\begin{figure}[htbp]
	\centering
		\includegraphics[scale=0.4]{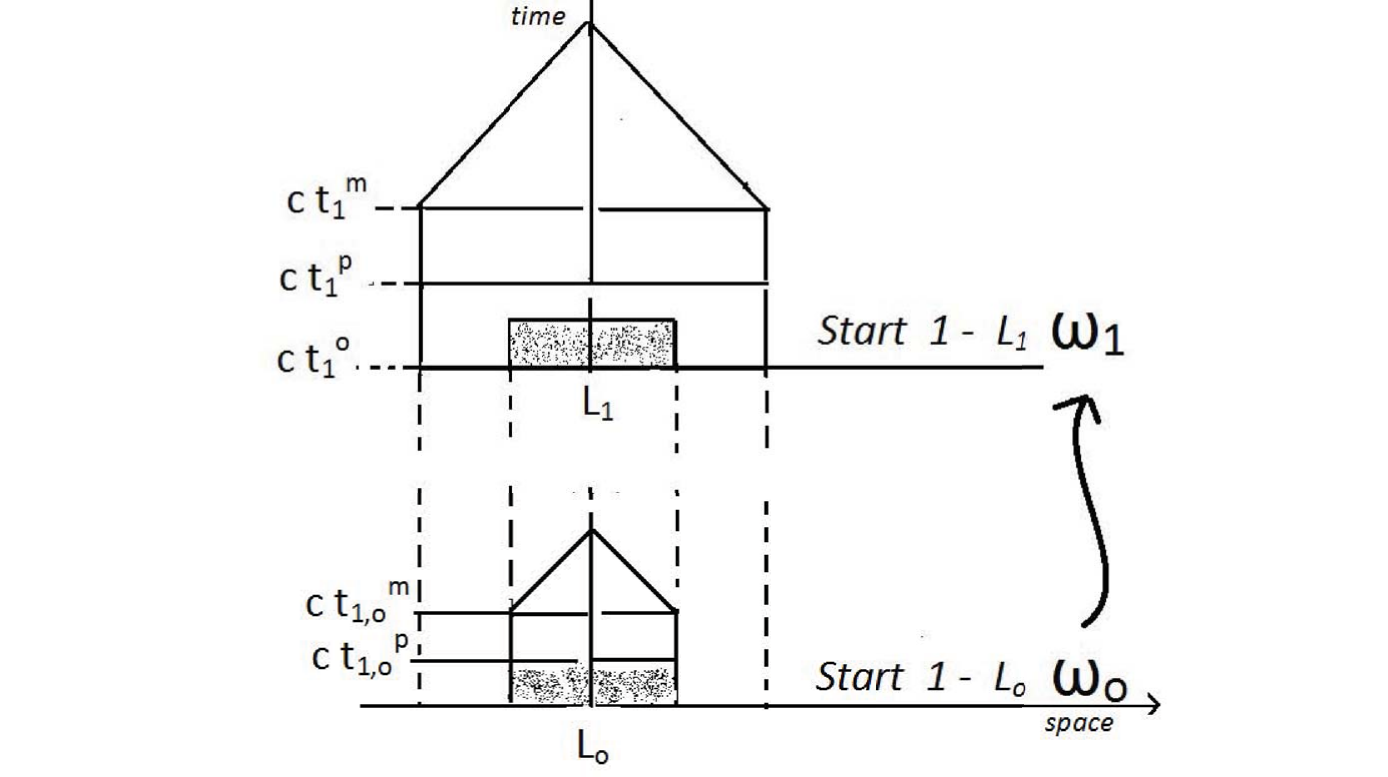}
	\caption{Centered Sublaboratory}
	\label{fig:09}
\end{figure}

\begin{figure}[htbp]
	\centering
		\includegraphics[scale=0.4]{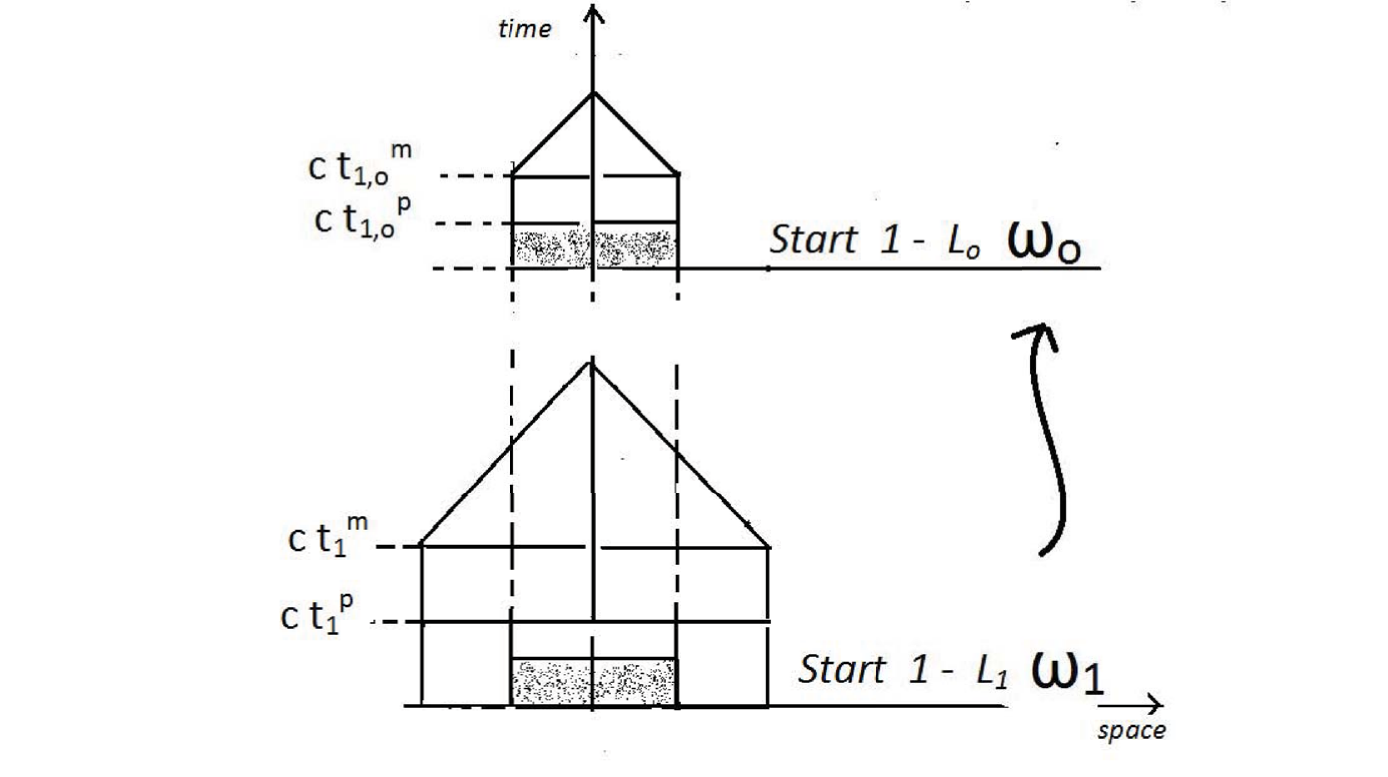}
	\caption{Sublaboratory in evidence}
	\label{fig:09b}
\end{figure}

In the previous sections we established that our experiments take place in a subset of space-time of the type $\mathcal O_o = L_o \times [0, t_0^p]$, where $L_o$ is our laboratory region and $t_o^p$ is the preparation time for the experiments. Let us see what happens if we consider a larger laboratory $L_1$ which appropriately includes our laboratory $L_o$ and a larger preparation time $t_1^p$. Appropriately, this means that
\begin{equation} 
\mathcal O_1 := L_1 \times [0, t_1^p] \ , \qquad t_o^p \leq t_1^p \ , \ L_o \subset L_1 \subset \mathfrak E
\end{equation}
therefore $L_o$ and $L_1$ are centered at the common origin where the eventual experimenter resides. In other words, we use the same clock positioned at $O$ and the same $K$ rulers for both laboratories (see Figure \ref{fig:09})\footnote{Thus we have the same reference $(K,O)$ for $L_o$ and $L_1$.}.
\\
We establish that our measurements take place at a fixed time $\tau \geq 0$ in both laboratories.
\\
In Figure \ref{fig:09} we have represented the first copy of the ensemble for both laboratories $L_o$ and $L_1$, where 
$$\tau = t_j^m - t_j^p = t_{j,o}^P - t_{j,o}^p \qquad \forall j = 1, 2, \ldots, n$$
\\
Furthermore, the way to operate is as follows: \textit{first perform all the measurements of all the $N$ copies in $L_o$ and then, after having rearranged the laboratory, perform the measurements on the $N$ copies of $L_1$.}
\begin{post}\label{subreg}\index{Postulate-Observables and Subregions I} 
Let $\mathcal O_o, \mathcal O_1$ be two laboratory-type regions. If $\mathcal O_o \subset \mathcal O_1$, then every observable measurable in $\mathcal O_o$ is measurable in $\mathcal O_1$:
$$\mathfrak X(\mathcal O_o) \subset \mathfrak X(\mathcal O_1)$$
\end{post}
This apparently banal statement has some critical issues; let us highlight them.
\\
If $a \in \mathfrak X(\mathcal O_o)$, then for every state $\omega_o \in \mathfrak S_a(\mathcal O_o)$ there must exist at least one state $\omega_1 \in \mathfrak S_a(\mathcal O_1)$\footnote{Otherwise $a$ cannot be a measurable quantity in $L_1$. 
\\ 
Warning: It is not being assumed that the set of states $\mathfrak S_a(\mathcal O_o)$ is contained in $\mathfrak S_a(\mathcal O_1)$.} such that
\begin{equation}\index{Axiom on states and subregions}
\label{restrizionestato}
P^{L_o}(a \in \Delta, \tau)_{\omega_o} = P^{L_1}(a \in \Delta, \tau)_{\omega_1} \ , \qquad \forall \Delta \in B(\mathbb R) 
\end{equation}
Indeed, the various devices we have in $L_o$ and the procedures that are carried out in $\mathcal O_o$, which establish $\omega_o$, can \textit{a fortiori} be carried out in a larger laboratory $L_1$ and with a longer preparation time $t_1^p$.
\\
This leads us to affirm the experimental validity of the following axiom, which generalizes property \ref{estensione_stato} of section \ref{stati_regioni}:

\begin{post}\label{subregB} \index{Postulate-Observables and States on Subregions II}
Let $\mathcal O_o, \mathcal O_1$ be two laboratory-type regions with $\mathcal O_o \subset \mathcal O_1$. For every state $\omega_o \in \mathfrak S(\mathcal O_o)$ there exists a state $\omega_1 \in \mathfrak S(\mathcal O_1)$ such that:
\begin{itemize}
\item With $\omega_1$ we can measure all observables of $\mathfrak X(\mathcal O_o)$ that are measurable in the state $\omega_o$; in formulas,
\begin{equation}\label{restrizionestatobis}
 \mathfrak X_{\omega_o}(\mathcal O_o) = \mathfrak X_{\omega_1}(\mathcal O_1) \cap \mathfrak X(\mathcal O_o)
\end{equation}
\item We obtain the same values for these observables in the two laboratories; in formulas,
\begin{equation}\label{restrizionestatotris}
P^{L_o}(x \in \Delta, \tau)_{\omega_o} = P^{L_1}(x \in \Delta, \tau)_{\omega_1} \ , \qquad \forall x \in \mathfrak X_{\omega_o}(\mathcal O_o)
\end{equation}
\end{itemize}
\end{post}
We emphasize that the state $\omega_1$ is \textit{not necessarily unique}, since we cannot rule out having two experimental procedures, with associated measuring instruments, etc., that "restricted" to the observables of $L_o$ result in exactly the same state $\omega_o$\footnote{Mathematically, this statement can be seen as the non-algebraic counterpart of the Hahn–Banach extension theorem for functionals on topological vector spaces.}.
\\
\\
More attention requires the converse of our statements\footnote{See also section \ref{nuovi_axiom} and \ref{pbinclus}, respectively on pages \pageref{nuovi_axiom} and \pageref{pbinclus}.}:
\\
Let us consider a state $\omega_1 \in \mathfrak S(\mathcal O_1)$ and ask whether it is possible to determine a state $\omega_o \in \mathfrak S_a(\mathcal O_o)$ for which relations \eqref{restrizionestatobis} and \eqref{restrizionestatotris} are valid.
\\
The general answer to this last statement \textit{is negative}, since to establish a state $\omega_o \in \mathfrak S(\mathcal O_o)$ we have a smaller geometric space available and a preparation time $t_o^p$ different from $t_1^p$, and it is not certain that this is sufficient to establish our state (see Figure \ref{fig:09b})\footnote{The assertion is positive if $\mathfrak X(\mathcal O_o) \subset \mathfrak X_{\omega_1}(\mathcal O_1)$ and property \ref{esistenza_stato_ristretto} is considered valid.}.
\\

Consider the following set of states:\index{$\mathfrak S(\mathcal O_1 \vert \mathcal O_o)$ }
\begin{equation}\label{sottostati}
\mathfrak S(\mathcal O_1 | \mathcal O_o) = \left\{ \omega \in \mathfrak S(\mathcal O_1) : \exists \omega_o \in \mathfrak S(\mathcal O_o) \text{ which satisfies \eqref{restrizionestatobis} and \eqref{restrizionestatotris}} \right\}
\end{equation}
\begin{proposition}\upshape\label{sottostatiprop}
The set $\mathfrak S(\mathcal O_1 | \mathcal O_o) \subset \mathfrak S(\mathcal O_1)$ is non-empty.
\end{proposition}
\begin{proof}
By definition, for each $a \in \mathfrak X(\mathcal O_o)$ there is at least one state $\omega_o \in \mathfrak S_a(\mathcal O_o)$, and therefore by Postulate \ref{subregB} we obtain the existence of a state $\omega_1 \in \mathfrak S(\mathcal O_1)$ which satisfies relations \eqref{restrizionestatobis} and \eqref{restrizionestatotris}. Therefore $a \in \mathfrak X_{\omega_1}(\mathcal O_1)$, and from this it follows easily that $\omega_1 \in \mathfrak S(\mathcal O_1 | \mathcal O_o)$.
\end{proof}
\begin{attenzione}\upshape
In this way, we also prove that for each $a \in \mathfrak X(\mathcal O_o)$ the set
\begin{equation}
\mathfrak S_a(\mathcal O_1 | \mathcal O_o) := \mathfrak S_a(\mathcal O_1) \cap \mathfrak S(\mathcal O_1 | \mathcal O_o)
\end{equation}
is non-empty.  
\end{attenzione}
Furthermore, we can write
\begin{equation} 
\mathfrak S(\mathcal O_1 | \mathcal O_o) = \bigcup_{a \in \mathfrak {X \ }({\mathcal O}_o)} \mathfrak S_a(\mathcal O_1 | \mathcal O_o)
\end{equation}
Indeed, if $\omega_1 \in \mathfrak S(\mathcal O_1 | \mathcal O_o)$, then by definition there exists $\omega_o \in \mathfrak S(\mathcal O_o)$ that satisfies relation \eqref{restrizionestatobis}; it follows that every $a \in \mathfrak X_{\omega_o}$ is also an element of $\mathfrak X_{\omega_1}$; in other words,
$\omega_1 \in \mathfrak S_a(\mathcal O_1)$.
We have the following proposition; the verification is a simple consequence of relations \eqref{restrizionestatobis}, \eqref{restrizionestatotris} and Axiom \ref{assio2}.
\begin{proposition}\upshape\label{mappaP}
Let $\mathcal O_o$ and $\mathcal O_1$ be two laboratory-type regions. If $\mathcal O_o \subset \mathcal O_1$, then there exists a surjective map\footnote{Which obviously depends on the two laboratory-type regions in question.} 
\begin{equation}
\texttt{P} : \mathfrak S(\mathcal O_1 | \mathcal O_o) \longrightarrow \mathfrak S(\mathcal O_o)
\end{equation}  
such that for every $\omega_1 \in \mathfrak S(\mathcal O_1 | \mathcal O_o)$ we have:
\begin{itemize}
\item [A.] $\qquad \mathfrak X_{\texttt{P}(\omega_1)} = \mathfrak X_{\omega_1} \cap \mathfrak X(\mathcal O_o)$;
\item [B.] $\qquad \mu^{L_1}_{\omega_1, x} = \mu^{L_o}_{\texttt{P}(\omega_1), x} \qquad , \qquad \forall x \in \mathfrak X_{\texttt{P}(\omega_1)}$.
\end{itemize}
\end{proposition}
Moreover, as mentioned, for every $a \in \mathfrak X(\mathcal O_o)$ the restriction of the map $\texttt{P}$ to the set $\mathfrak S_a(\mathcal O_1 | \mathcal O_o)$ determines a surjective map
\begin{equation}\label{mappastati}
\texttt{P}_a : \mathfrak S_a(\mathcal O_1 | \mathcal O_o) \rightarrow \mathfrak S_a(\mathcal O_o)
\end{equation}
which obviously satisfies the same relations as Proposition \ref{mappaP}.
\begin{remark}\upshape
If $x, y \in \mathfrak X(\mathcal O_o)$, then the sets $\mathfrak S_x(\mathcal O_1 | \mathcal O_o)$ and $\mathfrak S_y(\mathcal O_1 | \mathcal O_o)$ may have elements in common, and therefore from the surjectivity of the map $\texttt{P}$ it follows that
$$ \texttt{P}\left( \mathfrak S_x(\mathcal O_1 | \mathcal O_o) \cap \mathfrak S_y(\mathcal O_1 | \mathcal O_o) \right) = \mathfrak S_x(\mathcal O_o) \cap \mathfrak S_y(\mathcal O_o) $$     
\end{remark}
Intuitively, the set $\mathfrak S_a(\mathcal O_1)$ contains more elements than $\mathfrak S_a(\mathcal O_o)$, since having a more extensive region (spatial and temporal) in which to carry out our experiments increases the number of possible experiments and therefore increases the information about our observable $a$.
 \\ 
\\
In line with the topic covered in section \ref{inferenze}, we give the following
\begin{definition}\upshape\label{infogeo}\index{Sufficiently informative region}
The region $\mathcal O_o$ is said to be \textit{sufficiently informative} for the region $\mathcal O_1$ if for every $x \in \mathfrak X(\mathcal O_1)$ there exists a projector of the states
$$E_x : \mathfrak S_x(\mathcal O_1) \rightarrow \mathfrak S_x(\mathcal O_1 | \mathcal O_o) \subset \mathfrak S_x(\mathcal O_1)$$
\end{definition}
Therefore, if the region $\mathcal O_o$ is sufficiently informative, then from \eqref{mappastati} we have a surjective map 
$$\mu_{\omega,x}^{L_1} = \mu_{E_a(\omega), x}^{L_1} \ , \qquad \omega \in \mathfrak S_x(\mathcal O_1) $$ 
with $\widehat{E}_x = \texttt{P}_x \circ E_x$ such that
$$\mu_{\omega,x}^{L_1} = \mu_{\widehat{E}_x(\omega), x}^{L_o} \ , \qquad \omega \in \mathfrak S_x(\mathcal O_1) $$ 
\subsection{Localized Observables}
There are observables of the laboratory system such that even by increasing the size of the laboratory and the preparation times of the various procedures, \textit{we never obtain an increase in information about the observable itself}. This fact suggests introducing the following definition of geometric localization of an observable\footnote{See also remark \ref{misure_oss} on page \pageref{Reference Problems}.}:
\begin{definition}\upshape[\textbf{Localized Observable}]\index{Localized observable}
An observable $a \in \mathfrak X(\mathcal O_o)$ is said to be geometrically localized in $\mathcal O_o$ if for every laboratory-type region $\mathcal O_1$ containing $\mathcal O_o$ it holds that\footnote{In other words, every experiment carried out in $\mathcal O_1$ can always be traced back to one in $\mathcal O_o$, and the map $\texttt{P}_a$ is:
$$\texttt{P}_a : \mathfrak S_a(\mathcal O_1) \rightarrow \mathfrak S_a(\mathcal O_o)$$}:
$$\mathfrak S_a(\mathcal O_1 | \mathcal O_o) = \mathfrak S_a(\mathcal O_1)$$
 \end{definition}
Before introducing the notion of a global observable, we must make some simple observations.
\\
If $a \in \mathfrak X(\mathcal O_o)$, then for every pair of laboratory-type regions $\mathcal O_1$ and $\mathcal O_2$ with $\mathcal O_o \subset \mathcal O_1 \subset \mathcal O_2$ it turns out that:
$$a \in \mathfrak X(\mathcal O_o) \subset \mathfrak X(\mathcal O_1) \subset \mathfrak X(\mathcal O_2)$$
with 
$$ \mathfrak S_a(\mathcal O_1 | \mathcal O_o) \subset \mathfrak S_a(\mathcal O_1) \qquad , \qquad \mathfrak S_a(\mathcal O_2 | \mathcal O_1) \subset \mathfrak S_a(\mathcal O_2)$$
 
\begin{attenzione}\upshape
In general it is not necessarily true that the set $\mathfrak S_a(\mathcal O_2 | \mathcal O_o)$ is included in the set of states $\mathfrak S_a(\mathcal O_2 | \mathcal O_1)$.
\end{attenzione}
\begin{definition}[\textbf{Global Observable}]\upshape\index{Global observables}
The observable $a \in \mathfrak X(\mathcal O_o)$ is called global if for every pair of laboratory-type regions $\mathcal O_1, \mathcal O_2$ with $\mathcal O_o \subset \mathcal O_1 \subset \mathcal O_2$ and $\mathcal O_1 \neq \mathcal O_2$, it turns out that
$$ \mathfrak S_a(\mathcal O_2 | \mathcal O_1) \neq \mathfrak S_a(\mathcal O_2) $$ 
In other words, it is global if it can never be geometrically localized in any of the regions $\mathcal O_1$ that include $\mathcal O_o$. 
\end{definition}
The next postulate establishes the locality of the physical quantity time\footnote{See also postulate \ref{ipotesiorologiosincro} of section \ref{rif_pb}.}:
\begin{post}\index{Time observable localized}\index{Postulate-Time observable localized}
The time observable $\texttt{t}$ of our laboratory $L_o$ is an observable localized in any laboratory-type region $\mathcal O_o = L_o \times [0, t_o]$:
$$\mathfrak S_\texttt{t}(\mathcal O_1 | \mathcal O_o) = \mathfrak S_\texttt{t}(\mathcal O_1) \ , \qquad \forall \mathcal O_o \subset \mathcal O_1 $$  
\end{post}
We consider an observable $a \in \mathfrak X(\mathcal O_1)$ and ask ourselves whether an experimenter placed in the laboratory $L_o$ is somehow able to identify this physical quantity with the instruments that he has available in that laboratory\footnote{See also section \ref{nuovi_axiom}.}.  
\begin{definition}\upshape\label{restrizionegeo}\index{Geometrically detectable observable}
Let $\mathcal O_o \subset \mathcal O_1$. An observable $a \in \mathfrak X(\mathcal O_1)$ is geometrically detectable in $\mathcal O_o$ if there exists a state $\omega_o \in \mathfrak S(\mathcal O_o)$ suitable for $a$.
\end{definition}
We emphasize that we do not have sufficient knowledge to state that every observable $a \in \mathfrak X(\mathcal O_1)$ is detectable in $\mathfrak X(\mathcal O_o)$.

\begin{problem}\upshape
If the observable $a \in \mathfrak X(\mathcal O_1)$ is detectable in $\mathcal O_o$, what relationship exists between the set $\mathfrak S_a(\mathcal O_1 | \mathcal O_o) \subset \mathfrak S_a(\mathcal O_1)$ and the set $\mathfrak S_a(\mathcal O_o)$?
\end{problem}
\subsection{Geometric Markovianity}
We consider different preparation times\footnote{Not to be confused with the temporal evolution of the values of the quantities as the time $\tau$ changes.} for carrying out the experiments in the laboratory $L_o$ as shown in Figure \ref{fig:09ca}. We denote  
\begin{equation}\label{markovregione}
 \mathcal O_t := L_o \times [0, t]  
\end{equation}
As stated previously, for each $t_2 \geq t_1 \geq t_o$ we obtain the following relations:
$$ \mathfrak X(\mathcal O_{t_o}) \subset \mathfrak X(\mathcal O_{t_1}) \subset \mathfrak X(\mathcal O_{t_2}) $$
and for each $a \in \mathfrak X(\mathcal O_o)$ and $t \geq t_o$  
$$ \mathfrak S_a(\mathcal O_t | \mathcal O_o) \subset \mathfrak S_a(\mathcal O_t)$$
Furthermore, as we previously discussed, we have the existence of a surjective map $\texttt{P}^t_a : \mathfrak S_a(\mathcal O_t | \mathcal O_o) \rightarrow \mathfrak S_a(\mathcal O_o)$ such that for each $\omega \in \mathfrak S_a(\mathcal O_t | \mathcal O_o)$:
$$ P(a \in \Delta, \tau)_\omega = P(a \in \Delta, \tau)_{\texttt{P}^t_a(\omega)} \ , \ \forall \tau \geq 0 \ , \ t \geq t_o $$
\begin{definition}\upshape\index{Markov laboratory-type region}
If $\mathcal O_o$ is sufficiently informative for every region $\mathcal O_t$ with $t \geq t_o$, then the region $\mathcal O_o$ will be called a Markov laboratory-type region.
\end{definition}
Thus, if the region $\mathcal O_o$ is Markovian, then for each $t \geq t_o$ and $a \in \mathfrak X(\mathcal O_o)$ there exists a surjective map $E_a^t : \mathfrak S_a(\mathcal O_t) \rightarrow \mathfrak S_a(\mathcal O_o)$ such that for each $\omega \in \mathfrak S_a(\mathcal O_t)$ it turns out that
\begin{equation}
P(a \in \Delta, \tau)_\omega = P(a \in \Delta, \tau)_{E_a^t(\omega)} \ , \qquad \forall \tau \geq 0 \ , \ t \geq t_o 
\end{equation}
\begin{figure}[htbp] 
	\centering
		\includegraphics[scale=0.5]{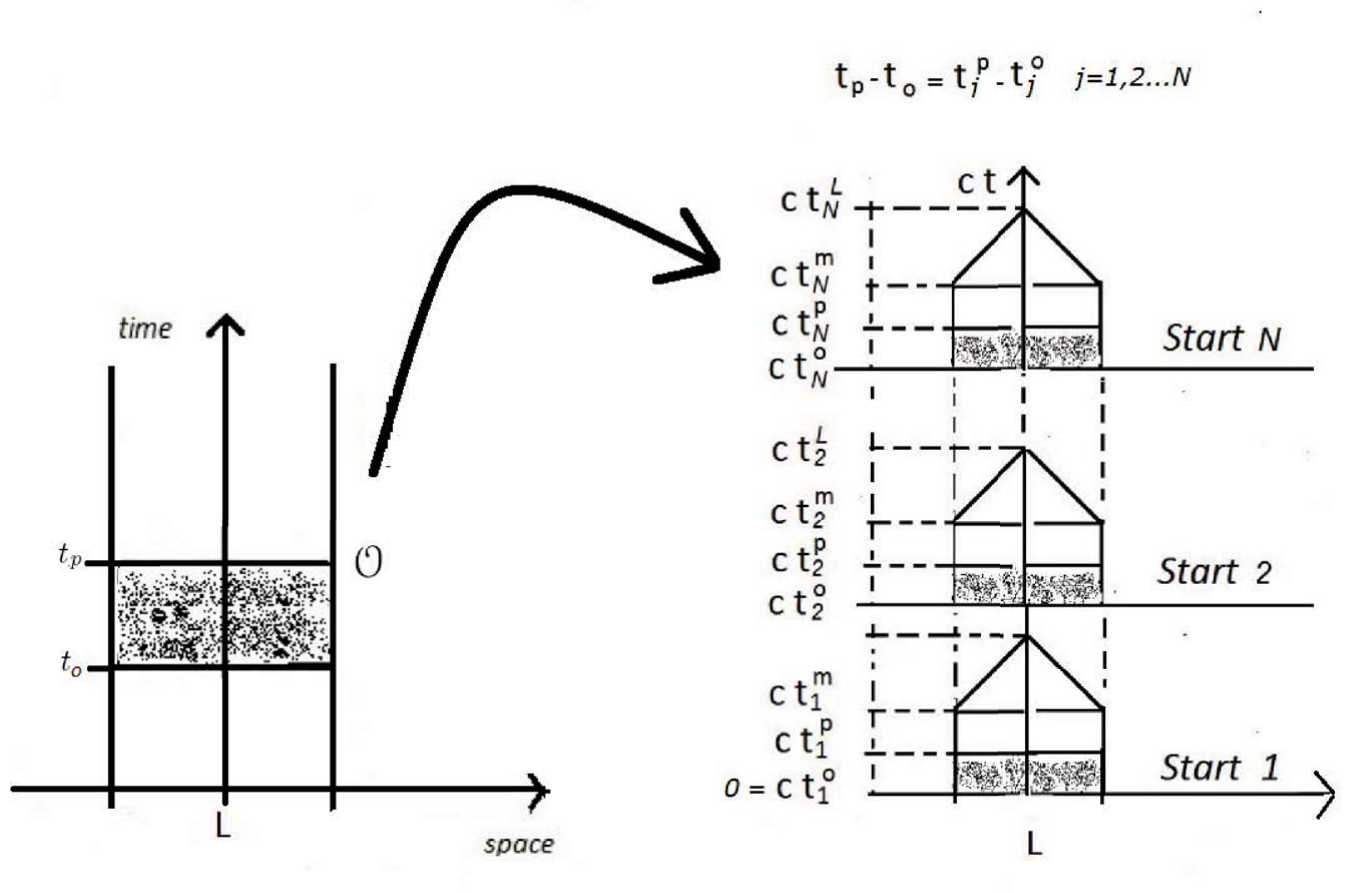}
	\caption{From the Regions to the Ensemble}
	\label{disle3bis}
\end{figure}

\section{Operational Space-Time} 
Consider the laboratory $L$ and its associated physical system $(\mathfrak X, \mathfrak S)$ and with it also the relevant region:
\begin{equation}\label{striscia}
\mathcal O_\infty = \bigcup_{t > 0} \mathcal O_t \ , \qquad \mathcal O_t = L \times [0, t] \subset \mathfrak E \times \mathbb R^+
\end{equation}

Let us examine all its sublaboratories $L_o$ of $L$ and all the possible start and preparation times of the various experiments:
\begin{equation}\label{set_striscia}
\mathfrak I = \left\{ \mathcal O = L_o \times [t_o, t_p] \ : \ L_o \subset L, \ t_o \geq 0, \ t_p > 0 \right\}
\end{equation}
By definition, the regions $\mathcal O \in \mathfrak I$ are associated with all the possible experiments that can potentially be carried out in the laboratory $L_o$ with an initial preparation time $t_o \geq 0$ and a final preparation time $t_p$. These experiments are labeled by the states $\omega \in \mathfrak S(\mathcal O)$ which allow us to measure the observables $a \in \mathfrak X(\mathcal O)$, and by definition it results that
$$ \mathfrak X(L) = \bigcup_{\mathcal O \in \mathfrak I} \mathfrak X(\mathcal O)$$
and\footnote{We recall that if $\mathcal O_o \subset \mathcal O_1$, then $\mathfrak X(\mathcal O_o) \subset \mathfrak X(\mathcal O_1)$, while for the states this is not true. We can say that for every $\omega_o \in \mathfrak S(\mathcal O_o)$ there exists $\omega_1 \in \mathfrak S(\mathcal O_1)$ as in Axiom \ref{subregB}.
\\
Furthermore, we note that 
$$ \bigcup_{t \geq 0} \mathfrak S(\mathcal O_t) \subset \mathfrak S(L) $$}
$$\mathfrak S(L) = \bigcup_{\mathcal O \in \mathfrak I} \mathfrak S(\mathcal O)$$
In the set $\mathcal M = \mathfrak E \times \mathbb R$, each laboratory-type region $\mathcal O \subset \mathcal M$ is linked with the ensemble of $N$ copies of the same trial that occurs in Minkowskian space-time, as shown in Figure \ref{disle3bis}\footnote{Remember that each laboratory $L \subset \mathfrak E$ is always associated with its own laboratory reference system $(K, O)$, where $O \in L$.
\\
Furthermore, even if the laboratory preparation begins at a non-zero time $t_o$, the first trial of the ensemble starts from zero, the actual start of the time counting of the experiment. In practice, the ensembles always start by resetting the chronometer present in our laboratory positioned at $O$.}:
$$\mathcal O \in \mathfrak I \ \Longrightarrow \ \text{Ensemble in } \mathcal M $$
\begin{figure}[htbp]
	\centering
		\includegraphics[scale=0.3]{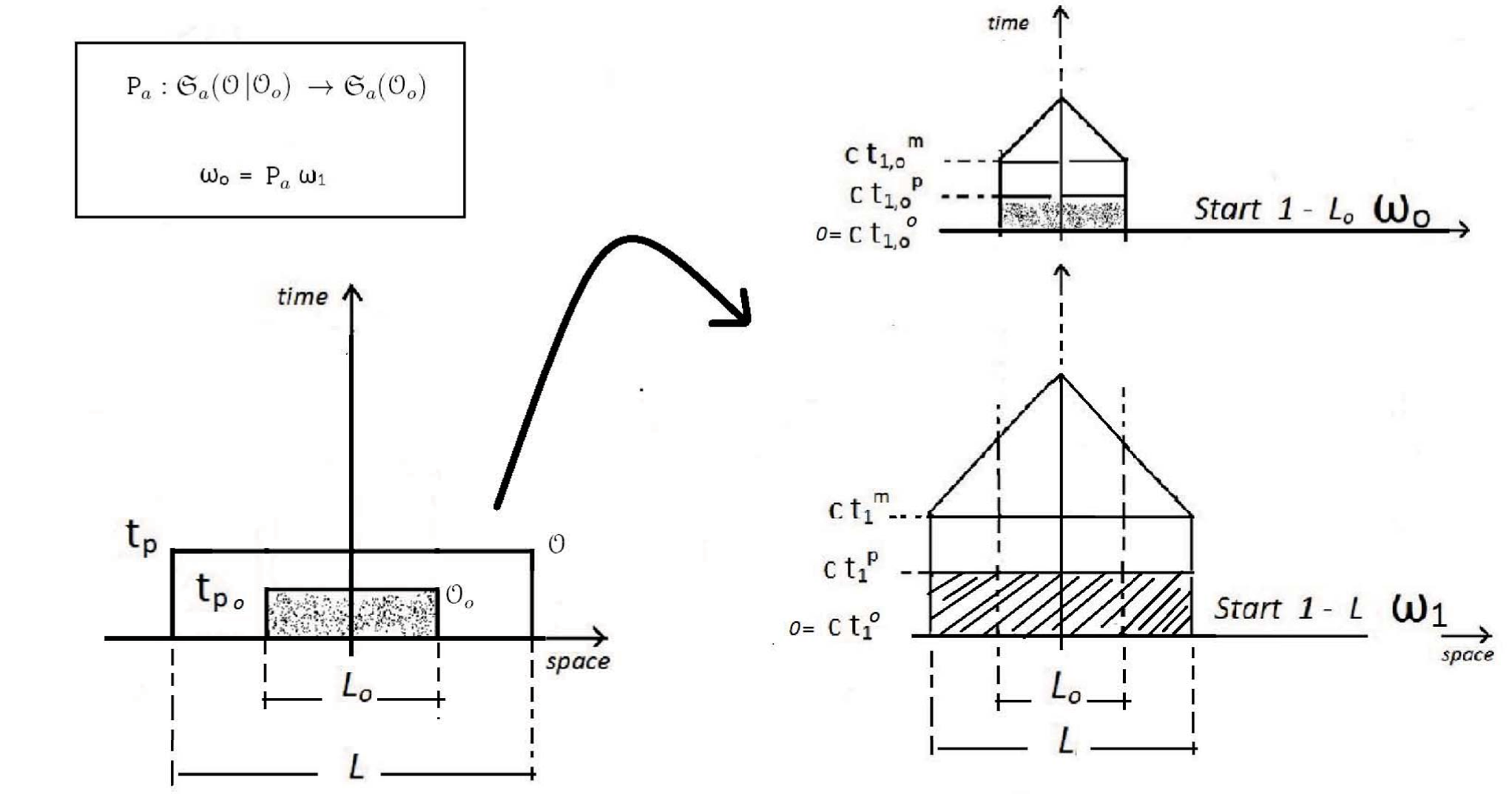}
	\caption{From Regions to Ensembles - Inclusions}
	\label{disle2bisens}
\end{figure}
Therefore, in $\mathcal O$ one can perform the operations, established by the state $\omega \in \mathfrak S(\mathcal O)$, to determine the probable values of the observable $a \in \mathfrak X(\mathcal O)$ at a given measurement time $\tau$, through the $N$ trials of the ensemble. In this way, we can think of the set $\mathcal M$ as an operational space-time, where the events of Minkowskian space-time are not contemplated, but only the laboratory-type regions are, where the potential experimental procedures for making the various measurements are associated.
\\
For example, let us consider two laboratories $L_o$ and $L$ with $L_o \subset L$ and two laboratory-type regions $\mathcal O_o$ and $\mathcal O$ as in Figure \ref{disle2bisens}, where we set the same start time of the experiment in the two laboratories but with different preparation times. 
\\
The $N$ measurement trials to establish the values of an observable $a \in \mathfrak X(\mathcal O_o)$ can be carried out first in $L$ and then in $L_o$, considered as two separate laboratories, as in Figure \ref{disle2bisens}\footnote{Obviously one can run the $N$ trials first for $L_o$ and then for $L$.}.
% 
 %%%
\begin{remark}\upshape
Operationally, when I perform a measurement in the sublaboratory $L_o$ of $L$, one might think that automatically the same measurement is carried out in the larger laboratory $L$, since $L_o$ is included in $L$, but this is not true (even if we have the same preparation time in both laboratories). In $L_o$ we make the measurement in the state $\omega_o$, which, as we have established, is not a state of the physical system of the laboratory $L$.
\end{remark}
\begin{figure}[htbp] 
	\centering
		\includegraphics[scale=0.5]{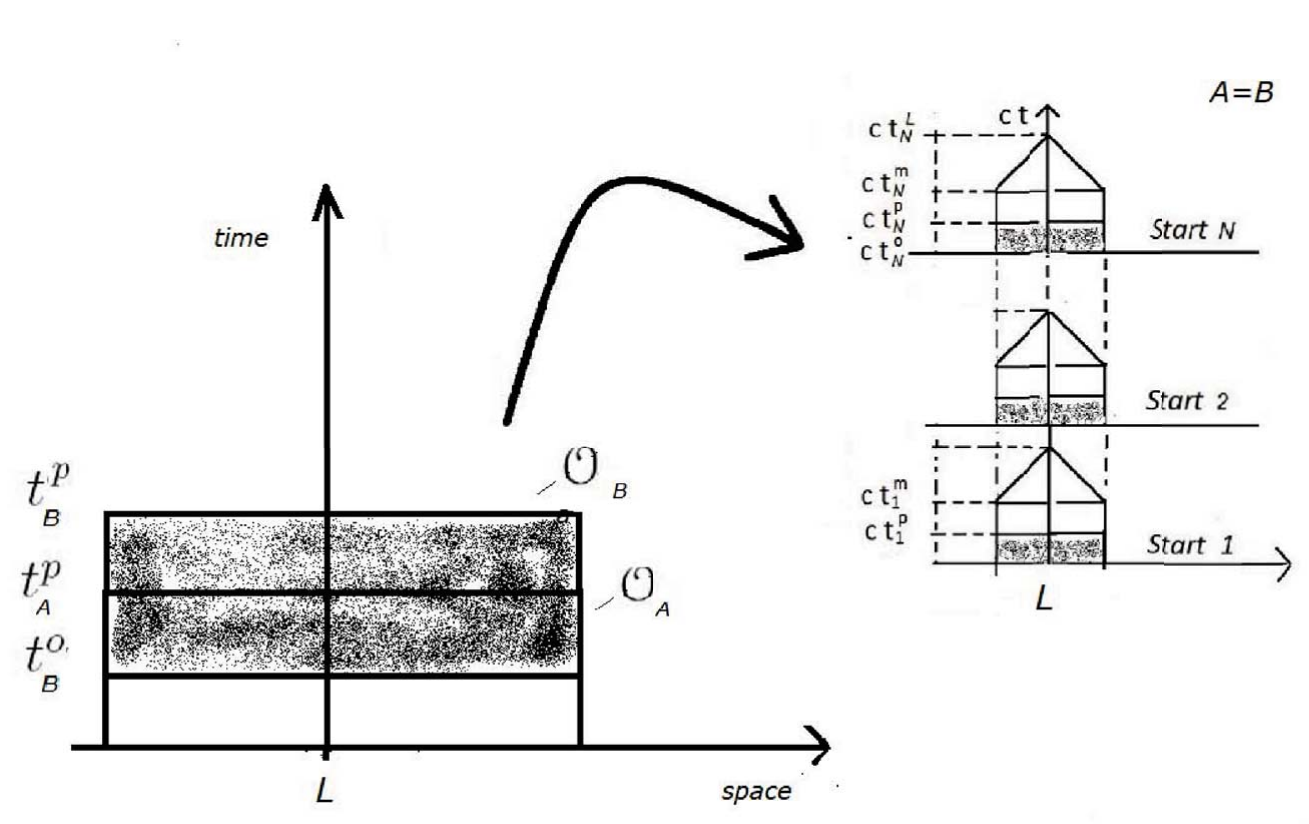}
	\caption{Time-shifted laboratories}
	\label{disle3traslibis}
\end{figure}
\subsection{Time-shifted regions}
If we consider two laboratory regions $\mathcal O_A$ and $\mathcal O_B$ 
$$ \mathcal O_A = L \times [0, t_A^P] \qquad , \qquad \mathcal O_B = L \times [t_B^o, t_B^P] $$
one shifted temporally with respect to the other, as shown in Figure \ref{disle3traslibis}, then we have that they have the same set of states:
$$\mathfrak S(\mathcal O_A) = \mathfrak S(\mathcal O_B)$$
since the various devices, equipment, etc., of the laboratory remain unchanged for all the initial moments of time that we consider. In other words, we are simply delaying the start of the preparation of the experiment. If we want to measure the values of $a \in \mathfrak X(\mathcal O_A) = \mathfrak X(\mathcal O_B)$ in the same state $\omega$ suitable for its measurement, we have that the graph of the $N$ copies of the ensemble is identical.
\\
In fact, having the two laboratory-type regions shifted temporally does not mean that we perform the $N$ trials of the ensemble relating to the two regions in temporal order; for example, first those of $\mathcal O_A$ and then those relating to $\mathcal O_B$. This does not happen. The departure of the ensembles always occurs by resetting the chronometer present in our laboratory $L$.
\\
We explicitly note that the set $\mathfrak S(\mathcal O_A \cap \mathcal O_B)$ is not contained in $\mathfrak S(\mathcal O_A)$. What we can say is that there exists a surjective map
$$\texttt{P} : \mathfrak S(\mathcal O_A | \mathcal O_A \cap \mathcal O_B) \longrightarrow \mathfrak S(\mathcal O_A \cap \mathcal O_B)$$
as described in Proposition \ref{mappaP}.
\section{Dislocated Laboratories} \label{disloc}
\begin{figure}[htbp] 
	\centering
		\includegraphics[scale=0.4]{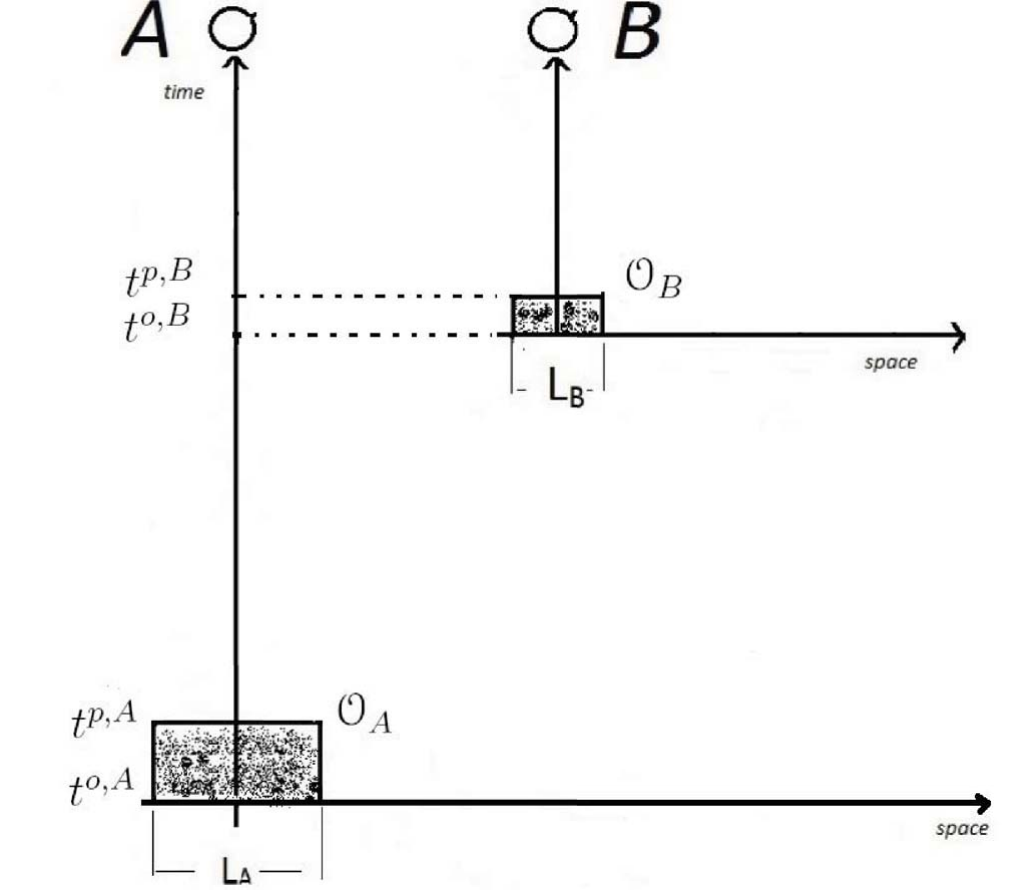}
	\caption{Dislocated laboratories}
	\label{fig:010.1}
\end{figure}%
Let us assume that we have two laboratories A and B located at two different points in space-time \textit{which are not moving with respect to each other}, and we denote by $A$ and $B$ their reference systems (and the associated clocks) on which the two experimenters rely for their measurements\footnote{Since they are not in motion with respect to each other, time flows in the same way in the clocks of $A$ and $B$, although their readings may be shifted.}
\\
These laboratories can be activated for various experimental verifications in different ways, for example:
\\
Prepare laboratory $A$ without activating laboratory $B$\footnote{In practice, without performing any experimental preparation in the set $L_B \subset \mathfrak E$} or vice versa, or activate both labs $A$ and $B$ but with different activation times, as shown in Figure \ref{fig:010.1}.
\\
In any case, two physical systems remain associated with the two laboratories $L_A$ and $L_B$, which we denote respectively by $(\mathfrak X_A, \mathfrak S_A)$ and $(\mathfrak X_B, \mathfrak S_B)$.
\\
We set in both laboratories the preparation time intervals for the experiments, which we denote by $s^{p,A}$ and $s^{p,B}$:
$$s^{p,A} = t^{p,A} - t^{o,A} \qquad , \qquad s^{p,B} = t^{p,B} - t^{o,B}$$
with $t^{o,A} = 0$.
\\
In this way, as shown in Figure \ref{fig:010.1}, we have the two laboratory-type regions 
\begin{equation}\label{dueregioni}
\mathcal O_A = L_A \times [0, t^{p,A}] \ , \qquad \mathcal O_B = L_B \times [t^{o,B}, t^{p,B}] \ , \qquad L_A, L_B \subset \mathfrak E 
\end{equation}
and the associated physical systems $(\mathfrak X(\mathcal O_A), \mathfrak S(\mathcal O_A))$ and $(\mathfrak X(\mathcal O_B), \mathfrak S(\mathcal O_B))$\footnote{In this case we decided to fix the preparation time in the respective laboratories, but this is an unnecessary condition; in fact, we could also act in the opposite way, not fixing a priori the preparation time of the various experiments but choosing a state of the laboratory and with it its relative preparation time and thus its laboratory-type region.}. 
\\
\begin{figure}[htbp] 
	\centering
		\includegraphics[scale=0.3]{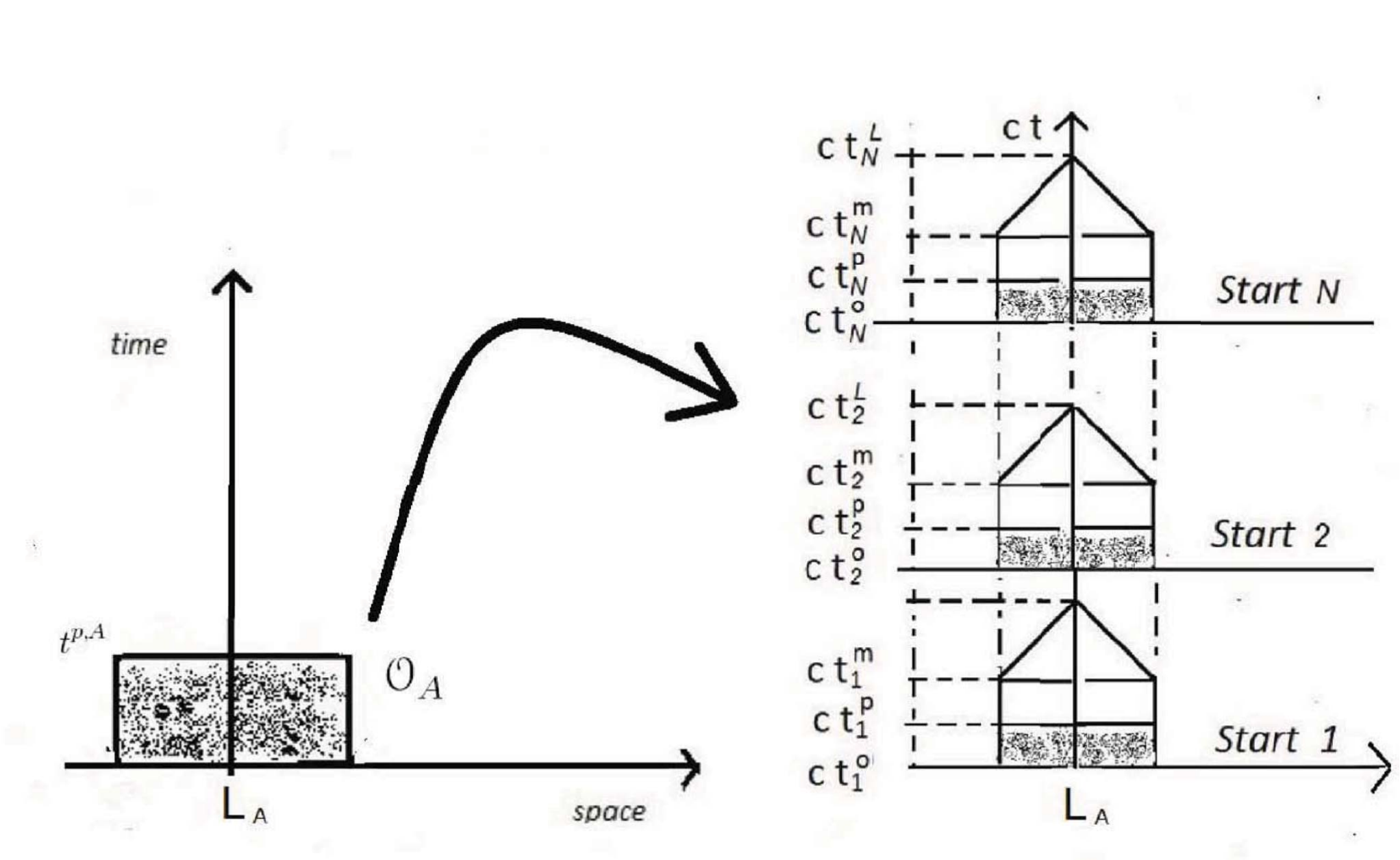}
	\caption{From Regions to Ensembles - Lab. A}
	\label{fig:010.2}
\end{figure}
In each of the two laboratories we can determine the distribution laws \eqref{distribuzio0} by applying the ensemble procedures as established in section \ref{proceduresperimentali}\footnote{The two laboratories can always exchange various information on the experimental procedures that determine their laboratory status; obviously this occurs with the necessary reception time due to the finite speed of the transmitted signals.} and, having two distinct systems in $A$ and $B$, we have the following possibilities:
\\

$\bullet$ First case: \textbf{Single measurement in the two laboratories}.\index{Single measurement in the two laboratories}
\\

Preparation of the ensembles occurs individually in each $A$ system and later in $B$ (or vice versa):
\\
Having fixed the observable $a \in \mathfrak X_A$ to be measured in the state $\omega^A_a \in \mathfrak S^A_a$, we associate with it an ensemble consisting of $N$ copies of the experiment, as shown in Figure \ref{fig:010.2}.
\\
Therefore, for each $a \in \mathfrak X_A$ and $\omega^A_a \in \mathfrak S^A_a$, we obtain, at the time $\tau_A$ established by the clock of laboratory A, the distribution law: 
\begin{equation}
\label{sistema-Asingolo}
\Delta \in B(\mathbb R) \longrightarrow P^{L_A}(a \in \Delta, \tau_A)_{\omega^A_a}
\end{equation}
and as regards laboratory B, we have the same considerations:
\\
For each $b \in \mathfrak X_B$ and $\omega^B_b \in \mathfrak S^B_b$, we obtain, at the time $\tau_B$ established by the clock of laboratory B, the distribution law:  
\begin{equation}
\label{sistema-Bsingolo}
\Delta \in B(\mathbb R) \longrightarrow P^{L_B}(b \in \Delta, \tau_B)_{\omega^B_b}
\end{equation}
\begin{figure}[htbp] 
	\centering
		\includegraphics[scale=0.5]{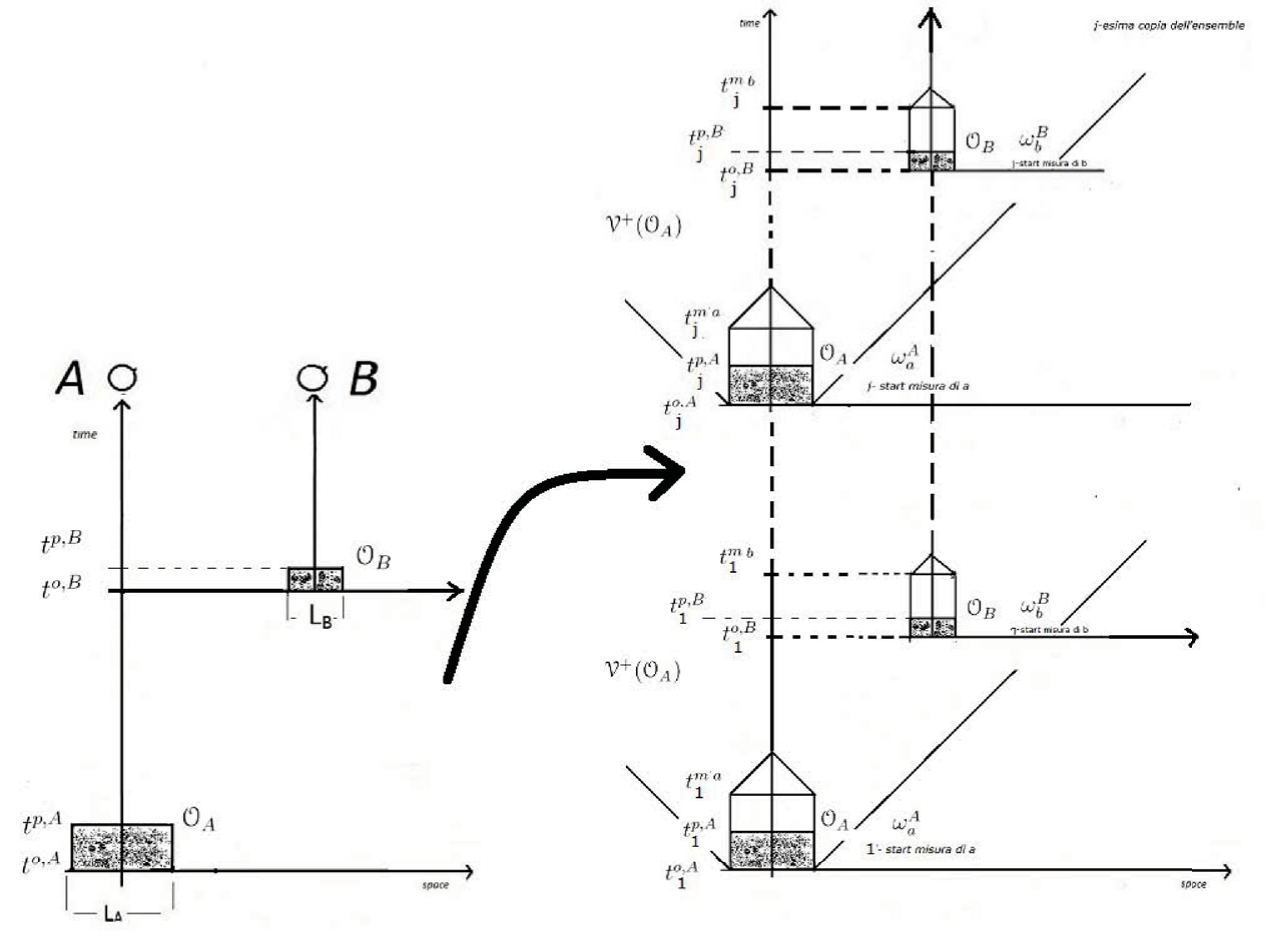}
	\caption{From Regions to Ensembles - Second case}
	\label{fig:010.3}
\end{figure}

$\bullet$ Second case: \textbf{Joint measurement in the two laboratories}.\index{Joint measurement in the two laboratories}
\\

\begin{figure}[htbp]
	\centering
		\includegraphics[scale=0.6]{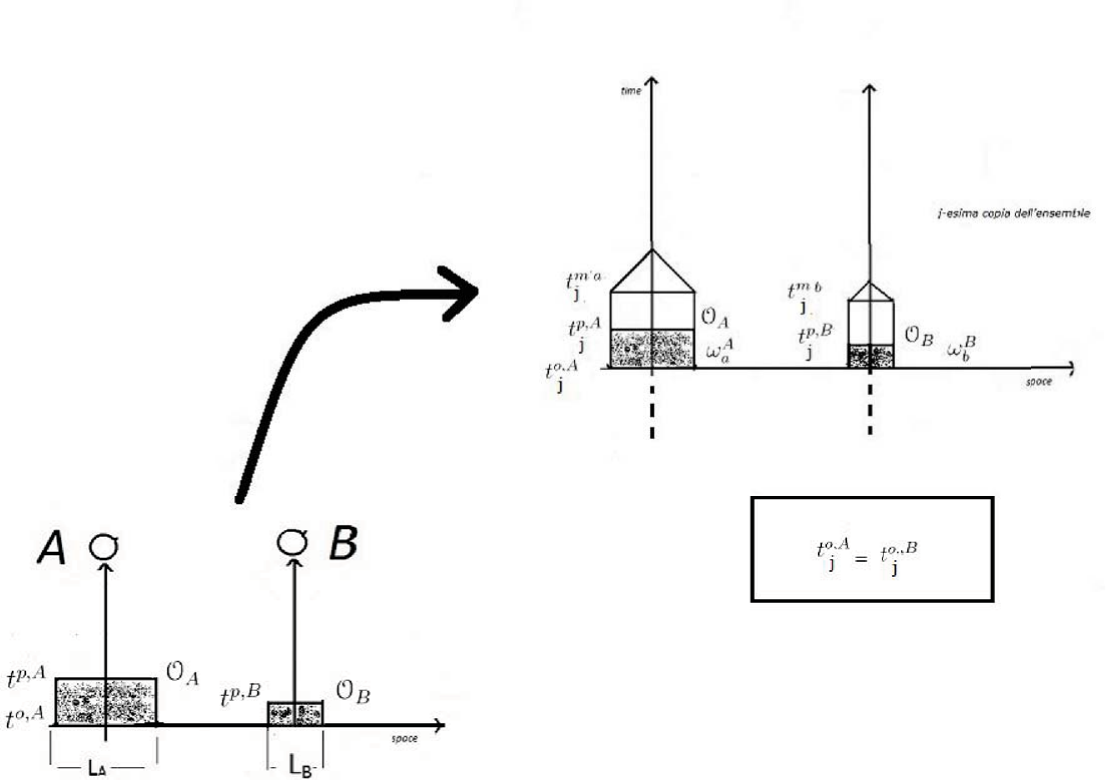}
	\caption{Simultaneous ensembles}
	\label{fig:09A.Bens}
\end{figure}

In this case, having two separate laboratories and thus \textit{two experimenters} who can act independently of each other, the preparation of the $N$ copies of the ensemble of experiments in the two systems can be carried out completely autonomously, preparing all $N$ copies in the same way, as shown in Figures \ref{fig:010.3} and \ref{fig:09A.Bens}.
\\
In the event that the initial times of the various preparations are the same in the two laboratories:
$$t_j^{o,A} = t_j^{o,B} \ , \qquad j = 1, 2, \ldots, N$$
we will speak of \textit{jointly simultaneous preparations separated in the two laboratories} $L_A$ and $L_B$\footnote{We observe from Figure \ref{fig:010.3} that the two ensembles start with an initial time equal to zero in the respective laboratory systems.}.
\\

As we discussed in section \ref{riproduzione}, everything that happens in the past light cone generated by the laboratory set $L_B$ has no effect on the preparation of the parametric state and on the experimental procedures carried out in the laboratory itself; \textit{they will only influence the act of measurement during the time interval between the end of the preparation and the measurement itself} — influences that we have established to be identical in all copies of our ensemble\footnote{This statement can be considered as a postulate of our model. 
\\
In practice, if the overall time of our measurements with respect to the measurement time $\tau$ is not very large, then we can assume that these perturbations do not change much across the various $N$ copies of the ensemble. }.
\\
In other words, even if the setup of the $L_B$ laboratory takes place \textit{after} that of the $L_A$ laboratory, as in Figure \ref{fig:010.3}, it is not influenced by having prepared and carried out a measurement in $L_A$, even if $O_A$ is in the past light cone of $O_B$. Therefore we assume that the preparations carried out in the region $\mathcal O_A$ cannot influence the preparations carried out in $B$, even if it turns out that\footnote{Recall that $\mathcal V^+(\mathcal O_A)$ is the future light cone generated by the region $\mathcal O_A$.}
$$ \mathcal O_B \subset \mathcal V^+(\mathcal O_A)$$
We observe that in $\mathfrak S(\mathcal O_B)$ there exist states that contain the information that the measurement of the observable $a$ in the state $\omega^A_a \in \mathfrak S^A$ has been carried out or not. This could happen if in $L_B$ there is a device capable of recording this event\footnote{The experimenter in $B$ records and prepares the experiment only with what he has available in his laboratory, with his equipment.} (with the fixed preparation time sufficient to operate such devices). We can prepare two states $\omega_B$ and $\omega_B'$ which have the same preparation time for the measurement at time $\tau = 0$, with the same fixed physical parameters, where $\omega_B$ has the same equipment/devices and procedures as $\omega_B'$, with the only difference that during the preparation of $\omega_B$ the measurement was not carried out in $L_A$, unlike $\omega_B'$ in which the measurement was carried out.
\\  
 
Therefore in $L_B$ we have two substantially identical states $\omega_B$ and $\omega_B'$ which differ only in the information of whether or not measurements have previously been carried out in the $L_A$ laboratory\footnote{This translates into the knowledge of variations in some physical parameters of the laboratory; the experimenter observes their change but does not act on the devices to control this variation. In practice, these parameters are not included in the parametric state.}.
\\
In this way, for the perturbations due to the measurement in $L_A$ we can write:
$$ P(b \in \Delta, 0)_{\omega_B} = P(b \in \Delta, 0)_{\omega_B'} $$
while if we consider their chronological states, for $\tau > 0$ we have:
 $$ P(b \in \Delta, \tau)_{\omega_B} \neq P(b \in \Delta, \tau)_{\omega_B'} $$
Let us now consider the case of the \textit{preparable conjunction in the two laboratories carried out simultaneously}. As mentioned, the various preparations of the experiments take place at the same time in the two laboratories, as shown in Figures \ref{fig:09A.Bens}\footnote{Warning: the measurement times could also be different:
$$\tau_A = t_j^{m,a} - t_j^{p,A} \ , \ \tau_B = t_j^{m,b} - t_j^{p,B} \ , \qquad \forall j = 1, 2, \ldots, N$$}. 
Since in this case the preparations take place simultaneously in the two laboratories, it cannot be ruled out that the two experimenters may not be able to take countermeasures to control the disturbances due to the individual preparations and measurements carried out in the two laboratories. The measurement of the observable $a$ in laboratory $A$ could make some states no longer suitable for the measurement of the observable $b$ in laboratory $B$ (and vice versa), which could be prepared individually in that laboratory. For example, the preparation carried out simultaneously in $A$ could put some equipment in laboratory $B$ offside and thus limit the preparations of any experiments for the measurement of $B$.
\begin{definition}\upshape\index{Jointly preparable observables in two distinct laboratories}
Two observables $a \in \mathfrak X_A$ and $b \in \mathfrak X_B$ are said to be jointly preparable in $L_A$ and $L_B$ in the respective states $\omega^A_a$ and $\omega^B_b$ if the preparation carried out in laboratory $A$ for the measurement of $a$ in the state $\omega^A_a \in \mathfrak S^A$ does not destroy the various operations carried out to prepare laboratory $B$ for the measurement of $b$ in the state $\omega^B_b \in \mathfrak S^B$, and vice versa.   
\end{definition}
The set of states of $L_A$ and $L_B$ for which $a$ and $b$ can be jointly prepared is denoted by\footnote{We did not use the notation $a:b$ because in this case there is no order of precedence in the preparation of the states, since we have two laboratories with two potential experimenters.}:
$$ \mathfrak S_{a \bowtie b}^{A \bowtie B} \subset \mathfrak S_a \qquad , \qquad \mathfrak S_{b \bowtie a}^{A \bowtie B} \subset \mathfrak S_b$$ 
and when we fix the laboratory-type regions we use the notation:\index{$\mathfrak S_{a \bowtie b}^{A \bowtie B}(\mathcal O_A)$}
$$ \mathfrak S_{a \bowtie b}^{A \bowtie B}(\mathcal O_A) \subset \mathfrak S_a(\mathcal O_A) \qquad , \qquad \mathfrak S_{b \bowtie a}^{A \bowtie B}(\mathcal O_B) \subset \mathfrak S_b(\mathcal O_B)$$ 
\begin{attenzione}\upshape
By definition of jointly preparable observables in the two laboratories, for every $\omega^A_a \in \mathfrak S_{a \bowtie b}^A(\mathcal O_A)$ there must exist a $\omega^B_b \in \mathfrak S_{b \bowtie a}^B(\mathcal O_B)$, and vice versa.
\\
Thus, if $\mathfrak S_{a \bowtie b}^A(\mathcal O_A) \neq \emptyset$, then it cannot be the case that $\mathfrak S_{b \bowtie a}^B(\mathcal O_B) = \emptyset$.
 \end{attenzione}
\begin{figure}[htbp]
	\centering
		\includegraphics[scale=0.5]{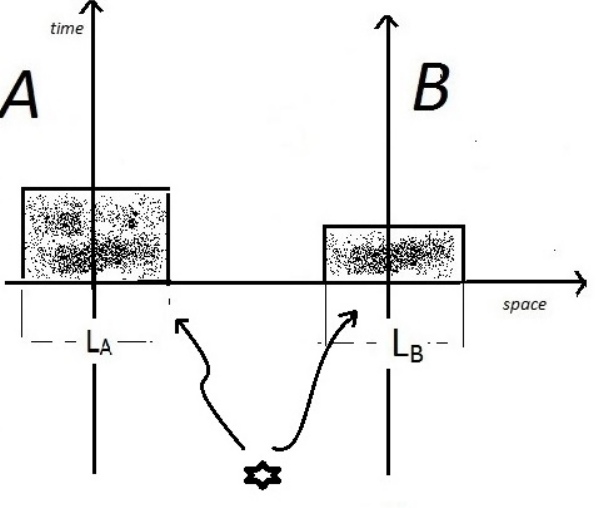}
	\caption{External source}
	\label{fig:09sorgentelab}
\end{figure}
\section{Experimental Invariance}\label{invarianza-sperimentale}
After these clarifications on measurement and ensemble modes, let us return to the study of observables in the two dislocated laboratories and ask the following question:
\\
Let $a \in \mathfrak X_A$; when can I say that $a \in \mathfrak X_B$?
\\
In other words, we ask whether the observable $a$ is measurable in laboratory $L_B$ with the preparation time $t^{p,B}$\footnote{It follows that it is measurable in the laboratory-type region $\mathcal O_B$.}, i.e., whether there exists a device capable of measuring it whose preparation requires a time $t^{p,B}$, i.e., whether there exists a state of $\mathfrak S^B$ suitable for $a$.
 \\
Therefore
$$a \in \mathfrak X_B \qquad \Longleftrightarrow \qquad \mathfrak S^B_a \neq \emptyset$$
Before proceeding with the discussion, we must make a banal but necessary remark:
\\
In the laboratory we have the instruments that are needed to measure our observables, and by definition they must be present only in the laboratory and not outside it. In fact, we can say that it is precisely the devices that determine the extent of the laboratory, unlike our source $S$ of the measurement, which is not necessarily contained in the laboratory.  
\\
Thus, the devices are internal to the laboratory, considered as a single measuring device, while the source $S$ is identified by the laboratory reference system considered.
\\

For example, if we need to measure the intensity of a certain wave frequency from a sound source $S$ in laboratories $A$ and $B$ (see Figure \ref{fig:09sorgentelab}), we have that $S$ is identified by the coordinates with respect to the two references\footnote{The source $S$ to be detected must be positioned in the past light cones generated by the laboratory-type regions of $A$ and $B$. Furthermore, it is assumed that in all $N$ copies of the ensembles the intensity of the sound source does not change.}, while the measuring devices are located in the respective laboratories with their preparation times, time that the two experimenters have available to prepare the measurement and which may not be sufficient for the measurement of this frequency (and therefore for its identification in the laboratory).
\\
Be careful: in this example, the experimenter in $A$ does not perform the same experiment as that in $B$, even if the observable to be measured is the same, since the sound source $S$ does not have the same spatial position with respect to the reference systems centered in the two laboratories (so we may not have the same situation in the two laboratories).
\\ 
We also observe that if $a$ is a physical quantity measurable in both laboratories $L_A$ and $L_B$, then it is possible to measure it jointly in the two laboratories, obtaining in this case the following sets of states:
$$ \mathfrak S_{a \bowtie a}^A \subset \mathfrak S_a^A \qquad , \qquad \mathfrak S_{a \bowtie a}^B \subset \mathfrak S_a^B $$ 
Physically we can set up two laboratories at rest with respect to each other, with the same geometric characteristics (therefore the possibility of having the same devices, the same experimental procedures, etc.) but positioned in different places in space. 
\\
What can we say about their observables and system states?
\\
Having the same devices/instruments in the two copy laboratories $A$ and $B$ does not ensure that the same observables can be detected in the two laboratories. Let us specify the issue better by focusing on the meaning of the newly introduced term detect\footnote{See also Definition \ref{rilevare}.}:
 \\
Since the two laboratories are identical with the same preparation times, etc., this means that for each device positioned in $L_A$ there corresponds an identical one positioned in $L_B$ with the same preparations, conditions, etc. Mathematically, this translates into stating that each state $\omega_A \in \mathfrak S^A$ identified through the reference system $(K_A, O_A)$ corresponds to one and only one copy state $\omega_B \in \mathfrak S^B$ identified through the reference system $(K_B, O_B)$, and vice versa.
\\
Therefore, if $a \in \mathfrak X^A$, then we have that the set $\mathfrak S_a^B \neq \emptyset$\footnote{ As discussed in section \ref{proceduresperimentali}, observables are physical quantities that remain stable in time and space; what changes are their values in time and space, but not their typology.
Thus, if the observable $a$ is in both $L_A$ and $L_B$, what changes is its value in those labs, which is underlined by the superscript $L_A$ and $L_B$ in their distribution law.} and one could have
$$P^{L_B}(a \in \left\{ 0 \right\}, \tau)_{\omega'} = 1 \ , \qquad \forall \omega' \in \mathfrak S_a(\mathcal O_B)$$
which means that $a \subset 0$.
\\
For example, in Figure \ref{fig:09sorgentelab}, if the sound source $S$ does not fall into the past light cone of laboratory $B$, then the laboratory, even having internal instruments for measuring sound frequencies, will not detect them.
\\
We are now ready to formalize the issue by introducing a new postulate relating to the two distinct laboratories at rest with respect to each other.
 
\begin{post}[\textbf{Isotropy of Space}]\label{isotropia-A}\index{Postulate-Isotropy of Space}
Let $L_A$ and $L_B$ be two laboratories at rest with respect to each other and let $(K_A, O_A)$ and $(K_B, O_B)$ be their respective laboratory reference systems. 
If the regions $L_A, L_B \subset \mathfrak E$ are superimposable through a space-time translation\footnote{\label{traslazione}In practice, it is the coordinate transformation of the passage from the reference systems centered in the two laboratories $(K_A, O_A) \longmapsto (K_B, O_B)$ given by:
$$ (\underline{x}, x^0) \rightarrow (\underline{x} + \underline{a}, x^0 + b) \ , \qquad \forall (\underline{x}, x^0) \in \mathbb R^3 \times \mathbb R $$
with $(\underline{a}, b) \in \mathbb R^3 \times \mathbb R$ the displacement vector.}, then the two systems $(\mathfrak X^A, \mathfrak S^A)$ and $(\mathfrak X^B, \mathfrak S^B)$ have the following properties:
\begin{itemize}
\item[1.] They have the same observables:
$$\mathfrak X_A = \mathfrak X_B $$
\item[2.] There is a one-to-one correspondence\footnote{Obviously this map will depend on the coordinate transformation $T^{K_B, O_B}_{K_A, O_A}$ to go from $(K_A, O_A) \longmapsto (K_B, O_B)$.}
\begin{equation}\label{copialab}
\Lambda^\natural : \mathfrak S_A \longrightarrow \mathfrak S_B  
\end{equation} 
such that   
$$ \Lambda^\natural(\mathfrak S_a^A) = \mathfrak S_a^B \ , \qquad \forall a \in \mathfrak X_A$$
and
$$ \mathfrak X_{\omega_A}^A = \mathfrak X_{\Lambda^\natural(\omega_A)}^B \ , \qquad \forall \omega_A \in \mathfrak S_A $$
\item[3.] The map $\Lambda^\natural$ sends jointly preparable observables of $(\mathfrak X^A, \mathfrak S^A)$ to jointly preparable observables of $(\mathfrak X^B, \mathfrak S^B)$.
\\
Furthermore, if $x, y \in \mathfrak X^A$ are jointly preparable in the order $x:y$, then it turns out that
$$ \Lambda^\natural (\mathfrak S_{x:y}^A) = \mathfrak S_{x:y}^B$$
\end{itemize}
In particular, if the laboratory-type regions $\mathcal O_A$ and $\mathcal O_B$ relating to the two laboratories are superimposable through a space-time translation, we have:
\begin{itemize}
\item[1.] For observables:
$$\mathfrak X(\mathcal O_A) = \mathfrak X(\mathcal O_B) $$
\item[2.] For states, the existence of a one-to-one correspondence:
\begin{equation} 
\Lambda^\natural : \mathfrak S(\mathcal O_A) \longrightarrow \mathfrak S(\mathcal O_B)  
\end{equation} 
such that 
$$ \Lambda^\natural(\mathfrak S_a(\mathcal O_A)) = \mathfrak S_a(\mathcal O_B) \ , \qquad \forall a \in \mathfrak X(\mathcal O_A)$$
and
$$ \mathfrak X_{\omega_A}(\mathcal O_A) = \mathfrak X_{\Lambda^\natural(\omega_A)}(\mathcal O_B) \ , \qquad \forall \omega_A \in \mathfrak S(\mathcal O_A) $$
In particular, if $x, y \in \mathfrak X^A$ are jointly preparable in the order $x:y$, it turns out that
$$ \Lambda^\natural (\mathfrak S_{x:y}(\mathcal O_A)) = \mathfrak S_{x:y}(\mathcal O_B)$$
\item[3.] For every $a \in \mathfrak X(\mathcal O_A)$ we have:
$$ Z_{\mathfrak S(\mathcal O_A)}(a) = Z_{\mathfrak S(\mathcal O_B)}(a)$$
\end{itemize}
\end{post}
\begin{remark}\upshape
For the two copy laboratories $L_A$ and $L_B$, for each $a \in \mathfrak X^A = \mathfrak X^B$, there exists a map, induced by the map given in \eqref{copialab}:
$$\Lambda_a : \mathbb{M}^A(a) \longrightarrow \mathbb{M}^B(a)$$
such that
$$ \Lambda_a (\mu^A_{\omega,a}) = \mu^B_{\Lambda^\natural(\omega), a} \qquad , \qquad \forall \omega \in \mathfrak S^A_a$$
\end{remark}
\section{Laboratory Inclusions}
\begin{figure}[htbp]
	\centering
		\includegraphics[scale=0.35]{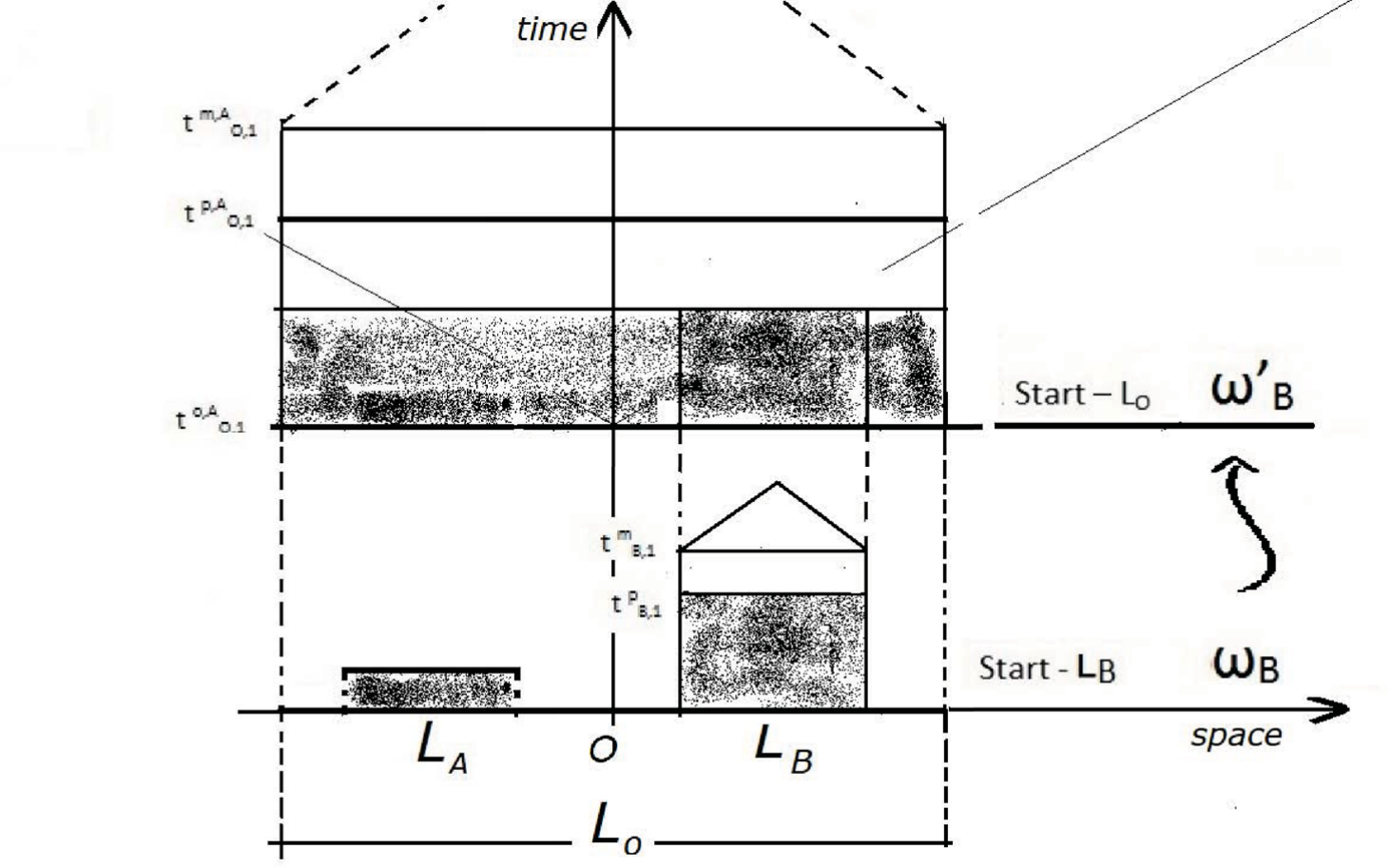}
	\caption{Sub-laboratory B}
	\label{fig:duelabBO}
\end{figure}

In our model we consider the laboratory system as a single body, where it is possible to prepare the laboratory by activating procedures and devices in sequential actions. This fact must be taken into account when you want to embed the two laboratories into a larger one. In fact, in the two laboratories $L_A$ and $L_B$ the actions undertaken to implement the experimental procedures can be performed autonomously from each other, which cannot happen when they are considered part of a larger laboratory $L_o$.
\begin{figure}[htbp]
	\centering
		\includegraphics[scale=0.3]{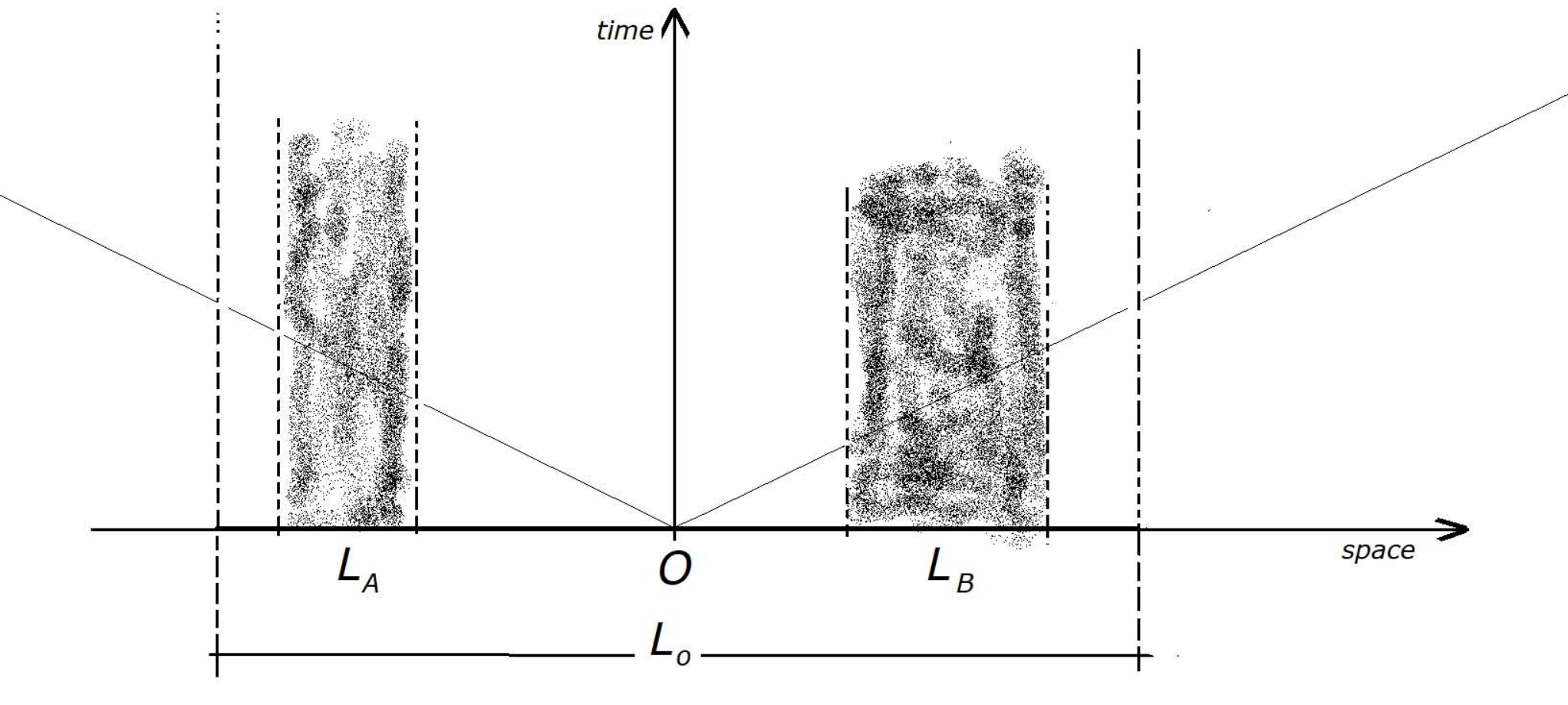}
	\caption{Sub-laboratories}
	\label{fig:duelab}
\end{figure}
Therefore, let us assume that we have two laboratories $L_A$ and $L_B$, \textit{at rest with respect to each other}. They can always be considered part of a larger laboratory $L_o$ centered at the point $O$ between the two reference systems, as in Figure \ref{fig:duelab}:
$$L_A, L_B \subset L_o \subset \mathfrak E \qquad , \qquad L_A \cap L_B = \emptyset$$
The observer of $L_o$ will keep the same times $\tau_A, \tau_B$ with respect to the clocks of $A$ and $B$, since \textit{the two laboratories are not moving with respect to $L_o$}. The synchronization of the clocks is possible to achieve because we know where $A$ and $B$ are positioned with respect to the laboratory system $(K,O)$ of $L_o$\footnote{The observer placed at the center of laboratory $L_o$ will read the measurements delayed with respect to $L_A$ and $L_B$ by $\tau_A^L = \tau_A - d_A/c$ and $\tau_B^L = \tau_B - d_B/c$, respectively, where $d_A$ and $d_B$ are the distances from the center of laboratory $A$ and $B$ to the origin of the $L_o$ system, respectively.
\\
We observe that laboratory $L_o$ is spatially limited but it does not necessarily mean that it is a room in a building; this happens, for example, when making astronomical measurements. In practice, we are stating that in $L_A$ and $L_B$ the passage of time could be different due to gravitational causes; we will assume that in $L_o$ these phenomena are negligible.
\\
Therefore we have $\tau = \tau_A = \tau_B$, which are the times at which the measurement takes place in the respective laboratories (and not the reading times).}.
\\
Let us consider the two laboratory-type regions $\mathcal O_A$ and $\mathcal O_B$ from equation \eqref{dueregioni}; let us see how we must choose the laboratory-type region $\mathcal O_o$ such that 
$$\mathcal O_A, \mathcal O_B \subset \mathcal O_o \qquad , \qquad \mathcal O_o = L_o \times [0, t^{p}] $$
is experimentally well-posed.
\\
Again we have two ways to include the two laboratories into a larger one.
\\
\begin{figure}[htbp]
	\centering
		\includegraphics[scale=0.35]{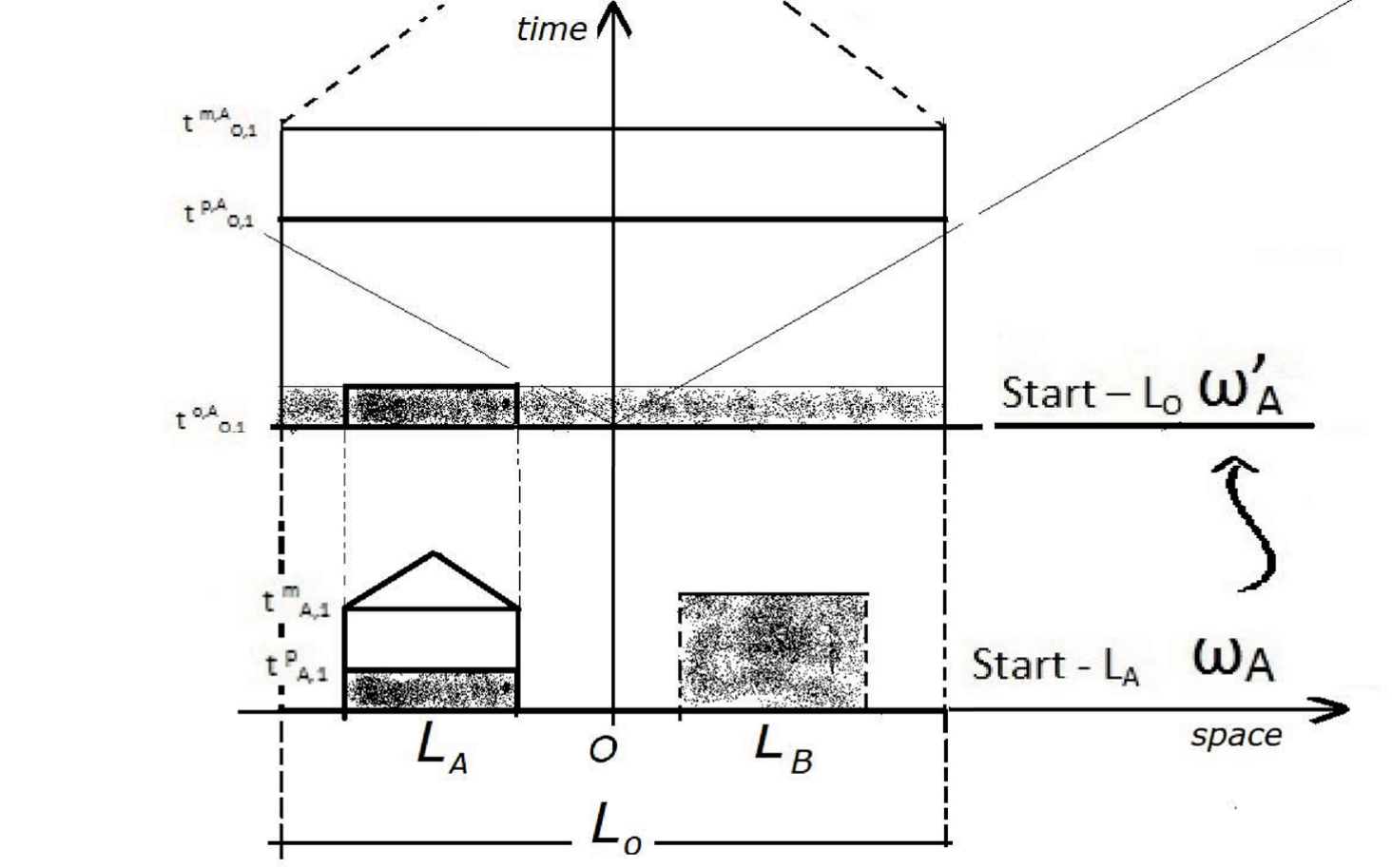}
	\caption{Sub-laboratory A}
	\label{fig:duelabAO}
\end{figure}
$\bullet$ First case: \textit{Single Inclusion}.\index{Single Inclusion}
\\
This is essentially the case discussed in section \ref{sottolaboratorio}. In fact, the matter does not change if laboratory $L_A$ is not centered in $L_o$ or if the preparations in $A$ and $B$ occur simultaneously or not (see Figures \ref{fig:duelabAO} and \ref{fig:duelabBO}). In this case, for the $L_o$ laboratory it is enough to fix a preparation time 
$$t^p \geq \max \left\{ t^{p,A}, t^{p,B} \right\}$$
in such a way as to illuminate from $O$ the entire laboratory $L_o$\footnote{See Definition \ref{illuminata}, on page \pageref{illuminata}.} and obtain the laboratory-type region $\mathcal O_o = L_o \times [0, t^p]$, obtaining the physical system $(\mathfrak X(\mathcal O_o), \mathfrak S(\mathcal O_o))$ associated with it.
\\
Therefore, if $a \in \mathfrak X_A \subset \mathfrak X(\mathcal O_o)$, as we said previously, for every $\omega_A \in \mathfrak S^A_a$ there exists $\omega_A' \in \mathfrak S_a(\mathcal O_o)$ such that  
\begin{equation}
P^{L_A}(a \in \Delta, \tau_A)_{\omega_A} = P^{L_o}(a \in \Delta, \tau_A)_{\omega_A'}
\end{equation}
and in a similar way we reason for system B:
\begin{equation}
P^{L_B}(b \in \Delta, \tau_B)_{\omega_B} = P^{L_o}(b \in \Delta, \tau_B)_{\omega_B'}
\end{equation}

\begin{figure}[htbp]
	\centering
		\includegraphics[scale=0.3]{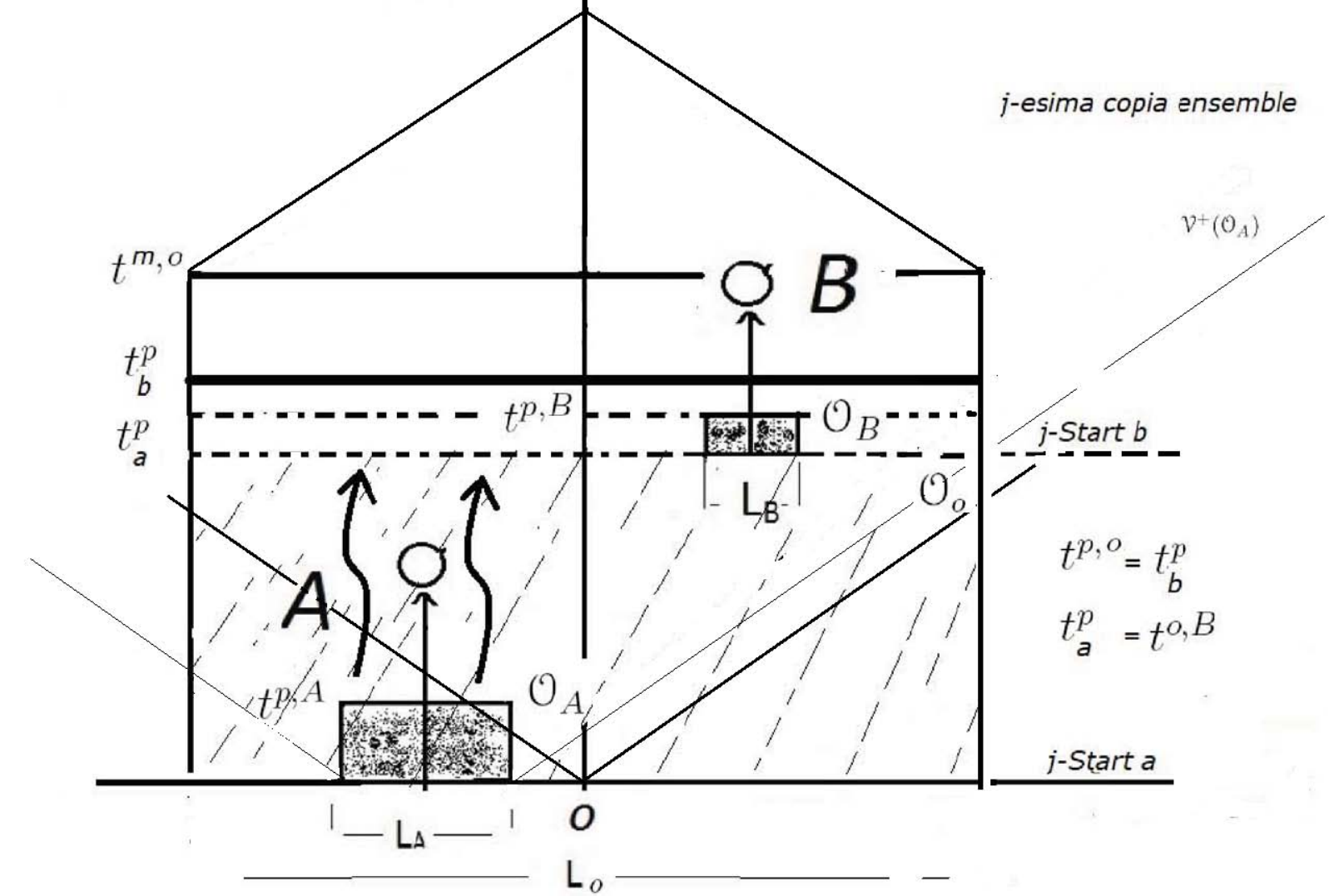}
	\caption{Non-Simultaneous Sub-laboratories}
	\label{fig:010.4}
\end{figure}

Also in this case the considerations made for the "centered" laboratory of the previous section can be applied\footnote{We reiterate that changing the reference system of the laboratory only has the effect of rearranging the state of the system. Moreover, we underline once again that we have no temporal problems, $\tau_A = \tau_B$, since all the laboratories are at rest with respect to each other.} and we can define the relations \eqref{restrizionestatobis} and \eqref{restrizionestatotris} for systems A and B and the related sets of states.
$$\mathfrak S_a(\mathcal O_o | \mathcal O_A) \subset \mathfrak S_a(\mathcal O_o) \qquad , \qquad \mathfrak S_b(\mathcal O_o | \mathcal O_B) \subset \mathfrak S_b(\mathcal O_o) $$
and, as discussed in section \ref{sottolaboratorio}, we have two surjective maps
$$ \texttt{P}^A_a : \mathfrak S_a(\mathcal O_o | \mathcal O_A) \longrightarrow \mathfrak S_a^A \qquad , \qquad \texttt{P}^B_a : \mathfrak S_a(\mathcal O_o | \mathcal O_B) \longrightarrow \mathfrak S_a^B $$
such that
$$ \mu_{\omega,a}^{L_o} = \mu_{\texttt{P}^A_a(\omega),a}^A \ , \qquad \forall \omega \in \mathfrak S_a(\mathcal O_o | \mathcal O_A)$$
and
$$ \mu_{\omega,a}^{L_o} = \mu_{\texttt{P}^B_a(\omega),a}^B \ , \qquad \forall \omega \in \mathfrak S_a(\mathcal O_o | \mathcal O_B)$$
\\
$\bullet$ Second case: \textit{Joint Inclusion}. \index{Joint Inclusion}
\\
Compared to the previous situation, the matter becomes more delicate, since what can be done simultaneously in two different laboratories, and therefore with two distinct experimenters, cannot necessarily be done in a single laboratory. In the preparable conjunction in $L_o$, the preparation of the observables is carried out sequentially. We can prepare in $L_o$ jointly, first $a$ and then $b$ (or vice versa) in the order $a<b$, or for the simultaneous measurement in the order $a:b$, a preparation which differs from the experimental procedures of the separate case, assuming that such preparations are experimentally feasible. In fact, we must remember once again that in $L_o$ we have only one experimenter who carries out all the experimental procedures.
\\
Thus, the problem is the following: include the preparation procedures of the two laboratories $L_A$ and $L_B$ into a larger laboratory.
\\
Let us ask ourselves the following problem:
\\
In order to include the two preparations of the two laboratories $L_A$ and $L_B$, constituted by the laboratory-type regions $\mathcal O_A$ and $\mathcal O_B$, into a single preparation of the laboratory $L_o$, what experimental properties must a laboratory-type region $\mathcal O_o = L_o \times [0, t^{P,o}]$ satisfy?
 \\
Let us address the problem by analyzing some of the possible situations that could occur experimentally.
 \\
Let us assume for the moment that the preparation of laboratory $L_A$ is carried out before that of $L_B$, as shown in Figure \ref{fig:010.4}.
\\

As we discussed, given an observable $a \in \mathfrak X_A$, it is also measurable in laboratory $L_o$ in any laboratory-type region $\mathcal O_o$ illuminated   from the point $O$ of the origin of our laboratory system of $L_o$ which contains the preparation of this observable in $L_A$, i.e., $\mathcal O_A \subset \mathcal O_o$, and its suitable states in $L_o$ induced by $\mathcal O_A$ of $L_A$ are given by
$$\mathfrak S_a(\mathcal O_o | \mathcal O_A) \subset \mathfrak S_a(\mathcal O_o)$$ 
Similarly, retracing the previous discussion for any observable $b \in \mathfrak X_B$, we have $\mathcal O_B \subset \mathcal O_o$, and its suitable states in $L_o$ induced by $\mathcal O_B$ of $L_B$ are given by the set
$$\mathfrak S_b(\mathcal O_o | \mathcal O_B) \subset \mathfrak S_b(\mathcal O_o)$$
This does not guarantee that, if $a \in \mathfrak X^A$ and $b \in \mathfrak X^B$ are \textit{jointly separately preparable} in $L_A$ and $L_B$ in their respective states $\omega^A_a$ and $\omega^B_b$, then there exists a laboratory-type region $\mathcal O_o$ of $L_o$ and a state $\omega_o$ of $\mathfrak S(\mathcal O_o)$ which contains the experimental procedures of $\omega^A_a$ and $\omega^B_b$ using their respective measuring instruments and devices, where they are jointly preparable for their subsequent or simultaneous measurement.
\\
Thus, if $\omega_A \in \mathfrak S_a^A$ and $\omega_B \in \mathfrak S_b^B$, it does not necessarily mean that there exists $\omega_o \in \mathfrak S^{a<b}(\mathcal O_o)$ such that
\begin{itemize}
\item [-] $\mathfrak X_{\omega_A}, \mathfrak X_{\omega_B} \subset \mathfrak X_{\omega_o}$;
\item [-] $P^A(a \in \Delta, \tau)_{\omega_A} = P^{L_o}(a \in \Delta | a < b, \tau)_{\omega_o}$;
\item [-] $P^B(b \in \Delta, \tau)_{\omega_B} = P^{L_o}(b \in \Delta | a < b, \tau)_{\omega_o}$
\end{itemize}
where we used the notation from relations \eqref{primamisura} and \eqref{secondamisura}.
\\
The same considerations apply for their simultaneous joint measurement in the order $a:b$:
\\
We cannot say that there exists a state $\omega_o \in \mathfrak S_{a:b}(\mathcal O_o)$ such that
\begin{property}\label{amalgama}
\begin{itemize}
\item [-] $\mathfrak X_{\omega_A}, \mathfrak X_{\omega_B} \subset \mathfrak X_{\omega_o}$;
\item [-] $P^A(a \in \Delta, \tau)_{\omega_A} = P^{L_o}(a \in \Delta : b \in \mathbb R, \tau)_{\omega_o}$;
\item [-] $P^B(b \in \Delta, \tau)_{\omega_B} = P^{L_o}(a \in \mathbb R : b \in \Delta, \tau)_{\omega_o}$.
\end{itemize}
 \end{property}

We now have the following definition:
\begin{definition}\upshape[\textbf{Amalgamated State}]\index{Amalgamated state}\index{$\omega_A \bowtie \omega_B$}
Let $a \in \mathfrak X^A$ and $b \in \mathfrak X^B$ be jointly observable and separately preparable in the two laboratories in the respective states $\omega_A \in \mathfrak S^A$ and $\omega_B \in \mathfrak S^B$, and suppose that $a$ and $b$ are compatible in the laboratory-type region $\mathcal O_o$ of $L_o$. 
\\
If there exists a state $\omega_o \in \mathfrak S_{a:b}(\mathcal O_o)$ such that the conditions in property \ref{amalgama} hold, then the state $\omega_o$ is called the amalgam of $\omega_A$ and $\omega_B$, in symbols:
$$ \omega_o = \omega_A \bowtie \omega_B $$
\end{definition}
Let us see what happens when we jointly include two laboratories where the preparations of the observables take place \textit{jointly simultaneously, separately} in the two laboratories $L_A$ and $L_B$, as shown in Figure \ref{fig:09A.Bens} of section \ref{disloc}.
\\
Can we include the two preparations in $L_o$ jointly for simultaneous or subsequent preparation?
\\
The answer is negative, for the same previous arguments: what can be done in two distinct laboratories cannot be done in a single laboratory that extends both laboratories; in $L_o$ we must first prepare $a$ and then $b$ or vice versa, and not simultaneously.
\\
Therefore, even if $a$ and $b$ are jointly preparable in $L_A$ and $L_B$, i.e.,
$$ \mathfrak S_{a \bowtie b}^A \neq \emptyset \qquad , \qquad \mathfrak S_{b \bowtie a}^B \neq \emptyset $$ 
it is not certain that they can be jointly prepared in $L_o$.
\\
Furthermore, we cannot even say the opposite: if $a \in \mathfrak X^A$ and $b \in \mathfrak X^B$ are jointly preparable in $L_o$ for their simultaneous measurement in the order $a:b$, it is not certain that they can be jointly prepared separately in the two laboratories due to some state of the system\footnote{We note that if $\omega \in \mathfrak S_{a:b}(\mathcal O_o) \subset \mathfrak S_b(\mathcal O_o)$, the observable $b$ belongs to $\mathfrak X_\omega$, but we cannot ensure that all of $\mathfrak X^B$ is contained in $\mathfrak X_\omega(\mathcal O_o)$, and thus guarantee at least the existence of a restricted state $\omega^B$ on $\mathfrak X^B$ of $\omega$.}.
\subsection{Independent Systems}
We adapt to our case the algebraic notion of statistical independence, a notion that can be found in \cite{Redei011, Summers90}, to physical systems associated with sublaboratories.
\\
The definitions we will give, differently from the algebraic case, depend heavily on the methods of preparing the experiments in the various laboratories\footnote{Let us recall the definition of independence in the context of operator algebras \cite{Summers90}:
\\
\textit{Let $\mathfrak A_A$ and $\mathfrak A_B$ be C*-subalgebras of $\mathfrak A$. The pair $(\mathfrak A_A, \mathfrak A_B)$ is said to be C*-independent if for every state $\varphi_A$ of $\mathfrak A_A$ and state $\varphi_B$ of $\mathfrak A_B$ there exists a state $\varphi$ of $\mathfrak A$ such that $\varphi_A = \varphi|_{\mathfrak A_A}$ and $\varphi_B = \varphi|_{\mathfrak A_B}$.} }.

\begin{figure}[htbp]
	\centering
		\includegraphics[scale=0.50]{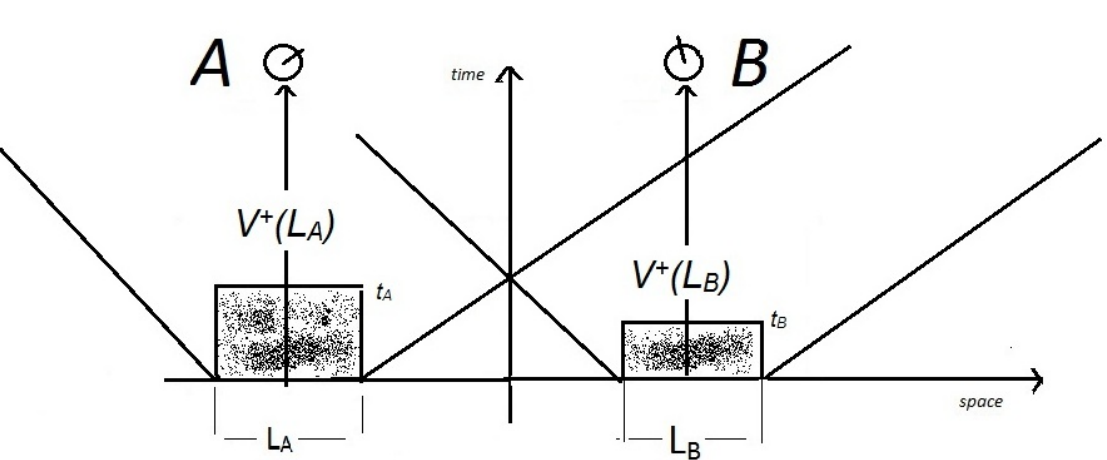}
	\caption{Measurement Disturbance}
	\label{fig:011}
\end{figure}

\begin{definition}[\textbf{Singular Independence}]\upshape\label{indp-noop}\index{Singular Independence of physical systems}
Two physical systems $(\mathfrak X_A, \mathfrak S_A)$ and $(\mathfrak X_B, \mathfrak S_B)$ associated with the two laboratory-type regions $\mathcal O_A$ and $\mathcal O_B$, respectively, are said to be singularly independent if, for any laboratory-type region $\mathcal O_o$ of $L_o$ which \textit{singularly includes} the two laboratories located at $A$ and $B$, then for every observable $x \in \mathfrak X_A$, $y \in \mathfrak X_B$ and $\omega_A \in \mathfrak S_x^A$, $\omega_B \in \mathfrak S_y^B$, there exists a state $\omega \in \mathfrak S(\mathcal O_o)$ such that: 
\begin{itemize}
\item $\omega$ is suitable for both $x$ and $y$, i.e., $\omega \in \mathfrak S_x(\mathcal O_o) \cap \mathfrak S_y(\mathcal O_o)$;
\item $P^{L_o}(x \in \Delta, \tau)_{\omega} = P^{L_A}(x \in \Delta, \tau)_{\omega_A}$;
\item $P^{L_o}(y \in \Delta, \tau)_{\omega} = P^{L_B}(y \in \Delta, \tau)_{\omega_B}$.
\end{itemize} 
\end{definition}
We have a second definition of independence: 
\begin{definition}\upshape[\textbf{Joint or Operational Independence}]\label{indip-op}\index{Joint Independence of physical systems} \index{Operational Independence}
Two physical systems\\
$(\mathfrak X_A, \mathfrak S_A)$ and $(\mathfrak X_B, \mathfrak S_B)$ associated with the two laboratory-type regions $\mathcal O_A$ and $\mathcal O_B$, respectively, are said to be operationally independent if, for any laboratory-type region $\mathcal O_o$ of $L_o$ which \textit{jointly includes} the two laboratories located at $A$ and $B$, then for every observables $x \in \mathfrak X_A$, $y \in \mathfrak X_B$ that are compatible observables of the system $(\mathfrak X, \mathfrak S)$ related to $\mathcal O_o$, and for states $\omega_A \in \mathfrak S^A_x$, $\omega_B \in \mathfrak S^B_y$, there exists a state $\omega \in \mathfrak S_{a:b}(\mathcal O_o)$ such that:
 $$P^{L_o}(x \in \Delta_0 : y \in \Delta_1, \tau)_{\omega} = P^{L_A}(x \in \Delta_0, \tau)_{\omega_A} \, P^{L_B}(y \in \Delta_1, \tau)_{\omega_B}$$
\end{definition}
\begin{remark}\upshape
If the laboratory systems $A$ and $B$ are operationally independent, then for every pair of observables $x \in \mathfrak X_A$ and $y \in \mathfrak X_B$ that are compatible in $(\mathfrak X, \mathfrak S)$, they are also singularly independent.
\\

Indeed, from operational independence we obtain: 
$$P^{L_o}(x \in \Delta_0 : y \in \mathbb R, \tau)_{\omega} = P^{L_A}(x \in \Delta_0, \tau)_{\omega_A} $$
and by compatibility,
$$P^{L_o}(x \in \Delta_0 : y \in \mathbb R, \tau)_{\omega} = P^{L_o}(x \in \Delta_0, \tau)_{\omega} $$
It follows that, for each $\tau \geq 0$, we obtain
$$P^{L_o}(x \in \Delta_0, \tau)_{\omega} = P^{L_A}(x \in \Delta_0, \tau)_{\omega_A} $$
Applying the same considerations to the observable $y$, we obtain:
$$P^{L_o}(x \in \Delta_0 : y \in \Delta_1, \tau)_{\omega} = P^{L_o}(x \in \Delta_0, \tau)_{\omega} \, P^{L_o}(y \in \Delta_1, \tau)_{\omega} $$   
In this case, if $\mu_{\omega, x:y}^{L_o}$ is the measure given in relation \eqref{misuracongiuntasimultanea}, then it follows from operational independence that
$$ \mu_{\omega, x:y}^{L_o} = \mu_{\omega_A, x}^{L_A} \otimes \mu_{\omega_B, y}^{L_B}$$
\end{remark}
 $$ \star \star \star $$
It is useful to make the following remarks.
\\
If the preparation of the ensembles is carried out in the $L_A$ laboratory as described in section \ref{proceduresperimentali}, and the total preparation time $t_A$, which is the sum of all the time intervals of all the $N$ copies that constitute the ensemble, is sufficiently small such that
$$ \mathcal O_A = L_A \times [0, t_A] \cap \mathcal V^+(L_B) = \emptyset $$
where $\mathcal V^+(L_B)$ is the future light cone generated by laboratory $L_B$ as shown in Figure \ref{fig:011}\footnote{where the dark region collects all $N$ copies of the ensemble for both laboratories}, then we can say that the experiments carried out in $L_B$ do not influence the experimental procedures carried out in $L_A$.
\\
We assume that the same considerations apply to the $L_B$ laboratory\footnote{In other words, $\mathcal O_A \subset \mathcal O_B^c$, where $\mathcal O^c$ denotes the causal complement of $\mathcal O$.}:
$$ \mathcal O_B = L_B \times [0, t_B] \cap \mathcal V^+(L_A) = \emptyset $$
Therefore, we can say that the experiments carried out in $L_A$ do not produce any effects on the experimental procedures of laboratory $L_B$\footnote{We underline that experimentally this can only happen by considering very short times $t_A$ and $t_B$ or astronomical distances between the two laboratories.}.
\\
In this way, can we say that all observables in $L_A$ and $L_B$ can be jointly prepared and measured simultaneously in the laboratory-type region $\mathcal O_o$?
\\
The answer is no!
\\
Indeed, we can only do this for observables located in $\mathcal O_A \subset \mathcal O_o$ with those located in $\mathcal O_B \subset \mathcal O_o$.
\\
In this way we obtain for every observable $a \in \mathfrak X_A$ localizable in $\mathcal O_A$ the following equality:
$$ \mathfrak S_{a \bowtie b}^A = \mathfrak S_a^A \qquad \forall b \in \mathfrak X_B \text{ localizable in } \mathcal O_B $$
and for every observable $b \in \mathfrak X_B$ localizable in $\mathcal O_B$ we have:
$$ \mathfrak S_{a \bowtie b}^B = \mathfrak S_b^B \qquad \forall a \in \mathfrak X_A \text{ localizable in } \mathcal O_A $$
In other words, we can say that every observable $x \in \mathfrak X_A$ and $y \in \mathfrak X_B$, localizable in their respective laboratory-type regions, are independent and therefore compatible in the physical system of laboratory $L_o$:
\begin{equation}
P(x \in \Delta_0 : y \in \Delta_1, \tau)_{\omega} = P(x \in \Delta_0, \tau)_{\omega} \, P(y \in \Delta_1, \tau)_{\omega}
\end{equation}
for each $\omega \in \mathfrak S_{x:y}(L_o) = \mathfrak S_{y:x}(L_o) \subset \mathfrak S_x(L_o) \cap \mathfrak S_y(L_o)$.
\begin{attenzione}
In this case, it is not certain that the two physical systems $(\mathfrak X_A, \mathfrak S_A)$ and $(\mathfrak X_B, \mathfrak S_B)$ corresponding to our two laboratories $L_A$ and $L_B$ are also operationally independent.  
\\
In fact, having a state $\omega_A$ in $L_A$ and $\omega_B$ in $L_B$ means having established experimental procedures and, with them, the various measuring instruments in these two laboratories. To achieve operational independence of the two systems, it will be necessary to identify a common state $\omega$ in $L_o$ that simultaneously carries out the experimental procedures contained in $\omega_A$ and $\omega_B$, using the respective measurement instruments, which is not always experimentally feasible.
\end{attenzione}
\section{Laboratories on the Move}
As mentioned in the introduction, here we will not deal with relativistic issues in detail; we will only address the meaning of invariance for laboratory-type regions.
\\
We will extend Axiom \ref{isotropia-A} to the case where the two laboratories $L_A$ and $L_B$ are not at rest with respect to each other, and their mutual motion is regulated by an element of a group of transformations $\mathcal G$, an element that establishes the change of coordinates between the two reference systems $(K_A, O_A)$ and $(K_B, O_B)$, centered respectively in our two laboratories as established in equation \eqref{cambiocoord}.
 \\ 
\textit{We will always assume the possibility that the two laboratories are physically capable of transmitting information about the instruments adopted\footnote{Obviously, the time delay of communications due to the finite speed of any signals used must be taken into account.}, of sending each other the various experimental protocols to be used in their respective laboratories for the use of instruments, etc.\footnote{In other words, we can exchange information about the respective states $\omega_A$ and $\omega_B$ established in the two laboratories.}, and obviously the respective results of the measurements of the various physical quantities carried out at given times $\tau_A$ in $L_A$ and $\tau_B$ in $L_B$, previously agreed upon.}
\\
Therefore, after having implemented the provisions of the experimental protocol and having set up our laboratory-type regions in the respective laboratories $L_A$ and $L_B$:
$$
 \mathcal O_A = L_A \times [0, t^p_A] \ , \qquad L_A \subset \mathfrak E 
$$
with respect to $(K_A, O_A)$ and 
$$ \mathcal O_B = L_B \times [0, t^p_B] \ , \qquad L_B \subset \mathfrak E $$ 
with respect to $(K_B, O_B)$, we measure the observables singly: $a \in \mathfrak X(\mathcal O_A)$ in state $\omega_A \in \mathfrak S_a(\mathcal O_A)$ and $b \in \mathfrak X(\mathcal O_B)$ in state $\omega_B \in \mathfrak S_b(\mathcal O_B)$, in their respective reference systems, obtaining the distributions\footnote{We reiterate that this means that the time $t^p_A$ is that indicated by the clock at $O_A$, while $t^p_B$ is that indicated by the clock at $O_B$; similarly, for the spatial regions, $L_A$ is determined by the oriented rulers $K_A$ centered at $O_A$, while $L_B$ is determined by the oriented rulers $K_B$ centered at $O_B$. 
Furthermore, the two laboratories $L_A$ and $L_B$ can mutually exchange information about their system states $\omega_A$ and $\omega_B$.}:
$$ P^{L_A}(a \in \Delta, \tau_A)_{\omega_A} \ , \ \text{relating to the laboratory system } (K_A, O_A) $$
$$ P^{L_B}(b \in \Delta, \tau_B)_{\omega_B} \ , \ \text{relating to the laboratory system } (K_B, O_B) $$
 $$ \star \star \star $$
Let us now focus our attention on the $L_A$ laboratory.
\\
Therefore, in laboratory $L_A$, for the various measurements we can only design the laboratory-type regions \eqref{regionelab0}, which we have denoted by $\mathcal O_A$.
\\
As we previously said, there are also other regions of space-time that are not of the laboratory type \eqref{regionelab0} which have experimental significance with respect to the laboratory system $(K_A, O_A)$; they are the $\mathcal G$-regions $\mathcal R$ of Definition \ref{G-regione}.
However, we want to underline that, physically, the measurements do not take place in the $\mathcal G$-regions $\mathcal R$ but in the laboratory $L_B$; \textit{it makes no experimental sense to consider states and observables relative to the $\mathcal G$-regions $\mathcal R$}.
\\ 
We remark that if the region $\mathcal R$ is also of laboratory type, i.e.,
$$ \mathcal R = L_A \times [0, t_A] $$
then the transformation $T^{K_B, O_B}_{K_A, O_A} \in \mathcal G$ is a translation as described in note \ref{traslazione} in section \ref{invarianza-sperimentale}, and in this way we obtain the laboratory-type region $\mathcal O_B$ centered at $O_B$\footnote{Recall that the clock positioned in laboratory $L_B$ is centered at the point $O_B \in \mathfrak E$; therefore it is a space-time event which, once the reference system of laboratory $(K_A, O_A)$ has been established, is identified by the relative coordinates $(x, x^0) \in \mathbb R^4$.
\\
Furthermore, if the region $\mathcal R$ is laboratory-type, then it means that $O_B$ is not moving relative to $O_A$.}.
\begin{figure}[htbp]
	\centering
		\includegraphics[scale=0.5]{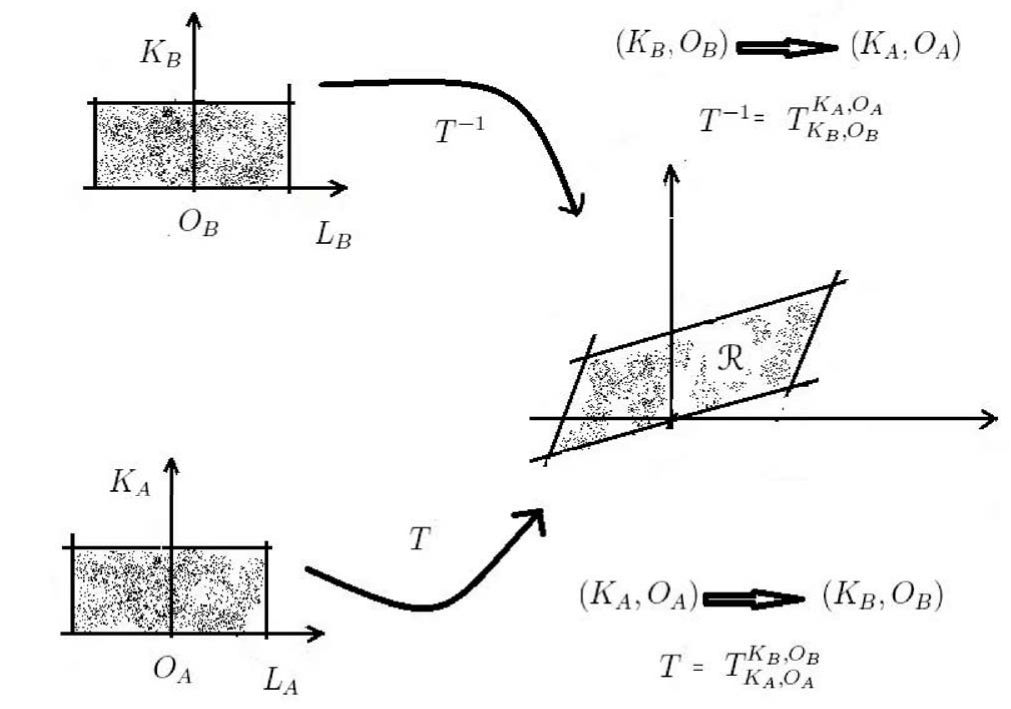}
	\caption{Equivalent Regions - Lab.}
	\label{fig:0111}
\end{figure}
\\
Let us now ask ourselves the following question:
\\
How do we determine when two laboratories are physically equivalent?
\\
Let us first give the following definition: 
\begin{definition}\upshape\index{$\mathcal G$-equivalent regions}
Two laboratory-type regions $\mathcal O_A$ and $\mathcal O_B$ are said to be $\mathcal G$-equivalent if there exists a $\mathcal G$-region $\mathcal R$ and an element $T$ of $\mathcal G$ such that:
$$ T \mathcal O_A = \mathcal R \qquad , \qquad T^{-1} \mathcal O_B = \mathcal R$$
\end{definition}
\begin{problem}\upshape
Verify that this definition is mathematically well posed, i.e., that it is independent of the two laboratory systems $(K_A, O_A)$ and $(K_B, O_B)$\footnote{We note that by changing the laboratory system, for example of $L_A$, from $(K_A, O_A)$ to $(K'_A, O'_A)$, by definition $O_A' \in L_A$; therefore the clock fixed at this point will mark the same time as that fixed at $O_A$. From this it follows that the transformation $T^{K'_A, O'_A}_{K_A, O_A} \in \mathcal G$ acts only on the spatial coordinates.}.
\end{problem}
The solution to our initial question is obtained by re-adapting Axiom \ref{isotropia-A} in an obvious way. In fact, it is sufficient to replace the space-time translation with the definition of $\mathcal G$-equivalence. In other words, if $\mathcal O_A$ and $\mathcal O_B$ are $\mathcal G$-equivalent, then the associated physical systems $(\mathfrak X^A, \mathfrak S^A)$ and $(\mathfrak X^B, \mathfrak S^B)$ possess the properties listed in Axiom \ref{isotropia-A}.

\begin{problem}\upshape
With the obvious meaning of the notation, let us ask ourselves what relations exist between the measures $\mu_{\omega_A, a}^{L_A}$ and $\mu_{\omega', a}^{L_B}$, where $\omega'$ is the image of $\omega$ via the map \eqref{copialab} on page \pageref{copialab}.
\end{problem}

\section{EPR Experiment and the Classical Analogy}

\begin{figure}[htbp]
	\centering
		\includegraphics[scale=0.5]{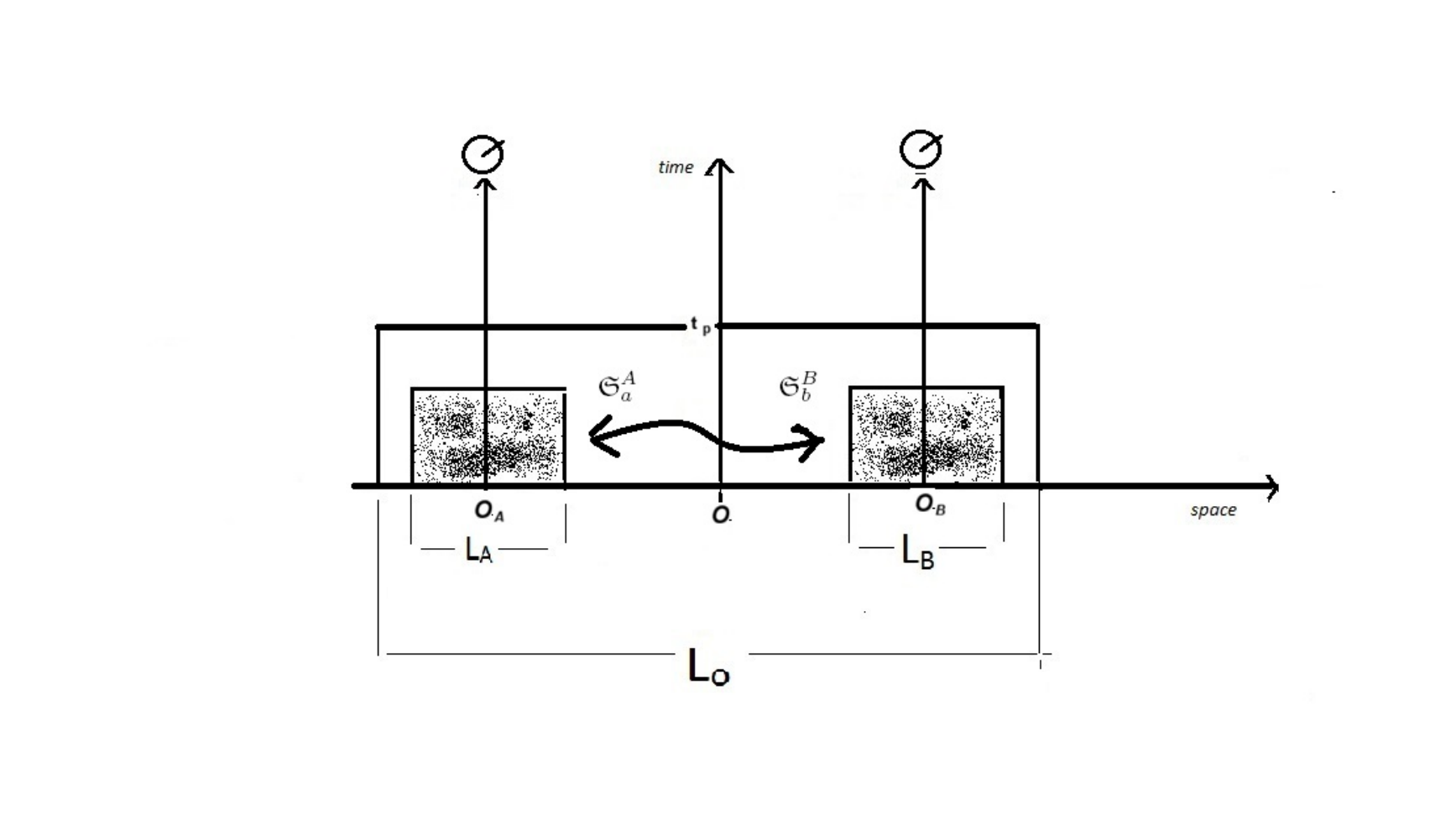}
	\caption{EPR}
	\label{fig:epr}
\end{figure}

Let $L_o$ be a laboratory containing two sub‑laboratories $L_A, L_B \subset L_o$, identical and positioned symmetrically with respect to the origin $O$ of the main laboratory, as shown in Figure~\ref{fig:epr}.

Assume that we have two compatible observables $a, b$ of the laboratory system $L_o$ that are correlated with each other in the chronological state $\omega \in \mathfrak S_{a:b}$; i.e., for every Borel set $\Delta$ we obtain
$$P^{L_o}(a \in \Delta : b \in \mathbb R)_{\omega^{\tau}} = P^{L_o}(a \in \mathbb R : b \in \Delta)_{\omega^{\tau}}, \qquad \forall \tau \in \mathbb R^+ 
$$
We also assume (see the set defined in equation \eqref{sottostati}):   
\begin{equation}\label{epr}
\omega \in \mathfrak S_{a}(L_o | L_A) \cap \mathfrak S_{b}(L_o | L_B) \subset \mathfrak S(L_o)
\end{equation}
Obviously, from what has been said in the previous sections, we cannot perform a simultaneous measurement of the observables in $L_o$ and in the two laboratories $L_A$ and $L_B$. However, we can establish, through ensemble procedures, the values of $a$ and $b$ in the laboratory $L_o$ and, once the statistical law
$$P^{L_o}(a \in \Delta : b \in \Delta_1)_{\omega^{\tau}}$$
has been determined, proceed to measure $a$ in $L_A$ and $b$ in $L_B$.
\\
Of course, the measurements of these two values must be carried out in the states $\omega_A \in \mathfrak S^{L_A}_{a}$ and $\omega_B \in \mathfrak S^{L_B}_{b}$, states that must “remember” the mother state $\omega$. These states are prepared through the procedures and devices employed in the laboratory $L_o$, i.e., they are obtained from hypothesis \eqref{epr} via Postulate~\ref{subregB}:
\[
\mathfrak X_{\omega_A}(L_A) = \mathfrak X_{\omega}(L_o) \cap \mathfrak X(L_A)
\]
and
\[
P^{L_A}(a \in \Delta, \tau)_{\omega_A} = P^{L_o}(a \in \Delta : b \in \mathbb R, \tau)_{\omega} 
\]
and similarly for the state $\omega_B$:
\[
\mathfrak X_{\omega_B}(L_B) = \mathfrak X_{\omega}(L_o) \cap \mathfrak X(L_B)
\]
and
\[
P^{L_B}(b \in \Delta, \tau)_{\omega_B} = P^{L_o}(a \in \mathbb R : b \in \Delta, \tau)_{\omega} 
\]

These considerations are always experimentally true if we prepare the two states $\omega_A$ and $\omega_B$ for a single measurement in the two laboratories. However, in EPR‑type experiments we are dealing with simultaneous measurements in both laboratories; therefore we must arrange everything for a joint preparation. As we saw in Section~\ref{disloc}, this is not always possible\footnote{This is possible precisely because the two laboratories are far apart and therefore do not perturb each other during the joint preparation.}. Thus we must assume that the preparation of the two states can be done jointly, i.e., using the definition such that
$$
\omega_A \in \mathfrak S_{a \bowtie b}^{A \bowtie B}(L_A), \qquad
\omega_B \in \mathfrak S_{b \bowtie a}^{A \bowtie B}(L_B)
$$
In this way we obtain that
$$
P^{L_A}(a \in \Delta, 0)_{\omega_A} = P^{L_B}(b \in \Delta, 0)_{\omega_B}
$$
so the two observers will obtain the same values simultaneously (i.e., at the same time $\tau = 0$ as shown by both clocks at $O_A$ and $O_B$), because both of their measurement states have a common ancestor: they derive from the mother state $\omega \in \mathfrak S_{a:b}(L_o)$.

This situation is operationally indistinguishable from the classical Regge ball example: two boxes, one containing a white ball and the other a black ball, placed in two separate laboratories. The correlation does not require any action at a distance; it is inherited from the initial preparation.

Hence, within our model, the EPR “paradox” does not arise. What is sometimes perceived as a quantum mystery is, in our framework, no more paradoxical than a classical correlation established by a common preparation.

\part{Algebraic Interpretation}

\chapter{The Algebra of Observables}
\begin{flushright}
\textit{... for a general observable, the choice of operator is as much of an art as a science, since none of the "rules of quantization" known is of universal validity. 
\\
— Muga et al. \cite{Muga2} }
\end{flushright}
In the previous sections we have not clearly specified the mathematical structure of the sets of states $\mathfrak{S}$ and observables $\mathfrak{X}$, so as to reveal only the basic framework of the theory. But this freedom of action is only apparent, since the main physical theories, such as classical mechanics and elementary quantum mechanics, require a less Spartan mathematical structure than the one exposed so far, equipped with a more elaborate mathematical formalism that has an effective exploratory function in understanding the various physical phenomena that occur in nature.
\\
We believe that the most suitable mathematical structure for this project is given by the algebraization of the set of observables, where the set $\mathfrak{X}$ is embedded, in an appropriate way, in an associative algebra, and its set of states is seen as a subset of its dual.
\\
We will see that we do not have a general rule for establishing an algebraization for a laboratory physical system, except in the particular case of classical or purely quantum systems given by the quantization of phase space, which historically takes the name of Weyl quantization\footnote{This topic will not be discussed in these notes, since there are many texts in the literature where it is exposed in an exhaustive way; e.g., for a rigorous mathematical treatment, we recommend Folland's book \cite{Folland2}.}.
%

%========================================

%
\section{Algebraization of a Physical System}\label{algebrizzazionesistema}
We will consider a mathematical model in which the set of observables is identified with the self-adjoint elements of a unital C*-algebra $\mathfrak{A}$, which we will generically refer to as the algebra of observables. Meanwhile, the set of physical states will be identified with a subset of the set of states $S(\mathfrak{A})$ of the algebra $\mathfrak{A}$.
\\
It should be noted that the existence of this algebra associated with the entire physical system of the laboratory is generally not guaranteed. To increase the likelihood of success, we must instead consider the physical subsystems of our laboratory. In fact, as we have seen, these subsystems are what hold real physical meaning, making the possibility of algebraization more likely to materialize.
\subsection{Representations on Associative Algebras}
An algebraic representation (briefly, an algebraization) of the physical system $(\mathfrak{X}, \mathfrak{S})$ is a triple $(\mathfrak{B}, \mathrm{J}, \mathrm{J}^{\natural})$ consisting of a unital real Banach algebra $\mathfrak{B}$ and two maps 
$$\mathrm{J}: \mathfrak{X} \rightarrow \mathfrak{B} \qquad , \qquad \mathrm{J}^{\natural}: \mathfrak{S} \rightarrow S(\mathfrak{B})$$
where $S(\mathfrak{B})$ denotes the space of linear functionals on $\mathfrak{B}$ with norm equal to $1$\footnote{If $\mathfrak{B}_1^*$ denotes the space of linear functionals on $\mathfrak{B}$ with norm less than or equal to $1$, we have
 $$S(\mathfrak{B}) \subset \mathfrak{B}_1^*.$$
By Alaoglu's theorem, $\mathfrak{B}_1^*$ is weak-star compact.
\\
By the Hahn–Banach theorem, for each $b \in \mathfrak{B}$ 
$$ \| b \| = \sup \left\{ |\varphi(b)| : \varphi \in \mathfrak{B}_1^* \right\}. $$} (The state space of $\mathfrak{B}$). \index{$S(\mathfrak{B})$} \index{$\mathfrak{B}_1^*$}. These maps satisfy the following Algebraic Representation on Banach Algebras (\texttt{ARBA}) conditions:
\\
\textbf{ARBA Conditions}\index{ARBA conditions}
\begin{itemize} 
\item[\textbf{a}] Polynomial property:
\\
For all $a \in \mathfrak{X}$ and $n \in \mathbb{N}_o := \mathbb{N} \cup \{ 0 \}$\index{$\mathbb{N}_o$},
\begin{equation}\label{Polynomial-property}
\mathrm{J}(a^n) = \mathrm{J}(a)^n 
\end{equation}
\item[\textbf{b}] Sum on compatibility:
\\
For every pair of \textit{compatible observables} $a, b \in \mathfrak{X}$,
\begin{equation}\label{Sum-compatibility}
\mathrm{J}(a + b) = \mathrm{J}(a) + \mathrm{J}(b)  
\end{equation}
\item[\textbf{c}] Product for a scalar:
\\
For every real number $r$ and observable $a$, we have
\begin{equation}\label{Product-scalar}
\mathrm{J}(r a) = r \mathrm{J}(a) 
\end{equation}

\item[\textbf{d}] Average value:
\\
For every observable $a \in \mathfrak{X}$ and state $\omega \in \mathfrak{S}_a$,  
\begin{equation}
\label{Valore medio}
\langle a \rangle_{\omega} = \mathrm{J}^{\natural}(\omega) \big( \mathrm{J}(a) \big)   
\end{equation}
\end{itemize}

The representation $(\mathfrak{B},\mathrm{J} ,\mathrm{J} ^{\natural })$ is called  \textit{minimal}  if the image $\mathrm{J} \left(\mathfrak{X}\right) $ generates the entire real Banach algebra $\mathfrak{B}$.
\\
This means $\mathfrak B$ is the norm closure of the real vector space spanned by elements of the form:
\begin{equation}\label{poly}  
\left\{ \mathrm{J} (a_1^{h_1} )\mathrm{J} (a_2^{h_2}) \cdots \mathrm{J} (a_{n}^{h_n} ) \ : \  a_i \in\mathfrak X  \  , \  h_i\in \mathbb N_o \  \    i=1,2\ldots n \right\}
\end{equation}
Let  $\mathcal P(\mathrm{J} (\mathfrak X))$  denote the unital algebra of non-commutative polynomials defined by the above relation  \eqref{poly}. Then
$$ \mathfrak B = \overline{\mathcal P(J(\mathfrak X))}^{\tau_N}$$

where $\tau_N$ is the operator norm topology of $\mathfrak B$.
\\
\textit{In other words $\mathfrak B$ is the smallest Banach algebra that contains the observables of the system}
\\

Let’s now make some simple observations:
\begin{itemize}
\item[I ] The observables $a^0$ and $b^0$   are different, because $\mathfrak S_a$ and $\mathfrak S_b$ generally  do not coincide. Furthermore, we recall that there may be incompatible observables.
\\
By  definition, we have
$$\mathrm{J} (a^0)= \mathrm{J} (b^0)= 1$$
so the map $\mathrm{J} $ is not injective map. 
\item[II ] From relation \eqref{Valore medio} and  axioms \ref{assio1} and \ref{assio2}, on page \pageref{assio1}, we have
$$ \mathrm{J} (a)=0 \qquad \Longrightarrow  \qquad  a\subset 0 $$
Indeed,
$$ [ \ \mathrm{J} ^{\natural }(\omega)(\mathrm{J} (a))=0 \ , \ \forall \omega\in\mathfrak S_a \ ] \ \Longrightarrow \ [ \  \left\langle a\right\rangle _{\omega }=0 \ , \ \forall \omega\in\mathfrak S_a \ ] \ \Longrightarrow  \ a\subset 0   $$
\item[III] Similarly to the previous case, if $\omega,\omega' \in\mathfrak S$ with $\mathfrak X_\omega \subset \mathfrak X_{\omega'}$ then we have  
$$ \mathrm{J} ^{\natural}(\omega)=\mathrm{J} ^{\natural}(\omega') \qquad \Longrightarrow  \qquad  \omega\subset \omega' $$
\item[IV] We recall that the observable $r a$ is the observable  $C(a) a$ where $C(t)=r$ for each $t \in\mathbb R$, as defined on page \pageref{Productscalar}.
\item[V] If $a,b$ are arbitrary  observables of $\mathfrak{X}$, then we can only write that 
\begin{equation*}
\left\langle a\right\rangle _{\omega }+\left\langle b\right\rangle _{\omega
}=\mathrm{J} ^{\natural }\left( \omega \right) \left( \mathrm{J} \left( a\right) +\mathrm{J} (b)\right)
\end{equation*} 
If they are compatible observables, then 
\begin{equation*}
\left\langle a\right\rangle _{\omega }+\left\langle b\right\rangle _{\omega
}=\mathrm{J} ^{\natural }\left( \omega \right) \left( \mathrm{J} \left( a+b\right) \right)=\left\langle a+b\right\rangle _{\omega }
\end{equation*}
\end{itemize}
\begin{attenzione}\upshape\label{oss-nulli}
If  $a\subset 0$,  we do not necessarily have $\mathrm{J}  \left( a\right)=0$\footnote{We recall that  two observables  $a\subset 0$ and $b\subset 0$ are not necessarily equal, since the set $\mathfrak S_a$ and $\mathfrak S_b$ may differ.}.
\\
Observables are not numbers; we cannot use property  \eqref{Product-scalar}  to assert this implication without an additional hypothesis:
 \\
Let $a\subset 0$. Suppose there exists an observable $b\in\mathfrak X$ such that:
\\
1 - $b$ is a non-null observable,
\\
2- $\mathfrak S_a =\mathfrak S_b$.
\\
Then, we can write $a= C(b)b$ with $C(t)=0$ for each $t\in\mathbb R$, so 
$$ \mathrm{J} \left( a \right) = \mathrm{J} \left( C(b)b \right)= 0 \mathrm{J} \left( b \right)=0$$
\textbf{We will always assume this hypothesis to hold.}
\\
Moreover, from relation \eqref{Sum-compatibility}, if $a,b$ are strongly compatible  observables with $\mathrm{J} \left( a \right)=\mathrm{J} \left( b \right)$ then $a=b$\footnote{if $a,b$ are merely compatible observables, this statement is not always true:
$$ \mathrm{J} \left( a \right)=\mathrm{J} \left( b \right) \ \Longrightarrow  \ \mathrm{J} \left(b- a \right)= 0  \ \Longrightarrow \ [ \ \left\langle b-a\right\rangle _{\omega }=0 \ \forall \omega \in \mathfrak S_{a-b} \ ]$$
and thus 
$$b- a \subset  0 $$ 
However, since $\mathfrak S_{a-b} =\mathfrak S_{a:b}$,  we cannot conclude that $b \subset a$ ( i.e., $\mathfrak S_b \subset \mathfrak S_a)$.}.
\end{attenzione}
\subsection{Convexity and algebrizations}\label{convex-algebre}
We must now make some simple but important observations, which will in the future identify what further properties the Banach algebra must have in order to achieve an optimal algebraization of the physical system.
\\
We would like to emphasize that it is not guaranteed that every positive functional of norm $1$ in the Banach algebra $\mathfrak{B}$ corresponds to a physically relevant state of the system.
\\
In other words, we cannot say that the map $\mathrm{J}^{\natural}: \mathfrak{S} \rightarrow S(\mathfrak{B})$ is surjective:
$$ \mathrm{J}^{\natural}(\mathfrak{S}) = \bigcup_{a \in \mathfrak{X}} \mathrm{J}^{\natural}(\mathfrak{S}_a) \subset S(\mathfrak{B}) $$
Another important consideration is the following:
\\ 
\textit{if $\varphi_1, \varphi_2 \in \mathrm{J}^{\natural}(\mathfrak{S})$, then it is not necessarily true that their algebraic mixture $t \varphi_1 + (1-t) \varphi_2$ is an element of the set $\mathrm{J}^{\natural}(\mathfrak{S})$}. 
\\
Let us properly frame the problem.
\\
We examine the relations between the $\mathbb{M}_k(a)$, the sectors in the measurement of an observable $a$, defined in Section \ref{sectors} on page \pageref{sectors}, and the set $\mathrm{J}^{\natural}(\mathfrak{S})$.
\\
Recall that for every observable $a$, we have the following set of physical system states:
$$\mathfrak{S}_a^k = \left\{ \omega \in \mathfrak{S}_a : \mu_{\omega,a} \in \mathbb{M}_k(a) \right\}$$
where $\mathbb{M}_k(a)$ is the $k$-th measurement sector of $a$, a convex subset of the set $\Pi$. 
\\
Furthermore 
$$ \mathfrak{S}_a = \bigcup_k \mathfrak{S}_a^k \qquad \text{with} \qquad \mathfrak{S}_a^h \cap \mathfrak{S}_a^k = \emptyset \quad (h \neq k).$$
The set $\mathrm{J}^{\natural}(\mathfrak{S}_a) \subset S(\mathfrak{B})$ is divided into multiple \textit{sectors}\footnote{These sectors are disjoint if the mapping $\mathrm{J}^{\natural}$ is injective:
	$$ \mathrm{J}^{\natural}(\mathfrak{S}_a^h \cap \mathfrak{S}_a^k) = \mathrm{J}^{\natural}(\mathfrak{S}_a^k) \cap \mathrm{J}^{\natural}(\mathfrak{S}_a^h) = \emptyset $$}:
\begin{equation}\label{sectors2}
\mathrm{J}^{\natural}(\mathfrak{S}_a) = \bigcup_k \mathrm{J}^{\natural}(\mathfrak{S}_a^k) \subset S(\mathfrak{B})
\end{equation} 
\begin{problem}\upshape
 If $(\mathfrak{B}, \mathrm{J}, \mathrm{J}^{\natural})$ is an algebraization (in the Banach algebra $\mathfrak{B}$) of the physical system $(\mathfrak{X}, \mathfrak{S})$, then for every observable $a$ and its $k$-th measurement sector, is the set $\mathrm{J}^{\natural}(\mathfrak{S}_a^k)$ a convex subset of $S(\mathfrak{B})$? 
\end{problem}
Let us analyze the problem.
\\
Take $\varphi_1, \varphi_2 \in \mathrm{J}^{\natural}(\mathfrak{S}_a^k)$ and $r \in [0,1]$, and consider the mixture:
$$ \varphi = r \varphi_1 + (1 - r) \varphi_2 \in S(\mathfrak{B})$$
By hypothesis, there exist $\omega_1, \omega_2 \in \mathfrak{S}_a^k$ such that
$$ \varphi = r \mathrm{J}^{\natural}(\omega_1) + (1 - r) \mathrm{J}^{\natural}(\omega_2) $$ 
Thus, for every $f \in C_o(\mathbb{R})$:
$$ \varphi(\mathrm{J}(f(a))) = r \mathrm{J}^{\natural}(\omega_1)(\mathrm{J}(f(a))) + (1 - r) \mathrm{J}^{\natural}(\omega_2)(\mathrm{J}(f(a))) $$
It follows that:
$$\varphi(\mathrm{J}(f(a))) = r \mu_{\omega_1, a}(f) + (1 - r) \mu_{\omega_2, a}(f) \ , \qquad \forall f \in C_o(\mathbb{R})$$  
and
$$ \nu_{\varphi, \mathrm{J}(a)} = r \mu_{\omega_1, a} + (1 - r) \mu_{\omega_2, a} \in \mathbb{M}_k(a) $$
since $\mu_{\omega_1, a}, \mu_{\omega_2, a} \in \mathbb{M}_k(a)$ and, by hypothesis, $\mathbb{M}_k(a)$ is convex.
\\
Therefore, there exists $\omega \in \mathfrak{S}_a^k$ such that:
$$\mu_{\omega, a} = r \mu_{\omega_1, a} + (1 - r) \mu_{\omega_2, a}$$
Hence:
$$\mathrm{J}^{\natural}(\omega)(\mathrm{J}(f(a))) = r \mathrm{J}^{\natural}(\omega_1)(\mathrm{J}(f(a))) + (1 - r) \mathrm{J}^{\natural}(\omega_2)(\mathrm{J}(f(a))) = \varphi(\mathrm{J}(f(a))) $$
However, this does not imply that:
$$ \mathrm{J}^{\natural}(\omega)(A) = \varphi(A) \ , \qquad \forall A \in \mathfrak{B}$$
In other words, we cannot conclude that $\mathrm{J}^{\natural}(\omega) = \varphi$ and thus that $\varphi \in \mathrm{J}^{\natural}(\mathfrak{S}_a^k)$, nor that the set $\mathrm{J}^{\natural}(\mathfrak{S}_a^k)$ is convex. 
\begin{attenzione}\upshape
The set $\mathrm{J}^{\natural}(\mathfrak{S}_a^k) \subset S(\mathfrak{B})$ is in general not a convex set. 
\end{attenzione}
 To summarize, we can only say that if $\omega \in \mathfrak{S}_a^k$ is a mixture in the measurement of $a$ of the states $\omega_1, \omega_2 \in \mathfrak{S}_a^k$, then by definition:
$$\mu_{\omega, a} = (1 - r) \mu_{\omega_1, a} + r \mu_{\omega_2, a} \ , \qquad \text{for some } r \in (0,1)$$
with $\mu_{\omega_1, a}, \mu_{\omega_2, a} \in \mathbb{M}_k(a)$.
\\  
It then follows directly that  
\begin{equation}\label{pseudoconvex}
\mathrm{J}^{\natural}(\omega)(\mathrm{J}(a)) = (1 - r) \mathrm{J}^{\natural}(\omega_1)(\mathrm{J}(a)) + r \mathrm{J}^{\natural}(\omega_2)(\mathrm{J}(a)) \qquad \forall a \in \mathfrak{X}
\end{equation}
In section \ref{Physic-math-states}, in the context of C*-algebras, we will try to answer the following question:
\begin{problem}\upshape\label{pure-states}
Study the relationship between purely informative states in the measurement of $a$ (or the pure states in the measurement of $a$), and the pure states of $S(\mathfrak{B})$ belonging to the set $\mathrm{J}^{\natural}(\mathfrak{S})$. 
\end{problem}
\subsection{Jordan product and algebraization}
We need to make an important consideration regarding our algebraizations over real Banach algebras.
\\
Given two compatible observables $a, b$ of the system, we deduce from the \texttt{ARBA} properties that
\begin{equation} 
\mathrm{J}(a \cdot b) = \frac{1}{2} \big[ \mathrm{J}(a) \mathrm{J}(b) + \mathrm{J}(b) \mathrm{J}(a) \big] 
\end{equation}
where $a \cdot b$ is the Jordan product defined in relation \eqref{prodotto jordan} on page \pageref{prodotto jordan}.
\\
This gives us a map $\mathrm{J}: \mathfrak{X} \longrightarrow \mathfrak{B}^{(+)}$ that preserves the Jordan algebra structure for compatible observables\footnote{For the definition of $\mathfrak{B}^{(+)}$, see relation \eqref{prodotto jordan2} on page \pageref{prodotto jordan2}.}:
\begin{equation}
\label{centrale1}
\mathrm{J}(a \cdot b) = \mathrm{J}(a) \circ \mathrm{J}(b) 
\end{equation}
where $a, b$ are compatible observables of the system, and the product $\mathrm{J}(a) \circ \mathrm{J}(b)$ is defined by relation \eqref{prodotto jordan2} on page \pageref{prodotto jordan2}.
\\
Furthermore, if $\{ a, b, c \}$ is a family of compatible observables of the system, then from relation \eqref{associaa} we have the associative property:
\begin{equation}
\label{centrale2}
\mathrm{J}(a) \circ \big( \mathrm{J}(b) \circ \mathrm{J}(c) \big) = \big( \mathrm{J}(a) \circ \mathrm{J}(b) \big) \circ \mathrm{J}(c)
\end{equation}
\begin{remark}\upshape
When considering the physical subsystems of the laboratory $(\mathfrak{X}_o, \mathfrak{S}_o)$, relations \eqref{centrale1} and \eqref{centrale2} remain valid, but they do not guarantee the validity of the inclusion\footnote{See relation \eqref{condizionecentrum0} on page \pageref{condizionecentrum0}.}:
\begin{equation} \label{condizionecentrum1}
\mathrm{J}\big( Z_{\mathfrak{S}_o}(\mathfrak{X}_o) \big) \subset \mathfrak{Z}\big( \mathfrak{B}^{(+)} \big)
\end{equation}
However, from relation \eqref{commutante-J} on page \pageref{commutante-J}, if for every pair of compatible observables $a, b$ of the system the algebraization satisfies
\begin{equation}
\mathrm{J}(a) \circ \big( X \circ \mathrm{J}(b) \big) = \big( \mathrm{J}(a) \circ X \big) \circ \mathrm{J}(b) \ , \qquad \forall X \in \mathfrak{B}
\end{equation}
then the central condition \eqref{condizionecentrum1} is satisfied.
\\ 
Moreover, \eqref{condizionecentrum1} does not ensure that $\mathrm{J}\big( Z_{\mathfrak{S}_o}(\mathfrak{X}_o) \big)$ lies in $\mathfrak{B}^c$, the commutant of the Banach algebra.
\\
Additionally, as we will see in Section \ref{centralcond}, satisfying this relation requires a stronger condition on the map $\mathrm{J}$:
\\
It must preserve the product of compatible observables in the associative algebra $\mathfrak{B}$, not just the product in the Jordan algebra $\mathfrak{B}^{(+)}$ as in \eqref{centrale1}.
\end{remark}
These observations lead us to consider possible algebraizations obtained through not necessarily associative algebras, such as real JB-algebras.
We will explore this possibility after introducing the algebraizations derived from complex C*-algebras, where the real JB-algebra arises from the set of its self-adjoint elements, as discussed in Section \ref{compatibili-algebre} on page \pageref{compatibili-algebre}.
\subsection{Spectral Connection}\label{spectral-connection}
We now want to study the connection between the spectrum of the observable $a$ and the spectrum of $\mathrm{J}(a)$ in the real Banach algebra $\mathfrak{B}$. Fundamental to this analysis is the embedding property introduced above. \textit{Of course, to have a good algebraization of the physical system, these two spectra should coincide.}
\\
As is known, discussing the spectrum of the real Banach algebra $\mathfrak{B}$ requires its complexification: 
$$\mathfrak{B}_{\mathbb{C}} = \mathfrak{B} \oplus j\mathfrak{B}$$
Indeed, if $A$ is an element of the algebra $\mathfrak{B}$, the set of $\lambda \in \mathbb{R}$ such that $A - \lambda I$ is non-invertible in $\mathfrak{B}$ could be empty\footnote{See Bingren \cite{Libin}, p. 7, and Kaniuth \cite{Kaniu} for further details on real operator algebras.}.
\\

Recall that each element $A \in \mathfrak{B}_{\mathbb{C}}$ is written as
$$A = B_1 + j B_2 \ , \qquad B_1, B_2 \in \mathfrak{B}$$
and there is a natural way to define an involution $^*: \mathfrak{B}_{\mathbb{C}} \rightarrow \mathfrak{B}_{\mathbb{C}}$:
\begin{equation}
\label{involuzione}
(B_1 + j B_2)^* = B_1 + j (-B_2) \ , \qquad B_1, B_2 \in \mathfrak{B} 
\end{equation}  
Obviously, the self-adjoint elements of $\mathfrak{B}_{\mathbb{C}}$ coincide with $\mathfrak{B}$\footnote{A non-negligible problem is the existence of an operator norm $\| \cdot \|_\bullet$ in $\mathfrak{B}_{\mathbb{C}}$, which makes the *-algebra $\mathfrak{B}_{\mathbb{C}}$ a C*-algebra with the following properties:
\begin{itemize}
\item [a.] $\| B + j 0 \|_\bullet = \| B \| \ , \qquad \forall B \in \mathfrak{B}$
\item [b.] $\| B_1 - j B_2 \|_\bullet = \| B_1 + B_2 \|_\bullet \ , \qquad \forall B_1, B_2 \in \mathfrak{B}$
\end{itemize}
This topic is already covered in the first chapter of Li Bingren's book \cite{Libin}, to which the interested reader is referred.}.
\\
The spectrum of an element $A \in \mathfrak{B}$ is defined as follows:
$$ \operatorname{Sp}(A) = \left\{ \lambda \in \mathbb{C} \ : \ \lambda I - A \text{ is not invertible in } \mathfrak{B}_{\mathbb{C}} \right\}$$
It can be proven that the spectrum is non-empty and compact (cf. Kaniuth \cite{Kaniu}, Proposition 1.2.8).\footnote{This fact helps explain why complex numbers are used in quantum mechanics.}.
\\
To determine the relationship between the spectrum $\sigma(a)$ of the observable and the spectrum $\operatorname{Sp}(\mathrm{J}(a))$ of its algebraization, we will use functional calculus as an investigative tool, applying it to both observables and associative algebras.
\\

Let us make some observations about spectral mapping, starting with a simple consequence of the properties of the map $\mathrm{J}$, as follows:
\begin{remark}\upshape\label{spec_poly}
For every real polynomial $P$, we have:
$$ \mathrm{J}(P(a)) = P(\mathrm{J}(a)) \ , \qquad \forall a \in \mathfrak{X}$$
\end{remark}
%%%%ù
Now, let $f: \mathbb{R} \rightarrow \mathbb{R}$ be a continuous function. Since $f(a) \in \mathfrak{X}$, it follows that $\mathrm{J}(f(a)) \in \mathfrak{B}$. However, \textit{we cannot yet assert} that
\begin{eqnarray}\label{spec_fuc}
\mathrm{J}(f(a)) = f(\mathrm{J}(a))
\end{eqnarray}
because $\mathfrak{B}$ has too weak an algebraic-topological structure to support continuous functional calculus. As previously noted, functional calculus is a useful tool for establishing an explicit connection between the spectrum of an observable $a$ and the spectrum of $\mathrm{J}(a)$ as an element of the algebra.
\\
This is one of the key motivations for enriching the mathematical structure of our algebraization of a physical system — a topic we will explore in the next section.
\\
Now we need to make a useful clarification:
\begin{remark} [\textbf{Associative real algebras vs. complex algebras}]\upshape
Let $\mathfrak{A}$ be a generic complex *-algebra. The set of its self-adjoint elements, denoted by $\mathfrak{A}_{s.a.}$, is not generally a true associative algebra. However, with the Jordan product given in \eqref{prodotto jordan2}, it becomes a real Jordan algebra (and thus a JC-algebra).
\\
The next sections will study algebraizations of self-adjoint elements in complex C*-algebras, which include the case of algebraizations on real Banach algebras.
\\
Indeed, given a real Banach algebra $\mathfrak{B}$, the self-adjoint elements of its complexification $\mathfrak{B}_{\mathbb{C}}$ coincide with $\mathfrak{B}$, which by initial hypothesis is a real associative algebra, unlike the set $\mathfrak{A}_{s.a.}$.
\\
Of course, we can further generalize the algebraizations over possible associative algebras by considering the real C*-algebras $\mathfrak{A}_{\mathbb{R}}$\footnote{Here too, to properly define the spectrum of an element, we must consider its complexification $\mathfrak{A}_{\mathbb{C}} = \mathfrak{A}_{\mathbb{R}} + j \mathfrak{A}_{\mathbb{R}}$, with the additional structure
$$ (A + j B)^* = A^* - j B^* \ , \qquad \forall A, B \in \mathfrak{A}_{\mathbb{R}}$$
In this case we have:
$$ (\mathfrak{A}_{\mathbb{R}})_{s.a.} = (\mathfrak{A}_{\mathbb{C}})_{s.a.} \subset \mathfrak{A}_{\mathbb{R}}$$}.
We will not consider real associative \textit{*-algebras} because, in our framework, the adjoint operation lacks operational meaning. A brief discussion of this possibility is given on page \pageref{aggiunto-accardi}.
\end{remark}
Before continuing, for notational simplicity, we adopt the following convention:
\begin{notation}\upshape
For each pair $(a, \omega) \in \mathfrak{X} \times \mathfrak{S}$ we set
$$ \widehat{a} = \mathrm{J}(a) \in \mathfrak{B} \qquad , \qquad \widehat{\omega} = \mathrm{J}^{\natural}(\omega) \in S(\mathfrak{B})$$
Thus, we can concisely express:
\begin{equation}
\label{v.m.}
\widehat{\omega}(\widehat{a}) = \langle a \rangle_{\omega}
\end{equation}
\end{notation}
\section{The C*-algebraic approach} \label{L'approccio C*-algebrico}
A physical system $(\mathfrak{X}, \mathfrak{S})$ admits a C*-algebraic representation if there exists a triple $(\mathfrak{A}, \mathrm{J}, \mathrm{J}^{\natural})$ consisting of a unital complex C*-algebra $\mathfrak{A}$, along with the maps
$$\mathrm{J}: \mathfrak{X} \rightarrow \mathfrak{A}_{s.a.} \qquad \text{and} \qquad \mathrm{J}^{\natural}: \mathfrak{S} \rightarrow S(\mathfrak{A})$$
which satisfy the same \texttt{ARBA} properties of the previous section. Here, $\mathfrak{A}_{s.a.}$ denotes the self-adjoint elements of the algebra $\mathfrak{A}$\footnote{Recall that a C*-algebra is generated by its self-adjoint elements. }.
\\
We will assume that the C*-algebra $\mathfrak{A}$ is a concrete algebra, i.e., that there exists a Hilbert space $\mathcal{H}$ (not necessarily separable) such that $\mathfrak{A} \subset \mathfrak{B}(\mathcal{H})$. We denote the bicommutant of $\mathfrak{A}$ by $\mathfrak{A}''$.
\\
Recall that $\mathfrak{A}''$ is a closed C*-algebra in the weak topology of $\mathfrak{B}(\mathcal{H})$.
\\
As in the previous case, we will only consider minimal algebraizations, i.e., algebraizations where $\mathfrak{A}$ \textit{is the norm closure of the *-algebra generated by $\mathrm{J}(\mathfrak{X})$}\footnote{From remark \ref{spec_poly} it follows that
$$ \mathcal{P}(\mathrm{J}(\mathfrak{X})) = \mathrm{J}(\mathcal{P}(\mathfrak{X}))$$}:
$$ \mathfrak{A} = \overline{\mathcal{P}(\mathrm{J}(\mathfrak{X}))}^{\tau_N} \subset B(\mathcal{H})$$
From the properties of the map $\mathrm{J}$ and the definition of the commutant, we have:
\begin{equation}\label{com-bic}
\mathrm{J}(\mathfrak{X})' = \mathrm{J}(\mathcal{P}(\mathfrak{X}))' 
\end{equation}
Therefore, 
$$ \mathrm{J}(\mathfrak{X})' = \mathcal{P}(\mathrm{J}(\mathfrak{X}))' \qquad \Longrightarrow \qquad \mathrm{J}(\mathfrak{X})'' = \mathcal{P}(\mathrm{J}(\mathfrak{X}))'' = \overline{\mathcal{P}(\mathrm{J}(\mathfrak{X}))}^w$$
Thus, we obtain the following result:
\begin{proposition}\upshape
For every minimal algebraization we obtain:
 $$ \mathfrak{A}' = \mathrm{J}(\mathfrak{X})'$$
\end{proposition}
\begin{proof}
Since $\mathrm{J}(\mathfrak{X}) \subset \mathfrak{A}$, it follows that $\mathfrak{A}' \subset \mathrm{J}(\mathfrak{X})'$.
\\
By the bicommutant property, we have $[\mathcal{P}(\mathrm{J}(\mathfrak{X}))']'' = \mathcal{P}(\mathrm{J}(\mathfrak{X}))'$, because $\mathcal{P}(\mathrm{J}(\mathfrak{X}))$ is a unital subalgebra of $\mathfrak{A}$.
\\
From the minimality condition we have $\mathfrak{A} \subset \mathcal{P}(\mathrm{J}(\mathfrak{X}))''$, therefore $\mathcal{P}(\mathrm{J}(\mathfrak{X}))' \subset \mathfrak{A}'$ and from the previous relation \eqref{com-bic}, $\mathrm{J}(\mathfrak{X})' \subset \mathfrak{A}'$.
\end{proof}
A simple consequence of the previous result:
\begin{equation}\label{SSR0}
\mathrm{J}(\mathfrak{X})' = \mathbb{C} I \qquad \Longleftrightarrow \qquad \mathfrak{A}'' = \mathfrak{B}(\mathcal{H})
\end{equation}
\subsection{Functional calculus and C*-algebraization}
As is well known, in a C*-algebra for each element $A \in \mathfrak{A}_{s.a.}$, the spectral mapping theorem holds for continuous functions\footnote{Cf. Blackadar \cite{Black} and Reed–Simon \cite{R_S}.}:
\begin{equation}\label{smcc}
 \operatorname{Sp}\big( f(A) \big) = f\big( \operatorname{Sp}(A) \big) \qquad , \qquad \forall f \in C(\mathbb{R})
\end{equation}
where $\operatorname{Sp}(A)$ denotes the spectrum of $A$, an element of the C*-algebra $\mathfrak{A}$.
\\
We recall that if $A \in \mathfrak{A}_{s.a.}$, then for every Borel function $f: \mathbb{R} \rightarrow \mathbb{R}$ bounded on the spectrum $\operatorname{Sp}(A)$, there exists an element $f(A) \in \mathfrak{A}''$ such that\footnote{Cf. Blackadar \cite{Black}, Proposition 1.6.2.4.}
\begin{equation}\label{funzborel}
\varphi(f(A)) = \int f(s) \, d \nu_{\varphi, A} \ , \qquad \forall \varphi \in S(\mathfrak{A}'')
\end{equation}
By the Stone–Weierstrass theorem\footnote{Cf. Reed–Simon \cite{R_S}.}, every real continuous function $f$ defined on a compact set $K \subset \mathbb{R}$ is the limit, in the uniform norm, of a net of polynomial functions $\{ P_\alpha \}_\alpha$ on $\mathbb{R}$\footnote{Although the relationship between the sets $\sigma(a)$ and $\operatorname{Sp}(\mathrm{J}(a))$ is not yet known, to apply the Stone–Weierstrass theorem it suffices to consider the compact set $K = \sigma(a) \cup \operatorname{Sp}(\mathrm{J}(a)) \subset \mathbb{R}$.}.
\\
Consequently, for every $\omega \in \mathfrak{S}_a$,
$$\left\langle P_\alpha(a) \right\rangle_{\omega} \longrightarrow \left\langle f(a) \right\rangle_{\omega}$$
since
\begin{equation}\label{poly-conv}
 \int P_\alpha(s) \, d \mu_{\omega, a} \longrightarrow \int f(s) \, d \mu_{\omega, a}  
\end{equation}
Let us return to the following question: 
\\
Given an algebraization $(\mathfrak{A}, \mathrm{J}, \mathrm{J}^{\natural})$ of our physical system, when does the equality\footnote{See page \pageref{spec_fuc}} 
\begin{equation}\label{strong-relation}
\mathrm{J}\big( f(a) \big) = f\big( \mathrm{J}(a) \big) \ , \qquad \forall f \in C(\sigma(a))
\end{equation}
hold?
\\
Here we have a first fundamental answer:
\begin{proposition}\upshape\label{weak-relation}
Let $f \in C(\mathbb{R})$ and $a \in \mathfrak{X}$. Then  
$$\mathrm{J}^{\natural}(\omega)\big( \mathrm{J}(f(a)) \big) = \mathrm{J}^{\natural}(\omega)\big( f(\mathrm{J}(a)) \big) \qquad \forall \omega \in \mathfrak{S}_a $$
\end{proposition}
\begin{proof}
The spectrum of the observable $a$ is a compact set; therefore we can use the Stone–Weierstrass theorem for continuous functions:
\\
There exists a net of real polynomials $P_\alpha$ such that $P_\alpha \longrightarrow f$ in the $\| \cdot \|_\infty$ topology.
\\

For the functional calculus of C*-algebras, we have:
$$ P_\alpha(\mathrm{J}(a)) \longrightarrow f(\mathrm{J}(a)) \ , \qquad \text{in the } \tau_n \text{ topology of } \mathfrak{A} $$
It follows that 
$$ \mathrm{J}^{\natural}(\omega)\big( P_\alpha(\mathrm{J}(a)) \big) \longrightarrow \mathrm{J}^{\natural}(\omega)\big( f(\mathrm{J}(a)) \big) \ , \qquad \forall \omega \in \mathfrak{S}_a $$
For the second term, we have
$$ \mathrm{J}^{\natural}(\omega)\big( \mathrm{J}(P_\alpha(a)) \big) = \left\langle P_\alpha(a) \right\rangle_{\omega} = \int P_\alpha(s) \, d \mu_{\omega, a}$$
and 
$$ \int P_\alpha(s) \, d \mu_{\omega, a} \longrightarrow \int f(s) \, d \mu_{\omega, a} = \left\langle f(a) \right\rangle_{\omega} = \mathrm{J}^{\natural}(\omega)\big( \mathrm{J}(f(a)) \big) $$
Therefore
$$ \mathrm{J}^{\natural}(\omega)\big( P_\alpha(\mathrm{J}(a)) \big) \longrightarrow \mathrm{J}^{\natural}(\omega)\big( \mathrm{J}(f(a)) \big) $$
Moreover 
$$ \mathrm{J}^{\natural}(\omega)\big( P_\alpha(\mathrm{J}(a)) \big) = \mathrm{J}^{\natural}(\omega)\big( \mathrm{J}(P_\alpha(a)) \big) $$
By the uniqueness of the limit, we obtain the thesis.
\end{proof}
\begin{remark}\upshape\label{separates-points}
Proposition \ref{weak-relation} does not ensure that equality \eqref{strong-relation} holds.
\\
Indeed, let $A, B \in \mathfrak{A}$ be such that
$$\mathrm{J}^{\natural}(\omega)(A) = \mathrm{J}^{\natural}(\omega)(B) \ , \qquad \forall \omega \in \mathfrak{S}$$
This does not guarantee that $A = B$.
 \end{remark}
\begin{proposition}\upshape\label{spectrum1}
For every $\omega \in \mathfrak{S}_a$, we have:
$$\mu_{\omega, a} = \nu_{\widehat{\omega}, \widehat{a}}$$
\end{proposition}
 \begin{proof}
Let $f \in C(\mathbb{R})$. By the definition of algebraization, we obtain
$$\mu_{\omega, a}(f) = \left\langle f(a) \right\rangle_{\omega} = \mathrm{J}^{\natural}(\omega)\big( \mathrm{J}(f(a)) \big)$$
Moreover, by Proposition \ref{weak-relation} and relation \eqref{funzborel}, we have:
$$ \mathrm{J}^{\natural}(\omega)\big( \mathrm{J}(f(a)) \big) = \mathrm{J}^{\natural}(\omega)\big( f(\mathrm{J}(a)) \big) = \nu_{\widehat{\omega}, \widehat{a}}(f)$$
\end{proof}
We emphasize that the measure $\nu_{\widehat{\omega}, \widehat{a}}$ is induced by the functional
\begin{equation}\label{misura_nu}
\nu_{\widehat{\omega}, \widehat{a}}(f) = \widehat{\omega}\big( f(\widehat{a}) \big) \ , \qquad f \in C_o(\mathbb{R}) 
\end{equation}
and these two measures have the same support:
$$\operatorname{Supp} \mu_{\omega, a} = \operatorname{Supp} \nu_{\widehat{\omega}, \widehat{a}}$$
\subsubsection{Spectrum of a self-adjoint element of a C*-algebra and support of spectral measures $\nu_{\varphi, A}$}
Let us recall some basic facts about the spectrum and spectral measures of a bounded self-adjoint operator.
\\
The starting point is relation \eqref{smcc}. The statements we will prove follow reasoning similar to the case of the observables of the physical system that we have previously discussed.
\\
We fix a self-adjoint element $A$ of the C*-algebra $\mathfrak{A}$. As we have already noted, for every state $\varphi$ on the algebra $\mathfrak{A}$, we obtain a regular Borel measure $\nu_{\varphi, A}$ defined as follows:
\begin{equation}
\label{misura_nu_1}
 \nu_{\varphi, A}(f) = \varphi(f(A)) \ , \qquad f \in C_o(\mathbb{R}) 
\end{equation}
Let us prove the following
 \\
\textbf{Step (A):}
\\
\textsl{If $\lambda \in \operatorname{Sp}(A)$, then for every open neighborhood $U$ of $\lambda$ there exists a state $\varphi \in S(\mathfrak{A})$ such that 
$$\nu_{\varphi, A}(U) \neq 0$$}
Let $U$ be an open neighborhood of $\lambda$. By Urysohn's lemma\footnote{See Folland \cite{Folland}.}, there exists a continuous function $f$ with the following properties: $0 \leq f \leq 1$, with $f(\lambda) = 1$ and $\operatorname{Supp} f \subset U$.
\\
From relation \eqref{smcc}, we obtain
$$\sup \operatorname{Sp}\big( f(A) \big) = \sup f\big( \operatorname{Sp}(A) \big) = 1$$
Thus, $1 \in \operatorname{Sp}\big( f(A) \big)$, which implies $f(A) \neq 0$. Therefore, there must exist at least one state $\varphi$ such that $\varphi(f(A)) \neq 0$.
\\ 
In other words:
$$0 < \int f(s) \, d \nu_{\varphi, A} < \int_U 1 \, d \nu_{\varphi, A} = \nu_{\varphi, A}(U) $$
Thus, $\nu_{\varphi, A}(U) > 0$.
\\
\textbf{Step (B):}
\\
\textsl{For every Borel function $f$ we have:
\begin{equation*}
\sigma\big( f(A) \big) \subset \overline{f\big( \operatorname{Sp}(A) \big)} 
\end{equation*}
}
The proof of this statement follows from Step (A) and retraces the same steps as in the proof of Theorem \ref{smb} on page \pageref{smb}.
\\
\textbf{Step (C):}
$$ \operatorname{Supp} \nu_{\varphi, A} \subset \operatorname{Sp}(A) \qquad \Longrightarrow \qquad \bigcup_{\varphi \in S(\mathfrak{A})} \operatorname{Supp} \nu_{\varphi, A} \subset \operatorname{Sp}(A) $$
The set $\mathbb{R} \setminus \operatorname{Sp}(A)$ is an open subset of $\mathbb{R}$. Take any compact $K$ contained in it, i.e., $K \subset \mathbb{R} \setminus \operatorname{Sp}(A)$.
\\
By Urysohn's lemma, there exists a continuous function $f$ with the following properties: 
\\
$0 \leq f \leq 1$, $f(K) = 1$, and $\operatorname{Supp} f \subset \mathbb{R} \setminus \operatorname{Sp}(A)$.
\\
From the spectral mapping theorem, we deduce that the element $f(A)$ is zero, since its spectrum $\operatorname{Sp}\big( f(A) \big) = \{ 0 \}$. This implies that the integral $\int f(s) \, d \nu_{\varphi, A} = 0$. It follows that $\nu_{\varphi, A}(K) = 0$.
\\
Due to the arbitrariness of the compact set $K$ and the regularity of our Borel measure, we conclude that $\nu_{\varphi, A}(\mathbb{R} \setminus \operatorname{Sp}(A)) = 0$, which proves the claim.
\\
\textbf{Step (D):}
\\
We have: 
$$\lambda \in \operatorname{Sp}(A) \ \Longleftrightarrow \ \exists \varphi \in S(\mathfrak{A}) \text{ such that } \varphi(A) = \lambda\footnote{More precisely, we can write 
$$ \operatorname{Sp}(A) = \left\{ \varphi(A) : \varphi \in PS(\mathfrak{A}) \right\} $$
where $PS(\mathfrak{A})$ denotes the pure states of the algebra $\mathfrak{A}$.
\\
For details, see Zhu's book \cite{Zhu}, p. 83.} $$
\subsubsection{Bounded Borel functions and algebraization}
Let us return to the initial discussion: the study of the relationship between the spectra $\sigma(a)$ and $\operatorname{Sp}(\mathrm{J}(a))$.
\begin{proposition}\upshape\label{spectrumm3}
If $(\mathfrak{A}, \mathrm{J}, \mathrm{J}^{\natural})$ is a C*-algebraization of the physical system $(\mathfrak{X}, \mathfrak{S})$, then
$$\sigma(a) \subset \operatorname{Sp}(\mathrm{J}(a)) \ , \qquad \forall a \in \mathfrak{X}$$
\end{proposition}
\begin{proof}
The proof follows from relation \eqref{spectrum1} on page \pageref{spectrum1}:
 \\
Indeed, if $\lambda \in \sigma(a)$, then there exists at least one $\omega \in \mathfrak{S}_a$ such that $\langle a \rangle_{\omega} = \lambda$. It follows that $\mathrm{J}^{\natural}(\omega)(\mathrm{J}(a)) = \lambda$ with $\mathrm{J}^{\natural}(\omega) \in S(\mathfrak{A})$, and from Step (D) we conclude $\lambda \in \operatorname{Sp}(A)$\footnote{Warning: The converse does not hold because $\mathrm{J}^{\natural}$ is not a surjective map.}.
\end{proof}
It is useful to emphasize the following relations:
$$ \bigcup_{\omega \in \mathfrak{S}_a} \operatorname{Supp} \mu_{\omega, a} = \bigcup_{\varphi \in \mathrm{J}^{\natural}(\mathfrak{S}_a)} \operatorname{Supp} \nu_{\varphi, \mathrm{J}(a)} \subset \operatorname{Sp}(\mathrm{J}(a)) $$
The next step is to study what happens if we obtain equality between the two spectra in the previous proposition.
\begin{proposition}\upshape
Equation \eqref{misura_nu} extends to all bounded Borel functions $F: \sigma(a) \rightarrow \mathbb{R}$:
\begin{equation}\label{spectrum01}
\mu_{\omega, a}(F) = \nu_{\widehat{\omega}, \widehat{a}}(F) 
\end{equation}
\end{proposition}
\begin{proof}
By Lusin's theorem (Proposition \ref{lusin}, on page \pageref{lusin}), for every state $\omega \in \mathfrak{S}_a$, there exists an equibounded net $\{ f^\omega_\alpha \}$ of functions in $C(\sigma(a))$ such that
 
$$f^\omega_\alpha \longrightarrow F \ , \qquad \mu_{\omega, a}\text{-}a.e.$$ 
Consequently,
$$\mu_{\omega, a}(f^\omega_\alpha) = \int f^\omega_\alpha(s) \, d \mu_{\omega, a} \longrightarrow \int F(s) \, d \mu_{\omega, a} = \mu_{\omega, a}(F)$$
and similarly,
$$\nu_{\widehat{\omega}, \widehat{a}}(f^\omega_\alpha) = \int f^\omega_\alpha(s) \, d \nu_{\widehat{\omega}, \widehat{a}}(s) \longrightarrow \int F(s) \, d \nu_{\widehat{\omega}, \widehat{a}}(s) = \nu_{\widehat{\omega}, \widehat{a}}(F)$$
Since $\mu_{\omega, a}(f^\omega_\alpha) = \nu_{\widehat{\omega}, \widehat{a}}(f^\omega_\alpha)$ by \eqref{misura_nu}, the limit yields \eqref{spectrum01}.
\end{proof}
  
Let $(\mathfrak{A}, \mathrm{J}, \mathrm{J}^{\natural})$ be a C*-algebraization of the physical system $(\mathfrak{X}, \mathfrak{S})$.
\\
For any observable $a \in \mathfrak{X}$ we have $\mathrm{J}(a) \in \mathfrak{A}_{s.a.}$, and for a bounded Borel function $F$, $\mathrm{J}(F(a)) \in \mathfrak{A}_{s.a.}$, but $F(\mathrm{J}(a))$ belongs to the von Neumann algebra $\mathfrak{A}''$. 
\\
Thus, while
$$ \mu_{\omega, a}(F) = \mathrm{J}^{\natural}(\omega)\big( \mathrm{J}(F(a)) \big)$$
the following \textbf{does not hold} in general:
$$ \nu_{\widehat{\omega}, \widehat{a}}(F) = \mathrm{J}^{\natural}(\omega)\big( F(\mathrm{J}(a)) \big)$$
since $F(\mathrm{J}(a)) \in \mathfrak{A}''$, it is not necessarily in $\mathfrak{A}$, and $\mathrm{J}^{\natural}(\omega) \in S(\mathfrak{A})$\footnote{See on page \pageref{Phys-Math-states}.}.
\\
Therefore, even if the spectral measures satisfy relation \eqref{spectrum01}, for the bounded Borel function $F$, we cannot generally assert the algebraic identity:
 \begin{equation}\label{funzborel00}
\mathrm{J}^{\natural}(\omega)\big( \mathrm{J}(F(a)) \big) = \mathrm{J}^{\natural}(\omega)\big( F(\mathrm{J}(a)) \big) \ , \qquad \forall \omega \in \mathfrak{S}_a 
 \end{equation}
\\

To clarify this distinction, consider the case of characteristic functions:
\\
For any $a \in \mathfrak{X}$ and Borel set $\Delta$:
\\
The element $\mathrm{J}(\mathbf{1}_\Delta(a)) \in \mathfrak{A}_{s.a.}$ is an orthogonal projection in $\mathfrak{A}$, since:
$$ \mathrm{J}(\mathbf{1}_\Delta(a))^2 = \mathrm{J}(\mathbf{1}_\Delta(a)^2) = \mathrm{J}(\mathbf{1}_\Delta(a))$$
The element $\mathbf{1}_\Delta(\mathrm{J}(a)) \in \mathfrak{A}''$ is also an orthogonal projection, but in the larger von Neumann algebra $\mathfrak{A}''$\footnote{Let $A \in B(\mathcal{H})$ be a bounded operator on a Hilbert space $\mathcal{H}$. We denote by $[A\mathcal{H}]$ the orthogonal projection onto the closed subspace $\overline{\operatorname{ran}(A)}$ (the closure of the range of $A$). The projection $[A\mathcal{H}]$ is the smallest projection $P$ satisfying $PA = A = AP$.
\\
If $A$ belongs to a unital C*-algebra $\mathfrak{A} \subset B(\mathcal{H})$, then $[A\mathcal{H}] \in \mathfrak{A}''$, where $\mathfrak{A}''$ is the double commutant (von Neumann algebra) of $\mathfrak{A}$.
 \\
Therefore, the orthogonal projection $\mathbf{1}_\Delta(\mathrm{J}(a)) \in \mathfrak{A}''$ while $\mathrm{J}(\mathbf{1}_\Delta(a)) \in \mathfrak{A}$.}.		
	\\
The fundamental obstruction is that these projections need not coincide:	
$$\mathrm{J}(\mathbf{1}_\Delta(a)) \overbrace{=}^{?} \mathbf{1}_\Delta(\mathrm{J}(a))$$		
This equality fails in general because:
\begin{itemize}
 \item[-] The left side is constructed via the functional calculus in the original system $(\mathfrak{X}, \mathfrak{S})$.
\item[-] The right side uses the von Neumann algebra functional calculus.
\item[-] The map $\mathrm{J}$ need not preserve spectral projections.	
\end{itemize}
\begin{remark}\label{vnborelfunc}
For von Neumann algebraizations $(\mathfrak{M}, \mathrm{J}, \mathrm{J}^{\natural})$ where $\mathfrak{M}$ is a von Neumann algebra (e.g., $\mathfrak{M} = \mathfrak{A}''$), with
$$ \mathrm{J}: \mathfrak{X} \longrightarrow \mathfrak{M}_{s.a.} \qquad , \qquad \mathrm{J}^{\natural}: \mathfrak{S} \longrightarrow S(\mathfrak{M})$$
the equality \eqref{funzborel00} holds for all states $\omega \in \mathfrak{S}_a$.
\\
In particular, for any Borel set $\Delta \in B(\mathbb{R})$ and any $\omega \in \mathfrak{S}_a$, we have:
$$\mathrm{J}^{\natural}(\omega)\big( \mathrm{J}(\mathbf{1}_\Delta(a)) \big) = \mathrm{J}^{\natural}(\omega)\big( \mathbf{1}_\Delta(\mathrm{J}(a)) \big)$$
\end{remark}
\subsection{Embedding Properties}
For every observable $a$ of $\mathfrak{X}$, we have the inequality:
\begin{equation}
\| a \| \leq \| \mathrm{J}(a) \|_{\mathfrak{A}}  
\end{equation}
which follows from the fundamental properties of C*-algebras. Specifically:
\\
The C*-norm is given by
$$\| \mathrm{J}(a) \|_{\mathfrak{A}} = \sup \left\{ |\varphi(\mathrm{J}(a))| : \varphi \in S(\mathfrak{A}) \right\}$$
while the observable norm satisfies:
\begin{eqnarray*}
\| a \| = \sup \left\{ |\langle a \rangle_{\omega}| : \omega \in \mathfrak{S}_a \right\} =  
\sup \left\{ |\mathrm{J}^{\natural}(\omega)(\mathrm{J}(a))| : \omega \in \mathfrak{S}_a \right\} \leq
\| \mathrm{J}(a) \|_{\mathfrak{A}} 
\end{eqnarray*}
since $\mathrm{J}^{\natural}(\mathfrak{S}_a) \subset S(\mathfrak{A})$. 
\\
At this stage, we cannot conclude that $\mathrm{J}: \mathfrak{X} \rightarrow \mathfrak{B}$ is isometric. This motivates the following key definition:
\begin{property}[\textbf{Embedding Property}] \label{prop_E}\index{Property of embedding} 
The C*-algebraization satisfies the embedding property if 
\begin{equation}
\| \mathrm{J}(a) \|_{\mathfrak{A}} = \| a \| \ , \qquad \forall a \in \mathfrak{X}
\end{equation}
\end{property}
 
The embedding property represents a subtle aspect of the algebraic formulation, due to the fundamentally different nature of the two norms involved:
\\
\textit{The physical norm $\| a \|$ is intrinsically tied to the set of states $\mathfrak{S}_a$, while the C*-norm $\| \mathrm{J}(a) \|_{\mathfrak{A}}$ depends only on the algebraic structure.}
\\
When the embedding property holds, the algebraic norm must necessarily concentrate on the image of physical states:
\begin{eqnarray}\label{prop_E1}\index{Embedding Properties}
 \| \mathrm{J}(a) \|_{\mathfrak{A}} = \sup \left\{ |\varphi(\mathrm{J}(a))| : \varphi \in \mathrm{J}^{\natural}(\mathfrak{S}_a) \right\}
\end{eqnarray}
\begin{proposition}\upshape\label{prop_density_embedding}
If for every observable $a$ of the physical system, the set $\mathrm{J}^{\natural}(\mathfrak{S}_a)$ is $w^*$-dense in the state space $S(\mathfrak{A})$, i.e.,
$$ \overline{\mathrm{J}^{\natural}(\mathfrak{S}_a)}^{w^*} = S(\mathfrak{A}) \ , \qquad \forall a \in \mathfrak{X}$$
then the embedding property (Property \ref{prop_E}) holds.
\end{proposition} 
\begin{proof}
By the definition of the supremum, for each $k \in \mathbb{N}$, there exists $\varphi_k \in S(\mathfrak{A})$ such that
$$ \| \mathrm{J}(a) \|_{\mathfrak{A}} \leq \varphi_k(\mathrm{J}(a)) + \frac{1}{k}$$
From the hypothesis, there exists $\omega_k \in \mathfrak{S}_a$ (depending on $k$) such that:
$$\varphi_k(\mathrm{J}(a)) \leq |\langle a \rangle_{\omega_k}| + \epsilon $$ 
Combining these inequalities, we obtain 
$$ \| \mathrm{J}(a) \|_{\mathfrak{A}} \leq |\langle a \rangle_{\omega_k}| + \epsilon + \frac{1}{k}$$
Since $|\langle a \rangle_{\omega_k}| \leq \| a \|$, it follows that:
$$ \| \mathrm{J}(a) \|_{\mathfrak{A}} \leq \| a \| + \epsilon + \frac{1}{k}$$ 
Taking the limit as $k \rightarrow \infty$, we conclude:
$$ \| \mathrm{J}(a) \|_{\mathfrak{A}} \leq \| a \| + \epsilon $$ 
Since $\epsilon > 0$ is arbitrary, the embedding property (Property \ref{prop_E}) holds.
\end{proof}
\begin{proposition}\upshape
If the embedding property is satisfied, then
\begin{eqnarray}\label{spec_fuc01}
\mathrm{J}\big( f(a) \big) = f\big( \mathrm{J}(a) \big) \ , \qquad \forall f \in C(\mathbb{R})
\end{eqnarray} 
\end{proposition}
\begin{proof}
By the Stone–Weierstrass theorem, there exists a net of real polynomials $P_\alpha$ such that $P_\alpha \rightarrow f$ uniformly (i.e., in the $\| \cdot \|_\infty$ topology).
\\
We estimate the norm difference as follows:
$$ \| f(\mathrm{J}(a)) - \mathrm{J}(f(a)) \|_{\mathfrak{A}} \leq \| f(\mathrm{J}(a)) - \mathrm{J}(P_\alpha(a)) \|_{\mathfrak{A}} + \| \mathrm{J}(f(a)) - \mathrm{J}(P_\alpha(a)) \|_{\mathfrak{A}} $$
For the first term, by remark \ref{spec_poly} on page \pageref{spec_poly} and the functional calculus for self-adjoint operators,
$$ \| f(\mathrm{J}(a)) - \mathrm{J}(P_\alpha(a)) \|_{\mathfrak{A}} = \| f(\mathrm{J}(a)) - P_\alpha(\mathrm{J}(a)) \|_{\mathfrak{A}} \leq \| f - P_\alpha \|_\infty $$ 
For the second term, by the embedding property\footnote{We recall that $\| a \| \leq r(a)$, and if the spectral property of the states \texttt{SPS} holds, then $\| a \| = r(a)$.},  
$$ \| \mathrm{J}(f(a)) - \mathrm{J}(P_\alpha(a)) \|_{\mathfrak{A}} = \| f(a) - P_\alpha(a) \| \leq \sup_{t \in \sigma(a)} |f(t) - P_\alpha(t)| = \| f - P_\alpha \|_\infty $$
Combining these estimates, we obtain
$$ \| f(\mathrm{J}(a)) - \mathrm{J}(f(a)) \|_{\mathfrak{A}} \leq 2 \| f - P_\alpha \|_\infty $$
Taking the limit as $P_\alpha \rightarrow f$ uniformly, the right-hand side vanishes, proving $\mathrm{J}(f(a)) = f(\mathrm{J}(a))$.
\end{proof}
We now have a statement that emphasizes the central role of the embedding properties:
\begin{proposition}\upshape\label{propos_E}
Let $(\mathfrak{A}, \mathrm{J}, \mathrm{J}^{\natural})$ be a C*-algebraization of the physical system $(\mathfrak{X}, \mathfrak{S})$.
If the physical system satisfies the state separation property \texttt{SPS}, then
$$ \sigma(a) = \operatorname{Sp}(\mathrm{J}(a)) \ , \qquad \forall a \in \mathfrak{X} $$
if and only if the C*-algebraization satisfies the embedding property.
\end{proposition}
\begin{proof}
$(\Longrightarrow)$
\\
Assume $\sigma(a) = \operatorname{Sp}(\mathrm{J}(a))$. Then the spectral radii coincide:
$$ r(\mathrm{J}(a)) = \| \mathrm{J}(a) \|_{\mathfrak{A}} = \| a \| = r(a)$$
where the last equality follows from the \texttt{SPS} property ($\| a \| = r(a)$). This implies the embedding property, as $\mathrm{J}$ preserves norms.
\\
$(\Longleftarrow)$
\\
Conversely, assume the embedding property holds. Suppose, for contradiction, that there exists $\lambda \in \operatorname{Sp}(\mathrm{J}(a))$ such that $\lambda \notin \sigma(a)$.
\\
Since $\{ \lambda \}$ is closed and $\sigma(a)$ is compact, Urysohn's lemma guarantees the existence of a continuous function $F \in C(\mathbb{R})$ satisfying:
$$ F(\lambda) = 1 \qquad \text{and} \qquad F|_{\sigma(a)} = 0 $$ 
By the embedding property, we have $\mathrm{J}(f(a)) = f(\mathrm{J}(a))$. Thus:
$$ F(a) = 0 \qquad \Longrightarrow \qquad \mathrm{J}(F(a)) = 0 \qquad \Longrightarrow \qquad F(\mathrm{J}(a)) = 0 $$  
Applying the spectral mapping theorem, we obtain:
$$ 0 = \operatorname{Sp}\big( F(\mathrm{J}(a)) \big) = F\big( \operatorname{Sp}(\mathrm{J}(a)) \big) $$ 
However, since $\lambda \in \operatorname{Sp}(\mathrm{J}(a))$ and $F(\lambda) = 1$, this yields $1 \in F\big( \operatorname{Sp}(\mathrm{J}(a)) \big)$, a contradiction\footnote{The \texttt{SPS} property was not used in the second part of the proof.}.
\end{proof}
\begin{attenzione}\upshape
While the embedding property guarantees
$$\mathrm{J}\big( F(a) \big) = F\big( \mathrm{J}(a) \big) \ , \qquad \forall F \in C(\mathbb{R})$$ 
this relation does not automatically extend to all bounded Borel functions $F$. 
\end{attenzione}

\subsection{Algebraic property of separation of physical states}
We now introduce a crucial selection criterion for possible algebraizations of a physical system — the \textit{algebraic state separation property} for the set $\mathrm{J}^{\natural}(\mathfrak{S}_a) \subset S(\mathfrak{A})$ (for more information see Bratteli–Robinson \cite{BR1}, Proposition 3.2.10)\footnote{
 Such selection rules are typically too strong for complete laboratory physical systems, but become applicable when considering properly selected physical subsystems.}.
\begin{property}[\texttt{ASSP}]\index{Property-Algebraic states separation property}\label{prop_S} \index{\texttt{ASSP}}
A C*-algebraization $(\mathfrak{A}, \mathrm{J}, \mathrm{J}^{\natural})$ of a physical system $(\mathfrak{X}, \mathfrak{S})$ satisfies the \textbf{Algebraic State Separation Property} if for every observable $a \in \mathfrak{X}$:
\begin{equation}
 \left[ \ A \in \mathfrak{A} \text{ such that } \mathrm{J}^{\natural}(\omega)(A) = 0 \ \ \forall \omega \in \mathfrak{S}_a \ \right] \qquad \Longrightarrow \qquad A = 0
\end{equation}
\end{property}
If the \texttt{ASSP} holds, then\footnote{Without using the hypothesis made in warning \ref{oss-nulli} on page \pageref{oss-nulli}.} 
$$ \mathrm{J}(a) = 0 \qquad \Longleftrightarrow \qquad a \subset 0 $$
This equivalence follows from the chain of implications:
$$ a \subset 0 \ \Longrightarrow \ [ \ \langle a \rangle_{\omega} = 0 \ , \ \forall \omega \in \mathfrak{S}_a \ ] \ \Longrightarrow \ [ \ \mathrm{J}^{\natural}(\omega)(\mathrm{J}(a)) = 0 \ , \ \forall \omega \in \mathfrak{S}_a \ ] \ \Longrightarrow \ \mathrm{J}(a) = 0 $$
This property is fundamental for the following spectral result:
\begin{proposition}\upshape
For any von Neumann algebraization $(\mathfrak{M}, \mathrm{J}, \mathrm{J}^{\natural})$ satisfying \texttt{ASSP}, we have spectral preservation:
$$ \sigma(a) = \operatorname{Sp}(\mathrm{J}(a)) \ , \qquad \forall a \in \mathfrak{X} $$
\end{proposition}
\begin{proof}
Since the von Neumann algebraization satisfies the \texttt{ASSP} property, by equation \eqref{funzborel00} for every Borel set $\Delta$, we obtain  
\begin{equation}\label{as}
\mathrm{J}\big( \mathbf{1}_\Delta(a) \big) = \mathbf{1}_\Delta\big( \mathrm{J}(a) \big)
\end{equation}

If $\lambda \in \operatorname{Sp}(\mathrm{J}(a))$, then for every $U_\epsilon = ]\lambda - \epsilon, \lambda + \epsilon[$ we have $\mathbf{1}_{U_\epsilon}(\mathrm{J}(a)) \neq 0$, so by equation \eqref{as} we obtain $\mathrm{J}\big( \mathbf{1}_{U_\epsilon}(a) \big) \neq 0$.
\\
Therefore, by the\texttt{ ASSP} property, $\mathbf{1}_{U_\epsilon}(a) \neq 0$, and then $\lambda \in \sigma(a)$\footnote{See Remark \ref{sp_risol} on page \pageref{sp_risol}.}.
\end{proof}
\begin{corollary}\upshape
For any von Neumann algebraization $(\mathfrak{M}, \mathrm{J}, \mathrm{J}^{\natural})$ satisfying the \texttt{ASSP} property, the embedding property (Property \ref{prop_E}) automatically holds:
$$\texttt{ASSP} \ \Longrightarrow \ \text{Embedding Property}$$
\end{corollary} 
\begin{proof}
This follows immediately from Proposition \ref{propos_E}, since \texttt{ASSP} guarantees the spectral equality $\sigma(a) = \operatorname{Sp}(\mathrm{J}(a))$ for all observables $a \in \mathfrak{X}$, which is equivalent to the embedding property.
\end{proof}
\begin{proposition}[\textbf{State Separation Property}] \upshape \index{Property of State separation}
Let $(\mathfrak{A}, \mathrm{J}, \mathrm{J}^{\natural})$ be a C*-algebraization of a physical system $(\mathfrak{X}, \mathfrak{S})$.
\\
Assume that for every observable $a \in \mathfrak{X}$, the set $\mathrm{J}^{\natural}(\mathfrak{S}_a)$ is weak*-dense in $S(\mathfrak{A})$:  
$$ \overline{\mathrm{J}^{\natural}(\mathfrak{S}_a)}^{w^*} = S(\mathfrak{A}), \quad \forall a \in \mathfrak{X}$$  
Then, the following separation property holds:  
$$\bigg[ \ \mathrm{J}^{\natural}(\omega)(A) = \mathrm{J}^{\natural}(\omega)(B) \quad \forall \omega \in \mathfrak{S} \ \bigg] \implies A = B$$
\end{proposition}
\begin{proof}
By assumption, for any state $\varphi \in S(\mathfrak{A})$, there exists a net $\{ \omega_\alpha \}_\alpha$ in $\mathfrak{S}_a$ such that:  
$$\mathrm{J}^{\natural}(\omega_\alpha) \overset{w^*}{\longrightarrow} \varphi$$  
For $A, B \in \mathfrak{A}$, if $\mathrm{J}^{\natural}(\omega)(A) = \mathrm{J}^{\natural}(\omega)(B)$ for all $\omega \in \mathfrak{S}_a$, then:  
$$ \varphi(A) = \lim_{\alpha} \mathrm{J}^{\natural}(\omega_\alpha)(A) = \lim_{\alpha} \mathrm{J}^{\natural}(\omega_\alpha)(B) = \varphi(B)$$  
Since $\varphi(A) = \varphi(B)$ holds for all $\varphi \in S(\mathfrak{A})$, and states separate points in $\mathfrak{A}$, it follows that $A = B$.  
\end{proof}

\begin{problem}\upshape
Determine whether a pair $(\mathfrak{A}, \Sigma)$ consisting of a unital C*-algebra and a set of states $\Sigma$ separating the points is an algebraization of a physical system $(\mathfrak{X}, \mathfrak{S})$.
\end{problem}

A first answer to this question is given by the following observations:
\\
We have stated that if $f$ is a Borel function and $A$ is a self-adjoint element of the algebra $\mathfrak{A}$, then there exists an operator $f(A) \in \mathfrak{A}''$ that satisfies relation \eqref{funzborel}.
\\
The map $\mathrm{J}: \mathfrak{X} \rightarrow \mathfrak{A}_{s.a.}$ of our C*-algebraization must have the property that $\mathrm{J}(\mathbf{1}_\Delta(a)) \in \mathfrak{A}_{s.a.}$ for every observable $a$ and Borel set $\Delta$ of $\mathbb{R}$.
\\
By Lusin's theorem, Proposition \ref{lusin} on page \pageref{lusin}, we obtain: 
$$ \widehat{\omega}\big( \widehat{\mathbf{1}_\Delta(a)} \big) = \widehat{\omega}\big( \mathbf{1}_\Delta(\widehat{a}) \big) \ , \qquad \forall \omega \in \mathfrak{S}_a$$
and by the property of separation of states \texttt{ASSP,} we can write that 
$$ \mathrm{J}\big( \mathbf{1}_\Delta(a) \big) = \mathbf{1}_\Delta\big( \mathrm{J}(a) \big) \in \mathfrak{A}''$$
To sum up:
\begin{remark}[\textbf{von Neumann algebra}]\upshape
To have a good algebraization, the C*-algebra $\mathfrak{A}$ must be closed in the weak topology; in other words, it must be a von Neumann algebra. C*-algebras, unlike von Neumann algebras, are too poor in orthogonal projectors to contain the 'questions' of the physical system\footnote{Recall that a von Neumann algebra is generated by its orthogonal projectors.}, and the projection lattice $\mathfrak{P}(\mathfrak{A})$ corresponds to idealized measurement questions. 
\end{remark}

While we have established that the C*-algebra $\mathfrak{A}$ should be weakly closed (making it a von Neumann algebra), we must still determine which specific von Neumann algebra most appropriately contains all observables of the physical system.
\\
Furthermore, we must characterize the image of the state map $\mathrm{J}^{\natural}: \mathfrak{S} \rightarrow S(\mathfrak{A})$ under our algebraization, particularly whether it consists entirely of normal states in $\mathfrak{A}^*$.
\\

Given these considerations regarding von Neumann algebras and normal states, Kastler's seminal observation provides crucial physical insight \cite{kastler}:
\begin{citazione}\index{Kastler}
It is important to realize that whilst the quasi-local algebra $\mathfrak A$ has elements corresponding to physical observables (procedures), this is no longer the case for the weak closure $\mathfrak M = \pi(\mathfrak A)''$ in some representation $\pi$. It is not correct, from a physical point of view, to consider global quantities (such as the bounded functions of the energy, or the number operator), as observables, although they belong to $\mathfrak M$ for certain representations. Indeed these cannot be observed locally, and \textbf{only local experiments are physically possible}. Thus the W*-systems\footnote{A W*-system $ \left\{\mathfrak M , G , \alpha\right\}$ is a triple of a von Neumann algebra $\mathfrak M$, a locally compact group $G$ and a morphism $t\rightarrow \alpha_t$  of $G$ into the automorphism group of $\mathfrak M$ such that for each $A\in\mathfrak M$  the map $t\rightarrow \alpha_t(A)$ is continuous from $G$ to $\mathfrak M$  with its $\sigma$-weak topology.} obtained by considering the weak closure in covariant representations together with the extended group action \textbf{should not be considered as physical systems, but as certain mathematical extensions of physical systems} pertaining to certain particular physical situations (physical states). This point is important for a correct realization of the respective roles of C* and W* algebras in algebraic field theory. 
\end{citazione}
The topic will be returned to next chapter.
\section{Guidelines for Construct a C*-algebraic framework-first Step} \label{guide_first_step}
Let us examine, \textit{in broad terms, } the steps required to construct a C*-algebraic framework for our physical system, while keeping in mind the  quote by Munga et al. referenced earlier in this chapter. 
\\
Consider the following families of observables of the system:  
$$
\left\{ \mathbf{1}_\Delta (a) : \Delta \in B(\mathbb{R}), \, a \in \mathfrak{X} \right\}  
$$  
which are sets of spectral families. 
\\
My approach to this problem is standard. It outlines a method for constructing a C*-algebraic framework for a physical system by associating spectral families with orthogonal projectors in a separable Hilbert space. 
\begin{itemize}
\item [I.] \textbf{Choosing the Hilbert Space $\mathcal H$}.
\\
Since the sets of observables and states are countable, we select a separable Hilbert space $\mathcal{H}$\footnote{See remark \ref{state-num} on page \pageref{state-num}.}.
\\
Because $\mathcal{H}$ is separable, its dimension is either finite or countably infinite. Moreover, all infinite-dimensional separable Hilbert spaces are isomorphic to $\ell^2(\mathbb{N})$, so the choice reduces to fixing the dimension.
\\
The dimension of $\mathcal{H}$ should reflect the \textit{degrees of freedom of the system.}
\item[II.] \textbf{Mapping Spectral Families to Orthogonal Projectors on $\mathcal{H}$.}
\\
We  associate  
$$\mathbf{1}_\Delta (a) \longrightarrow E_\Delta^a \in \mathcal{P}(\mathcal{H})$$ 
 thus defining a family of orthogonal projectors of $\mathfrak B(\mathcal H)$:  
\begin{equation}\label{project}
\left\{ E_\Delta^a : \Delta \in B(\mathbb{R}), \, a \in \mathfrak{X}, \ \Delta\in B(\mathbb R)  \right\}  
\end{equation} 
We  define  
$$
\mathrm{J}(\mathbf{1}_\Delta (a)) := E_\Delta^a  
$$  
and, using the spectral decomposition from Section \ref{Spectral Decomposition of an Observable}, we extend the map $\mathrm{J}$ as follows\footnote{In other words:
$$\mathrm{J} \left( \int t \   dF_t \right): = \int t \   dE_t^a $$  
}:  
$$
\mathrm{J}(a) : = \int t \, dE_t^a  
$$  
where  
$$
E^a_t = \mathrm{J} ( \mathbf{1}_{]-\infty, t]}(a) ) \ , \qquad \forall t \in \mathbb{R}
$$

\item[III.]   \textbf{The C*-algebra of $\mathfrak B(\mathcal H)$.} 
\\ 
We can consider the von Neumann algebra $\mathfrak M$  of $\mathfrak B(\mathcal H)$  generated by the family of orthogonal projectors given by relation \eqref{project}. 
\\
In this way $\mathfrak M$ encodes the observables and their commutation relations.
 \end{itemize}
The map  $\mathrm{J}:\mathfrak X \longrightarrow \mathfrak M$ must be carefully defined. For instance should preserve the algebraic relations between the observables discussed previously.
\\

In the next chapter we will discuss the construction and properties of the dual map $\mathrm{J}^\natural:\mathfrak S \longrightarrow S(\mathfrak M)$.
\chapter{Physical vs. Mathematical States}  
In this section, we resume the analysis begun in the previous chapter \ref{Pure States and Borel Measures}, with the aim of clarifying in a more structural way the relationship between physical states and algebraic states in the representation induced by $\mathrm{J}^{\natural}$.  
\\
We will see that there is a distinction between mathematical eigenstates and physical eigenstates: mathematical eigenstates, i.e., vectors in a Hilbert space, do not always correspond to pure physical states of the system. Even if a vector $\Psi_h$ is an eigenvector of the operator $\mathrm{J}(a)$, it is not guaranteed that there exists a physical state $\omega_h$ representing it in a pure way, that is, satisfying $\mathrm{J}^{\natural}(\omega_h)(A) = \langle \Psi_h | A \Psi_h \rangle$.  
\\
This observation has implications for the notion of mixed states understood as statistical mixtures. In other words, a physical state $\omega$ satisfying $P(a = \lambda)_\omega = 1$, and thus being an eigenstate of the observable $a$, can still be a mixed state in the algebraic representation. Its representation $\mathrm{J}^{\natural}(\omega)$ can indeed be a statistical mixture of mathematical pure states, each of which is an eigenvector of $\mathrm{J}(a)$.  
\\
Therefore, the image of physical eigenstates under the map $\mathrm{J}^{\natural}$ is not necessarily contained in the pure states of $\mathfrak{A}$. This means that physical eigenstates do not always correspond to pure states in the algebraic representation, and we can affirm that there exists a gap between the mathematical description, based on eigenvectors in $\mathcal{H}$, and the actual physical states of the system.

\section{States and Algebraic Normal States}\label{Phys-Math-states}
Let $\mathfrak{A}$ be a concrete C*-algebra on a Hilbert space $\mathcal{H}$. We denote by 
$$S_\sigma(\mathfrak{A}) \subset (\mathfrak{A}, \sigma)^*$$
the set of states on $\mathfrak{A}$ that are continuous in the $\sigma$-weak (ultraweak) topology\footnote{For a good summary of the various topologies induced by seminorms, see Bratteli–Robinson \cite{BR1}, \S 2.4.1.}.
\\
For a state $\varphi$ on a von Neumann algebra $\mathfrak{M} \subset B(\mathcal{H})$, the following are equivalent:
\begin{itemize}
\item[1.] $\varphi$ is normal (i.e., $\sigma$-weakly continuous).
\item[2.] $\varphi(A) = \operatorname{tr}(DA)$ for some density operator $D$ on $\mathcal{H}$.
\item[3.] $\varphi(A)$ is completely additive on orthogonal projections:
$$\varphi\Big( \bigvee_{i \in I} E_i \Big) = \sum_{i \in I} \varphi(E_i)$$
for any family $\{ E_i \}_{i \in I}$ of orthogonal projections in $\mathfrak{M}$\footnote{The expression $E = \bigvee_{i \in I} E_i$ represents the least upper bound of $F_n = \sum_{k=1}^n E_k$, and $F_n$ converges to $E$ in the \textit{strong operator topology}.  
\\
Since the strong topology $\tau_F$ coincides with the \textit{ultra-strong topology} $S$ on the unit ball $B(\mathcal{H})_1$, and the $\sigma$-weak topology is weaker than the $S$-topology, it follows that $F_n$ converges to $E$ in the $\sigma$-weak topology (and therefore also in the weak operator topology). }. 
\end{itemize} 
For a concrete C*-algebra $\mathfrak{A} \subset \mathfrak{B}(\mathcal{H})$, the bicommutant $\mathfrak{A}''$ is the smallest von Neumann algebra containing $\mathfrak{A}$.
\\

It is useful to recall that if the Hilbert space $\mathcal{H}$ is separable, then the von Neumann algebra $\mathfrak{A}''$ is $\sigma$-finite and therefore admits at least one faithful normal state\footnote{Cf. Bratteli–Robinson \cite{BR1}, Proposition 2.5.6.}.
\\
Furthermore, we have:
\begin{equation}\label{stati-normali}
S_\sigma(\mathfrak{A}) = \left\{ \varphi|_{\mathfrak{A}} : \varphi \in S(\mathfrak{A}'') \cap (\mathfrak{A}'')_* \right\}
\end{equation}
where we denote by $\mathfrak{M}_*$ the predual of the von Neumann algebra $\mathfrak{M} = \mathfrak{A}''$\footnote{We note that type III von Neumann algebras have no pure normal states.} .
\\
 
We recall that a linear functional $\varphi \in B(\mathcal{H})^*$ is said to be \textit{singular} if it vanishes on all compact operators, i.e.,
$$ \varphi(K) = 0 \qquad \forall K \in \mathcal{K}(\mathcal{H}) $$
where $\mathcal{K}(\mathcal{H})$ denotes the set of compact operators on the Hilbert space $\mathcal{H}$.
Such functionals cannot be represented by trace-class operators.
\\

Let us reconsider the topic discussed in Section \ref{questions-states} on page \pageref{questions-states}.  
\\
Let $\Delta$ be a Borel set and $\{ \Delta_k \}_{k \in \mathbb{N}}$ a disjoint Borel partition of $\Delta$.
\\
In this case, we can write:  
$$ \mathbf{1}_\Delta(a) = \lim_{N \rightarrow \infty} \sum_{k=1}^N \mathbf{1}_{\Delta_k}(a) $$  
where the convergence is established by relation \ref{proiettori}\footnote{This follows from the $\sigma$-additivity of our Borel measures.}:  
$$\left\langle \mathbf{1}_\Delta(a) \right\rangle_{\omega} = \lim_{N \rightarrow \infty} \sum_{k=1}^N \left\langle \mathbf{1}_{\Delta_k}(a) \right\rangle_{\omega}, \qquad \forall \omega \in \mathfrak{S}_a.$$  
Let $(\mathfrak{A}, \mathrm{J}, \mathrm{J}^{\natural})$ be a C*-algebraic representation of our physical system $(\mathfrak{X}, \mathfrak{S})$.  
We obtain that $\mathrm{J}(\mathbf{1}_{\Delta_k}(a))$ are orthogonal projectors in $\mathfrak{A} \subset \mathfrak{B}(\mathcal{H})$, and  
$$ \mathrm{J}\big( \mathbf{1}_\Delta(a) \big) = \bigvee_k \mathrm{J}\big( \mathbf{1}_{\Delta_k}(a) \big),$$  
with  
$$ \mathrm{J}^{\natural}(\omega) \left( \bigvee_k \mathrm{J}\big( \mathbf{1}_{\Delta_k}(a) \big) \right) = \sum_k \mathrm{J}^{\natural}(\omega)\big( \mathbf{1}_{\Delta_k}(a) \big).$$  
In this case, the state $\mathrm{J}^{\natural}(\omega)$ turns out to be completely additive.  
\\
Can we assert that $\mathrm{J}^{\natural}(\omega)$ must necessarily be $\sigma$-continuous?  
\\
No, because complete additivity only holds for specific projectors of the form $\mathrm{J}\big( \mathbf{1}_\Delta(a) \big)$. Not all projectors in $\mathfrak{A}$ are of this type.
\begin{property}[N1]\label{condizione N1}\index{Property N1}
The C*-algebraization $(\mathfrak{A}, \mathrm{J}, \mathrm{J}^{\natural})$ of $(\mathfrak{X}, \mathfrak{S})$ satisfies condition [\texttt{N1}] if
$$ \mathrm{J}^{\natural}(\mathfrak{S}) \subset S_\sigma(\mathfrak{A})$$
\end{property}
In this case, all physically realizable states in the laboratory are $\sigma$-continuous, and the singular states of the algebra are not physically realizable in our laboratory.
\begin{property}[N2]\label{condizione N2}\index{Property N2}
The C*-algebraization $(\mathfrak{A}, \mathrm{J}, \mathrm{J}^{\natural})$ of $(\mathfrak{X}, \mathfrak{S})$ satisfies condition [\texttt{N2}] if
$$ S_\sigma(\mathfrak{A}) \subset \mathrm{J}^{\natural}(\mathfrak{S})$$
\end{property}
In this case, all $\sigma$-continuous states are physically realizable, and even singular states may admit a physical realization in the laboratory.
\\
Furthermore, since $S_\sigma(\mathfrak{A})$ is a full subset of $S(\mathfrak{A})$\footnote{Cf. Bratteli–Robinson \cite{BR1}, Proposition 3.2.10.}, we have:
$$ \overline{S_\sigma(\mathfrak{A})}^{W^*} = S(\mathfrak{A}) \qquad \Longrightarrow \qquad \overline{\mathrm{J}^{\natural}(\mathfrak{S})}^{W^*} = S(\mathfrak{A})$$
Consequently, by Proposition \ref{prop_density_embedding} on page \pageref{prop_density_embedding}, the representation satisfies the embedding property. 
\begin{theorem}\upshape
Let $(\mathfrak{A}, \mathrm{J}, \mathrm{J}^{\natural})$ be a C*-algebraization of a physical system $(\mathfrak{X}, \mathfrak{S})$. Then there exists a von Neumann algebraization $(\mathfrak{R}, \mathtt{J}_o, \mathtt{J}_o^{\natural})$ such that 
\begin{itemize}
\item \textbf{Embedding of $\mathfrak{A}$ into $\mathfrak{R}$:}
\\
\qquad $\mathfrak{A}$ embeds into $\mathfrak{R}$ via an injective homomorphism: 
$$\mathfrak{A} \stackrel{\Lambda}{\hookrightarrow} \mathfrak{R}$$
where $\Lambda$ is $\sigma$-weakly continuous (i.e., continuous in the $\sigma$-weak topology).
\item \textbf{Condition [N1] holds:}
\\
\qquad The map $\mathtt{J}_o^{\natural}$ satisfies:
$$\mathtt{J}_o^{\natural}: \mathfrak{S} \longrightarrow S_\sigma(\mathfrak{R})$$ 
where $S_\sigma(\mathfrak{R})$ denotes the normal ($\sigma$-weakly continuous) states on $\mathfrak{R}$.
\item \textbf{Consistency on observables:}
\\
\qquad For every $a \in \mathfrak{X}$,
$$\mathtt{J}_o(a) = \Lambda(\mathrm{J}(a))$$
\item \textbf{Consistency on state evaluations:}
\\
\qquad For every $a \in \mathfrak{X}$ and $\omega \in \mathfrak{S}_a$,
$$\mathtt{J}_o^{\natural}(\omega)\big( \mathtt{J}_o(a) \big) = \mathrm{J}^{\natural}(\omega)\big( \mathrm{J}(a) \big)$$
\end{itemize}
\end{theorem}
The von Neumann algebraization $(\mathfrak{R}, \mathtt{J}_o, \mathtt{J}_o^{\natural})$ is called the \textit{dilation} of the C*-algebraization $(\mathfrak{A}, \mathrm{J}, \mathrm{J}^{\natural})$.
\begin{proof}
We divide the proof into four key steps:
\\

\textbf{1.} Canonical Embedding and Extensions:  
\\ 
Let $\mathrm{e}: \mathfrak{A} \hookrightarrow \mathfrak{A}^{**}$ denote the canonical embedding (an isometric $*$-homomorphism):  
$$\mathrm{e}(a)(x') = x'(a) \qquad \forall a \in \mathfrak{A}, \ x' \in \mathfrak{A}^*$$  
This embedding is continuous with respect to the weak topologies\footnote{Density property: $\overline{\mathrm{e}(\mathfrak{A})}^{w^{**}} = \mathfrak{A}^{**}$}:  
$$\mathrm{e}: (\mathfrak{A}, w) \to (\mathfrak{A}^{**}, w^{**})$$   
where $w = \sigma(\mathfrak{A}, \mathfrak{A}^*)$ and $w^{**} = \sigma(\mathfrak{A}^{**}, \mathfrak{A}^*)$.  
\\
Moreover, for every $\varphi \in S(\mathfrak{A})$, there exists a unique $\widehat{\varphi} \in S(\mathfrak{A}^{**})$ such that:  
$$\widehat{\varphi}\big( \mathrm{e}(a) \big) = \varphi(a) \qquad \forall a \in \mathfrak{A}$$  
where $\widehat{\varphi}$ is $\sigma(\mathfrak{A}^{**}, \mathfrak{A}^*)$-continuous.  
\\

\textbf{2.} Construction of the W*-Algebraization:
\\  
We define:  
$$\mathrm{J}_1: \mathfrak{X} \to \mathfrak{A}^{**}, \quad \mathrm{J}_1(a) := \mathrm{e}\big( \mathrm{J}(a) \big) \quad \forall a \in \mathfrak{X}$$  
and  
$$\mathrm{J}_1^{\natural}: \mathfrak{S} \to S(\mathfrak{A}^{**})$$  
as the $\sigma$-weakly continuous extension of $\mathrm{J}^{\natural}(\omega)$, satisfying\footnote{Cf. Sakai's book \cite{Sakai}, Proposition 1.21.13.}:  
$$\mathrm{J}_1^{\natural}(\omega)\big( \mathrm{e}(A) \big) = \mathrm{J}^{\natural}(\omega)(A) \quad \forall A \in \mathfrak{A}$$  
Consequently,  
$$
\mathrm{J}_1^{\natural}(\omega)\big( \mathrm{J}_1(a) \big) = \mathrm{J}^{\natural}(\omega)\big( \mathrm{J}(a) \big) = \langle a \rangle_\omega$$  
The bidual $\mathfrak{A}^{**}$ is the \textit{universal enveloping von Neumann algebra} of $\mathfrak{A}$, making it a W*-algebra. Thus, we obtain a W*-algebraization $(\mathfrak{A}^{**}, \mathrm{J}_1, \mathrm{J}_1^{\natural})$.  
\\

\textbf{3.} Universal Representation and von Neumann Algebraization:
\\  
Consider the \textit{universal representation} of $\mathfrak{A}$:  
$$\pi_u: \mathfrak{A} \to \mathfrak{B}(\mathcal{H}_u)$$  
which is isometric (by the well-known Sherman–Takeda Theorem\footnote{Sherman–Takeda Theorem: If $\pi_u: \mathfrak{A} \to \mathfrak{B}(\mathcal{H}_u)$ is the universal representation, then 
$$\mathfrak{A}^{**} \simeq \pi_u(\mathfrak{A})''$$
Moreover, if $\mathfrak{M}$ is a von Neumann algebra, then $\mathfrak{M} \simeq \mathfrak{M}^{**}$.}) and admits an isometric, $\sigma$-weakly continuous extension:  
$$\check{\pi}_u: (\mathfrak{A}^{**}, w^{**}) \to (\mathfrak{B}(\mathcal{H}_u), \sigma)$$  
where $\check{\pi}_u \circ \mathrm{e} = \pi_u$.  
\\
Now, define the concrete W*-algebra:  
$$ \mathfrak{R} := \check{\pi}_u(\mathfrak{A}^{**}) \subset \mathfrak{B}(\mathcal{H}_u)$$ 
We set:  
$$\mathrm{J}_o: \mathfrak{X} \to \mathfrak{R}, \quad \mathrm{J}_o(a) := \check{\pi}_u\big( \mathrm{J}_1(a) \big) \quad \forall a \in \mathfrak{X}$$ 
and  
$$\mathrm{J}_o^{\natural}: \mathfrak{S} \to S_\sigma(\mathfrak{R})$$  
such that for every $\omega \in \mathfrak{S}$,  
$$\mathrm{J}_o^{\natural}(\omega)(R) = \mathrm{J}_1^{\natural}(\omega)\big( \check{\pi}_u^{-1}(R) \big) \quad \forall R \in \mathfrak{R}$$  
This ensures:  
$$\mathrm{J}_o^{\natural}(\omega)\big( \mathrm{J}_o(a) \big) = \langle a \rangle_\omega$$  
Thus, we obtain a \textit{von Neumann algebraization} $(\mathfrak{R}, \mathrm{J}_o, \mathrm{J}_o^{\natural})$ of $(\mathfrak{X}, \mathfrak{S})$, satisfying condition [\texttt{N1}].  
\\

\textbf{4.} Density Operator Representation: 
\\  
By definition, for every $\omega \in \mathfrak{S}$, there exists a unique $D_\omega \in L^1(\mathcal{H}_u)$ such that:  
$$\mathrm{J}_o^{\natural}(\omega)(R) = \operatorname{tr}\big( D_\omega R \big) \qquad \forall R \in \mathfrak{R}$$  
In particular,  
$$\mathrm{J}_o^{\natural}(\omega)\big( \mathrm{J}_o(a) \big) = \operatorname{tr}\Big( D_\omega \, \pi_u\big( \mathrm{J}(a) \big) \Big) \quad \forall a \in \mathfrak{X}$$  
We thus obtain the following well-defined correspondence:
$$\omega \in \mathfrak{S}_a \longrightarrow D_\omega \in L^1(\mathcal{H}_u)$$
where\footnote{See section \ref{rep-states} and relation \eqref{stati-matrici}  on page \pageref{stati-matrici}.}
$$ \langle a \rangle_{\omega} = \operatorname{tr}\Big( D_\omega \, \pi_u\big( \mathrm{J}(a) \big) \Big) $$
\end{proof}
\section{Physical States and Mathematical States}\label{Physic-math-states}
The problem statement is as follows:
\begin{problem}\upshape
The process of preparing a physical state for measurement involves creating a well-defined physical state in the laboratory. The question is: how does the mathematical representation of these states correspond to the physical preparation procedure?
\end{problem}
Once the algebra of observables $\mathfrak{M} \subset \mathfrak{B}(\mathcal{H})$ has been established, the first step\footnote{See section \ref{guide_first_step} on page \pageref{guide_first_step}.} is to select a state $\varphi$ in $\mathfrak{M}$ such that  
$$\varphi(E_\Delta^a) = \langle \mathbf{1}_\Delta(a) \rangle_\omega, \quad \forall a \in \mathfrak{X}, \ \Delta \in B(\mathbb{R})$$    
and to associate  
$$ \omega \longrightarrow \varphi := \mathrm{J}^{\natural}(\omega)$$ 
As we have seen, for any disjoint partition $\{ \Delta_k \}_k$ of $\Delta$, the $\sigma$-additivity of the measure $\mu_{\omega,a}$ implies that  
$$ \varphi(E_\Delta^a) = \varphi\Big( \sum_k E_{\Delta_k}^a \Big) = \sum_k \varphi(E_{\Delta_k}^a)$$  

This suggests choosing the \textit{normal states} of $\mathfrak{M}$ as candidates for $\mathrm{J}^{\natural}(\omega)$:  
$$\mathrm{J}^{\natural}: \mathfrak{S} \longrightarrow S_\sigma(\mathfrak{M})$$  
 $$ \star \star \star $$
To be closer to physical reality, we must consider that the laboratory state $\omega$ is not defined for every observable $x$ of the physical system, but only for the observables suitable to it, i.e., on the set $\mathfrak{X}_\omega$. Thus, associating the algebraic state $\mathrm{J}^{\natural}(\omega)$ with the entire algebra $\mathfrak{A}$ is a questionable action. To be more precise, we should instead consider the following algebra:
\begin{equation}\label{algegra-ridotta}  
\mathfrak{A}_\omega := \overline{\mathcal{P}\big( \mathrm{J}(\mathfrak{X}_\omega) \big)}^{\tau_n}
\end{equation}    
and with  
$$\mathrm{J}^{\natural}(\omega) \in S(\mathfrak{A}_\omega)$$  
In other words, we can mathematically assert that  
\begin{equation}\label{stato-alg-ridotto}
\mathrm{J}^{\natural} \in \prod_{\omega \in \mathfrak{S}} S(\mathfrak{A}_\omega)
\end{equation}  
where the direct product is defined as:  
$$
\prod_{\omega \in \mathfrak{S}} S(\mathfrak{A}_\omega) := \left\{ \xi: \mathfrak{S} \rightarrow \bigcup_{\omega \in \mathfrak{S}} S(\mathfrak{A}_\omega) : \; \xi(\omega) \in S(\mathfrak{A}_\omega) \ \forall \omega \right\}.  
$$ 
This way, we will have a discrepancy between the element $\mathrm{J}(a) \in \mathfrak{A}$ and the algebraic state associated with $\mathrm{J}^{\natural}(\omega) \in S(\mathfrak{A}_\omega)$\footnote{Recall that in C*-algebras, for the spectrum of the element $\mathrm{J}(a) \in \mathfrak{A}_\omega \subset \mathfrak{A}$, we have  
$$ \operatorname{Sp}_{\mathfrak{A}_\omega}\big( \mathrm{J}(a) \big) = \operatorname{Sp}\big( \mathrm{J}(a) \big)$$}. 
\\
However, by the Hahn–Banach theorem, every state on $\mathfrak{A}_\omega$ extends (non-uniquely) to a state on $\mathfrak{A}$. Yet this mathematical step is entirely detached from the physical act, though it allows us to recover our original map $\mathrm{J}^{\natural}: \mathfrak{S} \rightarrow S(\mathfrak{A})$.
%g  
\begin{problem}\upshape\label{perditainformalgebrica}
What physical information do we lose in this extension?  
\end{problem}
\subsection{Physical Pure States and Mathematical Pure States}
Let us resume the discussion initiated in Section \ref{convex-algebre}, examining in detail the problem presented in Question \ref{pure-states} on page \pageref{pure-states}.
\\

We recall that a \textit{pure state} $\omega \in \mathfrak{S}_a$ in the measurement of an observable $a$ corresponds to an extreme point of the convex set of probability measures $\mathbb{M}_k(a)$, denoted by:
$$\mathrm{Ext}_k(a) \subset \mathbb{M}_k(a) \subset \mathbb{M}(a) \subset \Pi$$
Here:
\\
- $\mathbb{M}(a)$ is the total set of admissible probability measures for $a$,
\\
- $\Pi$ is a broader convex set of probability measures defined in equation \eqref{insieme-pi} on page \pageref{insieme-pi}.
\\  
The extreme points of $\mathbb{M}_k(a)$ (i.e., $\mathrm{Ext}_k(a)$) need not be extreme points of the larger convex set $\Pi$.  
\\
This implies the following structural relationship:
$$\Pi_p \cap \mathbb{M}_k(a) \subset \mathrm{Ext}_k(a) \quad \forall k $$
where $\Pi_p$ denotes the set of extreme points of $\Pi$.
\\
The inclusion $\mathrm{Ext}_k(a) \subset \mathbb{M}_k(a)$ reflects that "pure states" are a subset of all sectorial measures.
\\
Moreover, the fact that $\mathrm{Ext}_k(a) \not\subseteq \mathrm{Ext}(\Pi) = \Pi_p$ highlights that sectorial purity does not necessarily imply \textit{global purity}.
\\
Alongside the notion of pure states in the measurement of an observable $a$, we have introduced in Definition \ref{statopurodelsistema} on page \pageref{statopurodelsistema} the more physically relevant concept of \textit{purely informative states} for the measurement of an observable $a$, denoted by $\mathfrak{P}_a$.
\\
As established in Proposition \ref{stati-puri-relazione} on page \pageref{stati-puri-relazione}, we have the inclusion:
$$\mathfrak{P}_a \cap \mathfrak{S}^k_a \subset \mathrm{Ext}(\mathfrak{S}_a^k)$$
and consequently,
$$\mathrm{J}^{\natural} \left( \mathfrak{P}_a \cap \mathfrak{S}^k_a \right) \subset \mathrm{J}^{\natural} \left( \mathrm{Ext}(\mathfrak{S}_a^k) \right) \subset S(\mathfrak{A})$$
Moreover, from the inclusion \eqref{autostatifisici} on page \pageref{autostatifisici}:
$$ \texttt{V}_\lambda^{k}(a) \subset \mathfrak{P}_a \subset \mathrm{Ext}(\mathfrak{S}_a^k)$$
it follows that
\begin{equation}\label{autostatialgebr}
 \mathrm{J}^{\natural} \left( \texttt{V}^k_\lambda(a) \right) \subset \mathrm{J}^{\natural} \left( \mathfrak{P}_a \right) \subset \mathrm{J}^{\natural} \left( \mathrm{Ext}(\mathfrak{S}_a^k) \right)
\end{equation}
\begin{attenzione}\upshape
we  emphasize that the set $\mathrm{J}^{\natural} \left( \texttt{V}^k_\lambda(a) \right)$ need not consist of extreme points of $S(\mathfrak{A})$.
\end{attenzione}
\begin{remark}\upshape From an algebraic perspective, the pure states of $S(\mathfrak{A})$ are those states that cannot be expressed as non-trivial convex combinations of other states — in other words, they correspond to its extreme points.
\end{remark}
We recall that a representation $\pi$ is \textit{irreducible} if it admits no non-trivial invariant subspaces (i.e., the only closed subspaces $\mathcal{K} \subseteq \mathcal{H}$ satisfying $\pi(a)\mathcal{K} \subseteq \mathcal{K}$ for all $a \in \mathfrak{A}$ are $\{0\}$ and $\mathcal{H}$).
Consequently, for C*-algebras, we have:
$$\pi_\varphi(\mathfrak{A})' = \mathbb{C}I$$
Moreover, in the C*-algebra case $\mathfrak{A}$, if $\varphi$ is a pure state, then its associated GNS (Gelfand–Naimark–Segal) representation $\pi_\varphi: \mathfrak{A} \to \mathcal{B}(\mathcal{H}_\varphi)$ is irreducible.
\\
Conversely, if $\pi$ is an irreducible representation with a cyclic vector $\Omega$ of norm $1$, then the state
$$\varphi(a) := \langle \Omega, \pi(a)\Omega \rangle \quad \forall a \in \mathfrak{A}$$
is pure.
\begin{attenzione}\upshape
If $\mu_{\omega,a} \in \mathrm{Ext}_k(a)$, it does not necessarily follow that its GNS representation $\pi_{\mu_{\omega,a}}: C_o(\mathbb{R}) \rightarrow B(\mathcal{H}_{\mu_{\omega,a}})$ is irreducible. This holds true if and only if $\mu_{\omega,a} \in \Pi_p$.
\end{attenzione}
Let $\varphi \in S(\mathfrak{A})$ be a state on a C*-algebra $\mathfrak{A}$. Then for its GNS representation $(\pi_\varphi, \mathcal{H}_\varphi, \Omega_\varphi)$, there exists a density matrix $\rho \in B(\mathcal{H}_\varphi)$ such that:
\begin{equation}\label{GNS-density}
\varphi(A) = \operatorname{tr}\big( \rho \, \pi_\varphi(A) \big) \quad \forall A \in \mathfrak{A}
\end{equation}
If $\varphi \in \operatorname{Ext}\big( S(\mathfrak{A}) \big)$, where $\operatorname{Ext}\big( S(\mathfrak{A}) \big)$ denotes the set of pure states, then the density matrices satisfying equation \eqref{GNS-density} can only be of rank $1$. Consequently, their von Neumann entropy vanishes: $S(\rho) = 0$.

\subsection{Purity Index of a State in C*-algebras}

Let $\mathfrak{A}$ be a concrete C*-algebra on a Hilbert space $\mathcal{H}$. Given a density matrix $\rho$ on $\mathcal{H}$, the number $\operatorname{tr}(\rho^2)$ is the purity index of the corresponding normal state $\varphi$ associated with $\rho$.

Thus, the purity of a density matrix, and therefore of its associated normal state, is defined as follows:
$$p(\rho) := \operatorname{tr}(\rho^2)$$ 
Since $\rho \geq 0$, its eigenvalues $\lambda_k$ are real and non-negative, and they satisfy:
$$\operatorname{tr}(\rho) = \sum_k \lambda_k \qquad \text{and} \qquad \operatorname{tr}(\rho^2) = \sum_k \lambda_k^2$$
We recall that  
$$ \sum_{i=1}^n \lambda_i^2 \leq \left( \sum_{i=1}^n \lambda_i \right)^2 $$  
holds for every natural number $n$ and every sequence of non-negative real numbers $\lambda_1, \lambda_2, \dots, \lambda_n \geq 0$.  
\\
Proof of the inequality for finite sums:
\\ 
Expanding the square of the sum:  
$$\left( \sum_{i=1}^n \lambda_i \right)^2 = \sum_{i=1}^n \lambda_i^2 + 2 \sum_{1 \leq i < j \leq n} \lambda_i \lambda_j$$  
Since all $\lambda_i$ are non-negative, the cross term $2 \sum_{1 \leq i < j \leq n} \lambda_i \lambda_j$ is non-negative. Hence,  
$$ \sum_{i=1}^n \lambda_i^2 \leq \left( \sum_{i=1}^n \lambda_i \right)^2. $$  
Equality holds \textit{if and only if} the cross term is zero, i.e., if at most one of the $\lambda_i$ is positive (and the others are zero).  
\\
What happens for infinite sums?
\\  
If we consider an infinite sequence $\{ \lambda_i \}_{i=1}^\infty$ of non-negative numbers, the inequality continues to hold provided the sums converge.
\\ 
In particular: 
\\ 
- If $\sum_{i=1}^\infty \lambda_i$ converges, then $\sum_{i=1}^\infty \lambda_i^2$ also converges, since $\lambda_i^2 \leq \lambda_i$ for $0 \leq \lambda_i \leq 1$, and by elementary properties of numerical series.
\\  
- The inequality extends to the infinite case by taking limits:  
$$ \sum_{i=1}^\infty \lambda_i^2 \leq \left( \sum_{i=1}^\infty \lambda_i \right)^2$$  
In the context of density matrices:
\\ 
In quantum mechanics, the Hilbert space is often separable, and the density matrix $\rho$ is a trace-class operator.
\\ 
Its eigenvalues $\{ \lambda_i \}$ form a non-negative summable sequence with $\sum_i \lambda_i = 1$.
\\
Therefore,  
$$ \operatorname{tr}(\rho^2) = \sum_i \lambda_i^2 \leq \left( \sum_i \lambda_i \right)^2 = 1$$  
This inequality holds for both finite and infinite-dimensional spaces.
\\  
Thus, the density matrix $\rho$ is a Hilbert–Schmidt operator, i.e., $\rho \in L^2(\mathcal{H})$, and its Hilbert–Schmidt norm satisfies  
$$ \| \rho \|_2^2 = \operatorname{tr}(\rho^2) \leq 1$$  
It is straightforward to verify that  
$$ \operatorname{tr}(\rho^2) = 1 \iff \rho \text{ is a pure state} $$  
It follows that if $\operatorname{tr}(\rho^2) < 1$, the state $\rho$ is mixed.
\\ 
In other words, this norm measures purity:
\\  
$\| \rho \|_2 = 1$ for pure states, and $\| \rho \|_2 < 1$ for mixed states.  
$$ $$
Let us examine the relationship between purity and von Neumann entropy:
\begin{itemize}
\item[-] The purity $p(\rho)$ is a quadratic measure of mixedness: it is $1$ for pure states and decreases for mixed states.
\item[-] The von Neumann entropy $S(\rho) = -\operatorname{tr}(\rho \log \rho)$ is an entropic measure: it is $0$ for pure states and increases for mixed states.
\end{itemize}
Both quantities measure how "far" a state is from being pure. There is no direct, universal algebraic relationship between $p(\rho)$ and $S(\rho)$ because they depend on the eigenvalues of $\rho$ in different ways.
\\
However, they are both expressed in terms of the eigenvalues $\lambda_k$ of $\rho$:
\begin{itemize}
\item $p(\rho) = \sum_k \lambda_k^2$,
\item $S(\rho) = -\sum_k \lambda_k \log \lambda_k$.
\end{itemize}
\subsection{Purity for a Physical State}
We define the notion of a purity index for a physical state $\omega \in \mathfrak{S}^k_a$ of the laboratory system.
To do this, we must consider, as in the case of von Neumann entropy, the set of operationally realizable density matrices $\mathfrak{D}^k_{\omega,a}$ defined on page \pageref{Operationally-Realizable-Matrices}:
$$p(\omega, a) = \inf \left\{ \operatorname{tr}(\rho^2) \colon \rho \in \mathfrak{D}^k_{\omega,a} \right\}$$
The index $p(\omega, a) \in (0, 1]$ is called the \textit{purity degree} of the state $\omega$ with respect to the measurement of the observable $a$.\index{$p(\omega, a)$}\index{Purity degree}
$$p(\omega, a) = 1 \iff \mathfrak{D}^k_{\omega,a} \text{ is indecomposable } \iff S_k(\omega, a) = 0 \iff \omega \in \operatorname{Ext}(\mathfrak{S}_a^k)$$
The purity of a state $\omega$, to be a good quality index of the state itself, must be evaluated on every observable suitable for it; in other words, on every observable $x \in \mathfrak{X}_\omega$.
\\
To obtain a good purity index for the state, one should average the value $p(\omega, x)$ over the whole set $\mathfrak{X}_\omega$, which is experimentally unfeasible.
\\
We choose another path: we assume as the purity degree of $\omega$ its most ``mixed'' value:
$$p(\omega) = \inf \left\{ p(\omega, x) \colon x \in \mathfrak{X}_\omega \right\}$$
In this way, if $p(\omega) = 1$, then $p(\omega, x) = 1$ for every $x \in \mathfrak{X}_\omega$. This serves as a robust operational definition of a pure state. Conversely, if $p(\omega) < 1$, the state is mixed.
$$\circ\circ\circ$$ 
We now examine the problem posed in Question \ref{pure-states} on page \pageref{pure-states}:  
\\
Does the C*-algebraization $(\mathfrak{A}, \mathrm{J}, \mathrm{J}^{\natural})$ preserve purity?  
\\
But what does it mean for an algebraization to preserve purity?  
\\
Let us give a possible answer:  
\\
\textit{Purity preservation ensures that the algebraic states $\mathrm{J}^{\natural}(\omega)$ retain the physical interpretability of $\omega$ as an "extremal preparation".}  
\\
In physics, pure states represent the most precise possible preparation of a system.  
\\
If the map $\mathrm{J}^{\natural}$ preserves purity, then a perfectly prepared state in the physical laboratory system remains a perfectly prepared state in the algebraic representation.  
Without purity preservation, an extremal state in the physical laboratory system might become mixed (non-extremal) in the C*-algebraization $(\mathfrak{A}, \mathrm{J}, \mathrm{J}^{\natural})$, losing its interpretation as a "sharp" preparation. But is this question well-posed?  
\\
Thus, the following problem must be solved:  
\\
Given a state $\omega \in \mathfrak{S}_a^k$, determine the density matrix $\rho_\omega \in B(\mathcal{H})$ that implements the algebraic state $\mathrm{J}^{\natural}(\omega)$, i.e.,  
$$\mathrm{J}^{\natural}(\omega)(B) = \operatorname{tr}(\rho_\omega B), \qquad \forall B \in \mathfrak{A}$$  
such that the following equation holds:  
$$p(\rho_\omega) = p(\omega)$$
In other words, given $p(\omega) \in (0, 1]$, determine $\rho \in B(\mathcal{H})$ such that  
\begin{equation}\label{equazione-fondamentale}
\operatorname{tr}(\rho^2) = p(\omega).
\end{equation}  
Note that if \eqref{equazione-fondamentale} admits a solution when $p(\omega) = 1$, then the associated state $\mathrm{J}^{\natural}(\omega)$ is pure.  
\\
We will see later\footnote{See remark \ref{controesempio-eq-fond} on page \pageref{controesempio-eq-fond}.} that \textit{this equation does not always admit solutions}, and that purity cannot always be trivially preserved under algebraization.
\subsection{Deterministic States on C*-algebras}
Comparative definitions in the algebraic case:
\\
Let $\mathfrak{A}$ be a unital C*-algebra and $\varphi \in S(\mathfrak{A})$ a state. For a self-adjoint element $A \in \mathfrak{A}$, the \textit{variance} of $A$ in the state $\varphi$ is defined as:
$$\Delta_\varphi(A) = \varphi(A^2) - \varphi(A)^2$$
If $\Delta_\varphi(A) = 0$ for all $A \in \mathfrak{A}$, the state $\varphi$ is called \textit{deterministic}.
 \\

We recall that the spectral measure $\nu_{\varphi, A}$ associated with the self-adjoint operator $A$ is given by
$$ \varphi(f(A)) = \int f(s) \, d\nu_{\varphi, A}(s), \qquad \forall f \in C_o(\mathbb{R})$$
We have a result analogous to Proposition \ref{freeds1} on page \pageref{freeds1}\footnote{One proceeds in the same way as in the proof of Proposition \ref{freeds1}, with $m = \varphi(A)$ here.}:
\begin{equation}\label{freeds2}
\Delta_\varphi(A) = 0 \qquad \iff \qquad \nu_{\varphi, A} \in \Pi_p.
\end{equation}
Let us make the following observation:
\\
Let $A \in \mathfrak{A}_{s.a.}$. Denote by $C^*(A)$ the unital C*-algebra generated by $A$. Now, let $\varphi \in S(\mathfrak{A})$ and consider its restriction $\varphi_0$ to the commutative algebra $C^*(A)$.
\\
From the above, it follows that:
$$ \Delta_{\varphi_0}(A) = 0 \qquad \iff \qquad \nu_{\varphi_0, A} \in \Pi_p$$
We note that the two spectral measures $\nu_{\varphi, A}$ and $\nu_{\varphi_0, A}$ coincide\footnote{As is known from C*-algebra theory 
$$\operatorname{Sp}_{C^*(A)}(A) = \operatorname{Sp}_{\mathfrak{A}}(A)$$}. Furthermore, the respective variances are equal, $\Delta_{\varphi}(A) = \Delta_{\varphi_0}(A)$, and thus we can state that:
$$ \Delta_{\varphi}(A) = 0 \ \iff \ \nu_{\varphi_0, A} \in \Pi_p \ \iff \ \varphi_0 \text{ is a character} \ \iff \ \varphi_0 \text{ is a pure state}$$

\begin{proposition}\upshape[Deterministic States and Purity]\label{puritdeterm}
If $\Delta_\varphi(A) = 0$ for all $A \in \mathfrak{A}_{s.a.}$, then $\varphi$ is a pure state of $\mathfrak{A}$\footnote{\textit{The converse holds if $\mathfrak{A}$ is commutative.}}.
\end{proposition}
\begin{proof}
Assume $\varphi$ is not pure. Then there exist distinct states $\varphi_1, \varphi_2 \in S(\mathfrak{A})$ and $t \in (0,1)$ such that $\varphi = t\varphi_1 + (1 - t)\varphi_2$. 
\\ 
By distinctness, there exists $B \in \mathfrak{A}$ with $\varphi_1(B) \neq \varphi_2(B)$. 
\\
Let $A = B - \varphi(B)I$. Then:
$$\varphi(A) = \varphi(B) - \varphi(B) = 0$$
For any state $\gamma$, the Cauchy–Schwarz inequality gives $|\gamma(A^2)| \geq \gamma(A)^2$. 
\\
Applying this to $\varphi_1, \varphi_2$:
$$\varphi(A^2) = t\varphi_1(A^2) + (1 - t)\varphi_2(A^2) \geq t\varphi_1(A)^2 + (1 - t)\varphi_2(A)^2$$
Since $\varphi_1(A) = \varphi_1(B) - \varphi(B) \neq \varphi_2(B) - \varphi(B) = \varphi_2(A)$ (therefore $\varphi_1(A)^2 \neq \varphi_2(A)^2$), the inequality is strict:
$$ \varphi(A^2) > 0$$
But $\varphi(A) = 0$, so $\Delta_\varphi(A) = \varphi(A^2) > 0$, contradicting the hypothesis. Thus, $\varphi$ must be pure.
\end{proof}

In other words, the previous proposition states that if $\nu_{\varphi, A} \in \Pi_p$ for every $A \in \mathfrak{A}_{s.a.}$, then $\varphi$ is a pure state.
\\
Let us also recall that every element of a C*-algebra can be written as a linear combination of four self-adjoint elements of the algebra. Therefore,
$$ \left[ \Delta_\varphi(A) = 0 \quad \forall A \in \mathfrak{A}_{s.a.} \right] \ \iff \ \left[ \Delta_\varphi(A) = 0 \quad \forall A \in \mathfrak{A} \right]$$

\textit{Characterization of Deterministic States:}
\\
A state $\varphi$ is deterministic if and only if its spectral measure $\nu_{\varphi, A}$ is point-supported for all $A \in \mathfrak{A}_{s.a.}$, i.e., for each $A$, there exists $\lambda \in \mathbb{R}$ such that $\nu_{\varphi, A} = \delta_\lambda$:
$$ \Delta_\varphi(A) = 0 \quad \forall A \in \mathfrak{A}_{s.a.} \qquad \iff \qquad \nu_{\varphi, A} \in \Pi_p \quad \forall A \in \mathfrak{A}_{s.a.} $$

\subsection{Spectral Digression}\label{spectral-digre}
This section is intended for readers familiar with the Hilbert space formalism.  
We briefly recall some properties of self-adjoint operators and their spectral projections, in order to  compare  them with our notion of observables and spectral projections $\mathbf{1}_{\Delta}(a)$.  
The aim is to highlight  similarities and differences  between the two approaches, and to prepare the ground for the algebraic formulation.
\\

Let $A$ be a self-adjoint and unitary operator on a Hilbert space $\mathcal{H}$. We study the operator $\mathbf{1}_{\{\lambda\}}(A)$, the spectral projection associated with $\lambda$.  
\\ 
If $\mathbf{1}_{\{\lambda\}}(A) \neq 0$, then $\lambda \in \operatorname{Sp}(A)$ (the spectrum of $A$).
\\  
Conversely, if $\lambda \in \operatorname{Sp}(A)$, it may still happen that $\mathbf{1}_{\{\lambda\}}(A) = 0$.  
\\
If $\mathbf{1}_{\{\lambda\}}(A) \neq 0$, we have:  
$$ \mathbf{1}_{\{\lambda\}}(A) \mathcal{H} = \ker(\lambda I - A)$$  
Indeed, by the spectral theorem for normal operators, any $A \in B(\mathcal{H})$ can be written as:  
$$ A = \int_{\sigma(A)} z \, dE(z)$$ 
where $E$ is the spectral measure (a projection-valued measure on $\operatorname{Sp}(A)$). It follows that for any bounded Borel function $f$,  
$$f(A) = \int_{\sigma(A)} f(z) \, dE(z)$$ 

If $\Psi \in \mathcal{H}$ satisfies $A\Psi = \lambda \Psi$, then:  
$$ \mathbf{1}_{\{\lambda\}}(A) \Psi = \mathbf{1}_{\{\lambda\}}(\lambda) \Psi = \Psi$$ 
which implies:  
$$ \ker(\lambda I - A) \subset \mathbf{1}_{\{\lambda\}}(A) \mathcal{H}$$ 

For the reverse inclusion, take any $\Psi \in \mathcal{H}$. Then:  
$$ \left\langle \Psi \, \middle| \, (\lambda I - A) \mathbf{1}_{\{\lambda\}}(A) \Psi \right\rangle = \int_{\sigma(A)} (\lambda - z) \mathbf{1}_{\{\lambda\}}(z) \, d\nu_{\Psi,A}(z) = 0$$  
where $\nu_{\Psi,A}$ is the spectral measure associated with $\Psi$ and $A$. 
\\
This shows that:  
$$(\lambda I - A) \mathbf{1}_{\{\lambda\}}(A) = 0$$ 
and therefore:  
$$ \mathbf{1}_{\{\lambda\}}(A) \mathcal{H} \subset \ker(\lambda I - A)$$
\begin{remark}\upshape\label{spettropunpuro}  
If $\lambda$ is an eigenvalue of a self-adjoint operator $A$ on a Hilbert space, then the spectral projection $\mathbf{1}_{\{\lambda\}}(A)$ is the orthogonal projection onto the eigenspace $\ker(A - \lambda I)$:  
$$\mathbf{1}_{\{\lambda\}}(A) \mathcal{H} = \ker(A - \lambda I)$$ 
\end{remark}  
If $\mathbf{1}_{\{\lambda\}}(A) = 0$, then $\lambda$ is \textbf{not} an eigenvalue of $A$.

$$\star\star\star $$
Returning to the study of deterministic states (relation \eqref{autostatialgebr}). 
\\
Let $\varphi$ be a \textit{ normal state}  on $\mathfrak{A}$ such that  
$$ P(A \in \{\lambda \})_\varphi: = \nu_{A,\varphi}( \{\lambda\}) = 1$$
where $A \in \mathfrak{A}$ is a self-adjoint operator and $\lambda  \in Sp(A)$.  
\\
We then have two possibilities:
\begin{itemize}  
\item  $\varphi$ is  pure  if $\dim \ker(A - \lambda  I) = 1$.
\item $\varphi$ is a  mixture  if $\dim \ker(A - \lambda  I) > 1$.  
\end{itemize}
Proof of the pure/mixed state characterization:   
\\
Assume that $\varphi$ is a normal state on $\mathfrak{A}$ with density matrix representation  
$$\varphi(A) =  tr(\rho A), \quad \forall A \in \mathfrak{A}$$ 
where $\rho$ is a density operator of the form  
$$\rho = \sum_{j=1}^N p_j |\Psi_j\rangle \langle \Psi_j| $$  
with $\|\Psi_j\| = 1$ and $\{p_j\}$ a probability distribution ($p_j > 0$, $\sum_j p_j = 1$).  
\\
Suppose further that  
$$P(A \in \{\lambda\})_\varphi = \varphi\big(\mathbf{1}_{\{\lambda\}}(A)\big) = 1$$  
Expanding this expectation value, we obtain  
$$ \sum_{j=1}^N p_j \langle \Psi_j | \mathbf{1}_{\{\lambda\}}(A) \Psi_j \rangle = 1$$  
Since $\mathbf{1}_{\{\lambda\}}(A)$ is an orthogonal projection, we have  
$$ \langle \Psi_j | \mathbf{1}_{\{\lambda\}}(A) \Psi_j \rangle = \|\mathbf{1}_{\{\lambda\}}(A) \Psi_j\|^2$$  
and thus  
$$ \sum_{j=1}^N p_j \|\mathbf{1}_{\{\lambda\}}(A) \Psi_j\|^2 = 1$$  
Because $\sum_j p_j = 1$ and $\|\mathbf{1}_{\{\lambda\}}(A) \Psi_j\| \leq 1$, this equality holds  \textit{if and only if}   
$$\|\mathbf{1}_{\{\lambda\}}(A) \Psi_j\| = 1 \quad \text{for all } j$$  
Since $\|\Psi_j\| = 1$ and $\mathbf{1}_{\{\lambda\}}(A)$ is a projection, this implies  
$$\mathbf{1}_{\{\lambda\}}(A) \Psi_j = \Psi_j$$  
meaning $\Psi_j \in \ker(A - \lambda I)$ for all $j$.  
\\
Consequences:
\begin{itemize}  
\item[1.] Number of states $N$:
\\  
The $\{\Psi_j\}$ must be eigenvectors of $A$ with eigenvalue $\lambda$, so  
$$ N \leq \dim \ker(A - \lambda I)$$
\item [2.] Purity condition:
\\  
 - If $\dim \ker(A - \lambda I) = 1$, then $N = 1$ (only one $\Psi_j$ is possible), and $\varphi$ is pure.
\\  
 - If $\dim \ker(A - \lambda I) > 1$, then $N$ can be greater than 1, meaning $\varphi$ is a  mixture  of eigenstates.  
\end{itemize}
 $$ \star \star \star $$

Let us summarize the discussion up to this point:

Let $\varphi$ be a normal state on a C*-algebra $\mathfrak{A}$ and $A$ a self-adjoint element.
\\
As in the case of physical states, we define the following set:
$$ \mathcal{V}_\lambda(A) = \left\{ \varphi \in S(\mathfrak{A}) : P(A \in \{\lambda\})_\varphi = 1 \right\}$$
Unlike the case of physical states\footnote{See equation \eqref{autostatifisici} on \pageref{autostatifisici}.}, here we have:
$$\mathcal{V}_\lambda(A) \nsubseteq \mathrm{Ext}(S(\mathfrak{A}))$$
To summarize\footnote{$C^*(A)$ is the unitary C* algebra generated by $A$.}:
$$ \varphi \in \mathcal V_\lambda(A) \ \iff \ \Delta_\varphi(A) = 0 \ \iff \ \varphi_0 \text{ is a pure state on } C^*(A) $$

\begin{remark}\upshape
Let $\varphi$ be a normal state on a  C*-algebra $\mathfrak{A}$, we have:
$$ \nu_{A,\varphi} \in \Pi_p   \qquad \forall   A \in \mathfrak A_{s.a.}    \ \iff \ \Delta_\varphi(A) = 0   \qquad \forall   A \in \mathfrak A_{s.a.} $$
and by the Proposition \ref{puritdeterm}:
$$ [ \nu_{A,\varphi} \in \Pi_p    \qquad \forall   A \in \mathfrak A_{s.a.} ]  \ \Longrightarrow \   \varphi  \textit{ \ is a pure state} $$
If $\mathfrak{A}$ is an abelian algebra, the converse also holds, since in this case the pure states coincide with the characters, i.e., the multiplicative functionals.
\end{remark}

\subsubsection{Mathematical and Physical Eigenstates} \label{subsec:eigenstates}

We define $\omega \in \mathfrak{S}_a^k$ to be a $k$-eigenstate (or auto-state) with eigenvalue $\lambda$ relative to the observable $a$ if 
$P(a \in \{\lambda\})_\omega = 1$ and
$$\texttt{V}_\lambda^{k} (a) =   \left\{ \omega \in  \mathfrak{S}_a^k : P(a \in \{\lambda\})_\omega = 1 \right\}$$
Obviously, this not only fails to guarantee that 
$$ P(A \in \{\lambda\})_{\mathrm{J}^\natural (\omega)} = 1 \qquad \text{for all self-adjoint } A \in \mathfrak{A}$$
and thus that the normal state $\mathrm{J}^\natural (\omega)$ is deterministic, but it also does not ensure that it is pure, as established by the previous corollary. It could very well be a mixture with a density matrix 
\begin{equation}\label{matrixdens-A} 
\rho = \sum_{h=1}^N p_h |\Psi_h\rangle \langle \Psi_h|, \qquad \Psi_h \in \ker (\lambda I - \mathrm{J}(a)), \quad h=1,2,\ldots
\end{equation} 
In other words\footnote{We also recall that,
$$\pi_{\mathrm{J}^{\natural}(\omega)}(\mathfrak{A})' \neq \mathbb{C} I$$
Therefore, we may have
$$ \ker(\lambda I - \mathrm{J}(a)) = \mathbf{1}_{\{\lambda\}}(\mathrm{J}(a)) \mathcal{H}$$}
\begin{equation}\label{inclu-puro-stati-A}
 \mathrm{J}^{\natural}(\texttt{V}_\lambda^{k} (a)) \nsubseteq \mathrm{Ext}(S (\mathfrak{A})),
\end{equation}
Thus, if $\omega \in \texttt{V}_\lambda^{k} (a)$, then in general
$$ \mathrm{J}^{\natural}(\omega)(A) = \operatorname{tr}(\rho A) \qquad \forall A \in \mathfrak{A}$$
with $\rho$ being the density matrix given by \eqref{matrixdens-A}.
\\

So, what do the eigenstates $\Psi_h \in \ker (\lambda I - \mathrm{J}(a))$ represent physically? 
\\ 
If there exists a physical state $\omega_{h} \in \mathfrak{S}_a^k$ such that
\begin{equation}\label{matrixdens-B} 
 \mathrm{J}^{\natural}(\omega_h)(A) := \langle \Psi_h | A \Psi_h \rangle \qquad \forall A \in \mathfrak{A},
\end{equation}  
then by definition $\omega_h \in \texttt{V}_\lambda^{k} (a)$.  
Therefore, the mixed state $\mathrm{J}^{\natural}(\omega)$ takes the form:
$$ \mathrm{J}^{\natural}(\omega) = \sum_{h=1}^N p_h \mathrm{J}^{\natural}(\omega_h)$$
\textit{The problem is that relation \eqref{matrixdens-B} is not always true}; that is, the connection to a physical state of the laboratory system does not always exist.
\\

To summarize, from relation \eqref{inclu-puro-stati-A} we can say that in general
\begin{equation}\label{fundamental-rel}
\mathrm{J}^{\natural}(\mathrm{Ext}(\mathfrak{S}_a^k)) \nsubseteq \mathrm{Ext}(S (\mathfrak{A}))
\end{equation}
since
$$\texttt{V}_\lambda^{k} (a)\subset \mathrm{Ext}(\mathfrak{S}_a^k)$$
\begin{remark}\label{controesempio-eq-fond}
If $\omega\in \mathrm{Ext}(\mathfrak{S}_a^k)$, so that $p(\omega)=1$, it does not necessarily follow that $\mathrm{J}^{\natural}(\omega)$ is a pure state.  
\end{remark}

\subsection{Algebraic States and Operationally Realizable Density Matrices}

Let us consider the universal representation $(\pi_u, \mathcal{H}_u)$ of the algebra $\mathfrak{A}$.  
If $(\pi_\varphi, \mathcal{H}_\varphi)$ is the GNS representation of the state $\varphi \in S(\mathfrak{A})$, we have by definition:  
$$\pi_u : \mathfrak{A} \longrightarrow B(\mathcal{H}_u), \qquad \mathcal{H}_u = \bigoplus_{\varphi \in S(\mathfrak{A})} \mathcal{H}_\varphi, \qquad \pi_u = \bigoplus_{\varphi \in S(\mathfrak{A})} \pi_\varphi$$  
For every $\varphi \in S(\mathfrak{A})$, we define $D_\varphi$ as the following subset of $B(\mathcal{H}_u)$:\index{$D_\varphi$}  
the set of density matrices $\hat{\rho}$ in $B(\mathcal{H}_u)$ such that
\begin{equation}\label{traccia2}
 \varphi(A) = \operatorname{tr} (\hat{\rho} \, \pi_u(A)) \qquad \forall A \in \mathfrak{A}
\end{equation}
We recall that the  representation $(\pi_o, \mathcal{H}_o)$ is  equivalent to a subrepresentation of $(\pi_u, \mathcal{H}_u)$ if  there exists an isometry $V : \mathcal{H}_o \longrightarrow \mathcal{H}_u$ such that  
$$ V \pi_o(A) = \pi_u(A) V \qquad \forall A \in \mathfrak{A} $$
Meanwhile, $(\pi_o, \mathcal{H}_o)$ is a subrepresentation of $(\pi_u, \mathcal{H}_u)$ if there exists a projector $P \in \pi_u(\mathfrak{A})'$ such that  
$$ \pi_o(A) = \pi_u(A)|_{\mathcal{K}} \qquad \forall A \in \mathfrak{A} $$  
where $\mathcal{K} = P \mathcal{H}_u$, and the isometry $V$ is simply the embedding $\mathcal{H}_o \hookrightarrow \mathcal{H}_u$. 
 \\

Let us recall the following fact:  
\begin{lemma}\upshape
Let $V : \mathcal{K} \to \mathcal{H}$ be any isometry between Hilbert spaces. For every $A \in B(\mathcal{H})$, we have:  
$$ \operatorname{tr}(A) = \operatorname{tr}(V A V^*) $$
and  
$$
\operatorname{tr}(A^2) = \operatorname{tr}((V A V^*)^2)$$
\end{lemma}
\begin{proof}
Let $\{e_\alpha\}_\alpha$ be an orthonormal basis for the Hilbert space $\mathcal{K}$. Since $V$ is an isometry, $\{V e_\alpha\}_\alpha$ are orthonormal vectors in $\mathcal{H}$. They can be extended to an orthonormal basis $\{f_\beta\}_\beta$ of $\mathcal{H}$ via Gram–Schmidt, with $f_\alpha = V e_\alpha$ for every $\alpha$.  
Then:  
$$
\operatorname{tr}(V A V^*) = \sum_{\beta} \langle f_\beta | V A V^* f_\beta \rangle = \sum_{\alpha} \langle e_\alpha | A e_\alpha \rangle = \operatorname{tr}(A)$$
since  
$$
V^* f_\beta = 
\begin{cases}
e_\alpha & \text{if } \beta = \alpha, \\
0 & \text{if } \beta \neq \alpha.
\end{cases}
$$ 
Moreover,  
$$
\operatorname{tr}(A^2) = \operatorname{tr}(V A^2 V^*) = \operatorname{tr}((V A V^*)^2),
$$ 
because  
$$
(V A V^*)^2 = V A V^* V A V^* = V A^2 V^*.
$$
\end{proof}
Let $\pi_o \ll \pi_u$ and let $\rho \in B(\mathcal{H}_o)$ be a density matrix such that
$$ \varphi(A) = \operatorname{tr} ( \rho  \, \pi_o(A)) \qquad \forall A \ \in \mathfrak{A}$$
From the previous lemma, we can state that
$$ \varphi(A) = \operatorname{tr} ( V \rho V^*  \, \pi_u(A)) \qquad \forall A \in \mathfrak{A}$$
where $V:\mathcal{H}_o \rightarrow \mathcal{H}_u$ is the isometry that intertwines the two representations.
\\
Indeed, from the previous lemma we have:
$$ \operatorname{tr} ( V \rho V^*  \, \pi_u(A)) = \operatorname{tr} ( V \rho  \, \pi_o(A)V^*) = \operatorname{tr} ( \rho  \, \pi_o(A)).$$
It follows that $V \rho V^* \in D_\varphi$.
\\

We define $S(\varphi)$, the entropy of the state $\varphi$, as: \index{$S(\varphi)$} \index{Entropy of the algebraic state} 
$$S(\varphi) = \sup \left\{ S(\rho) : \rho \in D_\varphi \right\} $$  
As in the previous case regarding physical states, the set $D_\varphi$ is indecomposable if it contains only rank-1 density matrices.  
Moreover, the following holds:  
$$ \varphi \text{ is a pure state } \iff D_\varphi \text{ is indecomposable } \iff S(\varphi) = 0 $$  
Indeed, as before (see page \pageref{Operationally-Realizable-Matrices}), if $\varphi$ is not pure and is a mixture of two states $\varphi_1, \varphi_2$:  
$$ \varphi = t \varphi_1 + (1-t) \varphi_2 $$  
we can write (via the GNS representations of the individual states\footnote{After embedding the Hilbert spaces $\mathcal{H}_{\varphi_1}, \mathcal{H}_{\varphi_2}$ into $\mathcal{H}_u$.}):  
$$\rho = t | \Omega_{\varphi_1} \rangle \langle \Omega_{\varphi_1} | + (1-t) | \Omega_{\varphi_2} \rangle \langle \Omega_{\varphi_2} | \in B(\mathcal{H}_u)$$  
Thus, the set $D_\varphi$ will not be indecomposable.  
\\
An important role in our discussion is also played by the following set of density matrices, denoted $ D_\varphi(A) $ for $ A \in \mathfrak{A}_{s.a.} $:
the set of density matrices $\hat{\rho}$ in $B(\mathcal{H}_u)$ such that
\begin{equation}\label{traccia2bis}
\varphi(f(A)) = \operatorname{tr} (\hat{\rho}  \, \pi_u(f(A))) \qquad \forall f \in C_o(\mathbb{R}).
\end{equation}

Obviously, for every $ A \in \mathfrak{A}_{s.a.} $, we have the following inclusion:
$$ D_\varphi \subset D_\varphi(A)$$
Furthermore, it is easy to verify that we have the following set equality:
$$D_\varphi = \bigcap_{A \in \mathfrak{A}_{\ s.a.}} D_\varphi(A)$$ 
\begin{definition}[\textbf{A-Pure States}]\upshape
If $ D_\varphi(A) $ is indecomposable, i.e., it consists only of rank-1 density matrices, the functional $\varphi$ is said to be pure on $A$.
\end{definition}

We have the obvious implication: if there exists an element $ A \in \mathfrak{A}_{s.a.} $ such that $ D_\varphi(A) $ is indecomposable, then $ D_\varphi $ is also indecomposable. Therefore,
$$ \text{If } \varphi \text{ is pure on } A \qquad \Longrightarrow \qquad \varphi \text{ is a pure state}$$
The converse is not true.
\\

Now, let $\varphi_o$ be the restriction of $\varphi$ to $C^*(A)$, the unital  C*-algebra generated by $A \in \mathfrak{A}$, and let $( \pi^o_u , \mathcal{H}^o_u )$ be the universal representation of the algebra $C^*(A)$.
\\
There is an obvious embedding $\texttt{e}: \mathcal{H}^o_u \hookrightarrow \mathcal{H}_u$.
\\
If $\rho_o \in D_{\varphi_o} \subset B(\mathcal{H}^o_u)$, then $\texttt{e} \rho_o \texttt{e}^* \in D_\varphi(A)$.
\\
Indeed, for every $f \in C_o(\mathbb{R})$ we have:
$$\operatorname{tr} ( \texttt{e} \rho_o \texttt{e}^* \pi_u(f(A))) = \operatorname{tr} ( \texttt{e} \rho_o \pi^o_u(f(A)) \texttt{e}^*) = \operatorname{tr} ( \rho_o \pi^o_u(f(A))) = \varphi(f(A))$$
The commutative C*-algebra $C^*(A)$ is generated by all polynomials in $A$, so it suffices to consider continuous functions defined as polynomials on the spectrum $\text{Sp}(A)$, extended to functions in $C_o(\mathbb{R})$.
\\
Therefore,
$$\texttt{e} D_{\varphi_o} \texttt{e}^* \subset D_\varphi(A)$$
Hence, if $D_\varphi(A)$ is \textit{indecomposable}, it implies that $D_{\varphi_o}$ is also \textit{indecomposable} meaning $\varphi_o$ is a pure state (and thus a character) of $C^*(A)$.
\\
In this way, stating that $\varphi$ is pure on $A$ means:
\begin{equation} 
 \varphi \ \textit{ is pure on $A$} \ \Longrightarrow \ 
 \begin{cases}  
 \varphi \ \textit{is a pure state,}  &   \\  
 \varphi_o \ \textit{is a pure state on $C^*(A)$}  &   \textit{then}  \  \nu_{A,\varphi} \in \Pi_p      
\end{cases}
\end{equation}
Thus, the notion of \textit{A-purity }is stronger than purity alone.
\\

Let us now ask: given $\omega \in \mathfrak{S}_a^k$, what is the relationship between the set of operationally realizable density matrices $\mathfrak{D}^k_{\omega,a}$ and the set $D_{\mathrm{J}^{\natural}(\omega)}$?
\\  
We examine the relationship between the GNS representation $(\pi_\omega, \mathcal{H}_\omega, \Omega_\omega)$ of $\mu_{\omega,a} \in C_o(\mathbb{R})^*$ and the GNS representation $(\pi_\varphi, \mathcal{H}_\varphi, \Omega_\varphi)$ of the state $\varphi = \mathrm{J}^{\natural}(\omega) \in S(\mathfrak{A})$.  
\\
We define the following operator $W_\omega : \mathcal{H}_\omega \longrightarrow \mathcal{H}_\varphi$:  
$$ W_\omega \pi_\omega(f) \Omega_\omega = \pi_\varphi(f(\mathrm{J}(a))) \Omega_\varphi \qquad \forall f \in C_o(\mathbb{R})$$  
It is easy to verify that:  
\begin{itemize}
\item $W_\omega$ is an isometry,
\item $W_\omega W_\omega^*$ is the projector onto the Hilbert subspace $\mathcal{K}_{\omega} = \overline{\pi_\varphi(C^*(\mathrm{J}(a)) \Omega_\varphi} \subset \mathcal{H}_\varphi$, where $C^*(\mathrm{J}(a))$ is the unital C*-algebra generated by  $\mathrm{J}(a)$.
\\  
Note that the algebra $C^*(\mathrm{J}(a))$ is commutative.
\item $W_\omega \pi_\omega(f) = \pi_\varphi(f(\mathrm{J}(a))) W_\omega$ for every $f \in C_o(\mathbb{R})$.
\end{itemize}
In this way, we can define a new operator $W : \mathcal{H}_{a,k} \longrightarrow \mathcal{H}_u$ as:  
$$ W = \bigoplus_{\omega \in \mathfrak S_a^{ \ k}} W_\omega, \qquad \mathcal{H}_{a,k} = \bigoplus_{\omega \in \mathfrak S_a^{ \ k}} \mathcal{H}_\omega$$  
with the following properties:  
\begin{itemize}
\item $W$ is an isometry, 
\item $W W^*$ is the projector onto the Hilbert subspace  
$$ \mathcal{K}_o = \bigoplus_{\omega \in \mathfrak S_a^{ \ k}} \mathcal{K}_{\omega} \subset \mathcal{H}_u$$
\item $W \pi_{a,k}(f) = \pi_o(f(\mathrm{J}(a))) W$ for every $f \in C_o(\mathbb{R})$, where  
$$ \pi_{a,k} = \bigoplus_{\omega \in \mathfrak S_a^{ \ k}} \pi_\omega, \qquad \pi_o = \bigoplus_{\varphi \in \mathrm{J}^\natural(\mathfrak S_a^{ \ k})} \pi_\varphi, \qquad \pi_o \ll \pi_u$$  
In particular, by construction, $(\pi_o, \mathcal{K}_o)$ is a subrepresentation of $(\pi_u, \mathcal{H}_u)$.
\end{itemize}

\begin{proposition}\upshape
If $\rho \in \mathfrak{D}^k_{\omega,a}$, then $W \rho W^* \in D_{\mathrm{J}^{\natural}(\omega)}(\mathrm{J}(a))$.
\end{proposition}
\begin{proof}
From the previous lemma, we obtain:  
$$\operatorname{tr}(\rho) = \operatorname{tr}(W\rho W^*)$$
   
Now, by hypothesis, for every function $f \in C_o(\mathbb{R})$, we have:  
\begin{equation}\label{tr-A}
\mu_{\omega,a}(f) = \operatorname{tr}(\rho \, \pi_{a,k}(f)) = \mathrm{J}^{\natural}(\omega)(f(\mathrm{J}(a))).
\end{equation}
We have the following equality:  
$$\operatorname{tr}(W \rho W^* \, \pi_o(f(\mathrm{J}(a)))) = \operatorname{tr}(W \rho \, \pi_{a,k}(f) W^*)$$
and by the previous lemma:  
$$\operatorname{tr}(W \rho \, \pi_{a,k}(f) W^*) = \operatorname{tr}(\rho \, \pi_{a,k}(f)) \qquad \forall f \in C_o(\mathbb{R})$$
From equation \eqref{tr-A}:  
$$\mathrm{J}^{\natural}(\omega)(f(\mathrm{J}(a))) = \operatorname{tr}(W \rho W^* \, \pi_o(f(\mathrm{J}(a)))) \qquad \Longrightarrow \qquad W\rho W^* \in D_{\mathrm{J}^{\natural}(\omega)}(\mathrm{J}(a))$$
\end{proof}
Hence, for every observable $a$
$$ W \mathfrak{D}^k_{\omega,a} W^* \subset D_{\mathrm{J}^{\natural}(\omega)}(\mathrm{J}(a))$$
We emphasize that $W$ depends on $a$.
\\
%
%Attenzione
\begin{proposition}\upshape
If $\mathrm{J}^{\natural}(\omega)$ is pure, then $\omega \in \operatorname{Ext}(\mathfrak{S}_a^k)$\footnote{We have seen that the converse is not always true, and that  this state is not always deterministic.}:  
$$\mathrm{J}^\natural \left( \mathfrak{S}_a^k \right) \cap \operatorname{Ext} S(\mathfrak{A}) \subset \mathrm{J}^\natural \left( \operatorname{Ext}(\mathfrak{S}_a^k) \right)$$
\end{proposition}
\begin{proof}
By hypothesis, $\mathrm{J}^{\natural}(\omega) \in \operatorname{Ext} S(\mathfrak{A})$. Assume by contradiction that $\mathrm{J}^{\natural}(\omega) \notin \mathrm{J}^\natural \left( \operatorname{Ext}(\mathfrak{S}_a^k) \right)$.  
Then there exists at least one density matrix $\rho \in \mathfrak{D}^k_{\omega,a}$ with rank greater than 1, hence with purity $\operatorname{tr}(\rho^2) < 1$.  
By the previous proposition, there corresponds a density matrix $W_\omega \rho W_\omega^* \in D_\varphi$, and moreover:  
$$
\operatorname{tr}((W_\omega \rho W_\omega^*)^2) = \operatorname{tr}(W_\omega \rho^2 W_\omega^*) = \operatorname{tr}(\rho^2) < 1$$
It follows that $\mathrm{J}^{\natural}(\omega)$ cannot be pure.
\end{proof}

As discussed, the sectors $\mathrm{J}^{\natural }\left( \mathfrak S_a^k \right)$  may be non convex and not closed and non-disjoint intersections. This makes the problem highly non-trivial.  Based on the previous considerations, it becomes necessary to relax the requirements on the C*-algebraization of our physical laboratory system:
\begin{definition}[Purity-Preserving C*-Algebraization] \upshape\label{Purity-Preserving} 
The C*-algebraic realization $(\mathfrak{A}, \mathrm{J}, \mathrm{J}^\natural)$ preserves purity  if the following inclusion holds:
$$ \mathrm{J}^\natural \left( \mathrm{Ext}(\mathfrak{S}_a^k) \right) \subset \mathrm{Ext}(C(k,a)) \subset S(\mathfrak A)$$
where:
\\  
1. $C(k,a)$ is the Convex Hull, the  minimal convex subset of $ S(\mathfrak{A})$ containing $\mathrm{J}^\natural(\mathfrak{S}_a^k)$:
  $$   C(k,a) := \bigcap \left\{ C \subseteq S(\mathfrak{A}) \,\Big|\, \text{$C$ convex and } \mathrm{J}^\natural(\mathfrak{S}_a^k) \subset C \right\}$$
2. $\mathrm{Ext}(C(k,a))$ denotes the extremal state, i.e.,  extreme boundary (pure states) of $C(k,a)$.
\end{definition}
\begin{attenzione}\upshape
Even when a C*-algebraization satisfies the aforementioned property, this does \emph{not} guarantee that for every $\omega \in \mathrm{Ext}(\mathfrak{S}_a^k)$, the GNS representation associated to the functional $\mathrm{J}^\natural(\omega)$ will be irreducible.
\\
More precisely, the GNS representation:$$ \pi_{\mathrm{J}^{\natural}(\omega)}:\mathfrak A \longrightarrow \mathfrak B(\mathcal H_{\mathrm{J}^{\natural}(\omega)})$$
may remain reducible, since in general $\mathrm{J}^{\natural}(\omega)\notin \mathrm{Ext}(S(\mathfrak A))$\footnote{There is no general guarantee that $\mathrm{Ext}(C(k,a)) \subset \mathrm{Ext}(S(\mathfrak{A}))$.}.
\end{attenzione}
\begin{property}[Sectorial Rules for States] \index{Sectorial Rules for States  }
The C*-algebraic representation $(\mathfrak{A}, \mathrm{J}, \mathrm{J}^\natural)$ of a physical system $(\mathfrak{X}, \mathfrak{S})$ satisfies the Sectorial Rules for States if:  
\begin{itemize}  
\item[a.] For every observable $a$ and its measurement sector $k$, the set $\mathrm{J}^\natural \left( \mathfrak{S}_a^k \right)$ is a \textit{convex subset} of $S(\mathfrak{A})$.  
\item[b.] The family $\left\{ \mathrm{J}^\natural \left( \mathfrak{S}_a^k \right) \right\}_k$ forms a disjoint partition of $\mathrm{J}^\natural \left( \mathfrak{S}_a \right)$ for every observable $a$.\footnote{This holds trivially if the mapping $\mathrm{J}^\natural : \mathfrak{S}_a \rightarrow S(\mathfrak{A})$ is injective.}  
\end{itemize}  
\end{property}

\section{Guidelines for Construct a C*-algebraic framework II}
How can we associate a density matrix $\rho$ with the physical state of the laboratory $\omega \in \mathfrak{S}$ such that
\begin{equation}\label{identit}
  \mathrm{J}^{\natural}(\omega) (\mathrm{J}(a)) = \operatorname{tr}(\rho  \ \mathrm{J}(a)) \qquad \text{for all } a \in \mathfrak{X}_\omega?
\end{equation}
The answer to this question is not at all straightforward. In the literature, similar arguments are addressed using methods from \textit{Quantum Tomography}. The author of these notes, after reading some key references on the subject, does not believe the problem has been solved. What is presented here is an attempt to frame the problem, which will require further in-depth investigation in the future.
\\
\textit{First}, it should be noted that the identity given in equation \ref{identit} holds only on the subset $\mathrm{J}(\mathfrak{X}_\omega) \subset \mathfrak{A}$ and not on the entire algebra.
\\
\textsl{Second}, from an experimental standpoint, the set $\mathfrak{X}_\omega$ is \textit{finite cardinality}. Indeed, claiming that we can measure infinitely many distinct observables for a given preparation in $\omega$\footnote{Of course, we consider the set of states to be infinite; the human mind possesses the infinite capacity to imagine infinite preparations and thus system states, or so one hopes.} is not experimentally feasible.
\\
\textit{Third}, the experimenter has already chosen the Hilbert space $\mathcal{H}$ and the relevant algebra, as discussed in the initial phase of the algebraic construction.
\\
Therefore, once the Hilbert space $\mathcal H$ is fixed, we must determine a normal state on $B(\mathcal H)$ that satisfies equation \ref{identit}.
\\
We are not interested in the behaviour of the state $  \mathrm{J}^{\natural}(\omega)$ outside the C*-algebra generated by the set $\mathrm{J}(\mathfrak{X}_\omega)$, as it has no experimental validity.
\\
Let us briefly denote by $\mathcal P(\omega) $ the unital C*-algebra generated by $\mathrm{J}(\mathfrak{X}_\omega)$ and the unit of the algebra.\index{$\mathcal P(\omega) $}
\\
We choose a sequence of observables $ \mathcal F_n=\left\{x_1, x_2, \ldots, x_n \right\}$ from $\mathfrak X_\omega$, and consider the algebra generated by the family $ \left\{I , \mathrm{J}(x_1),\mathrm{J}( x_2) , \ldots,  \mathrm{J}(x_n)  \right\}$:
$$\mathfrak A (\mathcal F_n) \subset \mathcal P( \omega ) \subset \mathfrak A.$$
Now, we fix a family of orthonormal vectors in the Hilbert space $\left\{\Psi^n_1 , \Psi^n_2,  \ldots, \Psi^n_n \right\}$ and determine a matrix $\rho_n$ such that:
$$ \langle  x_k\rangle =   \operatorname{tr}(\rho_n\mathrm{J}(x_k))  \qquad \forall  k=1,2,\ldots, n $$
where
$$ \rho_n = \sum_{k=1}^n  \lambda^n_k  \  | \Psi^n_k   \rangle \langle \Psi^n_k  | $$
considering the $\lambda^n_k$ as unknowns with the constraint $\sum_{k=1}^n  \lambda^n_k=1$.
\\
It follows that
$$ \operatorname{tr}(\rho_n\mathrm{J}(x_k) ) =  \sum_{h=1}^n  \lambda^n_h  \  \langle \Psi^n_h  | \mathrm{J}(x_k)\Psi^n_h   \rangle = \sum_{h=1}^n  \lambda^n_h c_{ h,k}  $$
in other words,
$$ y_k = \langle  x_k\rangle =  \sum_{h=1}^n  \lambda^n_h c_{ h,k}$$
where $y_k$ and $c_{ h,k}$ are known numbers. In vector form:
 \begin{equation}\label{eq:system}
 \mathbf y^n  =
 \mathbf C_n \mathbf{ \lambda}^n,
 \quad \text{where } \mathbf C_n = \begin{bmatrix}
c_{1,1} & c_{1,2} & \cdots & c_{1,n} \\
c_{2,1} & c_{2,2} & \cdots & c_{2,n} \\
\vdots & \vdots & \ddots & \vdots \\
c_{n,1} & c_{n,2} & \cdots & c_{n,n}
\end{bmatrix}, \
\mathbf{\lambda}^n = \begin{bmatrix}
\lambda_1^n  \\ \lambda_2^n  \\ \vdots \\ \lambda_n^n
\end{bmatrix}.
\end{equation}
If the matrix $\mathbf C_n$ is invertible, we have
$$ \mathbf C_n^{-1}  \mathbf y^n = \mathbf{ \lambda}^n $$
and thus the solution.
\begin{problem}\upshape\label{pb-A1}
The invertibility of the matrix $\mathbf C_n$ depends on the arbitrary choice of the family of orthonormal vectors $\left\{\Psi^n_1 , \Psi^n_2,  \ldots, \Psi^n_n \right\}$. Therefore, choosing this family (besides verifying its existence) is a critical point of this method, a point to be analyzed in detail, which we will not do in these notes.
\end{problem}
We repeat the procedure by adding an observable $x_{n+1}$:
$$  \mathcal F_{n+1} =\left\{x_1, x_2, \ldots, x_n, x_{n+1} \right\} $$
and obtain the new equation
$$ \mathbf C_{n+1}^{-1}  \mathbf y^{n+1} = \mathbf{ \lambda}^{n+1} $$
where, by definition,
$$ y^{n+1}_j = y^{n }_j \qquad \forall j=1,2,\ldots, n$$
since \textit{$x^{n}_{j}$ is the same observable in both sets $\forall j=1,2,\ldots, n$.}
\\
In this way, we obtain a family of density matrices $\left\{ \rho_n \right\}$, which are operators in the Hilbert space $L^2(\mathcal H)$ with the scalar product induced by the $L^2$-norm:
$$ \langle T | X  \rangle_2 = \operatorname{tr}(T^* X  )  \qquad \forall  T, X \in L^2(\mathcal H) $$
Since the family of density matrices $ \left\{ \rho_n \right\}$ are vectors in the Hilbert space $L^2(\mathcal H)$, it has a limit point; thus, there exists a subnet $\left\{ \rho_{n_\alpha} \right\}$ convergent in the weak topology:
$$  \langle T |  \rho_{n_\alpha}   \rangle_2 \longrightarrow  \langle T |  \rho_{0}   \rangle_2   \qquad \forall  T \in L^2(\mathcal H).$$
If $x\in\mathfrak X_\omega$, the operator $\mathrm{J}(x)$ is not necessarily in $L^2(\mathcal H)$. However, since the equation holds for any $T \in L^2(\mathcal H)$, we can use the projector $ \mathbf{1}_{\{\Delta\}}(\mathrm{J}(x))$, which belongs to this space because it is a projector. Therefore,
$$  \langle \mathbf{1}_{\{\Delta\}}(\mathrm{J}(x)) |  \rho_{n_\alpha}   \rangle_2 \longrightarrow  \langle \mathbf{1}_{\{\Delta\}}(\mathrm{J}(x))|  \rho_{0}   \rangle_2  \qquad \forall x\in \mathfrak X_\omega $$
\\
Note that in our case, by how we have chosen the observables,
$$P(x_{j}\in \Delta )_\omega=  P(x_{n}\in \Delta )_\omega \qquad \forall  \  1\leq j\leq n, \text{ for all } n$$
and therefore, in general, for any $m \leq n$,
$$P(x^{ }_{j}\in \Delta )_\omega=  P(x^{ }_{m}\in \Delta )_\omega \qquad \forall  \ 1\leq j\leq m.$$
Since there exists an $\alpha_0$ such that $n_{\alpha_0} \geq m$\footnote{Recall that the subnet $\left\{ n_\alpha \right\}$ is increasing.}, we can assert that for every $\alpha \geq \alpha_0$,
$$P(x_{j}\in \Delta )_\omega = \operatorname{tr}(\rho_{m}  \mathbf{1}_{\{\Delta\}}(\mathrm{J}(x_{j})) ) \quad \text{for } 1\leq j\leq  m $$
It follows that
$$ \operatorname{tr}(\rho_{m}  \mathbf{1}_{\{\Delta\}}(\mathrm{J}(x_{j})) ) = \operatorname{tr}(\rho_{n_\alpha}  \mathbf{1}_{\{\Delta\}}(\mathrm{J}(x_{j})) )  $$
and therefore,
$$P(x_{j}\in \Delta )_\omega=    \operatorname{tr}(\rho_{n_\alpha}  \mathbf{1}_{\{\Delta\}}(\mathrm{J}(x_{j})) ) \longrightarrow \langle \mathbf{1}_{\{\Delta\}}(\mathrm{J}(x_{j}))|  \rho_{0}\rangle_2   \qquad \forall    1\leq j\leq m $$
In other words, for every $x\in \cup_n \mathcal F_n$ we have
$$ P(x\in \Delta )_\omega =\operatorname{tr}(\rho_{n_\alpha}  \mathbf{1}_{\{\Delta\}}(\mathrm{J}(x)) )\longrightarrow  \operatorname{tr}(\rho_0 \mathbf{1}_{\{\Delta\}}(\mathrm{J}(x)) ),$$
and thus,
$$ \mathrm{J}^{\natural}(\omega)(\mathbf{1}_{\{\Delta\}}(x)) = \operatorname{tr}(\rho_0 \mathbf{1}_{\{\Delta\}}(\mathrm{J}(x)) )   \qquad \forall x\in \cup_n \mathcal F_n$$

We still have other problems to consider:
\begin{itemize}
\item[--] The first is that we do not know if $\cup_n \mathrm{J}(\mathcal F_n)$ generates the entire algebra $\mathcal P(\omega)$.
\item[--] The second is that the initial net of density matrices may admit more than one limit point $\rho_0$. We propose to select this limit point based on the properties of our state $\omega$. Specifically, we would choose the $\rho_0$ that satisfies:
$$p(\omega)= \operatorname{tr} \rho_0^2 = \left\|\rho_0 \right\|^2_2 $$
i.e., the one that matches the purity. Furthermore, it should also satisfy the corresponding condition for the von Neumann entropy:
$$ S(\omega )= S(\rho_0) $$
\end{itemize}
Thus, our candidate matrix is constrained by these two equations.
\begin{attenzione}\upshape
As previously stated, the set $\mathfrak X_\omega$ is always finite, regardless of its size. Therefore, the first question presents a false theoretical problem, though it remains a technical one. Moreover, this finiteness also addresses the issue raised in problem \ref{pb-A1}.

The second question is more difficult to resolve. One should verify that there exists at least one limit point satisfying these two conditions, a task we will not undertake here. In fact, as we have seen, purity is not always preserved in the transition to the algebraic framework. Consequently, we are forced to make a mathematically convenient choice: to select the density matrix with the lowest possible purity index (even if it does not match the true purity of our physical state) or, alternatively, the one with the highest von Neumann entropy. In both cases, this leads to a discrepancy with physical reality.
\end{attenzione}

\section{Spectrum of Observables and C*-algebraization}
We will analyze the spectral types of observables and self-adjoint operators, along with their comparison.
\\
We have established that for a good algebraization, the spectral relation
$$ \sigma(a)= Sp(\mathrm{J} (a)) $$ 
must remain preserved for every observable $a\in\mathfrak X $. However, this relation alone\textit{ is not yet sufficient} to fully characterize our algebraization.
\\
Let $a$ be a non null observable of the system. On page \pageref{spettrolambda}, we defined the set $\sigma_{pd}(a)$ as the set of isolated points of the spectrum of $a$ and established the following equivalence:
\\
$\lambda\in \sigma_{pd}(a)$ if and only if $\lambda\in\sigma(a)$  and there exists an open neighborhood  $U_\lambda$  such that  $U_\lambda\cap\sigma(a)\neq\emptyset$\footnote{That is,  $\lambda$ is an isolated point of the spectrum}.
\\
Furthermore, we recall the implication:
$$\mathbf{1}_{ \left\{  \lambda   \right\}}(a)\neq 0 \qquad   \Longrightarrow \qquad \lambda \in \sigma(a)$$ 
whose converse is not true in general:
$$ \textit{ If }  \  \lambda \in \sigma(a) ,  \  \textit{it does not necessarily follow that} \  \mathbf{1}_{ \left\{  \lambda   \right\}}(a)\neq 0 $$
We now define the following new subsets of the spectrum:\index{Pure point spectrum od an observable}\index{Continuous spectrum od an observable}
\begin{itemize}
\item \textit{Pure point spectrum}:
$$ \sigma_{pp}(a) = \left\{  \lambda\in\sigma(a) \ : \   \mathbf{1}_{ \left\{  \lambda   \right\}}(a)\neq 0 \right\} $$
\item \textit{Continuous spectrum}:
$$ \sigma_{c}(a) = \left\{  \lambda\in\sigma(a) \ : \   \mathbf{1}_{ \left\{  \lambda   \right\}}(a)= 0 \right\} $$
\end{itemize}
Consequently, the spectrum decomposes into the following disjoint union:
$$  \sigma (a)=  \sigma_{pp}(a)\cup \sigma_{c}(a)$$
and
$$ \sigma_{pd}(a)\subset\sigma_{pp}(a)$$
We recall, that the spectrum of a  self-adjoint operator $A\in B(\mathcal H)$  satisfies the following relation\footnote{See Reed Simon \cite{R_S}, Chapter VII.3.
  
See also  the equation \eqref{puntispettro-2} on page \pageref{puntispettro-2}.  }:
$$Sp(A)= \left\{  \lambda\in \mathbb C  :     \mathbf{1}_{ ]\lambda-\epsilon ,\lambda+\epsilon[}(A)    \neq 0 \ \textit{ for any } \epsilon >0\right\}$$
This spectrum contains various types of spectral points (e.g., pure point spectrum, continuous and essential spectrum), all of w\textit{hich must be preserved }for the spectrum of an observable under algebraization.
\\
Let's see the fundamental spectral properties for self-adjoint operators in $B(\mathcal H)$:
\begin{itemize}
\item  The Pure Point spectrum $Sp_{p}(A)$,    consists of all $\lambda\in \mathbb C$ for which $A-\lambda I$  is not one-to-one:
$$Sp_{p}(A)= \left\{  \lambda\in Sp(A) :  \ker(A-\lambda I)\neq 0 \right\}$$
If $\mathcal H$ is separable, then $Sp_{p}(A)$  at most countable.
\item  The Continuous spectrum $Sp_{c}(A)$, consists of all $\lambda\in \mathbb C$  such that $A-\lambda I$ is a one-to-one mapping of $\mathcal H$ onto a dense proper subspace of $\mathcal H$:
\item  For self-adjoint operators, the spectrum partitions as:
$$ Sp(A)= \overline{Sp_{p}(A)} \cup  Sp_{c}(A) \qquad , \qquad Sp_{p}(A) \cap  Sp_{c}(A)=\emptyset  $$
where $\overline{Sp_{p}(A)}$  denotes the closure of the point spectrum.
\end{itemize}
As observed on page \pageref{spettropunpuro}, we can therefore state that:
\begin{itemize}
\item   $Sp_{p}(A)= \left\{  \lambda\in Sp(A) :   \mathbf{1}_{  \left\{\lambda \right\}}(A)\neq 0 \right\}$ 
\item  $Sp_{c}(A)= \left\{  \lambda\in Sp(A) :   \mathbf{1}_{  \left\{\lambda \right\}}(A) =  0 \right\}$
\end{itemize}
This leads us to the following proposition:

\begin{proposition}\label{spettro-alg.}
Let $(\mathfrak{M},\mathrm{J}, \mathrm{J}^{\natural})$ be a von Neumann algebrization of a physical system $\left( \mathfrak{X,S}\right)$. For every non-zero observable $a$, we have
$$ \sigma_{\mathrm{pp}}(a) \subset \mathrm{Sp}_{\mathrm{p}}(\mathrm{J}(a))$$
If the algebrization satisfies the following property\footnote{For instance, if it satisfies \texttt{ASSP}.}:
$$  \mathbf{1}_\Delta(\mathrm{J} (a)) = \mathrm{J} (\mathbf{1}_\Delta(a)) \quad \text{for every Borel set } \Delta$$
then
$$
\sigma_{\mathrm{pp}}(a) = \mathrm{Sp}_{\mathrm{p}}(\mathrm{J}(a)) \quad \Longrightarrow \quad \big[ \sigma_{\mathrm{c}}(a) = \mathrm{Sp}_{\mathrm{c}}(\mathrm{J}(a)) \big] $$
\end{proposition}

\begin{proof}
As previously discussed, we have the following equivalence:
$$ \mathbf{1}_{\{\lambda\}}(\mathrm{J} (a)) \neq 0 \quad \iff \quad \lambda \in \mathrm{Sp}_{\mathrm{p}}(\mathrm{J}(a))$$ 
If $\lambda \in \sigma_{\mathrm{pp}}(a)$, by definition $\mathbf{1}_{\{\lambda\}}(a) \neq 0$.
Thus, there exists at least one state $\omega_0$ of the laboratory system such that
$$\mu_{\omega_0, a} \{\lambda\} = \langle \mathbf{1}_{\{\lambda\}}(a) \rangle_{\omega_0} \neq 0$$
Since for any Borel set $\Delta \in B(\mathbb R)$ and any $\omega \in \mathfrak{S}_a$, we have:
\begin{equation}\label{eq:key_identity}
\mathrm{J}^{\natural}(\omega)(\mathrm{J} (\mathbf{1}_\Delta(a))) = \mathrm{J}^{\natural}(\omega)(\mathbf{1}_\Delta(\mathrm{J} (a))) 
\end{equation}
it follows that also
$$ \mathrm{J}^{\natural}(\omega)(\mathbf{1}_{\{\lambda\}}(a)) \neq 0 \quad \Longrightarrow \quad \sigma_{\mathrm{pp}}(a) \subset \mathrm{Sp}_{\mathrm{p}}(\mathrm{J}(a))$$
Now, if $\lambda \in \mathrm{Sp}_{\mathrm{p}}(\mathrm{J}(a))$, it follows that $\mathbf{1}_{\{\lambda\}}(\mathrm{J} (a)) \neq 0$.
Thus, there exists a state $\varphi$ of the algebra such that $\varphi(\mathbf{1}_{\{\lambda\}}(\mathrm{J} (a))) \neq 0$. However, this state is not necessarily derived from a physical state $\omega$, i.e., $\varphi = \mathrm{J}^{\natural}(\omega)$, so we cannot use identity \eqref{eq:key_identity}.

In this case, we must assume a stronger hypothesis and assert that
$$\mathrm{J}(\mathbf{1}_{\{\lambda\}}(a)) = \mathbf{1}_{\{\lambda\}}(\mathrm{J} (a))$$
Then, using the embedding property:
$$ \| \mathrm{J}(\mathbf{1}_{\{\lambda\}}(a)) \| = \| \mathbf{1}_{\{\lambda\}}(a) \| \quad \Longrightarrow \quad \| \mathbf{1}_{\{\lambda\}}(a) \| \neq 0 \quad \Longrightarrow \quad \lambda \in \sigma_{\mathrm{pp}}(a)$$
\end{proof}
To conclude this spectral discussion, recall that we have a second decomposition of the spectrum $Sp(A)$ of a self-adjoint operator:
\begin{itemize}
\item [$\circ$ ] The Essential spectrum $Sp_{ess}(A)$:
$$Sp_{ess}(A) = \left\{  \lambda\in Sp(A) :    \dim [ \mathbf{1}_{ ]\lambda-\epsilon ,\lambda+\epsilon[}(A)\mathcal H ]= \infty \  , \ \forall \epsilon >0    \right\}$$
The essential spectrum $Sp_{ess}(A)$ is always closed.
\item [$\circ$ ] The Discrete  spectrum $Sp_{dis}(A)$:
$$Sp_{dis}(A) = \left\{  \lambda\in Sp(A) :    \dim [ \mathbf{1}_{ ]\lambda-\epsilon ,\lambda+\epsilon[}(A)\mathcal H ]<\infty \  , \ \textit{ for some }  \  \epsilon >0  \right\}$$
 The discrete spectrum $Sp_{ess}(A)$ is not necessarily closed.
\item [$\circ$ ] For self-adjoint operators, the spectrum partitions as:
$$ Sp(A)=  Sp_{ess}(A)  \cup  Sp_{disc}(A) \qquad , \qquad Sp_{ess}(A)  \cap  Sp_{disc}(A) =\emptyset$$
\end{itemize}
We have the following statement:
\begin{equation} 
\lambda \in  Sp_{disc}(A)  \qquad \Longleftrightarrow  \qquad   \left\{   
\begin{array}{ccc}
\lambda \  \textit{is isolated point of spectrum}    \\ 
  \\
\dim \ker (\lambda I- A)<\infty      
\end{array}
\right.   
\end{equation}
\begin{attenzione}\upshape
Note that points in $Sp_{p}(A)$ are not necessarily isolated.
\end{attenzione}
In this case, we cannot simply transpose these definitions to the case of physical observables of the system, as these definitions make substantial use of spectral subspaces $[\mathbf{1}_{\Delta}(A)\mathcal{H}]$ which do not have a simple counterpart in the physical case.
\\
For example, we can introduce the following definition (which is similar to that given for self-adjoint operators on Hilbert spaces):
\\
Let $a$ be an observable and define its \textit{discrete spectrum}, denoted by $\sigma_d(a)$, as the set
\begin{equation}
\sigma_{\mathrm{disc}}(a) := \left\{ \lambda \in \sigma(a) : 
\begin{cases}
\lambda \text{ is an isolated point of the spectrum}, \\
\operatorname{card} V_\lambda(a) < \infty
\end{cases}
\right\} 
\end{equation}

\section{Equivalent Algebraic Representations}
Experimental information is intrinsically contained in the pair $(\mathfrak{X}, \mathfrak{S})$, which includes, for example, the spectrum of every observable of the system and its type. Therefore, the possible algebraic representations of our physical system must "reproduce" this information faithfully, making it mathematically more usable.
\\
For an algebrization  $(\mathfrak{A}, \mathrm{J}, \mathrm{J}^\natural)$ to be a good representation, it must necessarily preserve the spectrum of each observable and its type unchanged, as established by Proposition \ref{spettro-alg.}:
$$
\sigma_{\mathrm{pp}}(a) = \mathrm{Sp}_{\mathrm{p}}(\mathrm{J}(a)) \quad , \quad \sigma_{\mathrm{c}}(a) = \mathrm{Sp}_{\mathrm{c}}(\mathrm{J}(a)) 
$$
As for the dual representation $\mathrm{J}^\natural$, it should preserve as much as possible the degree of purity of the physical states, as discussed in the previous sections of this chapter.
\\
Two algebrizations $(\mathfrak{A}_0, \mathrm{J}_0, \mathrm{J}_0^\natural)$ and $(\mathfrak{A}_1, \mathrm{J}_1, \mathrm{J}_1^\natural)$ of the same physical system $(\mathfrak{X}, \mathfrak{S})$, to be considered \textbf{equivalent}, must satisfy for every observable $a \in \mathfrak{X}$ the following spectral property:
\begin{equation}\label{tipo-spettro-en}
\mathrm{Sp}_{\mathrm{ess}}(\mathrm{J}_0(a)) = \mathrm{Sp}_{\mathrm{ess}}(\mathrm{J}_1(a)), \qquad
\mathrm{Sp}_{\mathrm{dis}}(\mathrm{J}_0(a)) = \mathrm{Sp}_{\mathrm{dis}}(\mathrm{J}_1(a))
\end{equation}
Furthermore, for every $ \omega \in \mathfrak{S}_a $ it must:
\begin{itemize}
\item keep the purity of the algebraic states invariant:
$$p(\rho_0) = p(\rho_1)$$
\item and keep their entropy invariant:
$$S(\rho_0) = S(\rho_1) $$
\end{itemize}
where $\rho_i$ denote the density matrices associated with the normal states $\mathrm{J}_i^\natural(\omega)$, for $i=0,1$.
\\
In the case of concrete algebraic representations, i.e., when $\mathfrak{A}_i \subset B(\mathcal{H}_i)$, the definition of essential and discrete spectrum implies that the orthogonal projectors
$$ \mathbf{1}_{\Delta}(\mathrm{J}_i(a))\mathcal{H}_i, \quad i=0,1 $$
must have the same dimensions. This happens if there exists a unitary operator $U: \mathcal{H}_0 \to \mathcal{H}_1$ such that, for every observable $a\in \mathfrak{X}$, it holds\footnote{Thanks to the properties of functional calculus: $ U f(A) U^* = f(U A U^*) $ for every normal operator $A$ of $B(\mathcal{H}_1)$.} 
\begin{equation}\label{morf.interno-en}
U \mathrm{J}_0(a) U^* = \mathrm{J}_1(a).
\end{equation}
Consequently, for every $\omega \in \mathfrak{S}_a$ we will have:
$$ \mathrm{J}_1^\natural(\omega)(\mathrm{J}_1(a)) = \mathrm{J}_0^\natural(\omega)(U^* \mathrm{J}_1(a) U)$$
In other words, if there exists a *-algebra isomorphism $\Phi: \mathfrak{A}_0 \to \mathfrak{A}_1$  implemented by a unitary operator:
$$\Phi(A) = U A U^*, \quad \forall A \in \mathfrak{A}_0$$
then its dual map $\Phi^*: \mathfrak{A}_1^* \to \mathfrak{A}_0^*$ acts as:
$$ \Phi^*(\varphi) = \varphi \circ \Phi, \quad \forall \varphi \in \mathfrak{A}_1^*$$

\begin{definition}\upshape
Let $(\mathfrak{A}_1, \mathrm{J}_1, \mathrm{J}_1^\natural)$ and $(\mathfrak{A}_0, \mathrm{J}_0, \mathrm{J}_0^\natural)$ be two concrete algebraic representations on $ B(\mathcal{H}_1)$ and $B(\mathcal{H}_0)$, respectively\footnote{Recall that we can always view a C*-algebra as a concrete algebra by using its universal representation.}. They are said to be \textbf{equivalent} if there exists a unitary operator $U: \mathcal{H}_0 \to \mathcal{H}_1$ that satisfies equation \eqref{morf.interno-en}.
\end{definition}
In this way, the conditions on the spectral type from equation \eqref{tipo-spettro-en}, as well as those on purity and entropy, are automatically satisfied.
\begin{attenzione}\upshape
Nothing prevents us from thinking that there may exist representations that satisfy the conditions on the spectrum and on the states, but that are \textbf{not} equivalent, i.e., for which there is no unitary operator connecting them as in equation \eqref{morf.interno-en}.
\end{attenzione}

\begin{problem}\upshape\label{equivalenti-en}
If experimental information is intrinsically contained in the pair $(\mathfrak{X}, \mathfrak{S})$, what physical meaning do its possible \textbf{non-equivalent} algebraic representations have?
\end{problem}

\subsection{Kadison Theorem}
To establish the equivalence of two representations, one might consider using the following fundamental result due to Kadison in his work \cite{kad65}:

\begin{theorem}\upshape
If $\Phi$ is an affine mapping of the family $S_{\sigma}(\mathfrak{A})$ of weakly-continuous states of 
a von Neumann algebra $\mathfrak{A}$ acting on the Hilbert space $\mathcal{H}$ into the corresponding 
family $S_{\sigma}(\mathfrak{B})$ of another von Neumann algebra $\mathfrak{B}$ acting on the Hilbert 
space $\mathcal{K}$, then there is a weakly-continuous positive linear mapping 
$\alpha : \mathfrak{B}'' \to \mathfrak{A}''$ such that 
$$\widetilde{\omega} (\alpha(B)) = \widetilde{\Phi(\omega)}(B) \ , \ \forall B \in \mathfrak{B}'' \ , \ \omega \in S_{\sigma}(\mathfrak{A})$$ 
where $\widetilde{\omega}$ is the (unique) weakly-continuous (state) extension of $\omega$ to $\mathfrak{A}''$.
\\ 
If $\Phi$ is an affine isomorphism of $S_{\sigma}(\mathfrak{A})$ onto $S_{\sigma}(\mathfrak{B})$ then $\alpha$ 
is a $C^*$-isomorphism of $\mathfrak{B}''$ onto $\mathfrak{A}''$. 
\end{theorem}

Observe that if $\Phi$ is a bijective affine isomorphism between the state spaces of C*-algebras or von Neumann algebras, then it automatically preserves pure states, because pure states are exactly the extreme points of the compact convex set $S(\mathfrak{A})$, and a bijective affine isomorphism of convex sets maps extreme points to extreme points.

Given two algebraizations $(\mathfrak{A}_0, \mathrm{J}_0, \mathrm{J}_0^\natural)$ and $(\mathfrak{A}_1, \mathrm{J}_1, \mathrm{J}_1^\natural)$ of the same physical system $(\mathfrak{X}, \mathfrak{S})$, in order to apply the previous theorem one needs to identify an affine isomorphism $\Phi: S_{\sigma}(\mathfrak{A}_0) \to S_{\sigma}(\mathfrak{A}_1)$. One might consider extending the map defined by:
$$\mathrm{J}_0^\natural (\omega) \to \mathrm{J}_1^\natural (\omega) \ , \ \forall \omega\in \mathfrak{S}$$
and extending it by continuity (if possible) to the whole set of normal states. However, such a map is not always well-defined, since $\mathrm{J}_0^\natural, \mathrm{J}_1^\natural$ are not injective in general.

\chapter{Algebraizations and Compatibility}
In this section, we study the problem of algebraically representing the experimental measurements of two or more compatible observables of the system when they are measured simultaneously. We will see that, in order to identify the experimental measurement $\mu_{\omega, a:b}$ obtained from the experimental frequencies with a theoretical joint spectral measure $\mu_{\varphi, A:B}$, the self-adjoint operators $A, B$ associated with $a, b$ respectively via algebraization must necessarily commute.
\section{Commuting Self-Adjoint Operators}\label{section:joint-operator}
Let $A, B$ be self-adjoint operators in $\mathfrak B(\mathcal H)$ that \textit{commute}.  
For any Borel sets $\Delta_0, \Delta_1 \subset \mathbb R$, because $A$ and $B$ commute, it is easy to verify that the operator   
$$  \mathbf{1}_{\Delta_0}(A) \, \mathbf{1}_{\Delta_1}(B)$$   
is an orthogonal projection in $\mathfrak B(\mathcal H)$.
\\
For every state $\varphi$ on $\mathfrak B(\mathcal H)$ we can define the following map\footnote{Recall that $B(\mathbb R^2)$ denotes the Borel $\sigma$-algebra on $\mathbb R^2$. A set of the form $\Delta_0 \times \Delta_1$ with $\Delta_0,\Delta_1 \in B(\mathbb R)$ is called a measurable rectangle.  
The collection of all measurable rectangles is not a $\sigma$-algebra (it is not closed under arbitrary countable unions), but it generates $B(\mathbb R^2)$ as a $\sigma$-algebra. Hence every Borel set in $\mathbb R^2$ can be obtained from measurable rectangles by $\sigma$-algebraic operations, though it need not itself be a measurable rectangle:  
$$ B(\mathbb R^2)=B(\mathbb R)\otimes B(\mathbb R)$$  
where $B(\mathbb R)\otimes B(\mathbb R)$ is the product $\sigma$-algebra, i.e., the $\sigma$-algebra generated by the measurable rectangles $\Delta_0\times\Delta_1$ with $\Delta_0,\Delta_1\in B(\mathbb R)$.}:
$$ \Delta_0\times\Delta_1 \in B(\mathbb R^2) \;\longmapsto\; \varphi \bigg (\mathbf{1}_{\Delta_0}(A)\,\mathbf{1}_{\Delta_1}(B)\bigg ) \in [0,1] $$ 
Because the self-adjoint operators commute, the expression above defines a measure $\nu_{\varphi, A:B} \in \Pi(\mathbb R^2)$ that satisfies
\begin{equation}\label{condizione-spettrale}
\nu_{\varphi, A:B}(\Delta_0\times\Delta_1)=\varphi \bigg (\mathbf{1}_{\Delta_0}(A)\,\mathbf{1}_{\Delta_1}(B)\bigg ) 
\end{equation}
This measure is associated with the pair of operators $A,B \in \mathfrak B(\mathcal H)$,  which we denote  by the symbol $A:B$, and is called the \textit{joint spectral measure} of $A:B$.
Indeed, recall that in this case the Jordan product $\circ$ coincides with the ordinary operator product:
\begin{equation}\label{attenzione1}
\mathbf{1}_{\Delta_0}(A)\circ\mathbf{1}_{\Delta_1}(B)=\mathbf{1}_{\Delta_0}(A)\,\mathbf{1}_{\Delta_1}(B) 
\end{equation}
\begin{attenzione}
We have denoted the operator pair $(A,B)$ by the symbol $A:B$ in analogy with the notation for compatible observables. However, it should be stressed that $A:B$ is \textit{not} an element of $\mathfrak B(\mathcal H)$. 
\end{attenzione}

\subsection{Remarks on the joint spectral measure}
From Proposition \ref{markovkernel0} on page \pageref{markovkernel0} we obtain the existence of a Markov kernel $\big\{P_s^{A:B}\big\}_{s\in\mathbb R}$ and a measure $\nu\in\Pi(\mathbb R)$ such that
$$ \nu_{\varphi, A:B}(\Delta_0\times\Delta_1)=\int_{\Delta_0} P_s^{A:B}(\Delta_1)\,d\nu(s)$$
Since by definition
$$\nu_{\varphi, A:B}(\Delta_0\times\mathbb R)=\varphi\!\left(\mathbf{1}_{\Delta_0}(A)\,\mathbf{1}_{\mathbb R}(B)\right)
      =\varphi\!\left(\mathbf{1}_{\Delta_0}(A)\right)=\nu_{\varphi,A}(\Delta_0)$$
the measure $\nu$ in \eqref{condizione-spettrale} coincides with the spectral measure $\nu_{\varphi,A}$.

Similarly,
$$\nu_{\varphi,B}(\Delta_1)=\int_{\mathbb R} P_s^{A:B}(\Delta_1)\,d\nu_{\varphi,A}(s)$$
because
$$ \nu_{\varphi, A:B}(\mathbb R\times\Delta_1)=\varphi\!\left(\mathbf{1}_{\mathbb R}(A)\,\mathbf{1}_{\Delta_1}(B)\right)
      =\varphi\!\left(\mathbf{1}_{\Delta_1}(B)\right)=\nu_{\varphi,B}(\Delta_1)$$
From the commutativity of the operators we also obtain a relation completely analogous to~\eqref{a1} on page \pageref{a1}:
$$  \nu_{\varphi, A:B}(\Delta_0\times\Delta_1)=\int_{\Delta_0} P_s^{A:B}(\Delta_1)\,d\nu_{\varphi,A}(s)$$ 
Moreover, for every bounded Borel function $F:\mathbb R^{2}\to\mathbb R$ we have\footnote{Recall the identity  
$\mathbf{1}_{\Delta_0\times\Delta_1}(s,t)=\mathbf{1}_{\Delta_0}(s)\,\mathbf{1}_{\Delta_1}(t)$.}
$$ \nu_{\varphi, A:B}(F)=\int_{\mathbb R^{2}}F(s,t)\,dP_s^{A:B}(t)\,d\nu_{\varphi,A}(s)$$
Finally, for any state $\varphi$ on $\mathfrak B(\mathcal H)$,
\begin{align*}
\varphi(A)&=\int_{\mathbb R} s\,d\nu_{\varphi,A}(s)
          =\int_{\mathbb R^{2}} s\,d\nu_{\varphi, A:B}(s,t)\\[2mm]
\varphi(B)&=\int_{\mathbb R} t\,d\nu_{\varphi,B}(t)
          =\int_{\mathbb R^{2}} t\,d\nu_{\varphi, A:B}(s,t)
\end{align*}
We therefore define
\begin{equation}\label{expectation-pair}
\langle A:B \mid \varphi \rangle := \big(\varphi(A), \varphi(B)\big) \in  \mathbb R^2
\end{equation}
\subsection{The joint spectrum of $A:B$}
Using the measure defined in equation~\eqref{condizione-spettrale} for every pair $(A,B)$ of commuting self-adjoint operators in $\mathfrak B(\mathcal H)$, we can define their \textit{joint spectrum}, denoted by $\operatorname{Sp}(A:B) \subset \mathbb R^{2}$.  
Indeed, one can repeat step by step the reasoning carried out in Section~\ref{spettrocongiunto} on page~\pageref{spettrocongiunto} and arrive at completely analogous conclusions\footnote{For an overview of the joint spectrum see the works of Dash, in particular~\cite{Dash85}, although our notation and definitions differ from those of the author.}.

As in the case of physical observables, for each state $\varphi$ on $\mathfrak B(\mathcal H)$ we introduce the family of subsets of $\mathbb R^{2}$:
\[
\mathfrak F^{\varphi}(A:B) = \bigl\{ V \subseteq \mathbb R^{2} \text{ open} : \nu_{\varphi, A:B}(V) = 0 \bigr\}
\]
and the associated open set
\[
\rho^{\varphi}(A:B) = \bigcup \bigl\{ V : V \in \mathfrak F^{\varphi}(A:B) \bigr\}
\]
By definition, the support of the measure is
\[
\operatorname{Supp} \mu_{\varphi, A:B} := \mathbb R^{2} \setminus \rho^{\varphi}(A:B)
\]

We then set
\[
\mathfrak F^{\infty}(A:B) = \bigcap_{\varphi \in S(\mathfrak A)} \mathfrak F^{\varphi}(A:B)
\]
and define the \textit{joint resolvent} of $A:B$ as
\[
\rho^{\infty}(A:B) = \bigcup \bigl\{ V : V \in \mathfrak F^{\infty}(A:B) \bigr\}
\]
while its \textit{joint spectrum} is the set
\[
\operatorname{Sp}(A:B) = \mathbb R^{2} \setminus \rho^{\infty}(A:B)
\]

Repeating the arguments of Section~\ref{spettrocongiunto} on page~\pageref{spettrocongiunto}, we can also write
\[
\operatorname{Sp}(A:B) \subset \operatorname{Sp}(A) \times \operatorname{Sp}(B) \subset \mathbb R^{2}
\]

\subsection{Functional calculus for $A:B$}

It is well known (see e.g. Prugovečki~\S\,4.2~\cite{Prugo} or Schmüdgen~\S\,5.5~\cite{Schm12}) that for a family of commuting self-adjoint operators a \textit{joint functional calculus} exists\footnote{We stress once more that the commutativity of the operators is essential for the existence of such a calculus.}.  
In particular, for every bounded Borel function $F: \mathbb R^{2} \to \mathbb R$ there exists a unique self-adjoint operator $F(A:B) \in \mathfrak B(\mathcal H)$ such that for every state $\varphi$ on $\mathfrak B(\mathcal H)$,
\[
\varphi\big(F(A:B)\big) = \nu_{\varphi, A:B}(F) 
\]
where
\[
\nu_{\varphi, A:B}(F) = \int_{\mathbb R^{2}} F(s,t) \, d\nu_{\varphi, A:B}(s,t) 
\]

For instance, using the notation and calculations on page~\pageref{funzioneosservabile2} and considering again the function $\Theta_{m,n}(s,t) = s^{m} t^{n}$ for every $s,t \in \mathbb R$, we have
$$ \Theta_{1,1}(A:B) = A \circ B$$
but since the operators commute by hypothesis, 
$$\Theta_{1,1}(A:B) = AB$$
Furthermore, a direct consequence of the definition of the joint spectral measure for commuting operators is the relation
\begin{equation}\label{misura-spec-doppia}
\mathbf{1}_{\Delta_0}(A) \, \mathbf{1}_{\Delta_1}(B) = \mathbf{1}_{\Delta_0 \times \Delta_1}(A:B)
\end{equation}
In compact notation we may write, using the functional calculus,
$$ \langle A:B \mid \varphi \rangle := \big( \nu_{\varphi, A:B}(\Theta_{1,0}), \; \nu_{\varphi, A:B}(\Theta_{0,1}) \big) $$
%
%\section{Commutativity and Compatibility}
Consider a von Neumann algebraization $(\mathfrak M, \mathrm{J}, \mathrm{J}^{\natural})$ of a physical system $(\mathfrak X, \mathfrak S)$.
\\
Let $a, b$ be compatible observables of the system. As we have previously verified, in this case we have:
$$ \mathrm{J}(a \cdot b) = \mathrm{J}(a) \circ \mathrm{J}(b)$$
where $\cdot$ and $\circ$ are the respective Jordan products. Moreover, it is immediate to verify that 
$$ \mathrm{J}(a \cdot b) = \mathrm{J}(a) \mathrm{J}(b) \ \Longleftrightarrow \ [ \mathrm{J}(a), \mathrm{J}(b) ] = 0$$ 
\begin{property}[\textbf{Multiplicative Condition}]\index{Property of central condition} 
The algebraization satisfies the  multiplicative condition if for every pair of compatible observables $a, b$ of the system we have 
\begin{equation}\label{cond-centrale}
 [ \mathrm{J}(a), \mathrm{J}(b) ] = 0
\end{equation}
\end{property}
\textit{We assume that the multiplicative condition is always satisfied.}
\\

In this way, given compatible observables $a$ and $b$, we can define the joint operator $\mathrm{J}(a) : \mathrm{J}(b)$:
$$ a:b \ \longmapsto \ \mathrm{J}(a) : \mathrm{J}(b) $$ 
Recall that from relation \eqref{sestocaso} on page \pageref{sestocaso} we obtain 
$$ \mu_{\omega, a:b} (\Delta_0 \times \Delta_1) = \left\langle \mathbf{1}_{\Delta_0}(a) \cdot \mathbf{1}_{\Delta_1}(b) \right\rangle_{\omega} = \left\langle \mathbf{1}_{\Delta_0 \times \Delta_1}(a:b) \right\rangle_{\omega} $$
while from relation \eqref{attenzione1}:
 $$ \nu_{\varphi, A:B} (\Delta_0 \times \Delta_1) = \varphi \bigg ( \mathbf{1}_{\Delta_0}(A) \circ \mathbf{1}_{\Delta_1}(B) \bigg ) = \varphi \bigg ( \mathbf{1}_{\Delta_0 \times \Delta_1}(A:B) \bigg )$$
With this established, we can state that:
\begin{proposition}\label{misure-cong}
Given compatible observables $a, b \in \mathfrak X$ and a state $\omega \in \mathfrak S_{a:b}$, we have:  
$$ \nu_{\hat{\omega}, \hat{a}:\hat{b}} = \mu_{\omega, a:b} $$ 
\end{proposition}
\begin{proof}
By hypothesis, $\mathrm{J}(a)$ and $\mathrm{J}(b)$ commute; therefore we can write:
$$\nu_{\hat{\omega}, \hat{a}:\hat{b}}(\Delta_0 \times \Delta_1) = \mathrm{J}^\natural(\omega) \bigg ( \mathbf{1}_{\Delta_0}(\mathrm{J}(a)) \mathbf{1}_{\Delta_1}(\mathrm{J}(b)) \bigg ) = \mathrm{J}^\natural(\omega) \bigg ( \mathbf{1}_{\Delta_0}(\mathrm{J}(a)) \circ \mathbf{1}_{\Delta_1}(\mathrm{J}(b)) \bigg )$$
By the von Neumann algebraization, for any Borel set $\Delta \in B(\mathbb R)$ and any $\omega \in \mathfrak S_a$, we have:
$$\mathrm{J}^{\natural}(\omega) \big( \mathrm{J}(\mathbf{1}_\Delta(a)) \big) = \mathrm{J}^{\natural}(\omega) \big( \mathbf{1}_\Delta(\mathrm{J}(a)) \big)$$ 
Hence, also using relation \eqref{centrale1} on page \pageref{centrale1}, we obtain
$$\nu_{\hat{\omega}, \hat{a}:\hat{b}}(\Delta_0 \times \Delta_1) = \mathrm{J}^\natural(\omega) \bigg ( \mathrm{J}(\mathbf{1}_{\Delta_0}(a)) \circ \mathrm{J}(\mathbf{1}_{\Delta_1}(b)) \bigg ) = \mathrm{J}^\natural(\omega) \bigg ( \mathrm{J}(\mathbf{1}_{\Delta_0}(a) \cdot \mathbf{1}_{\Delta_1}(b)) \bigg ) $$  
In other words,
$$\nu_{\hat{\omega}, \hat{a}:\hat{b}}(\Delta_0 \times \Delta_1) = \langle \mathbf{1}_{\Delta_0}(a) \cdot \mathbf{1}_{\Delta_1}(b) \rangle_\omega = \mu_{\omega, a:b} (\Delta_0 \times \Delta_1) $$
\end{proof}
As a straightforward consequence of this proposition, if $F: \mathbb R^2 \to \mathbb R$ is a bounded Borel function, then for every $\omega \in \mathfrak S_{a:b}$ we obtain 
\begin{equation}
\mathrm{J}^\natural(\omega) \bigg ( \mathrm{J}(F(a:b)) \bigg) = \mathrm{J}^\natural(\omega) \bigg ( F(\mathrm{J}(a):\mathrm{J}(b)) \bigg )
\end{equation}
In this case as well, one could study the relationship between the joint spectrum $\sigma(a:b)$ of the two compatible observables $a, b$ and the joint spectrum $\operatorname{Sp}(\mathrm{J}(a):\mathrm{J}(b))$.
\\
From Proposition \ref{misure-cong} we obtain that 
\begin{equation}
\sigma(a:b) = \operatorname{Sp}(\mathrm{J}(a):\mathrm{J}(b))
\end{equation}
\subsection{The joint operator $A:B$}
As we have seen, the algebraization $\mathrm{J}: \mathfrak X \to \mathfrak A \subset \mathfrak B(\mathcal H)$ maps each observable to a self-adjoint operator. However, there is no natural way to define $\mathrm{J}(a:b)$ as an element of $\mathfrak A$ itself. To obtain an algebraic object that captures the joint spectral properties of two commuting self-adjoint operators $A = \mathrm{J}(a)$ and $B = \mathrm{J}(b)$, we introduce the \emph{joint operator} $A:B$.

Let $\mathfrak A \subset \mathfrak B(\mathcal H)$ be a $*$-algebra. Consider the $*$-algebra\footnote{See Kadison \& Ringrose \cite{kadring}, Vol. II, p. 881.} 
$$M_n(\mathfrak A) = \mathfrak A \otimes M_n(\mathbb C)$$ 
Every $X \in M_n(\mathfrak A)$ can be written uniquely as 
$$X = \sum_{i,j=1}^n X_{i,j} \otimes E_{i,j}, \qquad X_{i,j} \in \mathfrak A, \quad E_{i,j} \in M_n(\mathbb C)$$

For any state $\varphi$ on $\mathfrak A$, we define the \emph{induced $n$-state}\footnote{See \cite{kadring}, Ex. 11.5.21.}:
$$ \langle X \mid \varphi \rangle = \sum_{i,j=1}^n \varphi(X_{i,j}) E_{i,j} \in M_n(\mathbb C)$$

In the case $n = 2$, for commuting self-adjoint operators $A, B \in \mathfrak B(\mathcal H)_{\text{s.a.}}$, we define the joint operator
\begin{equation}\label{eq:joint-operator}
A:B = \begin{pmatrix} 0 & A \\ B & 0 \end{pmatrix} \in M_2(\mathfrak B(\mathcal H))
\end{equation}
For any state $\varphi$ on $\mathfrak B(\mathcal H)$, we then have
$$\langle A:B \mid \varphi \rangle := \begin{pmatrix} 0 & \varphi(A) \\ \varphi(B) & 0 \end{pmatrix} \in M_2(\mathbb R)$$
which we identify with the vector $(\varphi(A), \varphi(B)) \in \mathbb R^2$.

Observe that $(A:B)^* = B:A$ and that
$$(A:B)^*(A:B) = \begin{pmatrix} B^2 & 0 \\ 0 & A^2 \end{pmatrix}$$
Thus, the set of operators of the form $A:B$ forms a real $*$-vector space, but not a subalgebra of $M_2(\mathfrak B(\mathcal H))$.

\begin{remark}
The joint spectrum of $A:B$ is not the ordinary spectrum of $A:B$ as an element of the matrix algebra. Similarly, the joint functional calculus for the pair $(A,B)$ corresponds, via this construction, to an ordinary functional calculus on $A:B$, but one must be careful not to confuse the two.
\end{remark}
\subsection{Heisenberg's Theorem}
Let us revisit the discussion from Section \ref{FD-states}, adapting it to the algebraic case.
\\
Let $\mathfrak A$ be a C*-algebra, $\varphi$ a state on it. For every $A \in \mathfrak A$, we define:
$$ \Delta_{\varphi}(A) = \varphi(A^2) - \varphi(A)^2 \geq 0 $$
Note that 
$$ \Delta_{\widehat{\omega}}(J(a)) = \Delta_{\omega}(a) \ , \qquad \forall a \in \mathfrak X, \ \omega \in \mathfrak S_a $$
We have the following theorem:
\begin{theorem}\upshape[\textbf{Heisenberg Uncertainty}]
Let $\mathfrak A$ be a C*-algebra, $\varphi$ a state on it, and $A, B$ elements of $\mathfrak A_{s.a.}$. Then
\begin{equation*}
\Delta_{\varphi}(A) \ \Delta_{\varphi}(B) \geq \frac{1}{4} \ \varphi([A,B])^2
\end{equation*}%
\end{theorem}
\begin{proof}
Assume $\varphi(A) = \varphi(B) = 0$ and for every $t \in \mathbb R$ consider the positive element $(A + itB)(A - itB)$ of the algebra. We have 
$$t^2 \varphi(B^2) + i t \varphi([A,B]) + \varphi(A^2) \geq 0$$
Note that 
\begin{equation*}
\overline{\varphi([A,B])} = - \varphi([A,B]) \qquad \Longrightarrow \qquad \varphi([A,B]) \in i\mathbb R
\end{equation*}
It follows that 
$$ (-i \varphi([A,B]))^2 - 4 \varphi(B^2) \varphi(A^2) \leq 0$$
in other words
 $$\varphi(B^2) \varphi(A^2) \geq \frac{1}{4} \ \varphi([A,B])^2 $$
For the general case, consider the self-adjoint elements $A_o = A - \varphi(A) \mathbf{1}$ and $B_o = B - \varphi(B) \mathbf{1}$. We have 
$$ \varphi(A_o^2) = \Delta_{\varphi}(A_o) = \Delta_{\varphi}(A) \ , \qquad \varphi(B_o^2) = \Delta_{\varphi}(B_o) = \Delta_{\varphi}(B)$$
and also
$$ \varphi(B_o^2) \varphi(A_o^2) \geq \frac{1}{4} \ \varphi([A_o, B_o])^2 = \frac{1}{4} \ \varphi([A, B])^2$$
from which the thesis follows.
\end{proof}

Returning to the case of the C*-algebraization $(\mathfrak A, J, J^\natural)$, if $\omega \in \mathfrak S$ then for every observable $x, y \in \mathfrak X_\omega$ we have
$$ \Delta_{\omega}(x) \ \Delta_{\omega}(y) \geq \frac{1}{4} \ \widehat{\omega}([J(x), J(y)]) $$
The following result follows immediately:
\begin{proposition}
Given $\omega \in \mathfrak S$, for every $x, y \in \mathfrak X_\omega$ with $[J(x), J(y)] \neq 0$\footnote{Thus they are incompatible observables of the system.} we obtain
$$ \Delta_{\omega}(x) \ \Delta_{\omega}(y) > 0 $$
\end{proposition}

\section{Final discussion}

Let $J: \mathfrak{X} \to \mathfrak{A}$ be the map assigning to each experimental observable $a$ a self-adjoint operator $J(a)$ in the algebra $\mathfrak{A}$. As we have seen, to identify the experimental joint measure $\mu_{\omega, a:b}$ with the theoretical joint spectral measure $\mu_{\hat{\omega}, J(a):J(b)}$, it must hold that:
\[
\mu_{\omega, a:b} = \mu_{\hat{\omega}, J(a):J(b)}
\]
This identification is possible only if the joint spectral measure on the right-hand side exists, which requires that $J(a)$ and $J(b)$ commute. Hence, the map $J$ must preserve compatibility in the following sense:
\[
a, b \text{ compatible (experimentally simultaneously measurable)} \;\Rightarrow\; [J(a), J(b)] = 0 
\]
If this condition fails, the joint spectral measure $\mu_{\hat{\omega}, J(a):J(b)}$ is not defined within the usual spectral theory, and the equality above loses its meaning.
\\

We emphasize that the commutativity of $J(a)$ and $J(b)$ is a mathematical property of the representation, whereas experimental compatibility is a physical fact concerning the existence of an apparatus or procedure for simultaneous measurement.

\begin{remark}\upshape
If $a$ and $b$ are experimentally compatible observables, then in the theory they must be represented by commuting operators:
\[
[J(a), J(b)] = 0.
\]
The converse is not automatic, nor is it a general property. Commutativity alone does not guarantee that two observables can be measured simultaneously in the laboratory; it merely ensures that, within the mathematical model, a joint spectral measure exists.
\end{remark}

We are  convinced that many famous  paradoxes  of quantum mechanics arise precisely from confusing:
\begin{itemize}
  \item experimental compatibility (the existence of a simultaneous measurement procedure);
  \item mathematical commutativity (the vanishing of the commutator $[J(a), J(b)]$),
\end{itemize}
and from forgetting that the map $J$ must preserve compatibility, but that commutativity alone is not sufficient to determine whether two observables are compatible in the physical world.

In other words, in a rigorous treatment of quantum mechanics, one should always state:
\begin{itemize}
  \item that self-adjoint operators are not  the observables, but represent observables via a map $J$;
  \item that experimental compatibility implies $[J(a), J(b)] = 0$;
  \item that $[J(a), J(b)] = 0$ does not automatically imply that $a$ and $b$ are compatible.
\end{itemize}
Recognizing this distinction avoids many misunderstandings and restores conceptual clarity to the structure of quantum theory.

Another problem that arises in various cases is the following statement:
\\
\textit{If you measure $A$ and then $B$, the result is the same as measuring $B$ and then $A$ if $A$ and $B$ commute.}
\\
This is true, but there is a misinterpretation that is sometimes added:
\\
namely, that commuting $A$ and $B$ can nevertheless be seen as  one measured before the other  in an operational temporal sense, and that the  product  $AB$ corresponds to  first $A$ and then $B$.
\\
\textit{But this is not the case}:
\\
The operator product $AB$ (in our case $J(a)J(b)$) does not physically represent  measuring $A$ and then $B$  in a sequential sense, even if they commute; rather, it represents the product of the two operators in the algebraic sense, which is well-defined only because they commute.
\chapter{Some Algebraic Questions}
This chapter collects several algebraic issues that arise naturally   once a C* - algebraization $(\mathfrak A, J, J^\natural)$ of a physical system has been introduced. We examine how such an algebraization accommodates the description of time evolution (Schrödinger and Heisenberg pictures), the compatibility of algebraizations with nested laboratory regions (isotony), the lifting of geometric symmetries to algebraic symmetries, and finally a tentative algebraic model for sequential measurements via free algebras.

Each of these topics touches upon foundational questions at the interface between physics and operator algebras. The discussion is intentionally kept at a rather exploratory level; many of the constructions and definitions proposed here are only sketched, and their deeper implications, such as the precise conditions for isotony, the existence of algebraic time maps, the classification of algebraic symmetries, or the role of free algebras in non-commutative probability, are left for future investigation. This chapter therefore serves as a roadmap for further research, highlighting directions that deserve a more systematic treatment.

\section{Algebraization and Measurement Time}
Let us briefly review the practical-theoretical procedures that must be carried out to prepare measurements in a laboratory $L_o$:
\begin{itemize}
\item The laboratory is prepared to perform the experiment; this determines the laboratory-type region $\mathcal O_o$ and our \textit{chronological state}, which we still denote by $\omega$:
$$ t \in I \longrightarrow \omega^{(\tau)} \in \mathfrak S(\mathcal O_o)|\tau$$
\item The values of our observables at various instants of time are determined through the frequencies \eqref{freq} according to the procedures established in the first chapter.
\\
These yield, at each time, a probability measure 
$$ t \in \mathbb R^+ \longrightarrow \mu^t_{\omega,a} \in C_o(\mathbb R)^*$$
and an expectation value
$$t \in \mathbb R^+ \longrightarrow \left\langle a \right\rangle_{\omega}(t) = \int s \, d \mu^t_{\omega,a} (s)$$
\item It is assumed that the set of observables and states of our physical system can be endowed with a C*-algebraic structure, which is achieved through the algebraization of $\left( \mathfrak X , \mathfrak S \right)$: 
$$ J : \mathfrak{X} \rightarrow \mathfrak{A}_{\text{s.a.}} \qquad , \qquad J^{\natural} : \mathfrak{S} \rightarrow S(\mathfrak A) $$
\end{itemize} 
%*
%
As discussed in Section \ref{sezione_cronos}, given a state $\omega \in \mathfrak S_a |_{\tau=0}$ it is possible to determine a chronological state, which we still denote by $\omega$, such that for every $\tau \in I$ we have the relation: 
\begin{equation} \label{valoremedio-a}
J^{\natural}(\omega^{(\tau)})(a) = \left\langle a \right\rangle_{\omega}(\tau) \ , \qquad \forall a \in \mathfrak X_\omega
\end{equation}

We now ask whether it is possible to determine an element $S_\tau a \in \mathfrak A$ of the algebra such that 
\begin{equation} \label{valoremedio-b}
J^{\natural}(\omega)(S_\tau a) = \left\langle a \right\rangle_{\omega}(\tau) \ , \qquad \forall a \in \mathfrak X_\omega
\end{equation}
Historically, this problem is known as the Schrödinger and Heisenberg pictures. In the Schrödinger picture, the states associated with the algebraization of the physical system evolve in time according to
$$ \tau \in I \longrightarrow S^\natural_\tau \omega := J^{\natural}(\xi(\tau)) \in S(\mathfrak A) \ , \qquad \omega \in \mathfrak S_a$$
where
$$ \tau \in I \longrightarrow \xi(\tau) \in \mathfrak S^{\omega,a}_\tau \subset \mathfrak S_a|_{\tau'=0} \ , \qquad \xi(0) = \omega $$
is a possible evolution of the state in the measurement of $a$\footnote{Here it is assumed that the time evolution of the state can be established (see Remark \ref{esistenza-evol}).} with 
$$ J^{\natural}(\xi(\tau))(a) = \left\langle a \right\rangle_{\omega}(\tau)$$ 
while in the Heisenberg picture what changes in time are the \textit{representatives in the observable algebra}
$$ \tau \in I \longrightarrow S_\tau a \in \mathfrak A \ , \qquad S_0 a = a $$ 
which satisfy relation \eqref{valoremedio-b} while \textit{keeping the initial state fixed}. 
\\
In this case
\begin{equation}\label{valoremedio-c}
J^{\natural}(\omega^{(\tau)})(J(a)) = J^{\natural}(\omega)(S_\tau a) = \left\langle a \right\rangle_{\omega}(\tau) \ , \qquad \forall \tau \geq 0
\end{equation}
\begin{definition}\upshape
A triple $(\mathfrak A, S_\tau, S^\natural_\tau)$ that satisfies, for every observable $a$ of $\mathfrak X_\omega$, relation \eqref{valoremedio-c} is called an \emph{algebraic time map}\footnote{Obviously its existence is not guaranteed.}.
\end{definition} 
\subsection{Algebraizations and Preparation Time}
Consider now the laboratory-type regions $\mathcal O_t = L_o \times [0, t]$ as the preparation time $t > 0$ varies, as shown in Figure \ref{fig:09ca} on page \pageref{fig:09ca}, and the associated physical subsystem $\left( \mathfrak X(\mathcal O_t) , \mathfrak S(\mathcal O_t) \right)$. The possible algebraizations of this physical subsystem refer to a well‑defined instant of time:
\begin{equation}
\label{algebraztemp}
t \in \mathbb R^+ \longrightarrow ( \mathfrak A^t , J^{t}, J^{\natural}_{t})
\end{equation}
where
$$ J^{t} : \mathfrak{X}(\mathcal O_t) \rightarrow \mathfrak{A}^t_{\text{s.a.}} \qquad , \qquad J^{\natural}_{t} : \mathfrak{S}(\mathcal O_t) \rightarrow S(\mathfrak A^t) \ , \qquad \forall t > 0 $$
with the family of C$^*$-algebras $\left\{ \mathfrak A^t \right\}_{t \in \mathbb R^+}$ concretely represented on the same Hilbert space $\mathcal H$, i.e., $\mathfrak A^t \subset \mathfrak B(\mathcal H)$.
\\
By definition, the set $J^{t}(\mathfrak X(\mathcal O_t))$ generates the whole algebra $\mathfrak A^t$; it follows that 
$$ \mathfrak A^{t_1} \subset \mathfrak A^{t_2} \ , \qquad \forall t_2 > t_1$$
since the set of observables satisfies $\mathfrak X(\mathcal O_{t_1}) \subset \mathfrak X(\mathcal O_{t_2})$.
\\

The situation for states is more delicate. Recall that for every $a \in \mathfrak X(\mathcal O_{t_1})$ we have $\mathfrak S_a( \mathcal O_{t_2}| \mathcal O_{t_1}) \subset \mathfrak S_a(\mathcal O_{t_2})$ and there exists a map 
$$ P_a^{t_2, t_1} : \mathfrak S_a( \mathcal O_{t_2}| \mathcal O_{t_1}) \longrightarrow \mathfrak S_a(\mathcal O_{t_1})$$
such that
$$ \mu_{P_a^{t_2, t_1}(\omega), a} = \mu_{\omega, a} \ , \qquad \forall \omega \in \mathfrak S_a( \mathcal O_{t_2}| \mathcal O_{t_1}) $$
The family of algebraizations $\left\{ ( \mathfrak A^t , J^{t}, J^{\natural}_{t}) \right\}_{t \in \mathbb R^+}$ is called \textit{isotonic} if it satisfies the following properties:
\begin{itemize}
\item [1] For every $t_1 < t_2$ we must have
$$J^{t_1}(a) = J^{t_2}(a) \ , \qquad \forall a \in \mathfrak X(\mathcal O_{t_1})$$
\item [2] For every $t_1 < t_2$ and $a \in \mathfrak X(\mathcal O_{t_1})$
$$ Q^{t_2, t_1} \circ J^{\natural}_{t_2} = J^{\natural}_{t_1} \circ P_a^{t_2, t_1}$$
where the map $Q^{t_2, t_1}$ is the restriction of a generic state $\varphi \in S(\mathfrak A^{t_2})$ to the subalgebra $\mathfrak A^{t_1}$.
\\
In other words, the maps must make the following diagram commutative:
\begin{equation}
\begin{array}[c]{ccc}%
 \mathfrak S_a( \mathcal O_{t_2}| \mathcal O_{t_1}) & \overset{P_a^{t_2, t_1}}{\longrightarrow} & \mathfrak S_a(\mathcal O_{t_1}) \\
J^{\natural}_{t_2} \downarrow & & J^{\natural}_{t_1} \downarrow \\
S(\mathfrak A^{t_2}) & \overset{Q^{t_2, t_1}}{\longrightarrow} & S(\mathfrak A^{t_1})   
\end{array}
\label{commutative diagram}
\end{equation}
\end{itemize}
Given the physical system $\left( \mathfrak X , \mathfrak S \right)$ of the laboratory $L_o$, we assume it admits a C$^*$-algebraization $(\mathfrak A, J, J^{\natural})$. For every $t > 0$ we can define the following C$^*$-algebraization $(\mathfrak A^t, J^{t}, J^{\natural}_{t})$ of the subsystem $\left( \mathfrak X(\mathcal O_t) , \mathfrak S(\mathcal O_t) \right)$ by:
$$ J^t = J \circ i_t \ , \qquad \mathfrak X(\mathcal O_t) \overset{i_t}{\hookrightarrow} \mathfrak X$$
and
$$ J^{\natural}_{t} = J^{\natural} \circ i^{\natural}_t \ , \qquad \mathfrak S(\mathcal O_t) \overset{i^{\natural}_t}{\hookrightarrow} S(\mathfrak A)$$
with $\mathfrak A^t$ generated by the set $J_{t}(\mathfrak X(\mathcal O_t))$.
\\
These considerations lead to the following
\begin{definition}\upshape\label{isotonica}
A C$^*$-algebraization $(\mathfrak A, J, J^{\natural})$ of the system $\left( \mathfrak X , \mathfrak S \right)$ is called \emph{isotonic} if the family of algebraizations $(\mathfrak A^t, J^{t}, J^{\natural}_{t})$ of $\left( \mathfrak X(\mathcal O_t) , \mathfrak S(\mathcal O_t) \right)$ induced by it is isotonic.
\end{definition}
\subsection{Algebras and Regions}\label{isotonicabis}
We now extend the concept of isotony to any laboratory-type region.
\\
Let $\left( \mathfrak X , \mathfrak S \right)$ be the physical system associated with the laboratory $L$ and consider an algebraization $(\mathfrak A, J, J^{\natural})$ of it.
\\
For every $\mathcal O \in \mathfrak I$, where $\mathfrak I$ is the family of laboratory-type regions defined in relation \eqref{set_striscia}, we obtain its associated physical system $(\mathfrak X(\mathcal O), \mathfrak S(\mathcal O))$ and, since 
$$ \mathfrak X(\mathcal O) \subset \mathfrak X \qquad , \qquad \mathfrak S(\mathcal O) \subset \mathfrak S $$
we can define
\begin{equation}
\label{restrizio-alg}
 J_{\mathcal O}(x) = J(x) \ , \qquad \forall x \in \mathfrak X(\mathcal O)
\end{equation} 
with
$$\mathfrak A(\mathcal O) = J_{\mathcal O}(\mathfrak X(\mathcal O))''$$
\begin{equation}
\label{restrizio-alg-stato}
 J^{\natural}_{\mathcal O}(\omega) = J^{\natural}(\omega) \mid_{\mathfrak A(\mathcal O)} \ , \qquad \forall \omega \in \mathfrak S(\mathcal O)
\end{equation} 
In this way, the triple $(\mathfrak A(\mathcal O), J_{\mathcal O}, J^{\natural}_{\mathcal O})$ is an algebraization of the physical system $(\mathfrak X(\mathcal O), \mathfrak S(\mathcal O))$ induced by $(\mathfrak A, J, J^{\natural})$.
\\
Thus, given an algebraization $(\mathfrak A, J, J^{\natural})$ of the physical system $\left( \mathfrak X , \mathfrak S \right)$, we obtain 
$$\left\{ (\mathfrak A(\mathcal O), J_{\mathcal O}, J^{\natural}_{\mathcal O}) \right\}_{\mathcal O \in \mathfrak I}$$
algebraizations of the subsystems $(\mathfrak X(\mathcal O), \mathfrak S(\mathcal O))$.
\\
If for every $\mathcal O_o, \mathcal O_1 \in \mathfrak I$ with $\mathcal O_o \subset \mathcal O_1$, it holds that 
$$ Q^{\mathcal O_1, \mathcal O_0} \circ J^{\natural}_{\mathcal O_1}(\omega_1) = J^{\natural}_{\mathcal O_0} \circ P^{\mathcal O_1, \mathcal O_0}(\omega_1) \qquad , \ \forall \omega_1 \in \mathfrak S(\mathcal O_1 | \mathcal O_0) \subset \mathfrak S(\mathcal O_1),$$
in other words, if the following diagram
\begin{equation}
\begin{array}[c]{ccc}%
 \mathfrak S(\mathcal O_1 | \mathcal O_0) & \overset{P^{\mathcal O_1, \mathcal O_0}}{\longrightarrow} & \mathfrak S(\mathcal O_0) \\
J^{\natural}_{\mathcal O_1} \downarrow & & J^{\natural}_{\mathcal O_0} \downarrow \\
S(\mathfrak A(\mathcal O_1)) & \overset{Q^{\mathcal O_1, \mathcal O_0}}{\longrightarrow} & S(\mathfrak A(\mathcal O_0))   
\end{array}
\label{commutative diagram2}
\end{equation}
is commutative\footnote{Recall that we cannot assert that $J_{\mathcal O_0}(\mathfrak X(\mathcal O_0))$ is a dense set in $\mathfrak A(\mathcal O_0)$ with respect to the weak topology.}, then the algebraization $(\mathfrak A, J, J^{\natural})$ is called \textit{isotonic}.
\begin{problem}\upshape
Determine the conditions that the algebraization $(\mathfrak A, J, J^{\natural})$ must satisfy in order for the isotony property to hold.
\end{problem}
\section{Algebraizations and Symmetries}
Consider a physical system $(\mathfrak X(L_o), \mathfrak S(L_o))$ associated with the laboratory $L_o$ and let $(\mathfrak A, J, J^\natural)$ be a C$^*$-algebraization of it.
\\

Let $(\beta, \beta^\natural)$ be a pair of bijective maps
$$\beta : \mathfrak A_{\text{s.a.}} \longrightarrow \mathfrak A_{\text{s.a.}} \qquad , \qquad \beta^\natural : S(\mathfrak A) \longrightarrow S(\mathfrak A)$$
with the property that for every $A \in \mathfrak A_{\text{s.a.}}$ and $\varphi \in S(\mathfrak A)$ we have
\begin{equation}
\label{betanu}
\nu_{\beta^\natural(\varphi), \beta(A)} = \nu_{\varphi, A} 
\end{equation}
where $\nu_{\varphi, A}$ is the measure introduced in equation \eqref{misura_nu_1}.
\\
Obviously, in this case we also have
\begin{equation}
  \beta(f(A)) = f(\beta(A)) \ , \qquad \forall A \in \mathfrak A_{\text{s.a.}} 
\end{equation}
and from relation \eqref{betanu} it follows that
\begin{equation}
\beta^\natural(\varphi)(\beta(A)) = \varphi(A) \ , \qquad \forall A \in \mathfrak A_{\text{s.a.}} 
\end{equation}
If there exists a symmetry $(\alpha, \widehat{\alpha}) \in \operatorname{Sym}(L_o)$ of the system such that
\begin{equation}
\beta \circ J = J \circ \alpha \qquad , \qquad \beta^\natural \circ J^\natural = J^\natural \circ \alpha^\natural
\end{equation} 
the pair $(\beta, \beta^\natural)$ is called an \textit{algebraic symmetry} of the physical system $\left( \mathfrak X(L_o), \mathfrak S(L_o) \right)$.
\\
Moreover, for the probability measure defined in \eqref{misura_nu} it is easy to verify that for an algebraic symmetry we have, for every observable $a$ of the system,
\begin{equation}
\label{sym_alg}
\nu_{\beta^\natural(\hat{\omega}), \beta(\hat{a})} = \nu_{\hat{\omega}, \hat{a}} \ , \qquad \forall \omega \in \mathfrak S_a  
\end{equation}
\section{Free Algebras for Sequential Measurements}
Consider a laboratory region $\mathcal O_o$ and fix a sequence of observables that are jointly preparable in succession $ a_1 < a_2 < \cdots < a_n $, measurable at times $\left\{ \tau_1, \tau_2, \ldots, \tau_n \right\}$ in the state $\omega_n \in \mathfrak S_{a_1 < a_2 < \cdots < a_n}$.
\\
In this case, we can perform the following embedding:
$$ a_1, a_2, \ldots, a_n \in \mathfrak X \qquad \hookrightarrow \qquad a_1 \otimes a_2 \otimes \cdots \otimes a_n \in \mathbf{V}^{\otimes_n}$$
where $\mathbf{V}$ is the space generated by the set\footnote{Here "almost everywhere" is with respect to counting measure, so these are functions that are nonzero only on a finite number of points of $\mathfrak X$.
\\
As is well known, a Hamel basis for this space is given by the functions
$$ e_x(y) = \left\{
\begin{array}{c c}
0 & x \neq y \\
1 & x = y
\end{array} \right.
$$ 
while for $\mathbf{V}^{\otimes_n}$ a basis is given by the functions
$$\left\{ e_{a_1} \otimes e_{a_2} \otimes \cdots \otimes e_{a_n} \right\}$$ 
which we write in compact form as the tensor product 
$ a_1 \otimes a_2 \otimes \cdots \otimes a_n$.

 }
$$ \left\{ f : \mathfrak X \longrightarrow \mathbb R : \text{a.e. zero} \right\}$$  
and we can consider the graded algebra of noncommutative polynomials (with zero constant term):
$$ \mathbb{P}(\mathfrak X) := \bigoplus_{k=1}^{\infty} \mathbf{V}^{\otimes_k} = \left\{ \xi : \mathbb N \longrightarrow \bigcup_{n=1}^{\infty} \mathbf{V}^{\otimes_n} : \text{a.e. zero} \right\}$$ 
Obviously, for a single observable we trivially obtain
$$ a \in \mathfrak X \ \hookrightarrow \ a \in \mathbf{V}^{\otimes_1} \subset \mathcal{P}(\mathfrak X)$$
We wish to emphasize that if $a$ and $b$ are compatible observables, hence simultaneously measurable, their product $a \cdot b$ is still in $\mathbf{V}^{\otimes_1}$. The same holds for constants: if $a = r I$, we still have $a \in \mathbf{V}^{\otimes_1}$ and not in $\mathbb C = \mathbf{V}^{\otimes_0}$ as one might mistakenly think\footnote{Obviously, for the same reason, the observable $a^2$ does not correspond to the element $a \otimes a$ in the free algebra.}.
\\
Thus, each step in the tensor product corresponds to a step in the sequential measurement:
$$\begin{array}{c c c c c c c}
\tau_1 & & \tau_2 & & \ldots & & \tau_n \\
a_1 & \otimes & a_2 & \otimes & \cdots & \otimes & a_n
\end{array}$$
Recall that 
$$\mathfrak S_{a_1 < a_2 < \cdots < a_n} \subset \bigcap_{k=1}^n \mathfrak S_{a_k} \qquad \Longrightarrow \qquad \omega_n \in \bigcap_{k=1}^n \mathfrak S_{a_k} $$
and for the probability distribution of the individual observables measured at the prescribed times, we have
$$P(a_k \in \Delta_k \mid a_1 < a_2 < \cdots < a_n, \tau_k)_{\omega_n} \ , \qquad k = 1, 2, \ldots, n$$ 
and as we have seen previously, if the observables are Kolmogorovian, then there exists a Borel measure $\mu_{\omega_n, a_1 < a_2 < \cdots < a_n}$ such that for every Borel set $\Delta_j$ with $j = 1, 2, \ldots, n$ we have 
$$ P(a_1 \in \Delta_1, \tau_1 \wedge \cdots \wedge a_n \in \Delta_n, \tau_n)_{\omega_n} = \mu_{\omega_n, a_1 < a_2 < \cdots < a_n}(\Delta_1 \times \cdots \times \Delta_n) $$

\chapter{Algebras and State Selections}

In this section, we will analyze the algebraic realizations $(\mathfrak{A}, \mathrm{J}, \mathrm{J}^{\natural})$ of physical subsystems $(\mathfrak{X}_o, \mathfrak{S}_o)$ of the laboratory physical system $L_o$.
\\
We will study the role played by the center of its observables $\mathcal{Z}_{\mathfrak{S}_o}(\mathfrak{X}_o)$ in establishing the possible algebraic realizations of a physical system. To this end, we will need to introduce additional properties that the maps $\mathrm{J}$ and $\mathrm{J}^{\natural}$ must satisfy to obtain a physically suitable mathematical description of our laboratory system; in other words, we must specify to which \textit{selection rules} they must adhere.

\section{Jordan and Segal Algebraic Realizations}

The C*-algebraic approach we introduced is not the most general possible algebraic realization for a physical subsystem $(\mathfrak{X}_o, \mathfrak{S}_o)$.
\\
Let us therefore consider an algebraic realization $(\texttt{B}, \mathrm{J}, \mathrm{J}^{\natural})$ where $(\texttt{B}, \circ)$ \textit{is a real JB-algebra of Jordan} and $\mathrm{J}: \mathfrak{X}_o \to \texttt{B}$ and $\mathrm{J}^{\natural}: \mathfrak{S}_o \to S(\texttt{B})$\footnote{With $S(\texttt{B})$ we denote the state space of the JB-algebra; as in the case of Banach algebras, an element $\varphi$ of $\texttt{B}^*$ is a state if $\varphi(B^2) \geq 0, \ \forall B \in \texttt{B}$ and $\varphi(I) = 1$.} that always satisfy the \texttt{ARBA} conditions given in Section \ref{algebrizzazionesistema} on page \pageref{algebrizzazionesistema}, along with the embedding property \eqref{prop_E}.
\\
We emphasize that a JB-algebra is formally real (cf. \cite{Ol-St}, Corollary 3.3.8)\footnote{Therefore, we are in agreement with Proposition \ref{formalreal} on page \pageref{formalreal}, concerning the squares of compatible observables.}. Obviously, the C*-algebraic case treated in the previous section is a special case of Jordan algebraic realizations, precisely those obtained through JW-algebras.
\\
If $a, b \in \mathfrak{X}_o$ are $\mathfrak{S}_o$-compatible observables of the system, by the definition of the Jordan product given in \eqref{prodotto jordan} on page \pageref{prodotto jordan} and from the properties of the \texttt{ARBA} conditions of the Jordan algebraic realization, it is easy to verify that in this case too, relation \eqref{centrale1} is satisfied.
\\
As we have previously verified, if $\{a, b, c\}$ are compatible observables, we have $a \cdot (b \cdot c) = (a \cdot b) \cdot c$, and also in this case, we obtain that relation \eqref{centrale2} on page \pageref{centrale2} is satisfied.
\\
Recall that $(\mathfrak{Z}(\texttt{B}), \circ)$ defined in equation \eqref{centro} on page \pageref{centro} turns out to be an associative algebra, like the set $(\mathcal{Z}_{\mathfrak{S}_o}(\mathfrak{X}_o), \cdot)$ defined on page \pageref{commutanteobs}; therefore, to obtain a good algebraic realization of a physical subsystem, this property must be maintained. This leads to the following definition:
\begin{definition}\upshape[\textbf{Central Condition}]
The algebraic realization $(\texttt{B}, \mathrm{J}, \mathrm{J}^{\natural})$ of the physical system $(\mathfrak{X}_o, \mathfrak{S}_o)$ satisfies the central condition if
\begin{equation}
\label{condcentrale}
\mathrm{J}( \mathcal{Z}_{\mathfrak{S}_o}(\mathfrak{X}_o) ) \subset \mathfrak{Z}(\texttt{B})
\end{equation}
\end{definition}
\begin{remark}\upshape
Relation \eqref{centrale1} 
$$ \mathrm{J}(a \cdot b) = \mathrm{J}(a) \circ \mathrm{J}(b), \qquad \text{for all $a, b$ that are $\mathfrak{S}_o$-compatible}$$
does not guarantee the central condition.
\end{remark}
We note that from relation \eqref{commutante-J}, the central condition is satisfied if and only if $T_{\mathrm{J}(a)} T_B = T_B T_{\mathrm{J}(a)}$ for all elements $a \in \mathcal{Z}_{\mathfrak{S}_o}(\mathfrak{X}_o)$ and $B \in \texttt{B}$; therefore, if and only if
\begin{equation*}
( B \circ X ) \circ \mathrm{J}(a) = B \circ ( X \circ \mathrm{J}(a)) \ , \qquad \forall B, X \in \texttt{B}, \ a \in \mathcal{Z}_{\mathfrak{S}_o}(\mathfrak{X}_o).
\end{equation*}
\noindent
\\
We have the following question:
\\
 Why not directly use the more general real JB-algebras for the algebraic realization of a physical system, instead of resorting to the C*-algebraic approach? 
\\
An attempt to answer our question was provided by Horuzhy in \cite{Horu}, p. 10:
\begin{citazione}
Such noncanonical schemes have not been developed very far, however, as they involve a lot of mathematical difficulties which are not balanced by noticeable physical gains.
\end{citazione}
The mathematical difficulties mentioned in the previous statement lie in the fact that for generic JB-algebras, we have a spectral analysis that is more difficult to handle and a functional calculus that is, in practice, restricted to holomorphic functions (cf. \cite{Aupetit, Ol-St}), whereas this is not the case for JW-algebras. Moreover, for the latter, we have a connection, given by Topping's Proposition \ref{centro_commutante} on page \pageref{centro_commutante}, between the center of the Jordan algebra and its commutant.
\\

Historically, after the start of the joint work of Jordan, von Neumann, and Wigner in 1934 \cite{Jord_34}, applications of Jordan algebras to physics came to a halt. They were later proposed in another form by Segal in his \textit{postulates for quantum mechanics} \cite{Segal_47}, through the definition of what are now called \textit{Segal systems}\footnote{Segal's postulates were refined by Sherman in \cite{Sherm}.}.
\\

A Segal system consists of a real Banach space $\mathfrak{Q}$ and the existence, for each natural number $n$, of a continuous map in norm, called the $n$-th power,
$$A \in \mathfrak{Q} \longrightarrow A^n \in \mathfrak{Q}$$
such that for every $A, B \in \mathfrak{Q}$ we have:
\begin{itemize}
\item $A^0 = I$;
\item $\| A^2 - B^2 \| \leq \max \left\{ \| A^2 \|, \| B^2 \| \right\}$;
\item $\| A^2 \| = \| A \|^2$.
\end{itemize}
Now it is easy to verify that the space $\mathfrak{Q}$ with the formal product given in \eqref{prodotto jordan} becomes a real Jordan algebra.
\section{Central Connections}\label{centralcond}

Let us abandon the more general Jordan algebraic realizations and analyze the consequences of the central condition \eqref{condcentrale} on the mathematically more tractable algebraic realizations of our physical system.

\subsubsection{Case of associative algebras}

Consider the algebraic realization $(\mathfrak{B}, \mathrm{J}, \mathrm{J}^{\natural})$ of the physical system $(\mathfrak{X}_o, \mathfrak{S}_o)$ as defined in Section \ref{algebrizzazionesistema}, but now with $\mathfrak{B}$ being a \textit{unital real Banach algebra} (i.e., an associative real Banach algebra with unit).
\\ 
Assume that 
$$\mathcal{Z}_{\mathfrak{S}_o}(\mathfrak{X}_o) \neq \emptyset$$
and that it satisfies the additional condition:
\begin{property}[\textbf{Weak-Central Condition}]\index{Property of Weak-Central}
\begin{equation}\label{condcentraledebole}
 \mathrm{J}(\mathcal{Z}_{\mathfrak{S}_o}(\mathfrak{X}_o)) \subset \mathfrak{B}^c 
\end{equation}
\end{property}
where by definition the \emph{center} (or \emph{algebraic center}) of $\mathfrak{B}$ is:
$$ \mathfrak{B}^c = \{ B \in \mathfrak{B} : BX = XB \ \forall X \in \mathfrak{B} \} \subset \mathfrak{B} $$ 
Observe that from relation \eqref{prodotto jordan3} on page \pageref{prodotto jordan3}, we obtain $\mathfrak{B}^c \subset \mathfrak{Z}(\mathfrak{B}^{(+)})$, where $\mathfrak{B}^{(+)}$ denotes the \textit{Jordan algebra obtained from $\mathfrak{B}$ by symmetrization of the product \eqref{prodotto jordan2}.}
\\
Consequently, our algebraic realization automatically satisfies the central condition \eqref{condcentrale}.
\\
From \eqref{condcentraledebole} it follows trivially that
$$ \mathrm{J}(\mathcal{Z}_{\mathfrak{S}_o}(\mathfrak{X}_o)) \subset \mathfrak{B}^c \subset \mathfrak{B}
\quad \Longrightarrow \quad 
\mathfrak{B}^c \subset \mathfrak{B}^{cc} \subset \mathrm{J}(\mathcal{Z}_{\mathfrak{S}_o}(\mathfrak{X}_o))^c \subset \mathfrak{B}$$
Hence, the real Banach algebra $\mathfrak{B}$ is precisely\footnote{Recall that
$$ \mathrm{J}(\mathcal{Z}_{\mathfrak{S}_o}(\mathfrak{X}_o))^c = \{ B \in \mathfrak{B} : B \mathrm{J}(z) = \mathrm{J}(z) B, \ \forall z \in \mathcal{Z}_{\mathfrak{S}_o}(\mathfrak{X}_o) \}$$
and if $\mathfrak{B} \subset B(\mathcal{H}_{\mathbb{R}})$, then
$$ \mathrm{J}(\mathcal{Z}_{\mathfrak{S}_o}(\mathfrak{X}_o))^c = \mathfrak{B} \cap \mathrm{J}(\mathcal{Z}_{\mathfrak{S}_o}(\mathfrak{X}_o))'$$
therefore
$$ \mathfrak{B} \subset \mathrm{J}(\mathcal{Z}_{\mathfrak{S}_o}(\mathfrak{X}_o))'$$}
\begin{equation} \label{commutante1}
\mathfrak{B} = \mathrm{J}(\mathcal{Z}_{\mathfrak{S}_o}(\mathfrak{X}_o))^c 
\end{equation}

\begin{proposition}\upshape\label{commutanteprop}
The algebraic realization satisfies the weak-central condition \eqref{condcentraledebole} if and only if 
\begin{equation}\label{commutante2}
 \mathrm{J}(a \cdot z) = \mathrm{J}(a) \mathrm{J}(z), \qquad \forall a \in \mathfrak{X}_o, \ z \in \mathcal{Z}_{\mathfrak{S}_o}(\mathfrak{X}_o).
\end{equation}
\end{proposition}
\begin{proof}
Assume condition \eqref{condcentraledebole} holds. For $a \in \mathfrak{X}_o$, relation \eqref{commutante1} implies $ \mathrm{J}(a) \in \mathrm{J}(\mathcal{Z}_{\mathfrak{S}_o}(\mathfrak{X}_o))^c$, hence
$$ \mathrm{J}(a) \mathrm{J}(z) = \mathrm{J}(z) \mathrm{J}(a), \qquad \forall z \in \mathcal{Z}_{\mathfrak{S}_o}(\mathfrak{X}_o)$$   
By definition, the observable $a$ is compatible with every element of $\mathcal{Z}_{\mathfrak{S}_o}(\mathfrak{X}_o)$; therefore, using the Jordan product definition,
$$ \mathrm{J}(a \cdot z) = \frac{1}{2} \big[ \mathrm{J}(z) \mathrm{J}(a) + \mathrm{J}(a) \mathrm{J}(z) \big] = \mathrm{J}(a) \mathrm{J}(z)$$ 
in other words:
$$ \mathrm{J}(a) \circ \mathrm{J}(z) = \mathrm{J}(a) \mathrm{J}(z)$$
Conversely, assume \eqref{commutante2} holds for all $z \in \mathcal{Z}_{\mathfrak{S}_o}(\mathfrak{X}_o)$. Then
$$ \mathrm{J}(z) \mathrm{J}(x_1) \mathrm{J}(x_2) \cdots \mathrm{J}(x_n) = \mathrm{J}(x_1) \mathrm{J}(x_2) \cdots \mathrm{J}(x_n) \mathrm{J}(z), 
\qquad \forall x_1, x_2, \ldots, x_n \in \mathfrak{X}_o$$
From the minimality hypothesis of the algebraic realization (i.e., that $\mathfrak{B}$ is generated by $\mathrm{J}(\mathfrak{X}_o)$), it follows that $\mathrm{J}(z)$ commutes with every element of $\mathfrak{B}$, establishing \eqref{condcentraledebole}.
\end{proof}

We can extend our algebraic realization $(\mathfrak{B}, \mathrm{J}, \mathrm{J}^{\natural})$ of the physical system $(\mathfrak{X}_o, \mathfrak{S}_o)$ to the complex case.
\\
Define the complexification
$$ \mathfrak{B}_{\mathbb{C}} = \mathfrak{B} + i \mathfrak{B} $$
equipped with the natural involution $(A + iB)^* = A - iB$ for $A, B \in \mathfrak{B}$. The self-adjoint part of this complex $*$-algebra is then
$$ (\mathfrak{B}_{\mathbb{C}})_{\text{s.a.}} = \mathfrak{B} $$
The embedding $J_{\mathbb{C}} : \mathfrak{X} \longrightarrow (\mathfrak{B}_{\mathbb{C}})_{\text{s.a.}}$ remains unchanged: $J_{\mathbb{C}} = \mathrm{J}$, while the state extension $J^{\natural}_{\mathbb{C}} : \mathfrak{S} \longrightarrow S(\mathfrak{B}_{\mathbb{C}})$ is defined by
$$ J^{\natural}_{\mathbb{C}}(\omega)(A + iB) = \mathrm{J}^{\natural}(\omega)(A) + i \mathrm{J}^{\natural}(\omega)(B), \qquad \forall A, B \in \mathfrak{B} $$

As we have already discussed, this construction naturally leads us to consider C$^*$-algebraic realizations, which effectively absorb algebraic realizations over general associative real algebras.

\subsubsection{The Case of C*-algebraizations}
A physical subsystem $\left( \mathfrak X_o , \mathfrak S_o \right)$ admits a C*-algebraic representation if there exists a triple $(\mathfrak{A}, J, J^{\natural})$ consisting of a unital complex C*-algebra $\mathfrak A \subset B(\mathcal H)$, the maps 
$$\mathrm{J} : \mathfrak X_o \rightarrow \mathfrak A_{s.a.} \qquad \text{and} \qquad \mathrm{J}^{\natural} : \mathfrak S_o \rightarrow S(\mathfrak A)$$
satisfying the \textsl{ARBA} conditions given in section \ref{algebrizzazionesistema} with the embedding property \eqref{prop_E}.
\begin{remark}\upshape
The triple $(\mathfrak A, \mathrm{J}, \mathrm{J}^{\natural})$ establishes a Jordan algebraization on the JW-algebra $(\mathfrak A_{s.a.}, \circ)$ where the product $\circ$ is given by \eqref{prodotto jordan2}.
\end{remark}
By minimality, the elements of $\mathrm{J}(\mathfrak X_o)$ generate the whole algebra $\mathfrak A$:
\begin{equation}\label{minmil}
\bar{\mathfrak A}^w = \mathfrak A'' = \mathrm{J}(\mathfrak X_o)'' \qquad ( \ \Longrightarrow \mathfrak A' = \mathrm{J}(\mathfrak X_o)' \ ) 
\end{equation}
Recall that in general, if $S \subset B(\mathcal H)$ is any self-adjoint set, then its commutant $S'$ is a von Neumann algebra of $B(\mathcal H)$.
\\
If $\mathfrak A(S)$ is the von Neumann algebra of $B(\mathcal H)$ generated by the set $S$, we obtain\footnote{Here $\bar{S}^w$ denotes the closure of the set $S$ in the weak topology of $B(\mathcal H)$.}:
$$ S \subset \mathfrak A(S) \qquad \Longrightarrow \qquad S'' = \mathfrak A(S)'' = \mathfrak A(S) \neq \bar{S}^w $$ 
Our set $\mathrm{J}(\mathfrak X_o)$ is self-adjoint with identity, but this does not guarantee that its weak closure coincides with its double commutant\footnote{See Arveson \cite{Arveson76} Theorem 1.2.1.}.
\\

Since $(\mathfrak X_o , \mathfrak S_o)$ is a physical subsystem, the set $\mathcal{Z}_{\mathfrak S_o}(\mathfrak X_o)$ possesses a structure of a unital abelian normed real algebra with the Jordan product given in relation \eqref{prodotto jordan}.
\\
In our case we have 
$\mathrm{J}(\mathcal{Z}_{\mathfrak S_o}(\mathfrak X_o)) \subset \mathfrak A_{s.a.}$. Therefore $\mathrm{J}(\mathcal{Z}_{\mathfrak S_o}(\mathfrak X_o))$ possesses a structure of a real abelian operator algebra.
\\
 
Let us see what form the central condition takes in the case of concrete C*-algebras.
\\
In this case, by Topping's proposition \ref{centro_commutante}, we have\footnote{Recall that the real vector space $\mathfrak A_{s.a}$ with the Jordan product given in \eqref{prodotto jordan2} becomes a real Jordan algebra.}
$$ \mathfrak A_{s.a.}^c = \left\{ A \in \mathfrak A_{s.a.} : AX = XA \ , \ \forall X \in \mathfrak A_{s.a.} \right\} = \mathfrak Z(\mathfrak A_{s.a}) $$
hence for the central condition we have
\begin{equation}\label{condcentralebis}
 \mathrm{J}( \mathcal{Z}_{\mathfrak S_o}(\mathfrak X_o) ) \subset \mathfrak Z(\mathfrak A_{s.a}) = \mathfrak A_{s.a.}^c = \mathfrak A_{s.a.}' \cap \mathfrak A_{s.a.}
\end{equation}
since $\mathfrak A_{s.a.}'' = \bar{\mathfrak A}^w$ we can write 
$$ \mathrm{J}( \mathcal{Z}_{\mathfrak S_o}(\mathfrak X_o) ) \subset Z(\mathfrak A'') $$
From the validity of the central condition and from relation \eqref{commutante2} we obtain the following inclusions:
\begin{equation}\label{centroalg0}
\mathrm{J}(\mathcal{Z}_{\mathfrak S_o}(\mathfrak X_o)) \subset \mathrm{J}(\mathfrak X_o)' \subset B(\mathcal H) \ \Longrightarrow \ \mathrm{J}(\mathfrak X_o)'' \subset \mathrm{J}(\mathcal{Z}_{\mathfrak S_o}(\mathfrak X_o))' 
\end{equation}
Thus
\begin{equation}\label{centroalg}
 \mathrm{J}(\mathcal{Z}_{\mathfrak S_o}(\mathfrak X_o)) \subset \mathfrak A'' \subset \mathrm{J}(\mathcal{Z}_{\mathfrak S_o}(\mathfrak X_o))'
\end{equation}
it follows that
\begin{equation}\label{centroalgbis} \mathrm{J}(\mathcal{Z}_{\mathfrak S_o}(\mathfrak X_o)) \subset Z(\mathfrak A'') \subset \mathrm{J}(\mathcal{Z}_{\mathfrak S_o}(\mathfrak X_o))
\end{equation}
\begin{remark}\upshape\label{no_mqp}
If in our algebraization $\mathfrak A$ is the whole C*-algebra $B(\mathcal H)$, it must necessarily be that 
$$ B(\mathcal H) = \mathrm{J}(\mathcal{Z}_{\mathfrak S_o}(\mathfrak X_o))' \ \Longrightarrow \ \mathrm{J}(\mathcal{Z}_{\mathfrak S_o}(\mathfrak X_o)) = \mathbb R I \ \Longrightarrow \ \mathcal{Z}_{\mathfrak S_o}(\mathfrak X_o) = \mathbb R I$$
Therefore, for physical subsystems that are not quantum-pure, the C*-algebra $\mathfrak A$ of any algebraization satisfying the central condition cannot be $B(\mathcal H)$. 
\end{remark}
\subsection{Non-trivial center and algebraization}\label{centronontriviale}
Consider a physical subsystem $\left( \mathfrak X_o , \mathfrak S_o \right)$ and assume that $\mathcal{Z}_{\mathfrak S_o}(\mathfrak X_o)$ is non-empty and non-trivial.
\\
Take a question $p \in \mathcal{Z}_{\mathfrak S_o}(\mathfrak X_o)$. For every observable $a$ of the subsystem we obtain that $ p \cdot a = p \cdot a \cdot p $ and if $p^\bot$ is the question orthogonal to $p$ we can write
$$ a = p \cdot a \cdot p + p^\bot \cdot a \cdot p^\bot \qquad , \qquad \forall a \in \mathfrak X_o$$
since $p \cdot a \cdot p$ and $p^\bot \cdot a \cdot p^\bot$ are compatible observables with 
$$ (p \cdot a \cdot p) \cdot (p^\bot \cdot a \cdot p^\bot) = 0 $$ 
It follows that if $(\mathfrak{A}, J, J^{\natural})$ is an algebraization satisfying the central condition, then from relation \eqref{commutante2} we have
$$ \mathrm{J}(p \cdot a) = \mathrm{J}(p) \mathrm{J}(a) \mathrm{J}(p) $$
where $P = \mathrm{J}(p)$ is a non-trivial orthogonal projector\footnote{As we have shown, $\sigma(p) = \operatorname{Sp}(P)$ and hence if $p$ is non-trivial, so is $P$.} of the algebra $\mathfrak A \subset B(\mathcal H)$. Set $\mathcal K = P \mathcal H$.
In this way we obtain a new algebraization of the physical system $(\mathfrak A_\parallel, J_\parallel, J_\parallel^\natural)$\footnote{Attention: this algebraization does not satisfy the embedding property \ref{prop_E}; therefore we cannot claim that the spectrum of an observable $a \in \mathfrak X_o$ coincides with the corresponding algebraic spectrum of $J_\parallel(a)$.} where the algebra $\mathfrak A_\parallel$ is given by the elements
$$ \left\{ T \in B(\mathcal K) : \exists A \in B(\mathcal H) \text{ s.t. } T = P A \mid_{\mathcal K} \right\} $$
while $J_\parallel : \mathfrak X \longrightarrow \mathfrak A_{\parallel \text{s.a.}}$ with
$$ J_\parallel(a) = P \mathrm{J}(a) \mid_{\mathcal K} \ , \qquad \forall a \in \mathfrak X $$ 
and $J_\parallel^\natural : \mathfrak S \longrightarrow S(\mathfrak A_\parallel)$ with 
$$ J_\parallel^\natural(\omega)(T) = \frac{ \mathrm{J}^{\natural}(\omega)(P A P)}{ \mathrm{J}^{\natural}(\omega)(P)} \ , \qquad T = P A \mid_{\mathcal K} \ , \ A \in \mathfrak A $$
Thus
$$ J_\parallel^\natural(\omega)(J_\parallel(a)) = \frac{ \langle p a p \rangle_{\omega} }{ \langle p \rangle_{\omega} } \ , \qquad \forall a \in \mathfrak X $$
The same considerations apply to the orthogonal question $p^\bot$, yielding an algebraization $(\mathfrak A_\bot, J_\bot, J_\bot^\natural)$ of our physical system.
\\
Moreover, from relation \eqref{decomposizione} we have
$$ \mathrm{J}(a) = P \mathrm{J}(a) P + P^\bot \mathrm{J}(a) P^\bot $$
it follows that we can write:
$$ J_\parallel \oplus J_\bot : \mathfrak X \longrightarrow \mathfrak A_\parallel \oplus \mathfrak A_\bot $$ 
where
$$ \mathrm{J}(a) = J_\parallel(a) \oplus J_\bot(a) \ , \qquad \forall a \in \mathfrak X $$
while for every $\omega \in \mathfrak S$ we obtain:
$$ \mathrm{J}^{\natural}(\omega)(A) = r_0 J_\parallel^\natural(\omega)(P A \mid_{\mathcal K}) + r_1 J_\bot^\natural(\omega)(P^\bot A \mid_{\mathcal K^\bot}) \ , \qquad \forall A \in \mathfrak A$$
where $r_0 = \mathrm{J}^{\natural}(\omega)(P)$ and $r_1 = \mathrm{J}^{\natural}(\omega)(P^\bot)$ with $r_0, r_1 \in ]0, 1[$ and $r_0 + r_1 = 1$.
\\
From the minimality of the algebraization $(\mathfrak{A}, \mathrm{J}, \mathrm{J}^{\natural})$ we can assert that the projector $P$ is a central projector $P \in Z(\mathfrak A)$.
\\
Obviously these considerations for relation \eqref{decomposizionedue} can be easily extended to a countable family of mutually orthogonal questions $\left\{ p_k \right\}_{k \in \mathbb N}$ of the system.
\\

Recall that if $a \in \mathcal{Z}_{\mathfrak S_o}(\mathfrak X_o)$ and $\left\{ \Delta_k \right\}_k$ is a disjoint covering of $\mathbb R$, the projectors $p_k = \mathbf{1}_{\Delta_k}(a)$ form a family of orthogonal questions.
 $$ \star \star \star $$
We now perform a new selection among the possible algebraizations, extending the central condition \ref{condcentralebis} to all compatible observables:
\begin{definition}\upshape\label{strong-central-condition}
The algebraization satisfies the strong central condition if it preserves the product of $\mathfrak S_o$-compatible observables in the associative algebra:
\begin{itemize}
\item $\mathrm{J}(a \cdot b) = \mathrm{J}(a) \mathrm{J}(b)$
\end{itemize}
\end{definition} 
Obviously the strong central condition implies \ref{condcentralebis}:
$$ \mathrm{J}( \mathcal{Z}_{\mathfrak S_o}(\mathfrak X_o) ) \subset \mathfrak A_{s.a.}' \cap \mathfrak A_{s.a.} $$
\\
Moreover, as we discussed for the general case, if $a, b \in \mathfrak X_o$ are $\mathfrak S_o$-compatible observables, it follows that  
$$ [ \mathrm{J}(a), \mathrm{J}(b) ] = 0 $$
\begin{remark}\upshape
If $[\mathrm{J}(a), \mathrm{J}(b)] = 0$, it is not necessarily true that the observables $a, b$ are $\mathfrak S_o$-compatible.
\end{remark}

\section{Selection rules for algebraizations}

Unlike our approach, where the sets of observables and states are initial data of the model, in algebraic theory one starts from an operator algebra, often concrete, where observables are identified with its self-adjoint elements. This identification is \textit{generally not surjective}, since not all self-adjoint elements of the algebra are physically relevant. Hence the introduction of selection mechanisms on the algebras that attempt to delimit such observables within the algebra, historically known as superselection rules:
\begin{citazione}
In its most general form, a superselection rule for a quantum mechanical theory can be defined as any restriction on what is observable in the theory\footnote{Citation from Strocchi and Wightman \cite{strocchi}.}.
\end{citazione}
As already stated, in our model the physical quantities and the states of the system are given. The problem is therefore to determine an appropriate algebraization that adequately describes the physical phenomenon under consideration. We have already briefly indicated in section \ref{guide_first_step} the strategy to follow for this purpose. Here we intend to introduce additional properties that algebraizations ought to satisfy in order to better describe a physical subsystem of our quantum laboratory.
\\

Our starting point is the following:
\\
Let a C*-algebraization $(\mathfrak{A}, J, J^{\natural})$ be given, with $\mathfrak A$ a concrete C*-algebra on a Hilbert space $\mathcal H$ which is weakly closed, of the physical subsystem $(\mathfrak X_o, \mathfrak S_o)$ satisfying the central condition.
\\
Assume furthermore that $\mathcal{Z}_{\mathfrak S_o}(\mathfrak X_o) \neq \emptyset$.
\\
By Proposition \ref{commutanteprop}, for every $z \in \mathcal{Z}_{\mathfrak S_o}(\mathfrak X_o)$ we have
$$ \mathrm{J}(a) \mathrm{J}(z) = \mathrm{J}(z) \mathrm{J}(a) \ , \qquad \forall a \in \mathfrak X_o $$
hence 
\begin{equation}\label{centro_SR}
\mathrm{J}(\mathcal{Z}_{\mathfrak S_o}(\mathfrak X_o)) \subset \mathrm{J}(\mathfrak X_o)'
\end{equation}
and relation \eqref{centroalg} follows.
\\
We have a simple statement about the non-trivial center:
\begin{proposition}
$$ \mathcal{Z}_{\mathfrak S_o}(\mathfrak X_o) = \mathbb R I \qquad \Longleftrightarrow \qquad \mathrm{J}(\mathcal{Z}_{\mathfrak S_o}(\mathfrak X_o)) = \mathbb R I $$
\end{proposition}
\begin{proof}
The first implication is trivial; for the second we make use of the embedding property of the algebraization:
\\
If $A \in \mathrm{J}(\mathcal{Z}_{\mathfrak S_o}(\mathfrak X_o))$, by hypothesis $A = \texttt{r} I$ for some real $\texttt{r}$.
\\
Thus $\mathrm{J}(C_r(a)) = \mathrm{J}(A)$ where the observable $C_r(a) \subset \texttt{r} I$ with $C_r(t) = \texttt{r}$ for every $t \in \mathbb R$.
\\
The observables $C_r(a)$ and $a$ are compatible and therefore we can write $\mathrm{J}(a - C_r(a)) = 0$ and by the embedding property $a - C_r(a) \subset 0$.
\end{proof}
\begin{definition}\upshape [\textbf{Selection Rules}]
A C*-algebraization $(\mathfrak{A}, \mathrm{J}, \mathrm{J}^{\natural})$ of the physical system $(\mathfrak X_o, \mathfrak S_o)$ possesses selection rules (\textbf{SR}) if the algebra $\mathrm{J}(\mathfrak X_o)' \subset B(\mathcal H)$ is non-trivial, i.e.,
$$ \mathrm{J}(\mathfrak X_o)' \neq \mathbb C I $$
\end{definition}
From relation \eqref{centro_SR}, if $\mathrm{J}(\mathcal{Z}_{\mathfrak S_o}(\mathfrak X_o))$ is non-trivial then $\mathrm{J}(\mathfrak X_o)'$ will also be non-trivial and therefore we are in the presence of selection rules.
\\

Note that $\mathfrak A = \mathrm{J}(\mathfrak X_o)''$ and hence $\mathfrak A' = \mathrm{J}(\mathfrak X_o)'$; from this follows the double implication
$$ \mathfrak A = B(\mathcal H) \ \Longleftrightarrow \ \text{no selection rules are present}$$  
\begin{remark}\upshape
If the system $(\mathfrak X_o, \mathfrak S_o)$ is abelian, by definition the center $\mathcal{Z}_{\mathfrak S_o}(\mathfrak X_o) = \mathfrak X_o$ and from \eqref{centro_SR} it follows that 
$$ \mathrm{J}(\mathcal{Z}_{\mathfrak S_o}(\mathfrak X_o)) = \mathrm{J}(\mathfrak X_o) \ \Longrightarrow \ \mathfrak A \subset \mathrm{J}(\mathcal{Z}_{\mathfrak S_o}(\mathfrak X_o))' = \mathrm{J}(\mathfrak X_o)' = \mathfrak A' $$ 
Hence the algebra $\mathfrak A = Z(\mathfrak A)$ is abelian.
\end{remark}
We give another definition recurrent in the literature\footnote{See for instance the work of Jauch \cite{Jauch60}.}, here adapted to our model.
\\

Let a concrete C*-algebraization $(\mathfrak{A}, \mathrm{J}, \mathrm{J}^{\natural})$ of the laboratory physical system $(\mathfrak X, \mathfrak S)$ be given.
\\ 
Denote by $\mathfrak X_\oslash \subset \mathfrak X$ a family of $\mathfrak S$-mutually compatible observables of the physical system\footnote{Hence $\mathfrak X_\oslash \subset \mathcal C (\mathfrak X_\oslash)$.} and consider the following algebras:  
$$\mathfrak R := \mathrm{J}(\mathfrak X_\oslash)'' \subset \mathfrak A \qquad , \qquad \mathfrak R' = \mathrm{J}(\mathfrak X_\oslash)' \subset B(\mathcal H).$$
\begin{definition}\upshape
The set $\mathfrak X_\oslash$ of compatible observables of the system is \textit{complete with respect to the given algebraization} if  
\begin{equation}\label{completo_alg} \mathfrak R = \mathfrak R'. 
\end{equation}
Equivalently, $\mathfrak R$ is a maximal abelian subalgebra (masa) of $B(\mathcal H)$.
\end{definition}

Now it is useful to recall two well-known facts from operator algebra theory:
\begin{proposition}\upshape[Theorem 9.1.3 of \cite{kadring}] \label{kady1}
If a von Neumann algebra $\mathfrak M$ is of type I (resp. type II, type III), then the same holds for its commutant $\mathfrak M'$.  
\end{proposition}
\begin{proposition}\upshape[Problem 9.6.1 of \cite{kadring}] \label{kady2}
A von Neumann algebra $\mathfrak M$ is of type I if and only if $\mathfrak M$ is $*$-isomorphic to a von Neumann algebra with abelian commutant. 
\end{proposition}

From these two important statements, the following result follows:
\begin{proposition}\upshape
If our algebraization admits a complete system of compatible observables, then the algebra $\mathfrak A$ is of type I.  
\end{proposition}
\begin{proof}
Assume that $\mathfrak X_\oslash$ is a complete system of compatible observables. By definition, $\mathfrak R = \mathrm{J}(\mathfrak X_\oslash)''$ and $\mathfrak R' = \mathrm{J}(\mathfrak X_\oslash)'$, hence  
$$ \mathfrak R \subset \mathfrak A \qquad \Longrightarrow \qquad \mathfrak A' \subset \mathfrak R' = \mathfrak R \subset \mathfrak A.$$
Thus $\mathfrak A'$ is a subalgebra of an abelian algebra, and therefore it is itself abelian. By Proposition \ref{kady2}, we obtain that $\mathfrak A'' = \mathfrak A$ is of type I (and consequently, by Proposition \ref{kady1}, $\mathfrak A'$ is also of type I).
\end{proof}
\subsubsection{Strong Selection Property}
From \eqref{centro_SR} and the minimality of the algebraization we obtain that
$$ \mathfrak A \subset \mathrm{J}(\mathcal{Z}_{\mathfrak S_o}(\mathfrak X_o))' \subset \mathfrak B(\mathcal H) $$
from this relation we cannot deduce that every self-adjoint element of $\mathrm{J}(\mathcal{Z}_{\mathfrak S_o}(\mathfrak X_o))'$ is an element of the algebra $\mathfrak A$ and therefore a potential observable of the physical system.
\\

We introduce a new property of our algebraization that physical subsystems may have\footnote{For further considerations on the role of superselection rules and the type of the observable algebra $\mathfrak A$, the reader may consult the work of Earman \cite{Earman}.}, called the strong selection property of the system's observables:
\begin{property}[\textbf{StSR}]\label{prop_super} \index{Property of strong selection state} 
The physical subsystem $(\mathfrak X_o, \mathfrak S_o)$ admits a C*-algebraization $(\mathfrak A, J, J^\natural)$ that satisfies the strong selection property if
\begin{equation}
 \mathfrak B(\mathcal H)_{s.a.} \cap \mathrm{J}(\mathcal{Z}_{\mathfrak S_o}(\mathfrak X_o))' \subset \mathfrak A_{s.a.}
\end{equation}
\end{property}
If property [\textbf{StSR}] is satisfied together with the central condition, we have 
$$\mathrm{J}(\mathcal{Z}_{\mathfrak S_o}(\mathfrak X_o))' \subset \mathfrak A$$ 
and from \eqref{centroalg0} and minimality:   
 $$ \mathfrak A'' = \mathrm{J}(\mathfrak X_o)'' \subset \mathrm{J}(\mathcal{Z}_{\mathfrak S_o}(\mathfrak X_o))' $$
we obtain that $\mathfrak A$ is weakly closed (it must necessarily be a von Neumann algebra). 
\\
Thus
$$ \text{Strong Superselection Property} \qquad \Longrightarrow \qquad \mathfrak A = \mathrm{J}(\mathcal{Z}_{\mathfrak S_o}(\mathfrak X_o))' $$
we also have the converse implication:
$$ \mathfrak A = \mathrm{J}(\mathcal{Z}_{\mathfrak S_o}(\mathfrak X_o))' \qquad \Longrightarrow \qquad \text{Strong Superselection Property}   $$
since in this case
$$ \mathfrak B(\mathcal H)_{s.a.} \cap \mathrm{J}(\mathcal{Z}_{\mathfrak S_o}(\mathfrak X_o))' = \mathfrak B(\mathcal H)_{s.a.} \cap \mathfrak A = \mathfrak A_{s.a.}$$

Property [\textbf{StSR}] greatly restricts the possible types of the observable algebra $\mathfrak A$ of our algebraization.
\\
Indeed, from this we trivially obtain the following implication:
$$ \mathfrak A' \subset \mathfrak A \qquad \Longrightarrow \qquad \mathcal{Z}(\mathfrak A) = \mathfrak A' $$
and setting $\mathfrak M = \mathfrak A''$ we obtain that $\mathfrak M' = \mathcal{Z}(\mathfrak M)$ and therefore the algebra $\mathfrak M'$ is abelian, implying that $\mathfrak M$ is of Type I.
\\
Summarizing:
$$ \text{Strong Superselection Property} \qquad \Longrightarrow \qquad \mathfrak A'' = \mathfrak A \text{ of Type I} $$

$$\star\star\star$$ 

Let us consider two fundamental examples of C*-algebraization $(\mathfrak{A}, \mathrm{J}, \mathrm{J}^{\natural})$ with $\mathfrak A \subset \mathfrak B(\mathcal H)$ a von Neumann algebra: 

\begin{itemize}
\item A purely quantum system:
$$ \mathcal{Z}_{\mathfrak S_o}(\mathfrak X_o) = \mathbb R I$$
In this case  
$$\mathrm{J}(\mathcal{Z}_{\mathfrak S_o}(\mathfrak X_o)) = \mathbb R I \ \Longrightarrow \ \mathrm{J}(\mathcal{Z}_{\mathfrak S_o}(\mathfrak X_o))' = \mathfrak B(\mathcal H)$$
as we note, these conditions do not ensure that the C*-algebra $\mathfrak A$ coincides with $\mathfrak B(\mathcal H)$.
\item A classical system:
$$ \mathcal{Z}_{\mathfrak S_o}(\mathfrak X_o) = \mathfrak X_o$$
it follows
$$ \mathrm{J}(\mathcal{Z}_{\mathfrak S_o}(\mathfrak X_o)) = \mathrm{J}(\mathfrak X_o) \ \Longrightarrow \ \mathrm{J}(\mathfrak X_o)' = \mathrm{J}(\mathcal{Z}_{\mathfrak S_o}(\mathfrak X_o))' $$
From relation \eqref{centro_SR}:
$$ \mathfrak A \subset \mathrm{J}(\mathfrak X_o)' \ \Longrightarrow \ \mathfrak A \subset \mathfrak A' \ \Longrightarrow \ Z(\mathfrak A) = \mathfrak A $$
\end{itemize}
In physics there are many intermediate cases where the algebra $\mathrm{J}(\mathcal{Z}_{\mathfrak S_o}(\mathfrak X_o))$ is smaller than $\mathfrak A$ and contains non-trivial elements.
\section{Algebraization of Abelian Systems}
We study a possible algebraization for the abelian system $(\mathfrak X_o, \mathfrak S_o)$ given in section \ref{sist_abel} on page \pageref{sist_abel}.
\\
By property \ref{c.f.m.} of multivariable functional closure, in $\mathfrak X_o$ we have particular observables $c$ such that
$$ \left\langle c \right\rangle_{\omega} = \text{const} \qquad \forall \omega \in \mathfrak S_o$$
which play the role of constant observables of our subsystem, and hence there exists a null observable $o$ such that
$$ \left\langle o \right\rangle_{\omega} = 0 \qquad \forall \omega \in \mathfrak S_o $$ 

In this way we can state that 
$$ \text{if } \| a \| = 0 \ \Longrightarrow \ \left[ \left\langle a \right\rangle_{\omega} = \text{const} \ \forall \omega \in \mathfrak S_o \right] \ \Longrightarrow \ a = o $$
Thus by Proposition \ref{norma_banach} on norms, $(\mathfrak X_o, \| \cdot \|)$ with the \textit{Jordan product} turns out to be a unital abelian operator algebra. In this case the unit $I$ is given by the constant observable $1$:
$$ \left\langle I \right\rangle_{\omega} = 1 \qquad \forall \omega \in \mathfrak S_o$$
We denote by $\mathfrak B$ the real abelian Banach algebra, the completion of the normed algebra $(\mathfrak X_o, \| \cdot \|)$, and by $\mathfrak B_\mathbb C$ its complexification.
\\
We introduce the following norm on $\mathfrak B_\mathbb C$:
$$ \| a + i b \| = \sqrt{ \| a \|^2 + \| b \|^2 } \ , \qquad \forall a, b \in \mathfrak X_o $$
we verify that this norm satisfies the C*-property:
$$ \| a + i b \|^2 = \| (a+ib) \cdot (a-ib) \| \ , \qquad \forall a, b \in \mathfrak X_o $$
Observe that in the complexification we have the following product:
$$ (a+ib) \cdot (x+iy) = a \circ x - b \circ y + i(a \circ y + b \circ x) \ , \qquad \forall a, b, x, y \in \mathfrak X_o $$  
and hence
$$ (a+ib) \cdot (a-ib) = a^2 + b^2 \ , \qquad \forall a, b \in \mathfrak X_o $$
\begin{eqnarray*}
\| (a+ib) \cdot (a-ib) \| & = & \| a^2 + b^2 \| = \sup_{\omega \in \mathfrak S_o} | \left\langle a^2 + b^2 \right\rangle_{\omega} | = \\
& = & \sup_{\omega \in \mathfrak S_o} \left\langle a^2 \right\rangle_{\omega} + \sup_{\omega \in \mathfrak S_o} \left\langle b^2 \right\rangle_{\omega} = \| a^2 \| + \| b^2 \| = \\
& = & \| a \|^2 + \| b \|^2
\end{eqnarray*}
the last equality follows from observation \ref{C_star_norma}.
\\
The algebra $\mathfrak B_\mathbb C$ is a unital abelian C*-algebra; to obtain a von Neumann algebra we consider its bidual $\mathfrak B_{\mathbb C}^{**}$.
\\
Obviously, by the Gelfand transform, $\mathfrak B_\mathbb C$ is isomorphic to the algebra of continuous functions $C(\Omega)$ where 
$\Omega$ is the space of characters of the algebra, which we recall is a compact topological space\footnote{Thus the bidual of the algebra is isomorphic to the space $C(\Omega)^{**}$.}.  
\\ 
We have the following embeddings for observables:
$$ \mathfrak X_o \hookrightarrow \mathfrak B \hookrightarrow \mathfrak B_\mathbb C \hookrightarrow \mathfrak B_{\mathbb C}^{**}$$
while for states\footnote{We define 
$$\widehat{\omega}(a+ib) = \left\langle a \right\rangle_{\omega} + i \left\langle b \right\rangle_{\omega} \ , \ \forall a, b \in \mathfrak X_o$$
which is an element of $S(\mathfrak B_\mathbb C)$. Recall, moreover, that every state of a C*-algebra can be extended to a $\sigma$-continuous state of its bidual (cf. \cite{BR1} Proposition 5.2.10).}:
$$ \mathfrak S_o \hookrightarrow S(\mathfrak B) \hookrightarrow S(\mathfrak B_\mathbb C) \hookrightarrow S_w(\mathfrak B_{\mathbb C}^{**})$$ 
To obtain a concrete algebraization of our physical subsystem, it suffices to recall that the abelian von Neumann algebra $\mathfrak B_{\mathbb C}^{**}$ is isomorphic to the algebra $L^\infty(X, \nu)$ where $(X, \nu, \Sigma)$ is a $\sigma$-finite space. Moreover, the predual of $L^\infty(X, \nu)$ is given by the vector space $L^1(X, \nu)$, therefore the set of normal states $S_w(\mathfrak B_{\mathbb C}^{**})$ is in bijective correspondence with the set\footnote{Observe that this set is in bijective correspondence with the set:
$$ \hat S = \left\{ \theta \in \Pi(X) : \theta \ll \nu \right\} \subset \Pi(X)$$
via the map
$$ \theta \in \hat S \longrightarrow \frac{d\theta}{d\nu} \in S $$}
$$S = \left\{ \rho \in L^1(X, \nu) : \rho \geq 0 \text{ with } \| \rho \|_1 = 1 \right\} \subset L^\infty(X, \nu)_* $$
hence we have two maps\footnote{Obviously injective but not necessarily surjective.}, which establish the algebraization for the physical subsystem:
\begin{equation}\label{mappealgebriche}
a \in \mathfrak X_o \longrightarrow \widehat{a} \in L^\infty(X, \nu) \qquad , \qquad \omega \in \mathfrak S_o \longrightarrow \rho_\omega \in S 
\end{equation}
such that 
$$ \left\langle a \right\rangle_{\omega} = \int_X \rho_\omega(x) \ \widehat{a}(x) \ d \nu(x) $$ 
Observe that if for every $\rho \in S$ and $A \in L^\infty(X, \nu)$ we define:
\begin{equation} \label{termos}
\nu_{\rho, A}(f) = \int_X f(A(x)) \rho(x) \ d\nu(x) \ , \qquad \forall f \in C_o(\mathbb R)^*
\end{equation}
we obtain a map
$$\rho \in S \longmapsto \nu_\rho \in \Pi(X)$$
where 
$$ \nu_\rho(E) = \int_E \rho(x) \ d\nu \ , \qquad \forall E \in B(X)$$ 
It follows that we have the following identification:
\begin{equation}\label{misura_sistema_classico}
\mu_{\omega,a} (\Delta) = \nu_{\rho_\omega} ( \widehat{a}^{-1}(\Delta) ) \ , \qquad \forall \Delta \in B(\mathbb R)
\end{equation}  
Observe that the set $S$ is convex while the set $\widehat{\mathfrak S}_o \subset S$ is not necessarily convex, where $\widehat{\mathfrak S}_o$ is the image of the state map given in relation \eqref{mappealgebriche} and hence it is not a priori given that $\mathbb{M}_{\mathfrak S_o}(a)$ has a single sector in the measurement of $a$\footnote{In practice, for every $\omega_1, \omega_2 \in \mathfrak S_o$ and $r \in ]0,1[$ it is not guaranteed that there exists $\omega \in \mathfrak S_o$ such that: 
$$ r \rho_{\omega_1} + (1-r) \rho_{\omega_2} = \rho_{\omega} $$}.
\\
Observe that also in this case we obtain the following inclusion:
$$ \Pi_p \cap \operatorname{Ext}_{\mathfrak S_o}(a) \subset \operatorname{Ext}_{\mathfrak S_o}(a) \subset \mathbb{M}_{\mathfrak S_o}(a) $$
and that if $\mu_{\omega, a} \in \Pi_p$ it implies that there exists a real number $\lambda \in \mathbb R$, which will depend on the state $\omega$ and on the observable $a$, such that
$$\delta_{\lambda} = \mu_{\omega, a}$$ 
since it must be that $\mu_{\omega, a} \ll \nu$, it must necessarily be that $\nu(E) \neq 0$ where $E = \widehat{a}^{-1}(\{ \lambda \}) \in \Sigma$; in this case the density is given by 
$$ \rho_\omega = \frac{1}{\nu(E)} \ \mathbf{1}_E \in L^1(X, \nu)$$ 
\section{Algebrization for Mackey Systems*}
We resume the discussion on algebrizations by applying it to Mackey systems, as defined in \S \ref{Mack_syst}.
\\
Let $(\mathfrak X_o , \mathfrak S_o)$ be a Mackey system and $(\mathfrak{A}, J, J^\natural)$ a C*-algebrization of it. We assume that it satisfies the embedding property and the state separation property as established in the previous sections.
\\
In particular, the state separation property takes the following form for Mackey systems:
\begin{property} [\textbf{PS}]
\label{prop_S_bis} 
The algebrization $(\mathfrak A, J, J^\natural)$ of our Mackey system satisfies the state separation property if for every observable $a \in \mathfrak X_o$ it holds:
\begin{equation}
 \left[ \ A \in \mathfrak A_{s.a.} \ \text{such that} \ J^\natural(\omega)(A) = 0 \ \ \forall \omega \in \mathfrak S_o \ \right] \qquad \Longrightarrow \qquad A = 0
\end{equation}
\end{property}    
In addition to the properties listed in Section \ref{algebrizzazionesistema}, new properties arise for Mackey systems. For example, from the previous sections we derived relation \eqref{pseudoconvex}, which does not establish the actual convexity of the map $J^\natural$ but suggests it.  
\\
To this end, we can introduce the following further condition for Mackey systems:
\begin{itemize}
\item[I.] The map $J^\natural : \mathfrak S_o \rightarrow S(\mathfrak A)$ is an affine map, i.e.:
$$J^\natural((1-r)\omega_1 + r\omega_2) = (1-r) J^\natural(\omega_1) + r J^\natural(\omega_2), \qquad \forall \omega_1, \omega_2 \in \mathfrak S_o, \ r \in [0,1]$$
\end{itemize}
We also extend the continuity condition to the weak topologies:
\begin{itemize}
\item[II.] The map $J^\natural : \mathfrak S_o \rightarrow S(\mathfrak A)$ is weakly continuous:
\\
If $\omega_\alpha \rightarrow \omega_o$ in the $w*$-topology, then $J^\natural(\omega_\alpha) \rightarrow J^\natural(\omega_o)$ in the $w*$-topology.
\end{itemize}
From the affine property we obtain that the subset of states $J^\natural(\mathfrak S_o) \subset S(\mathfrak A)$ is convex.
\\
From the injectivity of the map $J^\natural$\footnote{If $\omega, \omega' \in \mathfrak S_o$ are such that
$$ \langle a \rangle_\omega = \langle a \rangle_{\omega'} \ \ \forall a \in \mathfrak X_o \qquad \Longrightarrow \qquad \omega = \omega' $$
hence in the Mackey case:
$$ J^\natural(\omega) = J^\natural(\omega') \qquad \Longrightarrow \qquad \omega = \omega' $$} we can assert that if $\omega \in \mathfrak S_o$ is a pure state of the system, \textit{then $J^\natural(\omega)$ is a pure algebraic state of the set $J^\natural(\mathfrak S_o)$.
}
\subsection{Maps Induced by State Transformations} 
Let us make some further observations on the maps introduced in defining a C*-algebrization $(\mathfrak A, J, J^\natural)$ of a Mackey system $(\mathfrak X_o, \mathfrak S_o)$:
$$J : \mathfrak X_o \longrightarrow \mathfrak A \qquad , \qquad J^\natural : \mathfrak S_o \longrightarrow S(\mathfrak A)$$
We have repeatedly stated that not all algebraic states in $S(\mathfrak A)$ are physical states of the system; in other words, $J^\natural(\mathfrak S_o) \neq S(\mathfrak A)$.
\\
To this end, we introduce mathematically a new property for our algebrization, called the full state property:
\begin{property}[\texttt{PF}]
\label{prop_F} 
The algebrization $(\mathfrak A, J, J^\natural)$ of our physical system satisfies the full state property if
\begin{equation}
 \left[ \ A \in \mathfrak A \ \text{such that} \ J^\natural(\omega)(A) > 0 \ \ \forall \omega \in \mathfrak S_o \ \right] \qquad \Longrightarrow \qquad A > 0
\end{equation}
\end{property}

We observe that if the algebrization $(\mathfrak A, J, J^\natural)$ satisfies this property, then the set of states $J^\natural(\mathfrak S_o)$ is $w*$-topologically dense in $S(\mathfrak A)$\footnote{Proposition 3.2.10 of \cite{BR1}}.
\\
This tells us that for every $\varphi \in S(\mathfrak A)$ there exists a net $\left\{ \omega_\alpha \right\}_\alpha$ in $J^\natural(\mathfrak S_o)$ such that 
\begin{equation}
\label{compact}
 \varphi(A) = \lim_\alpha J^\natural(\omega_\alpha)(A) \ , \qquad \forall A \in \mathfrak A 
\end{equation}
From this it follows that the full state property \ref{prop_F} implies the separability property \ref{prop_S_bis}:
\begin{equation}
\text{Property \texttt{PF}} \Longrightarrow \text{Property \texttt{PS}}
\end{equation}
In general, any map $F: \mathfrak S \rightarrow \mathfrak S$ is called a \textit{state transformation} of the system\footnote{As already recalled, such maps are called operations in the literature (cf. \cite{HaagKastler}).}.
\\
We ask when a state transformation determines a map 
$\hat{F} : S(\mathfrak A) \rightarrow S(\mathfrak A)$ such that 
\begin{equation}
\label{propindotta}
\hat{F} \circ J^\natural = J^\natural \circ F
\end{equation}
\begin{definition}\upshape
The transformation $F: \mathfrak S \rightarrow \mathfrak S$ is said to be \textit{$L$-uniformly continuous} (with respect to our algebrization) if for every $A \in \mathfrak X$ there exists an $L > 0$ such that for every $\omega, \omega_o \in \mathfrak S$ it holds:
\begin{equation}
\label{prop.L}
| J^\natural(F(\omega))(A) - J^\natural(F(\omega_o))(A) | \leq L \ | J^\natural(\omega)(A) - J^\natural(\omega_o)(A) |
\end{equation}
\end{definition}
\begin{proposition}\upshape
\label{mappeindotte}
Let an algebrization of our Mackey system $(\mathfrak X_o, \mathfrak S_o)$ be given that satisfies property \ref{prop_F}.
\\
If $F : \mathfrak S_o \rightarrow \mathfrak S_o$ is an $L$-uniformly continuous state transformation, then there exists a map $\hat{F} : S(\mathfrak A) \rightarrow S(\mathfrak A)$ satisfying relation \eqref{propindotta}.
\end{proposition}
\begin{proof}
Let $\varphi \in S(\mathfrak A)$ and let $\left\{ \omega_\alpha \right\}_\alpha$ be the net from relation \eqref{compact}. 
\\
The net $\left\{ J^\natural(F(\omega_\alpha)) \right\}_\alpha$ in $S(\mathfrak A)$ admits a unique limit point $\phi \in S(\mathfrak A)$ in the $w*$-topology.
 In this way we can define $\hat{F}(\varphi) = \phi$.
\\
The uniqueness of the limit point follows from $L$-uniform continuity.
\\
Indeed, assume the existence of two limit points $\phi$ and $\phi'$. Then there exist two subnets of the set $\left\{ J^\natural(F(\omega_\alpha)) \right\}_\alpha$ such that:
$$ J^\natural(F(\omega_{\alpha_\beta})) \stackrel{\beta}{\longrightarrow} \phi \qquad , \qquad J^\natural(F(\omega_{\alpha_{\beta'}})) \stackrel{\beta'}{\longrightarrow} \phi' $$
in the $w*$-topology.
\\
Note that for every $A \in \mathfrak A$,
\begin{eqnarray*}
|\phi(A) - \phi'(A)| & < & |\phi(A) - J^\natural(F(\omega_{\alpha_\beta}))(A)| + |\phi'(A) - J^\natural(F(\omega_{\alpha_{\beta'}}))(A)| \\
& & + | J^\natural(F(\omega_{\alpha_\beta}))(A) - J^\natural(F(\omega_{\alpha_{\beta'}}))(A) | \\
& < & |\phi(A) - J^\natural(F(\omega_{\alpha_\beta}))(A)| + |\phi'(A) - J^\natural(F(\omega_{\alpha_{\beta'}}))(A)| \\
& & + L \ | J^\natural(\omega_{\alpha_\beta})(A) - J^\natural(\omega_{\alpha_{\beta'}})(A) |
\end{eqnarray*}
Taking $\beta \rightarrow \infty$ we obtain:
\begin{eqnarray*}
|\phi(A) - \phi'(A)| & < & |\phi'(A) - J^\natural(F(\omega_{\alpha_{\beta'}}))(A)| \\
& & + L \ | \varphi(A) - J^\natural(\omega_{\alpha_{\beta'}})(A) |
\end{eqnarray*}
and when $\beta' \rightarrow \infty$ we have $|\phi(A) - \phi'(A)| = 0$.
\end{proof}

\subsection{Temporal Evolution for Mackey Systems}
Before introducing the temporal dynamics of a Mackey system, we must adapt the sets defined in \eqref{evol_temp_01}, \eqref{evol_temp_03} and in Definition \ref{evol_temp_02} on page \pageref{evol_temp_01}.
\\
We assume that every state $\omega \in \mathfrak S_o |_{\tau = 0}$ determines a chronological state\footnote{For simplicity we still denote it by $\omega$.}
$$ \omega : \tau \in I \longrightarrow \omega^{(\tau)} \in \mathfrak S_o | \tau \ , \qquad I = [0, \infty[ $$
We define the set:
 $$\mathcal S_o^\omega = \bigcap_{a \in \mathfrak X_o} \mathcal S_{a,\omega}^o$$
where\footnote{Recall that 
$$\mathfrak S_t^{a,\omega} \subset \mathfrak S_a |_{\tau = 0} \ , \ \forall t \geq 0 $$ }
$$ \mathcal S_{a,\omega}^o = \left\{ \xi: \mathbb R^+ \rightarrow \bigcup_{t \geq 0} \mathfrak S_t^{a,\omega} \ : \ \xi(t) \in \mathfrak S_t^{a,\omega} \cap \mathfrak S_o |_{\tau = 0}, \ \forall t \geq 0 \right\} 
$$
As we have already discussed, \textit{it is not guaranteed that a temporal evolution of the state $\omega \in \mathfrak S_o$ exists independently of the observables of the system; this statement continues to hold for Mackey systems as well.}
\\
In other words, it is not guaranteed that an element $\xi \in \mathcal S_o^\omega$ exists, but if it does, it is unique.
\\
Indeed, if $\xi, \xi' \in \mathcal S_o^\omega$, then for every $t \in \mathbb R^+$ we obtain:
$$ [ \ \mu_{\omega,a}^t = \mu_{\xi(t),a} = \mu_{\xi'(t),a} \ , \ \forall a \in \mathfrak X_o \ ] \qquad \Longrightarrow \qquad \xi(t) = \xi'(t)$$
In other words, we obtain a unique temporal evolution of the state $\omega$:
\begin{equation*}
\tau \in [0, +\infty[ \ \longrightarrow \xi(\tau) \in \mathfrak S_o |_{\tau' = 0}
\end{equation*}
that satisfies relation \eqref{evol_temp_04}. 
\bigskip

We make the following assumptions:
\begin{itemize}
\item[-] For every chronological state $\omega \in \mathfrak S_o$, the set $\mathcal S_o^\omega$ is non-empty.
\item[-] There exists a map\footnote{In agreement with Mackey in his famous book \cite{Mackey}.}
\begin{equation} \label{isometriatemp} 
V^t : \mathfrak S_o |_{\tau = 0} \longrightarrow \mathfrak S_o |_{\tau = 0}
\end{equation}
such that for every $\omega \in \mathfrak S_o |_{\tau = 0}$
$$ \mu_{V^t \omega, a} = \mu_{\omega^t, a} \ , \qquad \forall t \geq 0 $$
which is called the \textit{temporal evolution}, with the following obvious properties:
\begin{itemize}
\item [a.] For every $\omega \in \mathfrak S_o |_{\tau = 0}$ we have:
$$ V^0(\omega) = \omega $$
\item [b.] For every $\omega, \omega' \in \mathfrak S_o |_{\tau' = 0}$ and $t \geq 0$, we have the \textit{affine property}:
$$ V^t((1-r)\omega + r\omega') = (1-r) V^t(\omega) + r V^t(\omega'), \qquad r \in [0,1] $$
\item [c.] The map $V^t : \mathfrak S_o |_{\tau = 0} \longrightarrow \mathfrak S_o |_{\tau = 0}$ is $W^*$-topologically continuous.
\item [d.] For every state $\omega \in \mathfrak S_o |_{\tau = 0}$, the map
$$ t \in \mathbb R_+ \longrightarrow V^t(\omega) \in \mathfrak S_o |_{\tau = 0} $$
is a Borel map, where $\mathfrak S_o$ is equipped with the topological structure induced by \eqref{normstate}.
\end{itemize}
Property [a] is obvious, [b] follows from \eqref{amistura}, [c] follows from Axiom \ref{axio_evol_c}, while the last is a completely \textit{arbitrary} assumption introduced to avoid mathematical difficulties in the future.
\end{itemize}
\begin{remark}\upshape
In the definition of temporal evolution we did not assume that it satisfies the semigroup property (cf. relation \eqref{semigroup}):
\begin{equation}
\label{semigruppo}
V^s \circ V^t = V^{s+t} \ , \qquad t, s \geq 0 
\end{equation}
This property is assumed by Mackey in \cite{Mackey}, paragraphs 2–3.
\end{remark}
We now ask whether, starting from the temporal evolution $V^t$, it is possible to establish a temporal map $(\mathfrak A, S^t, S^\natural_t)$.
\\
For the map $S^\natural_t : \mathfrak S_o \rightarrow S(\mathfrak A)$ we can define it as 
\begin{equation}
\label{evol2}
  S^\natural_t = J^\natural \circ V^t \ , \qquad \forall t \geq 0
\end{equation}
while it is not possible to determine a natural candidate for the map $S_t$, since so far we have been working with objects that have too primitive a mathematical structure.
\\
Therefore, we take a step back and return to the basic discussion at the beginning of this topic, studying the possibility of introducing an algebrization of the system by assuming the following facts:
\begin{itemize}
\item[A1] We assume the existence of a C*-algebrization $(\mathfrak A, J, J^\natural)$ of a physical subsystem $(\mathfrak X_o, \mathfrak S_o)$ where $\mathfrak A$ is a von Neumann algebra satisfying the \texttt{ARBA} properties, the strong central condition of Definition \ref{strong-central-condition}, the embedding property, and the full state property \texttt{PF}.
\item[A2] We assume that the physical states are represented by the normal states of the von Neumann algebra, i.e.,
\begin{equation}
\label{evol3}
J^\natural : \mathfrak S_o \rightarrow S_w(\mathfrak A)
\end{equation}
\item[A3] We assume the existence of a temporal evolution map $V^t : \mathfrak S_o |_{\tau = 0} \rightarrow \mathfrak S_o |_{\tau = 0}$ with the properties [a], [b], [c], [d] listed above, and that it is $L$-uniformly continuous:
$$ |S^\natural_t(\omega)(A) - S^\natural_t(\omega_o)(A) | < L \ | J^\natural(\omega)(A) - J^\natural(\omega_o)(A) | \ , \qquad \forall \omega, \omega_o \in \mathfrak S_o |_{\tau = 0} $$
where $S^\natural_t : \mathfrak S \rightarrow S_w(\mathfrak A)$ is the map defined in \eqref{evol2}.
\end{itemize}
By Proposition \ref{mappeindotte}, the temporal map induces a map, which in this case will be affine, $\Phi^\natural_t : S_w(\mathfrak A) \rightarrow S_w(\mathfrak A)$ such that
\begin{equation}
\Phi^\natural_t \circ J^\natural = S^\natural_t \ , \qquad \forall t \geq 0
\end{equation}
By a well-known result of Kadison found in \cite{kad65}\footnote{See also \cite{BR1}, Proposition 3.2.8}, there exists a Jordan morphism 
$\Phi_t : \mathfrak A \rightarrow \mathfrak A$ 
such that for every $\varphi \in S_w(\mathfrak A)$,
\begin{equation}
\Phi^\natural_t(\varphi) = \varphi \circ \Phi_t \ , \qquad \forall t \geq 0
\end{equation}
and setting $S^t = \Phi_t \circ J$, we obtain the following relations:
\begin{eqnarray*} 
S^\natural_0(\omega)(S^t(a)) & = & J^\natural(\omega)(\Phi_t(J(a))) = (\Phi^\natural_t \circ J^\natural)(\omega)(J(a)) \\
& = & (J^\natural \circ V^t)(\omega)(J(a)) = \langle a \rangle_{V^t \omega}
\end{eqnarray*}
while
\begin{eqnarray*} 
S^\natural_t(\omega)(S^0(a)) & = & (J^\natural \circ V^t)(\omega)(J(a)) = \Phi^\natural_t J^\natural(\omega)(J(a)) = \langle a \rangle_{V^t \omega}
\end{eqnarray*}
Therefore, the triple $(\mathfrak A, S_t, S^\natural_t)$ defined in this way is an algebraic temporal map.
\\

In conclusion, the existence — in a non operational sense — of a canonical evolution in the sense of Mackey cannot be considered a simple assumption of the model. It is realized only for very restricted classes of physical systems, namely those for which it is possible to construct a C*-algebrization endowed with particularly strong structural properties, given by assumptions [A1], [A2], [A3].
  
\chapter*{Final Remarks}

\begin{flushright}
\textit{Talking once about a colleague who had spent a few years in Germany, Wigner said: 
``... and when he came back he was transformed into a German physicist.'' 
Puzzled, I asked: ``What is a German physicist?'' 
The answer: ``Well, an American physicist, if he has no ideas, makes himself useful — perhaps he calculates something. 
A German physicist, if he has no ideas, simply does nothing.''}
\\
Rudolf Haag — Personal discussion with Wigner, 1957–59 \cite{Doebner}.
\end{flushright}

In these notes, I have chosen to return to the laboratory itself: to the measurement instruments, the preparation protocols, and the operations that define what can be done and what can be observed. From this operational context, I have attempted to reconstruct the quantum formalism - not by postulating a Hilbert space or an algebraic structure from the outset. Only in the final section did we explore the possibility of embedding a physical system into an operator algebra, while emphasizing the distinction between the mathematical framework and the operationalist perspective adopted here.
\\
A natural continuation of these notes would be to analyze in greater detail the relation between the physical subsystems of a laboratory (as discussed in the previous sections) and their possible algebrizations - a topic closely connected to the notion of independence of local algebras in the algebraic approach (cf.\ \cite{Summers90}).
\\
These notes are not a closed system.\\
They are, rather, an invitation to rethink the foundations of quantum physics starting from the only things we truly possess: a laboratory, some instruments, and the ability to repeat experiments.
\\ 
I am not a German physicist.  
\\
So I wrote a book.

\begin{center}
\textbf{Comments, suggestions, criticisms, and reports of errors are most welcome.}
\end{center}

\begin{flushright}
\texttt{pandiscia.carlo@gmail.com}
\end{flushright}

\printindex

\end{document}